\pdfoutput=1

\documentclass[12pt]{tcibook}

\usepackage{fancyhea}
\usepackage{work}
\usepackage[hmargin=3cm,vmargin={2.5cm,3cm},includefoot]{geometry}
\usepackage{graphicx}
\usepackage{relsize}
\usepackage{enumerate}
\usepackage{array}
\usepackage{tabularx}
\usepackage{subfigure}
\usepackage{amstext}
\usepackage{amsfonts}
\usepackage{amsmath}
\usepackage{lineno}
\usepackage{cite,varioref,psfrag,multirow,dcolumn,hhline,amssymb,threeparttable}
\usepackage{setspace}
\usepackage{epsfig,rotating}
\usepackage{longtable}
\usepackage{morefloats}
\usepackage{afterpage}

\input babarsym
\def\tenTo#1#2{\ensuremath{10^{#1#2} \cms}\xspace}
\newcommand{\EE}[1] {\ensuremath{\times 10^{#1}}\xspace}

\newcommand{\PM}[2] {\ensuremath{^{+#1}_{-#2}}\xspace}

\def\ze#1   {\ensuremath{\zeta_{#1}}\xspace}
\newcommand{\VE}[1]  {\ensuremath{\vec{\mathrm{#1}}}\xspace}

\def\amp    {\ensuremath{{\rm \,A}}\xspace}
\def\mA     {\ensuremath{{\rm \,mA}}\xspace}
\def\nC     {\ensuremath{{\rm \,nC}}\xspace}
\def\pC     {\ensuremath{{\rm \,pC}}\xspace}
\def\Ohm    {\ensuremath{{\,\rm \Omega}}\xspace}
\def\mOhm   {\ensuremath{{\rm \,m\Omega}}\xspace}
\def\kOhm   {\ensuremath{{\rm \,k\Omega}}\xspace}
\def\MOhm   {\ensuremath{{\rm \,M\Omega}}\xspace}
\def\V     {\ensuremath{{\rm \,V}}\xspace}
\def\kV     {\ensuremath{{\rm \,kV}}\xspace}
\def\MV     {\ensuremath{{\rm \,MV}}\xspace}
\def\Tesla  {\ensuremath{{\rm \,T}}\xspace}
\def\Gauss  {\ensuremath{{\rm \,G}}\xspace}
\def\kGauss  {\ensuremath{{\rm \,kG}}\xspace}
\def\dB     {\ensuremath{{\rm \,dB}}\xspace}

\def\J      {\ensuremath{{\rm \,J}}\xspace}
\def\kJ     {\ensuremath{{\rm \,kJ}}\xspace}

\def\W      {\ensuremath{{\rm \,W}}\xspace}
\def\MW     {\ensuremath{{\rm \,MW}}\xspace}
\def\kW     {\ensuremath{{\rm \,kW}}\xspace}

\def\Hz     {\ensuremath{{\rm \,Hz}}\xspace}
\def\MHz    {\ensuremath{{\rm \,MHz}}\xspace}
\def\GHz    {\ensuremath{{\rm \,GHz}}\xspace}
\def\kHz    {\ensuremath{{\rm \,kHz}}\xspace}

\def\msec   {\ensuremath{{\rm \,ms}}\xspace}
\def\nsec   {\ensuremath{{\rm \,ns}}\xspace}

\def\Torr  {\ensuremath{{\rm \,Torr}}\xspace}
\def\nTorr  {\ensuremath{{\rm \,nTorr}}\xspace}
\def\Pa     {\ensuremath{{\rm \,Pa}}\xspace}

\def\picom  {\ensuremath{{\rm \,pm}}\xspace}
\def\km  {\ensuremath{{\rm \,km}}\xspace}

\def\mbarn  {\ensuremath{{\rm \,mbarn}}\xspace}

\def\pepii    {PEP-II\xspace}
\def\kekb     {KEKB\xspace}
\def\KEKB     {\kekb}

\def\daphne   {DA$\Phi$NE \xspace}
\def\KLOE     {KLOE \xspace}
\def\phifactory{$\phi$~Factory \xspace}
\def\ilc      {ILC\xspace}
\def\pps      {PPS\xspace}

\def\Npos   {\ensuremath{{N^{+}}}\xspace}
\def\Nele   {\ensuremath{{N^{-}}}\xspace}

\def\fcoll  {\ensuremath{{f_{c}}}\xspace}

\def\bets  {\ensuremath{{\beta^{\star}}}\xspace}
\def\betxs  {\ensuremath{{\beta_{x}^{\star}}}\xspace}
\def\betys  {\ensuremath{{\beta_{y}^{\star}}}\xspace}

\def\betx   {\ensuremath{{\beta_{x}}}\xspace}
\def\bety   {\ensuremath{{\beta_{y}}}\xspace}

\def\epsx   {\ensuremath{{\varepsilon_{x}}}\xspace}
\def\epsy   {\ensuremath{{\varepsilon_{y}}}\xspace}

\def\xix  {\ensuremath{{\xi_{x}}}\xspace}
\def\xiy  {\ensuremath{{\xi_{y}}}\xspace}

\def\thx  {\ensuremath{{\theta}}\xspace}
\def\piwi {\ensuremath{{\varphi}}\xspace}
\def\CW   {CW\xspace}
\def\br   {\ensuremath{B\rho}\xspace}
\def\thr  {\ensuremath{\theta_{r}}\xspace}

\def\Be   {Be\xspace}
\def\Cu   {Cu\xspace}
\def\Au   {Au\xspace}
\def\CO2  {$\mathrm{CO}_2$\xspace}

\def\Ngacoll       { \ensuremath{N_{\gamma \; \mathrm coll}}\xspace}
\def\Moller        { M{\o}ller\xspace}

\def\CrabbedWaist {{\it Crabbed waist}\xspace}

\usepackage[colorlinks=true, linkcolor=blue, linktocpage=true,
            a4paper=false, citecolor=blue, urlcolor=blue]{hyperref}

\begin{document}

%
\graphicspath{{covers/}}
\thispagestyle{empty}
\begin{figure}[!h]
\vskip-32.8mm
\hskip-34.3mm
\includegraphics[height=279.4mm]{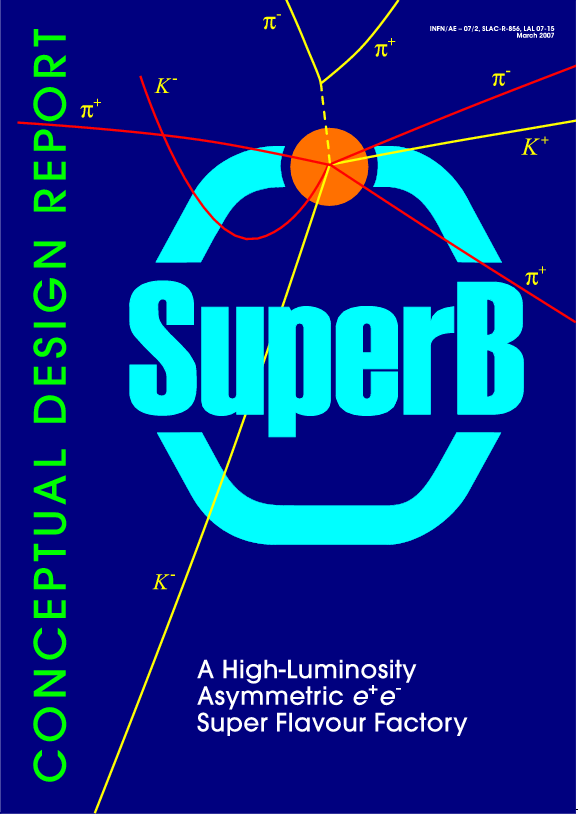}
\end{figure}
\clearpage

\newpage
\thispagestyle{empty}
\phantom{test}
\newpage

\hoffset-.1mm

\thispagestyle{empty}

\long\def\boxit#1{\quad\quad\quad\emspace\emspace\vbox{\hrule\hbox{\vrule
                   \kern8pt\vbox{\kern3pt#1\kern3pt}
                    \kern3pt\vrule}\hrule}}

 \hfill  \bigskip\bigskip\bigskip\bigskip\bigskip\bigskip

\begin{center}
 {
 \bfseries
     {\Huge   Super$B$\\} {\vskip 1.3cm}
         {\Huge  A High-Luminosity\\}
          {\huge Heavy Flavour Factory\\}  {\vskip 9cm}
             {\Huge  Conceptual Design Report\\
          }}

\bigskip\bigskip\bigskip
 {\Large  Presented to INFN}
 \par
  {\Large  March, 2007}

\end{center}

\begin{center}

\ \ Available on the World Wide Web, with figures in full color, at\break
{\underline{http://www.pi.infn.it/SuperB/?q=CDR}}

\end{center}


\newpage

\thispagestyle{empty}

\vspace*{1.5in}

\begin{center}
{\Large INFN/AE - 07/2, SLAC-R-856, LAL 07-15}\\
\smallskip
{\Large March, 2007}
\end{center}

\vspace*{1.5in}


\begin{center}
\vspace*{1.8in}
Available from INFN Publishing Services, INFN-Pisa, L.go Pontecorvo, 3, I-56127 Pisa, Italy.
\end{center}
\vfill

\newpage

\raggedbottom


\pagenumbering{roman}
{\normalsize \thispagestyle{empty}
\begin{center}
 The people on the following list have indicated interest in
 and support for the \superb\ project. It includes a subset
 who have been directly involved with the preparation of this
 document, as well as individuals who have contributed in other
 ways or plan on future involvement.
\\ 
\rule[.4mm]{3cm}{.2mm} \rule{1mm}{1mm}  \rule[.4mm]{3cm}{.2mm}\
\end{center}

\begin{center}

\mbox{ M.~Bona}\\*
 {\bf Laboratoire d'Annecy-le-Vieux de Physique des Particules,}\\*
 {\sl  Universit\'e de Savoie,  IN2P3/CNRS,  F-74941 Annecy-Le-Vieux,  France }\end{center}

\begin{center}

\mbox{ J.~Garra Tic\'o,}
\mbox{ E.~Graug\'es Pous}\\*
 {\bf Universitat de Barcelona,}
 {\sl  Facultat de Fisica,  Departament ECM,\\*  E-08028 Barcelona,  Spain }\end{center}

\begin{center}

\mbox{ P.~Colangelo,}
\mbox{ F.~De Fazio,}
\mbox{ A.~Palano}\\*
 {\bf Universit\`a di Bari,}
 {\sl  Dipartimento di Fisica and INFN,\\*  I-70126 Bari,  Italy }\end{center}

\begin{center}

\mbox{ M.~Manghisoni,}
\mbox{ V.~Re,}
\mbox{ G.~Traversi}\\*
 {\bf Universit\`a di Bergamo,}
 {\sl  Dipartimento di Ingegneria Industriale,  I-24129 Bergamo,  Italy}\end{center}

\begin{center}

\mbox{ G.~Eigen}\\*
 {\bf University of Bergen,}
 {\sl  Institute of Physics,  N-5007 Bergen,  Norway }\end{center}

\begin{center}

\mbox{ M.~Venturini}\\*
 {\bf Lawrence Berkeley National Laboratory,}
 {\sl  University of California,  Berkeley,  California 94720,  USA }\end{center}

\begin{center}

\mbox{ N.~Soni}\\*
 {\bf University of Birmingham,}
 {\sl  Birmingham,  B15 2TT,  United~Kingdom }\end{center}

\begin{center}

\mbox{ M.~Bruschi,}
\mbox{ S.~De Castro,}
\mbox{ P.~Faccioli,}
\mbox{ A.~Gabrielli,}
\mbox{ B.~Giacobbe,}
\mbox{ N.~Semprini Cesari,}
\mbox{ R.~Spighi,}
\mbox{ M.~Villa,}
\mbox{ A.~Zoccoli}\\*
 {\bf Universit\`a di Bologna,}
 {\sl  Dipartimento di Fisica and INFN,  I-40127 Bologna,  Italy }\end{center}

\begin{center}

\mbox{ C.~Hearty,}
\mbox{ J.~McKenna}\\*
 {\bf University of British Columbia,}
 {\sl  Vancouver,  British Columbia,  Canada V6T 1Z1 }\end{center}

\begin{center}

\mbox{ A.~Soni}\\*
 {\bf Brookhaven National Laboratory,}
 {\sl  Department of Physics,  Upton,\\*  NY 11973-5000,  USA}\end{center}

\begin{center}

\mbox{ A.~Khan}\\*
 {\bf Brunel University,}
 {\sl  Uxbridge,  Middlesex UB8 3PH,  United~Kingdom }\end{center}

\begin{center}

\mbox{ A.Y.~Barniakov,}
\mbox{ M.Y.~Barniakov,}
\mbox{ V.E.~Blinov,}
\mbox{ V.P.~Druzhinin,}
\mbox{ V.B.~Golubev,}
\mbox{ S.A.~Kononov,}
\mbox{ I.A.~Koop,}
\mbox{ E.A.~Kravchenko,}
\mbox{ E.B.~Levichev,}
\mbox{ S.A.~Nikitin,}
\mbox{ A.P.~Onuchin,}
\mbox{ P.A.~Piminov,}
\mbox{ S.I.~Serednyakov,}
\mbox{ D.N.~Shatilov,}
\mbox{ Y.M.~Shatunov,}
\mbox{ Y.I.~Skovpen,}
\mbox{ E.P.~Solodov}\\*
 {\bf Budker Institute of Nuclear Physics,}
 {\sl  Novosibirsk 630090,  Russia }\end{center}

\begin{center}

\mbox{ C.-H.~Cheng,}
\mbox{ B.~Echenard,}
\mbox{ F.~Fang,}
\mbox{ D.G.~Hitlin,}
\mbox{ F.C.~Porter}\\*
 {\bf California Institute of Technology,}
 {\sl  Pasadena,  California 91125,  USA }\end{center}

\begin{center}

\mbox{ D.M.~Asner}\\*
 {\bf Carleton University,}
 {\sl  Ottawa,  Ontario,  Canada K1S 5B6}\end{center}

\begin{center}

\mbox{ T.N.~Pham}\\*
 {\bf Centre de Physique Th\'eorique,}
 {\sl  CNRS/MP,  Ecole Politechnique,\\*  F- 91128 Palaiseau,  France}\end{center}

\begin{center}

\mbox{ R.~Fleischer,}
\mbox{ G.F.~Giudice,}
\mbox{ T.~Hurth,}
\mbox{ M.~Mangano}\\*
 {\bf CERN,}
 {\sl  TH-Division,  CH-1211 Geneva 23,  Switzerland}\end{center}

\begin{center}

\mbox{ G.~Mancinelli,}
\mbox{ B.T.~Meadows,}
\mbox{ A.J.~Schwartz,}
\mbox{ M.D.~Sokoloff}\\*
 {\bf University of Cincinnati,}
 {\sl  Cincinnati,  Ohio 45221,  USA }\end{center}

\begin{center}

\mbox{ A.~Soffer}\\*
 {\bf Colorado State University,}
 {\sl  Fort Collins,  Colorado 80523,  USA }\end{center}

\begin{center}

\mbox{ C.D.~Beard}\\*
 {\bf ASTeC,}
 {\sl  Daresbury Laboratory,  Daresbury Warrington,  Cheshire,  WA4 4AD,  United~Kingdom}\end{center}

\begin{center}

\mbox{ T.~Haas,}
\mbox{ R.~Mankel}\\*
 {\bf DESY,}
 {\sl  D-22607 Hamburg,  Germany}\end{center}

\begin{center}

\mbox{ G.~Hiller}\\*
 {\bf Universit\"at Dortmund,}
 {\sl  Institut f\"ur Physik,  D-44221 Dortmund,  Germany}\end{center}

\begin{center}

\mbox{ P.~Ball,}
\mbox{ M.~Pappagallo,}
\mbox{ M.R.~Pennington}\\*
 {\bf Durham University,}
 {\sl  IPPP,  Physics Department,  Durham DH1 3LE,  United~Kingdom }\end{center}

\begin{center}

\mbox{ W.~Gradl,}
\mbox{ S.~Playfer}\\*
 {\bf University of Edinburgh,}
 {\sl  Edinburgh EH9 3JZ,  United~Kingdom }\end{center}

\begin{center}

\mbox{ A.~Abada,}
\mbox{ D.~Becirevic,}
\mbox{ S.~Descotes-Genon,}
\mbox{ O.~P\`ene}\\*
 {\bf Laboratoire de Physique Th\'eorique,}
 {\sl  CNRS/MP Universit\'e Paris Sud,  F-91405 Orsay,  France}\end{center}

\begin{center}

\mbox{ D.~Andreotti,}
\mbox{ M.~Andreotti,}
\mbox{ D.~Bettoni,}
\mbox{ C.~Bozzi,}
\mbox{ R.~Calabrese,}
\mbox{ A.~Cecchi,}
\mbox{ G.~Cibinetto,}
\mbox{ P.~Franchini,}
\mbox{ E.~Luppi,}
\mbox{ M.~Negrini,}
\mbox{ A.~Petrella,}
\mbox{ L.~Piemontese,}
\mbox{ E.~Prencipe,}
\mbox{ V.~Santoro,}
\mbox{ G.~Stancari}\\*
 {\bf Universit\`a di Ferrara,}
 {\sl  Dipartimento di Fisica and INFN,  I-44100 Ferrara,  Italy  }\end{center}

\begin{center}

\mbox{ F.~Anulli,}
\mbox{ R.~Baldini-Ferroli,}
\mbox{ M.E.~Biagini,}
\mbox{ M.~Boscolo,}
\mbox{ A.~Calcaterra,}
\mbox{ A.~Drago,}
\mbox{ G.~Finocchiaro,}
\mbox{ S.~Guiducci,}
\mbox{ G.~Isidori,}
\mbox{ S.~Pacetti,}
\mbox{ P.~Patteri,}
\mbox{ I.M.~Peruzzi,}
\mbox{ M.~Piccolo,}
\mbox{ M.A.~Preger,}
\mbox{ P.~Raimondi,}
\mbox{ M.~Rama,}
\mbox{ C.~Vaccarezza,}
\mbox{ A.~Zallo,}
\mbox{ M.~Zobov,}
\mbox{ R.~de Sangro}\\*
 {\bf Laboratori Nazionali di Frascati dell'INFN,}
 {\sl  I-00044 Frascati,  Italy }\end{center}

\begin{center}

\mbox{ A.~Buzzo,}
\mbox{ M.~Lo Vetere,}
\mbox{ M.~Macr\'i,}
\mbox{ M.R.~Monge,}
\mbox{ S.~Passaggio,}
\mbox{ C.~Patrignani,}
\mbox{ E.~Robutti,}
\mbox{ S.~Tosi}\\*
 {\bf Universit\`a di Genova,}
 {\sl  Dipartimento di Fisica and INFN,  I-16146 Genova,  Italy }\end{center}

\begin{center}

\mbox{ J.~Matias}\\*
 {\bf IFAE,}
 {\sl  Universitat Autonoma de Barcelona,  E-08193 Bellaterra,  Barcelona,  Spain}\end{center}

\begin{center}

\mbox{ W.~Panduro Vazquez}\\*
 {\bf Imperial College London,}
 {\sl  London,  SW7 2AZ,  United~Kingdom }\end{center}

\begin{center}

\mbox{ F.~Borzumati}\\*
 {\bf International Centre for Theoretical Physics ICTP,}
 {\sl  I-34014 Trieste,  Italy}\end{center}

\begin{center}

\mbox{ V.~Eyges,}
\mbox{ S.A.~Prell}\\*
 {\bf Iowa State University,}
 {\sl  Ames,  Iowa 50011-3160,  USA }\end{center}

\begin{center}

\mbox{ T.K.~Pedlar}\\*
 {\bf Luther College,}
 {\sl  Department of Physics,  Decorah,  Iowa 52101,  USA}\end{center}

\begin{center}

\mbox{ S.~Korpar,}
\mbox{ R.~Pestotnik,}
\mbox{ M.~Stari\v c}\\*
 {\bf J. Stefan Institute,}
 {\sl  SI-1000 Ljubljana,  Slovenia}\end{center}

\begin{center}

\mbox{ M.~Neubert}\\*
 {\bf Johannes Gutenberg-Universit\"at,}
 {\sl  Institut f\"ur Physik (ThEP),  D-55099 Mainz,  Germany}\end{center}

\begin{center}

\mbox{ A.G.~Denig,}
\mbox{ U.~Nierste}\\*
 {\bf Karlsruhe Institut f\"ur Technologie (KIT),}
 {\sl  D-76131 Karlsruhe,  Germany}\end{center}

\begin{center}

\mbox{ T.~Agoh,}
\mbox{ K.~Ohmi,}
\mbox{ Y.~Ohnishi}\\*
 {\bf KEK,}
 {\sl  High Energy Accelerator Research Organization,  1-1 Oho,  Tsukuba,  Ibaraki 305-0801 Japan}\end{center}

\begin{center}

\mbox{ J.R.~Fry,}
\mbox{ C.~Touramanis}\\*
 {\bf University of Liverpool,}
 {\sl  Liverpool L69 7ZE,  United~Kingdom }\end{center}

\begin{center}

\mbox{ A.~Wolski}\\*
 {\bf University of Liverpool,}
 {\sl  Liverpool L69 7ZE and Cockcroft Institute,  Daresbury Science and Innovation Campus,  Warrington,  WA4 4AD,  United~Kingdom}\end{center}

\begin{center}

\mbox{ B.~Golob,}
\mbox{ P.~Kri\v zan}\\*
 {\bf University of Ljubljana and J. Stefan Institute,}
 {\sl  SI-1000 Ljubljana,  Slovenia}\end{center}

\begin{center}

\mbox{ H.~Flaecher}\\*
 {\bf University of London,}
 {\sl  Royal Holloway and Bedford New College,  Egham,  Surrey TW20 0EX,  United~Kingdom }\end{center}

\begin{center}

\mbox{ A.J.~Bevan,}
\mbox{ F.~Di Lodovico,}
\mbox{ K.A.~George}\\*
 {\bf Queen Mary,}
 {\sl  University of London,  London E1 4NS,  United~Kingdom}\end{center}

\begin{center}

\mbox{ R.~Barlow,}
\mbox{ G.~Lafferty}\\*
 {\bf University of Manchester,}
 {\sl  Manchester M13 9PL,  United~Kingdom }\end{center}

\begin{center}

\mbox{ A.~Jawahery,}
\mbox{ D.A.~Roberts,}
\mbox{ G.~Simi}\\*
 {\bf University of Maryland,}
 {\sl  College Park,  Maryland 20742,  USA }\end{center}

\begin{center}

\mbox{ P.M.~Patel,}
\mbox{ S.H.~Robertson}\\*
 {\bf McGill University,}
 {\sl  Montr\'eal,  Qu\'ebec,  Canada H3A 2T8 }\end{center}

\begin{center}

\mbox{ A.~Lazzaro,}
\mbox{ F.~Palombo}\\*
 {\bf Universit\`a di Milano,}
 {\sl  Dipartimento di Fisica and INFN,  I-20133 Milano,  Italy }\end{center}

\begin{center}

\mbox{ A.~Kaidalov}\\*
 {\bf ITEP,}
 {\sl  Institute for Theoretical and Experimental Physics,  RU-117218 Moscow,  Russia }\end{center}

\begin{center}

\mbox{ A.J.~Buras,}
\mbox{ C.~Tarantino}\\*
 {\bf Technische Universit\"at M\"unchen,}
 {\sl  Physik Department,  D-85748 Garching,  Germany }\end{center}

\begin{center}

\mbox{ G.~Buchalla}\\*
 {\bf Ludwig-Maximilians-Universit\"at M\"unchen,}
 {\sl  Arnold Sommerfeld Center for Theoretical Physics,  D-80333 M\"unchen,  Germany}\end{center}

\begin{center}

\mbox{ A.I.~Sanda}\\*
 {\bf Nagoya University,}
 {\sl  Department of Physics,  Nagoya 464-8602,  Japan}\end{center}

\begin{center}

\mbox{ G.~D'Ambrosio,}
\mbox{ G.~Ricciardi}\\*
 {\bf Universit\`a di Napoli Federico II,}
 {\sl  Dipartimento di Scienze Fisiche and INFN,  I-80126,  Napoli,  Italy }\end{center}

\begin{center}

\mbox{ I.~Bigi,}
\mbox{ C.P.~Jessop,}\\*
\mbox{ J.M.~Losecco}\\*
 {\bf University of Notre Dame,}
 {\sl  Notre Dame,  Indiana 46556,  USA }\end{center}

\begin{center}

\mbox{ K.~Honscheid}\\*
 {\bf Ohio State University,}
 {\sl  Columbus,  Ohio 43210,  USA }\end{center}

\begin{center}

\mbox{ N.~Arnaud,}
\mbox{ R.~Chehab,}
\mbox{ Y.~Fedala,}
\mbox{ F.~Polci,}
\mbox{ P.~Roudeau,}
\mbox{ V.~Sordini,}
\mbox{ V.~Soskov,}
\mbox{ A.~Stocchi,}
\mbox{ A.~Variola,}
\mbox{ A.~Vivoli,}
\mbox{ G.~Wormser,}
\mbox{ F.~Zomer}\\*
 {\bf Laboratoire de l'Acc\'el\'erateur Lin\'eaire,}
 {\sl  Universit\'e Paris-Sud 11,  IN2P3/CNRS,  F-91898 Orsay,  France }
\end{center}
\begin{center}
% ...
\end{center}

\begin{center}

\mbox{ H.~Stoeck}\\*
 {\bf University of Sydney,}
 {\sl  School of Physics,  Sydney,  NSW 2006,  Australia}\end{center}

\begin{center}

\mbox{ H.-Y.~Cheng,}
\mbox{ H.-N.~Li}\\*
 {\bf Academia Sinica,}
 {\sl  Institute of Physics,  Nankang,  Taipei Taiwan 11529,  Republic of China  (R.O.C.)}\end{center}

\begin{center}

\mbox{ Y.-Y.~Keum}\\*
 {\bf National Taiwan University,}
 {\sl  Department of Physics,  Taipei Taiwan 10617,  Republic of China  (R.O.C.)}\end{center}

\begin{center}

\mbox{ M.~Gronau,}
\mbox{ Y.~Grossman}\\*
 {\bf Technion - Israel Institute of Technology,}
 {\sl  Haifa 32000,  Israel}\end{center}

\begin{center}

\mbox{ F.~Bianchi,}
\mbox{ D.~Gamba,}
\mbox{ P.~Gambino,}
\mbox{ F.~Marchetto,}
\mbox{ E.~Menichetti,}
\mbox{ R.~Mussa,}
\mbox{ M.~Pelliccioni}\\*
 {\bf Universit\`a di Torino,}
 {\sl  Dipartimento di Fisica Sperimentale and INFN,\\*  I-10125 Torino,  Italy }\end{center}

\begin{center}

\mbox{ G.F.~Dalla Betta}\\*
 {\bf Universit\`a di Trento,}
 {\sl  ICT Department,  I-38050 Trento,  Italy }\end{center}

\begin{center}

\mbox{ M.~Bomben,}
\mbox{ L.~Bosisio,}
\mbox{ C.~Cartaro,}
\mbox{ L.~Lanceri,}
\mbox{ L.~Vitale}\\*
 {\bf Universit\`a di Trieste,}
 {\sl  Dipartimento di Fisica and INFN,  I-34127 Trieste,  Italy }\end{center}

\begin{center}

\mbox{ V.~Azzolini,}
\mbox{ J.~Bernabeu,}
\mbox{ N.~Lopez-March,}
\mbox{ F.~Martinez-Vidal,}
\mbox{ D.A.~Milanes,}
\mbox{ A.~Oyanguren,}
\mbox{ P.~Paradisi,}
\mbox{ A.~Pich,}
\mbox{ M.A.~Sanchis-Lozano}\\*
 {\bf IFIC,}
 {\sl  Universitat de Valencia-CSIC,  E-46071 Valencia,  Spain }\end{center}

\begin{center}

\mbox{ R.~Kowalewski,}
\mbox{ J.M.~Roney}\\*
 {\bf University of Victoria,}
 {\sl  Victoria,  British Columbia,  Canada V8W 3P6 }\end{center}

\begin{center}

\mbox{ J.~Back,}
\mbox{ T.J.~Gershon,}
\mbox{ P.F.~Harrison,}
\mbox{ T.E.~Latham,}
\mbox{ G.B.~Mohanty}\\*
 {\bf University of Warwick,}
 {\sl  Department of Physics,  Coventry CV4 7AL,  United~Kingdom }\end{center}

\begin{center}

\mbox{ A.A.~Petrov}\\*
 {\bf Wayne State University,}
 {\sl  Department of Physics and Astronomy,  Detroit,  Michigan 48202,  USA}\end{center}

\begin{center}

\mbox{ M.~Pierini}\\*
 {\bf University of Wisconsin,}
 {\sl  Madison,  Wisconsin 53706,  USA }\end{center}

}

\tableofcontents
\newpage

\parindent=0pt
\parskip=8pt

%
\renewcommand{\chapname}{chap:intro_}
\renewcommand{\chapterdir}{.}
\renewcommand{\arraystretch}{1.25}
\addtolength{\arraycolsep}{-3pt} \pagestyle{empty}
\pagenumbering{arabic}

\pagestyle{fancyplain} \pagestyle{fancyplain}
\addtolength{\headwidth}{\marginparsep}
\addtolength{\headwidth}{\marginparwidth}
\renewcommand{\chaptermark}[1]%
                  {\markboth{#1}{#1}}
\renewcommand{\sectionmark}[1]%
                 {\markright{\thesection\ #1}}

%
\graphicspath{{Intro/figures/}}
\chapter{Introduction}

Elementary particle physics in the next decade will be focused on the
investigation of the origin of electroweak symmetry breaking and the
search for extensions of the Standard Model (SM) at the TeV scale. The
discovery of New Physics will likely produce a period of excitement
and progress recalling the years following the discovery of the
\jpsi. In this new world, attention will be riveted on the detailed
elucidation of new phenomena uncovered at the LHC; these discoveries
will also provide strong motivation for the construction of the ILC. 
High statistics studies of heavy quarks and leptons will have a
crucial role to play in this new world. 

The  two asymmetric $B$ Factories, \pepii~\cite{ref:PEP-II} and
KEKB\cite{ref:KEKB}, and their companion detectors,
\babar\cite{ref:babar} and Belle\cite{ref:belle}, have over the last
seven years produced a wealth of flavour physics results, subjecting
the quark and lepton sectors of the Standard Model to a series of
stringent tests, all of which have been passed. With the much larger
data sample made possible by a Super~$B$~Factory, qualitatively new
studies will be possible. These studies will provide a uniquely
important source of information about the details of the New Physics
uncovered at hadron colliders in the coming decade. 

We thus believe that continued detailed studies of heavy quark and
heavy lepton (henceforth {\it heavy flavour}) physics will not only be
pertinent in the next decade; they will be central to understanding
the flavour sector of New Physics phenomena. A Super Flavour Factory
such as \superb\ will, perforce, be a partner, together with LHC, and
eventually, ILC, experiments, in ascertaining exactly {\it what kind}
of New Physics has been found. The capabilities of \superb\ in
measuring \CP-violating asymmetries in very rare $b$ and $c$ quark
decays, accessing branching fractions of heavy quark and heavy lepton
decays in processes that are either extremely rare or forbidden in the
Standard Model, and making detailed investigations of complex
kinematic distributions will provide unique and important constraints
in, for example, ascertaining the type of supersymmetry breaking or
the kind of extra dimension model behind the new phenomena that many
expect to be manifest at the LHC. 

This \superb\ Conceptual Design Report is the founding document of a
nascent international enterprise aimed at the construction of a very
high luminosity asymmetric \epem Flavour Factory.  A possible location
for \superb\ is the campus of the University of Rome ``Tor Vergata'',
near the INFN National Laboratory of Frascati. This report has been
prepared by an international study group set up by the President of
INFN at the end of 2005, with the charge of studying the physics
motivation and the feasibility of constructing a Super Flavour Factory
that would come into operation in the first half of the next decade
with a peak luminosity in excess of \hbox{$10^{36}$ cm$^{-2}$
  s$^{-1}$} at the \FourS resonance. This report is the response to
that charge.

We discuss herein the exciting physics program that can be
accomplished with a very large sample of heavy quark and heavy lepton
decays produced in the very clean environment of an \epem collider; a
program complementary to that of an experiment such as \lhcb at a
hadronic machine. It then presents the conceptual design of a new type
of \epem collider that produces a nearly two-order-of-magnitude
increase in luminosity over the current generation of asymmetric $B$
Factories. The key idea is the use of low emittance beams produced in
an accelerator lattice derived from the ILC Damping Ring Design,
together with a new collision region, again with roots in the ILC
final focus design, but with important new concepts developed in this
design effort.  Remarkably, \superb\ produces this very large
improvement in luminosity with circulating currents and wallplug power
similar to those of the current $B$ Factories. There is clear synergy
with ILC R\&D; design efforts have already influenced one another, and
many aspects of the ILC Damping Rings and Final Focus would be
operationally tested at \superb.  Finally, the design of an
appropriate detector, based on an upgrade of \babar\ as an example, is
discussed in some detail. A preliminary cost estimate is presented, as
is an example construction timeline.

\section{The Physics}

\begin{figure}[htbp]
 \begin{center}
  \includegraphics[width=0.71\textwidth]{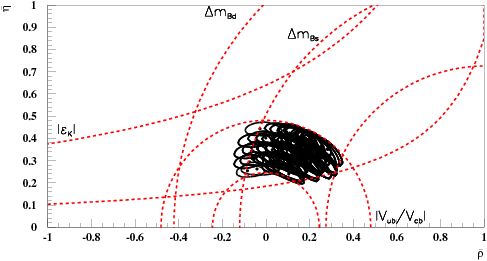}
  \caption{Status of the Unitarity Triangle before the $B$ Factories,
  with allowed regions in the ($\bar{\rho}$--$\bar{\eta}$) plane
  delineated only by determinations of the sides of the triangle
  (dashed lines), which were, in general, dominated by theoretical
  uncertainties.  The ellipses show representative statistical errors
  for various choices of theoretical parameters within the allowed
  region.\cite{ref:UT}} 
  \label{fig:ligeti}
  \bigskip\bigskip\bigskip
  \includegraphics[width=0.75\textwidth]{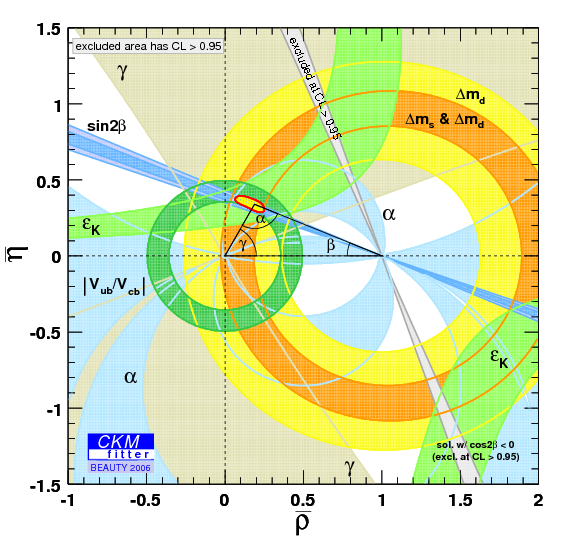}
  \caption{Global fit of the Unitarity Triangle construction as of the
  Beauty 2006 conference.} 
  \label{fig:beauty06}
 \end{center}
\end{figure}

By measuring mixing-dependent $C\!P$-violating asymmetries in the $B$
meson system for the 
first time, \pepii/\babar\ and KEKB/Belle have shown that the CKM
phase accounts 
for all observed $C\!P$-violating phenomena in $b$ decays.
The Unitarity Triangle construction provides a set of unique
overconstrained precision tests of the self-consistency of the three
generation Standard Model. Figure~\ref{fig:ligeti} shows the status of
knowledge of the Unitarity Triangle in 1998, before the new series of
tests made possible by the measurement of \CP-violating asymmetries in
$B^0$ decay at the $B$ Factories.  Figure~\ref{fig:beauty06} shows the
current status of the Unitarity Triangle construction, incorporating
measurements from \babar\ and Belle, as well as the $B_s$ mixing
measurement of CDF; the addition of \CP asymmetry measurements,
together with the improvement in the precision of \CP-conserving
measurements, has made this uniquely precise set of Standard Model
tests possible.

The fact that the CKM phase has now been shown to be consistent with
all observed \CP-violating phenomena is both a triumph and an
opportunity. In completing the experimentally-verified Standard Model
{\it ansatz} (except, of course, for the Higgs), it intensifies the
mystery of the creation of the baryon-antibaryon asymmetry of the
universe: the observed \CP-violation is too small for the Standard
Model to account for electroweak baryogenesis. 
This intriguing result opens the door to two possibilities: the matter
antimatter asymmetry is produced by another mechanism, such as
leptogenesis, or baryogenesis proceeds through the additional
\CP-violating phases that naturally arise in many extensions of the
Standard Model. 
These extra phases produce measurable effects in the weak decays of
heavy flavour particles. The detailed pattern of these effects, as
well as of rare decay branching fractions and kinematic distributions,
is, in fact, diagnostic of the characteristics of New Physics at or
below the TeV scale, 

By the end of this decade, the two $B$ Factories will have accumulated
a total of \hbox{$\sim 2$ \ab}. Even at this level, most important
measurements pertinent to the Unitarity Triangle construction will
still be statistics limited: an even larger data sample would provide
increasingly stringent tests of three-generation CKM unitarity. There
are two main thrusts here. The first is the substantial remaining
improvement that can still be made in the Unitarity Triangle
construction. 
Here measurements in $B$, $D$ and $\tau$ decay play an important role,
as do improvements in lattice QCD calculations of hadronic matrix
elements. This important physics goal is {\bf NOT}, however, the sole,
or even the primary, motivation for a \sbf. The precision of our
knowledge of the Unitarity Triangle will perforce improve to the limit
allowed by theoretical uncertainties as we pursue the primary goal:
improving the precision of the measurement of \CP\ asymmetries, rare
decay branching fractions, and rare decay kinematic distributions in
penguin-dominated $b\rightarrow s$ transitions, to a level where there
is substantial sensitivity to New Physics effects. 
This requires data samples substantially larger than the
current $B$ Factories will provide. Some of these measurements are
accessible at the LHC~\cite{ref:LHCb},
but the most promising approach to this
physics is \superb, a very high luminosity \abf, which is also, of
course, a Super Flavour Factory, providing large samples of $b$ and
$c$ quark and $\tau$ lepton decays. 

\superb, having an initial luminosity of $10^{36}$ cm$^{-2}$s$^{-1}$,
will collect 15 ab$^{-1}$ in a New Snowmass
Year~\cite{ref:snowmassyear}, or 75 ab$^{-1}$ in five years. 
A data sample this large will make the Unitarity Triangle tests, in
their manifold versions, the ultimate precision test of the flavour
sector of the Standard Model, and open up the world of New Physics
effects in very rare $B$, $D$, and $\tau$ decays 

A primary tool for isolating new physics is the time-dependent
\CP\ asymmetry in decay channels that
proceed through penguin diagrams, such as
the $b\rightarrow s\bar{s}s$ processes $B^0_d \rightarrow \phi K^0$ and $B^0_d \rightarrow (K\bar{K})_{C\!P} K^0$
or similar transitions such as
$B^0_d \rightarrow \eta 'K^0$,  $B^0_d \rightarrow f_0K^0$,  $B^0_d \rightarrow \pi^0 K^0$,
$B^0_d \rightarrow \rho^0 K^0$, $B^0_d \rightarrow \omega K^0$, and $B^0_d \rightarrow \pi^0\pi^0 K^0$.
The dominant contribution to these decays is the combination of CKM elements $V_{tb}V_{ts}^{*}$; these amplitudes have the
same phase as
the charmonium channels $b \rightarrow c\bar{c}s$,
up to a small phase shift of $V_{ts}$ with respect to $V_{cb}$. New heavy particles contribute new loop amplitudes, with new phases that
can contribute to the 
$C\!P$ asymmetry and the $S$ coefficient of the time-dependent analysis,
so that the measured \CP violation parameter
could be substantially different from $\sin 2\beta$.

Physics beyond the Standard Model can affect rare $B$ decay modes, through observables such as branching fractions, $C\!P$-violating asymmetries and kinematic distributions.
These decays do not typically occur at tree level,
and thus their rates are strongly suppressed in the Standard Model. Substantial enhancements in the
rates and/or variations in angular distributions of final state particles could result from the presence of new
heavy particles in loop diagrams, resulting in clear evidence of New Physics. Moreover, because the pattern
of observable effects is highly model-dependent, measurements of several rare decay modes can provide
information regarding the source of the New Physics. An extended run at the \FiveS is also contemplated; such a run would yield a wealth of interesting new $B_s^0$ decay results.

The \superb\ data sample will also contain unprecedented numbers of charm quark and $\tau$ lepton decays. This data is also of great interest, both for its capacity to improve the precision of existing measurements and for its sensitivity to New Physics. This interest extends beyond weak decays; the detailed exploration of new charmonium states is also an important objective. Limits on rare $\tau$ decays, particularly lepton-flavour-violating decays, already provide important constraints on New Physics models. \superb\ may have the sensitivity to actually observe such decays. The accelerator design will allow for longitudinal polarization of the $e^-$ beam, making possible uniquely sensitive searches for a $\tau$ electric dipole moment, as well as for \CP-violating $\tau$ decays.

Some measurements in charm and $\tau$ physics are best done near threshold. \superb\ also has the capability of running in the 4 \gev region. Short runs at specific center-of-mass energies in this region, representing perhaps 10\% of data taking time, would produce data samples substantially larger than those currently envisioned to exist in the next decade.

\section{The Super$B$ Design}

Given the strong physics motivation, there has been a great deal of activity over the past few years aimed at designing an $e^+e^-$ $B$ Factory that can produce samples of $B$ mesons 50 to 100 times larger than will exist when the current $B$ Factory programs end. Several approaches were tried before the design presented here was developed.

Upgrades of \pepii~\cite{ref:SuperPEPII} and
KEKB~\cite{ref:SuperKEKB} to Super $B$ Factories that accomplish this goal have been proposed
at SLAC and at KEK. These machines are extrapolations of the existing $B$ Factories, with higher currents, more bunches, and smaller $\beta$ functions (1.5 to 3 mm). They also use a great deal of power ($\ge 100$ MW), and the high currents (as much as 10A) pose significant challenges for detectors. To minimize the substantial wallplug power, the SuperPEP-II design doubled the current RF frequency, to 958 MHz. In the case of SuperKEKB, a factor of two increase in luminosity is assumed for the use of crab
crossing, which will soon be tested at KEKB.

SLAC has no current plans for an on-site accelerator-based
high energy physics program, so the SuperPEP-II proposal
is moribund. As of this writing, no decision has been made on SuperKEKB.
In the interim, the problematic power consumption and background issues associated with the SLAC and KEK-based Super $B$ Factory designs stimulated a new approach, using low emittance beams, to constructing a Super $B$ Factory with a luminosity of $10^{36}$, but with reduced power consumption~\cite{ref:linac1}.

We first turned to a colliding linac approach, but this proved to be a difficult design that also had high power consumption. We then developed the current concept, which has roots in ILC R\&D: a very low emittance storage ring, based on the ILC damping ring minimum emittance growth lattice and final focus, that incorporates several novel accelerator concepts and appears capable of meeting all design criteria, while reducing the power consumption, which dominates the operating costs of the facility, to a level similar to that of the current $B$ Factories. Due to similarities in the design of the low emittance rings and the final focus, operation of \superb\ can serve as a system test for these linear collider components

By utilizing concepts developed for the ILC damping rings and final focus in the design of the Super$B$ collider, it is possible to produce a two-order-of-magnitude increase in luminosity with beam currents that are comparable to those in the existing asymmetric $B$ Factories.  Background rates and radiation levels associated with the circulating currents are comparable to current values; luminosity-related backgrounds such as those due to radiative Bhabhas, increase substantially. With careful design of the interaction region, including appropriate local shielding, and straightforward revisions of detector components, upgraded detectors based on \babar\ or Belle are a good match to the machine environment:  in this discussion, we use \babar\ as a specific example.  Required detector upgrades include: reduction of the radius of the beam pipe, allowing a first measurement of track position closer to the vertex and improving the vertex resolution (this allows the energy asymmetry of the collider  to be reduced to 7 on 4 GeV); replacement of the drift chamber, as the current chamber will have exceeded its design lifetime; replacement of the endcap calorimeter, with faster crystals having a smaller Moli{\`e}re radius, since there is a large increase in Bhabha electrons in this region.

The \superb\ design has been undertaken subject to two important
constraints: 1) the lattice is closely related to the ILC
  Damping Ring lattice, and 2) as many \pepii components as possible
have been incorporated into the design. A large number of \pepii
components can, in fact, be reused: The majority of the HER and LER
magnets, the magnet power supplies, the RF system, the digital
feedback system, and many vacuum components. This will reduce the cost
and engineering effort needed to bring the project to fruition. 

The Super$B$ concept is a breakthrough in collider design. The invention of the ``crabbed waist'' final focus can, in fact, have impact even on the current generation of colliders. A test of the crabbed waist concept is planned to take place at Frascati in 2007; a positive result of this test would be an important milestone as the \superb\ design progresses. The low emittance lattice, fundamental as well to the ILC damping ring design, allow high luminosity with modest power consumption and demands on the detector.

\superb\ appears to be the most promising approach to
producing the very high luminosity \abf\ that is required to observe
and explore the contributions of physics beyond the Standard Model
to heavy quark and $\tau$ decays.

\section{The Opportunity}

There is substantial international interest in both the experimental and theoretical communities in studying heavy flavour physics with a very large data sample. This is reflected in the large number of Super $B$ Factory workshops that have been held by the \pepii/\babar\ and KEK $B$/Belle groups, jointly and separately, as well as by workshops specifically oriented to the study of the physics capabilities, such as the Workshop on the {\sl Discovery Potential of a Super~$B$~Factory}~\cite{ref:disc} and the {\sl Workshop on Flavour Physics in the LHC Era}. These workshops have clearly demonstrated the importance of heavy flavour studies to arriving at an understanding of New Physics beyond the Standard Model.

The $B$~Factories, building on more than thirty years of work in heavy flavour studies, have developed an extraordinarily vibrant and productive physics community. They have produced more than four hundred refereed publications on mixing-induced and direct \CP\ violation, improved the measurements of leptonic, semileptonic and hadronic decays and discovered a series of surprising charmonium states. The $B$~Factories have also been an excellent training ground for hundreds of graduate students and postdoctoral fellows. \superb\ will no doubt be similarly productive.

INFN has formed an International Review Committee to critically examine this \superb\ Conceptual Design Report and give advice as to further steps, which include submission of this CDR to the CERN Strategy Group, requests for funding to the Italian government, and application for European Union funds.

Should the proposal process move forward, it is expected that the collider and detector projects will be realized as an international collaborative effort. Members of the \superb\ community will apply to their respective funding agencies for support, which will ultimately be recognized in Memoranda of Understanding. A cadre of accelerator experiments must be assembled to detail the design of \superb, while an international detector/physics collaboration is formed. The prospect of the reuse of substantial portions of \pepii and \babar\ raises the prospect of a major in-kind contribution from the US DOE and/or other agencies that contributed to \babar\ construction; support of the project with other appropriate in-kind contributions is also conceivable. It is anticipated that the bulk of the US DOE contribution would be in kind, in the form of \pepii components made available with the termination of the SLAC heavy flavour program. These include the HER and LER magnets, the RF and digital feedback systems, power supplies and vacuum components and the \babar\ detector as the basis for an upgraded \superb\ detector.

\babar\ is generally recognized as a successful example of an international collaboration formed to design, build and operate an HEP detector and to produce physics. The \babar\ model was based on experience gained at CERN and other major laboratories in building and managing international collaborations over the past several decades; it is expected to serve as a model for the \superb\ effort~\cite{ref:web}. The funding agencies of the participating countries will have a role, together with the host agency and host laboratory, in the management of the enterprise, as well as a fiscal role through an International Finance Committee and various review committees.  As with \babar, the international character of the enterprise will be reflected in the governance of the collaboration and in participation in the operating expenses of the experiment, which include the substantial offline computing required.

\section{Conclusions}

The two first generation asymmetric $B$ Factories, \pepii and KEKB,
were built after consideration of a wide variety (more than twenty) of technical
options for achieving very high luminosity with asymmetric energies.
Both $B$ Factories have been very successful, exceeding
design luminosity in a short time, and performing very reliably.

The associated detectors, \babar\ and Belle have utilized the very
large data samples provided by \pepii and KEKB to provide a
cornucopia of new heavy quark and heavy lepton physics measurements.
These have subjected the Standard Model to new and stringent tests,
all of which have thus far been passed.

With much larger data samples of 50-100 ab$^{-1}$, new physics
effects in $B$, $D$, and $\tau$ decays should be readily measurable, and
will play a crucial and complementary role with the LHC and ILC in
deciphering the details of, for example, supersymmetry breaking.
Samples of this size require the construction of a machine like Super$B$
to provide data at a rate exceeding 15 ab$^{-1}$ per year.

\bibliography{sample}

\begin{thebibliography}{9}
\bibitem{ref:UT}
{Differences in the treatment of theoretical uncertainties in fits to extract Unitarity Triangle parameters have led to several distinct approaches. There is a Bayesian contingent~\cite{ref:UTfit} that treats theoretical uncertainties as if they followed a probability distribution. There are also frequentist treatments~\cite{ref:CKMfit},\cite{ref:unfit} that make no assumptions about the distribution of theoretical uncertainties.}
\bibitem{ref:snowmassyear}
{The New Snowmass Year is an updating of the convention that multiplying peak luminosity by a ``year'' containing $10^7$ seconds provides a good measure of actual running time, the effects of accelerator and detector down time, dead time effects and the difference between peak and average luminosity. \pepii/\babar\ experience has shown that a ``New Snowmass Year'' with $1.5\times 10^7$ seconds is a more precise estimator of actual performance at a $B$ Factory.}
\end{thebibliography}

\afterpage{\clearpage}

\graphicspath{{Physics/figures/}}
\graphicspath{{./}{../}{figures/}{Physics/}{Physics/figures/}}


\def\superb{\mbox{\normalfont Super$B$\xspace}}
\def\sbf{\mbox{\normalfont \superb\ Factory}\xspace}
\def\sff{\mbox{\normalfont Super Flavour Factory}\xspace}
\def\babar{\mbox{\slshape B\kern-0.1em{\smaller A}\kern-0.1em
    B\kern-0.1em{\smaller A\kern-0.2em R}}\xspace}
\def\belle{\mbox{\normalfont Belle}\xspace}

\def\etal{{\it et al.}}
\def\ie{{\it i.e.}}
\def\eg{{\it e.g.}}
\def\etc{{\it etc.}}
\def\cf{{\it cf.}}
\def\vs{{\it vs.}}

\def\B{{\ensuremath{{\cal B}}\xspace}}

\newcommand{\subsubsubsection}[1]{\vspace{2ex}\par\noindent {\bf\boldmath\em #1} \vspace{2ex}\par}
\newcommand{\mysection}[1]{\section[#1]{\boldmath #1}}
\newcommand{\mysubsection}[1]{\subsection[#1]{\boldmath #1}}
\newcommand{\mysubsubsection}[1]{\subsubsection[#1]{\boldmath #1}}
\newcommand{\mysubsubsubsection}[1]{\subsubsubsection{\boldmath #1}}


\renewcommand{\Re}{{\rm Re}\,}
\renewcommand{\Im}{{\rm Im}\,}

\newcommand{\GeV}{\,\mbox{GeV}}
\newcommand{\MeV}{\,\mbox{MeV}}
\newcommand{\aver}[1]{\langle #1\rangle}
\newcommand{\matel}[3]{\langle #1|#2|#3\rangle}
\newcommand{\state}[1]{|#1\rangle}
\newcommand{\ve}[1]{\vec{\bf #1}}

\newcommand{\note}[1]{\marginpar{\hspace*{-2mm}\tiny #1}}


\def\mtiny{\vrule width 0pt}
\def\mrm#1{\mathrm{#1}}
\def\DZ{\relax\ifmmode{D^0}\else{$\mrm{D}^{\mrm{0}}$}\fi}
\def\KZ{\relax\ifmmode{K^0}\else{$\mrm{K}^{\mrm{0}}$}\fi}
\def\BZ{\relax\ifmmode{B^0}\else{$\mrm{B}^{\mrm{0}}$}\fi}
\def\BZS{\relax\ifmmode{B_s^0}\else{$\mrm{B_s}^{\mrm{0}}$}\fi}
\def\DZB{\relax\ifmmode{\overline{D}\mtiny^0}
        \else{$\overline{\mrm{D}}\mtiny^{\mrm{0}}$}\fi}
\def\KZB{\relax\ifmmode{\overline{K}\mtiny^0}
        \else{$\overline{\mrm{K}}\mtiny^{\mrm{0}}$}\fi}
\def\BZB{\relax\ifmmode{\overline{B}\mtiny^0}
        \else{$\overline{\mrm{B}}\mtiny^{\mrm{0}}$}\fi}
\def\BZBS{\relax\ifmmode{\overline{B_s}\mtiny^0}
        \else{$\overline{\mrm{B_s}}\mtiny^{\mrm{0}}$}\fi}

\newcommand{\MSbar}{\hbox{$\overline{MS}$\ }}
\newcommand{\ket}[1]{\left| {#1} \right\rangle}
\newcommand{\bra}[1]{\left\langle {#1}\right|}
\newcommand{\braket}[2]{\left\langle {#1} \right| \left. {#2} \right\rangle}
\newcommand{\too}{\mathop{\ \longrightarrow}\ }
\newcommand\eqn[1]{\label{eq:#1}} 
\newcommand\eq[1]{eq.~(\ref{eq:#1})} 
\newcommand\tab[1]{{\footnotesize {\bf Table}~[{\bf\ref{#1}}]}} 
\newcommand\sect[1]{sec.~\ref{sec:#1}} 
\newcommand{\Det}{\mathop{\rm Det}}
\newcommand{\bfn}{{\bf n}}
\newcommand{\bfr}{{\bf r}}
\newcommand{\bfw}{{\bf w}}
\newcommand{\bfm}{{\bf m}}
\newcommand{\bfe}{{\bf  e}}
\newcommand{\xh}{{\bf \hat x}}
\newcommand{\yh}{{\bf \hat y}}
\newcommand{\zh}{{\bf \hat z\,}}
\newcommand{\Tr}{\mathop{\rm Tr}}
\newcommand{\sla}[1]%
        {\kern .25em\raise.18ex\hbox{$/$}\kern-.75em #1}
\newcommand{\mybar}[1]%
        {\kern 0.8pt\overline{\kern -0.8pt#1\kern -0.8pt}\kern 0.8pt}
\newcommand{\cs}{charged scalar}
\newcommand{\css}{charged scalars}
\newcommand{\nfcl}{\mbox{$95\%~{\rm CL}$}}
\newcommand{\ncl}{\mbox{$90\%~{\rm CL}$}}
\newcommand{\brsm}{\mbox{${\rm BR}^{SM}$}}
\newcommand{\brmh}{\mbox{${\rm BR}^{MHDM}$}}
\newcommand{\bxtaunu}{\mbox{$B \rightarrow X \tau \nu_{\tau}$}}

\newcommand{\bbr}{$B${\footnotesize $A$}$B${\footnotesize $AR$} }
\newcommand{\ov}[1]{\overline{#1}}

\newcommand{\msbar}{\overline{\rm MS}}
\newcommand{\bea}{\begin{eqnarray}}
\newcommand{\eea}{\end{eqnarray}}
\newcommand{\beq}{\begin{equation}}
\newcommand{\eeq}{\end{equation}}
\newcommand{\kkbar}{K^0-\overline K^0}
\newcommand{\bbbard}{\overline B_d^0-B_d^0}
\newcommand{\bbbars}{\overline B_s^0-B_s^0}
\newcommand{\dmd}{\Delta m_d}
\newcommand{\dms}{\Delta m_s}

\def\simge{\mathrel{\rlap{\raise 0.511ex \hbox{$>$}}{\lower 0.511ex
\hbox{$\sim$}}}}
\def\simle{\mathrel{\rlap{\raise 0.511ex \hbox{$<$}}{\lower 0.511ex
\hbox{$\sim$}}}}

\chapter{The Physics}
\label{sec:physics}


The search for evidence of physics beyond the Standard Model will be
the main objective of elementary particle physics in the coming
decade. The LHC at CERN will soon commence a search for the Higgs
boson, the missing building block of the Standard Model. It will also
begin an intensive search for New Physics beyond the Standard Model, a
search motivated by the expectation that a new scale is expected make
an appearance at energies around 1~TeV, which will be accessible to
the LHC. 

The production and observation of new particles is not, however, the
only way to look for New Physics. New particles can reveal themselves
through virtual effects in decays of Standard Model particles such as
$B$ and $D$ mesons and $\tau$ leptons. 
Since quantum effects typically become smaller as the mass of the
virtual particles increases, high-precision measurements are required
to have an extended mass reach. In some instances, in fact,
high-precision measurements of heavy flavour decays allow us to probe
New Physics energy scales inaccessible at present and next-generation
colliders. 

Flavour physics is fertile ground for indirect New Physics searches
for several reasons. Flavour Changing Neutral Currents (FCNC), neutral
meson-antimeson mixing and $\CP$ violation occur only at the loop
level in the Standard Model and  are therefore potentially subject to
${\cal O}(1)$ New Physics virtual corrections. In addition, quark
flavour violation in the Standard Model is governed by the weak
interaction and suppressed by the small Cabibbo-Kobayashi-Maskawa
(CKM) mixing angles. These features are not necessarily shared by New
Physics, which could, therefore, produce very large effects in
particular cases. Indeed, the inclusion in the Standard Model of
generic New Physics  flavour-violating terms with natural ${\cal
O}(1)$ couplings is known to violate present experimental constraints
unless the New Physics scale is pushed up to $10$--$100$ TeV,
depending on the flavour sector. The difference between the New
Physics scale emerging from flavour physics and that suggested by
Higgs physics could be a problem for model builders, but it clearly
indicates that flavour physics has either the potential to push the
explored New Physics scale in the $100$ TeV region or, if the New
Physics scale is indeed close to $1$ TeV, that the flavour structure
of New Physics is non-trivial and the experimental determination of
the flavour-violating couplings is particularly interesting. 

On quite general grounds, indirect New Physics searches in
flavour-changing processes explore a parameter space including the New
Physics scale and the New Physics flavour- and $\CP$-violating
couplings. In specific models, these are related to fundamental
parameters, such as the masses and couplings of new particles. In
particular, an observable New Physics effect could be generated by
small New Physics scales and/or large couplings. Conversely, small
effects in the flavour sector could be due to large New Physics scales
and/or small couplings. The question of whether or not New Physics is
flavour-blind is therefore crucial; if so, New Physics searches in
flavour physics would be unfeasible. Fortunately, the concept of
Minimal Flavour Violation (MFV) provides a negative answer: even if
New Physics did not contain new sources of flavour and $\CP$
violation, the flavour-violating couplings present in the Standard
Model are enough to produce a new phenomenology that makes flavour
processes sensitive to the presence of new particles. In other words,
MFV puts a lower bound on the flavour effects generated by New Physics
at a given mass scale, a sort of ``worst case'' scenario for the
flavour-violating couplings.  Thus the MFV concept is extremely useful
to exclude New Physics flavour-blindness and to assess the ``minimum''
performance of flavour physics in searching for New Physics, keeping
in mind that larger effects are quite possible and easily produced in
many scenarios beyond MFV. 

\begin{figure}[htb!]
  \begin{center}
    \includegraphics[width=0.45\textwidth]{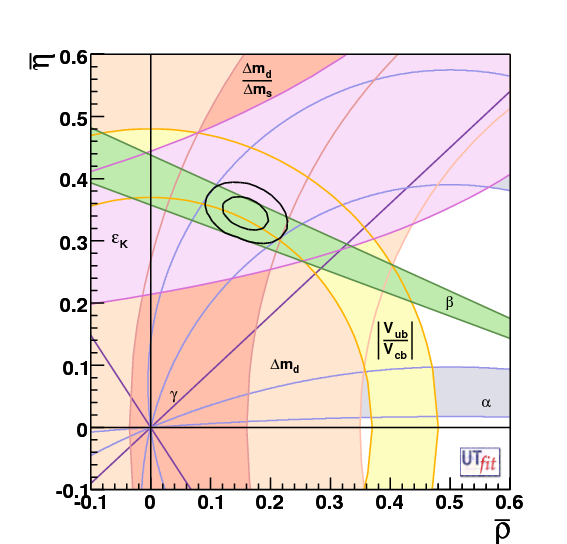}
    \includegraphics[width=0.45\textwidth]{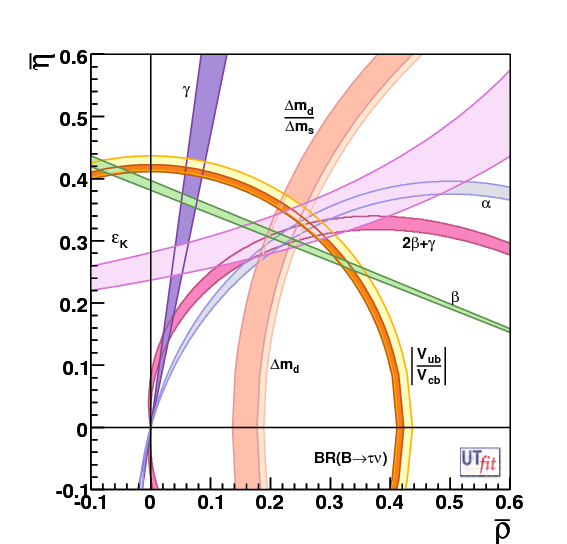}
    \caption{
      Regions corresponding to $95\%$ probability for $\rhobar$
      and $\etabar$ selected by different constraints, assuming
      present central values with present errors (left) or with
      errors expected at \superb\ (right).
    }
    \label{fig:sm2015}
  \end{center}
\end{figure}

The effectiveness of flavour physics in constraining New Physics has
already been demonstrated by the $B$ Factories, whose superb
performance in measuring the parameters of the CKM matrix, together
with new results from the Tevatron on $B_s$ physics, already allow
interesting bounds on New Physics.  A few discrepancies exist in the
current data, although several measurements alone do not approach
$10\%$ accuracy. One lesson from the $B$ Factories is that precision
is crucial in these kind of studies, as are redundant measurements of
the same underlying quantity. 
In Fig.~\ref{fig:sm2015} we show the regions on the
$\rhobar$-$\etabar$ plane selected by different constraints assuming
the current measurement precision, and that 
expected at \superb. With the precision reached at \superb, the
current discrepancies would clearly indicate the presence of New
Physics in the flavour sector! 

In light of these considerations, it is clear that a Super Flavour Factory can provide unique evidence for New Physics in the heavy flavour sector
by searching for virtual effects that induce deviations from Standard Model predictions at the percent level, and for processes that are highly suppressed, or even forbidden, in the Standard Model, but can be enhanced by New Physics.  Two features of the \sff\ are appealing from an experimental point of view: the possibility of measuring dozens of New Physics-sensitive observables with unprecedented precision, thanks to the high luminosity and the very clean experimental environment; and the ability to change the center-of-mass energy to produce well-defined particle-antiparticle pairs of
$B^+$, $B_d$, $B_s$, $D^0$, $D^+$, $D_s$ mesons and $\tau$ leptons,
exploiting the quantum-coherence inherent in production via resonances $e^+e^-$ annihilation.

Physics at \superb\ could begin around 2012. An obvious question is then how the \sff\ physics program fits into the program of particle physics early in the next decade? Several scenarios are conceivable, but the most pertinent is whether the LHC will have produced non-standard (possibly flavoured) particles with masses below 1 to 2 TeV or not.

If New Physics has been found elsewhere, the importance of flavour
physics studies becomes twofold: not only could the open window on
much larger scales extend the New Physics mass spectrum found at the
LHC, but a detailed study of the flavour- and $\CP$-violating
couplings of newly discovered particles could be carried out even in
the unfavourable MFV case, taking advantage of the crucial information
on the New Physics scale provided by the LHC. 
Although \lhcb, ATLAS or CMS
could be the first to observe flavour-related effects in new particle
production or decay, only with the \sff\  would we be able to perform
a systematic analysis of their flavour- and $\CP$-violating couplings
in processes involving the second and third generations of quarks and
leptons. These studies have a unique capability to reconstruct the New
Physics Lagrangian from the observed phenomenology. A typical example
is supersymmetry (SUSY): most of the couplings appearing in the soft
SUSY-breaking sector of the Lagrangian could be measured at the
\sff. In this scenario, high $p_T$ and flavour physics observations
would both be required to understand the nature of New Physics. 

If physics beyond the Standard Model is not found at the LHC, indirect
searches in flavour-changing processes become of the utmost importance
to probe New Physics scales in the $10$--$100$ TeV region. After all,
the $1$ TeV New Physics scale naturally required in order to stabilize
the Fermi scale could be somewhat higher, without invalidating the
concept of naturalness. Yet an acceptable upward shift of the New
Physics scale would put LHC out of the game, and leave the task of
discovering New Physics to indirect searches. Flavour physics would be
able to probe the interesting mass range, giving naturalness a second
chance before discarding it in favour of more exotic explanations of
the Fermi scale. Unfortunately, given the presence of the unknown
flavour couplings, there is no guarantee that the virtual effects of a
new particle with a mass of $100$ TeV are observable even at the
\sff\. Still, values of the New Physics scale in the $10$--$100$ TeV
range can be naturally reached in most New Physics models, including,
for example, the Minimal Supersymmetric Standard Model, and even
models with MFV are sensitive to scales larger than $1$ TeV in the
large $\tan\beta$ regime. Notice that \lhcb\ and the \sff, which find
their strengths in measuring different decay processes, are
complementary in the effort to observe New Physics effects from large
scales.

In any case, regardless of whether or not New Physics has already been
found, it is crucial to exploit the full richness of the phenomenology
accessible at the Super Flavour Factory in order to increase the
chances of observing 
New Physics flavour effects and to study the New Physics flavour
structure. 

Another anticipated result related to \sff\ physics is
the search for lepton flavour violation (LFV) in the decay
$\mu\to e\gamma$ by the MEG collaboration.
Indeed, searches for LFV in the transitions between the
second and third generations, the golden mode being $\tau\to\mu\gamma$, are
a centerpiece of the \sff\ physics program.
The observation of $\tau\to\mu\gamma$ with a branching ratio
around $10^{-9}$, an unmistakable signal of New Physics, is accessible at \superb.
\superb\ will probe values of $\BR(\tau\to\mu\gamma)$
an order of magnitude smaller than previous experiments;
this is the range predicted by most New Physics models. 
For example, within Grand-Unified models, MEG and \superb\
sensitivities 
are such that the pattern of LFV observations (and non-observation) 
can identify the dominant source of LFV and distinguish whether it 
is governed by the CKM or the PNMS matrix. 
Other topics in $\tau$ physics can be studied at the \sff\ as well, 
in particular, the precise determination of $\tau$ production and
decay properties, including $\CP$-violating observables, 
such as the $T$-odd triple products which benefit from the 
polarized $\tau$ leptons that \superb\ can produce with a polarized
electron beam. 
 
New Physics searches with $B_d$ and $B^+$ decays proceed along the
lines already begun at the $B$ Factories. The full set of $B$ Factory
measurements can be addressed, improving the accuracy of several
observables, \eg\ CKM angles, $b\to s$ penguin transitions,
$\B(B^+\to\tau^+\nu_\tau)$, \etc\ down to ${\cal O}(1\%)$. Additional 
New Physics-sensitive measurements such as the $\CP$ asymmetry in
$B\to X_s\gamma$ or the forward-backward asymmetry in $B\to X_s l^+
l^-$ become possible with the \superb\ dataset. Any of these
measurements could show a clear deviation from the Standard Model or
be used to feed more sophisticated New Physics analyses. Notice that,
in this sector, the overlap with the \lhcb physics program is rather
limited and the \sff\ performance is, typically, superior. 

It is worth noting that while some New Physics analyses depend only on
measured quantities, others require theoretical information on
hadronic parameters. The only approach that can, in principle, achieve
the required theoretical accuracy is lattice QCD, where the limiting
factor is likely to be uncontrolled systematic uncertainties. From
this point of view, it is reassuring that lattice simulations have
already begun to go beyond the quenched approximation. Extrapolations
based on computing power foreseen in 2015, taking into account
different sources of systematics (chiral extrapolation, heavy mass
extrapolation, continuum limit, finite-size effects, \etc), indicate
that an accuracy of ${\cal O}(1\%)$ is achievable on the hadronic
parameters of interest for the \sff\ physics program, even without
considering progress in theory and in algorithms, which are likely to
occur, but difficult to anticipate. 

The case of $B_s$ studies is somewhat different. The high oscillation
frequency makes it impossible to perform fully time-dependent
measurements at \superb. In addition, most of the interesting
observables, such as the phase $\phi_{B_{s}}$ of the $B_s$ mixing
amplitude or $\B(B_s\to\mu^+\mu^-$), will have been measured with high
precision by \lhcb\ (and possibly by Belle running at the
$\Upsilon(5{\rm S})$) before \superb\ begins. Nevertheless, a short
run at the $\Upsilon(5{\rm S})$ would suffice to accurately measure
New Physics-sensitive quantities, such as the semileptonic $\CP$
asymmetry $a_{sl}^s$, which cannot be observed at hadronic colliders. 
It is interesting to note that, thanks to the quantum coherence of the
$\BsBsb$ pairs and the (limited) time sensitivity achievable at
\superb, it would be possible 
to measure $CP$ violating phases through terms in the time-dependent
decay rates that depend on $\Delta \Gamma_s$. 
That is, the same quantities that can be extracted from the full
time-dependent analysis can still be determined. 
Using this method and the full \superb\ statistics, it should be
possible not only to measure $\phi_{B_{s}}$ with an accuracy
competitive with \lhcb, but also to access other CKM angles with $B_s$
decays. A similar consideration applies to $B_s\to\mu^+\mu^-$, where,
with the full statistics, one could hope to probe the Standard Model
value of this branching ratio. However, gains in $B_s$ physics would
be paid for with statistics potentially available for $B_d$/$B^+$
physics. It is not clear at this point whether this would be
worthwhile in the first few years of operation of
\superb. Nevertheless, it seems prudent to maintain this unique
capability. 

Finally, it is important to note that a large numbers of charmed
particles are produced at the \superb\ while running on the $\Upsilon$
resonances; this sample would be $10^4$ times the statistics of
existing charm factories and would still be much larger than samples
at future dedicated facilities. It is clear that the next generation
physics program of a charm factory could be carried out at
\superb. Some studies, for instance those related to the calibration
of lattice QCD, could benefit from a short run at the $D\Dbar$
threshold. Others, such as mixing studies based on quantum coherence,
can only be done at threshold. In any case, a run of 1 to 2 months at
threshold would produce a $D\Dbar$ sample ten times larger than that
available at the conclusion of running at the new charm
factories. With these statistics, interesting New Physics-related
measurements in the $D$ sector become possible, in particular $\CP$
violation in $D$ decay and improved measurements of $D\Dbar$
oscillation parameters.

In this chapter
we will identify those measurements that can be
performed at \superb\ which constitute clear motivation for its construction.
In Section~\ref{sec:bd} we discuss physics with $B^\pm$ and $B^0_d$ mesons,
that is $B$ physics at the $\FourS$ resonance;
in Section~\ref{sec:tau} we discuss $\tau$ physics,
with particular emphasis on searches for lepton flavour violation;
in Section~\ref{sec:bs} we discuss measurements
in the $B_s$ sector that can be made at the $\FiveS$ resonance;
in Section~\ref{sec:charm} we discuss the charm physics reach,
including a discussion of the case for running at charm threshold.
We briefly mention some of the other physics topics
that can be tackled at \superb\ in Section~\ref{sec:other},
before summarizing the physics potential in Section~\ref{sec:summary}.
We also include an Appendix on the expected improvement in
lattice QCD calculations, and how these can affect the \superb\ program.

The discussions in this chapter take as their starting point the results
from the current generation of asymmetric $B$ Factory experiments,
\babar\ and \belle.
We emphasize those measurements that are unique to the \superb\
program, and which cannot be accessed at hadronic machines.
For each measurement, we consider the current precision, the
precision at the end of the
$B$ Factory programs with $2 \ {\rm ab}^{-1}$ of integrated luminosity, and results with $75 \ {\rm ab}^{-1}$, the integrated luminosity that would be
collected in five years of data taking at a peak luminosity of $10^{36}\ {\rm cm}^{-1} {\rm sec}^{-2}$.
Note that the integrated luminosity profiles shown in Section~\ref{section:Overview} are based on more detailed scenarios for the progress of peak luminosity with time. We take particular care to consider potentially limiting systematic
or theoretical uncertainties, and how these may be reduced.
We also consider how changes in the detector and operating conditions
(energy asymmetry, acceptance (hermeticity), vertex resolution, \etc)
may affect the precision.

\clearpage

\mysection{$B$ Physics at the $\Upsilon(4{\rm S})$}
\label{sec:bd}
\mysubsection{The Angles of the Unitarity Triangle}
\label{ss:angles}

\begin{figure}[!h]
  \begin{center}
    \resizebox{0.55\textwidth}{!}{\includegraphics{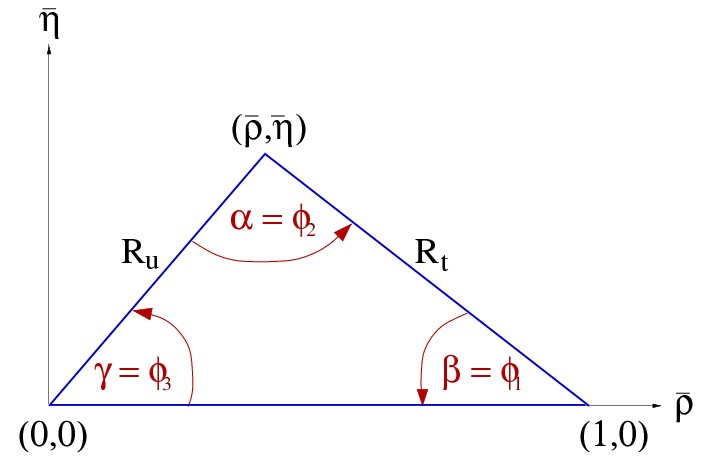}}
    \caption{The Unitarity Triangle.}
    \label{fig:cp_uta:ut}
  \end{center}
\end{figure}

The Unitarity Triangle (UT), shown in Fig.~\ref{fig:cp_uta:ut}, is a convenient graphical representation
of one of the unitarity conditions of the CKM matrix~\cite{Cabibbo:1963yz,Kobayashi:1973fv} given by
{\small
\begin{equation}
  \label{eq:cp_uta:ut}
  V_{ud}V^*_{ub} + V_{cd}V^*_{cb} + V_{td}V^*_{tb} = 0 \, .
\end{equation}
}

Two popular naming conventions for the UT angles exist in the literature:
\begin{equation}
  \label{eq:cp_uta:abc}
  \alpha  \equiv  \phi_2  =
  \arg\left[ - \frac{V_{td}V_{tb}^*}{V_{ud}V_{ub}^*} \right],
  \hspace{0.2cm}
  \beta   \equiv   \phi_1 =
  \arg\left[ - \frac{V_{cd}V_{cb}^*}{V_{td}V_{tb}^*} \right],
  \hspace{0.2cm}
  \gamma  \equiv   \phi_3  =
  \arg\left[ - \frac{V_{ud}V_{ub}^*}{V_{cd}V_{cb}^*} \right].
  \nonumber
\end{equation}
We use the $\left( \alpha, \beta, \gamma \right)$ convention.

A significant fraction of $B$ physics is centered on measuring the
properties of this triangle
and, by overconstraining it, searching for New Physics effects.
This will continue to be a prominent part of the physics program of \superb.

Interpreting measurements in terms of properties of the UT
is not always completely straightforward.
In order to estimate \superb\ measurement sensitivities, we have used the results of the UTfit
Collaboration~\cite{Ciuchini:2000de,Bona:2005vz,Bona:2005eu,Bona:2006ah}.
For a different approach producing similar results, see~\cite{Charles:2004jd}.

\noindent
\mysubsubsection{Measurement of $\beta$}
\label{sss:beta}

\mysubsubsubsection{$\beta$ in
charmonium-kaon final states}

The measurement of $\sin(2\beta)$
through mixing-induced $\CP$ violation in the decay $B^0 \to J/\psi\,K^0$,
one of the theoretically cleanest measurements
that can be made in flavour physics~\cite{Carter:1980tk,Bigi:1981qs},
was the {\it raison d'\^etre} of the current generation of $B$ Factories.
The most recent measurements give a
world average~\cite{hfag,Aubert:2006aq,Chen:2006nk}
\begin{equation}
  \sin(2\beta) = 0.675 \pm 0.026 \, ,
\end{equation}
yielding a solution consistent with the Standard Model of
$\beta = \left( 21.2 \pm 1.0 \right)^\circ$
(alternative solutions for $\beta$ are strongly disfavoured by other
measurements).
This result provides one of the tightest constraints on the
Standard Model parameters in the $\bar \rho$-$\bar \eta$ plane.

Further reduction of the error is straightforward,
since the statistical error is almost twice the systematic uncertainty.
However, beyond an integrated luminosity of a few ${\rm ab}^{-1}$
the measurements approach the systematics-dominated regime.
The systematic error budget of both \babar\ and \belle\ analyses
suggest that the limit due to systematic uncertainties
(in vertexing algorithms, beam spot position and
tag-side interference~\cite{Long:2003wq}) is about $0.010$.
The statistical precision will reach this limit
with about $10 \ {\rm ab}^{-1}$.
Nevertheless, this channel is an important benchmark for any
$B$ physics experiment, and improved understanding of
detector-related systematic effects will have benefits for all analyses.
Using the very high statistics control samples that will be available at \superb\ to improve the understanding of the detector,
it may be possible to reduce this error to $\sim 0.005$.

This level of precision is still above the size of the
expected theoretical uncertainty~\cite{Boos:2004xp,Li:2006vq,Ciuchini:2005mg}.
At \superb, it is possible to control the
the possible penguin contribution
using a data-driven approach
that employs the experimental measurement of time-dependent $\CP$
asymmetry
in $B^0 \rightarrow J/\psi\,\pi^0$~\cite{Ciuchini:2005mg}.
Although experiments in a hadronic environment should be able to
measure $\sin(2\beta)$ from $B^0 \rightarrow J/\psi\,\KS$
to the same level of accuracy as \superb,
they would not be able to perform this kind of cross-check,
nor can they check the consistency of the values obtained with different
final states such as $\eta_c \KS$, $\chi_{c1} \KS$, \etc~\cite{Atwood:2003tg}.

Decays with charmonium-kaon final states offer a number of
additional important observables.
Direct $\CP$ violation in $B^+ \to J/\psi\,K^+$
would be a clear New Physics signal~\cite{Hou:2006du};
\superb\ can probe for this effect to the limit of detector systematics,
expected to be $\sim 0.4\%$.
Furthermore, time-dependent studies of $B^0 \to J/\psi\,K^{*0}$,
with $K^{*0} \to \KS\pi^0$, provide sensitivity to $\cos(2\beta)$
~\cite{Dunietz:1990cj,Dighe:1998vm,Aubert:2004cp,Itoh:2005ks},
which could be measured to a precision of 0.05 at \superb.

\mysubsubsubsection{Complementary measurements of $\beta$}

The value of $\sin(2\beta)$ can also be measured from mixing-induced
$\CP$
asymmetries in several other $B^0$ decays, and the consistency of
these results with
those obtained from $b \to c \bar c s$ transitions
provides a powerful
test of New Physics effects.
Foremost among these are decays dominated by the $b \to s$ penguin
amplitude,
discussed below, but there are also others.
Decays such as $B^0 \to J/\psi\,\pi^0$ and $B^0 \to D^+D^-$
are expected to be dominated by the $b \to c \bar c d$ tree diagram,
although contributions from the $b \to d$ penguin amplitude are also allowed.
Sizeable deviations from the Standard Model predictions may suggest
New Physics enhancements in the $b \to d$ penguin topology.
Decays such as $B^0 \to D \pi^0$, where the $D$ meson
is reconstructed in a final state accessible to both
$\Dz$ and $\Dzb$ decay,
such as a $\CP$ eigenstate (\eg\ $K^+K^-$)
or a multibody final state (\eg\ $\KS \pi^+\pi^-$),
are dominated by the $b \to c \bar u d$ tree diagram, with negligible
Standard Model backgrounds
~\cite{Grossman:1996ke,Fleischer:2003ai,Fleischer:2003aj,Aubert:2007mn}.
The $\KS \pi^+\pi^-$ channel also allows $\cos(2\beta)$ to be
cleanly determined~\cite{Bondar:2005gk}.
With $75 \ {\rm ab}^{-1}$, these channels will yield
measurements of $\sin(2\beta)$ and $\cos(2\beta)$
with precision of about $0.02$ and $0.04$ respectively.
Experiments in hadronic environments are not competitive
for these measurements.

\mysubsubsubsection{Measurement of $\beta$ with $b \to s$
penguins}

Perhaps the most interesting channels to search for New Physics effects
in mixing-induced $\CP$ violation are those dominated by the
$b \to s$ penguin
transition~\cite{Grossman:1996ke,London:1997zk,Ciuchini:1997zp}. In the Standard Model,
these decays should measure \stwob, up to small corrections.
New Physics particles in the loops
can cause deviations from Standard Model predictions.
The potential of this approach to search for New Physics depends on
the precision of the Standard Model predictions for individual channels;
estimates of these hadronic uncertainties
necessarily rely on models or symmetries.
Recent calculations indicate that the modes with the
smallest theoretical uncertainties are
$\Bz \to \phi \Kz$, $\Bz \to \eta^\prime \Kz$ and
$\Bz \to \Kz\Kzb\Kz$ (the latter reconstructed as $\KS \KS \KS$)
and that these have uncertainties of $\sim 0.02$--$0.05$ on $\sin(2\beta)$
(see, for example,~\cite{Grossman:2003qp,Gronau:2003kx,Gronau:2004hp,
  Cheng:2005bg,Cheng:2005ug,Gronau:2005gz,Beneke:2005pu,Engelhard:2005hu,
  Buchalla:2005us,Gronau:2006qh}, and references therein).
Model-independent data-driven analyses find
larger uncertainties~\cite{pieriniSLAC,pieriniCKM}
but these will decrease as data become more precise.

\begin{figure}[!h]
  \begin{center}
    \resizebox{0.7\textwidth}{!}{\includegraphics{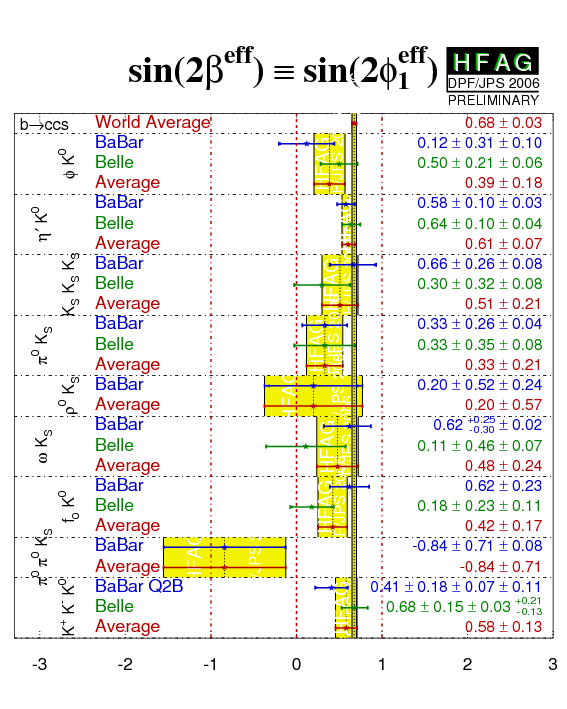}}
    \caption{HFAG compilation of measurements of $\sin(2\beta^{\rm eff})$
      in decays dominated by $b \to s$ penguin amplitudes.}
    \label{fig:hfag_btos}
  \end{center}
\end{figure}

The current world averages of $\sin(2\beta)$ measured using
$\Bz \to \phi \Kz$, $\Bz \to \eta^\prime \Kz$ and $\Bz \to \Kz
\Kzb\Kz$
have uncertainties of $0.18$, $0.07$ and $0.21$
respectively~\cite{hfag,Chen:2006nk,Aubert:2006av,Aubert:2006wv,Aubert:2006ar};
see Fig.~\ref{fig:hfag_btos}.
By the end of this decade,
these errors will be reduced by a factor of $\sim\sqrt{2}$;
\footnote{Note however that the relative sizes of the errors in
$B^0 \to \phi K^0$ differ between \babar\ and \belle, due to different treatment of the $K^+K^-$ $S$-wave under the $\phi$ peak.
If this contribution is large, as suggested by the \babar\ results,
the uncertainty on the average will be larger than expected.}
the precision will still be
much worse than most estimates of the theoretical uncertainties.

A number of studies~\cite{superkekb,superpep,Gershon:2006mt} have shown
that
approximately $75 \ {\rm ab}^{-1}$ are necessary to reduce
the experimental error to the level of the theoretical precision
in the channels $\Bz \to \phi \Kz$, and $\Bz \to \Kz \Kzb\Kz$.
For $\Bz \to \eta^\prime \Kz$,
the results will become theory-dominated much earlier.
However, in this case (and indeed in others)
it may be possible to
use data-driven techniques to gain better control over
the theoretical errors~\cite{Grossman:2003qp,Gronau:2004hp,Gronau:2006qh}.
These analyses obtain constraints on the hadronic parameters by using as input
a large number of branching fractions of charmless hadronic $B$
decays.
The complete set of these measurements can only be obtained
at a Super $B$ Factory.
Indeed, experiments at hadronic machines have limited
capability
in measuring time-dependent $\CP$ asymmetries in $B^0$ decays
dominated by the $b \to s$ penguin amplitudes,
whereas  \superb\ can study not only the channels discussed above,
but also $B^0 \to \pi^0 K^0$, $B^0 \to \rho^0 K^0$,
$B^0 \to \omega K^0$, $B^0 \to f_0 K^0$,
\etc~\cite{Aubert:2006ad,Abe:2006gy,
  Aubert:2006xu,Aubert:2006ai,Aubert:2004gk,Gershon:2004tk}.
While hadronic machines are well-suited to the study of time-dependent
$\CP$ asymmetries in $B_s \to \phi\phi$ decay,
\superb running at the $\Upsilon(5{\rm S})$ can complement these
results with measurements of related channels such as $B_s \to K^0 \overline{K}^0$
$B_s \to \phi\eta$, $B_s \to \eta\eta^\prime$.
For more details on the physics opportunities
at the $\Upsilon(5{\rm S})$, see Section~\ref{sec:bs}.


\noindent
\mysubsubsection{Measurement of $\gamma$}
\label{sss:gamma}

%
Many different processes are sensitive to the UT angle $\gamma$;
consequently a large number of techniques have been proposed to measure $\gamma$. While the majority of these methods suffer from
hard-to-quantify hadronic uncertainties,
there is a method that provides
a theoretically clean
measurement.
Using $B \to DK$ decays, this method exploits the fact that
the neutral $D$ meson decay product can be either a $D^0$
(from a $b \to c\bar{u}s$ transition),
or a $\Dzb$ (from a $b \to u\bar{c}s$ transition;
or {\it vice versa} for $\bar{b}$ decays).
If the final state is chosen such that both $\Dz$ and $\Dzb$
can contribute, the interference between these amplitudes
is sensitive to the phase $\gamma$,
allowing $\gamma$ to be determined with essentially
no theoretical assumptions.
Choices for the final state include $D^0$ meson decays to
$\CP$ eigenstates~\cite{Gronau:1990ra,Gronau:1991dp},
doubly-Cabibbo suppressed states~\cite{Atwood:1996ci,Atwood:2000ck}
or self-conjugate multibody states~\cite{Giri:2003ty}.
The sensitivity to $\gamma$ for each depends on
the (unknown) ratio of the magnitudes of the $b \to u$ and $b \to c$
decay amplitudes, denoted $r_B$, as well as on the structure of the $D$ decay.
Both $B \to D^{*}K$ and
$B \to D K^*$ decays can be employed in addition to $B \to DK$;
both neutral and charged $B$ decays can be used~\cite{Gronau:2004gt}.
In the case of $D^*K$, it is particularly important
to distinguish $D^* \to D\pi^0$ and $D^* \to D\gamma$
decays~\cite{Bondar:2004bi},
which may be difficult in a hadronic environment.
The value of $r_B$ must, in general, be measured, for each $B$
decay channel.

Both \babar\ and \belle\ have made measurements for each of the $D$
decays cited above~\cite{Aubert:2005rw,Abe:2006hc,Aubert:2004hu,
  Aubert:2005cc,Aubert:2005pj,Abe:2005gi,Aubert:2005cr,Aubert:2006ga,
  Aubert:2006am,Poluektov:2006ia,Aubert:2005yj}.
No statistically significant $\CP$-violating effect has yet been observed
in $B \to DK$ decays.
The most precise constraints on $\gamma$:
\begin{equation}
  \babar: \
  \gamma = (92 \pm 41 \pm 11 \pm 12)^\circ
  \hspace{5mm}
  \belle: \
  \gamma = (53 \, ^{+15}_{-18} \pm 3 \pm 9)^\circ,
\end{equation}
currently come from
analyses of the multibody decay $D \to \KS \pi^+\pi^-$
~\cite{Aubert:2006am,Poluektov:2006ia}.

The three sources of error are
statistical, systematic and uncertainty related to the
hadronic structure of the $D \to \KS\pi^+\pi^-$ Dalitz plot.
The smaller statistical error of the \belle\ result
is a consequence of the larger central value obtained for $r_B$.

Combining all the available measurements provides a determination of
$\gamma$ with an error of about 20$^{\circ}$.
The central values of $r_B$ (for each $B$ decay)
are found to be around 0.08~\cite{Bona:2005vz},
slightly smaller than the expectation.

As we extrapolate to high luminosity,
we find that the $D$ decay model uncertainty can become a limiting
factor for the multibody analysis.
Recent studies~\cite{ref:p5} have shown that with $2 \ {\rm ab}^{-1}$,
assuming a Dalitz plot model error of $6^{\circ}$ and $r_B = 0.10$,
the uncertainty on $\gamma$ can be reduced to $\sim 6.4^{\circ}$.
This can in principle be reduced with
a better understanding of the model describing the $D$ resonance
substructure.
A model-independent approach,
using $\CP$-tagged neutral $D$ mesons
collected at an $e^+e^-$ machine operating at the $\psi(3770)$~\cite{Bondar:2005ki}, can also reduce the uncertainty (see Section~\ref{sec:charm}).
Different multibody final states can also be used
(\eg\ $D \to \KS K^+K^-$, $D \to \pi^+\pi^-\pi^0$);
these channels have model uncertainties that are, in general, uncorrelated.
With very high luminosity, it should be possible to use a large number of
states, including singly-Cabibbo suppressed decays
such as $D \to \KS K^\pm \pi^\mp$~\cite{Grossman:2002aq},
and even four-body decays such as $D \to \KS\pi^+\pi^-\pi^0$ and
$D \to K^+ K^-\pi^+\pi^-$~\cite{Rademacker:2006zx}.
Finally, note that approaches using $D$ decays to $\CP$ eigenstates
and doubly-Cabibbo suppressed final states do not suffer from this error.
These methods contribute significantly to the overall constraint
at high luminosities, with the latter particularly important when $r_B$ is small.
Self-tagging $B^0$ meson decays, such as
$B^0 \to D K^{*0}$~\cite{Pruvot:2007yd} can also be used;
this mode may have a larger value of $r_B$ ($\sim 0.4$),
and thus could contribute significantly to a precise
determination of $\gamma$ at high luminosity.

With $75 \ {\rm ab}^{-1}$,
it should be possible to determine $\gamma$ with an uncertainty of
$2$--$3^\circ$ using decays to $\CP$ eigenstates and
doubly-Cabibbo suppressed states alone.
Assuming that $D$ decay model uncertainties can be tamed,
and exploiting the large variety of $D$ decays that can be
reconstructed at \superb,
an uncertainty of $1^{\circ}$ may be possible.

Note that this discussion has neglected the possibility of
mixing and $\CP$ violation in the neutral $D$ meson system.
It is, however, straightforward to take these effects into account
in the analysis, if necessary~\cite{Grossman:2005rp}.
Finally, it is interesting to observe that the determination of
$\gamma$ discussed in this section is not affected by New Physics
under the assumption that the New Physics does not change tree-level processes.
This assumption is expected to be valid at the subpercent level
in most models, and, in any case, would produce observable effects
in the decay branching ratios.
Thus, together with the measurement of $|V_{ub}/V_{cb}|$,
a precise measurement of $\gamma$ provides a significant constraint on the
$\bar{\rho}$--$\bar{\eta}$ plane that must be met by any New Physics model.

\noindent
\mysubsubsection{Measurement of $2\beta+\gamma$}
\label{sss:2b+g}

Interference effects between $b \to c$ and $b \to u$ decay amplitudes in
$B^0$ decays to $D^{(*)\pm}\pi^\mp$ and $D^{(*)\pm}\rho^\mp$ final states
allow the determination of the combination of UT angles $2\beta+\gamma$.
From this analysis the quantities
$r \, \sin(2\beta +\gamma \pm \delta)$ can be determined,
where $r$ is the absolute ratio of the $b \to u$ and $b \to c$ decay
amplitudes
and $\delta$ is their strong phase difference.
Since there are two observables and three unknowns
($r$, $\delta$ and $2\beta+\gamma$),
additional information is needed to extract the weak phase.
Measurements have been performed by both
\babar~\cite{Aubert:2006tw,Aubert:2005yf}
and
\belle~\cite{Ronga:2006hv}
in the channels $D^\pm\pi^\mp$, $D^{*\pm}\pi^\mp$
(the two experiments using both full and partial reconstruction techniques)
and $D^\pm\rho^\mp$ (\babar\ only).
The most precise constraint is provided by the measurement of
$a_{D^*\pi} = 2 \, r_{D^*\pi} \sin(2\beta+\gamma)\cos(\delta_{D^*\pi})
= -0.037 \pm 0.011$~\cite{hfag}.
Using flavour SU(3) symmetry to estimate the size of the $r$ parameters,
allowing for breaking effects as large as 100\%,
and combining measurements in the three modes,
the current world average is
$2\beta+\gamma = \pm (90 \pm 33)^\circ$~\cite{Bona:2005vz}.

A simple extrapolation of these numbers suggests that
precise constraints on $2\beta+\gamma$ can be obtained with
\superb\ luminosities.
However, there are two complications.
The first is that the experimental measurements,
while still dominated by statistical errors, are already rather
precise;
reduction of the systematic errors
much below the $0.01$ level will be challenging.
Secondly, the theoretical uncertainty
related to SU(3) breaking in the estimation of the $r$ parameters
does not allow to extrapolate the error simply scaling
with the statistics.
Such effects are hard to quantify, although there are means to address
them using data.

Both problems can be circumvented by using a channel in which
the value of $r$ is much larger
(and hence can be determined directly from the data).
A good example is
$B^0 \to D^\pm \KS \pi^\mp$~\cite{Charles:1998vf,Aleksan:2002mh},
for which the ratio $r$ is expected to be of order 0.4.
It has been shown~\cite{Cavoto:2006um,Polci:2006aw}
that all the amplitudes and the strong phases of the intermediate
states contributing to this decay can be
determined using a time-dependent Dalitz plot analysis, together with
$2\beta+\gamma$.
The uncertainty on $2\beta+\gamma$ from this mode, with the statistics
available at \superb, could be better than $5^{\circ}$.
The channel $B^0 \rightarrow D K^0$
with $D$ reconstructed into $\CP$ and three-body final states
should also allow a comparable precision in the determination of $2\beta+\gamma$.

\noindent
\mysubsubsection{Measurement of $\alpha$}
\label{sss:alpha}

Experimental information on the angle $\alpha$ derives from the
interference between $\BzBzb$ mixing and
decays dominated by the $b \to u$ amplitude ($T$),
\eg, from the charmless decays
$B \to \pi \pi$, $B \to \rho \pi$ and $B \to \rho \rho$.
In the absence of contributions from top-penguin diagrams,
the $\CP$ asymmetries in these decays provide a measurement of $\sin(2\alpha)$.
Penguin diagrams
introduce an additional amplitude ($P$) with a different weak phase.
In this case, the experimentally measured quantity is
$\sin(2\alpha_{\rm eff})$,
which is a function of $\alpha$ but also of unknown hadronic
parameters.
Several strategies have been proposed to remove this
so-called ``penguin pollution''.
Note that, generically, the uncertainty on $\alpha$
due to the penguin pollution depends on the ratio $|P/T|$.

\begin{figure}[htb]
  \begin{center}
    \resizebox{0.55\textwidth}{!}{\includegraphics{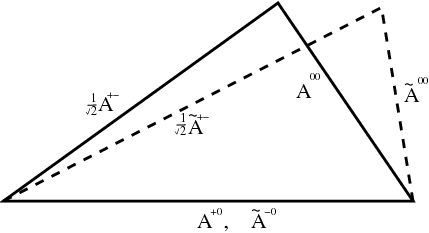}}
    \caption{Isospin triangles in the $B \to \pi\pi$ system.}
    \label{fig:GLtriangle}
  \end{center}
\end{figure}

The archetypal method to extract $\alpha$ uses an isospin analysis of
$B \to \pi\pi$~\cite{Gronau:1990ka};
the same method can also be employed for $B \to \rho\rho$ decays.
This method makes use of the fact that,
due to Bose-Einstein statistics,
the two pions produced in $B$ decay can only have isospin $I=0$ or $I=2$.
Since $B$ mesons have $I=1/2$,
the physical amplitudes can be decomposed in terms of isospin amplitudes
with $\Delta I = 1/2, \, 3/2, \, 5/2$.
On the other hand,
the $\Delta B = 1$ weak effective Hamiltonian only contains
operators that contribute to $\Delta I = 1/2$ and $\Delta I = 3/2$ transitions.
It follows that the physical amplitudes
for the $B^0 \to \pi^+\pi^-$, $B^+ \to \pi^+\pi^0$ and $B^0 \to \pi^0\pi^0$
decays
can be written respectively as:
\begin{equation}
  \begin{array}{ccccc}
    A^{+-}
    & = & \sqrt 2 \times [A_{I=2} - A_{I=0}] \\
    A^{+0}
    & = & \!\!\!\!\!\!\!\!\!\!\!\!\!\!\!\!\!\!\! 3 \times  A_{I=2} \\
    A^{00}
    & = & \ \ 2 \times [A_{I=2} + A_{I=0}], \\
  \end{array}
  \label{eq:su2analysis}
\end{equation}
yielding the relation
$A^{+-} + \sqrt{2} A^{00} = \sqrt{2} A^{+0}$.
An equivalent expression holds for the $\bar{b}$ decay amplitudes.
These relations can be represented in terms of isospin triangles, as shown in Fig.~\ref{fig:GLtriangle}.
In writing the above relations,
we have used the fact that penguin operators mediate only $\Delta I = 1/2$
transitions (since the gluon has $I=0$),
and have thus assumed that $A_{I=2}$ receives contributions
only from tree operators.
Hence $|A^{+0}| = |\bar{A}^{+0}|$,
and the triangles can be rotated to be drawn with a common base.

Construction of the isospin triangles allows
the penguin contribution to be isolated,
and therefore $\alpha$ can be extracted~\cite{Gronau:1990ka} from the measured $\CP$ asymmetry.
The experimental inputs required are the $\CP$-averaged branching fractions
for the three modes ($+-$, $+0$ and $00$), the direct $\CP$ asymmetries for
neutral $B$ decays, together with the mixing-induced $\CP$ asymmetry parameter $S^{+-}$.
Measurements of all of these quantities have been made in both the
$B \to \pi\pi$ and $B \to \rho\rho$
systems~\cite{Aubert:2006ap,Abe:2006cc,Aubert:2006af,Abe:2007ez}.
The latter is a vector-vector final state; contributions from different
helicity amplitudes could, in principle, complicate matters.
However, measurements show that the decay is almost completely longitudinally polarized,
and the analysis is performed on that component only.
Note that when and if measurements of $S^{00}$ become available,
they can be included in the analysis.

The situation for $B \to \rho\pi$ is more complicated.
Since the Bose-Einstein statistics argument no longer applies,
five amplitudes can contribute to the decays and the triangle becomes an isospin pentagon.
There is, however, alternative approach
that exploits the interference of the $\rho^+\pi^-$, $\rho^-\pi^+$ and
$\rho^0\pi^0$ amplitudes in the
$B^0 \to \pi^+\pi^-\pi^0$ Dalitz plot~\cite{Snyder:1993mx}.
Interference between resonances on the Dalitz plot
provides additional information, so that
isospin is only needed to relate the penguin contributions.
Results for this analysis are available from both
\babar~\cite{Aubert:2006fg} and
\belle~\cite{Abe:2006yg,Kusaka:2007dv}.

By combining all the available analyses, $\alpha$ can determined quite precisely.
Considering the solution which is compatible with the Standard Model value,
the current data give $\alpha = (92 \pm 7) ^{\circ}$~\cite{Bona:2005vz}.
Several extrapolations into the multi ${\rm ab}^{-1}$ regime
exist~\cite{ref:p5,superkekb,superpep,Bevan:2007ds},
showing that the precision on $\alpha$ could reach a few degrees.
To improve the precision to or below the degree level,
one must understand the size of the genuine isospin breaking effects
and the effect of a penguin amplitude in $A^{+0}$,
since electroweak penguins (EWP), which are usually neglected, contribute to $\Delta I = 3/2$ transitions.
The consequence is that the middle relation of Eq.~\ref{eq:su2analysis}
receives an additional contribution containing two additional parameters.
In this case the SU(2) triangular relations still hold true,
but the condition $|A^{+0}| = |\bar{A}^{+0}|$ could be invalid.
However it has been shown that the effect of the dominant EWP operators
can be included in the isospin analysis~\cite{Gronau:1998fn}.
Their effect has recently been estimated to produce a shift in the
extracted value of $\alpha$ equal to $(1.5 \pm 0.3 \pm 0.3)^\circ$,
where the first error is experimental and the second comes from
neglected subdominant EWP operators~\cite{Bona:2006ah,Zupan:2007fq}.
Note that the uncertainty on $\alpha$ induced by the
EWP correction to the isospin analysis can be reduced at \superb.


The second important ingredient in the isospin analysis is that
the effective Hamiltonian does not mediate $\Delta I = 5/2$ transitions.
Nevertheless $\Delta I = 5/2$ amplitudes can be generated by genuine
isospin-breaking effects such as the $u$-$d$ quark mass difference,
$\pi^0$-$\eta$-$\eta^\prime$ mixing,
or by electromagnetic interactions~\cite{Gardner:2005pq}.
All these effects break the triangular relation, causing
the number of free parameters to exceed the number of observables.
Some of these contributions have been estimated and found to induce an
uncertainty on $\alpha$ of about $1^\circ$~\cite{Zupan:2007fq}.
The presence of a $\Delta I = 5/2$ amplitude can be tested by measuring
the $\CP$ asymmetry in $B^+\to\pi^+\pi^0$
(which can also possibly be generated by subdominant EWP operators)
and experimentally testing the triangular relation~\cite{Botella:2006zi}.


There is an additional complication in the $B \to \rho\rho$ system,
where the non-negligible width of the $\rho$ allows $I = 1$ final states  to contribute~\cite{Falk:2003uq}.
The effect of $I=1$ amplitudes can be tested by measuring $\alpha$ as
a function of $\pi^\pm\pi^0$ mass.
Estimates of the size of the isospin breaking effects are
generally around $1$--$2^\circ$.


We note that once again the strength of the \superb\ program is that
multiple approaches are possible.
Since the size of $|P/T|$ is quite different in the
$B \to \pi\pi$, $B \to \rho\pi$ and $B \to \rho\rho$ systems,
the consistency between the results for $\alpha$
obtained in the different channels will allow to test with high statistics
the theoretical assumptions used to extract $\alpha$.
Finally, note that since all measurements of $\alpha$ require
reconstruction of modes containing neutral particles,
it is extremely unlikely that experiments in a hadronic environment
will provide competitive measurements.

\mysubsection{Measurement of the CKM Elements $|V_{ub}|$ and $|V_{cb}|$}
\label{ss:sides}

The determination of the magnitude of the CKM matrix elements $V_{cb}$ and $V_{ub}$
from inclusive and exclusive semileptonic $B$ meson decays requires knowledge of
the absolute branching fractions as well as an absolute prediction of
QCD corrections that relate quark level processes to meson decays.
Improvements in the measured branching fractions can only be realized
by improved understanding of the impact of the detector on the detection of both
signal and backgrounds. Since all semileptonic decays involve an undetected neutrino,
improvements to the detector acceptance and detection efficiency for
charged and neutral particles are very important.
The reduced beam-energy asymmetry at \superb\ leads to an increase in solid-angle
coverage. Detailed studies of
detection efficiencies and misidentification rates with large control samples
selected from data, which will become available at \superb,
will be critical to achieving simulations accurate 
to better than one percent.
This includes production rates of kaons in $B$ meson
decays and continuum events, as well as the interactions of neutral kaons
in the detector.
For the study of Cabibbo-suppressed $B\to X_u \ell \nu$ decays,
improved understanding of the background from
exclusive $B\to X_c \ell \nu$ decays is becoming critical.
Likewise, background contributions from $B\to X_u \ell \nu$ decays
involving higher-mass mesons, or mesons with strangeness, limit  
the current precision.

Theoretical understanding of QCD effects
for both inclusive and exclusive $B$ meson decays is expected to improve with time,
since experimental errors are in many cases smaller than uncertainties in
form factor normalization and Operator Product Expansion (OPE) 
and Heavy Quark Symmetry (HQS) corrections to the inclusive
decay rates. The impact of effects such as weak annihilation must also be
assessed, both experimentally and theoretically.

\mysubsubsection{Perspectives on Exclusive Semileptonic Measurements}

Measurements of exclusive decays such as $B \to \pi \ell \nu$
can yield a very precise
determination of $|V_{ub}|$, provided that the theoretical precision in
the determination of the form factor has a comparable accuracy
(see Section~\ref{sec:ph_lattice}).
From the experimental point of view, we should provide precise
measurements of $\delta \Gamma / \delta q^2$,
where $q^2$ is the invariant mass-squared of the lepton-neutrino pair.
Different tagging techniques have been developed,
using full or partial reconstruction of one of the two $B$ mesons
in the event to reduce background and 
improve the determination of kinematic quantities,
but untagged events are also employed, 
and at present still provide the most precise results. 
The study of exclusive charmless decays, making use of
tagging techniques, will become more powerful with significantly
larger data samples.
The various methods have quite different efficiencies
and background contamination.
Analyses using full-reconstruction techniques select about
130 events per ${\rm ab}^{-1}$
while those using untagged events select about 22,000 events per ${\rm ab}^{-1}$.
The signal-to-background ratio is about 20 times higher for
the tagged approach than for the untagged approach.
Recent \babar\ studies show that
full-reconstruction technique yields an error on the branching fraction of
$29\%$, dominated by the $25\%$ statistical uncertainty.
In the untagged analysis the precision has reached $8\%$ and the
systematic and statistical uncertainties are of comparable size.
In this case the dominant systematic uncertainties come from neutrino reconstruction
(due to missing particles, unidentified $\KL$ mesons, \etc),
non-$\BzBzb$ background,
and from contributions of other $B \to X_u \ell \nu$ decays.
These systematic uncertainties could be reduced by
improving the reconstruction of charged and neutral particles, and by performing more
measurements of resonant and non-resonant $B \to X_u \ell \nu$ decays.
At \superb, the branching fraction can be measured with a precision of
a few percent.
The large data samples will also allow a similarly precise determination of the
$q^2$ dependence of the $B \to \pi$ form factor through measurements of
partial branching fractions in many different $q^2$ intervals.
Similar analyses could be performed by reconstructing
$\rho$, $\omega$, $\eta$, $\eta'$, and higher-mass mesons in the final state.
Final states with a vector meson, $\rho$ or $\omega$, will especially
benefit from the large \superb\ data samples. Here three form factors
are used to describe the decay kinematics, which requires a detailed
study of angular distributions.
The contribution to the uncertainty on $|V_{ub}|$
from the partial branching fraction measurements could reach a level
of $1$--$2\%$.
The crucial point is that the form factors must be determined
at the same level of precision (see Appendix \ref{sec:ph_lattice}).
A total error of $3$--$4\%$ on $|V_{ub}|$ from exclusive analyses
appears possible with \superb\ data samples.

Estimates for improvements on $|V_{cb}|$ from measurements of
$\BR(B \to D^{(*)} \ell \nu)$ are given in Table~\ref{tab:inputs2}.
The experimental and theoretical uncertainties are currently
of comparable size ($\sim$ 3\%). It will be difficult to improve
the experimental uncertainties, which are mainly due to detector
effects, such as the reconstruction of the low-momentum pion from
the decay of the $D^*$ meson, below the $1$--$2\%$ level. A simultaneous
measurement of $B \to D \ell \nu$ and $B \to D^* \ell \nu$ decays
would be useful to better control the background due to feed-down
from $D^*$ decays for the analysis of the $B \to D \ell \nu$ mode.
The form-factor calculations for these decays could reach a
precision of better than 1\% (see Appendix~\ref{sec:ph_lattice}).
Thus a total error on $|V_{cb}|$ of $1$--$2\%$ from exclusive analyses
can be expected at \superb\ . In addition, a much improved understanding
of semileptonic decays with higher-mass ($D^{**}$) 
and non-resonant ($D^{(*)}\pi$) states, 
which are curently not well-understood, will become feasible.

\mysubsubsection{Perspectives on Inclusive Semileptonic Measurements}

The total decay rate and lepton spectra for inclusive semileptonic $B$~meson decays to charmed final states
have been measured with great accuracy at the current generation
of $B$ Factories,
allowing the determination of $|V_{cb}|$ with a precision of $1.5\%$.
Theoretical uncertainties already dominate this error.
These are mostly uncalculated perturbative corrections,
of ${\cal O}(\alpha_s^2)$ and ${\cal O}(\alpha_s \Lambda/m_b)$,
to the Wilson coefficients of the OPE used to compute the rate.
Though difficult, the required calculations are feasible
with present techniques, and are likely to be
available by the start of \superb,
where $|V_{cb}|$ can be expected to be determined inclusively
with a total error below 1$\%$.
Theoretical rate calculations for these decays
also include non-perturbative OPE parameters, which are obtained from a fit
to the moments of the electron energy or hadronic mass spectra in
$B \to X_c \ell \nu$ decays, or from the photon-energy spectrum in $B \to X_s \gamma$ decays.
With this method, a determination of $m_b$ and $m_c$ with a precision of less than 30 and 50 MeV,
respectively, and of $\mu_\pi^2$ and $\mu_G^2$ with better than 10$\%$ precision
should be possible. These measurements allow crosschecks of
lattice calculations of, for instance, quark masses, and are important
inputs for the determination of $|V_{ub}|$, as outlined below, or for studies
of rare $B$~meson decays.

The situation is more complicated in the case of charmless inclusive semileptonic
decays which play a crucial role in the determination of $|V_{ub}|$.
Experimentally, the separation of the $B\to\X_u\ell\nu$ signal from the
overwhelming $B\to\X_c\ell\nu$ background, the most challenging task,
requires harsh kinematical cuts.
Even with current statistics, the theoretical error has
begun to dominate here as well.
The kinematical cuts
needed to suppress the $B\to\X_c\ell\nu$ background make the
measurements particularly sensitive
to non-perturbative effects, and spoil the convergence of the OPE.
A resummation of the non-perturbative contributions into a so-called
shape function becomes necessary.
The non-perturbative dynamics relevant to these decays
cannot be simulated on the lattice. 
For future inclusive determinations of $|V_{ub}|$, it is important
to minimize the dependence on the shape function by avoiding overly stringent
kinematical cuts at the expense of higher backgrounds.
Here the large data samples available at \superb\ will help, because they allow
an improved determination of the backgrounds.
In the case of the lepton-endpoint analysis, the lepton momenta could
be extended well below the currently used minimum momentum of 2~GeV, which will
significantly reduce the shape function and resummation effects, or even
make them irrelevant.
The most important remaining source of theoretical error on $|V_{ub}|$
from inclusive measurements would then be the mass of the bottom quark.
An inclusive determination of $|V_{ub}|$ with a precision of about $2\%$
might then be possible.
New measurements, such as precise measurements of kinematic spectra in
$B \to X_u \ell \nu$ decays, will also become feasible at \superb.
These will yield significant information on the leading
and subleading shape functions, as well as on contributions from weak annihilation,
and will thus reduce the theoretical uncertainties and test the
validity of the theoretical framework.
Some of these studies are possible at the current $B$ Factories,
but they will become far more stringent at \superb.
Recent efforts aim at the elimination of the dependence on the
shape function (to first order), using theoretical calculations
that relate the differential rate for charmless semileptonic decays to the
photon-energy spectrum measured in $B\to\X_s\gamma$ decays through weighting
functions. These studies will also benefit from much larger samples of both
$B\to\X_u\ell\nu$ and $B\to\X_s\gamma$ decays.

\mysubsubsection{Measurement of $\BR(B \to D^{(*)} \tau \nu$)}

The decays $B \to D \tau \nu$ and $B \to D^{*} \tau \nu$ 
are sensitive to New Physics through virtual exchange of charged
Higgs bosons.
The branching fractions are expected to be of the order of $8 \times 10^{-3}$
in the Standard Model.
Because of the presence of at least two neutrinos in the final state,
the reconstruction of these modes requires the reconstruction of
the other $B$ meson in the event, and hence requires
a larger data sample with respect to that used to measure
$\B(B \to D^{(*)} \mu \nu)$ and $\B(B \to D^{(*)} e \nu)$.
Simulations show that by combining the hadronic
and leptonic $\tau$ decays in final states containing a $D^0$ meson, a relative precision of $\sim 10\%$
can be reached with $2 \ {\rm ab}^{-1}$.
In the hadronic $\tau$ analysis, the most important backgrounds
are the decays $B^+ \to D^{*-} \ell^+ \nu \pi^+$ with a
missing $\ell^+$ and a soft pion from $D^{*+}$ and
$B^+ \to D^{*0} \ell^+ \nu$ with misidentification of $\ell^+$ as $\pi^+$
and missing slow pion.
There is cross-feed between these decays, and the channels with a
$\tau$ decaying into leptons can be considered as backgrounds.
The background situation in the leptonic $\tau$ channel shows
similar patterns.
The systematics uncertainties attached to this measurement
are of quasi-statistical origin:
the efficiency and the purity of the $B$ recoil sample,
the particle identification and the reconstruction efficiency for slow pions.
For this reason, the precision of this measurement can be improved
to $2\%$ using \superb\ statistics.
It is clear that improvement of detector hermiticity and PID
can improve the sensitivity of the analysis.
Final states containing a $D^-$ have also been studied.
In this case, the efficiency is lower;
a precision of about $30\%$ ($6\%$) can be reached
with $2 \ {\rm ab}^{-1}$ (or the statistics available at \superb).
To fully exploit the experimental precision in this channel
the form factors must be known at the percent level.
It should be stressed that, while $\BR(B \to D \mu \nu)$ and $\BR(B \to D e \nu)$
depend on a single form factor, $\BR(B \to D \tau \nu)$
is also sensitive to a second form factor, since
the $\tau$ lepton mass is not negligible compared to the $B$ meson mass.

\afterpage{\clearpage}
\mysubsection{Rare Decays}
\label{sec:U4s_rare}


Rare $B$ decays provide a powerful window into New Physics.
Decays that are highly suppressed within the Standard Model
may not suffer the same constraints when New Physics is introduced,
and hence significant effects can be observed.
Due to the clean environment and excellent particle identification
capabilities of \superb,
a large number of rare decay channels can be studied,
with rates covering several orders of magnitude,
down to as low as ${\cal O}(10^{-10})$ for the cleanest channels.
We provide herein a brief summary of the reach
for some of the most interesting channels.
A key strength of the \superb\ program
is that the abundance of New Physics-sensitive measurements
allows the diagnosis of the origin of the New Physics.


\mysubsubsection{Leptonic Decays : $\BR(B^+ \rightarrow \ell^+ \nu_\ell (\gamma))$ and $\BR(B^0 \rightarrow \ell^+\ell^-)$ }
\label{sss:leptonic}

\begin{figure}[t]
  \begin{center}
    \begin{tabular}{cc}
      \resizebox{0.45\textwidth}{0.27\textwidth}{\includegraphics{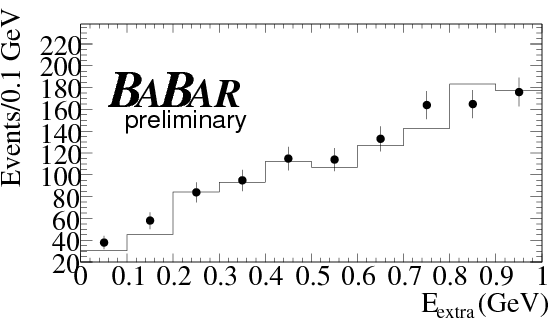}}
      &
      \resizebox{0.45\textwidth}{0.33\textwidth}{\includegraphics{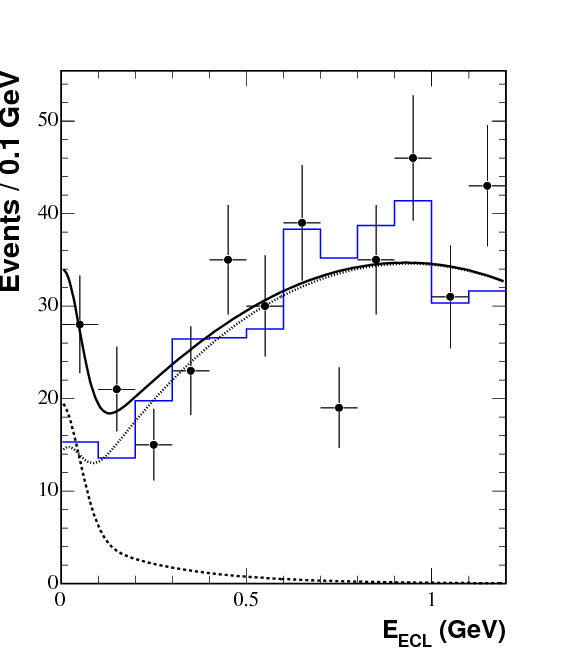}}
    \end{tabular}
    \caption{
      Signals for $B^+ \to \tau^+\nu_\tau$ from
      (left) \babar~\cite{Aubert:2006fk} and
      (right) \belle~\cite{Ikado:2006un}.
      Signal candidates are shown as data points,
      the expectations for background, obtained from Monte Carlo simulation,
      are shown as histograms.
      For the figure from \belle,
      the fit result is shown as a solid curve,
      with signal and background components
      shown as dashed and dotted curves respectively.
      \belle\ uses the variable $E_{\rm ECL}$,
      which is essentially equivalent to $E_{\rm extra}$ used by \babar
      and defined in the text.    }
    \label{fig:taunu}
  \end{center}
\end{figure}

Leptonic decay processes are described by annihilation diagrams.
The rates of leptonic decays of the $B^+$ meson are therefore
proportional to $f_B^2 \left| V_{ub} \right|^2$,
where $f_B$ is the same pseudoscalar constant that enters
the determination of $\Delta m_d$ (assuming isospin symmetry).
Leptonic decay rates are helicity suppressed;
the branching fraction is given by:
\begin{eqnarray}
  \BR(B^+ \to \ell^+\nu) & = &
  \frac{G_F^2}{8\pi} f_{B}^2 \left| V_{ub} \right|^2 \tau_{B^{+}}
  M_{B^+} m_{\ell}^2 \left( 1- \frac{m_{\ell}^2}{M_{B^+}^2} \right)^2 \, ,
\end{eqnarray}
where $G_F$ is the Fermi constant, and
$M_{B^+}$ and $m_{\ell}$ are the masses of the $B^+$ meson and the lepton $\ell$ respectively.
The branching fractions are expected to be about
         $10^{-4}$  for $\BR(B^+ \to \tau^+ \nu_\tau)$,
$5 \times 10^{-7}$  for $\BR(B^+ \to \mu^+  \nu_\mu)$ and
$         10^{-11}$ for $\BR(B^+ \to e^+    \nu_e)$.
Evidence for the $\tau$ mode has recently been reported~\cite{Ikado:2006un}
(see Fig.~\ref{fig:taunu}).
The world average for the branching ratio is
$\BR(B^+ \to \tau^+ \nu_\tau) = (1.3 \pm 0.5)
\times 10^{-4}$~\cite{hfag,Aubert:2006fk}.
The upper limits on the muonic decay are approaching the
Standard Model expectation~\cite{Satoyama:2006xn},
while those for the highly suppressed $B^+ \to e^+\nu_e$ decay are still far away from the Standard Model value.

The measurements of leptonic decay branching fractions can be interpreted in various ways.
If the value of $f_B$ is taken from lattice QCD calculations,
then a determination of $\left| V_{ub} \right|$ can be obtained from
the branching fraction,
allowing a consistency check with other approaches used to measure
this quantity, as described in Section~\ref{ss:sides}.
Alternatively, one can take the value of $\left| V_{ub} \right|$
and use the leptonic decay rate
to check the consistency of the lattice calculations.
Finally, if one takes known values of $f_B$ and $\left| V_{ub} \right|$,
the Standard Model expectation can be compared to the branching fraction measurement.
This is particularly interesting, since the leptonic decay processes
are sensitive to New Physics,
in particular to charged Higgs exchange in a scenario with large $\tan \beta$.
For example, in the two Higgs doublet model (2HDM),
the effect of the charged Higgs is that the branching fraction is
scaled by a factor $(1 - \tan^2 \beta ( M_B^2 / M_{H^+}^2 ))^2$~\cite{Hou:1992sy},
where $M_{H^+}$ is the mass of the charged Higgs boson,
and $\tan \beta$ is the ratio of Higgs expectation values, and
is not related to the UT angle.
Considering the leading corrections in the large $\tan\beta$ limit,
the charged Higgs contribution in the MSSM is rather similar,
the scaling factor becoming
$(1 - \tan^2 \beta( M_B^2 / M_{H^+}^2 ) / ( 1 + \epsilon_0\tan\beta ))^2$
where $\epsilon_0 \sim 10^{-2}$~\cite{Isidori:2007zm}.

We discuss the potential of \superb\ to measure these
leptonic modes in turn.
While we focus our attention on the precision with which the
branching ratios can be measured,
it is worth mentioning that any direct $\CP$ violation
in these channels would be an unmistakable signal of New Physics.

\vspace{1ex}
\noindent
$\BR(B^+ \to \tau^+ \nu_\tau)$ 
\par\noindent
Since the decay of the $\tau$ lepton necessarily involves at least one
neutrino, there are multiple sources of missing energy in the decay,
rendering conventional reconstruction techniques impossible.
However, the signal can be isolated by taking advantage of
$B^+B^-$ production at the $\FourS$ resonance.
The analysis technique proceeds by reconstructing either
exclusively or partially one $B$ meson in the event (the {\it tag}),
and then compares the remainder of the event with the signature
for the signal decay.
The distribution of the variable $E_{\rm extra}$, defined as
$E_{\rm extra} = E_{\rm total} - \sum E_{\rm tag} - \sum E_{\rm signal}$
peaks near zero for the signal, while
the backgrounds tend to take higher values,
as shown in Fig.~\ref{fig:taunu}.
This analysis is clearly highly sensitive to quantities that
depend on neutral particle detection.
Therefore, a detailed simulation of the calorimeter response and
knowledge of the beam backgrounds are important
for a realistic estimate of the sensitivity at \superb.
It is important to reduce and understand background sources,
such as unreconstructed tracks and undetected $\KL$ mesons.
The current analyses assign total systematic errors of more than $10\%$.
With $75 \ {\rm ab}^{-1}$,
the statistical error will be of the order of $3$--$4\%$, so
systematic effects will have to be much better controlled to match this
statistical precision.

Many of the systematic effects can, in fact, be reduced by
careful studies of control samples (such as $B \to D^{(*)}\ell\nu_\ell$),
and further reduction of the error may be possible
if the detector performance can be improved.
The reduced energy asymmetry of \superb\ will improve the detector solid angle coverage, and hence the hermeticity,
leading to better control of backgrounds.
Studies also show that the contemplated addition of more iron in the flux return
leads to an improved $\KL$ meson detection efficiency
(see Section~\ref{sec:det:IFR}),
which will directly benefit this analysis.
We conclude, therefore, that the $\BR(B^+ \to \tau^+ \nu_\tau)$ branching fraction can be measured with a total error of $\sim 4\%$ with $75 \ {\rm ab}^{-1}$.


\noindent
$\BR(B^+ \to \mu^+ \nu_\mu)$ 
\par\noindent
In contrast to the tauonic decay,
the muonic decay mode has a very distinctive signature:
a high transverse momentum muon and
a missing energy vector that balances the momentum of the lepton.
Kinematic constraints on the companion $B$ in the event
allow the mono-energetic final state lepton to
be reconstructed with little or no background.
Due to this clean signature,
we expect a statistical error of about
$5\%$ on the measured branching fraction at the Standard Model value.
Since backgrounds are small, it should be possible to
control systematic uncertainties to a similar level.

\noindent
$\BR(B^+ \to e^+ \nu_e)$ 
\par\noindent
The case when the lepton in the decay is an electron is as clean
as the muonic mode, but due to the small electron mass,
the helicity suppression is severe,
and the Standard Model rate is below the sensitivity of \superb.
The expected upper limit would be at the level of ${\cal O}(10^{-9})$.

\vspace{1ex}

\noindent
$\BR(B^+ \rightarrow \ell^+ \nu_\ell \gamma)$ and $\BR(B^0 \rightarrow \ell^+\ell^-)$
\par\noindent
The radiative leptonic decays do not suffer the same degree of helicity suppression as the
purely leptonic decays.
\superb\ has excellent sensitivity for,
\eg, $B^+ \to \ell^+ \nu_\ell \gamma$~\cite{Abe:2004dm}.
The theoretical branching ratio for this mode is model-dependent~\cite{Korchemsky:1999qb}, but, if a value for $f_B$ is taken from other measurements, this mode can be used to determine $\lambda_B$,  the first
inverse moment of the $B$ light-cone distribution amplitude, a quantity that enters into calculations of the branching fraction of hadronic $B$ decays such as $B\to\pi\pi$.

Finally, the neutrinoless leptonic decays
$B^0 \to e^+e^-$, $B^0 \to \mu^+\mu^-$ and $B^0 \to \tau^+\tau^-$
can also be studied at \superb,
together with their lepton flavour-violating
counterparts~\cite{Aubert:2004gm,Aubert:92qw}
(see the discussion of leptonic $B_s$ decays in Section~\ref{sss:bsmumu}).
With $75 \ {\rm ab}^{-1}$,
the sensitivity would reach the $10^{-10}$ level
for the $e^+e^-$ and $\mu^+\mu^-$ final states,
which is close to the Standard Model expectation for the muon mode.
The very clean experimental signatures
of these channels make them well-suited for experiments
in a hadronic environment,
in stark contrast to channels involving $\tau$ leptons, or neutrinos, or both.
The $B^0 \to \tau^+\tau^-$ decay can only be studied at a Super $B$ Factory,
although the sensitivity will still be far above the Standard Model expectation.
For New Physics searches, lepton flavour-violating decays such as $B^0 \to \ell^+\tau^-$
may, however, reach an interesting level of precision.

\mysubsubsection{Radiative Decays : $b \to s \gamma$ and $b \to d\gamma$}
\label{sss:radiative}

The radiative FCNC decays
$b \to s \gamma$ and $b \to d\gamma$
are very sensitive probes of New Physics.
Since these decays occur only at loop level, and
furthermore are CKM-suppressed in the Standard Model,
the rates for these transitions alone provide
severe constraints for New Physics model builders.
Indeed, early measurements of the rate of the $b \to s\gamma$
decay~\cite{Alam:1994aw}
have been very highly cited, due to their phenomenological impact.
To fully take advantage of these decays, however,
several other observables,
such as $\CP$ asymmetries and the polarization of the photon, must be measured.
These measurements, which can be performed at \superb\,
also provide clean tests of the Standard Model,
It is important to note that \superb\ can make these measurements in the theoretically cleaner
inclusive modes, and is not restricted only to exclusive channels.

\mysubsubsubsection{$b \to s\gamma$ : Exclusive}

The primary focus of exclusive measurements
is on $\CP$ asymmetries,
which have comparatively small theoretical uncertainties,
in contrast to the rates.
Indeed, studies of direct $\CP$ asymmetries in radiative penguin decays
are among the golden modes for \superb.
Direct $\CP$ violation in these decays is expected to be $\simeq 0.5\%$
in the Standard Model, but could be an order of magnitude larger
if there are New Physics contributions in the penguin loops.
Experimentally, the most accessible channel is $B^0 \to K^{*0}\gamma$.
One can also make an average with the $B^+ \to K^{*+}\gamma$ decay,
and search for isospin violation,
which could be caused by New Physics, in the rates and asymmetries.
The current experimental average is
$A_{\CP}( B \to K^{*}\gamma ) = -0.010 \pm 0.028$~\cite{Aubert:2004te,Nakao:2004th,hfag}.

With $75 \ {\rm ab}^{-1}$, the limiting factor in this measurement
will be systematic uncertainty due to asymmetries in the detector
response to positive and negative kaons.
Such effects are under study at the $B$ Factories,
with the residual errors already below the $1\%$ level.
Further reduction of the uncertainty should be possible,
but will require highly detailed studies of both Monte Carlo simulation
and data control samples.
We estimate the ultimate precision to be $\sim 0.4\%$.

\mysubsubsubsection{$b \to d\gamma$ : Exclusive}

The ratio of rates of $b \to d \gamma$ and $b \to s \gamma$ decays
can give a precise determination of $\left| V_{td}/V_{ts} \right|$,
complementing the information obtained from
the ratio of oscillation frequencies $\Delta m_d/\Delta m_s$.
In the ratio of exclusive decay branching fractions
$\BR(B \to \rho\gamma) / \BR(B \to K^*\gamma)$ many theoretical uncertainties cancel, allowing
a measurement of the same combination of CKM matrix elements.

The theoretically cleanest case is for the neutral modes,
since in the charged modes there is the possibility of
non-negligible weak annihilation contributions~\cite{Ali:1995uy,Beyer:2001zn}.
We therefore concentrate on the neutral modes in the following.

The ratio of decay rates can be written as~\cite{Ali:2001ez,Ball:2006nr}
\begin{equation}
  R = \frac{\BR(B^0 \to \rho^0\gamma)}{\BR(B^0 \to K^{*0}\gamma)} =
  \frac{1}{2}
  \left(
    \frac{1 - m_\rho^2/M_B^2}{1 - m_{K^*}^2/M_B^2}
  \right)^3
  \left|
    \frac{V_{td}}{V_{ts}}
  \right|^2
  \xi^2 [1 + \Delta R]
\label{eq:rhogKsg}
\end{equation}

which contains a factor due to isospin ($1/2$),
a kinematic factor,
the ratio of the CKM matrix elements squared,
the ratio of form factors squared ($\xi^2$)
and a term containing additional hadronic effects
($\Delta R$ contains non-factorizable SU(3)-breaking effects
and also accounts for annihilation contributions).

\begin{figure}[t]
  \begin{center}
    \begin{tabular}{cc}
      \resizebox{0.45\textwidth}{0.35\textwidth}{\includegraphics{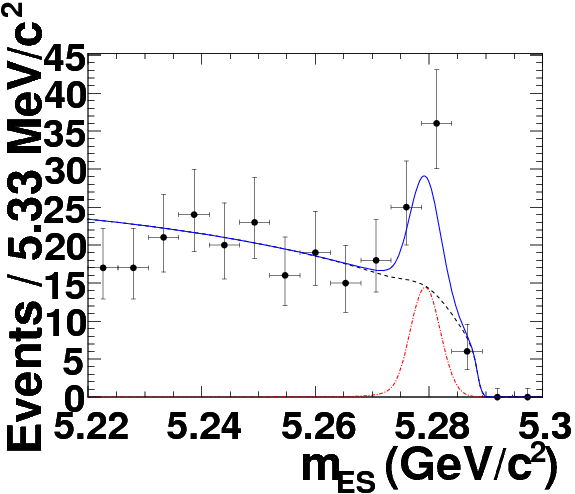}}
      &
      \resizebox{0.45\textwidth}{!}{\includegraphics{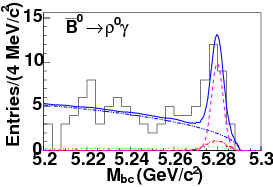}}
    \end{tabular}
    \caption{
      Signals for $B^0 \to \rho^0 \gamma$ from
      (left) \babar~\cite{Aubert:2006pu} and (right) \belle~\cite{Abe:2005rj}.
      The variables $m_{\rm ES}$ (\babar) and $m_{\rm bc}$ (\belle)
      are essentially equivalent.
    }
    \label{fig:btorhogamma}
  \end{center}
\end{figure}

The current experimental world averages are
$\BR(B^0 \to \rho^0 \gamma) = ( 0.91 \pm 0.19) \times 10^{-6}$~\cite{Abe:2005rj,Aubert:2006pu,hfag} and
$\BR(B^0 \to K^{*0} \gamma) = (40.1  \pm 2.0 ) \times 10^{-6}$~\cite{Aubert:2004te,Nakao:2004th,hfag}.
The signals for $B^0 \to \rho^0 \gamma$ are shown in Fig.~\ref{fig:btorhogamma}.
The limiting factor is currently the statistical precision on
$\BR(B^0 \to \rho^0 \gamma)$.
With \superb\ statistics of $75 \ {\rm ab}^{-1}$ this can be reduced
to a level of about $2\%$.
Systematic uncertainties will not be negligible at that scale.
In particular, good control of the particle identification performance
will be necessary to understand possible feed-across from $K^*\gamma$
with a misidentified kaon.
Nonetheless, even without any improvements in analysis techniques and
detector performances, the experimental precision should reach about $3\%$.
It is then crucial to control the form factors and the  SU(3)-breaking
terms at a similar level of accuracy to extract the
$V_{td}/V_{ts}$ ratio with ${\cal O}(1\%)$ uncertainty (see discussion in Section
\ref{sec:lattice}).


\superb\ will also be able to measure
direct $\CP$ asymmetries in $b \to d \gamma$ processes,
which have not been seen at the current $B$ Factories, to $\sim 10\%$ precision,
which is the Standard Model expectation for this asymmetry.

\mysubsubsubsection{$b \to s\gamma$ : Inclusive}

Measurements of the inclusive branching fraction of $b \to s\gamma$
provide powerful, theoretically clean constraints on New Physics.
Theoretical calculations of the Standard Model prediction
have recently been advanced to next-to-next-to-leading order (NNLO),
giving the value
${\cal B}(B \to X_s \gamma$, $E_\gamma \gsim 1.6 \ {\rm GeV}) =
(3.15 \pm 0.23) \times 10^{-4}$~\cite{Misiak:2006zs}
(see also~\cite{Misiak:2006ab,Becher:2006pu,Andersen:2006hr}),
which is in good agreement with the latest
experimental determination,
$(3.55 \pm 0.24 \pm 0.10 \pm 0.03) \times 10^{-4}$~\cite{Aubert:2005cu,
  Koppenburg:2004fz,Chen:2001fj,hfag}.
Although further reduction of the theoretical error will be
difficult,
improved measurements of the total rate with larger statistics
will greatly improve our knowledge of the photon energy spectrum,
and will allow the minimum energy requirement to be moved
to smaller values.
Consequently, the theoretical error associated with the extrapolation
required to obtain the total branching fraction will be reduced.
This approach also provides the most accurate determination
of $\left| V_{ts} \right|$.

\begin{figure}[t]
  \begin{center}
    \begin{tabular}{cc}
      \resizebox{0.45\textwidth}{0.35\textwidth}{\includegraphics{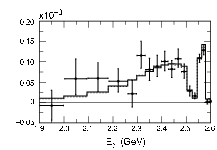}}
      &
      \resizebox{0.45\textwidth}{0.40\textwidth}{\includegraphics{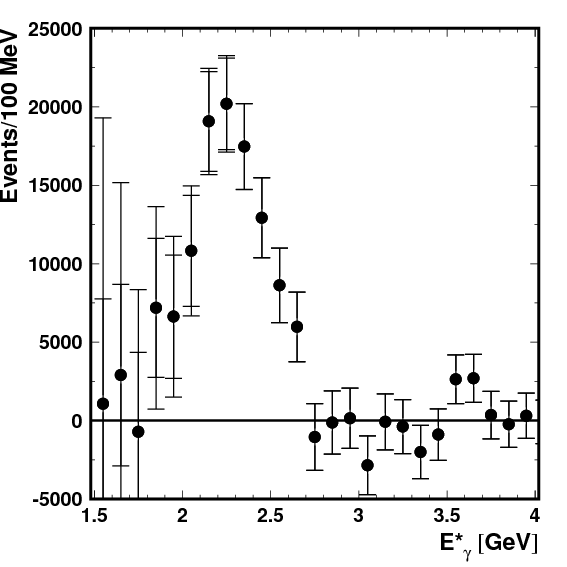}}
    \end{tabular}
    \caption{
      Photon energy spectra in inclusive $b \to s\gamma$ analyses from
      (left) \babar~\cite{Aubert:2005cu},
      using a sum of exclusive channels, and
      (right) \belle~\cite{Koppenburg:2004fz},
      using a fully inclusive analysis.
      In the \babar\ plot the data points are compared to
      theoretical predictions using the shape function~\cite{Neubert:1993um,Kagan:1998ym,Bosch:2004cb,Neubert:2004cu,Neubert:2004sp}
      (solid histogram) and the kinetic~\cite{Bigi:2002qq,Benson:2004sg,Buchmuller:2005zv}
      (dashed histogram) schemes.
      In the \belle\ plot the two sets of error bars show
      the statistical and total uncertainties.
    }
    \label{fig:bsgamma}
  \end{center}
\end{figure}

As in the exclusive case, the
inclusive $\CP$ asymmetry has less theoretical uncertainty
than the rate~\cite{Hurth:2003dk}.
The current world average is
$A_{\CP}(B \to X_s \gamma) = 0.004 \pm 0.037$~\cite{hfag},
using results in which both \babar~\cite{Aubert:2004hq}
and \belle~\cite{Nishida:2003yw}
reconstruct the $X_s$ system as a sum of exclusive final states
and correct for the missing fraction.
Such analyses suffer from the same source of systematic error
in particle identification as the exclusive modes.
Therefore, with $75 \ {\rm ab}^{-1}$,
the level of precision is likely to be limited at the same level
as the exclusive modes, about $0.004$.

\mysubsubsubsection{$b \to d\gamma$ : Inclusive}

Inclusive studies of the $b \to d\gamma$ transition have not yet been carried out at the $B$~Factories.
The analysis, using a sum of exclusive final states, is quite challenging,
since the $b \to s\gamma$ amplitude becomes a background
that can only be reduced using particle identification.
Nonetheless, preliminary studies show that such an analysis can
be done with around $10 \ {\rm ab}^{-1}$;
the results would clearly become very interesting with the
full \superb\ statistics.

It is interesting to note that, in the Standard Model,
the partial width differences in $b \to s\gamma$ and $b \to d\gamma$
should cancel, so that $A_{\CP}(B \to X_{s+d}\gamma)$ is predicted to be zero.
This prediction is exact in
the $U$-spin limit~\cite{Soares:1991te}, and flavour-breaking effects
have been calculated to be small~\cite{Hurth:2001yb,Hurth:2001ja,Chay:2006ve},
giving a very precise null test~\cite{Gershon:2006mt}.
A fully inclusive approach can be used;
particle identification systematics do not then contribute.
The price, however, is a very large background that must be controlled.
With \superb\ statistics it will be possible to do so
using full reconstruction of the other $B$ meson in the event.
A first measurement of $A_{\CP}(B \to X_{s+d}\gamma)$ has already been
carried out by \babar~\cite{Aubert:2006gg},
suggesting that this asymmetry can also be measured to subpercent
precision at \superb.

\mysubsubsubsection{Photon polarization measurements}

Within the Standard Model,
photons emitted in radiative $b$ decay are predominantly left-handed,
while those emitted in $\bar{b}$ decay are predominantly right-handed.
Based on the leading order effective Hamiltonian,
the amplitude for the emission of wrong-helicity photons is suppressed
by a factor $\propto m_q / m_b$~\cite{Atwood:1997zr},
where $m_q = m_s$ for $b \to s\gamma$ transitions
and $m_q = m_d$ for $b \to d\gamma$ transitions.
More detailed treatments, including QCD corrections, give
a suppression as ${\cal O}(\Lambda_{\rm QCD}/m_b)$~\cite{Grinstein:2004uu}.

New Physics
can modify this suppression
without introducing any new $\CP$ violating phase.
Measurements of the photon polarization therefore provide
an approach to search for New Physics that is complementary
to those based on rates and $\CP$ asymmetries.

Several different methods of measuring the photon polarization have been
suggested.
The only approach that has been attempted to date
uses mixing-induced $\CP$ asymmetries to probe the
level of interference between $b$ and $\bar{b}$ decays~\cite{Atwood:1997zr,Atwood:2004jj}.
Even with the inclusion of the QCD corrections discussed above
(see also~\cite{Grinstein:2005nu}),
recent calculations show that the Standard Model prediction
is below $5\%$~\cite{Matsumori:2005ax,Ball:2006cv,Ball:2006eu}.
The current experimental world average is
$S(\KS\pi^0\gamma) = -0.09 \pm 0.24$~\cite{Aubert:2005bu,Ushiroda:2006fi,hfag}.
The statistical uncertainty dominates the error.

Some care is required to extrapolate this result to \superb\ luminosities.
The critical feature is that the location of the $B$ decay vertex
is reconstructed from the $\KS$ (decaying via $\KS \to \pi^+\pi^-$)
using constraints on the beam spot position.
This is only effective if the $\KS$ decay occurs inside the vertex detector,
a larger silicon detector therefore results in better precision and efficiency.
This effect, already clearly visible in a comparison of the
\babar~\cite{Aubert:2005bu} and \belle~\cite{Ushiroda:2006fi} results,
should also be taken into account when considering other modes
from which time-dependent information is extracted in a similar manner,
particularly $B^0 \to \KS \pi^0$ and $B^0 \to \KS\KS\KS$.

Since the \superb\ vertex detector is likely to have a similar radius
to the current \babar\ detector (see Section~\ref{sec:det:SVT}),
we estimate the precision that will be reached by extrapolating
the most recent \babar\ results
($S(\KS\pi^0\gamma) = -0.06 \pm 0.37$,
obtained from $\sim 220 \ {\rm fb}^{-1}$~\cite{Aubert:2005bu}).
This suggests that a precision of $0.02$ can be reached
with $75 \ {\rm ab}^{-1}$, close to the expected systematic limit,
and also at a level where theoretical uncertainties become important.
It is interesting to note that a data-driven method to control
the theoretical errors exists~\cite{Atwood:2004jj}.

It will also be possible to apply the same approach
with different final states.
Measurements of the mixing-induced $\CP$ violation parameters $S$
will be done with additional $b \to s\gamma$ exclusive channels,
such as $B^0 \to \KS \eta \gamma$~\cite{Nishida:2004fk,Aubert:2006vs} and
$B^0 \to \KS \phi \gamma$~\cite{Drutskoy:2003xh,Aubert:2006he}.
In addition, the exclusive $b \to d\gamma$ channels
$B^0 \to \rho^0\gamma$ and $B^0 \to \omega\gamma$ can be used.
Note that these channels do not rely on the $\KS$ vertexing
discussed above.
Estimates are that a precision of $\sim 0.10$ on
$S(\rho^0\gamma)$ can be achieved with $75 \ {\rm ab}^{-1}$.
This quantity is predicted to be unobservably small in the Standard Model,
since it is suppressed not only by the photon polarization,
but also by the cancellation of the weak phase in
${\Bz}_d{\Bzb}_d$
mixing with that in the $b \to d\gamma$ decay.
A measurement of a non-zero $S(\rho^0\gamma)$ would be an unmistakable
sign of New Physics.

There are also other techniques to probe photon polarization
with \superb\ luminosity.
These include approaches in which interference between
different resonances~\cite{Gronau:2002rz,Gronau:2001ng,Lee:2003ci}
or different helicity states~\cite{Atwood:2007qh}
of the hadronic recoil system provide sensitivity to the polarization,
and also those in which the photon converts to
an $e^+e^-$ pair~\cite{Grossman:2000rk,Sehgal:2004xy}.
Although the current $B$ Factory data has not yet yielded results
using these approaches, various studies indicate that these
methods will provide competitive and complementary measurements
of the photon polarization with \superb\
luminosity~\cite{superkekb,superpep,bnm}.


\mysubsubsection{Radiative Decays : $b \to s \ell \ell$ and $b \to d \ell \ell$}
\label{sss:radiative_ll}

The electroweak penguin decays $b \to s \ell \ell$ and $b \to d \ell \ell$
are also highly sensitive to New Physics.
The phenomenology is different, however, since different operators contribute to the decay amplitude.
Furthermore, different observables are available:
in addition to the rates and direct $\CP$ asymmetries,
the forward-backward asymmetry ($A_{\rm FB}$)
is known to be particularly sensitive to the presence of new particles in the loops
~\cite{Burdman:1995ks,Ali:1998sf,Lee:2006gs}.
Within the Standard Model,
$A_{\rm FB}$ is caused by electroweak effects,
and its shape as a function of the dilepton invariant mass-squared $q^2$
is predicted with quite small theoretical uncertainty,
particularly at low values of $q^2$.
Notably, $A_{\rm FB}$ is expected to have a zero at
$q^2 \approx 3 \ {\rm GeV}^2/c^4$,
while in Standard Model extensions this zero can be at a different position or even
absent~\cite{Lunghi:1999uk,Ali:1999mm,Cornell:2005kb,Buras:2003mk,Hovhannisyan:2007pb}.
As for the $b \to s\gamma$ case, theoretical calculations
for $b \to s \ell^+\ell^-$ have recently been advanced to
NNLO~\cite{Asatryan:2001zw,Ghinculov:2002pe,Ghinculov:2003qd,Asatrian:2002va,Bobeth:1999mk,Bobeth:2003at}
(see~\cite{Hurth:2007xa} for a recent review).

\begin{figure}[t]
  \begin{center}
    \begin{tabular}{cc}
      \resizebox{0.45\textwidth}{0.30\textwidth}{\includegraphics{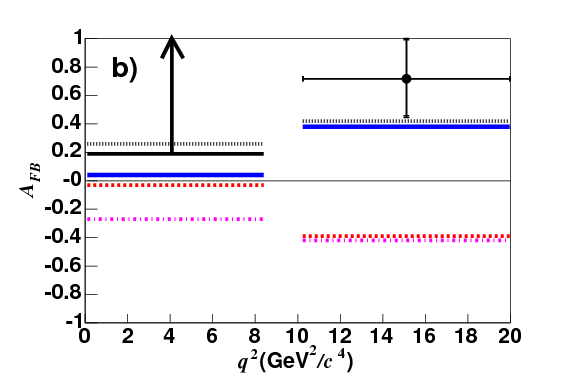}}
      &
      \resizebox{0.45\textwidth}{0.30\textwidth}{\includegraphics{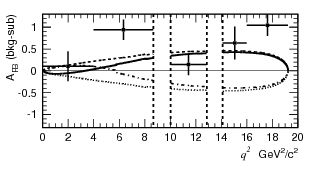}}
    \end{tabular}
    \caption{
      Measurements of $A_{\rm FB}$ from
      (left) \babar~\cite{Aubert:2006vb} and
      (right) \belle~\cite{Ishikawa:2006fh}.
      The forward-backward asymmetry is shown plotted in bins
      of the dilepton invariant mass-squared $q^2$.
      The Standard Model prediction is shown as (left) solid lines
      (right) solid curve.
      The other lines and curves show the predictions for values of
      various effective Wilson coefficients
      with the same magnitude but opposite signs to the Standard Model.
      These alternative values are not ruled out by any other measurement.
    }
    \label{fig:afb_kstarll}
  \end{center}
\end{figure}

Considering first the exclusive channels with charged lepton pairs,
the current situation is that the exclusive modes $B \to K\ell^+\ell^-$
and $B \to K^*\ell^+\ell^-$ (for $\ell = e, \mu$)
have been used to study rates,
direct $\CP$ asymmetries and the forward-backward asymmetry
(which is zero for $K\ell^+\ell^-$).
The results from \babar~\cite{Aubert:2006vb} and \belle~\cite{Ishikawa:2006fh}
on $A_{\rm FB}$ are shown in Fig.~\ref{fig:afb_kstarll}.
The current results show some enticing hints of New Physics effects,
but the precision of the $B$ Factories is not sufficient to make
the required stringent tests.
The first limit on an exclusive $b \to d \ell^+\ell^-$ mode
was recently announced by \babar~\cite{Aubert:2006ah}.
If the rates are at the Standard Model level,
detailed studies of these channels can be made at \superb.

The expected precision on observables of interest
is shown in Table~\ref{tab:inputs2}.
It should be noted that the decays $B^+ \to K^+\ell^+\ell^-$ and
$B^0 \to K^{*0}\ell^+\ell^-$ are among the modes
that can be studied at a hadron machine (with $\ell = \mu$).
Again, \superb\ can study a larger set of interesting channels,
and can also measure the parameters for the inclusive decays,
which have smaller theoretical uncertainties.
Initial studies of the exclusive $b \to s\ell^+\ell^-$ process
carried out at the $B$ Factories~\cite{Aubert:2004it,Iwasaki:2005sy},
indicate that \superb\ will be able to probe the asymmetries
down to the phenomenologically interesting percent level.

\mysubsubsection{Radiative decay : $b \to s \nu \overline{\nu}$}

It is also interesting to study channels
in which the emitted leptons are neutrinos --
modes closely related to the oft-cited
$K^+ \to \pi^+\nu\bar{\nu}$ and $K_L \to \pi^0\nu\bar{\nu}$ decays
~\cite{Littenberg:1989ix,Hagelin:1989wt,Rein:1989tr,
  Buchalla:1993wq,Buras:1994ec,Buchalla:1996fp,Buras:1997ij}.
Since there are two neutrinos in the final state,
studies of these $B$ decays are extremely challenging:
they can only be done at a \sff.
Branching fractions measurements can be done using
the technique described for $B^+ \to \tau^+\nu_\tau$ above.
Both \babar~\cite{Aubert:2004kt} and \belle~\cite{Abe:2005bq}
have used this strategy to obtain limits of
$\BR(B^+ \to K^+\nu\bar{\nu})$ at the level of $40 \times 10^{-6}$,
about an order of magnitude above the Standard Model expectation.
As is the case for other analyses using a similar reconstruction technique,
there are significant backgrounds from events with
undetected soft particles or $\KL$ mesons.
Improvements in detector hermeticity
and in the ability to better veto events containing $\KL$ mesons
can improve the sensitivity.
We therefore expect that with \superb\ statistics,
the branching fraction $\BR(B^+ \to K^+ \nu \bar{\nu})$
can be measured with about $20\%$ relative error.
A cut on the momentum of the kaon
would significantly reduce the background,
but could complicate the interpretation of the results.

Other exclusive $b \to s \nu\bar{\nu}$ channels,
such as $B \to K^* \nu\bar{\nu}$, can also be studied at \superb;
the sensitivity is such that
it should also be possible to observe $B^+ \to \pi^+ \nu\bar{\nu}$.
However, it is worth noting that these channels suffer
irreducible backgrounds from $B^+ \to \tau^+ \nu_\tau$
with hadronic decays of the $\tau$ lepton.

It will also be interesting to pursue the analysis
of $B \to {\rm invisible}$, where the observation of a signal
would be a clear sign of New Physics (see also Section~\ref{ss:lowUpsilon}).

\mysubsubsection{Charmless Hadronic $B$ Decays}
\label{sss:hadronic}

Studies of charmless hadronic $B$ decays are
in principle highly sensitive to New Physics contributions,
yet in practice many effects,
both in rates and in direct $\CP$ asymmetries,
are obscured by theoretical uncertainties.

A notable class of measurements are the sum rules
between rates and asymmetries in $B^+$ and $B_d$ decays to $K\pi$ final
states~\cite{Atwood:1997iw,Lipkin:1998ie,Gronau:2005kz,
  Matias:2001ch,Gronau:2006xu,Gronau:2006ha}.
These are sensitive to New Physics effects, in, for example, the EWP sector.
The complete set of measurements necessary for this kind
of test can only be performed at a Super $B$ Factory.
With $75 \ {\rm ab}^{-1}$, the precision of these tests
can reach $1$--$2\%$~\cite{Gershon:2006mt}.
Yet, for these, as well for other hadronic decays,
the meaningful extraction of information on fundamental Standard Model and New Physics parameters
requires a model-independent check, preferably on data,
of the approximations on which theoretical predictions are based.
To this end, general parameterizations of the hadronic
amplitudes~\cite{Gronau:1994bn,Hernandez:1994rh,Ciuchini:1997hb},
together with the full set of measurements of non-leptonic decays
performed at \superb, may prove useful.

Decays to multibody final states contain additional information
in the distribution of the final state particles.
Measurements of vector-vector final states can probe
the Standard Model predictions for the polarization~\cite{Kagan:2004uw,Kagan:2004ia,Baek:2005jk}.
Although there may, in general, still be significant hadronic uncertainties,
there are classes of observables that are relatively clean,
such as $T$-odd triple product asymmetries~\cite{Datta:2003mj,London:2000zi,London:2004ws,London:2003rk}.
Studies of the Dalitz plot distributions of
three-body charmless $B$ decays allow the relative phases
between the interfering resonant intermediate states to be determined,
allowing measurements that cannot be achieved for the equivalent
two-body decays~\cite{Ciuchini:2006kv,Ciuchini:2006st,Gronau:2006qn}.

Finally, it should be pointed out that measurements of rates
and asymmetries in charmless hadronic $B$ decays provide
an excellent testing ground for theoretical models~\cite{Li:1994iu,Keum:2000wi,Beneke:2000ry,Ciuchini:2001gv,Beneke:2003zv,Bauer:2000ew,Bauer:2000yr,Buras:1998ra}.
The comprehensive measurements that can be performed at a
Super $B$ Factory will be essential for efforts to improve our
understanding of various theoretical issues related to hadronic amplitudes (factorization, power corrections, flavour-symmetry breaking, \etc) and reduce the associated theoretical uncertainties.

\mysubsection{Other Measurements}
\label{sec:U4s_DB2}

We will not attempt to describe all the measurements
that can be performed by a \sff\ running at the $\Upsilon(4{\rm S})$,
but several important modes have not yet been mentioned.

\mysubsubsection{Semileptonic \CP Asymmetry $A_{\rm SL}$}

The semileptonic asymmetry $A_{\rm SL}$ probes $\CP$ violation
in $\BzBzb$ mixing,
giving the $B$ system equivalent of
the kaonic $\CP$ violation parameter $\epsilon_K$.
Within the Standard Model $\epsilon_B$ is expected to be
${\cal O}(10^{-3})$~\cite{Mohapatra:1998ai,Cahn:1999gx,Ciuchini:2003ww,Beneke:2003az}.
A larger value would be indicative of New Physics.
Since this measurement directly probes $\CP$ violation
in $\Delta B = 2$ transitions,
it has a large phenomenological impact.

Two different methods have been exploited to measure $A_{\rm SL}$
at the current $B$~Factories.
In the first, $\FourS \to B\Bbar$ events
with two high momentum leptons are selected,
to obtain a sample in which both $B$ mesons decayed semileptonically.
The background from charged $B$ pairs can be separated
using the vertex information of the two leptons
to evaluate the proper time difference $\Delta t$
between the decays of the two $B$ mesons.
Corrections for other background events, typically from misidentification of hadronic tracks,
or from events in which one lepton originates from a charm decay,
can be made based on Monte Carlo and data control samples.
After further corrections for possible differences in efficiencies
for reconstruction of positive and negative leptons,
the semileptonic asymmetry is obtained from the difference
in the numbers of events with two like-charged leptons:
\begin{equation}
  A_{\rm SL} =
  \frac{
    N(\ell^+\ell^+) - N(\ell^-\ell^-)
  }{
    N(\ell^+\ell^+) + N(\ell^-\ell^-) \, .
  }
\end{equation}


Measurement of $A_{\rm SL}$ allows the extraction of
the $\BzBzb$ mixing parameters $|q/p|$ and $\epsilon_{B}$:
\begin{equation}
  A_{\rm SL} = \frac{1 - |q/p|^4}{1 + |q/p|^4} =
  \frac{4\, \Re{\epsilon_{B}}}{1 + |\epsilon_{B}|^2}
\end{equation}

The second method is similar,
but requires one of the leptons to originate from a
partially or fully reconstructed semileptonic or hadronic $B$ decay.
This reduces the backgrounds, and in principle allows
finer modelling of detector and background charge asymmetries,
resulting in a reduced systematic error, at some cost to the statistics.
However, it must be noted that highly detailed studies on
large data and Monte Carlo samples are necessary in order to
control systematic effects.
Moreover, with the \superb\ data set,
a similar analysis could be performed
where both $B$ decays are partially or fully reconstructed.
This approach would carry a lower efficiency,
but also a reduced background and a potentially smaller systematic uncertainty.

The most recent experimental results give
$A_{\rm SL} = (-1.1 \pm 7.9 \pm 7.0) \times 10^{-3}$~\cite{Nakano:2005jb},
$A_{\rm SL} = ( 0.8 \pm 2.7 \pm 1.9) \times 10^{-3}$~\cite{Aubert:2006nf}
and
$A_{\rm SL} = (-6.5 \pm 3.4 \pm 2.0) \times 10^{-3}$~\cite{Aubert:2006sa}.

Clearly, controlling the systematic uncertainties will be
the most difficult issue for this measurement at \superb.
However, the capability to use different experimental approaches
provides a significant handle on systematic effects.
It may be possible to push the precision to the $10^{-3}$ or below,
potentially allowing the
Standard Model $\CP$ violation in $\BzBzb$ mixing to be seen.
This would be an impressive achievement
that would constrain many New Physics models~\cite{Laplace:2002ik}.

\mysubsubsection{Tests of Fundamental Symmetries}

The analyses described above for the measurement of
$\CP$ violation in $\BzBzb$ mixing
can be extended to include additional free parameters.
These can lead to precise constraints on the
neutral $B$ meson lifetime difference parameter $\Delta \Gamma_d$,
and can further be used to search for $CPT$ violation effects~\cite{Ren:2007dz}.
The precision achieved by the current
$B$~Factories~\cite{Aubert:2006nf,Hastings:2002ff}
could be improved by an order of magnitude
assuming a slight improvement in the systematic uncertainties.
Other fundamental tests of quantum mechanics,
such as the Bell inequality~\cite{Bellineq,Clauser:1969ny}
test of the Einstein-Podolsky-Rosen locality principle~\cite{Einstein:1935rr}
are also possible~\cite{Datta:1986ut,Go:2007ww}.

\afterpage{\clearpage}

\mysubsection{Summary of Experimental Reach}
\label{ss:u4s_summary}

As described in this section,
\superb, with an expected integrated luminosity of $75 \ {\rm ab}^{-1}$
can perform a wide range of important measurements
and dramatically improve upon the results from the
current generation of $B$~Factories.
The expected sensitivities for some of the most important measurements
are summarized in Tables~\ref{tab:inputs1} and~\ref{tab:inputs2}.
Many of these measurement cannot be made in a hadronic environment,
and are unique to \superb.

\begin{table}[!hp]
  \caption{
    The expected precision of some of the most important measurements
    that can be performed at \superb.
    For comparison, we put the reach of the $B$~Factories at $2 \ {\rm ab}^{-1}$.
    Numbers quoted as percentages are relative precisions.
    Measurements marked ($\dagger$) will be systematics limited;
    those marked ($\ast$) will be theoretically limited, with $75 \ {\rm ab}^{-1}$.
    Note that in many of these cases, there exist data driven methods of reducing the errors.
    See the text for further discussion of each measurement.
  \smallskip}
  \label{tab:inputs1}
  \begin{center}
    \begin{tabular}{l@{\hspace{-5mm}}cc}
      \hline \hline
      Observable                      & $B$~Factories ($2 \ {\rm ab}^{-1}$) & \superb\ ($75 \ {\rm ab}^{-1}$)  \\
      \hline
      $\sin(2\beta)$ ($J/\psi\,\Kz$)           &       0.018               &        0.005 ($\dagger$) \\
      $\cos(2\beta)$ ($J/\psi\,K^{*0}$)        &       0.30                &        0.05 \\
      $\sin(2\beta)$ ($Dh^0$)                  &       0.10                &        0.02              \\
      $\cos(2\beta)$ ($Dh^0$)                  &       0.20                &        0.04              \\
      $S(J/\psi\,\pi^0)$                       &       0.10                &        0.02              \\
      $S(D^+D^-)$                              &       0.20                &        0.03              \\
      $S(\phi K^0)$               &       0.13                &        0.02 ($\ast$)     \\
      $S(\eta^\prime K^0)$        &       0.05                &        0.01 ($\ast$)     \\
      $S(\KS\KS\KS)$              &       0.15                &        0.02 ($\ast$)     \\
      $S(\KS\pi^0)$               &       0.15                &        0.02 ($\ast$)     \\
      $S(\omega \KS)$             &       0.17                &        0.03 ($\ast$)     \\
      $S(f_0 \KS)$                &       0.12                &        0.02 ($\ast$)     \\
 & \\
      $\gamma$ ({\small $B \to DK$, $D \to$ $\CP$ eigenstates})  &  $\sim 15^\circ$   &   $2.5^\circ$   \\
      $\gamma$ ({\small $B \to DK$, $D \to$ suppressed states}) &  $\sim 12^\circ$   &   $2.0^\circ$   \\
      $\gamma$ ({\small $B \to DK$, $D \to$ multibody states})  &  $\sim  9^\circ$   &   $1.5^\circ$   \\
      $\gamma$ ($B \to DK$, combined)                  &  $\sim  6^\circ$   &     $1$--$2^\circ$  \\
 & \\
      $\alpha$ ($B \to \pi\pi$)                 &   $\sim 16^\circ$         &     $3^\circ$               \\
      $\alpha$ ($B \to \rho\rho)$               &   $\sim  7^\circ$         &     $1$--$2^\circ$ ($\ast$) \\
      $\alpha$ ($B \to \rho\pi)$                &   $\sim 12^\circ$         &     $2^\circ$               \\
      $\alpha$ (combined)                       &   $\sim  6^\circ$         &     $1$--$2^\circ$ ($\ast$) \\
 & \\
      $2\beta+\gamma$ ($D^{(*)\pm}\pi^\mp$, $D^\pm \KS \pi^\mp$) & $20^\circ$  &  $5^\circ$        \\
      \hline
    \end{tabular}
  \end{center}
\end{table}

\begin{table}[!hp]
  \caption {
    The expected precision of some of the most important measurements
    that can be performed at \superb.
    For comparison we put the reach of the $B$~Factories at $2 \ {\rm ab}^{-1}$.
    Numbers quoted as percentages are relative precisions.
    Measurements marked ($\dagger$) will be systematics limited,
    and those marked ($\ast$) will be theoretically limited, with $75 \ {\rm ab}^{-1}$.
    Note that in many of these cases, there exist data driven methods of reducing the errors.
    See the text for further discussion of each measurement.
 \smallskip }
  \label{tab:inputs2}
  \begin{center}
    \begin{tabular}{lcc}
      \hline \hline
      Observable                      & $B$~Factories ($2 \ {\rm ab}^{-1}$) & \superb\ ($75 \ {\rm ab}^{-1}$)  \\
      \hline
      $\left| V_{cb} \right|$ (exclusive)       &      $ 4\%$ ($\ast$)      &       $1.0\%$ ($\ast$)   \\
      $\left| V_{cb} \right|$ (inclusive)       &      $ 1\%$ ($\ast$)      &       $0.5\%$ ($\ast$)   \\
      $\left| V_{ub} \right|$ (exclusive)       &      $ 8\%$ ($\ast$)      &       $3.0\%$ ($\ast$)   \\
      $\left| V_{ub} \right|$ (inclusive)       &      $ 8\%$ ($\ast$)      &       $2.0\%$ ($\ast$)   \\
      \\
      $\BR(B \to \tau \nu)$                      &      $20\%$               &       $ 4\%$ ($\dagger$) \\
            $\BR(B \to \mu \nu)$                       &      visible              &       $5\%$             \\
      $\BR(B \to D \tau \nu)$                    &      $10\%$               &       $ 2\%$             \\
 & \\
      $\BR(B \to \rho \gamma$)                   &      $15\%$               &       $  3\%$ ($\dagger$) \\
      $\BR(B \to \omega \gamma$)                 &      $30\%$               &       $  5\%$             \\
      $A_{\CP}(B \to K^* \gamma)$                &      $0.007$ ($\dagger$)  &       $0.004$ ($\dagger$ $\ast$) \\
      $A_{\CP}(B \to \rho \gamma)$               &      $\sim 0.20$          &        0.05               \\
      $A_{\CP}(b \to s \gamma)$                  &      $0.012$ ($\dagger$)  &       $0.004$ ($\dagger$) \\
      $A_{\CP}(b \to (s+d) \gamma)$              &      $0.03$               &       $0.006$ ($\dagger$)  \\
      $S(\KS\pi^0\gamma)$                       &       0.15                &       $0.02$ ($\ast$)     \\
      $S(\rho^0\gamma)$                         &      possible             &       $0.10$              \\
 & \\
      $A_{\CP}(B \to K^* \ell \ell) $                  &      $7\%$                &         $1\%$             \\
      $A^{FB}(B \to K^* \ell \ell) s_0$               &      $25\%$               &         $9\%$             \\
      $A^{FB}(B \to X_s \ell \ell) s_0$               &      $35\%$               &         $5\%$             \\

      $\BR(B \to K \nu \overline{\nu})$          &      visible              &         $20\%$            \\
      $\BR(B \to \pi \nu \bar{\nu})$             &        --                 &        possible           \\
      \hline
    \end{tabular}
  \end{center}
\end{table}

It can be useful to schematically classify
the various results in two categories:

\begin{itemize}

\item
  \underline{Searching for New Physics} \\
  Many of the measurements that can be made at \superb\
  are highly sensitive to New Physics effects,
  and those with precise Standard Model predictions are potential discovery channels.
  As an example: the mixing-induced $\CP$ asymmetry parameter
  for $B^0 \to \phi \Kz$ decays 
  can be measured to a precision of $0.02$,
  as can equivalent parameters for numerous hadronic decay channels
  dominated by the $b \to s$ penguin transition.
  These constitute very stringent tests of any New Physics scenario
  which introduces new $\CP$ violation sources, beyond the Standard Model.
  Similarly, direct $\CP$ asymmetries can be measured to the fraction of a percent level in $b \to s \gamma$ decays,
  using both inclusive and exclusive channels,
  and $b \to s \ell^+\ell^-$ can be equally thoroughly explored.
  At the same time, \superb\ can access channels that are sensitive to New Physics
  even when there are no new sources of $\CP$ violation phases,
  such as the photon polarization in $b \to s\gamma$,
  and the branching fractions of $B^+ \to \ell^+ \nu_\ell$,
  the latter being sensitive probes of New Physics in MFV scenarios
  with large $\tan \beta$.
  Any of these measurements constitutes clear motivation for \superb.

\item
  \underline{Future metrology of the CKM matrix} \\
  As discussed further in Section~\ref{sec:U4s_pheno} below,
  there are several measurements which are unaffected by New Physics
  in many likely scenarios, and which allow the extraction
  of the CKM parameters even in the presence of such New Physics effects.
  Among these, the angle $\gamma$ can be measured with a precision
  of $1$--$2{^\circ}$,
  where the precision is limited only by statistics,
  not by systematics or by theoretical errors.
  By contrast,
  the determination of the elements $|V_{ub}|$ and $|V_{cb}|$
  will be limited by theory,
  but the large data sample of \superb\ will allow
  many of the theoretical errors to be much improved.
  With anticipated improvements in lattice QCD calculations
  (as discussed in Section~\ref{sec:ph_lattice}),
  the precision on $|V_{ub}|$ and $|V_{cb}|$ can be driven
  down to the percent level, and to a fraction of a percent, respectively.
  These measurements could allow
  tests of the consistency of the Standard Model at a few per mille level and provide the
  New Physics phenomenological analyses with a determination of the CKM
  matrix at the percent level.
\end{itemize}

\mysubsubsection{Comparison with \lhcb}
\label{sss:u4s_compLHCb}

Since \superb\ will take data in the LHC era,
it is reasonable to ask how the physics reach
compares with the $B$ physics potential of the LHC experiments,
most notably \lhcb.
By 2014, the \lhcb\ experiment is expected to have accumulated
$10 \ {\rm fb}^{-1}$ of data
from $pp$ collisions at a luminosity of
$\sim 2 \times 10^{32} \ {\rm cm}^{-2} {\rm s}^{-1}$.
We use the most recent estimates of \lhcb\
sensitivity with that data set~\cite{schneider} in the following.

The most striking outcome of any comparison between \superb\ and \lhcb\
is that the strengths of the two experiments are largely complementary.
For example, the large boost of the $B$ hadrons produced at \lhcb\
allows studies of the oscillations of $B_s$ mesons
(see the discussion in Section~\ref{sec:bs}).
It is particularly important to stress that
many of the measurements that constitute the primary
physics motivation for \superb\ cannot be performed
in the hadronic environment.
For example, modes with missing energy,
such as $B^+ \to \ell^+\nu_\ell$ and $B^+ \to K^+\nu\bar{\nu}$,
measurements of the CKM matrix elements $|V_{cb}|$ and $|V_{ub}|$,
and inclusive analyses of processes such as $b \to s\gamma$
are unique to \superb.
\lhcb\ has limited capability for channels containing
neutral particles,
or in studies where the analysis requires that the $B$ decay vertex
be determined from a $\KS$ meson,
precluding measurements of photon polarization
via mixing-induced $\CP$ violation in $B^0 \to \KS \pi^0 \gamma$.
Furthermore, the sensitivity of \lhcb\ to possible New Physics effects in
hadronic $b \to s$ penguin decays is seriously compromised,
since none of $\phi \Kz$, $\eta^\prime \Kz$, $\KS\KS\KS$ or $\KS \pi^0$
can be well studied.
\superb, on the other hand, can measure the $\CP$ asymmetries in all of these modes and more.

Where there is overlap,
the strength of the \superb\ program in its ability to use multiple approaches
to reach the objective becomes apparent.
For example, \lhcb\ may potentially be able to measure
$\alpha$ to about $5^\circ$ precision using $B \to \rho\pi$,
but would not be able to access the full information in the
$\pi\pi$ and $\rho\rho$ channels, which is necessary to drive
the uncertainty down to the $1$--$2^\circ$ level of \superb.
Similarly, \lhcb\ can certainly measure $\sin(2\beta)$
through mixing-induced $\CP$ violation in $B^0 \to J/\psi\KS$ decay
to high accuracy (about 0.01),
but will not be able to make the complementary measurements
({\it e.g.}, in $J/\psi\,\pi^0$ and $Dh^0$)
that help to ensure that the theoretical uncertainty is under control.
\superb\ is likely to have an advantage of a factor of two to three in
the precision for the angle $\gamma$ with respect to
\lhcb, with further improvements possible.

\lhcb\ can make a precise measurement of the zero of the
forward-backward asymmetry in $B^0 \to K^{*0}\mu^+\mu^-$,
but \superb\ can also measure the equivalent mode
for charged $B$ decay, as well as the corresponding mode with an $\epem$ pair,
and the inclusive channel $b \to s \ell^+\ell^-$.
The broader program of \superb\ thus provides a more comprehensive set of measurements.
As discussed in more detail below,
this will be of great importance for the study of
flavour physics in the LHC era.

The comparison with \lhcb\ for some specific topics on $B_s$
physics is given in Section \ref{sec:bs} which discusses the
physics case for running at the $\Upsilon(5S)$.

\mysubsection{Phenomenological Impact}
\label{sec:U4s_pheno}

\mysubsubsection{Determination of UT Parameters at \superb}
\label{sec:ckmfit}

In this section we discuss the determination of the CKM parameters
$\bar\rho$--$\bar\eta$ at \superb.

\begin{figure}[htb!]
  \begin{center}
    \includegraphics[width=0.85\textwidth]{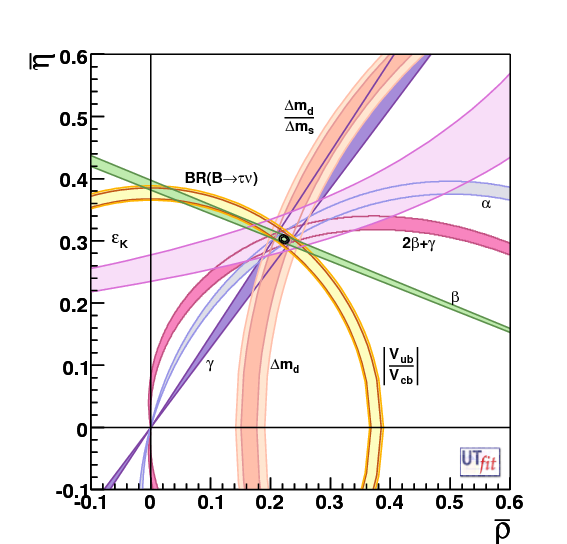}
    \caption{
      Allowed regions for $\rhobar$ and $\etabar$ using some of the parameters
      listed in Tables~\ref{tab:inputs1},~\ref{tab:inputs2}
      and~\ref{tab:lattice}.
      The closed contours at $68\%$ and $95\%$ probability are shown.
      The full lines correspond to $95\%$
      probability regions for each of the constraints.
    }
    \label{fig:smfit}
  \end{center}
\end{figure}

We start by assuming the validity of the Standard Model.
Most of the measurements described in the previous section
can be used to select a region in the $\rhobar$--$\etabar$ plane
as shown in Fig.~\ref{fig:smfit}.
The corresponding numerical results are given in Table~\ref{tab:smfit}.

\begin{table}[ht]
  \caption{
    Uncertainties of the CKM parameters obtained from the Standard Model fit
    using the experimental and theoretical information available today (left)
    and at the time of \superb\ (right) as given in
    Tables~\ref{tab:inputs1},~\ref{tab:inputs2} and \ref{tab:lattice}.
  }
  \begin{center}
    \begin{tabular}{lll}
      \hline\hline
      Parameter               &   SM Fit today        &  SM Fit at \superb \\
      \hline
      $\overline {\rho}$      &   $0.163\pm 0.028$    &  $\pm 0.0028$      \\
      $\overline {\eta}$      &   $0.344\pm 0.016$    &  $\pm 0.0024$      \\
      $\alpha$ ($^{\circ}$)   &   $92.7\pm4.2$        &  $\pm 0.45$        \\
      $\beta$ ($^{\circ}$)    &   $22.2\pm0.9$        &  $\pm 0.17$        \\
      $\gamma$ ($^{\circ}$)   &   $64.6\pm4.2$        &  $\pm 0.38$        \\
      \hline
    \end{tabular}
  \end{center}
  \label{tab:smfit}
\end{table}

The results shown in Fig.~\ref{fig:smfit} and in Table~\ref{tab:smfit}
indicate that a precision of a fraction of a percent can be reached,
significantly improving the current situation,
and providing a generic test of the presence of
New Physics at that level of precision.

This is done assuming the validity of the Standard Model.
Many of the measurements used for the Standard Model determination of
$\rhobar$--$\etabar$ can, however, be affected by the presence of New Physics.
Unambiguous New Physics searches require a determination of $\rhobar$ and $\etabar$ in
the presence of arbitrary New Physics contributions.

It is straightforward to generalize the Standard Model analysis including generic
New Physics effects in $\Delta F = 2$ processes.
In fact, those processes can be described by a single amplitude
and parameterized, without loss of generality, in terms of two New Physics parameters,
that quantify the ratio of the full amplitude
to the Standard Model amplitude~\cite{Soares:1992xi,
  Deshpande:1996yt,Silva:1996ih,Cohen:1996sq,Grossman:1997dd}.
Thus, for instance, in the case of $B^0_q$--$\Bbar^0_q$ mixing we define
\begin{equation}
  C_{B_q} \, e^{2 i \phi_{B_q}} =
  \frac{
    \langle B^0_q | H_{\rm eff}^{\rm full} | \Bbar^0_q \rangle
  }{
    \langle B^0_q | H_{\rm eff}^{\rm SM}   | \Bbar^0_q \rangle
  } \,, \qquad (q=d,s)
  \label{eq:paranp}
\end{equation}
where $H_{\rm eff}^{\rm SM}$ includes the Standard Model box diagrams only,
while $H_{\rm eff}^{\rm full}$ includes also the New Physics contributions.
In the absence of New Physics effects, by definition $C_{B_q}=1$ and $\phi_{B_q}=0$.
A subset of the \superb\ measurements, those of tree level and mixing-related processes, can be used.

For instance, the experimental quantities determined from
$B^0_d$--$\Bbar^0_d$ mixing are related to their Standard Model counterparts
and the New Physics parameters by the following relations:
\begin{equation}
  \Delta m_d^{\rm exp}   = C_{B_d} \Delta m_d^{\rm SM} \,,\;
  \sin 2 \beta^{\rm exp} = \sin (2 \beta^{\rm SM} + 2\phi_{B_d}) \,,\;
  \alpha^{\rm exp}       = \alpha^{\rm SM} - \phi_{B_d}\,.
  \label{eq:NPangles}
\end{equation}
The use of $\alpha$ in this context deserves explanation.
In principle, the extraction of $\alpha$ from
$B \to \pi \pi,~\rho \pi,~\rho \rho$ decays is affected by
New Physics effects in $\Delta F = 1$ transitions.
However, New Physics effects can be reabsorbed in a redefinition of the hadronic
parameters, as long as they induce only $\Delta I = 1/2$ transitions.
In this case they do not prevent the
extraction of $\alpha$~\cite{Grossman:1997gd,Botella:2005ks}.
Thus, information on $\alpha$ can be used to constrain
$\rhobar$ and $\etabar$ independently of
$\Delta I=1/2$ $\Delta F = 1$ New Physics contributions~\cite{Bona:2005eu}.
In the case of large $\Delta I = 3/2$ New Physics contributions, however,
such as large EWP-like New Physics,
the measurement of $\alpha$ cannot be used as a New Physics-independent constraint.

The numerical results are given in Table~\ref{tab:deltaf2fit}.
The precision of the CKM parameters obtained in the presence of generic New Physics
is not drastically worse than that of the Standard Model fit (Table~\ref{tab:smfit}),
and remains at the subpercent level.
This is a good starting point for New Physics analyses,
which require the model-independent determination of the CKM parameters
as an input.

\begin{table}[ht]
  \caption {
    Uncertainty of the CKM parameters obtained from the UT fit
    with generic New Physics contributions in $\Delta F = 2$ processes.
    The fits are performed using the experimental and theoretical information
    available today (left) and at the time of \superb\ (right)
    as given in Tables~\ref{tab:inputs1},~\ref{tab:inputs2}
    and~\ref{tab:lattice}.
  }
  \label{tab:deltaf2fit}
  \begin{center}
    \begin{tabular}{lll}
      \hline\hline
      Parameter             &  New Physics fit today     &  New Physics fit at \superb \\
      \hline
      $\overline {\rho}$    &  $0.187\pm0.056$  &  $\pm 0.005$       \\
      $\overline {\eta}$    &  $0.370\pm0.036$  &  $\pm 0.005$       \\
      $\alpha$ ($^{\circ}$)   &  $92\pm9$         &  $\pm 0.85$        \\
      $\beta$ ($^{\circ}$)    &  $24.4\pm1.8$     &  $\pm 0.4$         \\
      $\gamma$ ($^{\circ}$)    &  $63\pm8$         &  $\pm 0.7$         \\
      \hline
    \end{tabular}
  \end{center}
\end{table}

\mysubsubsection{New Physics Contributions in $\Delta F = 2$ Processes}
\label{sec:deltaf2}

\begin{figure}[t]
  \begin{center}
    \begin{tabular}{cc}
      \includegraphics[width=0.45\textwidth]{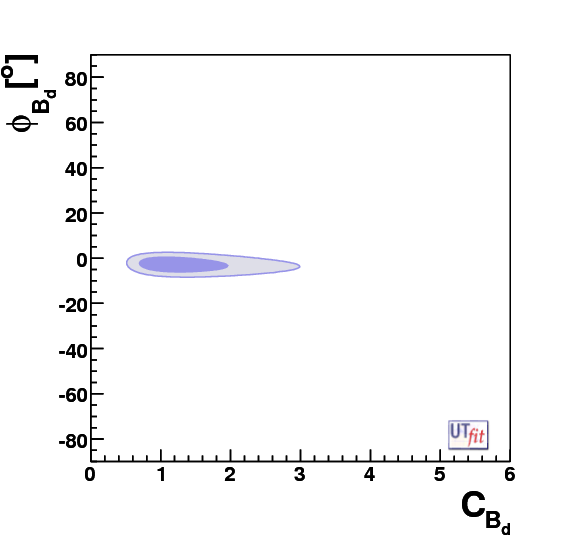} &
      \includegraphics[width=0.45\textwidth]{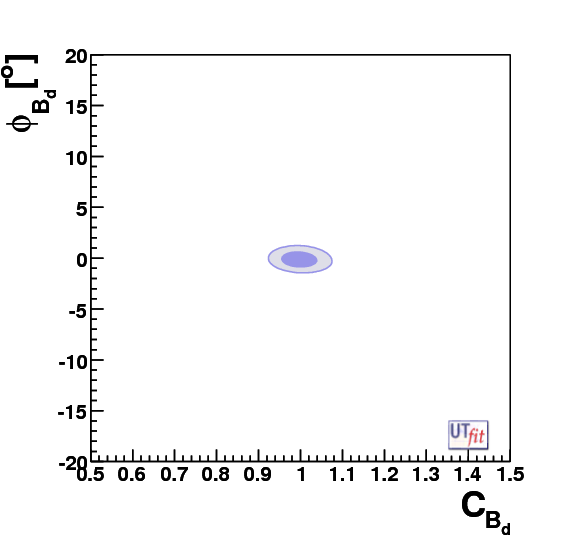}
    \end{tabular}
    \caption{
      Allowed regions in the $C_{B_d}$--$\phi_{B_d}$ plane
      given by the current data (left) and at the time of \superb\ (right).
      Note that the scales for the axes are different in the two cases.
    }
    \label{fig:bdphid}
  \end{center}
\end{figure}

The fit using $\Delta F = 2$ amplitudes
with generic New Physics contributions also allows us to obtain constraints
on the New Physics parameters $C_{d}$ and $\phi_{d}$,
which in turn provide information on the extent to which the experimental data
allow for New Physics in $\Delta F = 2$ amplitudes~\cite{Bona:2005eu}.
The numerical results are given in Table~\ref{tab:cvsphi}.
To illustrate the impact of the measurements at \superb,
in Fig.~\ref{fig:bdphid} we show the allowed regions in the
$C_{B_d}$--$\phi_{B_d}$ plane, as compared to the current situation.

\begin{table}[ht]
  \caption{
    Uncertainties on the New Physics parameters $C_{B_d}$ and $\phi_{B_d}$
    obtained using the experimental and theoretical information
    available today (left) and at the time of \superb\ (right),
    see Tables~\ref{tab:inputs1}~\ref{tab:inputs2} and~\ref{tab:lattice}.
  }
  \label{tab:cvsphi}
  \begin{center}
    \begin{tabular}{llll}
      \hline\hline
      Parameter              &  New Physics fit today      &  New Physics fit at \superb \\
      \hline
      $C_{B_d}$              &  $1.24 \pm 0.43$   &  $\pm 0.031$       \\
      $\phi_{B_d}$ ($^{\circ}$) &  $-3 \pm 2$        &  $\pm 0.4$         \\
      \hline
    \end{tabular}
  \end{center}
\end{table}

It is important to note that the reduction of the uncertainty on
the parameter $C_{B_d}$ is largely due to the expected improvement
of lattice calculations discussed in the Section~\ref{sec:lattice}.
By contrast, the impressive progress in the determination of
$\phi_{B_d}$ is due to the improved precision of the
experimental quantities measured at \superb.

\mysubsubsection{New Physics in Models with Minimal Flavour Violation}
\label{sec:mfv}

We now specialize to the case of Minimal Flavour Violation
(MFV)~\cite{Gabrielli:1994ff,Buras:2000dm,D'Ambrosio:2002ex}.
The basic assumption of MFV is that New Physics does not introduce
new sources of flavour and $\CP$ violation.
Hence the only flavour-violating couplings are the Standard Model Yukawa couplings.
In the simplest case with one Higgs doublet
(or two Higgs doublets with small $\tan \beta$),
one can safely assume that the top Yukawa coupling is dominant, so that
all New Physics effects amount to a real contribution added to the Standard Model
loop function generated by virtual top exchange.
In particular, considering the $\Delta B = 2$ amplitude,
MFV New Physics can be parameterized as
$$S_0(x_t) \to S_0(x_t) + \delta S_0$$
where the function $S_0(x_t)$ represents the top contribution
in the box diagrams and $\delta S_0$ is the New Physics contribution.
Therefore, in this class of MFV models,
the New Physics contribution to all $\Delta F = 2$ processes is universal,
and the effective Hamiltonian retains the Standard Model structure.


Following ref.~\cite{D'Ambrosio:2002ex},
this value can be converted into a New Physics scale using
\begin{equation}
  \delta S_0 =  4 a \left( \frac{\Lambda_0}{\Lambda}\right)^2\,,
\end{equation}
where $\Lambda_0=Y_t \sin^2 \theta_W M_W/\alpha \approx 2.4 \ {\rm TeV}$
is the Standard Model scale, $Y_t$ is the top Yukawa coupling,
$\Lambda$ is the New Physics scale and $a$ is an unknown (but real)
Wilson coefficient of ${\cal O}(1)$.

The UT analysis can constrain the value of the New Physics parameter $\delta S_0$ together
with $\rhobar$ and $\etabar$.
In the absence of a New Physics signal, $\delta S_0$ is distributed around zero.
From this distribution, we can obtain a lower bound on the New Physics scale $\Lambda$.

For the 1HDM and 2HDM in the low $\tan \beta$ regime,
the combination of \superb\ measurements and the improved lattice results gives
\begin{equation}
  \Lambda > 14 \ {\rm TeV}~@~95\% \ {\rm CL}
\end{equation}


\begin{figure}[t]
  \begin{center}
   \begin{tabular}{cc}
    \includegraphics[width=0.45\textwidth]{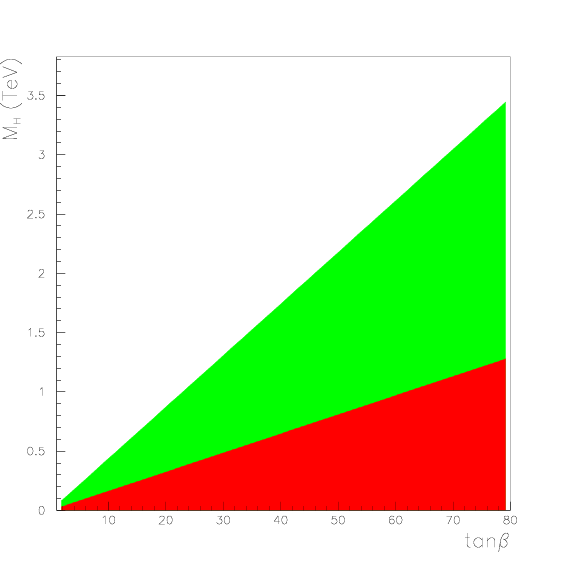} &
    \includegraphics[width=0.45\textwidth]{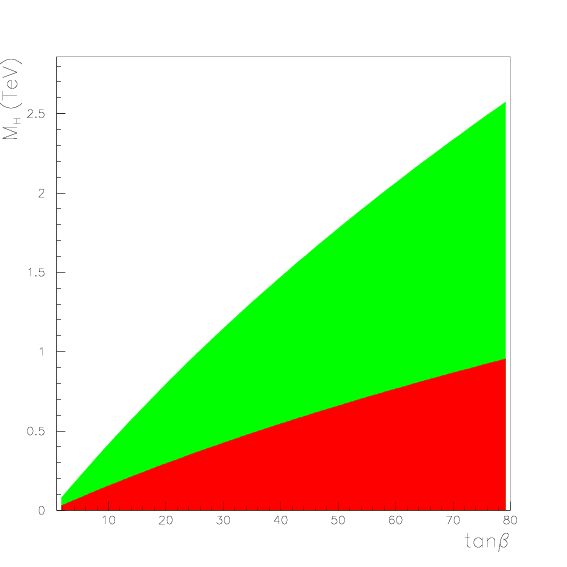}
    \end{tabular}
    \caption{%
      Exclusion regions at $95\%$ CL
      in the $M_{H^\pm}$--$\tan \beta$ plane for the 2HDM-II
      (left) and the MSSM (right),
      assuming the Standard Model value of $\B(B\to\ell\nu)$ measured with
      $2 \ {\rm ab}^{-1}$ (dark area) and
      $75 \ {\rm ab}^{-1}$ (dark+light area). In the MSSM case, we have used
      $\epsilon_0 \sim 10^{-2}$~\cite{Isidori:2007zm} (see Section \ref{sec:U4s_rare}
      for definitions).
    }
    \label{fig:btaunu}
  \end{center}
\end{figure}

These bounds are a factor of three larger than
those available today~\cite{Bona:2005eu}.
This means that even in the ``worst case'' scenario,
{\it i.e.},~in models with MFV at small $\tan \beta$,
the sensitivity of flavour-violating processes to New Physics
is strong enough to allow for the study of the flavour-violating
couplings of new particles with masses up to $600$ GeV.
This conversion to a New Physics scale in the MFV case deserves explanation.
We should consider that the Standard Model reference scale
corresponds to virtual $W$-exchange in the loops. As MFV
has the same flavour violating couplings as the Standard Model, the
MFV-New Physics scale is simply translated in a new virtual particle
mass as $\Lambda/\Lambda_0\times  M_W$.
It must be noted, however, that as soon as one considers large $\tan \beta$,
or relaxes the MFV assumption in this kind of analysis,
the New Physics scale is raised by at least a factor of $3$,
covering the whole range of masses accessible at the LHC.
In fact the RGE-enhanced contribution of the scalar operators
(absent in the small $\tan \beta$ MFV case)
typically sets bounds an order of magnitude stronger than that
on the Standard Model current-current operator,
correspondingly increasing the lower bound on the New Physics scale.
This is the case, for instance, of the NMFV models discussed in
ref.~\cite{Agashe:2005hk} as described in the analysis of ref.~\cite{df2gen}.

The large $\tan \beta$ scenario offers additional opportunities
to reveal New Physics by enhancing flavour-violating couplings in
$\Delta B = 1$ processes with virtual Higgs exchange.
This can be the case in decays such as $B\to\ell\nu$ or $B\to D\tau\nu$
whose branching ratios are strongly affected by a charged Higgs
for large values of $\tan\beta$.
In Fig.~\ref{fig:btaunu} we show the region excluded in the
$M_{H^\pm}$--$\tan \beta$ plane by the measurement of $\B(B\to\ell\nu)$
with the precision expected at the end of the current $B$~Factories
and at \superb, assuming the central value given by the Standard Model.
It is apparent that \superb\ pushes the lower bound on $M_{H^\pm}$,
corresponding, for example, to $\tan \beta \sim 50$ from the hundreds of GeV region
up to about 2 TeV, both in the 2HDM-II and in the MSSM.
Another interesting possibility is looking at LFV by measuring the ratio
$R_B^{\mu/\tau} = \B(B\to\mu\nu)/\B(B\to\tau\nu)$,
which could have a ${\cal O}(10\%)$ deviation from its Standard Model value
at large $\tan \beta$~\cite{Isidori:2006pk,Masiero:2005wr},
whereas the relative error on the individual branching fraction measurements
at \superb\ is expected to be $5\%$ or less, see Table~\ref{tab:inputs2}.

\mysubsubsection{MSSM with Generic Squark Mass Matrices}
\label{sec:MSSM}

We now discuss the impact of \superb\ on
the parameters of the MSSM with generic squark mass matrices
parameterized using the mass insertion (MI) approximation~\cite{Hall:1985dx}.
In this framework, the New Physics flavour-violating couplings are the complex MIs.
For simplicity, we consider only the dominant gluino contribution.
The relevant parameters are therefore the gluino mass $m_{\tilde g}$,
the average squark mass $m_{\tilde q}$ and the MIs $(\delta^{d}_{ij})_{AB}$,
where $i,j=1,2,3$ are generation indices and
$A,B=L,R$ refer to the helicity of the SUSY partner quarks.
For example, the parameters relevant to $b \to d$ transitions are
the two SUSY masses and the four MIs $(\delta^{d}_{13})_{LL,LR,RL,RR}$.
To simplify the analysis, we consider the
contribution of one MI at a time.
This is justified to some extent by the hierarchy of
the present bounds on the MIs.
Barring accidental cancelations,
the contributions from two or more MIs would produce larger New Physics effects
and thus make the detection of New Physics easier,
while simultaneously making the phenomenological analysis more involved.
The analysis presented here profits from results and techniques
developed in Refs.~\cite{Gabbiani:1996hi,Becirevic:2001jj,Ciuchini:2002uv}.

\begin{figure}[t]
  \begin{center}
    \begin{tabular}{cc}
      \includegraphics[width=0.48\textwidth]{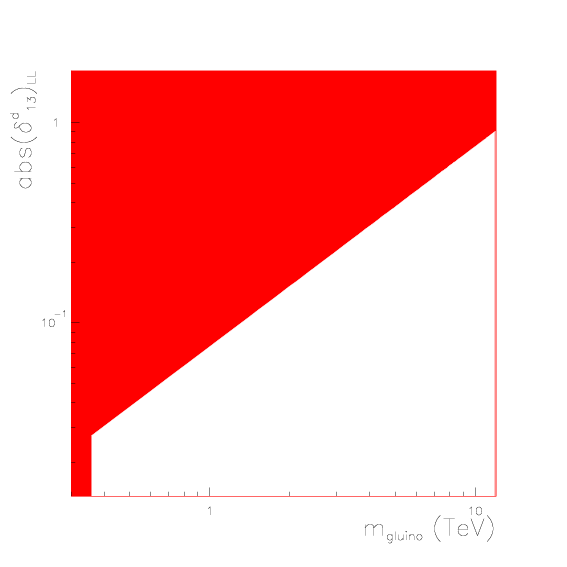} &
      \includegraphics[width=0.48\textwidth]{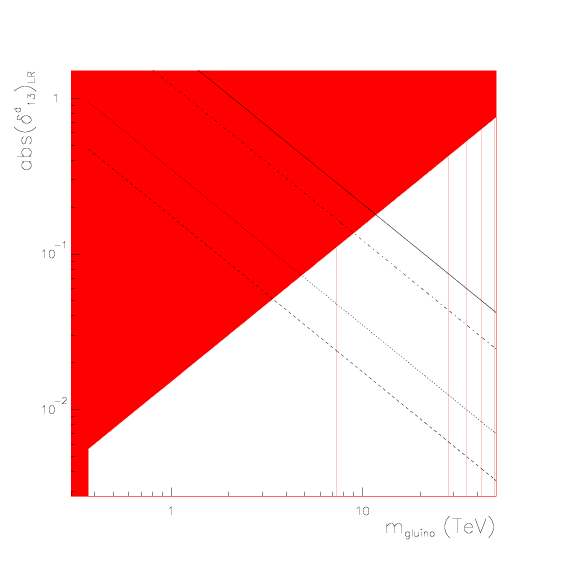} \\
      \includegraphics[width=0.48\textwidth]{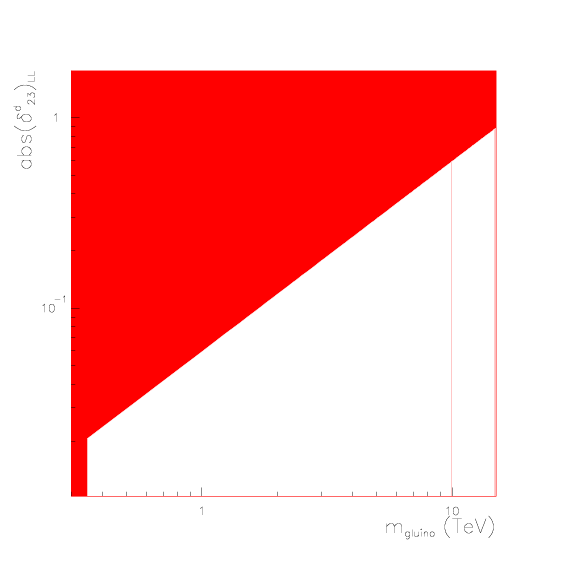} &
      \includegraphics[width=0.48\textwidth]{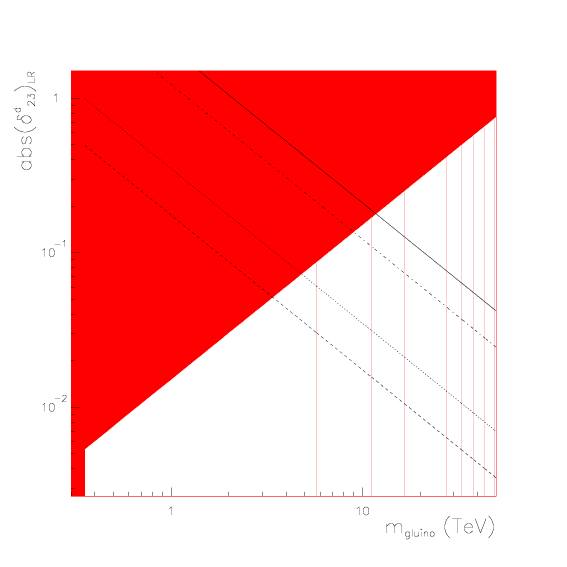}
    \end{tabular}
    \caption{
      Sensitivity region of \superb\ in the
      $m_{\tilde g}$--$\vert(\delta^d_{ij})_{AB}\vert$ plane.
      The region is obtained by requiring that the reconstructed
      MI is $3\sigma$ away from zero.
      The cases of $(\delta^d_{13})_{LL}$ (upper left),
      $(\delta^d_{13})_{LR}$ (upper right),
      $(\delta^d_{23})_{LL}$ (lower left) and
      $(\delta^d_{23})_{LR}$ (lower right) are shown.
      For LR MIs the theoretical upper bound (allowed parameter region is below these lines)
      discussed in the text is also shown for $\tan \beta= 5, 10, 35, 60$
      (dashed, dotted, dot-dashed, solid line respectively).
    }
    \label{fig:MIvsMg}
  \end{center}
\end{figure}

The aim of this analysis is twofold.
On the one hand, we want to show the bounds on the MSSM parameter space
as they would appear at \superb.
For this purpose, we first simulate the signals produced by the MSSM
for a given value of one MI.
Then we check how well we are able to determine this value
using the constraints coming from \superb.
In particular, we are interested in ascertaining the range of masses
and MIs for which clear New Physics evidence,
indicated by a non-vanishing value of the extracted MI, can be obtained.
In Fig.~\ref{fig:MIvsMg} we show for some of the different MIs,
the observation region in the plane $m_{\tilde g}$--$\vert\delta^d\vert$
obtained by requiring that the absolute value of the reconstructed MI
is more than $3\sigma$ away from zero.
For simplicity we have taken $m_{\tilde q}\sim m_{\tilde g}$.
From these plots, one can see that \superb\ could detect New Physics effects
caused by SUSY masses up to $10$--$15$ TeV,
corresponding to $(\delta^d_{13,23})_{LL}\sim 1$.
Even larger scales could be reached by $LR$ MIs,
although overly large LR MIs are known to produce
charge- and colour-breaking minima in the MSSM potential~\cite{Casas:1996de}.
This can be avoided by imposing the bounds
shown in the LR plots of Fig.~\ref{fig:MIvsMg}.
These bounds decrease as $1/m_{\tilde q}$ and
increase linearly with $\tan \beta$.
Taking them into account,
we can see that LR MIs are still sensitive to gluino masses
up to $5$--$10$ TeV for $\tan \beta$ between 5 and 60.

Fig.~\ref{fig:MIvsMg} shows the values of the MI
that can be reconstructed if SUSY masses are below $1$ TeV.
In the cases considered we find
$(\delta^{d}_{13})_{LL}=2$--$ 5\times 10^{-2}$,
$(\delta^{d}_{13})_{LR}=2$--$15\times 10^{-3}$,
$(\delta^{d}_{23})_{LL}=2$--$ 5\times 10^{-1}$ and
$(\delta^{d}_{23})_{LR}=5$--$10\times 10^{-3}$.
These value are typically one order of magnitude smaller than
the present upper bounds on the MIs~\cite{silvckm}.

\begin{figure}[t]
  \begin{center}
    \begin{tabular}{cc}
      \multicolumn{2}{c}{
      \includegraphics[width=0.8\textwidth]{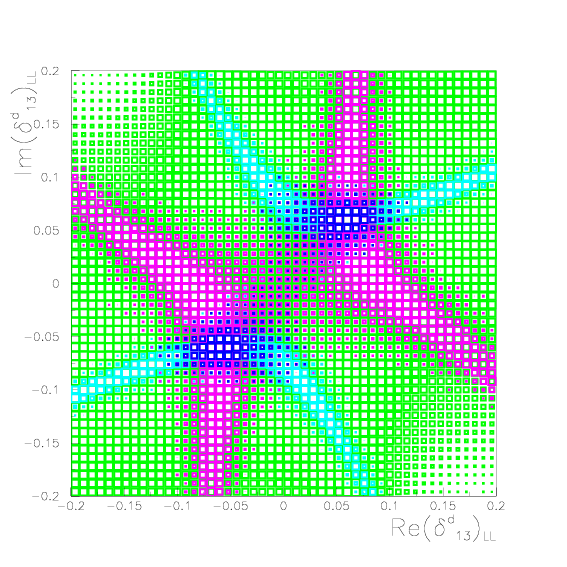}} \\
      \includegraphics[width=0.4\textwidth]{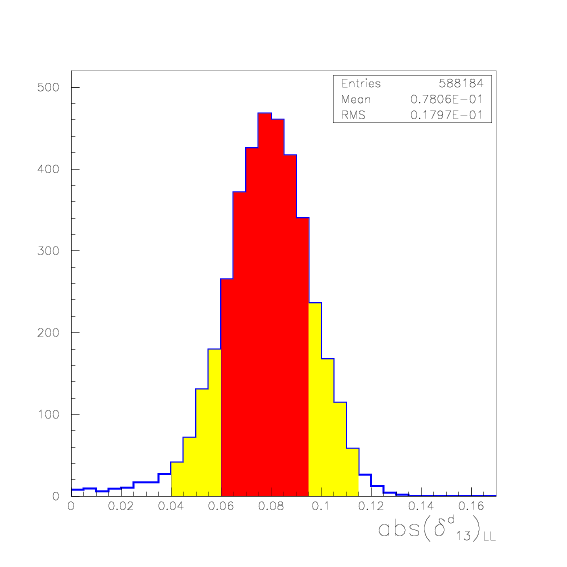} &
      \includegraphics[width=0.4\textwidth]{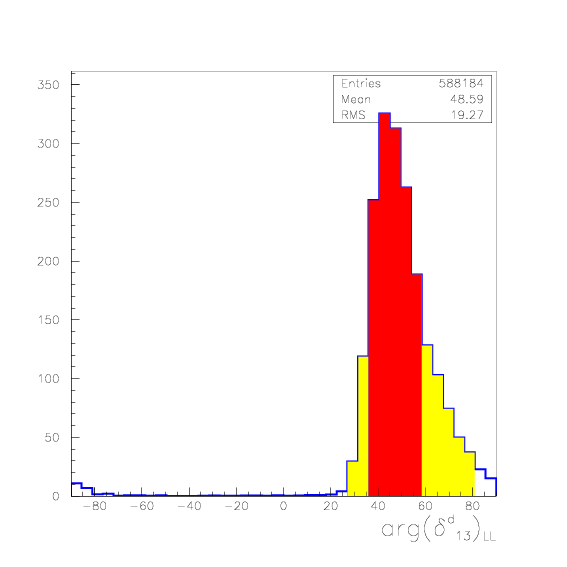} \\
    \end{tabular}
    \caption{
      Density plot of the selected region in the
      $\Re(\delta^d_{13})_{LL}$--$\Im(\delta^d_{13})_{LL}$ for
      $m_{\tilde q}=m_{\tilde g}=1 \ {\rm TeV}$ and
      $(\delta^d_{13})_{LL}=0.085 e^{i \pi/4}$
      using \superb\ measurements. Different colours correspond to different
      constraints: $A_{\rm SL}^d$ (green), $\beta$ (cyan),
      $\Delta m_d$ (magenta), all together (blue).
      On the lower line the distributions of the modulus (left) and
      phase (right) of the reconstructed MI are also shown.
    }
    \label{fig:MI13LL}
  \end{center}
\end{figure}

\begin{figure}[t]
  \begin{center}
    \begin{tabular}{cc}
     \multicolumn{2}{c}{
      \includegraphics[width=0.8\textwidth]{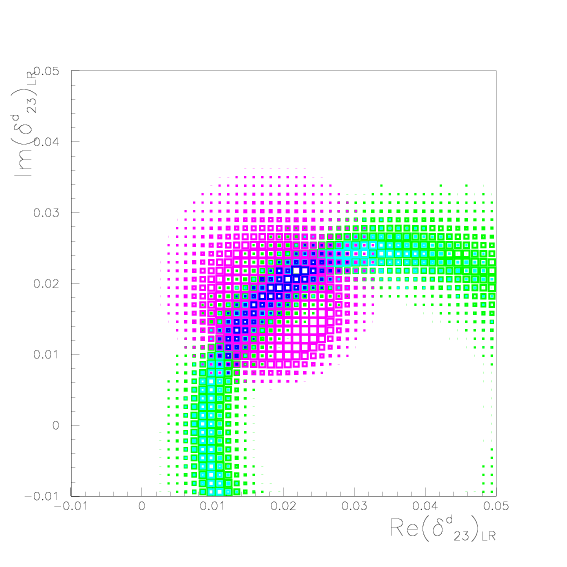}} \\
      \includegraphics[width=0.4\textwidth]{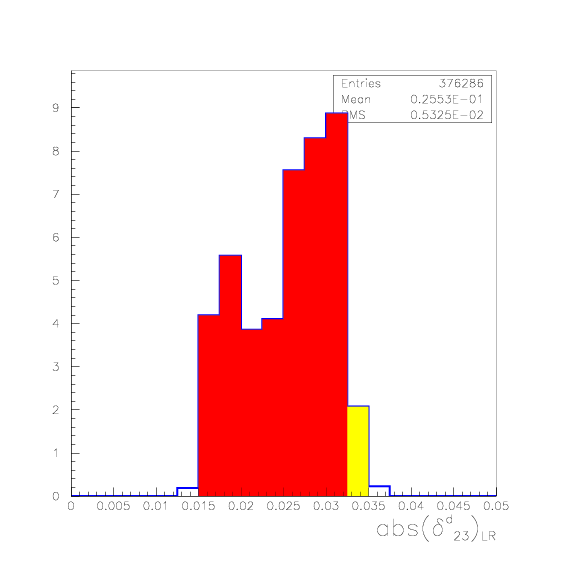} &
      \includegraphics[width=0.4\textwidth]{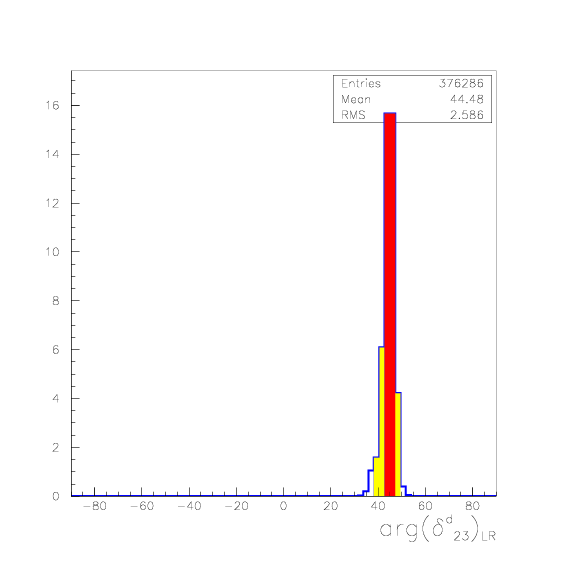} \\
    \end{tabular}
    \caption{
      Density plot of the selected region in the
      $\Re(\delta^d_{23})_{LR}$--$\Im(\delta^d_{23})_{LR}$ for
      $m_{\tilde q}=m_{\tilde g}=1 \ {\rm TeV}$ and
      $(\delta^d_{13})_{LL}=0.028 e^{i\pi/4}$
      using \superb\ measurements. Different colours correspond to different
      constraints: $\BR(B\to X_s\gamma)$ (green),
      $\BR(B\to X_s\ell^+\ell^-)$ (cyan), $A_{\CP}(B\to X_s\gamma)$ (magenta),
      all together (blue).
      On the lower line the distributions of the modulus (left) and
      phase (right) of the reconstructed MI are also shown.
    }
    \label{fig:MI23LR}
  \end{center}
\end{figure}

Figures~\ref{fig:MI13LL} and~\ref{fig:MI23LR} display examples of
the allowed region in the plane
$\Re(\delta^d_{ij})_{AB}$--$\Im(\delta^d_{ij})_{AB}$
with a value of $(\delta^d_{ij})_{AB}$
allowed from the present upper bound,
$m_{\tilde g}=1$ TeV and using the \superb\ measurements as constraints.
In particular, Fig.~\ref{fig:MI13LL} shows the selected region in the
$\Re(\delta^d_{13})_{LL}$--$\Im(\delta^d_{13})_{LL}$
using the measurements of $\Delta m_{B_d}$, $\beta$ and $A^d_{\rm SL}$
as constraints, together with the distributions of the reconstructed value
of the modulus and the phase of the MI.
In this case, the CKM angle $\beta$ and the mass difference
are the crucial constraints,
although accurate determination of the CKM parameters
$\bar\rho$ and $\bar\eta$ is crucial in order to separate New Physics contributions.

The plots for the case $(\delta^d_{23})_{LR}$ are shown
in Fig.~\ref{fig:MI23LR}.
Here the relevant constraints come from
$\BR(b\to s\gamma)$, $A_{\CP}(b\to s\gamma)$, $\BR(b\to s\ell^+\ell^-)$,
$A_{\CP}(b\to s\ell^+\ell^-)$, $\Delta m_{B_s}$ and $A^s_{\rm SL}$.
It is apparent the key role of $A_{\CP}(b\to s\gamma)$
together with the branching ratios of $b\to s\gamma$ and $b\to s\ell^+\ell^-$.
The zero of the forward-backward asymmetry
in $b\to s\ell^+\ell^-$, missing in the present analysis,
is expected to give an additional strong constraint,
further improving the already excellent extraction of
$(\delta^d_{23})_{LR}$ shown in Fig.~\ref{fig:MI23LR}.

\afterpage{\clearpage}

\mysection{$\tau$ Physics}
\label{sec:tau}
\mysubsection{Lepton Flavour Violation in $\tau$ Decays}
The search for FCNC transitions
of charged leptons is one of the most promising directions
to search for physics beyond the Standard Model.
Neutrino physics has provided
unambiguous evidence for non-conservation of lepton flavour.
We therefore expect this phenomenon to also occur in the charged lepton sector,
although, if the light neutrino mass matrix ($m_{\nu}$)
is the only source of lepton flavour violation (LFV),
FCNC transitions of charged leptons occur
well below any realistic experimental sensitivity.
However, in many realistic extensions of the Standard Model this is not the case.
In particular, the small value of $m_{\nu}$ is naturally explained by
a strong suppression associated with the breaking of total lepton number (LN),
which is not directly related to the size of LFV interactions.
As a result, there exist various well-motivated scenarios where LFV decays
of charged leptons occur just below the present experimental bounds
(given in Table~\ref{lfvtable}).

Rare FCNC decays of the $\tau$ lepton are particularly interesting since
LFV sources involving the third generation are naturally the largest.
In particular, searches for $\tau \to \mu \gamma$
at the $10^{-8}$ level or below are extremely interesting, even
taking into account the present stringent bounds on $\mu \to e \gamma$.
In the following we will illustrate this point
both within the explicit example of the MSSM with right-handed neutrinos,
and by means of a general effective-theory approach.
In all cases, the comparison of experimental results on
$\tau \to \mu \gamma$ with those for $\mu \to e \gamma$
and other LFV rare decays provides a unique tool to identify in detail
the nature of the New Physics model.

\mysubsubsection{Low-energy Supersymmetry}

A generic low-energy SUSY model with arbitrary
mixing in the soft-breaking parameters would induce unacceptably large
flavour-violating effects.
Limits on departures from Standard Model expectations in quark FCNC transitions motivate the assumption of flavour-universality of the SUSY breaking mechanism.
Even with this assumption, however, sizable flavour-mixing effects
can be generated at the weak scale by the running of the soft-breaking
parameters from the (presumably high) scale of SUSY-breaking mediation.
In the leptonic sector, the relevance of such effects strongly depends
on the assumptions about the neutrino sector.
If the light neutrino masses are obtained via a see-saw mechanism,
then the induced flavour-mixing coupling relevant to LFV rates
are naturally large~\cite{Borzumati:1986qx}.

Assuming a see-saw mechanism with three heavy right-handed neutrinos,
the effective light-neutrino mass matrix obtained by integrating
out the heavy fields is:
\begin{equation}
  \label{see-saw}
  m_\nu = - Y_\nu \hat{M}^{-1}_R Y_\nu^T \langle H_u \rangle^2~,
\end{equation}
where $\hat{M}_R$ is the $3\times 3$
right-handed neutrino mass matrix (which breaks LN),
$Y_{\nu}$ are the $3\times 3$ Yukawa couplings between
left- and right-handed neutrinos (the potentially large sources of LFV),
and $\langle H_u \rangle$ is the vacuum expectation value of the up-type Higgs
boson.
The LFV effects on charged leptons originate from any misalignment
between fermion and sfermion mass eigenstates.
Taking into account the renormalization-group evolution (RGE),
the slepton mass matrix $(m^2_{\tilde{L}})_{ij}$ acquires LFV entries
given by
\begin{equation}
  \label{mLfromseesaw}
  (m^2_{\tilde{L}})_{i\neq j} \approx
  - \frac{3m^2_0}{8\pi^2} (Y_{\nu} Y_{\nu}^\dagger)_{i\neq j}
  \ln \left(\frac{M_X}{M_{R}} \right)\,,
\end{equation}
where $M_X$ denotes the scale of SUSY-breaking mediation and
$m_0$ the universal SUSY breaking scalar mass.
Since the see-saw equation~\ref{see-saw}
allows large $(Y_\nu Y_\nu^\dagger)$ entries,
sizable effects can result from this running.

A complete determination of $(m^2_{\tilde{L}})_{i\neq j}$
would require a complete knowledge of the neutrino Yukawa matrix $(Y_\nu)_{ij}$,
which is not possible using only low-energy observables
from the neutrino sector.
This is in contrast with the quark sector,
where similar RGE contributions are completely determined in
terms of quark masses and CKM matrix elements.
As a result, the predictions of FCNC effects in the lepton
sector usually have sizable uncertainties.

More stable predictions can be obtained by embedding the SUSY model within
a Grand Unified Theory (GUT), such as $SO(10)$, where the see-saw mechanism can naturally
arise.
In this case, the GUT symmetry allows us to obtain some hints
about the unknown neutrino Yukawa matrix $Y_{\nu}$.
Moreover, in GUT scenarios there are other contributions
stemming from the quark sector~\cite{Barbieri:1994pv,Barbieri:1995tw}.
These effects are completely independent of the structure of
$Y_{\nu}$ and  can be regarded as new irreducible LFV contributions
within SUSY GUTs.
For instance, within $SU(5)$,
as both $Q$ and $e^c$ are hosted in the {\bf 10} representation,
the CKM matrix mixing of the left handed quarks
gives rise to off-diagonal entries in the running of the right-handed
slepton soft masses~\cite{Barbieri:1994pv,Barbieri:1995tw}.

\begin{table}[htb]
  \caption{
    \label{lfvtable}
    Present experimental bounds on some LFV decays of $\tau$ and $\mu$ leptons.
    Note that CR($\mu \to e$ in Ti) is a limit on the rate of conversions,
    $\sigma( \mu^- {\rm Ti} \to e^- {\rm Ti}) /
    \sigma( \mu^- {\rm Ti} \to {\rm capture})$.
    For more limits, see~\cite{Yao:2006px}.
  }
  \begin{center}
    \begin{tabular}{lr}
      \hline
      \hline
      Process & Present bound \\
      \hline
      $\BR(\tau \to \mu\gamma$) & $6.8 \times 10^{-8}$~\cite{Aubert:2005ye} \\
      $\BR(\tau \to e\gamma$)   & $1.1 \times 10^{-7}$~\cite{Aubert:2005wa} \\
      $\BR(\tau \to \mu\mu\mu$) & $1.9 \times 10^{-7}$~\cite{Aubert:2003pc} \\
      $\BR(\tau \to \mu\eta$)   & $1.5 \times 10^{-7}$~\cite{Enari:2005gc} \\
      & \\
      $\BR(\mu \to e\gamma$)    & $1.2 \times 10^{-11}$~\cite{Brooks:1999pu} \\
      $\BR(\mu \to eee$ )       & $1.0 \times 10^{-12}$~\cite{Bellgardt:1987du} \\
      CR($\mu \to e$ in Ti)    & $4.3 \times 10^{-12}$~\cite{Dohmen:1993mp} \\
      \hline
    \end{tabular}
  \end{center}
\end{table}

\begin{figure}
  \centering
  \includegraphics[angle=-90, width=0.80\textwidth]{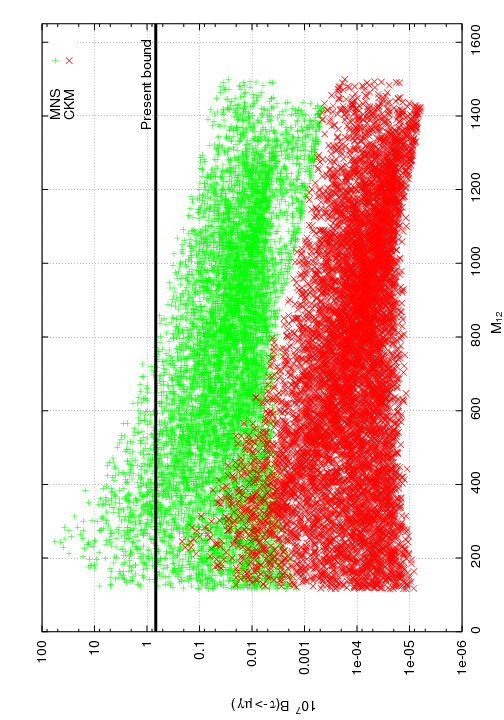}
  \caption{
    $\BR(\tau\to \mu\gamma)$ in units of $10^{-7}$ {\it vs.}
    the high energy universal gaugino mass ($M_{1/2}$)
    within a $SO(10)$ framework~\cite{Calibbi:2006nq}.
    The plot is obtained by scanning the LHC accessible parameter space
    $m_0\leq 5 \ {\rm TeV}$ for $\tan \beta = 40$.
    Green (red) points correspond to the PMNS (CKM) case,
    namely the scenario where $Y_{\nu} = U_{\rm PMNS}$
    ($Y_{\nu} = V_{\rm CKM}$).
    The thick horizontal line denotes the present experimental sensitivity.
  }
  \label{masieroLFV06}
\end{figure}

Once non-vanishing LFV entries in the slepton mass matrices
are generated, LFV rare decays are naturally induced by
one-loop diagrams with the exchange of gauginos and sleptons
(gauge-mediated LFV amplitudes). An order-of-magnitude
approximation for $\B(l_{i} \to l_{j} \gamma)$ is
\begin{equation}
  \frac{\BR(\ell_{i}\to\ell_{j}\gamma)}{\BR(\ell_{i}\to\ell_{j}\nu\bar{\nu})}
  \approx 4.5 \times 10^{-6}
  \left( \frac{500 \ {\rm GeV}}{m_{\rm SUSY}}\right)^4
  (\delta_{LL})^2_{ij} \left(\frac{\tan \beta}{10}\right)^2\,,
  \label{BrLFV}
\end{equation}
where $m_{\rm SUSY}$ is a typical SUSY mass running in the loop,
$(\delta_{LL})^2_{ij} = (m^2_{\tilde{L}})_{ij}/\langle m^2_{\tilde{L}} \rangle$
and, as usual, $\tan \beta$ denotes the ratio of the two MSSM
Higgs vacuum expectation values.
Besides $\ell_{i}\to\ell_{j}\gamma$, there are also other
promising LFV channels, such as $\ell_i\to\ell_j\ell_k\ell_k$,
that could be measured with the upcoming experimental sensitivities.
However, in the case of gauge-mediated LFV amplitudes
the $\ell_{i}\to\ell_{j}\ell_{k}\ell_{k}$ decays
are dominated by the $\ell_{i}\to\ell_{j}\gamma^{*}$
dipole transition, which leads to the unambiguous prediction:
\beq
\mathcal{BR}(\ell_{i}\to \ell_{j}\ell_{k}\ell_{k}) ~\sim~
\alpha_{\rm em} \mathcal{BR}(\ell_{i}\to \ell_{j}\gamma)~.
\label{eq:dipole}
\eeq
In Fig.~\ref{masieroLFV06}, we show the prediction
for $\BR(\tau\to\mu\gamma)$ within a SUSY SO(10) framework
for the accessible LHC SUSY parameter space $M_{1/2}\leq 1.5 \ {\rm TeV}$,
$m_0 \leq 5 \ {\rm TeV}$ and $\tan \beta = 40$~\cite{Calibbi:2006nq}.
Note that the scenarios where $Y_{\nu}=U_{\rm PMNS}$
and where $Y_{\nu} = V_{\rm CKM}$
can be distinguished by the measurement of $\BR(\tau \to \mu\gamma)$ at \superb.

An independent (and potentially large) class of LFV
contributions to rare decays comes from the Higgs sector:
if the slepton mass matrices have LFV entries and
the effective Yukawa interaction includes non-holomorphic
couplings, Higgs-mediated LFV amplitudes are necessarily
induced~\cite{Babu:2002et}.
Interestingly enough, gauge- and Higgs-mediated LFV amplitudes
lead to very different correlations
among LFV processes~\cite{Paradisi:2005tk,Paradisi:2006jp}.
In particular, the relation (Eq.~\ref{eq:dipole}) does not hold
for Higgs-mediated LFV amplitudes.
Thus, if several LFV transitions are observed,
a correlated analysis could shed light on the underlying
mechanism of LFV~\cite{Paradisi:2005tk,Paradisi:2006jp,Blanke:2007db}.

\mysubsubsection{Effective-theory Approaches}

All Standard Model extensions having new degrees of freedom at the TeV scale
and carrying flavour quantum numbers face severe constraints
from low-energy FCNC measurements.
An economical and elegant solution to this {\em flavour problem}
is provided by the MFV hypothesis,
namely by the assumption that the irreducible sources of
flavour symmetry breaking are minimally linked to the
fermion mass matrices observed at low energy.
This hypothesis guarantees the suppression of FCNC rates
to a level consistent with experimental constraints,
without resorting to unnaturally high scales of New Physics, and
allows the description of New Physics effects
in low-energy observables by means of a general
and falsifiable effective theory approach.

The MFV hypothesis has a strong motivation and a unique realization in
the quark sector~\cite{Chivukula:1987py,Hall:1990ac,D'Ambrosio:2002ex}:
the Standard Model Yukawa couplings are the only sources of breaking of the
$SU(3)^3$ quark-flavour symmetry group
(see Section~\ref{sec:U4s_pheno} for tests of MFV in the quark sector).
Apart from arguments based on the analogy with quarks,
the introduction of a Minimal Lepton Flavour
Violation (MLFV) principle~\cite{Cirigliano:2005ck}
is motivated by a severe fine-tuning problem in the lepton sector:
within a generic effective theory approach,
the non-observation of $\mu \to e\gamma$ implies
an effective New Physics scale above $10^5 \ {\rm TeV}$
unless the LFV coupling of the corresponding
operator is suppressed by some symmetry principle.

The implementation of a MFV principle in the lepton sector is
not as simple as in the quark sector, since
the neutrino mass matrix itself cannot be accommodated
within the renormalizable part of the Standard Model Lagrangian.
This implies that we need to employ an additional hypothesis
to identify the irreducible flavour-symmetry breaking structures.
The resulting effective theory can be viewed as a general tool to exploit
the observable consequences of a specific (minimal) hypothesis about
the irreducible sources of lepton-flavour symmetry breaking.

The most interesting MLFV ansatz is based on a see-saw mechanism with
three heavy right-handed neutrinos and $O(3)_{\nu_R}$ flavour symmetry.
The latter forces $M_R$ to be proportional to the identity matrix
in flavour space.
As a result, the irreducible LFV sources are only Yukawa couplings
(similar to the quark sector)
and their structure can be determined by the lepton and neutrino mass matrices.
The general consequences of this hypothesis
can be summarized as follows~\cite{Cirigliano:2005ck}:
\begin{itemize}
\item
  The overall normalization of LFV decay rates is largely unknown,
  being proportional to $M_R^2$.
  Results close to the present exclusion bounds (Table~\ref{lfvtable})
  could arise in the presence of a large hierarchy between the
  scale of the new flavour degrees of freedom ($\Lambda$) and
  the scale of lepton number violation: $M_R > 10^{9} \times \Lambda$.
\item
  Ratios of similar LFV decay rates 
  are free from the normalization ambiguity and can be
  predicted in terms of neutrino masses and PMNS angles:
  violations of these predictions would unambiguously signal
  the presence of additional sources of lepton-flavour symmetry breaking.
  One of these predictions is the ratio
  $\BR(\mu \to e\gamma)/\BR(\tau\to \mu\gamma) \sim 10^{-2}$
  (with dependence on the PMNS mixing angle $\theta_{13}$
  and $\CP$-violating phase $\delta$)
  Given the present bounds on $\mu \to e\gamma$, this implies
  $\BR(\tau \to \mu\gamma) \lsim 10^{-9}$ within the MLFV framework.
\end{itemize}

Once we accept the idea that flavour dynamics obeys a MFV principle,
both in the quark and in the lepton sector,
it is interesting to ask whether and how this is compatible with a
grand-unified theory (GUT),
where quarks and leptons are in the same representations
of a unified gauge group.
This question has recently been addressed in ref.~\cite{Grinstein:2006cg},
considering the case of $SU(5)_{\rm gauge}$ as an example.
Within this framework, the largest group of flavour transformations
commuting with the gauge group is
${\mathcal G}_{\rm GUT} = SU(3)_{\bar 5} \times SU(3)_{10}\times SU(3)_1$,
which is smaller than the direct product of the quark- and lepton-flavour
groups allowed in a non-GUT framework~\cite{D'Ambrosio:2002ex}.
We should therefore expect some violations of the MFV+MLFV predictions,
either in the quark or in the lepton sector or in both.
As far as LFV rates are concerned, the following
phenomenological conclusions can be drawn~\cite{Grinstein:2006cg}:
\begin{itemize}
\item
  Contrary to the non-GUT MFV framework,
  the rate of LFV decays cannot be arbitrarily suppressed by
  lowering the average mass $M_R$ of the heavy $\nu_R$.
  This fact can easily be understood by looking at the flavour structure
  of the relevant effective couplings.
  For $l_i \to l_j \gamma$ transitions, the following combination appears:
  \begin{equation}
    \label{eq:MFVgut}
    c_{1}~(Y_E Y_\nu^\dagger Y_\nu)_{ij} ~+~
    c_{2}~(Y_U Y_U^\dagger Y_E)_{ij}     ~+~
    c_{3}~(Y_U Y_U^\dagger Y_D^T)_{ij}\, +~\ldots
  \end{equation}
  where the $c_i$ are ${\cal O}(1)$ couplings.
  In addition to the terms involving $Y_\nu \sim \sqrt{M_R}$,
  already present in the non-unified case,
  the GUT group allows also $M_R$-independent terms
  involving the quark Yukawa couplings.
  The latter become competitive for $M_R \lsim 10^{12}$ GeV
  and their contribution is such that for $\Lambda \lsim 10$ TeV
  the $\mu \to e \gamma$ rate is above $10^{-13}$
  ({\it i.e.}~within the reach of MEG~\cite{Grassi:2005ac}).
\item
  Improved experimental information on $\tau \to \mu \gamma$ and
  $\tau \to e \gamma$ is a key tool.
  Their comparison with $\mu \to e \gamma$ is the best way
  to compare the relative size of the MLFV contributions
  with respect to the GUT-MFV contributions.
  In particular, if the quark-induced terms turn out to be dominant,
  then $\BR(\tau\to\mu\gamma) \propto (\lambda^2)^2$ and
  $\BR(\mu\to e\gamma) \propto (\lambda^5)^2$,
  where $\lambda \approx 0.22$.
  This implies a $\BR(\mu\to e\gamma)/\BR(\tau\to\mu\gamma)$
  ratio of ${\cal O}(10^{-4}$), which allows $\tau\to\mu\gamma$
  to be just below the present exclusion bounds.
\end{itemize}

\mysubsubsection{Little Higgs Models}
\label{sec:LHM}

Little Higgs Models address the tension between the naturalness of the electroweak scale and the precision electroweak measurements showing no eveidence for new physics up to $5-10$ TeV.

The Littlest Higgs model~\cite{Arkani-Hamed:2002qy} is based on a $SU(5)/SO(5)$ non-linear sigma model. It is strongly constrained by the electroweak precision data due to tree-level contributions of the new particles.

Implementing an additional discrete symmetry, the so-called T-parity~\cite{Cheng:2003ju}, makes the new particles contributing at the loop-level only and allows for a new-physics scale around $500$ GeV. It also calls for additional (mirror) fermions providing an interesting flavour phenomenology.

The high sensitivity for lepton flavour violation serves as an important tool
to test the littlest Higgs model with T-parity (LHT), in particular
to distinguish it from the MSSM~\cite{Blanke:2007db}.
Upper bounds on the branching ratio of lepton flavour violating $\tau$ decays are given in Table~\ref{tab:LHbounds}. Most of them could be within the reach of \superb.

However, large LFV branching ratios are not a specific feature of the LHT but a general property of many new physics model including the MSSM. Nevertheless, as Table~\ref{tab:LHratios} clearly shows, specific corrrelations are
very suitable to distinguish between the LHT and the MSSM.
The different ratios are a consequence of the fact that in the MSSM
the dipole operator plays the crucial role in these observables, while
in the LHT the $Z_0$ penguin and the box diagram contributions
are dominant.

\begin{table}
{
\begin{center}
\caption{
Upper bounds on LFV decay branching ratios in the LHT model with a new physics scale  $f=500 \ {\rm GeV}$, after imposing the constraints on $\mu\to e\gamma$ and $\mu^-\to e^-e^+e^-$.}
\begin{tabular}{|c|c|c|c|}
\hline
decay & $f=500 \ {\rm GeV}$  \\\hline\hline
$\tau\to e\gamma$ & ${1\cdot 10^{-8}}$\\
$\tau\to \mu\gamma$ &$2\cdot 10^{-8}$\\
$\tau^-\to e^-e^+e^-$ & ${2\cdot10^{-8}}$ \\
$\tau^-\to \mu^-\mu^+\mu^-$ & ${3\cdot10^{-8}}$ \\
$\tau^-\to e^-\mu^+\mu^-$ & ${2\cdot10^{-8}}$\\
$\tau^-\to \mu^-e^+e^-$ & ${2\cdot10^{-8}}$ \\
$\tau^-\to \mu^-e^+\mu^-$ & ${2\cdot10^{-14}}$ \\
$\tau^-\to e^-\mu^+e^-$ &${2\cdot10^{-14}}$ \\
$\tau\to\mu\pi$ & ${5.8\cdot10^{-8}}$ \\
$\tau\to e\pi$ & ${4.4\cdot10^{-8}}$ \\
$\tau\to\mu\eta$ & ${2\cdot10^{-8}}$ \\
$\tau\to e\eta$ & ${2\cdot10^{-8}}$\\
$\tau\to \mu\eta'$ & ${3\cdot10^{-8}}$ \\
$\tau\to e\eta'$ & ${3\cdot10^{-8}}$\\\hline
\end{tabular}
\end{center}
}
 \label{tab:LHbounds}
\end{table}

\begin{table}
{
\begin{center}
\caption{
Comparison of various ratios of branching ratios in the LHT model and in the MSSM without and with significant Higgs contributions.}
\vspace*{3mm}
\begin{tabular}{|c|c|c|c|}
\hline
ratio & LHT  & MSSM (dipole) & MSSM (Higgs) \\\hline\hline
$\frac{\BR(\tau^-\to e^-e^+e^-)}{\BR(\tau\to e\gamma)}$   & 0.4\dots2.3     &$\sim1\cdot10^{-2}$ & ${\sim1\cdot10^{-2}}$\\
$\frac{\BR(\tau^-\to \mu^-\mu^+\mu^-)}{\BR(\tau\to \mu\gamma)}$  &0.4\dots2.3     &$\sim2\cdot10^{-3}$ & $0.06\dots0.1$ \\
$\frac{\BR(\tau^-\to e^-\mu^+\mu^-)}{\BR(\tau\to e\gamma)}$  & 0.3\dots1.6     &$\sim2\cdot10^{-3}$ & $0.02\dots0.04$ \\
$\frac{\BR(\tau^-\to \mu^-e^+e^-)}{\BR(\tau\to \mu\gamma)}$  & 0.3\dots1.6    &$\sim1\cdot10^{-2}$ & ${\sim1\cdot10^{-2}}$\\
$\frac{\BR(\tau^-\to e^-e^+e^-)}{\BR(\tau^-\to e^-\mu^+\mu^-)}$     & 1.3\dots1.7   &$\sim5$ & 0.3\dots0.5\\
$\frac{\BR(\tau^-\to \mu^-\mu^+\mu^-)}{\BR(\tau^-\to \mu^-e^+e^-)}$   & 1.2\dots1.6    &$\sim0.2$ & 5\dots10 \\\hline
\end{tabular}
\end{center}
}

\label{tab:LHratios}
\end{table}

\mysubsubsection{Experimental Reach of LFV Decays}
\label{sec:tau_exp_reach_LFV}

 Experimentally, LFV decays can be conveniently classified as
$\tau\to\ell\gamma$, $\tau\to\ell_1\ell_2\ell_3$ and
$\tau\to\ell h$ where $\ell$ is either an electron or muon and
$h$ represents a hadronic system
({\it e.g.}, $\pi^0$, $\eta$, $\eta'$, $\KS$, {\it etc.})
The results of searches for LFV decays in data from \babar\ and \belle\
are summarized in Table~\ref{lfvtable}.
There is no evidence for LFV violating $\tau$ decays
and the individual experiments have each set
$90\%$ confidence level (CL) limits of order $10^{-7}$
using data sets of in the range of $100$ to $500 \ {\rm fb}^{-1}$.

The considerable experience developed in searching for these decays
in large data sets enables us to make projections of the sensitivities
to these decays with \superb\ delivering roughly
a 100-fold increase in the data set.
We express the experimental reach in terms of
``the expected $90\%$ CL upper limit'' that can be reached assuming no signal,
and, for brevity's sake, refer to this as the ``sensitivity''.
In the absence of signal,
for large numbers of background events $N_{\rm bkd}$,
the $90\%$ CL upper limit for the number of signal events can be given as
$N^{UL}_{90} \sim 1.64 \sqrt{N_{\rm bkd}}$,
whereas for small $N_{\rm bkd}$ a value for $N^{UL}_{90}$
is obtained using the method described in~\cite{Cousins:1991qz},
which gives, for  $N_{\rm bkd} \sim 0$, $N^{UL}_{90} \sim 2.4$.
Schematically, the $90\%$ CL branching ratio upper limit is then
\begin{equation}
  \B^{UL}_{90} =
  \frac{N^{UL}_{90}}{2 N_{\tau\tau} \epsilon} =
  \frac{N^{UL}_{90}}{2 \cal{L}\sigma_{\tau\tau} \epsilon}\,,
\end{equation}
where  $N_{\tau\tau}=\cal{L}\sigma_{\tau\tau}$
is the number of $\tau$-pairs produced in $e^+e^-$ collisions;
$\cal{L}$ is the integrated luminosity,
$\sigma_{\tau\tau}$ is the $\tau$-pair production cross section and
$\epsilon$ is the reconstruction efficiency.
We have based our projections on \babar\ analyses of
$\tau\to \mu\gamma$, $\tau\to e\gamma$, $\tau\to\ell_1\ell_2\ell_3$ and
$\tau\to\ell h h'$~\cite{Aubert:2005ye,
  Aubert:2005wa,Aubert:2003pc,Aubert:2005tp}
and \belle\ analyses of $\tau\to\ell\pi^0$, $\tau\to\ell\eta$,
$\tau\to\ell\eta^\prime$ and $\tau\to\ell \KS$~\cite{Enari:2005gc,Yusa:2002ff}.

The experimental signature for
LFV $\tau$ decays is extremely clean.
In $e^+e^- \to \tau^+\tau^-$ events at $\sqrt{s} \sim m_{\Upsilon(4{\rm S})}$,
the event can be divided into hemispheres in the center-of-mass frame,
each containing the decay products of one $\tau$ lepton.
Furthermore, unlike Standard Model $\tau$-decays, which contain at least one neutrino,
the LFV decay products have a combined energy in the center-of-mass frame
equal to $\sqrt{s}/2$ and a mass equal to that of the $\tau$.
A two dimensional signal region in the $E_{\ell X}$--$M_{\ell X}$ plane
therefore provides a powerful tool to reject background,
which usually arise from well-understood Standard Model $\tau$ decays.
Consequently, residual background rates and distributions
can be reliably estimated from Monte Carlo.

The estimated physics reach of \superb\ based on projections
from existing analyses depends on how the background is treated.
A ``worst-case scenario'' is obtained if identical analyses to those published
by \babar\ and \belle\ are repeated on a sample with more data:
the expectations then simply scale as
$\sim \sqrt{N_{\rm bkd}}/{\cal L}$,
which for large $N_{\rm bkd}$ scales as $1/\sqrt{{\cal L}}$.
A ``best case'' scenario would take the current expected limit
and scale linearly with the luminosity.
This is equivalent to a statement that analyses can be developed
maintaining the same efficiency and backgrounds as the current analyses.

For $\tau\to\ell\gamma$, there is an ``irreducible background'' from
$\tau\to\ell\nu\nu +  \gamma ({\rm ISR})$
in which the photon from initial state radiation can be combined
with a lepton to form a candidate that accidentally overlaps
with the signal region in the $E_{\ell X}$--$M_{\ell X}$ plane.
In the existing \babar\ analyses,
these events account for approximately one fifth of the total background.
We therefore consider a ``realistic'' scenario,
in which this source of background is present
at the rate determined with the existing analyses,
while all other backgrounds are suppressed with
minimal cost to the signal efficiency.
Note, however, that improvements on this ``realistic'' scenario
are possible if the $\ell\gamma$ mass resolution is improved,
which could be achieved by improving the spatial resolution
of the electromagnetic calorimeter (see Section~\ref{sec:det:EMC}).
Additional signal-to-background gains can be made by restricting
the polar-angle acceptance of the $\gamma$
thereby reducing initial state radiation (ISR)-related backgrounds, at the cost of efficiency.

The situation for the other LFV decays,
$\tau\to\ell_1\ell_2\ell_3$\ and $\tau\to\ell h$,
is even more promising, since these modes do not suffer
from the aforementioned backgrounds from ISR.
In this case, one can project sensitivities assuming $N_{\rm bkd}$
comparable to backgrounds in existing analyses
for approximately the same efficiencies.
Table~\ref{LFVExptSensitivities} summarizes the sensitivities
for various LFV decays.

\begin{table}[thb]
  \caption{
    \label{LFVExptSensitivities}
    Expected $90\%$ CL upper limits on representative LFV $\tau$ lepton decays
    with $75 \ {\rm ab}^{-1}$.
  }
  \begin{center}
    \begin{tabular}{ll}
      \hline \hline
      Process &  Sensitivity \\
      \hline
      $\BR(\tau \to \mu\,\gamma)$          &  $2 \times 10^{-9}$     \\
      $\BR(\tau \to e\,\gamma)$            &  $2 \times 10^{-9}$     \\
      $\BR(\tau \to \mu\, \mu\, \mu)$      &  $2 \times 10^{-10}$  \\
      $\BR(\tau \to e e e )$               &  $2 \times 10^{-10}$  \\
      $\BR(\tau \to \mu \eta)$             &  $4 \times 10^{-10}$    \\
      $\BR(\tau \to e \eta)$               &  $6 \times 10^{-10}$    \\
      $\BR(\tau \to \ell \KS)$           &  $2 \times 10^{-10}$    \\
      \hline
    \end{tabular}
  \end{center}
\end{table}

\mysubsection{Lepton Universality in Charged Current $\tau$ Decays}

Precise tests of lepton-flavour universality (LFU) in charged-current
interactions (CCI) represent a complementary window on New Physics.
In fact, within the Standard Model, possible departures from the LFU in $\tau$ decays,
described by $R_{\tau}^{\mu/\tau} \equiv
\Gamma(\tau\to \mu\nu\bar{\nu})/\Gamma(\mu\to e\nu\bar{\nu})$
are predicted to be
\begin{equation}
  |R_{\tau}^{\mu/\tau} - (R_{\tau}^{\mu/\tau})_{\rm SM}| =
  {\cal O}[(\alpha/4\pi)\times(m^{2}_{\tau}/M^{2}_{W})]\,,
\end{equation}
and thus completely negligible.

Violations of LFU in CCI can be classified as:

i) corrections to the strength of the effective $(V-A)\times(V-A)$
four-fermion interaction,
ii) four-fermion interactions with new Lorentz structures.
\par\noindent
As an example of class i), we mention the $W\ell\nu_{\ell}$
vertex correction through a loop of New Physics particles:
the induced effect is of order
$(\alpha/4\pi) \times (M^{2}_{W}/M^{2}_{NP}) < 10^{-4}$,
which is hardly measurable.
Class ii) is definitely more promising:
the typical example is the scalar current induced by
tree level Higgs exchange, with mass-dependent coupling
($H\ell\nu \sim m_{\ell} \tan\beta$).
In this case, it has been shown that~\cite{Krawczyk:1987zj},
\begin{equation}
  \begin{array}{rcl}
    R_{\tau}^{\mu/\tau} & \approx &
    (R_{\tau}^{\mu/\tau})_{\rm SM} \times
    \left[
      1-2 \frac{m^{2}_{\mu}}{M_{H^\pm}^2}\tan^{2}\beta
    \right] \\
    & \approx &
    (R_{\tau}^{\mu/\tau})_{\rm SM} \times
    \left[
      1-10^{-3}
      \left(\frac{200 \ {\rm GeV}}{M_{H^\pm}}\right)^2
      \left(\frac{\tan \beta}{50}\right)^2
    \right]\,.
  \end{array}
  \label{LFU}
\end{equation}
Note that the same relative effect can be seen in the ratio
$\Gamma(\tau\to \mu\nu\bar{\nu})/\Gamma(\tau\to e\nu\bar{\nu})$,
which can be determined at the ${\cal O}(10^{-3})$ level
at \superb.
A non-Standard Model effect at this level would have
a rather precise interpretation within the MSSM:
large $\tan\beta\geq 40$ and small $M_{H^{\pm}} \sim 200-300 \ {\rm GeV}$.
On the other hand, it must be stressed that a detailed re-analysis
of the Standard Model predictions of such ratios within the Standard Model
(which are fine with present techniques)
would be necessary in view of very precise measurements.

As pointed out in Refs.~\cite{Masiero:2005wr,Isidori:2006pk},
precise tests of LFU in CCI represent a complementary window on
Higgs-mediated LFV amplitudes.
These effects are particularly interesting within the MSSM at large
$\tan\beta$, and could rise to visible effects both in
$K_{\ell 2}$~\cite{Masiero:2005wr} and in
$B_{\ell 2}$ decays~\cite{Isidori:2006pk}.
However, while in the $K_{\ell 2}$ and $B_{\ell 2}$ cases
sizable LFU breaking effects can be induced with only
LFV couplings, in $\tau$ decays LFU breaking effects
are mainly generated by LF-conserving (but mass-dependent) couplings,
such as in Eq.~(\ref{LFU}), while LFV effects provide only
a second order correction. From this point of view,
the study of LFU breaking in $K_{\ell 2}$, $B_{\ell 2}$
and $\tau$ decays can be regarded as complementary tools to shed
light on New Physics effects, given their sensitivity to different New Physics contributions.

\mysubsubsection{
  Charged Current Universality Measurements}
\label{sec:tau_exp_reach_CCUniv}

Charged current universality is probed in $\tau$ decays via:
\begin{equation}
  \label{eq:CCUniv1}
  \tau_{\tau} =
  \tau_{\mu} \frac{g_{\mu}^2}{g_{\tau}^2} \frac{m_{\mu}^5}{m_{\tau}^5}
  \BR(\tau^- \to e^- \bar{\nu_e} \nu_{\tau} )
  \frac{f(m_{e}^2/m_{\mu}^2)r^{\mu}_{RC}}{f(m_e^2/m_{\tau}^2)r^{\tau}_{RC}}
\end{equation}
\begin{equation}
  \label{eq:CCUniv2}
  \tau_{\tau} =
  \tau_{\mu} \frac{g_{e}^2  }{g_{\tau}^2} \frac{m_{\mu}^5}{m_{\tau}^5}
  \BR(\tau^- \to \mu^- \bar{\nu_{\mu}} \nu_{\tau} )
   \frac{f(m_{e}^2/m_{\tau}^2)r^{\mu}_{RC}}
        {f(m_{\mu}^2/m_{\tau}^2)r^{\tau}_{RC}}\,,
\end{equation}
where the $g_{e}$, $g_{\mu}$ and $g_{\tau}$ are CC couplings,
all equal to unity in the Standard Model,
but different from each other in extensions of the Standard Model;
$f(x)=1-8x+8x^3-x^4-12x\ln(x)$ is a phase-space factor and
$r^{\ell}_{RC}$ are radiative corrections.
Equation~\ref{eq:CCUniv1} indicates that $g_{\mu}/g_{\tau}$
can be determined from measurements of the masses and lifetimes
of the $\tau$ and $\mu$ and the electronic branching fraction of the $\tau$.
A precise determination of $g_{\mu}/g_{\tau}$ is currently limited
by the errors on the measurements of
$\BR(\tau^- \to e^- \bar{\nu_e} \nu_{\tau}$) and of the $\tau$ lifetime.
Using the world average values~\cite{Lusiani:2005sy,Yao:2006px}
($\BR(\tau^- \to e^-   \bar{\nu_e}     \nu_{\tau}) = (17.824 \pm 0.052)\%$,
 $\BR(\tau^- \to \mu^- \bar{\nu_{\mu}} \nu_{\tau}) = (17.331 \pm 0.048)\%$,
 $\tau_{\tau} = (290.15 \pm 0.77) \ {\rm fs}$)
the ratio of the $\mu$ to $\tau$ charged current coupling constants
is found to be $g_{\mu}/g_{\tau}=0.9982\pm 0.0021$.
The ratio of Eq.~\ref{eq:CCUniv1} to Eq.~\ref{eq:CCUniv2}
indicates that $g_{\mu}/g_{e}$ only requires measurements
of the two leptonic branching fractions.
From world averages of these branching ratios,
the ratio of the $\mu$ to $e$ charged current coupling constants
is found to be $g_{\mu}/g_e = 0.9999 \pm 0.0020$.

Charged current universality can be probed at the $0.05\%$ level
if measurements of leptonic branching ratios
and the lifetime are controlled at better than $0.1\%$.
A determination of $g_{\mu}/g_e$ will require accurate control over:
the differences in the trigger and filter efficiencies between
events with an electron compared to those with a muon;
differences in tracking efficiency for electrons and muons;
the electron particle ID; the muon particle ID.
The latter two can be determined from control samples in the data from
$\mu$-pair and radiative Bhabha events.
As there will be cuts placed on the momentum spectra of the leptons,
there will be some sensitivity to assumptions of the
Lorentz structure of the decays that would have to be taken into account.
Because this will depend on a relative measurement,
the large data sample at \superb\ will make it possible
to trade-off considerable numbers of events to bring systematic errors
under control.
One can therefore expect $g_{\mu}/g_e$ to be determined to
better than $0.05\%$ at \superb.

More challenging will be a determination of $g_{\mu}/g_{\tau}$, which will
require an absolute measurement of the electronic branching fraction,
as well as the $\tau$ lifetime.
The absolute branching fraction measurement will require
(in addition to the absolute trigger, filter, tracking,
and particle ID efficiencies)
the absolute luminosity to be known with precision better than $0.1\%$.
The LEP experiments, using specialized luminosity detectors,
achieved a precision of $0.05\%$,
which was dominated by the theory cross section uncertainty.
The cross sections will have to be calculated to a comparable precision.
To improve the lifetime measurement, one must overcome the challenges of
backgrounds and detector alignment, as well as selection biases.
It is possible to approach a precision of $0.10\%$.
Consequently, it may be possible for $g_{\tau}/g_{\mu}$
to be determined to a precision of ${\cal O}(10^{-3})$ at \superb.

\mysubsubsection{
$\CPT$ Tests with the $\tau$ Lepton}
\label{sec:tau_exp_reach_CPT}

The \superb\ data set opens up a new window on $\CPT$ tests of
the third generation charged lepton from measurements of
the difference between the lifetimes and masses of the $\tau^-$ and $\tau^+$.
The difference in lifetimes,
$\frac{\tau_{\tau-} - \tau_{\tau+}}{\tau_{\tau-} + \tau_{\tau+}}$,
currently has a value of $(0.12 \pm 0.32)\%$ (the error is statistical only)
from a preliminary \babar\ result
that uses less than $100 \ {\rm fb}^{-1}$~\cite{Lusiani:2005sy}.
With \superb\ data, the statistical precision will reach $10^{-4}$.
Most systematic errors cancel in this test
but care is needed in the selection process in order to avoid
effects of known differences in hadronic interaction cross sections
for $\pi^+$ and $\pi^-$.
A reach of $10^{-4}$ compares favourably with the analogous test
for muons, which currently has a measured value of
$\frac{\tau_{\mu-} - \tau_{\mu+}}{\tau_{\mu-} + \tau_{\mu+}} =
(2 \pm 8) \times 10^{-5}$~\cite{Bardin:1984ie,Yao:2006px}.

A similar test related to the $\tau$ mass has been performed by
\belle~\cite{Abe:2006vf} using $414 \ {\rm fb}^{-1}$
and a pseudo-mass observable:
$m_{\tau-} - m_{\tau+} = (0.05 \pm 0.23 \pm 0.14) \ {\rm MeV}/c^2$.
This gives a $90\%$ CL limit on $\CPT$ violation of
$|\frac{m_{\tau-} - m_{\tau+}}{m_{\tau}}| < 2.8 \times 10^{-4}$.
The $0.14 \ {\rm MeV}/c^2$ systematic error is dominated
by assessments of potential charge asymmetries in the detector
using charmed meson control samples, such as $D^0\to K^-\pi^+$.
At \superb\ one expects a statistical error of $\sim 0.025 \ {\rm MeV}/c^2$.
In order to fully exploit the statistical power of such a test,
charge asymmetric momentum scales would have to be controlled
at the $10^{-5}$ level,
which is a challenging detector systematics problem.
Nonetheless, a $\CPT$ test at such precision would represent
one of the most precise fundamental fermion mass tests of $\CPT$ available.

\mysubsection{New Physics from $\CP$ Violation in the $\tau$ System}
\label{sec:tau_exp_reach_CPV}

The quantitative confirmation of the Standard Model mechanism for $\CP$ violation in both the kaon and $B$ meson systems means that there must be additional sources of $\CP$ violation beyond the Standard Model, if we are
to explain the dominance of matter over antimatter in the universe.
The origin of the non-Standard Model $\CP$ violation remains one of the most important
problems in physics;
it is thus important to look for the phenomenon in as many systems as possible.
Searches for $\CP$ violation in $\tau$ decays have been proposed~\cite{Tsai:1994rc,Tsai:1996bu,Kuhn:1996dv,Kuhn:1997wr,Datta:2006kd},
as the observation of a non-zero $\CP$ asymmetry in $\tau$ decays
would be a clear and unambiguous signature for New Physics.
Since all $\CP$-violating effects result from
the interference of at least two amplitudes with a relative phase and,
since in the Standard Model the $\tau$ decays via a single decay amplitude,
there can be no $\CP$-violating asymmetries in Standard Model $\tau$ decays
(apart from a Standard Model $\CP$ asymmetry of ${\cal O}(10^{-3})$
in $\tau\to\pi \KS \nu_{\tau}$ arising from
$\CP$ violation in the neutral kaon system~\cite{Bigi:2005ts}).

With unpolarized $\tau$ leptons, for example,
one can measure the branching ratios of $\tau$ decays
to at least two hadrons and determine if, for example,
$\BR(\tau^- \to K^-\pi^0 \bar{\nu}_{\tau})$ is equal to
$\BR(\tau^+ \to K^+\pi^0 \nu_{\tau})$.
However, such a simple asymmetry is not expected
in many conventional New Physics models.

There could be a $\CP$-violating asymmetry in multi-Higgs doublet models
in which the $\tau$ decay can proceed through charged Higgs exchange
in addition to the Standard Model decay via
a virtual $W$~\cite{Grossman:1994jb,Kuhn:1996dv,Kuhn:1997wr,Datta:2006kd}.
In these scenarios, $\CP$ violation arises from the interference
between the $W$ vector boson and the scalar charged Higgs amplitudes.
One of the most promising $\tau$ decay modes in multi Higgs doublet models
is  $\tau^{\pm} \to K^{\pm}\pi^0\nu_{\tau}$~\cite{Kuhn:1996dv,Kuhn:1997wr},
where charged Higgs exchange would
modify the scalar form factor in the hadronic matrix element.
Transitions from the QCD vacuum to two pseudoscalar mesons,
$h_1 = K^{\pm}$ and $h_2 = \pi^0$,
can proceed only through the vector and scalar currents.
The hadronic matrix elements can be expanded along
the set of independent momenta,
$(q_1-q_2)_{\beta}$ and $Q_{\beta}=(q_1+q_2)_{\beta}$:
\begin{equation}
  \langle h_1(q_1)h_2(q_2) \mid \bar{u}\gamma_{\beta}d \mid 0 \rangle =
  (q_1-q_2)^{\alpha}T_{\alpha\beta}F(Q^2)+Q_{\beta}F_S(Q^2)\, ,
\end{equation}
where $T_{\alpha\beta}=g_{\alpha\beta} - (Q_{\alpha}Q_{\beta}/Q^2)$,
$F(Q^2)$ is the vector form factor associated with
the $J^P=1^-$ component of the weak charged current and
$F_S(Q^2)$ is the scalar form factor corresponding to the $J^P=0^+$ component.
A charged Higgs exchange contribution is introduced as
a term proportional to $\eta_S F_H(Q^2)$ where:
\begin{equation}
  F_H(Q^2) = \langle h_1(q_1)h_2(q_2) | \bar{u} d | 0 \rangle \, .
\end{equation}
The complex parameter $\eta_S$ transforms under $\CP$ as $\eta_S \to \eta_S^*$
thereby allowing for the parametrization of possible $\CP$ violation.
The general amplitude for the decay of a $\tau$ with spin $s$,
$\tau(l,s) \to \nu(l',s') + h_1(q_1,m_1) + h_2(q_2,m_2)$,
can be written as:
\begin{equation}
  {\cal M} =
  \sin\theta_C \frac{G}{\sqrt{2}} \bar{u}(l',s')\gamma_\alpha(1-\gamma_5)u(l,s)
  [(q_1-q_2)_\beta T^{\alpha\beta}F+Q^\alpha\tilde{F_S}]
\end{equation}
where
\begin{equation}
  \tilde{F_S} = F_S + \frac{\eta_S}{m_{\tau}}F_H \, .
\end{equation}
One searches for the presence of a $\CP$-violating phase by
comparing the structure functions $W_{SF}$ and $W_{SG}$~\cite{Kuhn:1992nz}
measured in $\tau^+$ and $\tau^-$ decays,
\begin{equation}
  \Delta W_{SF} =
  \frac{1}{2}(W_{SF}[\tau^-] - W_{SF}[\tau^+]) =
  4\sqrt{Q^2}\frac{|q_1|}{m_{\tau}} \Im(FF_H^*)\Im(\eta_S)
\end{equation}
\begin{equation}
  \Delta W_{SG} =
  \frac{1}{2}(W_{SG}[\tau^-] - W_{SG}[\tau^+]) =
  4\sqrt{Q^2}\frac{|q_1|}{m_{\tau}} \Re(FF_H^*)\Im(\eta_S) \, .
\end{equation}
The structure functions $W_X$
are defined from the hadronic tensor $H^{\mu\nu}=J^{\mu}J^{\nu *}$
in the hadronic rest frame~\cite{Kuhn:1992nz}.
A non-zero value for $\Delta W_{SF}$ or $\Delta W_{SG}$
signals $\CP$ violation, and hence New Physics.

$\Delta W_{SF}$ is obtained from an analysis of the difference
in the correlated energy distribution of the charged $K$ and $\pi^0$
in $\tau^+$ and $\tau^-$ decays in the lab.
It can be determined by studying single unpolarized $\tau$ decays
produced at \superb. where the precision will be limited by the understanding
of the charge asymmetry in the detector response, which will be dominated
by the differences in the interaction of $K^-$ and $K^+$
in the material of the detector.
This can be controlled by studying the response of $K^{\pm}$ from,
for example, $\tau^-\to K^-K^+\pi^-\nu_{\tau}$
or from $K^{\pm}$ in charm decays.

The measurement of $\Delta W_{SG}$ requires knowledge of
the full kinematics and polarization of the $\tau$.
The full kinematics can be determined using vertex detectors.
The component of the $\tau$ polarization along the $\tau^-$ direction
can be obtained from the longitudinal beam polarizations ($w_{e^-}$ and $w_{e^+}$)
as a function of the $\tau^-$ production angle and energy,
 $E_{\tau}$~\cite{Tsai:1994rc}.
It is important to recognize, however, that only one of the beams needs to
 be polarized to obtain a non-zero $\tau$ polarization.

Searching for $\CP$ violation via $\Delta W_{SG}$
is similar to the methods proposed in~\cite{Tsai:1994rc,Tsai:1996bu},
which suggest using a $T$-odd rotationally invariant product,
such as   $P_Z^{\tau} \cdot (\vec{p}_{K^+} \times \vec{p}_{\pi^0} )$,
where $P_Z^{\tau}$ is the component of the $\tau$ polarization along the beam axis
 averaged over the production angle:
\begin{equation}
  P^{\tau}_Z =
  \frac{w_{e^-}+w_{e^+}}{1+w_{e^-}w_{e^+}}
  \frac{1+2m_{\tau}/E_{\tau}}{2 + m_{\tau}^2/E_{\tau}} \, .
\end{equation}

Other multi-meson decay modes should also be considered, and at \superb\
the large data sample will offset the smaller branching ratio to these modes.
In particular, $\tau^-\to a_1^- \pi^0 \nu_{\tau}$ decays could well be a
fruitful mode for $\CP$ violation searches~\cite{Datta:2006kd}.
In this case a polarization-dependent rate asymmetry is the most
sensitive observable.
The asymmetry is present with no polarization but can grow by large factors
as a function of $Q^2$ as the polarization changes from zero to one.
This not only gives an enhancement of an $\CP$ violating signal,
but with tunable polarization,
provides for a powerful systematic control over the $\CP$ asymmetry observable.

A polarized beam also provides for a very sensitive probe
of the electric dipole moment (EDM) of the $\tau$.
This observable is sensitive to $\CP$ violation in $\tau$ production and,
with $75 \ {\rm ab}^{-1}$ it is estimated that an upper limit
on the EDM of the $\tau$ of
$7.2 \times 10^{-20} e \, {\rm cm}$ can be achieved~\cite{Bernabeu:2006wf}.
This represents an improvement on the current limits~\cite{Inami:2002ah}
of about three orders of magnitude.
In light of existing limits on the EDM of the electron,
it would be surprising to observe such a large value and therefore
searches for a non-zero EDM at \superb\ would be probes for
non-``standard'' New Physics $\CP$ violation.

\afterpage{\clearpage}

\mysection{$B_s$ Physics at the $\Upsilon(5{\rm S})$}
\label{sec:bs}
\mysubsection{Running at the $\Upsilon(5{\rm S})$}

Measurement of CKM- and New Physics-related
quantities in the $B_s$ sector is a natural extension of the traditional $B$~Factory program.
In some cases, studies of $B_s$ mesons allow the extraction of
the same fundamental quantities accessible at a $B$~Factory
operating at the $\Upsilon(4{\rm S})$ resonance,
but with reduced theoretical uncertainty.
Experiments running at hadronic machines are expected
to be the main source of $B_s$-related measurements.
In particular, in the near future,
the increased dataset of the Tevatron experiments and the start
of the \lhcb, ATLAS, and CMS programs will surely yield important new results.

It is also worth noting, however, that despite the rapid $\BsBsb$ oscillation frequency, it is
also feasible to carry out $\Bs$ studies in the very clean environment of
$e^+e^-$ annihilation machines by running at the $\FiveS$ resonance,
where it is possible to perform measurements involving neutral particles
({\it e.g.}, $\pi^0$, $\eta$ and $\eta^\prime$ mesons,
radiative photons, {\it etc.})
CLEO~\cite{Artuso:2005xw,Bonvicini:2005ci,Huang:2006em}
and \belle~\cite{Drutskoy:2006fg,Abe:2006xc}
have had short runs at the $\FiveS$,
measuring the main features of this resonance.
The results clearly indicate the potential for an $e^+e^-$ machine
to contribute to this area of $B$ physics,
and have inspired the work in this section,
and elsewhere~\cite{Hou:2006mx,Hou:2007ps,Baracchini:2007ei}.
Note that, in contrast to much of the remainder of this chapter, there are
no experimental analyses for many of the measurements of interest,
and therefore our studies are based on Monte Carlo simulations.

The production of $B_s$ mesons at the $\FiveS$ allows comprehensive
studies of the decay rates of the $B_s$ with a completeness
and accuracy comparable to that currently available for $B_d$ and $B_u$ mesons,
thereby improving our understanding of $B$ physics and helping to reduce
the theoretical uncertainties related to
New Physics-sensitive $B_d$ quantities.
Moreover, $B_s$ physics provides additional methods
to probe New Physics effects in $b \to s$ transitions.
In the following, we concentrate on this second point,
providing examples of some of the highlight measurements
that could be performed by \superb\
operating at the $\FiveS$ resonance.

The $\FiveS$ resonance is a $J^{PC} = 1^{--}$ state
of a $b\bar{b}$ quark pair, having an invariant mass of
$m_{\FiveS} = (10.865 \pm 0.008) \ {\rm GeV}/c^2$~\cite{Besson:1984bd,Lovelock:1985nb,Yao:2006px}.
The cross section of $\FiveS$ production in $e^+e^-$ collisions is
$\sigma(e^+e^- \to  \FiveS =
0.301 \pm 0.002 \pm 0.039 \ {\rm nb}$~\cite{Huang:2006mf}, which corresponds to about one third of the $\FourS$ one.
Unlike the $\FourS$ state, this resonance is sufficiently massive to
decay into several $B$ meson states:
vector-vector ($B^* \bar{B}^*$),
pseudoscalar-vector ($B \bar{B}^*$),
and pseudoscalar-pseudoscalar ($BB$) combinations of charged $B$ mesons,
as well as neutral $B_d$ and $B_s$ mesons, as well as into
$B^{(*)}\bar B^{(*)} \pi$ states.
Tab.~\ref{tab:Y5sinput} shows the current experimental status
of $B$ pair production rates,
along with the values used in the study presented in this section.

\begin{table}[htb]
    \caption{
      \label{tab:Y5sinput}
      $\Upsilon(5{\rm S})$ decay branching ratios as measured by
      CLEO~\cite{Huang:2006mf} and \belle~\cite{Drutskoy:2006dw}.
      The last column shows the values used throughout this section.
    }
  \begin{center}
    \begin{tabular}{lccc}
      \hline
      \hline
      $\Upsilon(5{\rm S})$ Decay Modes & CLEO & \belle & This Study \\
       \hline
      $B_s^{(*)} \bar B_s^{(*)}$  (\%) & $26^{+7}_{-4}$ & $21^{+6}_{-3}$ & 26 \\
      $(B_s^* \bar B_s^*)/(B_s^{(*)}\bar B_s^{(*)})$ & $ - $ & $0.94^{+0.06}_{-0.09}$ & 0.94 \\
      $(B_s^* \bar B_s + B_s \bar B_s^*)/(B_s^{(*)}\bar B_s^{(*)})$ & $ - $ & $ - $ & 0.03 \\
      $(B_s \bar B_s)/(B_s^{(*)}\bar B_s^{(*)})$ & $ - $ & $ - $ & 0.03 \\
 & \\
      $B_d^* \bar B_d^*$ (\%) & $ 43.6 \pm 8.3 \pm 7.2 $ & $ - $ & 44 \\
      $B_d \bar B_d^* + B_d^* \bar B_d $ (\%) & $ 14.3 \pm 5.3 \pm 2.7 $ & $ - $ &  7 \\
      $B_d \bar B_d$ (\%) & $ <13.8 $ & $ - $ & 7 \\
 & \\
      $B_d \bar B_d^{(*)} \pi + B_d^{(*)} \bar B_d \pi$ (\%) & $ <19.7 $ & $ - $ & 16 \\
      $B_d \bar B_d \pi \pi$ (\%) & $< 8.9$ & $ - $ & - \\
      \hline
    \end{tabular}
  \end{center}
\end{table}

\begin{figure}[!htbp]
  \begin{center}
    \includegraphics[width=9cm]{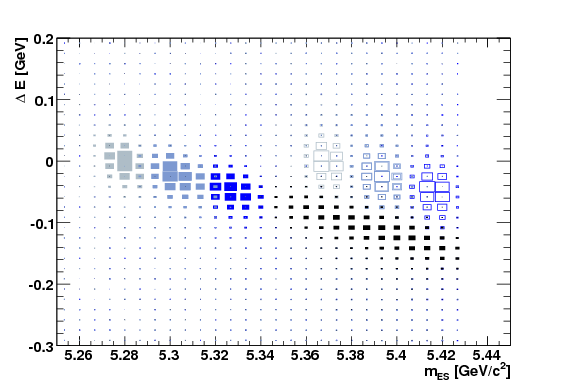}
    \caption{
      Distribution of $\Delta E$ {\it vs.} $m_{\rm ES}$
      for a sample of simulated $B_{d,s}$ mesons
      produced at the $\FiveS$ resonance and
      decaying into $J/\psi\,\phi$ final states.
      Events coming from $B_q^{(*)}B_q^{(*)}$ ($q=d,s$)
      are all generated with the same relative rate.
      We use full boxes for $q=d$ and empty boxes for $q=s$.
      The colour scale identifies VV, VP and PP events
      (from the darker to the lighter).
      Events from $B_d B_d \pi$ events are also shown (black boxes).
    }
    \label{fig:mesvsde}
  \end{center}
\end{figure}

The multiplicity of possible final states implies different momenta
for the produced $B \overline B$ pairs and affects the reconstruction methods.
In particular, the distribution of the usual discriminating variables
$m_{\rm ES}$ and $\Delta E$ is different depending on the final state,
as shown in Fig~\ref{fig:mesvsde}.
This feature is extremely helpful in isolating
the different final states in the $(m_{\rm ES},\Delta E)$ plane.
With the small beam energy spread of \superb,
the resolution of $m_{\rm ES}$ will be comparable to the current $B$~Factories,
resulting in almost negligible crossover between
$\BsBsb$ and $B\Bbar \pi$ states.
We have taken this small effect into account in our simulations.

\mysubsection{Measurement of $B_s$ Mixing Parameters}
\label{sec:U5s_timeind}

The absolute value and the phase of the $\BsBsb$ mixing amplitude
can be used to test for the presence of New Physics in
$\Delta B=2$ $b \to s$ transitions.
These measurements can be made at hadronic colliders~\cite{Bona:2006sa}.
The recent measurement of
$\Delta m_s$~\cite{Abulencia:2006ze,Abulencia:2006mq,Abazov:2006dm}
provides the first milestone in this physics program.
These studies exploit the high Lorentz boost
$\beta \gamma$ of $B_s$ mesons produced at high energy hadronic colliders;
the rapid $B_s$ oscillations can be resolved, with
current vertex detector spatial resolution ($\sim 100\,\mu{\rm m}$),
only with a large boost.

Similar tests for New Physics effects can be made by measuring
quantities such as $\Delta \Gamma_s$ and the $\CP$ asymmetry in
semileptonic decays $A^s_{\rm SL}$,
which can be done at \superb,
taking advantage of the large statistics,
high efficiency of lepton reconstruction, and low backgrounds.
These measurements do not require the $B_s$ oscillations to be resolved.

In a generic New Physics scenario, the effect of $\Delta B = 2$ New Physics contributions
can be parameterized in terms of two quantities, $C_{B_s}$ and $\phi_{B_s}$,
given by the relation (see also Section~\ref{sec:U4s_pheno}):
\begin{equation}
  C_{B_s} \, e^{2 i \phi_{B_s}} =
  \frac{
    \langle B_s | H_{\rm eff}^{\rm full} | \overline{B}_s \rangle
  }{
    \langle B_s | H_{\rm eff}^{\rm SM}   | \overline{B}_s \rangle
  }
  \,.
  \label{eq:paranp_bs}
\end{equation}

In the absence of New Physics effects, $C_{B_s}=1$ and $\phi_{B_s}=0$, by definition.
The measured values of $\Delta m_s$ and $\sin 2\beta_s$
(discussed in Section~\ref{sec:U5s_timeind}) are related to Standard Model
quantities through the relations :
\begin{equation}
  \Delta m_s^{\rm exp} = C_{B_s} \cdot \Delta m_s^{\rm SM}
  ~~~;~~~
  \sin 2\beta_s^{\rm exp} = \sin (2\beta_s^{\rm SM} + 2\phi_{B_s}) \,.
\end{equation}

The semileptonic $\CP$ asymmetry~\cite{Laplace:2002ik}
and the value of $\Delta \Gamma_s/\Gamma_s$~\cite{Dunietz:2000cr}
are sensitive to New Physics contributions to the $\Delta B=2$ effective Hamiltonian,
and can be expressed in terms of the parameters $C_{B_s}$ and $\phi_{B_s}$.

\begin{figure}[!htbp]
  \begin{center}
    \includegraphics[width=9cm]{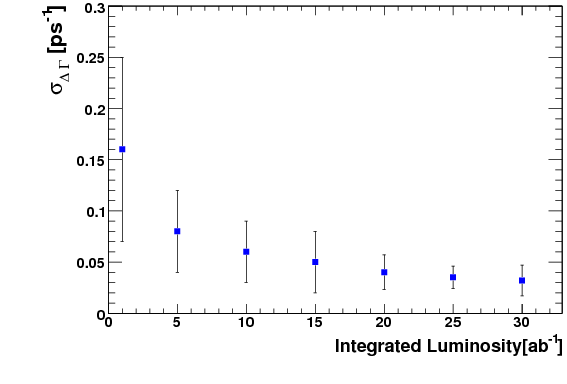}
    \includegraphics[width=9cm]{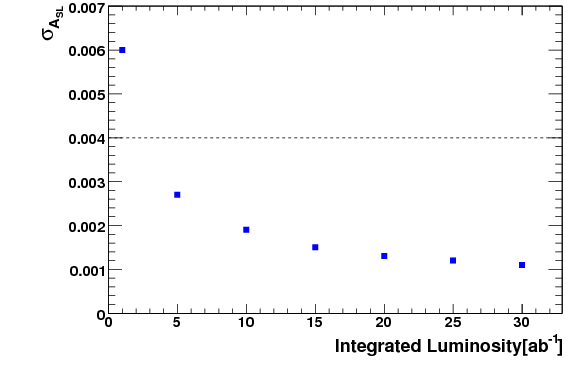}
    \includegraphics[width=9cm]{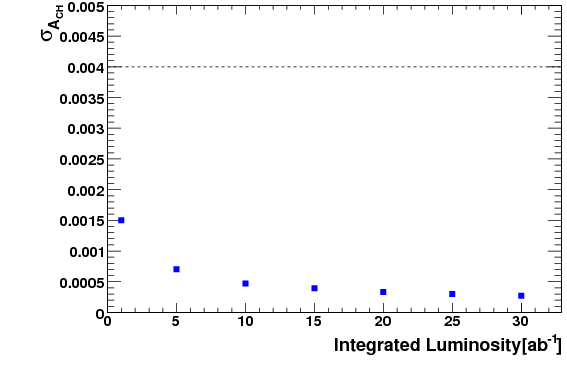}
    \caption{
      Trend of the error on $\Delta \Gamma_s$, $A_{\rm SL}$ and $A_{\rm CH}$
      as a function of the integrated luminosity.
      The error bars show the {\it rms} of the error distribution
      in the toy Monte Carlo experiments.
      The dashed line in the last two plots represents
      the systematic error on the current measurements
      at the $\FourS$ resonance, shown for comparison.
      \label{fig:dg}
    }
  \end{center}
\end{figure}

Different experimental methods have been proposed to extract the
lifetime difference $\Delta \Gamma_s$~\cite{Dighe:1998vk}.
For instance, $\Delta \Gamma_s$ can be obtained from the angular distribution
of untagged $B_s \to J/\psi \phi$ decays.
This angular analysis allows separation of the $\CP$ odd and $\CP$ even
components of the final state, which have a distinct time evolution,
given by different combinations of the
two exponential factors $e^{-\Gamma_{L,H}t}$.
This allows the extraction of the two parameters $\Gamma_{L,H}$
or, equivalently, $\Gamma_s$ and $\Delta \Gamma_s$.
The weak phase of the mixing amplitude, $\beta_s$, also
appears in this parametrization, and a constraint on this phase
can be extracted along with the other two parameters
(see Eq.~\ref{eq:BsPDFuntagged} below).
Measurements of $\Delta \Gamma_s$ have been performed
by CDF~\cite{Acosta:2004gt} and D\O~\cite{Abazov:2007tx};
D\O\ also obtains a constraint on $\beta_s$.
We have performed a simulation based on toy Monte Carlo experiments to
evaluate the sensitivity of this measurement at \superb.
An example of the evolution of the precision on $\Delta \Gamma_s$
as a function of the integrated luminosity is shown in Fig.~\ref{fig:dg}.
We see that with a few ${\rm ab}^{-1}$ of data accumulated
at the $\FiveS$ it will be possible to improve upon
the current experimental precision.
Clearly, \lhcb\ also has the potential to improve this measurement.

We have also studied the performance of two different experimental techniques
that can be used to to extract the semileptonic asymmetry $A^s_{\rm SL}$,
defined as (see also Section~\ref{sec:U4s_DB2}):
\begin{eqnarray}
  A_{\rm SL}^s & = &
  \frac{
    \BR(     B_s \to \overline {B}_s \to D_s^{(*)-} {\it l}^+ \nu_{\it l}) -
    \BR(\overline {B}_s \to      B_s \to D_s^{(*)+} {\it l}^- \nu_{\it l})
  }{
    \BR(     B_s \to \overline {B}_s \to D_s^{(*)-} {\it l}^+ \nu_{\it l}) +
    \BR(\overline {B}_s \to      B_s \to D_s^{(*)+} {\it l}^- \nu_{\it l})
  } \nonumber \\
  & = & \frac{1-|q/p|^4}{1+|q/p|^4} \, .
\end{eqnarray}

The first technique consists of exclusively reconstructing one of the two $B$
mesons into a self-tagging hadronic final state
(such as $B_s \to D_s^{(*)} \pi$)
and looking for the signature of a semileptonic decay (high momentum lepton) in
the rest of the event.
The second approach is more inclusive,
using all events with two high momentum leptons.
In this case, contributions from $B_s$ and $B_d$ decays
cannot be separated, and a combined asymmetry, $A_{\rm CH}$ is measured.
Results from this type of analysis are available from D\O~\cite{Abazov:2007nw}.
Fig.~\ref{fig:dg} shows the statistical errors
we expect on $A^s_{\rm SL}$ and $A_{\rm CH}$.
Notice that, in both cases, the error becomes systematics dominated
after a relatively small period of data taking.
Nonetheless, a clear improvement on the current
experimental situation is possible.
Since measurements in a hadronic environment generally suffer
from larger systematic effects;
\superb\ appears better-suited to obtain precise measurements
of the semileptonic asymmetries.

\begin{figure}[!htbp]
\begin{center}
\includegraphics[width=0.48\textwidth]{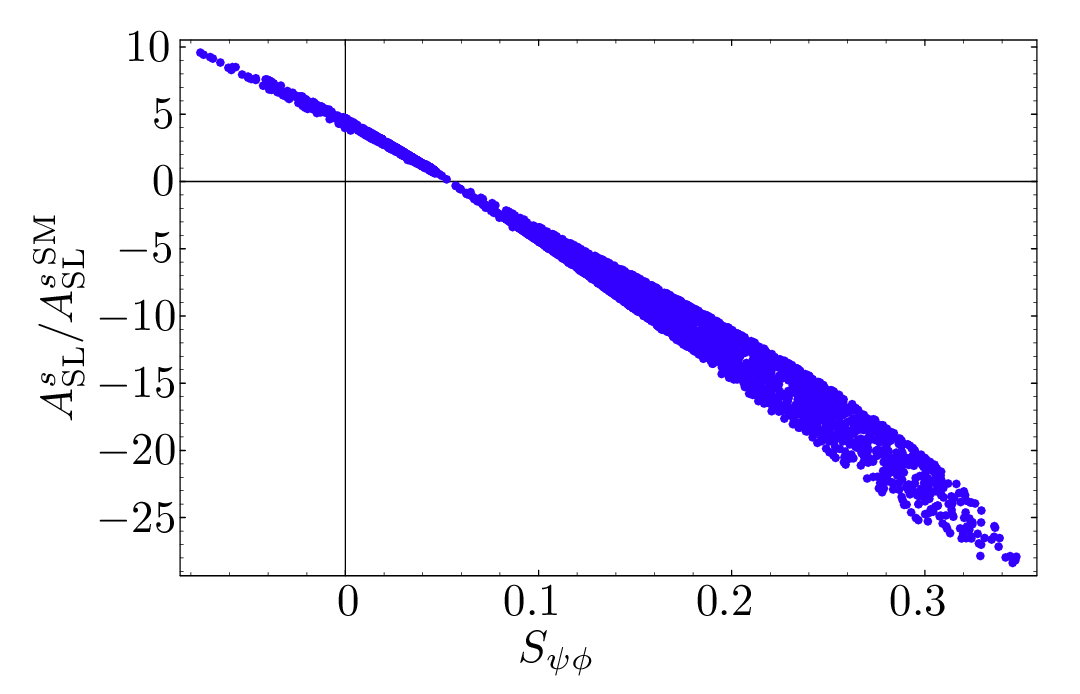}
\includegraphics[width=0.48\textwidth]{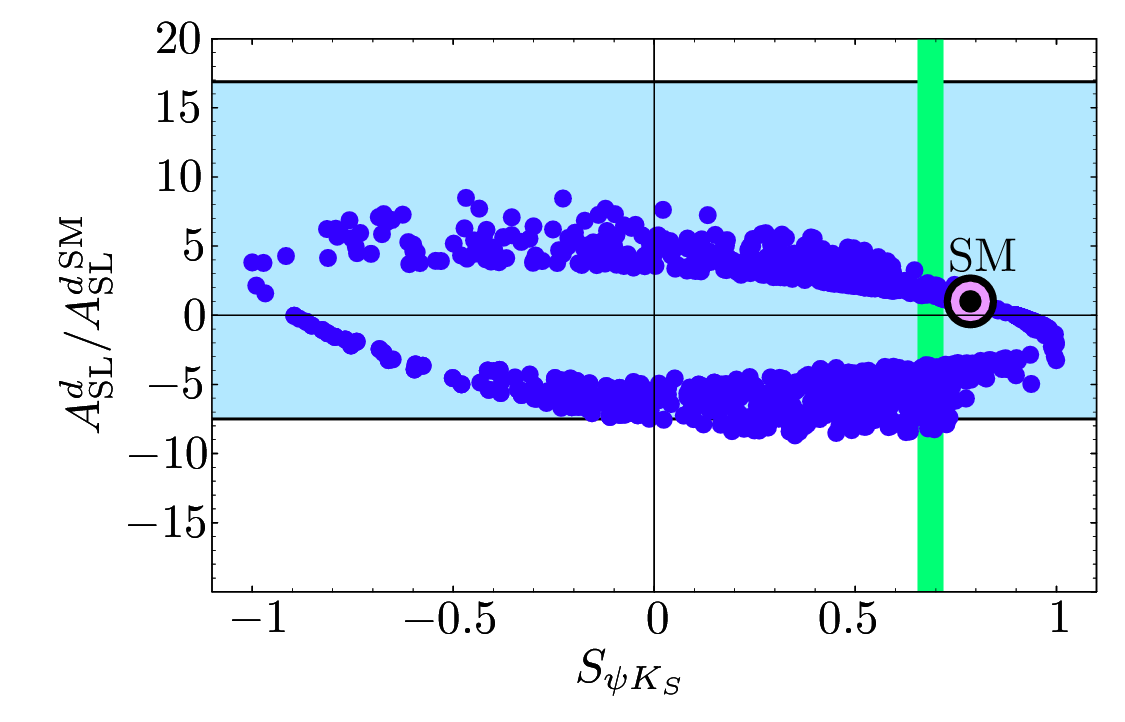}
\caption{Left (right) plot shows the correlation between
  $A_\mathrm{SL}^s$ ($A_\mathrm{SL}^d$) and $S_{J/\psi\phi}$ ($S_{J/\psi K_S}$)
  computed in the Littlest Higgs Model with T-parity (see text). The
  shaded areas represent the present experimental constraints.}
\label{fig:LHASL}
\end{center}
\end{figure}

It is interesting to mention that in the Littlest Higgs Model with T-parity introduced in Section~\ref{sec:LHM} one finds large and correlated corrections
to the $\CP$ asymmetries $S_{J/\psi\phi}$ and $A_\mathrm{SL}^s$ (and, to a lesser extent, also to $A^d_\mathrm{SL}$), as shown in Fig.~\ref{fig:LHASL}).
Note that all these $\CP$ asymmetries, in contrast to many other flavour observables,  are not sensitive to the UV completion of the model and, thus allow for more reliable
theoretical predictions.

\mysubsection{Time Dependent $\CP$ Asymmetries}
\label{sec:U5s_timedep}

Let us consider a $B_s$ pair produced at the $\FiveS$ resonance,
through a $B_s^* \overline {B}_s^*$ state.
If one of the two $B_s$ mesons decays into a $\CP$ eigenstate $f$
and the other to a flavour-tagging final state,
the decay rates as a function of the proper time difference $\Delta t$
can be written in terms of the parameter
$\lambda_f = \frac{q}{p}\frac{\bar{A}_f}{A_f}$ as~\cite{Dunietz:2000cr}:
\begin{equation}
  \begin{array}{rcl}
    \multicolumn{2}{l}{
      \Gamma_{\Bsb \to f} (\Delta t) =
      {\cal N} 
      \frac{e^{-| \Delta t | / \tau(\Bs)}}{4\tau(\Bs)}
      \Big[
      \cosh(\frac{\Delta \Gamma_s \Delta t}{2}) +
    } & \hspace{10mm} \\
    \phantom{.} \hspace{20mm} \phantom{.} &
    \multicolumn{2}{r}{
      \frac{2\, \Im(\lambda_f)}{1 + |\lambda_f|^2} \sin(\Delta m_s \Delta t) -
      \frac{1 - |\lambda_f|^2}{1 + |\lambda_f|^2} \cos(\Delta m_s \Delta t) -
      \frac{2\, \Re(\lambda_f)}{1 + |\lambda_f|^2} \sinh(\frac{\Delta \Gamma_s \Delta t}{2})
      \Big],
    } \\
    \multicolumn{2}{l}{
      \Gamma_{\Bs \to f} (\Delta t) =
      {\cal N} 
      \frac{e^{-| \Delta t | / \tau(\Bs)}}{4\tau(\Bs)}
      \Big[
      \cosh(\frac{\Delta \Gamma_s \Delta t}{2}) -
    } & \hspace{10mm} \\
    \phantom{.} \hspace{20mm} \phantom{.} &
    \multicolumn{2}{r}{
      \frac{2\, \Im(\lambda_f)}{1 + |\lambda_f|^2} \sin(\Delta m_s \Delta t) +
      \frac{1 - |\lambda_f|^2}{1 + |\lambda_f|^2} \cos(\Delta m_s \Delta t) -
      \frac{2\, \Re(\lambda_f)}{1 + |\lambda_f|^2} \sinh(\frac{\Delta \Gamma_s \Delta t}{2})
      \Big].
    } \\
  \end{array}
  \label{eq:BsPDFtagged}
\end{equation}
giving an {\it untagged} time-dependent decay rate of
\begin{equation}
  \Gamma_{\Bsb \to f} (\Delta t) + \Gamma_{\Bs \to f} (\Delta t)
  =
  {\cal N} 
  \frac{e^{-| \Delta t | / \tau(\Bs)}}{2\tau(\Bs)}
  \Big[
  \cosh(\frac{\Delta \Gamma_s \Delta t}{2}) -
  \frac{2\, \Re(\lambda_f)}{1 + |\lambda_f|^2} \sinh(\frac{\Delta \Gamma_s \Delta t}{2})
  \Big] \, .
  \label{eq:BsPDFuntagged}
\end{equation}
With the requirement
$\int_{-\infty}^{+\infty} \Gamma_{\Bsb \to f} (\Delta t) + \Gamma_{\Bs \to f} (\Delta t) d(\Delta t) = 1$,
the normalization factor ${\cal N}$
is fixed to $1 - (\frac{\Delta \Gamma_s}{2\Gamma_s})^2$.
In this formulation, we have neglected effects due to
$\CP$ violation in mixing.


We have investigated the possibility of performing a similar
time-dependent analysis to that for the case of $B_d \to J/\psi\Kz$ decays,
despite the very fast $B_s$ oscillations.
We performed a toy simulation to find the sensitivity to the
time dependent $\CP$ asymmetry in the decay $B_s \to J/\psi \phi$,
and found that in order to measure the $\CP$ violation parameters
it would be necessary to achieve a resolution
$\sigma (\Delta t) < 0.11 \ {\rm ps}$,
which does not appear to be possible --
improvements coming from new technology,
together with the possibility of adding a layer of silicon detectors
close  to the beam pipe (see Section~\ref{sec:det:SVT}),
can only reduce the resolution $\sigma (\Delta t)$ to $\sim 0.4 \ {\rm ps}$
with a Lorentz boost of $\beta\gamma \sim 0.3$.

However, since $\Delta \Gamma_s \neq 0$,
the untagged time-dependent decay rate also allows $\lambda_f$ to be probed,
through the $\Re(\lambda_f)$-dependence of the coefficient of the
$\Delta t$-odd $\sinh(\frac{\Delta \Gamma_s \Delta t}{2})$ term.
Such an analysis has been performed by D\O~\cite{Abazov:2007tx,Abazov:2007zj}.
We have explored the possibility of taking advantage of this,
using a ``two-bin'' time-dependent analysis.
We have carried out toy simulations in which we perform a simultaneous fit
to extract the yields in four categories
(for different signs of $\Delta t$ and tag flavour).
These yields can then be used to constrain $\lambda_f$.

For instance, considering the $B_s \to J/\psi \phi$ decay, and assuming,
for simplicity, that it is a pure $\CP$-even eigenstate
(in the general case, an angular analysis can be used to isolate
$\CP$-even and $\CP$-odd contributions),
this technique can be used to extract a constraint on
the weak phase of the mixing $2\beta_s$.
A precision on $\beta_s$ of $\sim 10^\circ$ and $\sim 3^\circ$ can be achieved,
with $1 \ {\rm ab}^{-1}$ and $30 \ {\rm ab}^{-1}$ of integrated luminosity,
respectively. Anyway, a two-fold ambiguity between $\beta_s$ and $-\beta_s$ can produce a 
(almost) two-times larger resolution in the total pdf, when the value of
$\beta_s$ is close to zero (as it should be in the SM). 
On the other side, this measurement is not limited by systematics,
and the precision can be readily improved by collecting more data.

While the precision that can be achieved in the
$B_s \to J/\psi \phi$ channel is not fully competitive with
that possible at \lhcb, where a tagged analysis can be done,
the success of this technique opens the possibility of using several
other channels, not accessible at hadronic machines,
that are sensitive to the weak phase of the $B_s$ mixing amplitude.
Among the many interesting final states \superb\ could study are
$B_s \to J/\psi \eta$, $B_s \to J/\psi \eta^\prime$,
$B_s \to D^{(*)+}_s D^{(*)-}_s$, $B_s \to D^{(*)} \KS$,
$B_s \to D^{(*)} \phi$, $B_s \to J/\psi \KS$,
$B_s \to \phi \eta^\prime$ and $B_s \to \KS \pi^0$.
We have performed a study on the particularly interesting channel
$B_s \to \Kz \bar {K}^0$,
which is a pure $b \to s$ penguin transition,
complementary to those that can be studied in $B_d$ decays
(see Section~\ref{sss:beta}).
With $30 \ {\rm ab}^{-1}$ accumulated at the $\FiveS$,
we can reach an error on $\beta_s$ of $11^\circ$.

\mysubsection{Rare Decays}
\label{ss:u5s_rare}

\mysubsubsection{Leptonic Decays}
\label{sss:bsmumu}

In the Standard Model
$\BR(B_s \to \mu^+\mu^-) = (3.35 \pm 0.32) \times 10^{-9}$~\cite{Buchalla:1998ba,Blanke:2006ig};
this decay is chirally suppressed, and proceeds in the Standard Model
through loop diagrams, which makes it particularly sensitive to
New Physics contributions~\cite{Choudhury:1998ze,Babu:1999hn,Buras:2002wq,Buras:2002ej,Buras:2003td,Dermisek:2003vn,Dermisek:2005sw,Auto:2003ys,Blazek:2003hv,Misra:2006xt}.
A combined analysis of $B$ and $K$ rare decays~\cite{Bobeth:2005ck}
has recently studied this decay in the context of MFV models with one Higgs doublet
or two Higgs doublets at small $\tan\beta$,
finding $\BR(B_s \to \mu\mu)<7.42 \times 10^{-9}$ at $95\%$ probability:
this decay rate requires large $\tan \beta$
to receive significant New Physics contributions in MFV models.

Indeed, in a very large $\tan\beta$ scenario,
Yukawa couplings contribute, resulting in a sizable enhancement
of the decay rate~\cite{D'Ambrosio:2002ex,Isidori:2006pk,Lunghi:2006uf}.
The current experimental limit is
$\BR(B_s \to \mu^+\mu^-) < 1.0 \times 10^{-7}$
at $90\%$ confidence level~\cite{Abulencia:2005pw,Abazov:2004dj}.

We have estimated the \superb\ sensitivity to the branching ratio for this decay.
The numbers of expected events
($6$ signal events and $960$ background events in $30 \ {\rm ab}^{-1}$)
suggest that \superb\ would not be competitive for this measurement;
indeed, this is one of the primary motivations of the LHC $B$ physics program.


\mysubsubsection{Radiative Decays}

An independent measurement of $|V_{td}/V_{ts}|$,
to be compared with the information coming from the $\Delta m_s$ measurement,
can be provided by  $\Delta B=1$ $b \to s$ transitions,
which can be sensitive to New Physics in a different way than $\Delta m_s$.

Such a test is provided by the ratio
$R = \BR(B_d^0 \to \rho^0 \gamma)/\BR(B_d \to K^{*0}\gamma)$
(see Section~\ref{sec:U4s_rare}),
which allows a measurement of $|V_{td}/V_{ts}|$,
with an uncertainty that is expected to be ultimately limited by the
presence of the power-suppressed correction term $\Delta R$ in Eq.~\ref{eq:rhogKsg}.
In particular, a significant contribution is expected to come from the
$W$-exchange diagram, which contributes to $B_d^0 \to \rho^0 \gamma$
but not to $B_d^0 \to K^{*0}\gamma$.
This contribution is of order $\Lambda_{\rm QCD}/m_b$
and is CKM suppressed in the Standard Model.
Beyond the Standard Model, however,
the CKM suppression may no longer be present.
It is therefore interesting to look for a similar observable
that is not affected by the presence of the $W$-exchange term,
to be sure that no hadronic uncertainty
is introduced going from the Standard Model to New Physics scenarios.
There is such an observable:
the ratio $R_s = \BR(B_s^0 \to K^{*0} \gamma)/\BR(B_d^0 \to K^{*0} \gamma)$.
These two decays are not affected by $W$-exchange
and $\Delta R$ is expected to be small even in the presence of New Physics.
The ratio $R_s$ is given again by Eq.~\ref{eq:rhogKsg}
where $\Delta R$, and the isospin, kinematic and form factor terms
are appropriately replaced.

We have performed toy simulations to estimate our sensitivity,
with an assumption of $\BR(B_s^0 \to K^{*0} \gamma) = 1.54 \times 10^{-6}$.
The results have been combined with the lattice QCD prediction
for the form factor ratio $\xi$
to extract the corresponding determination of $|V_{td}/V_{ts}|$.
The error on this determination is fully dominated by the experimental
statistical error, even assuming the present error on $\xi$.
Thus the ratio of $R_s$ can be thought of as a golden method
for a clean determination of the ratio $|V_{td}/V_{ts}|$
from radiative $B$ decays.
As shown in in Table~\ref{tab:inputsbs},
$|V_{td}/V_{ts}|$ can be measured to a precision of a few percent
with a multi-ab$^{-1}$ data sample accumulated by \superb\
at the $\FiveS$.

\mysubsubsection{Measurement of $B_s \to \gamma \gamma$}

For several years, $b \to s \gamma$ has been considered
the golden mode to probe New Physics in the flavour sector.
Indeed, branching ratios and $\CP$ asymmetries of $b \to s \gamma$ provide
significant constraints on the mass insertion parameters of the mass
matrix (see Sections~\ref{sss:radiative} and~\ref{sec:U4s_pheno}).
It is important to look for other channels of this type that can play a similar r\^ole.
An interesting candidate is the decay $B_s \to \gamma \gamma$.
The final state contains both $\CP$-odd and $\CP$-even components,
allowing for the study of $\CP$-violating effects with $B$~Factory tagging techniques.
The Standard Model expectation for the branching ratio is
$\BR(B_s \to \gamma \gamma) \sim (2-8) \times 10^ {-7}$~\cite{Reina:1997my}.
New Physics effects are expected to give sizable contributions to the decay rate
in certain scenarios~\cite{Aliev:1993ea,Devidze:1998cf}.
For instance, in R-parity-violating SUSY models,
neutralino exchange can enhance the branching ratio up to
$\BR(B_s \to \gamma \gamma)\simeq 5 \times 10^ {-6}$~\cite{Gemintern:2004bw}.
On the other hand, in R-parity-conserving SUSY models,
in particular in softly broken supersymmetry,
$\BR(B_s \to \gamma \gamma)$ is found to be highly correlated
with $\BR(b \to s \gamma)$~\cite{Bertolini:1998hp}.

From the experimental point of view, the exclusive measurement
of $B_s \to \gamma \gamma$ is very similar to other measurements
already performed at $B$~Factories (such as $B_d^0 \to \pi^0 \pi^0$).
The presence of two high-energetic photons
presents a clear signature for signal events, particularly with a
recoil technique. 
Both \babar~\cite{Aubert:2001fm} and \belle~\cite{Abe:2005bs}
have published results of searches for $B_d^0 \to \gamma \gamma$,
setting the current experiment upper limit at
$\BR(B_d \to \gamma \gamma) < 6.2 \times 10^{-7}$.
These results are encouraging for the study of
$B_s \to \gamma \gamma$ at \superb,
though they show that considerable effort will be necessary
to control systematic uncertainties.  The limiting systematic is knowledge of the
efficiency for photon reconstruction, which can be reduced with
dedicated studies on control samples with similar photon energy range.

A dedicated simulation shows that
we expect $14$ signal events and $20$ background events
in a sample of $1 \ {\rm ab}^{-1}$,
indicating that the decay could be observed if it
has the Standard Model branching fraction.
With $30 \ {\rm ab}^{-1}$,
one can achieve a statistical error of $7\%$
and a systematic error smaller than $5\%$.
Using tagging information, direct CP asymmetry can also be measured with
good accuracy, as already done at the B factories for neutral decays.

\mysubsection{Summary of Experimental Reach}

The results presented in this section are summarized
in Table~\ref{tab:inputsbs}
for the case of either a short ($1 \ {\rm ab}^{-1}$)
or a long ($30 \ {\rm ab}^{-1}$) run at the $\FiveS$.
Collecting $1 \ {\rm ab}^{-1}$
takes less than one month at a
design peak luminosity of $10^{36} \ {\rm cm}^{-2} \ {\rm sec}^{-1}$.

\begin{table}[htb]
  \caption{
    Summary of the expected precision of some of the most important
    measurements that can be performed at \superb\
    operating at the $\FiveS$ resonance,
    with an integrated luminosity of
    $1 \ {\rm ab}^{-1}$ and $30 \ {\rm ab}^{-1}$.
  }
  \label{tab:inputsbs}
  \begin{center}
    \begin{tabular}{lcc}
      \hline
      \hline
      Observable    & Error with $1 \ {\rm ab}^{-1}$ & Error with $30 \ {\rm ab}^{-1}$ \\
      \hline
      $\Delta \Gamma$                 & $0.16 \ {\rm ps}^{-1}$ & $0.03 \ {\rm ps}^{-1}$ \\
      $\Gamma$                        & $0.07 \ {\rm ps}^{-1}$ & $0.01 \ {\rm ps}^{-1}$	\\
      $\beta_s$ from angular analysis & $20^\circ$             & $8^\circ$ \\
      $A^s_{\rm SL}$                  & 0.006                  & 0.004 \\
      $A_{\rm CH}$                    & 0.004                  & 0.004 \\
      $\BR(B_s \to \mu^+ \mu^-)$       &  -                     & $<8 \times 10^{-9}$ \\
      $|V_{td}/V_{ts}|$               & 0.08                   & 0.017 \\
      $\BR(B_s \to \gamma \gamma)$     & $38 \%$                & $7 \%$ \\
      $\beta_s$ from $J/\psi \phi$    & $10^\circ$             & $3^\circ$ \\
      $\beta_s$ from $B_s \to K^0 \bar{K}^0$ & $24^\circ$      & $11^\circ$ \\
      \hline
    \end{tabular}
  \end{center}
\end{table}

It is fortunate for experiments in the hadronic environment
that many of the most interesting $B_s$ decay channels
contain dileptons in the final state.
It is clear that \superb\ cannot compete with hadronic experiments
on modes such as $B_s \to \mu^+\mu^-$ and $B_s \to J/\psi \phi$.
It is also clear that many important
channels that are not easily accessible
at hadronic experiments such as \lhcb,
among them $B_s \to \gamma \gamma$ and $B_s \to K^0 \bar{K}^0$.
Therefore, \superb\ will complement the results from \lhcb,
and enrich its own physics program,
by accumulating several ${\rm ab}^{-1}$ at the $\FiveS$.

\mysubsection{Phenomenological Implications}

The experimental measurements of $\Delta \Gamma$, $A^{s}_{\rm SL}$, $A_{\rm CH}$
and $\CP$ violation parameters described in the previous sections
can be used to determined the $\Delta B = 2$ New Physics contributions
in the $B_s$ sector.
The knowledge of $\overline{\rho}$ and $\overline{\eta}$ is assumed to come
from studies at the $\FourS$.

\begin{figure}[t]
  \begin{center}
    \includegraphics[width=0.45\textwidth]{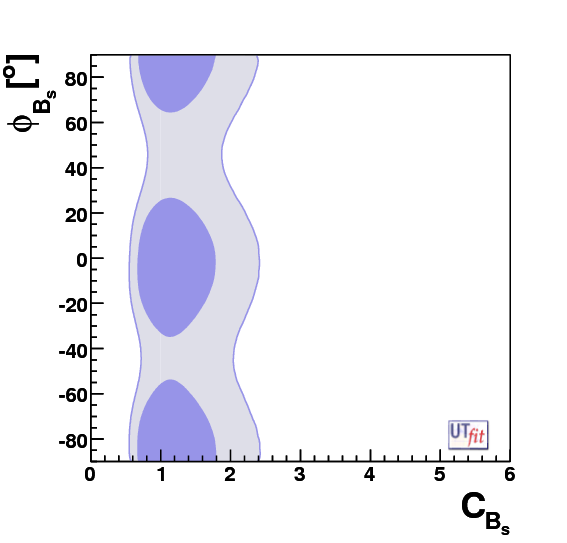}
    \includegraphics[width=0.45\textwidth]{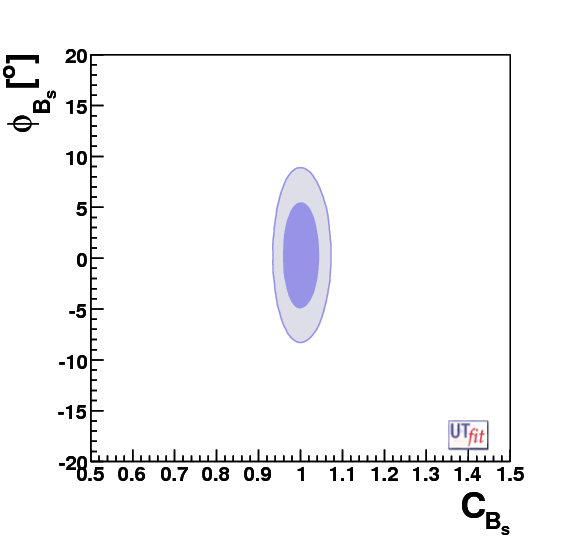}
    \caption{
      Allowed regions in the $C_{B_s}$--$\phi_{B_s}$ plane
      given by the current data (left) and at the time of \superb\ (right).
      Note that the scales for the axes are different in the two cases.
    }
    \label{fig:cvsphi}
  \end{center}
\end{figure}

To illustrate the impact of the measurement at \superb at the $\FiveS$,
we show in Fig.~\ref{fig:cvsphi} selected regions in the
$\phi_{B_s}$--$C_{B_s}$ plane (right),
compared to the current situation (left).
Corresponding numerical results are given in Table~\ref{tab:deltaf2sfit}.

\begin{table}[ht]
  \caption{
    Uncertainty of New Physics parameters $\phi_{B_s}$ and $C_{B_s}$
    using the experimental and theoretical information available
    at the time of \superb\ and given in Tables~\ref{tab:inputsbs}
    ($30 \ {\rm ab}^{-1}$) and~\ref{tab:lattice}.
    These uncertainties are compared to the present determination.
  }
  \label{tab:deltaf2sfit}
  \begin{center}
    \begin{tabular}{lll}
      \hline\hline
      Parameter    & Today        & At \superb\ ($30 \ {\rm ab}^{-1}$) \\
      \hline
      $\phi_{B_s}$ & $(-3 \pm 19)^\circ \cup (94 \pm 19)^\circ$ & $\pm 1.9^{\circ}$ \\
      $C_{B_s}$    & $1.15 \pm 0.36$     & $\pm 0.026$ \\
      \hline
    \end{tabular}
  \end{center}
\end{table}

It is important to note that the uncertainty on the parameter $C_{B_s}$
is largely dominated by the uncertainty on the related hadronic quantity,
namely $f_{B_s}$ and bag parameters.
The error on $\phi_{B_s}$ is not limited by systematics and theory,
and can be improved to $1$--$2^\circ$ with a
longer dedicated run at the $\FiveS$.

\lhcb\ will also measure the New Physics phase $\phi_{B_s}$.
With the final available statistics ($\sim 10 \ {\rm fb}^{-1}$),
the uncertainty on $\phi_{B_{s}}$ is estimated to be less than $1^\circ$.

\afterpage{\clearpage}

\mysection{Charm Physics}
\label{sec:charm}

It is a truth universally accepted that charm studies played
a seminal role in the evolution and acceptance of the Standard Model.
Yet the continuing importance of this sector is not widely appreciated,
since the Standard Model electroweak phenomenology for charm decays is on the dull side:
the CKM parameters are known, $\DzDzb$ oscillations are slow,
$\CP$ asymmetries are small or absent and loop-driven decays are extremely rare.

Yet on closer examination, a strong case emerges in two respects,
both of which derive from this apparent dullness:
\begin{itemize}
\item
  Detailed and comprehensive analyses of charm transitions
  will continue to provide us with new insights into QCD's
  nonperturbative dynamics, and advance us significantly
  towards establishing theoretical control over them.
  Beyond the intrinsic value of such lessons, they will also
  calibrate our theoretical tools for $B$ studies;
  this will be essential to saturate the discovery potential for
  New Physics in $B$ transitions.
\item
  Charm decays constitute a novel window into New Physics.
\end{itemize}
Lessons from the first item will have an obvious impact on the tasks
listed under the second.
They might actually be of great value even beyond QCD,
if the New Physics anticipated for the TeV scale
is of the strongly interacting variety.

The capabilities of a \sff\ are well matched to these goals.
It allows uniquely clean determinations of CKM parameters,
with six of the nine matrix elements impacted by charm measurements.
New Physics signals can easily exceed Standard Model predictions by considerable factors
such that there will be no ambiguity in interpreting them,
yet they are unlikely to be large;
this again requires the clean environment and huge statistics
that a \sff\ can provide.

A number of other facilities either currently running or
soon to commence operation provide competition in the area of charm physics.
The current $B$ Factory program is expected to produce a sample of about
$10^{10}$ charm hadrons from operation at or near the
$\FourS$ resonance.
The CLEO$c$ experiment at CESR
is operating in the charm threshold region,
and anticipates collecting a total of $5 \times 10^6$ $\DzDzb$ pairs
and about $7 \times 10^5$ $D_s^{*+}D_s^- + D_s^{+}D_s^{*-}$
through coherent production.
The BESIII experiment at BEPCII expects first $e^+e^-$ collisions in 2008,
and will collect large charmonium samples,
in addition to exceeding the CLEO$c$ data set in open charm production.
Although there will be no successors to the Fermilab fixed target
charm production experiments,
the LHC will produce copious quantities of charm
(notably, charm physics forms a part of the \lhcb\ physics program);
these are expected to result in very large samples of charmed hadrons
in final states with reconstructible topologies.

Most of the benchmark charm measurements will still be statistics-limited
after the CLEO$c$, BESIII and $B$ Factory projects,
and many will not be achievable in a hadronic environment.
\superb\ is the ideal machine with which to pursue
these measurements to their ultimate precision.
Operation near the $\FourS$ will provide
enormous samples of charm hadrons,
in a clean environment and with a detector well-suited for charm studies.
The charm physics program would benefit further from the ability
to operate in the threshold region, in order to
exploit the quantum correlations associated with coherent production.
The expected lower luminosity at threshold would be partly compensated by the
higher production cross-section,
resulting in a comparable charm production rate.
To estimate the reach of \superb\ from operation at the charm threshold,
we have assumed a simple dependence of the luminosity
on the center-of-mass energy: ${\cal L}_{\rm peak} \propto s$.
Thus, we expect that \superb\ (which will integrate
$\sim 15 \ {\rm ab}^{-1}$ per year operating at the $\Upsilon(4{\rm S})$)
can accumulate $\sim 150 \ {\rm fb}^{-1}$ per month
when operated at the $\psi(3770)$.

\mysubsection{Lessons on Strong Dynamics}

Detailed analyses of (semi)leptonic decays of charm hadrons
provide a challenging test bed for validating lattice QCD (LQCD),
which is the only known framework with realistic promise for a truly
quantitative treatment of charm hadrons that can be systematically improved .
Such studies form the core of the ongoing CLEO$c$
and the nascent BESIII programs;
they are also pursued very profitably at the $B$~Factories.
Central goals are measuring the decay constants $f_{D^+}$ and $f_{D_s}$
and going beyond total rates for semileptonic $D^+$, $D^0$ and $D_s^+$ decays.
Such high quality studies will greatly improve
our understanding of hadronization and provide an even richer test bed
for LQCD with the lessons to be learned
of crucial importance for extracting $V_{ub}$ from semileptonic $B$ decays.
Our knowledge of charm baryon decays is also rather limited;
\eg, no precision data on absolute
branching ratios or semileptonic decay distributions exist.
CLEO$c$ will not run above the charm baryon threshold, and BESIII cannot.

\mysubsubsection{Leptonic Charm Studies}
\label{LEPCHARM}

In the Standard Model the leptonic decay width is given by~\cite{Silverman:1988gc}:
\begin{eqnarray}
  \Gamma(D^+ \to \ell^+\nu) & = &
  \frac{G_F^2}{8\pi} f_{D^+}^2 m_{\ell}^2 M_{D^+}
  \left( 1 - \frac{m_{\ell}^2}{M_{D^+}^2} \right)^2 \left|V_{cd}\right|^2
  \nonumber \\
  \Gamma(D_s^+ \to \ell^+\nu) & = &
  \frac{G_F^2}{8\pi} f_{D^+_s}^2 m_{\ell}^2 M_{D^+_s}
  \left( 1 - \frac{m_{\ell}^2}{M_{D^+_s}^2}\right)^2 \left|V_{cs}\right|^2\,.
  \label{eq:equ_rate}
\end{eqnarray}
Taking $|V_{cd}|$ and $|V_{cs}|$ from elsewhere, one uses Eq.(\ref{eq:equ_rate})
to extract $f_{D^+}$ and $f_{D^+_s}$.
The ratio $R_\ell$ of the leptonic decay rates of the $D_s^+$ and the $D^+$
is proportional to $(f_{D^+_s}/f_{D^+})^2$,
for which the lattice calculation is substantially more precise.
A significant deviation from its predicted value would be a clear
sign of New Physics,
probably in the form of a charged Higgs exchange~\cite{Akeroyd:2003jb}.
On the other hand, the ratio of the rates of tauonic and muonic decays
for either $D^+$ or $D^+_s$ is independent of both form factors
and CKM elements, and serves as a useful cross-check in this context.

CLEO$c$ has published a measurement of
$f_{D^+}$~\cite{Artuso:2005ym,Bonvicini:2004gv},
and several measurements of
$f_{D_s^+}$~\cite{Rubin:2006nt,Artuso:2006kz}.
These measurements have benefitted from a ``double-tag'' method
uniquely possible at threshold,
where a $D^+_{(s)}D^-_{(s)}$ pair is produced with no extra particles.
The latest results are
\begin{equation}
  f_{D^+} = (222.6 \pm 16.7 \,^{+2.8}_{-3.4})~{\rm MeV}~.
\end{equation}
\begin{equation}
  f_{D_s} = 280.1 \pm 11.6 \pm 6.0 {~\rm MeV}
  \; , \;
  f_{D_s^+}/f_{D^+} = 1.26 \pm 0.11 \pm 0.03~.
\end{equation}
The central values for $f_{D_s^+}$ and $f_{D_s^+}/f_{D^+}$
are slightly above, but consistent with, the present LQCD calculations.
It is important to note that the desired $1$--$3\%$ accuracy level
has not yet been reached on either the experimental or theoretical side.
While LQCD practitioners expect to reach this level over the next decade,
the experimental precision is likely to fall significantly short
of this goal, even after BESIII.
Since larger statistics will certainly allow reduction
of the systematic errors in the current results,
it is clear that data accumulated by \superb\
from a relatively short run ($\sim 1$ month) at charm threshold
would allow the desired improvement of the experimental precision
(see discussion below, and Table~\ref{tab:statl}).
Validating LQCD on the ${\cal O}(1\%)$ level will have
important consequences for $B_d$ and $B_s$ oscillations,
since it would give us demonstrated confidence in the theoretical
extrapolation to $f_{B_d}$ and $f_{B_s}/f_{B_d}$.

\mysubsubsection{Semileptonic Charm Studies}

In the area of semileptonic decays,
CLEO$c$ has made the most accurate measurements for
the inclusive $D^0$ and $D^+$
semileptonic branching fractions --
${\cal B}(D^0 \rightarrow X \ell \nu_\ell)= (6.46 \pm 0.17 \pm 0.13)\%$ and
${\cal B}(D^+ \rightarrow X \ell \nu_\ell)=(16.13 \pm 0.20 \pm 0.33)\%$~\cite{cleoc:incl} --
and expects to do the same for $D_s^+$.
Such data provide important ``engineering input''
for other $D$ and $B$ decay studies.
However, a central goal must be to go beyond the total rates for these decays
and to extract the form factors \etc\
In order to do so, it is essential to analyze lepton spectra
and perform ``meaningful'' Dalitz plot studies.
To quantify ``meaningful'', it is instructive to
compare to analyses on $K_{e4}$ decays.
With a sample size of 30,000 events which became available in 1977,
one was able to begin extracting dynamical information.
Precise measurements are now possible,
with NA48/2 and E685 each having accumulated
400,000 events~\cite{Pislak:2001bf,Batley:2004cp}.
For charm we are nowhere near that level:
CLEO$c$ will have about 10,000 semileptonic charm decays --
comparable to kaon studies in the late 1970s.
Since for charm the phase space is larger,
thereby opening more domains of interest,
a reasonable target sample size is $10^6$ events,
which is far beyond the reach of CLEO$c$, and most probably, of BESIII.

Three-family unitarity constraints on the CKM matrix yield
rather precise values for $|V_{cs}|$ and $|V_{cd}|$.
Using these, one can extract the form factors from analyses of
exclusive semileptonic charm decays.
Both the normalization and $q^2$ dependence must be measured.
Existing LQCD studies do not allow us to
determine the latter from first principles;
instead a parametrization originally proposed by Becirevic and Kaidalov ($BK$)
is used~\cite{Becirevic:1999kt}.
Recent and forthcoming results from CLEO$c$, \babar\ and
\belle~\cite{AsnerCKM06} are expected to be statistics limited,
and will not reach the desired $1$--$3\%$ level.

The current status can be characterized by comparing the measured value
of the ratio $R_{sl}$,
which is independent of $|V_{cd}|$,
to that inferred from a recent LQCD calculation~\cite{Aubin:2005ar}:
\begin{equation}
  R_{sl} = \sqrt{
    \frac{
      \Gamma(D^+ \rightarrow \mu^+ \nu_\mu)
    }{
      \Gamma(D \rightarrow \pi e \nu _e)
    }
  } =
  \left\{
    \begin{array}{c@{\hspace{3mm}}c}
      0.237 \pm 0.019 & ({\rm exp}) \\
      0.212 \pm 0.028 & ({\rm theo})\, . \\
    \end{array}
  \right.
\end{equation}
The values are nicely consistent,
yet both are still far from the required level of precision.


While operation in the $\Upsilon$ region will produce
large quantities of charm hadrons,
there are significant backgrounds and
one pays a price in statistics when using kinematic
constraints to infer neutrino momenta, \etc.
On the other hand, even a limited run at charm threshold
will generate the statistics required to
study (semi)leptonic decays with the desired accuracy.
Assuming that systematic uncertainties in tracking and muon identification
will provide a limit to the precision at the $0.5\%$ level,
we estimate the integrated luminosity from threshold running required
to achieve a similar statistical uncertainty.
As shown in Table~\ref{tab:statl} we expect to be able to measure
$f_{D^+}$, $f_{D_s}$ and their ratio with better than $0.5\%$
statistical uncertainty with integrated luminosities
of at least $100 \ {\rm fb}^{-1}$.
\begin{table}[!htbp]
    \caption{
      Statistics required to obtain $0.5\%$ statistical uncertainties
      on corresponding branching fractions using double-tagged events,
      when running at threshold.
    }
    \label{tab:statl}
  \begin{center}
    \begin{tabular}{cc}
      \hline \hline
      Channel & Integrated luminosity  \\
      & (${\rm fb}^{-1}$)  \\
      \hline
      $D^+ \rightarrow \mu^+ \nu_{\mu}$  & 500  \\
      $D_s^+ \rightarrow \mu^+ \nu_{\mu}$  & 100  \\
      \hline
    \end{tabular}
  \end{center}
\end{table}

For semileptonic decays, a case-by-case study is necessary.
One also has to distinguish between merely determining the branching ratio
and performing a ``meaningful'' Dalitz plot analysis, as discussed above.
The required integrated luminosities are given in Table~\ref{tab:statsl}.
It is clear that the $\sim 150 \ {\rm fb}^{-1}$ anticipated
from one month of running in the threshold region
would provide the desired statistics for most measurements.
Note that while $D_s$ mesons are not produced at the $\psi(3770)$,
short runs at other energies are possible.
\bigskip
\begin{table}[!b]
    \caption{
      Statistics required to obtain $0.5\%$ statistical uncertainties
      on corresponding branching fractions (column 2) or
      one million signal events (column 3) using double tagged events,
      when running at threshold.
    }
  \begin{center}
    \begin{tabular}{lcc}
      \hline \hline
      Channel & Integrated luminosity & Integrated luminosity \\
      & (${\rm fb}^{-1}$)  & (${\rm fb}^{-1}$)  \\
      \hline
      $D^0 \to K^- e^+ \nu_e$     & 1.3 &   33 \\
      $D^0 \to K^{*-} e^+ \nu_e$  & 17  &  425 \\
      $D^0 \to \pi^- e^+ \nu_e$   & 20  &  500 \\
      $D^0 \to \rho^- e^+ \nu_e$  & 45  & 1125 \\
      $D^+ \to \KS e^+ \nu_e$   &  9  &  225 \\
      & \\
      $D^+ \to \bar{K}^{*0} e^+ \nu_e$  &    9 &   225 \\
      $D^+ \to \pi^{0} e^+ \nu_e$       &   75 &  1900 \\
      $D^+ \to \rho^{0} e^+ \nu_e$      &  110 &  2750 \\
      & \\
      $D_s^+ \to \phi e^+ \nu_e$        &   85 &  2200 \\
      $D_s^+ \to \KS e^+ \nu_e$       & 1300 & 33000 \\
      $D_s^+ \to K^{*0} e^+ \nu_e$      & 1300 & 33000 \\
      \hline
    \end{tabular}
    \label{tab:statsl}
  \end{center}
\end{table}




\mysubsection{Precision CKM Measurements}
\label{CKMPREC}

Studies of leptonic decay constants and semileptonic form factors
will yield a set of measurements,
including $\left| V_{cd}\right|$ and $\left| V_{cs}\right|$,
at the few percent level.
These measurements will constrain theoretical calculations,
and those that survive will be validated for use in a variety of areas
in which interesting physics cannot be extracted without theoretical input.
This broader impact of charm measurements extends beyond
those measurements that can be performed directly at charm threshold, and
has a large impact on the precision determination of CKM matrix elements.

The determination of $\left| V_{td}\right|$ and $\left| V_{ts}\right|$
is limited by ignorance of $f_B\sqrt{B_{B_d}}$ and $f_{B_s}\sqrt{B_{B_s}}$;
improved determinations of $f_B$ and $f_{B_s}$ are required.
Precision measurements of $f_D$ and $f_{D_s}$ can validate
the theoretical treatment of the analogous quantities for $B$ mesons.
Similarly, improved form factor calculations in the decays
$D \to \pi \ell \nu$ and $D \to \rho \ell \nu$
and inclusive semileptonic charm decays will enable
improved precision in $\left|V_{ub}\right|$ and $\left|V_{cb}\right|$.

The precision measurement of the UT angle $\gamma$
depends on decays of $B$ mesons to final states
containing neutral $D$ mesons (see Section~\ref{ss:angles}).
A variety of charm measurements impact these analyses, including:
improved constraints on charm mixing amplitudes,
measurements of relative rates and strong phases
between Cabibbo-favoured and -suppressed decays
(\eg, between $\Dz \to K^+\pi^-$ and $\Dzb \to K^+\pi^-$),
and studies of charm Dalitz plots tagged by hadronic flavour or
$\CP$ content~\cite{Giri:2003ty,Bondar:2005ki}.
Note that the latter two measurements
can only be performed with data from charm threshold.

\mysubsubsection{Overconstraining the Unitarity Triangle}

At present three-family unitarity constraints yield more precise values
for $|V_{cs}|$ and $|V_{cd}|$ than direct measurements.
Since it is conceivable that a fourth family exists
(with neutrinos so heavy that the $Z^0$ could not decay into them),
one would like to obtain more accurate direct determinations.
This should be possible if LQCD is indeed validated
at the ${\cal O}(1\%)$ level through its predictions
on form factors and their ratios.

From four-family unitarity,
and using current experimental constraints~\cite{Yao:2006px}
we can infer for a fourth quark doublet $(t^{\prime},b^{\prime})$:
\begin{eqnarray}
  |V_{cb'}| & = & \sqrt{1 - |V_{cd}|^2 - |V_{cs}|^2 - |V_{cb}|^2} \lsim 0.5~,\\
  |V_{t's}| & = & \sqrt{1 - |V_{us}|^2 - |V_{cs}|^2 - |V_{ts}|^2} \lsim 0.5~.
\end{eqnarray}
These loose bounds are largely due to the $10\%$ error on $|V_{cs}|$.

\bigskip

\mysubsection{Charm Decays as a Window to New Physics}
\label{HYPOGEN}

While significant progress can be guaranteed for the
Standard Model studies outlined above,
the situation is much less certain concerning the search for New Physics.
No sign of it has yet been seen,
but we have only begun to approach the regime of experimental sensitivity
in which a signal for New Physics could realistically emerge in the data.
The interesting region of sensitivity extends
several orders of magnitude beyond the current status.

New Physics scenarios in general induce flavour-changing neutral currents
that {\em a priori} have no reason to be as strongly suppressed as in the Standard Model.
More specifically, they could be substantially stronger
for up-type than for down-type quarks;
this can occur in particular in models that
reduce strangeness-changing neutral currents
below phenomenologically acceptable levels through an alignment mechanism.

In such scenarios,
charm plays a unique role among the up-type quarks $u$, $c$ and $t$;
for only charm allows the full range of probes for New Physics.
Since top quarks do not hadronize~\cite{Bigi:1986jk},
there can be no $T^0\bar{T}^0$ oscillations
(recall that hadronization, while hard to bring under theoretical control,
enhances the observability of $\CP$ violation).
As far as $u$ quarks are concerned, $\pi^0$, $\eta$ and $\eta ^{\prime}$
do not oscillate, and decay electromagnetically, not weakly.
$\CP$ asymmetries are mostly ruled out by $\CPT$ invariance.
Our basic contention can then be formulated as follows:
charm transitions provide a unique portal for a novel access
to flavour dynamics with the experimental situation being {\em a priori}
quite favourable.
The aim is to go beyond ``merely'' establishing the existence of
New Physics around the TeV scale --
we want to identify the salient features of this New Physics as well.
This requires a comprehensive study,
\ie, that we also search in unconventional areas such as charm decays.

\mysubsubsection{On New Physics scenarios}
\label{CHARMNP}

In a scenario in which the LHC discovers direct evidence of SUSY
via observation of sleptons or squarks,
the \sff\ program becomes even more important.
The sfermion mass matrices are a new potential source of flavour mixing
and $\CP$ violation and contain information about the SUSY-breaking mechanism.
Direct measurements of the masses can only constrain
the diagonal elements of this matrix.
However, off-diagonal elements can be measured through the study
of loop-mediated heavy flavour processes.
As a specific example, a minimal flavour violation scenario such as mSUGRA
with moderate $\tan \beta$, could result in a SUSY partner mass spectrum
that is essentially indistinguishable from an SU(5) GUT model
with right-handed neutrinos.
However the mSUGRA scenario would be expected to yield no observable effects
in the heavy flavour sector,
whereas the SU(5) model is expected to produce measurable effects
in time-dependent $\CP$ violation in penguin-mediated hadronic
and radiative decays.

While there is no compelling scenario that would generate observable effects
in charm, but not in beauty and strange decays,
it is nevertheless reassuring that such scenarios do exist.
One should keep in mind that New Physics signals in charm $\CP$ asymmetries
are particularly clean,
since the Standard Model background (which often exists in $B$ decays) is largely absent.
The consequence is that New Physics could produce signals
that exceed Standard Model predictions by an order of magnitude or more --
something that is of great help in interpreting the signals.
We will focus on the most promising areas;
more details can be found in several
recent reviews~\cite{Burdman:2001tf,Bianco:2003vb,Burdman:2003rs}.

\mysubsubsection{$\DzDzb$ oscillations}
\label{DOSC}

Oscillations of neutral $D$ mesons driven by the two quantities
$x_D = \Delta M_D/\Gamma_D$ and $y_D = \Delta \Gamma_D/2\Gamma_D$
lead to an effective violation of the Standard Model
$\Delta C = \Delta Q$ and $\Delta C = \Delta S$
rules in semileptonic and nonleptonic channels.
The status of the Standard Model prediction can be summarized as~\cite{Bianco:2003vb}:
while one predicts $x_D \sim {\cal O}(10^{-3}) \sim y_D$,
at present one cannot rule out $x_D, \, y_D \sim 0.01$.

Many different charm decay modes can be used to search for charm mixing.
The appearance of ``wrong-sign'' kaons in semileptonic decays
would provide direct evidence for $\DzDzb$ oscillations
(or another process with origin beyond the Standard Model).
The wrong-sign hadronic decay $D^0 \to K^+\pi^-$
is sensitive to linear combinations of the mass and lifetime differences,
denoted $x_D^{\prime2}$ and $y_D^\prime$.
The relation of these parameters to $x_D$ and $y_D$ is controlled
by a strong phase difference.
Direct measurements of $x_D$ and $y_D$
independent of unknown strong interaction phases,
can also be made using time-independent studies of amplitudes present
in multi-body decays of the $\Dz$, for example, $\Dz\to\KS \pip\pim$.
Direct evidence for $y_D \ne 0$ can also appear through lifetime differences
between decays to $\CP$ eigenstates.
The measured quantity in this case, $y_{\CP}$, is equivalent to $y_D$
in the absence of $\CP$ violation.
Another approach is to study quantum correlations near
threshold~\cite{Bianco:2003vb,Bigi:1989ah,Asner:2005wf}
in $e^+e^- \to \Dz\Dzb (\pi^0)$ and in $e^+e^- \to \Dz \Dzb \gamma$,
which yield $C$-odd and $C$-even $\Dz\Dzb$ pairs, respectively.




Very recently, several new results have suggested that charm mixing may
be at the upper end of the range of Standard Model predictions.
\babar\ finds evidence for oscillations in $D^0 \to K^+\pi^-$
with $3.9\sigma$ significance~\cite{Aubert:2007wf},
while \belle\ sees a $3.2\sigma$ effect in $D^0 \to K^+K^-$,
with results using $D^0 \to \KS\pi^+\pi^-$
supporting the claim~\cite{Abe:2007dt}.
These results are consistent with previous measurements,
some of which had hinted at a mixing
effect~\cite{Godang:1999yd,Abe:2003ys,Zhang:2006dp,Aubert:2006kt,Asner:2005sz}.
The results are not systematics limited,
and further improvements are anticipated.

\begin{figure}[t]
  \begin{center}
    \includegraphics[angle=-90, width=0.7\textwidth]{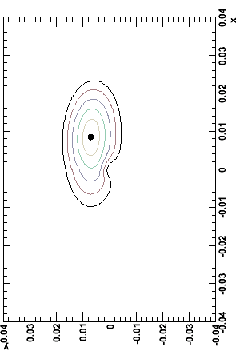}
    \caption{
      Likelihood contours in the $\left( x_D, y_D \right)$ plane
      from HFAG~\cite{asnerFLAVLHC}.
      These preliminary world averages use all available charm mixing results.
    }
    \label{fig:HFAG_charm}
  \end{center}
\end{figure}

The charm decays subgroup of the Heavy Flavour Averaging Group~\cite{hfag}
is preparing world averages of all the charm mixing measurements,
taking into account correlations between the measured quantities.
A very preliminary average is available~\cite{asnerFLAVLHC}, giving:
\begin{center}
  $x_D = \left( 8.5^{+3.2}_{-3.1} \right) \times 10^{-3}$
  \hspace{5mm} and \hspace{5mm}
  $y_D = \left( 7.1^{+2.0}_{-2.2} \right) \times 10^{-3}$\,.
\end{center}
Contours in the $\left( x_D, y_D \right)$ plane are shown
in Fig.~\ref{fig:HFAG_charm}.
The significance of the oscillation effect in
the preliminary world averages exceeds $5\sigma$.

The interpretation of these new results in terms of New Physics is inconclusive.
For one thing, it is not yet clear whether the effect is caused by
$x_D \neq 0$ or $y_D \neq 0$ or both,
though the latter is favored and this point may be clarified soon.
As shown in Table~\ref{tab:charm},
\superb\ will be able to observe both lifetime and mass differences
in the $\Dz$ system, if they lie in the range of Standard Model predictions.
It should be noted that the full benefit of measurements
in the $\Dz \to K^+\pi^-$ system (and indeed for other hadronic decays)
can only be obtained if the strong phases are measured.
This can be achieved with a short ($\sim 1$ month) period
of data taking at charm threshold.

A serious limitation in the interpretation of charm oscillations
in terms of New Physics is the theoretical uncertainty on
the Standard Model prediction.
Nonetheless, if oscillations indeed occur at the level suggested
by the latest results, this will open the window to searches for
$\CP$ asymmetries that do provide unequivocal New Physics signals.

\begin{table}[htb]
  \caption{
    Summary of the expected precision on charm mixing parameters.
    For comparison we put the reach of the $B$~Factories at $2 \ {\rm ab}^{-1}$.
    The estimates for \superb\ assume that systematic uncertainties
    can be kept under control.
  }
  \label{tab:charm}
  \begin{center}
    \begin{tabular}{llcc}
      \hline \hline
      Mode & Observable                      & $B$~Factories ($2 \ {\rm ab}^{-1}$) & \superb\ ($75 \ {\rm ab}^{-1}$)  \\
      \hline
      $\Dz \to K^+K^-$   & $y_{\CP}$       & $2$--$3 \times 10^{-3}$ & $5 \times 10^{-4}$ \\
      $\Dz \to K^+\pi^-$ & $y_D^\prime$    & $2$--$3 \times 10^{-3}$ & $7 \times 10^{-4}$ \\
                         & $x_D^{\prime2}$ & $1$--$2 \times 10^{-4}$ & $3 \times 10^{-5}$ \\
      $\Dz \to \KS\pi^+\pi^-$ & $y_D$      & $2$--$3 \times 10^{-3}$ & $5 \times 10^{-4}$ \\
                              & $x_D$      & $2$--$3 \times 10^{-3}$ & $5 \times 10^{-4}$ \\
      \hline
      Average            & $y_D$           & $1$--$2 \times 10^{-3}$ & $3 \times 10^{-4}$ \\
                         & $x_D$           & $2$--$3 \times 10^{-3}$ & $5 \times 10^{-4}$ \\
      \hline \hline
    \end{tabular}
  \end{center}
\end{table}


\mysubsubsection{$\CP$ Violation With and Without Oscillations}
\label{CPV}

Several factors favor dedicated searches for $\CP$ violation
in charm transitions:

\textbullet\   Within the Standard Model, the effective weak phase is highly diluted,
namely $\sim {\cal O}(\lambda ^4)$,
and can arise only in singly-Cabibbo-suppressed transitions,
where one expects asymmetries to reach the ${\cal O}(0.1 \%)$ level;
significantly larger values would signal New Physics.
Any asymmetry in Cabibbo-allowed or -doubly suppressed channels
requires the intervention of New Physics --
except for $D^{\pm}\to \KS\pi ^{\pm}$~\cite{Bianco:2003vb}
where the $\CP$ impurity in $\KS$ induces an asymmetry of $3.3 \times 10^{-3}$.
One should keep in mind that in going from Cabibbo-allowed to
Cabibbo-singly and -doubly suppressed channels,
the Standard Model rate is suppressed by factors of about
twenty and four hundred, respectively.
One would expect that this suppression will enhance
the visibility of New Physics.

\textbullet\   Strong phase shifts required for direct $\CP$ violation to emerge
are, in general, large, as are the branching ratios into relevant modes.
Although large final state interactions complicate the interpretation of
an observed signal in terms of the microscopic parameters
of the underlying dynamics, they enhance its observability.

\textbullet\   With the Standard Model providing one amplitude,
observable $\CP$ asymmetries can be linear in New Physics amplitudes --
unlike the case for rare decays -- thus increasing the sensitivity.

\textbullet\   Decays to multibody final states contain more dynamical information
than given by their widths;
their decay distributions as described by Dalitz plots or $T$-odd moments
can exhibit $\CP$ asymmetries that might be considerably larger
than those for the width.
Final state interactions,
while not necessary for the emergence of such effects,
can produce a signal that can be disentangled from New Physics effects
by comparing $T$-odd moments for $\CP$ conjugate modes~\cite{Link:2005th}.

\textbullet\   The distinctive channel $D^{*\pm} \to D \pi^{\pm}$ provides a powerful tag
on the flavour identity of the neutral $D$ meson.

The notable ``fly in the ointment'' in searching for $\CP$ violation in
the charm sector is that $\DzDzb$ oscillations are slow.
Nevertheless one should accept this challenge:
$\CP$ violation involving $\DzDzb$ oscillations
is a reliable probe of New Physics: the asymmetry is controlled by
$\sin(\Delta m_D t) \, \times \, \Im (q/p)\bar{\rho} (D \to f)$.
In the Standard Model both factors are small, namely $\sim {\cal O}(10^{-3})$,
making such an asymmetry unobservably tiny --
unless there is New Physics (see, \eg,~\cite{Grossman:2006jg,Agashe:2004cp}).
$\DzDzb$ oscillations, $\CP$ violation and New Physics
might thus be discovered simultaneously in a transition.
Such effects can be searched for in final states common to
$\Dz$ and $\Dzb$ such as $\CP$ eigenstates (\eg,~$\Dz \to K^+K^-$)
doubly Cabibbo suppressed modes (\eg,~$\Dz \to K^+\pi^-$)
or three-body final states (\eg\ $\Dz \to \KS\pi^+\pi^-$).
Undertaking time-dependent Dalitz plot studies requires a
high initial overhead,
yet in the long run this should pay handsome dividends,
since Dalitz plot analyses can invoke many internal correlations
that, in turn, serve to control systematic uncertainties.
Such analyses may allow the best sensitivity to New Physics.

\mysubsubsubsection{Experimental Status and Future Benchmarks}
\label{BENCH}

Time-integrated $\CP$ asymmetries have been searched for
and sensitivities of order $1\%$ [several $\%$]
have been achieved for Cabibbo-allowed and -singly suppressed modes with two
[three] body final states~\cite{Shipsey:2006gf}.
Time-dependent $\CP$ asymmetries
(\ie, those involving $\DzDzb$ oscillations)
are still largely {\it terra incognita}.

Since the primary goal is to establish the intervention of New Physics,
one ``merely'' needs a sensitivity level above the reach of the Standard Model;
``merely'' does not mean this can easily be achieved.
As far as direct $\CP$ violation is concerned,
this means asymmetries down to the $10^{-3}$ or $10^{-4}$ level in
Cabibbo-allowed channels
and down to the $1\%$ level or better in doubly Cabibbo-suppressed modes.
In Cabibbo-singly-suppressed decays one wants to reach the $10^{-3}$ range
(although CKM dynamics can produce effects of that order,
future advances might sharpen the Standard Model predictions).
For  time-dependent asymmetries in
$D^0 \to \KS\pi^+\pi^-$, $K^+K^-$, $\pi^+\pi^-$ \etc,
and in $D^0 \to K^+\pi^-$,
one should strive for the
${\cal O}(10^{-4})$ and ${\cal O}(10^{-3})$ levels, respectively.

When striving to measure asymmetries below the $1\%$ level,
one has to minimize systematic uncertainties.
There are at least three powerful weapons in this struggle:
i) resolving the time evolution of asymmetries that are controlled by
$x_D$ and $y_D$, which requires excellent vertex detectors;
ii) Dalitz plot consistency checks;
iii) quantum statistics constraints on distributions, $T$-odd moments,
\etc~\cite{Bigi:1989ah}.

\mysubsubsection{Experimental reach of New Physics searches}

In this section we briefly summarize the experimental reach
of \superb\ for New Physics sensitive channels in the charm sector.
Table~\ref{tab:charm_rare} shows the expected $90\%$ confidence level
upper limits that may be obtained on various important rare $D$ decays,
including suppressed flavour-changing neutral currents,
lepton flavour-violating and lepton number-violating channels,
from one month of running at the $\psi(3770)$.
It is expected that the results from running at the $\FourS$
will be systematics limited before reaching this precision.

For studies of $\DzDzb$ mixing,
running in the $\Upsilon$ region appears preferable,
and, if the true values of the mixing parameters are unobservably small,
the upper limits on both $x_D$ and $y_D$ can be driven to below $0.1\%$
in several channels ($D^0 \to K^+\pi^-$, $K^+K^-$, $\KS\pi^+\pi^-$, \etc)
Therefore, \superb\ can study charm
mixing if $x_D$ and $y_D$ lie within the ranges predicted by the Standard Model, and recently observed.
The sensitivity to mixing-induced $\CP$ violation effects
obviously depends strongly on the size of the mixing parameters.
If one or both of $x_D$ and $y_D$ are ${\cal O}(1\%)$,
as indicated by the most recent results,
\superb\ will be able to make stringent tests of
New Physics effects in this sector.

The situation for searches of direct $\CP$ violation is clearer:
the \superb\ statistics will be sufficient to observe the Standard Model effect
of $\sim 3 \times 10^{-3}$ in $D^+ \to \KS\pi^+$~\cite{Bianco:2003vb},
and other channels can be pursued to a similar level.
Within three body modes, uncertainties in the Dalitz model are likely
to become the limiting factor.
However, model-independent $T$-odd moments can be constructed
in multibody channels,
and limits in the $10^{-4}$ region appear obtainable.

\begin{table}[!t]
    \caption{
      Expected $90\%$ confidence level upper limits that may be obtained on
      various important rare $D$ decays,
      from 1 month of \superb\ running at the $\psi(3770)$.
    }
    \label{tab:charm_rare}
  \begin{center}
    \begin{tabular}{lc}
      \hline \hline
      Channel & Sensitivity \\
      \hline
      $D^0 \to e^+e^-$,       $D^0 \to \mu^+\mu^-$      & $1 \times 10^{-8}$ \\
      $D^0 \to \pi^0 e^+e^-$, $D^0 \to \pi^0 \mu^+\mu^-$& $2 \times 10^{-8}$ \\
      $D^0 \to \eta  e^+e^-$, $D^0 \to \eta  \mu^+\mu^-$& $3 \times 10^{-8}$ \\
      $D^0 \to \KS   e^+e^-$, $D^0 \to \KS   \mu^+\mu^-$& $3 \times 10^{-8}$ \\
      $D^+ \to \pi^+ e^+e^-$, $D^+ \to \pi^+ \mu^+\mu^-$& $1 \times 10^{-8}$ \\
& \\
      $D^0 \to e^\pm\mu^\mp$        & $1 \times 10^{-8}$ \\
      $D^+ \to \pi^+ e^\pm\mu^\mp$  & $1 \times 10^{-8}$ \\
      $D^0 \to \pi^0 e^\pm\mu^\mp$  & $2 \times 10^{-8}$ \\
      $D^0 \to \eta  e^\pm\mu^\mp$  & $3 \times 10^{-8}$ \\
      $D^0 \to \KS   e^\pm\mu^\mp$  & $3 \times 10^{-8}$ \\
& \\
      $D^+ \to \pi^- e^+e^+$,       $D^+ \to K^-   e^+e^+$       & $1 \times 10^{-8}$ \\
      $D^+ \to \pi^- \mu^+\mu^+$,   $D^+ \to K^-   \mu^+\mu^+$   & $1 \times 10^{-8}$ \\
      $D^+ \to \pi^- e^\pm\mu^\mp$, $D^+ \to K^-   e^\pm\mu^\mp$ & $1 \times 10^{-8}$ \\
      \hline
    \end{tabular}
  \end{center}
\end{table}

\mysubsection{Summary}
\label{SUMMCHARM}

One does not have to be an incorrigible optimist to argue
that the best might still be ahead of us in the exploration
of the weak decays of charm hadrons.
Detailed studies of leptonic and semileptonic charm decays will allow
experimental verification of improvements in lattice QCD calculations,
down to the required ${\cal O}(1\%)$ level of precision.
This will result in significant improvements in the precision of
CKM matrix elements.
The possibility to operate with $e^+e^-$ collision energies
in the charm threshold region further extends the physics reach
and the charm program of the \sff.

While no evidence for New Physics has yet been found in charm decays,
the searches have only recently entered a domain
where one could realistically hope for an effect.
New Physics typically induces flavour-changing neutral currents.
Those could be considerably less suppressed for
up-type than for down-type quarks.
Charm quarks are unique among up-type quarks in the sense that
only they allow to probe the full range of phenomena induced by
flavour changing neutral currents,
including $\CP$ asymmetries involving oscillations.

There is little Standard Model background to New Physics signals in charm $\CP$ asymmetries,
and what there is will probably be under good control
by the time \superb\ starts operating.
Baryogenesis -- necessary to explain the observed matter-antimatter asymmetry
in our Universe --
requires a new source of $\CP$ violation beyond that of the Standard Model.
Such new sources can be probed in charm decays on
three different Cabibbo levels, differing in rates
by close to three orders of magnitude.
With the Standard Model providing one amplitude, observable $\CP$ asymmetries
can be linear in a New Physics amplitude, thus
greatly enhancing their sensitivity.
Finally, as stated repeatedly,
the goal has to be to identify salient features of the anticipated New Physics
beyond ``merely'' ascertaining its existence.
This will require probing channels with
one or even two neutral mesons in the final state --
something that is possible only in an $e^+e^-$ production environment.
CLEO$c$ and BESIII are unlikely to find $\CP$ asymmetries in charm decays,
and the $B$ Factory results will still be statistics limited.
A \sff\ would allow conclusive measurements.
\superb, with data taken at the $\FourS$ and near threshold, will complete the charm program down to the Standard Model level.

\afterpage{\clearpage}

\mysection{Other Topics}
\label{sec:other}
\def\cleo{\mbox{\normalfont CLEO}\xspace}

\subsection{Spectroscopy}

The recent experience of the $B$~Factories shows that many of the
most exciting results that can be produced by an $e^+e^-$ Super $B$ Factory
cannot be predicted in advance.
A brief inspection of the most cited papers from \babar\ and \belle\
provides evidence of the renaissance of hadronic spectroscopy
that has been stimulated by measurements such as:
\begin{itemize}\setlength{\itemsep}{0.2ex}
\item
  Discoveries of excited $D_s$ mesons by \babar~\cite{Aubert:2003fg}
  and \cleo~\cite{Besson:2003cp},
  and subsequent studies of their properties~\cite{Krokovny:2003zq,Abe:2003jk}.
\item
  Studies of the properties of $D^{**}$ states~\cite{Abe:2003zm,Kuzmin:2006mw}.
\item
  Observation of the $\eta_c(2S)$ in $B$ decay~\cite{Choi:2002na}.
\item
  Observation of a $\chi_{c2}^\prime$ candidate in
  $\gamma \gamma \to D\Dbar$~\cite{Uehara:2005qd}.
\item
  Discovery of a narrow charmonium-like state,
  denoted $X(3872)$~\cite{Choi:2003ue},
  and subsequent studies of its
  properties~\cite{Aubert:2004ns,Aubert:2005zh,Aubert:2006aj,Gokhroo:2006bt}.
\item
  Observation of an excited charm baryon $\Omega_c^*$~\cite{Aubert:2006je}.
\item
  Observation of double $c\bar{c}$ production in $e^+e^-$
  annihilation~\cite{Abe:2002rb},
  and subsequent studies of the
  production mechanism~\cite{Abe:2004ww,Abe:2005hd}.
\item
  Observation of a broad structure around $4.26 \, {\rm GeV}/c^2$
  in $\pi^+ \pi^- J/\psi$ produced by $e^+e^-$ collisions after
  initial state radiation~\cite{Aubert:2005rm},
  and subsequent studies of related processes~\cite{Aubert:2006ge}.
\item
  Studies of the hadronic structure of charmless three-body
  hadronic $B$ decays~\cite{Garmash:2004wa,Aubert:2005sk,Aubert:2005ce,Aubert:2006nu,Garmash:2006fh}.
\end{itemize}

Although past performance provides no guarantee of future success,
new particles have been discovered by the $B$~Factories at a rate of
more than one per year, and there is no reason to believe that this
should not continue into multi-${\rm ab}^{-1}$ territory.
Furthermore, it is clear that the clean $e^+e^-$ environment is ideal for
the complicated analyses necessary to pin down the
nature of these new hadrons.
Many different types of production, such as initial state radiation,
$\gamma\gamma$ cross sections, production in $B$ decay or in the $e^+e^-$
continuum, can be probed.
The possibility of running at different center-of-mass energies
extends the reach of this branch of the physics program.
Particles can be searched for in exclusive decays,
or using inclusive techniques, such as recoil analysis.
Amplitude analyses of multibody decays allow
further probes of resonant states.
Thus \superb\ is the ideal machine with which to study
hadronic spectroscopy over a large mass range,
and to discover new particles, both conventional and exotic.

\mysubsection {Studying Lower $\Upsilon$ Resonances}
\label{ss:lowUpsilon}

A high luminosity $B$ Factory with flexible center-of-mass energy
opens possibilities for the study of lower $\Upsilon$ resonances.
Such studies would allow tests of extensions of the Standard Model
in a complementary manner
to the physics program of a classic $B$ Factory and to the LHC.

Taking a non-minimal supersymmetric
standard model (like the NMSSM) as an example, it will be possible to detect the
presence of a light pseudoscalar Higgs produced in the decay
$\Upsilon(n{\rm S}) \to ll \gamma$ ($n=1,2,3$)
as an intermediate (perhaps quite  broad) state.
In this scenario, the first experimental evidence for New Physics
would be likely to appear in the breaking of lepton universality,
since the tauonic branching fraction would
be enhanced with respect to the electronic and muonic ones~\cite{Sanchis-Lozano:2003ha}.

Indeed, a light non-standard Higgs has not been excluded by LEP in
certain scenarios~\cite{Accomando:2006ga},
including the MSSM with explicit $\CP$ violation~\cite{Carena:2000ks}.
From LEP data, a window in the Higgs mass {\it vs.} $\tan\beta$ plane
is still allowed, which would approximately match the values
needed to explain the hint of lepton universality breaking observed
by CLEO in $\Upsilon(n{\rm S}) \to ll$ decays~\cite{Besson:2006gj}.
In this class of models, the lightest pseudoscalar Higgs
(often denoted as $a_1$ or $A_1$) would be, in fact,
a mixture of singlet and MSSM-like components~\cite{Gunion:2005rw}.

In addition, 
the study of
$\Upsilon(n{\rm S}) \to$ invisible decays allows independent constraints
on models with light dark matter (LDM) to be obtained.
It is important to remark that the searches for LDM
in invisible decays of the $\Upsilon$ carried out by CLEO~\cite{Rubin:2006gc}
and \belle~\cite{Tajima:2006nc}
only put limits on a vector mediator of the decay.
To extend such searches, one should look at the decay
$\Upsilon(n{\rm S}) \to \gamma +$ invisible~\cite{McElrath:2005bp}.

From an experimental point of view, the analysis can be performed
using the production of a light $\Upsilon$ resonance through ISR radiation,
or with short runs at threshold~\cite{Sanchis-Lozano:2006gx}.
The experimental challenge for studies of $\Upsilon \to \ell \ell$
lies in the ability to detect the breaking of lepton universality,
which demands a detailed understanding of trigger efficiencies, and a proper simulation of
QED effects and the EMC response. In this picture, statistics is not a
limiting factor; even short runs of one month or less
at very high luminosity will close
this currently open window or lead to the discovery of a light Higgs.

In addition, the possibility of running at the
$\Upsilon(n{\rm S})$ ($n=1,2,3$) resonances should be considered.
This would provide a very clean environment for
$\Upsilon(n{\rm S})\to$ invisible decays,
allowing additional searches for LDM.
Several other studies possible at
the $\Upsilon(n{\rm S})$ resonances have been considered
in the literature~\cite{Li:2006wx,Brambilla:2004wf,Artuso:2004fp,Zambetakis:1983te,Godfrey:1985xj,Godfrey:2001eb,Lahde:2002wj}.

On the experimental side, the improved hermeticity of the \superb\ detector, and
the lower boost compared to the existing asymmetric $B$~Factories
will improve the detector performance.
Consequently, \superb\ should provide an
increase in the reconstruction efficiency of ISR photons 
and a reduction in the impact of QED and machine background.

\subsection{Studies with Light Quarks}

A Super Flavour Factory can also address several interesting open questions
concerning light quarks that exist in low energy $e^+e^-$ annihilation data;
these questions are otherwise likely to remain unanswered.
This can be done by exploiting initial state radiation,
as has been done at the present $B$ Factories running at the $\FourS$,
but it can be done better if \superb\ runs at
lower center-of-mass energies, with the expected very high luminosity,
and the planned reduced energy asymmetry, which
improves the detection efficiency for processes at threshold.

The number of unanswered questions remaining to be addressed will
depend on developments at the new low energy $e^+e^-$ colliders:
BEPCII~\cite{Harris:2006qh},
the $\tau$/charm factory in Beijing (if BEPCII collects data below the $\jpsi$);
VEPP2000~\cite{Serednyakov:2004px,Serednyakov:2004me},
the VEPP2M machine in Novosibirsk renewed and upgraded in energy
up to 2 GeV in the center-of-mass (if VEPP2000 has enough luminosity above the injection energy, 1.9 GeV); and
DANAE~\cite{Zobov:2005ne},
the present DA$\Phi$NE in Frascati upgraded in energy and luminosity (if DANAE is funded).

In the following, some of the open questions
are very briefly discussed:
$\sigma_{\rm tot} (e^+e^- \rightarrow {\rm hadrons})$,
$\gamma \gamma$ interactions, light quark hadronic chemistry,
and baryon time-like form factors.
More details on these topics can be found in ref.~\cite{Ambrosino:2006gk}.

The measurement of the muon anomalous magnetic moment,
$(g-2)_\mu$, does not agree with the QED prediction at the $\sim 3\sigma$ level~\cite{Davier:2003pw,Passera:2005mx}.
The total cross-section $\sigma_{\rm tot} (e^+e^- \rightarrow {\rm hadrons})$,
integrated over the total center-of-mass energy and
weighted by a function strongly peaked at low energies,
contributes heavily to the $(g-2)_\mu$ calculation~\cite{hadro,Gourdin:1969dm}.
It also contributes, with less weight, to
$\alpha(M^2_Z)$~\cite{alpha,Geshkenbein:1994ks}.
Present measurements of
$\sigma_{\rm tot} (e^+e^- \rightarrow {\rm hadrons})$ up to $\sim$ 1 GeV,
(mostly around the $\rho$(770)),
are performed with a claimed $\sim 1\%$ accuracy,
but actually disagree at the $\sim 5\%$
level~\cite{Franzini:2006aa,Sibidanov:2006ts,Aloisio:2004bu,Davier:2002dy,Cirigliano:2002pv}.
Should a future new measurement of $(g-2)_\mu$ be performed,
its interpretation would require improved measurements of
$\sigma_{\rm tot} (e^+e^- \rightarrow {\rm hadrons})$ at these energies,
as well as at energies in the $1$--$2$ GeV range,
where existing data are quite old and scanty~\cite{Jegerlehner:2006ju}.
These measurements show a violation of local duality,
as in this energy range they are always lower than the pQCD expectation.
Many hadronic final states, detected with different detection efficiencies,
contribute to the total cross section, and a measurement at $\sim$ 1\% level
may not be done in the near future.
%
%
%
%
%

The process $\gamma \gamma \rightarrow {\rm hadrons}$ is an important tool
to test chiral perturbation theory at
threshold~\cite{Bochicchio:1985xa,Donoghue:1988ee,Pennington:2006qi},
as well as being complementary to $e^+e^-$ interactions for
studies of electromagnetic form factors of hadrons.
In spite of the large number of events collected
at the present $B$~Factories,
severe cuts are necessary in their selection
(namely an invariant mass $> 700\; {\rm MeV}$
for pion pairs~\cite{Mori:2006jj}
and an angle in the center-of-mass frame $|\cos \theta^*|< 0.6$,
for pion as well proton pairs~\cite{Kuo:2004kz}.
Hence chiral perturbation theory at threshold has not been tested.
Furthermore at higher center-of-mass energies
the angular distributions are not at all isotropic.
For instance,
$\frac{d\sigma}{d\cos\theta} (\gamma \gamma \rightarrow p \bar p)$ at
$\sim 2\ {\rm GeV}$ is consistent with $|Y^0_2|^2$,
which means that $\sim 50\%$ of these events are
lost because of the angular cuts.

Hadron spectroscopy will remain interesting in the future,
whether or not it is fashionable.
Evidence for non $q\overline q$ states in the charm sector has recently
been observed.
In particular, a candidate $D D^*$ molecular state, the $X(3872)$
meson~\cite{Bander:1975fb,Voloshin:1976ap,DeRujula:1976qd,Manohar:1992nd,Tornqvist:1993ng},
has been found at the expected mass and with a vanishing width,
with the expected quantum numbers and the expected properties,
such as isospin violating decays~\cite{Choi:2003ue}.
In the light and strange quark sector, it has been suggested that the
$f_0(980)$ meson,
which has vacuum quantum numbers, is not a $q\bar q$ state.
Recently KLOE and CMD2 data on $\phi \rightarrow f_0 \gamma$
and \babar\ data on $e^+e^- \rightarrow \phi f_0$
strongly support the hypothesis that the $f_0(980)$ meson
is a tetraquark state~\cite{f0-noi}.
Most of the non $q\bar{q}$ vector mesons suggested by QCD
are expected to be isoscalar, but the luminosity produced
at the current $B$~Factories by means of ISR
is not high enough to explore isoscalar channels, which are weakly coupled to $e^+e^-$.
There are clear hints of unexpected structures in some isoscalar channels,
for instance $\phi f_0$,
and in some isovector channels, 6$\pi$
or $\phi \pi$.
In conclusion, it is very likely that hadronic chemistry will still be
topical in the next decade.
\begin{figure}[!ht]
    \begin{flushright}
    \includegraphics[height=70mm]{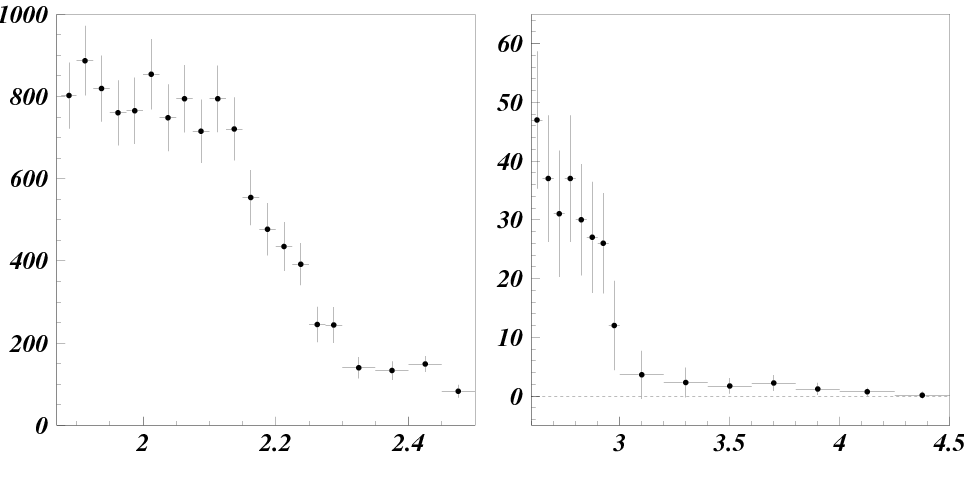}
    %
    \put(-60,0){$M_{p\bar{p}}$(GeV)}
    \put(-260,0){$M_{p\bar{p}}$(GeV)}
    \put(-420,105){\rotatebox{90}{$\sigma(e^+e^-\to p \bar{p})[{\rm nb}]$}}
    \caption{
      \label{pp-babar}
      Stepwise behaviour of the $e^+e^-\to p \overline{p}$ cross-section
      observed by \babar~\cite{Aubert:2005cb,BaldiniFerroli:2005ap}.
    }
  \end{flushright}
\end{figure}

\begin{figure}[!t]
  \begin{minipage}{160mm}
      %
      \begin{minipage}{75mm}
      \centering
          \includegraphics[width=70mm]{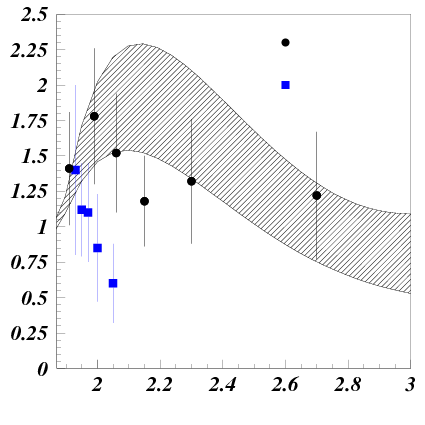}
          \put(-61,0){$M_{pp}(GeV)$}
          \put(-210,145){\rotatebox{90}{$|G_E/G_M|$}}
          \put(-60,176){\babar}
          \put(-60,156){Lear}
          \put(-205,-50){%
            \begin{minipage}{75mm}
              \caption{
                \label{ratio}
                Ratio of form factors $|G_E/G_M|$,
                compared between
                \babar~\cite{Aubert:2005cb,BaldiniFerroli:2005ap}
                and Lear data~\cite{Bardin:1994am}.
                The hatched area is a description based on
                analyticity~\cite{Baldini:2006bh}.
              }
            \end{minipage}
          }
      \end{minipage}
     \hspace{2mm}
      \raisebox{0mm}{%
        \begin{minipage}{75mm}
            \includegraphics[width=70mm]{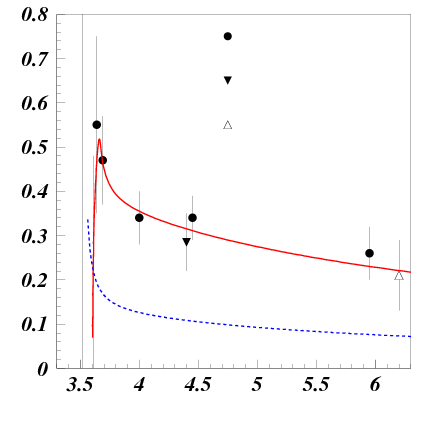}
            \put(-61,0){$M_{nn}(GeV)$}
            \put(-210,145){\rotatebox{90}{$\left|G_M^n(Q^2)\right|$}}
            \put(-87,178){FENICE}
            \put(-87,156){DM2}
            \put(-87,135){DM2 from $G^\Lambda$}
            \put(-205,-50){%
              \begin{minipage}{75mm}
                \caption{
                  \label{gmn}%
                  $\left|G_M^n (Q^2)\right|$ as measured by FENICE,
                  compared to pQCD expectations~\cite{Blunden:2003sp}
                  (dashed line) and as suggested by the
                  average angular distribution (solid line).
                }
              \end{minipage}
            }
        \end{minipage}
      }
      %
  \end{minipage}
\end{figure}

Nucleon space-like form factors were supposed to be well-established
thirty years ago.
In many textbooks,
a scaling law between electric and magnetic form factors was discussed.
However, a few years ago, this result was shown to be
completely wrong~\cite{Jones:1999rz,Gayou:2001qt,Gayou:2001qd,Milbrath:1997de},
although in agreement with early predictions~\cite{iachello,Iachello:1972nu}.
\babar\ has recently made the best measurement of
$\sigma (e^+e^- \rightarrow p \bar p)$,
showing hints of unexpected behavior, with plateaux and drops
(see Fig.~\ref{pp-babar})
and electric/magnetic form factors ratios varying with
center-of-mass energy~\cite{Aubert:2005cb,BaldiniFerroli:2005ap}
(see Fig.~\ref{ratio}),
but much more data is needed to draw a conclusion.
The neutron time-like form factors
(data on $|G_M^n|$ is shown in Fig.~\ref{gmn}),
have been measured only once, nevertheless showing
$\sigma(e^+e^- \to n \bar n) \sim  2 \sigma(e^+e^- \to p \bar p)$,
that is $\sim$ 8 times the na\"{\i}ve $(Q_u/Q_d)^2$
expectation~\cite{Antonelli:1998fv}.
No data exist on strange baryon form factors;
baryon time-like form factors are an open
field.


\afterpage{\clearpage}

\mysection{Summary}
\label{sec:summary}

This chapter has described the rich physics opportunities available at \superb.
The unique ability of high statistics studies of $b$, $c$ and $\tau$ decays at a high
luminosity \abf\ to
clarify the flavour structure of physics beyond the Standard Model is the flagship of the
program, but the wealth of precision measurements in weak decays and new discoveries in spectroscopy are also of clear interest in the coming decade.

Signals for the $\CP$ asymmetry $A_{\CP}(B\to X_{s+d}\gamma)$,
violations of lepton flavour or lepton universality symmetries
in $\tau$ or $B$ decays,
a shift in the dilepton invariant mass at which
$A_{\rm FB}(B\to X_s\ell^+\ell^-)=0$,
$\CP$ violation in $\tau$ decays,
or in Cabibbo-allowed or doubly-Cabibbo-suppressed charm decays,
would be unmistakable signs of New Physics.
These are all processes whose Standard Model contribution is
either vanishingly small or already very well known,
and they can only be probed at an interesting level by a Super $B$ Factory, with
data samples 50 to 100 times that which will exist at the end of the \babar\ and Belle programs.

At the same time, many other measurements can be performed
that are sensitive to the CKM parameters in the Standard Model:
they allow for a redundant determination of the UT sides and angles
which could display inconsistencies signaling the presence of New Physics .
To this end, precise CKM metrology,
which is interesting by itself within the Standard Model,
is a crucial ingredient for New Physics searches.
\superb\ can determine the UT parameters with percent errors,
allowing for New Physics tests in $\Delta B=1$ and $\Delta B=2$ processes,
involving all species of $B$ mesons, at the same level of accuracy.
For example, a well-known class of measurements
is the time-dependent $\CP$ asymmetries of non-leptonic decays
involving $b\to s$ transitions.
These are sensitive to $\sin2\beta$ in the Standard Model, but are expected
to receive substantial corrections in several New Physics scenarios.
\superb\ can reduce the error of the $B$ Factory measurements for these modes
by more than a factor of five, which is crucial to achieving a New Physics sensitivity at
the highest possible mass scale.

Another notable set of measurements are those sensitive to Higgs exchange
in models with two Higgs doublets in the large $\tan\beta$ regime.
These include $\BR(B\to\tau\nu)$, $\BR(B\to D\tau\nu)$
and LUV in $B$ and $\tau$ decays.
These measurements are a probe of Higgs-mediated New Physics effects,
which can be large even in the unfavourable MFV scenario.

To take full advantage of the \superb\ statistics,
improvements in non-perturbative calculations will be necessary.
Lattice QCD seems capable of achieving the required accuracy
by the time \superb\ is operational.
Inclusive techniques,
mainly the OPE and the heavy-quark expansion,
will also profit from the large statistics of \superb, which, for instance,
allows unprecedentedly accurate measurements of hadronic spectra.
On the other hand, in spite of recent theoretical progress,
predictions for non-leptonic decays still rely to some extent on models.
While the main theoretical ideas (factorization, flavour symmetries, \etc)
would benefit from the improved measurements
of non-leptonic modes performed at \superb,
New Physics searches with non-leptonic decay modes are likely to require methods
allowing data-driven control over the theoretical uncertainties.

We have presented estimates of the potential for \superb\
to measure these processes, as well as many others.
We find typical improvements over previous measurements,
either in error or in sensitivity, of factors between five and ten.
We have given arguments to support our estimates, considering
experimental systematic as well as theoretical uncertainties.
The main results are collected in
Tables~\ref{tab:inputs1} and~\ref{tab:inputs2}.

In addition, we have tried to assess the
theoretical interest of these measurements.
We have clearly demonstrated that flavour physics at \superb\
is an open window on virtual effects induced by new particles
with masses up to hundreds of TeV,
and that even in the MFV scenario there are processes
that probe New Physics scales beyond 1 TeV.
\superb\ provides a unique opportunity to extend
the New Physics reach of the LHC into the multi-TeV region.
This is important,
whether or not LHC finds direct evidence of New Physics.
In the likely scenario that new flavoured particles are discovered at LHC,
measuring the flavour-violating couplings of these new particles
will become a high-priority task that can be
systematically carried out only at \superb.
In addition, the \superb\ sensitivity to LFV in $\tau$ decays
reaches the theoretically interesting range, and will
effectively complement the MEG result on muon LFV
in helping to ascertain the underlying mechanism of LFV.

While several other interesting measurements have been discussed,
from QCD-related studies with charm quarks to hadronic spectroscopy,
we have certainly not touched on every aspect of heavy flavour phenomenology
that can be studied at \superb.
There is little doubt that the richness of this phenomenology,
and the window of opportunity that it offers for New Physics studies,
is as unprecedented in flavour physics as is
the planned peak luminosity of the accelerator and
its unique capability to run at charm and $\tau$ threshold as well as in the $\Upsilon$ region.
Indeed, the potential of the \superb\ physics program goes beyond
the traditional flavour physics domain, and could have a
profound impact on the future progress of particle physics.

\afterpage{\clearpage}



\def\B       {\ensuremath{B}\xspace}


%
\afterpage{\clearpage}

\graphicspath{{Accel/figures/}}

\chapter{The Accelerator}

\section{Overview}
\label{section:Overview}
\subsection{A History of $B$ Factories}
\label{sec:accel-overview}

A Super $B$ Factory, an asymmetric energy \epem collider with a
luminosity of order \tenTo36, can provide a uniquely sensitive probe of
New Physics in the flavour sector of the Standard Model. 

The \pepii and \kekb
asymmetric colliders \cite{KEKB_SR,PEP2_SR} have produced unprecedented
luminosities, above
\tenTo34, taking our understanding of the accelerator physics and engineering demands of asymmetric
\epem colliders to a new parameter regime.
This very high luminosity, coupled with the innovation of continuous injection
and the high efficiency of the accelerators and
detectors, will allow each of these machines to
produce 1000\invfb\ or more by the end of this decade.
The study of New Physics
effects in the heavy quark and heavy lepton sectors,
however, requires a data sample two orders of magnitude larger, hence the
luminosity target of $10^{36}$ for \superb\!.

Attempts to design a Super $B$~Factory date to 2001.
The initial approach at SLAC and KEK had much in common: they were
extrapolations of the very successful $B$~Factory designs,
with increased bunch charge, more bunches, somewhat reduced
$\beta_y^*$ values, and crab cavities.
These proposed designs reached luminosities of 5 to $7\times 10^{35}$,
but had wall plug power of the order
of 100\MW.

This daunting power consumption motivated us to adapt linear collider concepts
from the SLC and ILC to the regime of high
luminosity storage ring colliders. The low emittance design presented
herein reaches the desired luminosity regime with beam
currents and wall plug power comparable to those in the current $B$ Factories.

The parameters for a Flavor Factory based on
an asymmetric-energy \epem\ collider
operating at a luminosity of
order \tenTo36 at the \FourS resonance and \tenTo35 at $\tau$
production threshold are described below.
Such a collider would produce an integrated
luminosity of at least 15,000\invfb (15\invab) in a running year
($10^7 \sec$) at the \FourS resonance.

The construction and operation of
modern multi-bunch \epem colliders have brought about many advances in
accelerator physics in the area of high currents, complex interaction
regions, high beam-beam tune shifts, high-power RF systems, control of
beam instabilities, rapid injection rates, and reliable up-times
(90\%). The successful operation of the currently operating $B$ Factories
has proven the validity of their design concepts:
\begin{itemize}
\item Colliders with asymmetric energies work;
\item Beam-beam energy transparency conditions provide only weak constraints;
\item Interaction regions with two energies can be built for both head-on and small angle collisions;
\item IR backgrounds can be handled successfully;
\item High-current RF systems can be operated with excellent efficiency;
\item Beam-beam tune shift parameters can reach 0.06 to 0.09;
\item Good injection rates can be sustained. Continuous injection is now in routine operation, largely removing the distinction between peak and average luminosity;
\item The electron cloud effect (ECI) can be managed; and
\item Bunch-by-bunch feedback works well with 4\ns bunch spacing.
\end{itemize}

Lessons learned from SLC, and more recent ILC studies and
experiments (FFTB, ATF, ATF2), have also produced and proven new concepts:
\begin{itemize}{

\item Small horizontal and vertical emittances can be produced in a
damping ring having a short damping time.
\item Very small beam spot sizes
and beta functions can be produced at the interaction region; and
}
\end{itemize}
The design of the \superb\ \epem collider combines extensions of the
design of the current $B$ Factories with new linear collider concepts to produce
an extraordinary leap in $B$ Factory luminosity without increasing beam currents or power consumption.

The luminosity \lum of an
\epem collider is given by the expression
\begin{alignat}{1}
&\lum =  \frac
{ \Npos  \Nele  }{ 4 \; \pi \; \sy \sqrt{(\sz\tan \thx/2)^2 + \sx^2}} \,\fcoll\\
&\sigma_{x,y} =  \sqrt{ \beta_{x,y} \;  \varepsilon_{x,y}\;}\,,
\end{alignat}
where  \fcoll is the frequency of collision of each bunch,
$N$ is the number of particles in the positron $(+)$ and electron $(-)$
bunches, $\sigma$ is the beam size in the horizontal ($x$),
vertical ($y$) and longitudinal ($z$) directions,
$\varepsilon$ is the beam emittance, $\beta$ is the beta function
(in \cm) at the collision point in each plane and \thx is the
crossing angle between the beam lines at the interaction point (IP).

In this chapter we describe the principles of the
design of a new asymmetric collider that can reach
a peak luminosity of \tenTo36 with beam currents and bunch lengths
similar to those of the currently operating \epem factories,
through the use of smaller emittances and a new scheme of crossing angle
collision.

\subsection{Key Issues for a Super $B$ Factory}
\label{sec:Key_issues}

Our design is based on a new collision scheme,
that we call a ``crabbed waist''. This new scheme will
allow \superb\ to reach a luminosity of the order of \tenTo36 by overcoming
some of the issues that have plagued earlier super \epem collider designs, such as very high
beam currents and very short bunches. In this section we will review
the crabbed waist concept and address key issues related to high
luminosity colliders, such as luminosity with a crossing angle, beam lifetime
and injection, backgrounds, beam emittances and stability, polarization, power
and costs.


\subsubsection {The crabbed waist collision scheme}

In high luminosity colliders, one of the key requirements is very
short bunches, since this allows a decreased \betys at the IP, thereby
increasing the luminosity. However, \betys cannot be made much smaller
than the bunch length without incurring an ``hourglass'' effect.
Moreover, high luminosity requires small
vertical emittance, together with large horizontal beam size and horizontal emittance,
to minimize the beam-beam effect. It is, unfortunately, very difficult
to shorten the bunch length \sz in a ring.

This problem can be overcome with the recently proposed
crabbed waist scheme \cite{bib:Panta} for beam-beam collisions, which can
substantially increase luminosity without having to decrease the
bunch length, since it combines several potentially advantageous
ideas.

The first idea is the use of a large Piwinski angle: for collisions at a
crossing angle \thx, the luminosity \lum, the horizontal
\xix and the  vertical \xiy tune shifts scale according to \cite{bib:G58}:
\begin{alignat}{1}
\label{eq:lumi_tune}
\lum & = \frac{ \gamma^+ \xiy\, \Npos\, \fcoll}{ 2\, r_e\, \bety}
\left( 1 + \frac{\sy}{\sx}\right) \propto \frac { \Npos \; \xiy}{\bety}\\
\label{eq:tuney}
\xiy & = \frac{r_e \Nele }{2 \pi \gamma^+}
\frac {\bety}{\sy \left( \sx \sqrt {1 + \piwi^2} +\sy \right)}
\propto \frac{ \Nele \sqrt{\bety}}{\sy \sz \thx}\\
\label{eq:tunex}
\xix & =  \frac{r_e \Nele }{2 \pi \gamma^+}
\frac{\betx}{\sx^2 \left[ \left(1+\piwi^2\right) +
\frac{\sy}{\sx} \sqrt{1+\piwi^2}  \right]}
\propto \frac{ \Nele \betx}{ \left( \sz \thx \right)^2}\,.
\end{alignat}

The Piwinski angle \piwi is defined as:
\begin{equation}
  \piwi = \frac{\sz}{\sx}\; \tan \frac{\thx}{2} \sim
\frac{\sz}{\sx}\;  \frac{\thx}{2}
\end{equation}
where \sx the horizontal {\it rms} bunch size, \sz the {\it rms}
bunch length, \Nele(\Npos) the number of electron per HER (LER) bunch, and
$\gamma^+$ the Lorentz factor for the positrons in the LER.
We consider here the case of flat beams, small horizontal crossing
angle $\thx \ll 1$ and large Piwinski angle $\piwi \gg1$.

The idea of colliding with a large Piwinski angle is not new
(see for example \cite{bib:Hirata}). It has been also proposed for
the LHC upgrade \cite{bib:LHC}, to increase the bunch length and
the crossing angle. In such a case, if it were possible to increase
$N$ in proportion to $\sx \thx$, the vertical tune shift \xiy would
indeed remain constant, while the luminosity would grow
proportional to $\sz\thx$
(Eqs.~\ref{eq:lumi_tune} and \ref{eq:tuney}).
Moreover, the horizontal tune shift \xix drops like $1/(\sz\thx)^2$
(Eq.~\ref{eq:tunex}), so that for very large \piwi the
beam-beam interaction can be considered, in some sense,
one-dimensional, since the horizontal footprint in the tune
plane shrinks. However, as distinct from \cite{bib:LHC}, in the
crabbed waist scheme described here, the Piwinski angle is increased
by decreasing the horizontal beam size and increasing the crossing
angle. In this way, the luminosity is increased, and the
horizontal tune shift due to the crossing angle decreases. The most important effect is
that the overlap area of colliding bunches
is reduced, as it is proportional to $\sx/\thx$. Thus, if the vertical
beta function \bety can be made comparable to the overlap area size:
$$
\bety \sim \frac{\sx}{\thx} \ll \sz,
$$
several  advantages are gained:
\begin{itemize}
\item Small spot size at the IP, \ie, higher luminosity (Eq.~\ref{eq:lumi_tune})
\item Reduction of the vertical tune shift (Eq.~\ref{eq:tuney}); and
\item Suppression of vertical synchrobetatron resonances \cite{bib:Pestrikov}.
\end{itemize}

There are additional advantages in such a collision scheme:
there is no need to decrease the bunch length to increase the
luminosity, as proposed in standard upgrade plans for $B$ and
$\phi$ Factories \cite{bib:SuperPEP,bib:SuperKEKB,bib:SuperDafne}.
This will certainly ease the problems of HOM heating, coherent
synchrotron radiation of short bunches, excessive power consumption,
\etc. Moreover the problem of parasitic collisions (PC) is
automatically solved by the  higher crossing angle and smaller
horizontal beam size, which makes the beam separation at the PC large in
terms of \sx.

\begin{figure}[h!tb]
\begin{center}
\psfragscanon
\psfrag{sz}{$2 \sz$}
\psfrag{Sx}{$2 \sx$}
\psfrag{bx}{\betx}
\psfrag{th}{$2 \thx$}
\psfrag{th}{\thx}
\psfrag{s}{$2 \frac{\sx}{\thx}$}
\psfrag{S}{$  \frac{\sz}{\thx}$}
\includegraphics[width=15cm]{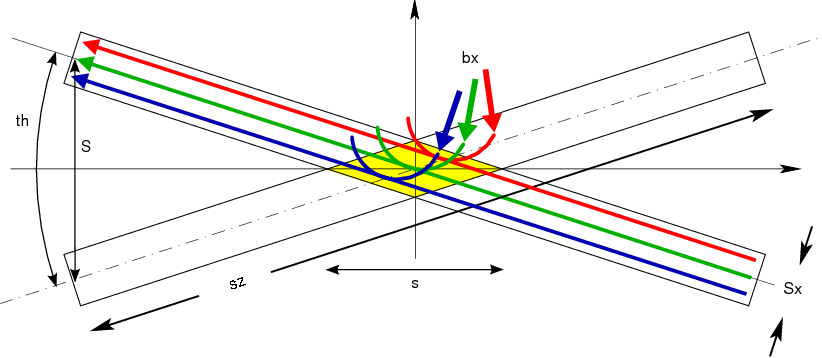}
\caption{\label{fig:Piwinski_CW} Large Piwinski angle and crabbed waist
scheme.
The collision area is shown in yellow.}
\end{center}
\end{figure}

However, a large Piwinski angle itself introduces new beam-beam
resonances and may strongly limit the maximum achievable tune shifts
(see for example \cite{bib:Ohmi}). This is where the crabbed waist innovation is required.
The crabbed waist
transformation boosts the luminosity,
mainly by suppression of betatron (and
synchrobetatron) resonances that usually arise (in collisions without the
crabbed waist) through the vertical motion modulation by horizontal beam
oscillations \cite{bib:CrabWaist}.
A sketch of the crabbed waist scheme is shown in Fig.~\ref{fig:Piwinski_CW}.

\begin{figure}[h!tb]
\begin{center}
\psfragscanon
\psfrag{IP}[][]{IP}
\psfrag{T}[]{$T_{x,y}$}
\psfrag{Tt}[]{$\bar{T}_{x,y}$}
\psfrag{Sex}[]{\parbox{2cm}{\begin{center} Sextupole  $+K_s$\end{center}}}
\psfrag{Asex}[]{\parbox{3cm}{\begin{center} Anti-sextupole  $-K_s$\end{center}}}
\psfrag{Dm}[]{\parbox{2cm}{\begin{center} $\Delta \mu_x = \pi$
$\Delta \mu_y = \pi/2$\end{center}}}
\includegraphics[width=15cm]{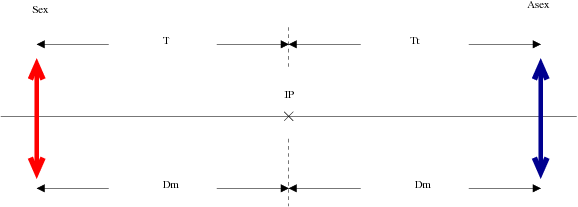}
\caption{\label{fig:Sextupoles} \CrabbedWaist correction by sextupole lenses.}
\end{center}
\end{figure}

The crabbed waist correction scheme can easily be realized in
practice with two sextupoles magnets in phase with the IP in the $x$
plane and at $\pi/2$ in the $y$ plane, on both sides of the IP, as
shown in Fig.~\ref{fig:Sextupoles}.

\subsubsection{Review of the key issues}

The \superb\ design aims to achieve a luminosity in excess of \tenTo36. Some
of the key design elements that need consideration in realizing
this goal are described in this section, beginning with the
luminosity-{\it vs.}-power issue.

\begin{enumerate}
\item {\bf Luminosity}: For very flat beams, luminosity can be written as:
$$
    \lum = \frac{ 2.17\EE{34}}{  \gev\cm\rm{\, C}}
\frac{ E \, I\, \xiy}{\bety}\,,
$$
where $I$ is the LER beam current, $E$ is the LER energy,
\xiy the vertical tune shift, \bety is the vertical beta at the
Interaction Point.
\item {\bf Synchrotron radiation power}: Power dissipation is related to the
beam current $I$ and to the energy loss per turn $U_o$ via:
$$
    P = I U_o.
$$
All colliders aim to maximize \lum while keeping $P$ as small as
possible. The \superb\ design is based on a ``large Piwinski
angle'' and crabbed waist scheme as described above. This allows us to lower
\bety to $0.2\mm$ and increase \xiy to $0.2$. These values should
be compared with the present \KEKB parameters of $\bety = 6\mm$ and $\xiy
= 0.06$. The \superb\ parameters result in a luminosity about two orders
of magnitude
larger than that achieved at \KEKB, with beam currents
and power consumption essentially unchanged.

\item {\bf Detector backgrounds}: Maintaining beam power as low as
possible is important to minimize backgrounds, which
scale with the beam currents. The interaction region (IR)
design also plays a fundamental role. The combination of large
crossing angle and small beam sizes, emittances and beam angular
divergences at the IP in the \superb\ design is very effective in
further decreasing the absolute background levels with respect to the
current $B$ Factories. These same factors also relax design requirements
for the IR. Luminosity-related backgrounds must, of course, be taken into
account, and can impose serious shielding requirements.

\item {\bf Beam lifetime}: In the current \epem\ factories, beam lifetime is
determined mainly by ring characteristics such as vacuum quality, dynamic
aperture, \etc. In \superb, beam lifetime is instead almost entirely dominated
by the luminosity itself: radiative Bhabhas limit the lifetime to a
few minutes for both rings. All other contributions are much smaller,
except for the Touschek lifetime of the low energy beam, which
causes a worsening by about a factor 1.3.
Given the short beam lifetime, the injection system must
be able to provide particles at a rate about 10 times larger than those
for the present factories.

\item {\bf Beam emittance}: The horizontal emittance \epsx is
 determined mainly by the ring lattice optics; the vertical emittance
\epsy is dominated by ring imperfections, which must be tightly controlled
to reach the design value. The current factories, and most
of the other \epem colliders, have achieved vertical/horizontal
emittance ratios similar to the  \superb\ design. However, the
absolute values for \superb\ are much smaller; they are similar to those
at the test damping ring for the \ilc
project, the ATF~\cite{bib:ATF}.
Thus, tolerances, stability levels and tuning constraints are also tighter than
those for the current factories. Instead, they are very similar to those for 
the ATF and the design values for the ILC Damping Rings, which will produce 
beams very similar to those of \superb\!.

\item {\bf Polarization}: \superb\ can provide collisions with
longitudinally polarized electrons by using a polarized electron gun
and spin rotators in the ring. Polarized positrons could be provided as well, but further study is required to evaluate whether the additional physics
benefit outweighs the added complexity.
A vigorous R\&D program (see references
in Sec.~\ref{sec:Posipol})
is being pursued by the \ilc community to provide a polarized positron
source. Production rates required by \superb\ are 100 times less 
demanding than those for the
\ilc, so such a source could be feasible by
the time \superb\ is funded. Preliminary considerations for a
polarized positron source are discussed in Sec.~\ref{sec:Posipol}.

\item {\bf Cost}: In the conventional Super $B$ Factory designs, the cost is
dominated by the requirements for dealing with higher currents and shorter bunches: for example,
substantial additions
to the RF system, engineering design for larger HOM power due to shorter
bunches, and the cooling and vacuum challenges posed by larger synchrotron radiation power.
Most of these problems do not exist in the
\superb\ design; the absolute cost of \superb\ is
therefore very similar to the present machines. In addition, the \superb\ design allows
the reuse of a great deal of \pepii hardware, resulting in substantial savings 
for the project, even at a new site.
\end{enumerate}


\subsection{Parameters}
\subsubsection{Nominal parameters for \tenTo36}

The IP and ring parameters have been optimized based on several
constraints. The most significant are:
\begin{itemize}
\item Maintaining wall plug power, beam currents, bunch lengths, and RF requirements comparable to present $B$ Factories;
\item Planning for the reuse as much as possible of the \pepii hardware;
\item Requiring ring parameters as close as possible to those already
  achieved in the $B$ Factories, or under study for the ILC
  Damping Ring or achieved at the ATF ILC-DR test facility~\cite{bib:ATF};
\item Simplifying the IR Design as much as possible. In particular, reduce
  the synchrotron radiation in the IR, reduce the HOM power and increase the
  beam stay-clear. In addition, eliminate the effects of the parasitic
  beam crossings;
\item Relaxing as much as possible the requirements on the beam
  demagnification at the IP; and
\item Designing the final focus system to follow as closely as possible 
  already tested systems, and integrating the system as much as possible into the ring
  design.
\end{itemize}

Column 1 of Table~\ref{tab:params} shows a parameter set that closely matches
these criteria. Further details on beam-beam
simulations and lattice design will be presented in the following
sections.

\subsubsection{Upgrade parameters}

Many of the nominal \superb\ design parameters could, in principle, be pushed
further to increase performance. This provides a excellent upgrade path after
experience is gained with the nominal design. The upgrade
parameters are based on the following assumptions:

\begin{itemize}
\item Beam currents could be raised to the levels that
  \pepii should deliver in 2008;
\item Vertical emittance at
  high current could be reduced to the ATF values;
\item Lattice properties support a further reduction in \betx and \bety; and
\item Beam-beam effects are still far from saturating the
  luminosity.
\end{itemize}

In principle, the design supports these improvements, so a luminosity
higher than nominal may well be feasible. In addition, it should be
pointed out that, since the nominal design parameters are not pushed to
maximum values, there is flexibility in obtaining the design
luminosity by relaxing certain parameters, if they prove
more difficult to achieve, and pushing others.
Columns 2 and 3 of Table~\ref{tab:params} show two potential upgrade paths.

{ \setlength{\tabcolsep}{5pt}  
\begin{table}[p]
  \caption{
  \label{tab:params}{\superb\ beam parameters.}
}
\vspace*{2mm}
\setlength{\extrarowheight}{-1pt}
\centering
\begin{tabular}{lcccccc}
\hline
\hline
  &
\multicolumn{2}{c}{Nominal} &
\multicolumn{2}{c}{Upgrade} &
\multicolumn{2}{c}{Ultimate} \\
  &
\multicolumn{2}{c}{Parameters} &
\multicolumn{2}{c}{Parameters} &
\multicolumn{2}{c}{Parameters} \\
             {Parameter}               &     {LER}      &  {HER}   &  {LER}   &   {HER}    &  {LER}   & {HER}\\
\hline
              Particle type                &       $\ep$        &    $\en$     &     $e^+$      &       $e^-$       &      $e^+$      & $e^-$ \\
               Energy (GeV)                &         4          &      7       &      4       &       7        &      4       & 7\\
      Luminosity ($\cm^{-2}\s^{-1}$)       &     \multicolumn{2}{c}{1.0\EE{36}}& \multicolumn{2}{c}{2.4\EE{36}}& \multicolumn{2}{c}{3.4\EE{36}}\\
            Circumference (m)              &        \multicolumn{2}{c}{2250}   &    \multicolumn{2}{c}{2250}   &     \multicolumn{2}{c}{2250}\\
          Revolution freq. (MHz)           &        \multicolumn{2}{c}{0.13}   &    \multicolumn{2}{c}{0.13}   &     \multicolumn{2}{c}{0.13}\\
         Long. polarization (\%)           &         0          &      80      &      0       &       80       &      0       & 80\\
            RF frequency (MHz)             &        \multicolumn{2}{c}{476}    &    \multicolumn{2}{c}{476}    &     \multicolumn{2}{c}{476}\\
             Harmonic number               &        \multicolumn{2}{c}{3570}   &    \multicolumn{2}{c}{3570}   &     \multicolumn{2}{c}{3570}\\
             Momentum spread ($\EE{-4}$)   &        8.4         &     9.0      &     10       &       10       &     10       & 10  \\
           Mom. compaction ($\EE{-4}$)     &        1.8         &     3.0      &     1.8      &      3.0       &     1.8      & 3.0 \\
             RF voltage (MV)               &         6          &      18      &      6       &       18       &     7.5      & 18  \\
          Energy loss/turn (MeV)           &        1.9         &     3.3      &     2.3      &      4.1       &     2.3      & 4.1 \\
            Number of bunches              &        \multicolumn{2}{c}{1733}   &    \multicolumn{2}{c}{3466}   &     \multicolumn{2}{c}{3466}\\
     Particles/bunch $\times 10^{10}$      &        6.16        &     3.52     &     5.34     &      2.94      &     6.16     & 3.52\\
             Beam current (A)              &        2.28        &     1.30     &     3.95     &      2.17      &     4.55     & 2.60\\
               \betys (mm)                 &        \multicolumn{2}{c}{0.30}   &    \multicolumn{2}{c}{0.20}   &     \multicolumn{2}{c}{0.20}\\
               \betxs (mm)                 &        \multicolumn{2}{c}{20}     &    \multicolumn{2}{c}{20}     &     \multicolumn{2}{c}{20}\\
               \epsy (pm-rad)                 &        \multicolumn{2}{c}{4}      &    \multicolumn{2}{c}{2}      &     \multicolumn{2}{c}{2}\\
               \epsx (nm-rad)                 &        \multicolumn{2}{c}{1.6}    &    \multicolumn{2}{c}{0.8}    &     \multicolumn{2}{c}{0.8}\\
            $\sy^\star$ (nm)              &        \multicolumn{2}{c}{35}     &    \multicolumn{2}{c}{20}     &     \multicolumn{2}{c}{20}\\
            $\sx^\star$ ($\mu$m)             &        \multicolumn{2}{c}{5.657}  &    \multicolumn{2}{c}{4.000}  &     \multicolumn{2}{c}{4.000}\\
            Bunch length (mm)              &        \multicolumn{2}{c}{6}      &    \multicolumn{2}{c}{6}      &     \multicolumn{2}{c}{6}\\
        Full Crossing angle (mrad)         &        \multicolumn{2}{c}{34}     &    \multicolumn{2}{c}{34}     &     \multicolumn{2}{c}{34}\\
              Wigglers (\#)                &         4          &      2       &      4       &       4        &      4       & 4\\
$\tau_{\mathrm{Damping}}$ (trans/long)(ms) &        \multicolumn{2}{c}{32/16}  &    \multicolumn{2}{c}{25/12.5}&     \multicolumn{2}{c}{25/12.5}\\
        Luminosity lifetime (min)          &        10.3        &     5.7      &     7.4      &      4.1       &     6.1      & 3.5\\
         Touschek lifetime (min)           &        5.5         &      38      &     2.9      &       19       &     2.3      & 15\\
       Total beam lifetime (min)            &        3.6         &     5.0      &     2.1      &      3.4       &     1.7      & 2.8\\
        Inj. rate pps (100\%) \EE{11}      &        4.9         &    2.0       &     15       &      5.0       &     21       & 7.2\\
   \xix (from Eq.~\ref{eq:tunex})          &        \multicolumn{2}{c}{0.004}  &    \multicolumn{2}{c}{0.007}  &    \multicolumn{2}{c}{0.009}\\
   \xiy (from Eq.~\ref{eq:tuney})          &        \multicolumn{2}{c}{0.17}   &    \multicolumn{2}{c}{0.16}   &    \multicolumn{2}{c}{0.2}\\
              RF Power (MW)                &        \multicolumn{2}{c}{ 17}    &    \multicolumn{2}{c}{35}     &    \multicolumn{2}{c}{44}\\ \hline
\end{tabular}
\end{table}
} 

\subsubsection{Projected \superb\ integrated luminosity}

We project the performance of \superb\ as a function of time using assumptions based on
present experience. The construction of the \superb\ collider will take approximately four
years.  The injector will be commissioned over a period of a few
months approximately six months prior to the commissioning of the collider. The
injector performance is assumed to be adequate for the needs of the 
collider at the start of
collider commissioning.  To calculate the integrated
luminosity expected, we assume that the collider will operate for ten months each year, with
a two month maintenance and upgrade period each year.  During the first
few months of each run, the collider luminosity will be unsteady and
the luminosity will be incrementally increased. After about two months,
the luminosity will reach the previous run's luminosity level. It is
then slowly increased during the rest of the run, while nearly
continuous luminosity is being provided to the detector and particle
physics data is recorded.

\begin{figure}[htb]
\centering
\includegraphics[width=\textwidth]{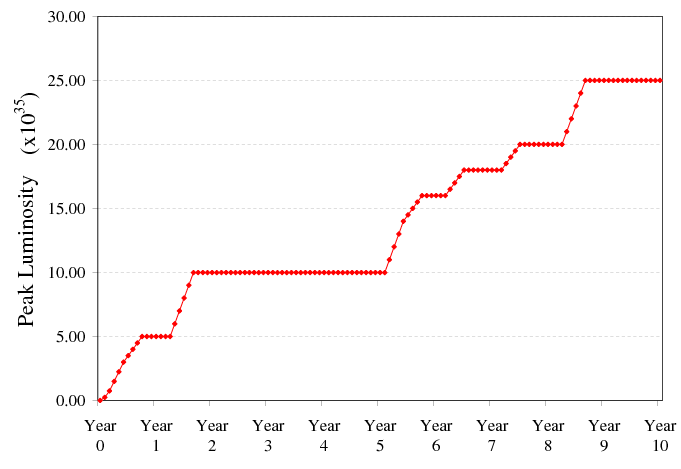}
\caption{\label{fig:peaklumi} Peak luminosity projection over 10 years.}
\end{figure}

\begin{figure}[htb]
\centering
\includegraphics[width=\textwidth]{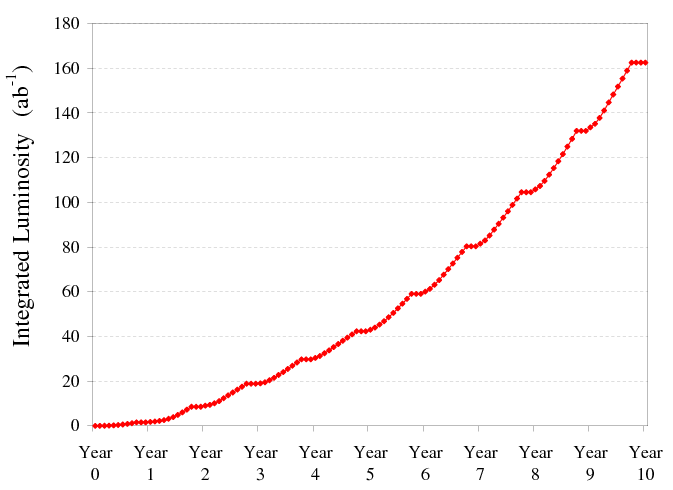}
\caption{\label{fig:intlumi} Integrated luminosity projection over 10 years.}
\end{figure}

We envision that the peak luminosity will reach 50\% of the design ($\approx
5\times10^{35}$) during the first year of operation, and full design value after two years of operation. The
luminosity is then held constant for about three years. After five
years the luminosity is increased by hardware upgrades by a factor
of $\sim$2.5 to a peak of $2.5\times10^{36}$.  The peak luminosity for each
month over a ten year span is shown in Fig.~\ref{fig:peaklumi}.
The resulting integrated luminosity over the ten year period is shown in
Fig.~\ref{fig:intlumi}.  With this model, more than $150 \invab$
will be delivered to the detector in the ten years.

\subsubsection{Energy asymmetry at the \FourS}

The energy asymmetry plays an important role in the design of \superb\ and the optimization
of parameters. It is not straightforward to
quantify precisely the ultimate luminosity achievable at a given
asymmetry, but simple scaling of some fundamental machine
parameters clearly shows a more than linear dependence with
respect to the boost (greater than $\gamma{^2}$). In addition, there
are other consequences for the design and practical limits. A list of
some of the more significant dependences follows; the choice of the
\superb\ energy asymmetry, 4 on 7\gev, has been based on these considerations.

Factors leading to a LER energy of 4\gev:
\begin{itemize}
\item Because of the transparency condition, the bunch charge of the
LER beam increases as $1/\gamma{^{1/2}}$. This provides a direct
limit to the luminosity, whose ultimate value depends on the
maximum beam colliding currents;

\item The Touschek lifetime scales as $\gamma{^4}$, the additional
factor in the exponent being due to the fact that the bunch charge must
also be increased. This causes a factor of two decrease in the Touschek
lifetime in going from $4$ to $3.5\gev$ for the LER (from
$4\min$ to $2\min$ for the present beam parameters);

\item Intrabeam scattering (IBS) instead scales as $1/\gamma{^4}$. Since the emittance
growth in the LER due to IBS is already of the order of $30\%$, it is
not possible to obtain the design horizontal emittance at lower energies
without significant changes in the present ring design. In particular,
the bends would have to be weakened and lengthened. Wigglers would have to be
added to reduce the intrinsic horizontal emittance;

\item All collective effects, such as instabilities, electron
cloud, \etc, increase at least with $1/\gamma{^{3/2}}$, since the LER becomes
weaker because of the reduced energy and the increased bunch charge;

\item The wiggler length required to produce the desired damping time nearly doubles if the energy is reduced from 4 to $3.5 \gev$. The magnet costs and the
overall ring length would increase correspondingly.

\end{itemize}

Factors leading to a HER Energy of 7\gev:

\begin{itemize}
\item The equilibrium emittance increases with energy. To produce the design parameters at higher energy,
we would have to make softer and longer bends, compared to the \pepii values,
or add more cells. For the latter solution, the sextupoles would become
stronger, and the \pepii components could not be reused without modification.
The ring dynamic aperture would also decrease considerably,
especially if we add more cells;

\item The final focus chromaticity increases linearly with energy. The
ultimately achievable $\bety$ scales directly with the chromaticity.
This additional factor adds to the luminosity dependence on
the boost;

\item The final focus emittance growth increases with energy. At a higher 
energy the final focus bends become
incompatible with the reuse of the \pepii components. Other solutions to
reduce the emittance growth would damage the optical and chromatic
properties of the final focus;

\item The IR design becomes more difficult with larger energy asymmetry. The final quadrupoles would be much closer
to the IP and stronger in comparison to those at \pepii or \kekb.
The value of $L^\star$, the distance of the first quadrupole from
the IP, would increase, worsening the final focus properties;

\item IR synchrotron radiation increases proportional to
$\gamma{^2}$. The related backgrounds in the detector would correspondingly worsen;

\item The site power consumption would increase by about 10\% if the HER
energy were to be increased from 7 to 8\gev; and

\item The cost of the injector linac increases linearly with the HER
energy.

\end{itemize}

\subsubsection{Energy scaling for operation at the $\tau$/charm threshold}

\superb\ can operate at a lower center-of-mass energy with a
somewhat reduced luminosity. In order to operate at $\tau$/charm
threshold energies (in the vicinity of $3.8\gev$) with minimal modifications to the
machine, beam energies will be scaled,  maintaining the nominal
energy asymmetry ratio used for operation at the center-of-mass energy of the \FourS. All
magnet currents will be rescaled accordingly, except for the wigglers, which
will be kept at the same field to provide maximum damping. If the IR magnets are permanent magnets, they will be
replaced with weaker versions. The main differences in the ring
properties will be:

\begin{itemize}
  \item Lower energy by a factor of about 2.78 per ring;
  \item Longer damping time by a factor of about 4.3 per ring;
  \item Decreased Touschek lifetime by a factor of $(2.78){^3}$ for a given current and
emittance;
  \item Rematched optics in the wiggler section to maintain (or increase if necessary)
  the beam emittance of the rings at the nominal energy; and
  \item Increased sensitivity to collective effects by a factor of about 2.78 per
ring.
\end{itemize}

Luminosity should scale linearly with energy (see formula
in Sec.~\ref{sec:Key_issues}). However, the damping time and
collective effects will result in a
further decrease the luminosity. In general, the luminosity
dependence is less then linear with respect to the damping time
(about $1/\tau^{0.3-0.5}$). However, given decreased Touschek lifetime
and increased collective effects, we expect that operations at lower 
energy will require a
decrease of the beam current and/or an increase of the beam
emittance. It is thus reasonable to expect a luminosity about
10 times smaller than that at $10.58\gev$.

It should also be noted that the beam polarization scheme,
described in Sec.~\ref{section:SpinPolarization},
does not work at lower energy. For a given running period and polarization, however, $\tau$ polarization
studies are best done at high energy.

\subsubsection{Synergy between \superb\ and \ilc}
There are significant similarities between the \superb\ storage
rings and the \ilc damping rings \cite{ilc_1}: Table
\ref{table:ILC_superb} compares some of the important parameters.
Beam energies and beam sizes are similar.  The \ilc damping
rings have a circumference three times larger than the \superb\
rings (because of the need to store a long train of bunches with
bunch spacing sufficiently large to allow injection and extraction
of individual bunches); the nominal bunch charge is smaller in the
\ilc damping rings than in the \superb\ storage rings, leading to a
lower average current.  Nevertheless, one may expect the
overall beam dynamics in the two facilities to be in comparable
regimes.  
A similar lattice design is used in both cases, the main difference being
a reduction in circumference and the insertion of an interaction region 
in the case of \superb.

\begin{table}[htb]
\caption{\label{table:ILC_superb}
Comparison between parameters for the \superb\ storage rings
and the \ilc damping rings.}
\vspace*{2mm}
\centering
\setlength{\extrarowheight}{2pt}
\begin{tabular}{lccc}
\hline
\hline
                    Unit & \superb\ & \superb\ & ILC \\
                    & LER      & HER      & DRs \\
\hline
Beam energy         (GeV) & 4          & 7  & 5\\
Circumference       (m)   & 2249       & 2249     & 6695\\
Particles per bunch        & $6.16\times10^{10}$& $3.52\times 10^{10}$&$2\times10^{10}$\\
Number of bunches          & 1733       & 1733    & 2767\\
Average current     (A) & 2.28       & 1.30   & 0.40\\
Horizontal emittance (nm) & 1.6        & 1.6 & 0.8 \\
Vertical emittance  (pm) &  4 & 4 & 2 \\
Bunch length        (mm)  &  6 & 6 & 9 \\
Energy spread       (\%)   & 0.084   & 0.09  & 0.13\\
Momentum compaction   &
                     $1.8\times 10^{-4}$&$3.1\times 10^{-4}$&$4.2\times 10^{-4}$\\
Transverse damping time (ms) & 32  & 32 & 25 \\
RF voltage          (MV)  & 6 & 18 & 24 \\
RF frequency        (MHz) & 476 & 476  & 650 \\
\hline
\end{tabular}
\end{table}

The \ilc damping rings and the \superb\ storage rings will face similar
demands on beam quality and stability: the \superb\ rings for direct
production of luminosity, and the \ilc damping rings for reliable
tuning and operation of the downstream systems, to ensure efficient luminosity production
from the extracted beams.  Significant issues common to both the \superb\
rings and the \ilc damping rings include:
\begin{itemize}
\item Alignment of the magnets, including orbit and coupling corrections, with the
precision needed to produce vertical emittances of a few picometers on a
routine basis;
\item Reduction of magnet vibration to a minimum, to ensure beam orbit
stability at the level of a few microns;
\item Optimization of lattice design and tuning to ensure sufficient dynamic
aperture for good injection efficiency (for both \superb\ and the \ilc damping
rings) and lifetime (particularly for the \superb\ low energy ring); and
\item Control of beam instabilities, including electron cloud and ion effects.
\end{itemize}

These are all active areas of research and development for
the \ilc damping rings.  For example, there has been significant progress in
recent years in the development of techniques for suppression of the electron cloud instability (including low secondary
yield vacuum chamber coatings; use of grooved chamber surfaces; and clearing improved
electrodes), that could have a major impact
on the performance of the \ilc damping rings.  While small-scale tests of these
techniques in the laboratory are essential, the experience of operating a
full-scale facility in the regime of the \superb\ storage rings or the
\ilc damping rings would be beneficial whether the facilities are 
constructed and
commissioned sequentially or in parallel.

In general, the similarity of the proposed operating regimes for the \ilc
damping rings and the \superb\ storage rings presents an opportunity for a
well-coordinated program of activities that could yield much greater benefits
than would be achieved by separate, independent research and development
programs.

The \superb\ baseline design includes a polarized electron beam. The
addition of a polarized positron beam would increase the effective luminosity 
for polarization studies since, in the unpolarized case, chirality 
conservation in QED processes acts as a ``filter'' on half of the $B$ and 
$\tau$ production channels. Thus, for example, in the limiting case with 
both beams fully polarized, the same production rate would be achieved with
 half the luminosity, allowing some relaxation in machine parameters.

A polarized positron source (\pps) is included in the \ilc baseline. 
The design envisages the
installation of a superconducting helical undulator in
the $150\gev$ electron beamline, a solution that is clearly
not applicable to \superb. An
alternative solution for the \ilc is a \pps based on Compton
scattering~\cite{ilc_2}. This
solution presents the important advantage that the electron
beam energy is in the 1--2\gev\ range; this approach could potentially 
be adapted for the
\superb\ project. Different \ilc R\&D
programs are in place to develop the associated technologies. Of
particular note are the efforts at LAL Orsay and KEK to develop
and test optical resonators with very high gain and very small
waists to improve the Compton cross section. With the rapid
development of high power pulsed lasers, and of high current electron
guns, the \ilc Compton scheme could become the basis
for upgrading \superb\ to a fully polarized configuration. A
polarized positron source for the \superb\ upgrade, could even prove to be
an interesting test facility for the \ilc.
The details of such an approach are
discussed in Sec.~\ref{sec:Posipol}.

The choice of wiggler technology, either permanent or
superconducting magnets, is also of interest for both the \ilc Damping
Rings and \superb.  The permanent magnet solution
seems less costly for \superb. The low
$\beta$ quadrupoles of the final focus will, however, likely employ superconducting
technology, in order to accommodate the detector and allow for changing the beam energy.



\afterpage{\clearpage}

\section{Layout}
\label{section:Layout}
\subsection{The Rings}
The lattices for the \superb\ rings must satisfy several requirements:
\begin{itemize}
\item Very small emittances;
\item Asymmetric energies ($4\times 7 \gev$);
\item Insertion of a final focus with very small \bets;
\item Good dynamic aperture and lifetimes; and
\item Reuse of available \pepii hardware as much as possible.
\end{itemize}

A crossing angle with the crabbed waist scheme relaxes
the requirements on the bunch lengths and beam currents, compared to the older
high-current designs. However, the objective remains to design a lattice that could
deliver a peak luminosity of \tenTo36 while keeping the wall plug power
requirements as low as possible. By adapting the approach of the
\ilc damping ring design, we have developed a \superb\ design for
low-emittance rings that reuses all the available \pepii magnets. 
Since the RF
requirements for \superb\ are also fully satisfied by the present 
\pepii RF system, consdierable cost savings are possible.

The $7\gev$ high energy ring (HER) and the $4\gev$ low energy ring (LER)
will be built on the same horizontal plane, with a horizontal crossing
angle of $2\times 17\mrad$. The beams will travel together only over a short
section (about $1.2\m$) of the interaction region (IR), where they will
collide at the interaction point (IP). On the opposite side of the IP the
beams will be vertically separated in order not to collide, and the rings
will be horizontally separated by a magnetic chicane. The design uses
1733 bunches with a 5\% ion gap.

The IP collision parameters listed in Table~\ref{tab:IPParam} 
have been chosen based on beam-beam
simulations, which also show that the requirements on 
damping times can be relaxed. The crossing angle has been fixed by optimization of the 
interaction region design.

\begin{table}[h]
\caption{\label{tab:IPParam}IP Parameters.} \centering \vspace*{2mm}
\begin{tabular}{ll} \hline\hline
         IP horizontal \betx           & 20\mm    \\
          IP vertical \bety            & 0.2\mm   \\
       Horizontal beam size \sx        & 4\mm    \\
Horizontal beam divergence $\sigma_{x'}$ & 200\mrad \\
        Vertical beam size \sy         & 20\nm    \\
 Vertical beam divergence $\sigma_{y'}$  & 100\mrad  \\
           Bunch length \sz            & 7\mm     \\
         Crossing angle \thx           & $2\times 17$\mrad  \\
\hline
\end{tabular}
\end{table}

\begin{figure}[htb]
\centering
\includegraphics[width=90mm]{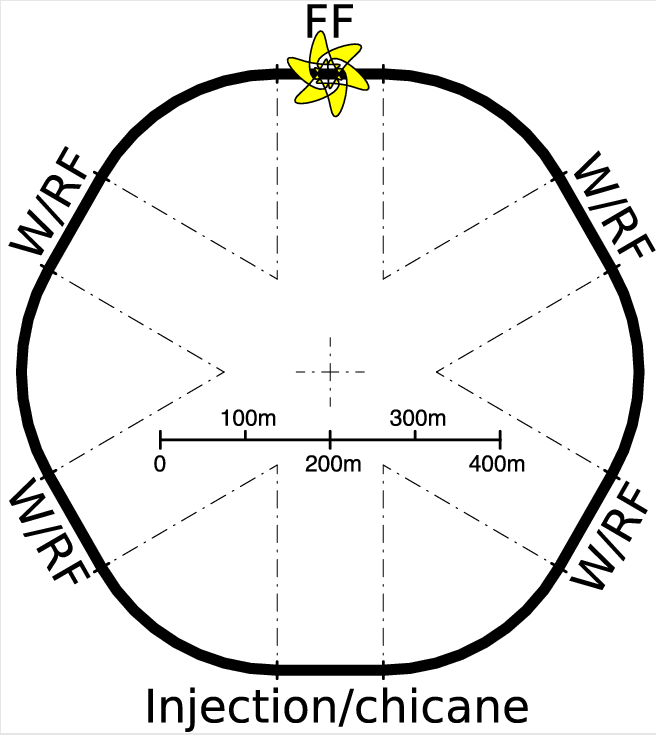}
\caption{ \label{fig:footprint} Footprint of one ring.}
\end{figure}

The magnetic layouts of the two rings are identical. Each ring has a 6-fold
symmetry, with 6 arcs separated by 6 long straight sections, similar to the
\pepii design. Wiggler magnets will be installed in some of the straight 
sections in order to control emittance and damping time.
A sketch of the rings is shown in Fig.~\ref{fig:footprint}.

Each ring is about 2249\m long, corresponding to a harmonic number of 3568
for the \pepii RF frequency of 476\MHz, although the circumference can 
easily be adjusted if needed.

A special final focus (FF) section brings the beams together,
focuses them to the very small $\beta$ functions required by the
design, and separates them after the collision. The FF has been designed as part of
the final arc and straight section. A magnetic chicane on the
opposite straight section will be used to adjust the ring length and
to inject both beams. Details of the ring lattices are
provided in Sec.~\ref{sec:lattice}. Table~\ref{tab:LatticeParam} 
summarizes the upgrade lattice parameters for both rings.

\begin{table}[htbp]
\setlength{\extrarowheight}{2pt}
    \caption{Lattice parameters for HER and LER rings.}
    \label{tab:LatticeParam}
    \vspace*{2mm}
    \centering
\begin{tabular}{lcc}
\hline
\hline
& \multicolumn{1}{p{1.2in}}{\centering LER} &
 \multicolumn{1}{p{1.2in}}{\centering HER} \\
\hline
   Energy (GeV)                            &       4       & 7 \\
   C (m)                                   &     2250      & 2250 \\
   Bw (T)                                  &     1.00      & 0.83 \\
   $L_{\textrm{bend}}$ (m) (Arc/FF)        & 0.45/0.75/5.4 & 5.4/5.4 \\
   Number of Bends (Arc/FF)                &  120/120/16   & 120/16 \\
   $U_0$ (MeV/turn)                        &     1.9       & 3.3 \\
   Wiggler sections: Number                &       4       & 2   \\
   Wiggler sections: $L_{\textrm{tot}}$(m) &     100       & 50  \\
   $\sz$ (mm)                              &     4.07      & 5.00 \\
   $\tau_{s}$ (ms)                         &     16.00     & 16.00 \\
   $\epsx$  (nm-rad)                       &     0.8       & 0.8 \\
   Emittance ratio                         &    0.25\%     & 0.25\% \\
   $\sigma_{E}$                            &   1.\EE{-3}   & 1.\EE{-3} \\
   Momentum compaction                     &  1.8\EE{-4}   & 3.\EE{-4} \\
   $\nu_{s}$                               &     0.011     & 0.02 \\
   $V_{\textrm{RF}}$ (MV),$N_{\textrm{cav}}$&     6, 8      & 18, 24 \\
   $N_{part}$ $(\times 10^{10})$           &     6.16      & 3.52 \\
   $I_{\textrm{beam}}$ (A)                 &     2.3       & 1.3 \\
   $P_{\textrm{beam}}$ (MW)                &     4.4       & 4.3 \\
   $f_{rf}$ (MHz)                          &      476      & 476 \\
   $N_{\textrm{bunches}}$                  &     1733      & 1733  \\
   Ion gap                                 &      5\%      & 5\% \\
\hline
\end{tabular}
\end{table}

\subsection{Interaction Region}
The \superb\ interaction region (IR) has been designed with the following
design constraints in mind:
\begin{itemize}
\item Very small spot sizes at the IP;
\item Local correction for the very high chromaticity due to the
highly focused beam, keeping geometric aberrations small;
\item Separation of the LER and HER beams as soon as possible;
\item Preventing synchrotron radiation (SR) production from hitting the beam pipe
and the detector;
\item Compatibility with a beam pipe of minimum radius and thickness; and
\item Maintenance of the largest possible angular acceptance for the detector.
\end{itemize}

The study of beam trajectories has led to the introduction of a small dipole
between the first two low-$\beta$ quadrupoles in each beam in order to redirect
the SR coming from the focusing element. The crossing angle has been chosen to be  $2\times 17 \mrad$.
A detailed description of the IR and SR backgrounds can be found in
Sec.~\ref{section:InteractionRegion}.

\subsection{Injector}

A possible injection scheme for \superb\ is the one presently used for the \daphne$\phi$-Factory at Frascati. This scheme inlcudes an electron gun, a linac for positron production, a
positron converter and a linac for electron and positron acceleration to operational
energies. Separate transfer lines bring the two beams into the rings. Details of this simple scheme are described in Sec.~\ref{section:InjectionSystem}.
Alternatively, a design incorporating two damping-rings 
could offer advantages.
Other schemes, including a recirculating linac, are under consideration.



\afterpage{\clearpage}

\section{Interaction Region}
\label{section:InteractionRegion}
\subsection{Geometry}
\label{subsec:FF_geom}
The final focus of the \superb\ design calls for a small \betxs ($20\mm)$
and a very small \betys ($0.2\mm$).  These small beta functions require
the final focus magnets to be as close to the interaction point (IP) as
possible in order to keep the maximum beta values as low as possible and
minimize the chromaticity generated in the final focus.
Table~\ref{table:IR_Table_1} lists the accelerator parameters that
are important for the interaction region (IR) design. In the table and throughout this section we 
assume the higher and more challenging beam currents from the \superb\ upgrade.

\begin{table}[ht]
\caption{\label{table:IR_Table_1}
\superb\ parameters that influence the design of the interaction
region. The $\beta$ functions and emittances define the beam size, and
thus set the
beam-stay-clear dimensions. The beam currents are taken from the upgrade scenario and not the baseline design values, in order to confront backgrounds and synchrotron radiation
power for this more challenging case.
}
\vspace*{2mm}
\centering
\setlength{\extrarowheight}{2pt}
\begin{tabular}{lcc}
\hline
\hline
                     & LER   & HER \\\hline
    Energy (GeV)     & 4.0   & 7.0 \\
  Beam current (A)   & 3.95  & 2.17 \\
   No. of bunches    & \multicolumn{2}{c}{ 1733 } \\
 Bunch spacing (m)   & \multicolumn{2}{c}{ 1.26 } \\
    \betxs (mm)      &  20   & 20 \\
    \betys (mm)      & 0.2   & 0.2 \\
   \epsx (nm-rad)    & 1.6   & 1.6 \\
   \epsy (pm-rad)    &  4    & 4 \\
Crossing angle (mrad)& \multicolumn{2}{c}{ 34 } \\\hline
\end{tabular}
\end{table}

We have adopted a beam-stay-clear (BSC) envelope that is similar to that
used in the \pepii design~\cite{bib:ir1}. The $x$ stay-clear is defined as 15 uncoupled
beam $\sx + 1\mm$ for closed orbit distortion (COD). The $y$ stay-clear is
defined as 15 fully coupled beam $\sy + 1\mm$ COD.

With these parameters in mind, we have positioned the first quadrupole
magnet (QD0) to start $0.3\m$ away from the IP. A collision crossing angle
of $\pm 17\mrad$ ($\pm 1$ degree) means that the beam centers are
$5.1\mm$ apart at this location, while the two
BSC envelopes are only $1.8\mm$. The small separation distance precludes
the use of separate initial quadrupole magnets, so QD0 is shared by the two 
beams. In order
to produce similar final focus beta functions for both beams, we would like to have set
the gradient of QD0 by the requirements for the high-energy beam (HEB), resulting in a
magnet length of $0.75\m$. However, this is too strong for the low-energy
beam (LEB), so we shorten the length to $0.46\m$ to obtain the
correct integrated strength for the LEB. The beams therefore need to be
separate enough at $0.76\m$ from the IP ($0.30 + 0.46\m$) to be able to
place an additional magnet that continues the vertical focusing for the HEB, while providing
a field free region for the LEB. We label this $0.29\m$-long magnet QD0H.
The two beams enter separate beam pipes at this
location.

\subsection{The QD0H magnet}
\label{sec:QD0}

The beam center separation at $0.76\m$ is $31.9\mm$  for the incoming
LEB side and $36.4\mm$  for the incoming HEB side. The difference is due to
the fact that the LEB is easier to bend than the HEB. If we include the BSC
envelopes we have $25.6\mm$  of clearance on the incoming LEB side and
$30.1\mm$  of clearance on the incoming HEB side. If we assume a beam pipe
radius of $10\mm$  for each beam at $0.76\m$  from the IP, we then have 11.9
and $16.4\mm$  of space for a beam pipe and magnet design. Using permanent
magnet (PM) material with a remnant field of $1.4\Tesla$, we can construct a
cylinder with an inner radius of $12\mm$  and an outer radius of $20\mm$
($8\mm$  thick) that has enough strength to satisfy the HER gradient
requirements. The PM blocks have a very low residual field beyond the outer
radius of the material and hence make a good field free region for the LEB.
This leaves enough room for a $2\mm$  thick beam pipe for each beam at this
narrow location. Beyond $0.76\m$  from the IP the distance between the beams
grows rapidly, and it is relatively easy to accommodate separate beam pipes and
magnets for the two beams.

The next quadrupole magnet (QF1) from the IP is an $x$ focusing magnet that
is $0.4\m$  long and is located between $1.45\m$  and $1.85\m$  from the IP.
There are two separate magnets at this location, one for each beam; beyond
QF1 the beam lines for both beams have the same layout with counterpart
magnets at each $z$ location.

\subsection{Synchrotron Radiation Fans and Backgrounds }
\label{subsec:sync_rad_fan}

We distinguish between synchrotron radiation (SR) in the form of ``bending radiation'', 
which are ``fans'' of radiation generated by bending the entire beam, and
``focusing radiation'', which is SR that results from a beam that travels
through a quadrupole magnet on axis. The SR power levels from focusing
radiation are, in general, about 100 times lower than the power levels from
bending radiation. For high current storage rings, the power levels for SR
fans can easily be several kilowatts. The power levels for both types of SR
are too high to allow these sources to strike directly the detector
beam pipe.

\begin{figure}[htbp]
\centering
\includegraphics[width=0.75\textwidth]{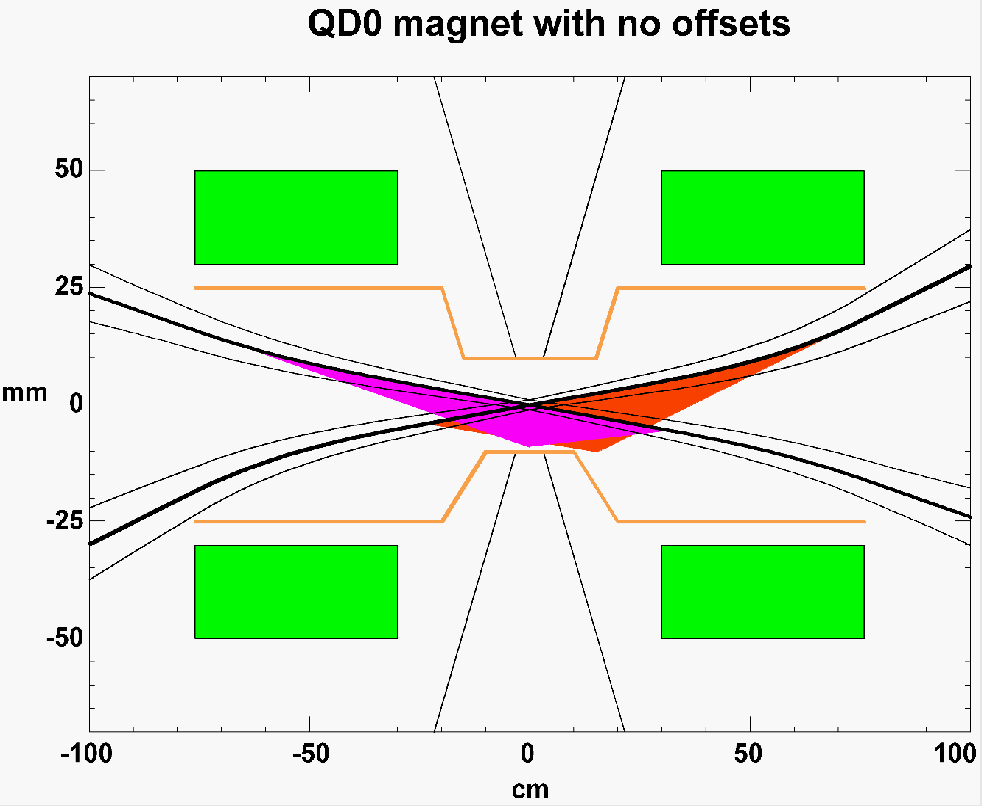}
\caption{
Layout of an IR design where the QD0 magnets have no $x$ offsets. 
The incoming beams are both off-axis, and hence produce SR fans that would,
in the absence of masks, directly strike the detector beam pipe.. 
The radiation fans are shown as
shaded triangles. The background rates from these direct hits would be much too
high for the detector to tolerate.
}
\label{fig:IR_Fig_1}
\end{figure}

The non-zero collision crossing angle means the two beams
enter the shared QD0 magnet at different $x$ locations. If we center the two QD0
magnets along the axis of the detector beam pipe, the vertically focusing, and
hence horizontally defocusing, QD0 magnet starts to bend the two beams away
from each other in the $x$ plane. The incoming beam trajectories then
produce SR fans that strike the detector beam pipe located at the collision
point, as illustrated in Fig.~\ref{fig:IR_Fig_1}. In order to protect the
detector beam pipe from this radiation without increasing the radius of the
beam pipe, masks are placed on either side of the central pipe to
shadow this incoming radiation. These masks, which are quite close to the 
beams, intercept high levels of SR radiation on the inside surfaces 
near the central beam pipe.

\begin{figure}[htbp]
\vspace*{5mm}
\centering
\includegraphics[width=0.8\textwidth]{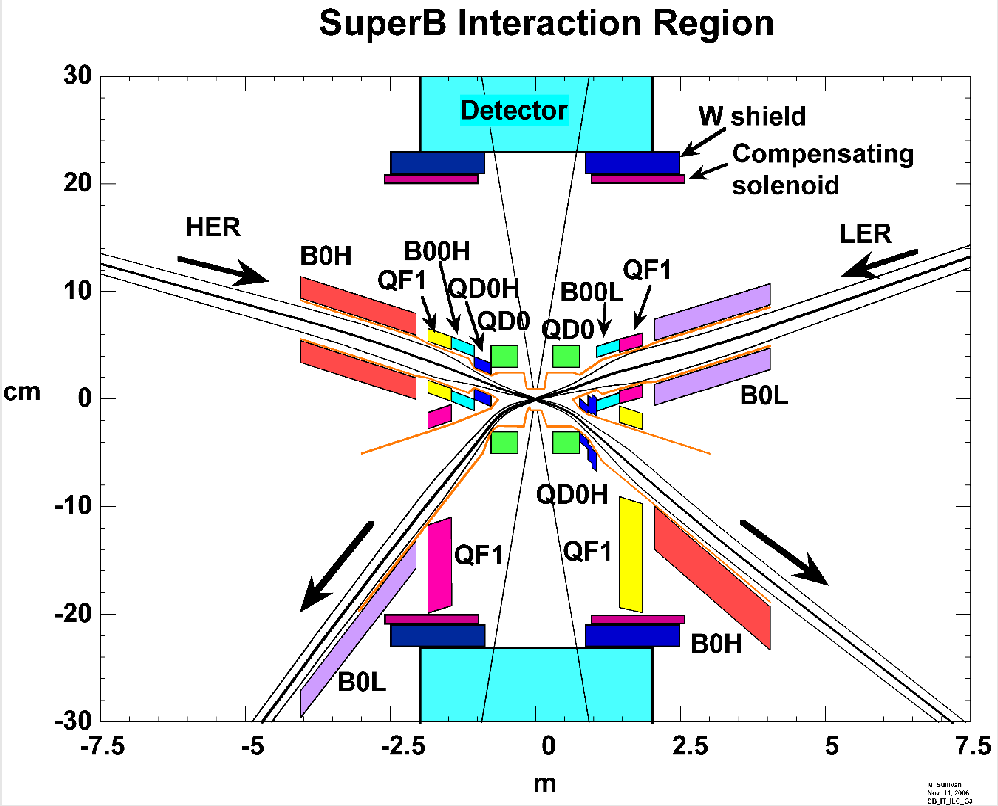}
\caption{
Layout of the interaction region. Note the asymmetric scales for the two axes.}
\label{fig:IR_Fig_2}
\end{figure}

These surfaces have a relatively large solid angle acceptance for
backscattered photons to strike the central \Be beam pipe, causing
unacceptably high backgrounds. Increasing the central
detector radius would reduce this problem, 
but compromise the detector physics performance
and, in a sense, reduce the effective luminosity of the collider.

We have therefore adopted an alternative solution of offsetting the magnetic axis of
QD0 to a value that is closer to the trajectory of the incoming beams.
The axis is still parallel to the detector beam pipe. This eliminates the
incoming SR fans from the QD0 magnets and also directs the focusing radiation
from QD0 away from the detector beam pipe. Consequently, the detector
background from SR has no component from QD0. The QD0 offsets are not
identical, because we are partially compensating for the fact that the LEB is
easier to bend than the HEB.

The next most important background source after the QD0 is the focusing
radiation coming from the incoming beams as they travel through the QF1 magnets.
This radiation comes from the horizontal over-focusing of the beam and, as a consequence, 
the photon trajectories are steeper, making
it more difficult to shield the detector beam pipe from this source. In order
to control this background rate, we introduce small bending magnets between the
QD0 and QF1 magnets on the incoming beam lines. These bending magnets redirect
the focusing radiation from the QF1 magnets away from the central \Be beam pipe.
Figure~\ref{fig:IR_Fig_2} shows a layout of the interaction region design and
Table~\ref{table:IR_Table_2} lists the magnet parameters for the magnets in the IR.

\begin{table}[htb]
\caption{\label{table:IR_Table_2}
Strengths and locations of the magnets around the IR. The offsets
of the QD0 magnets are produced by a dipole field winding in
the super-conducting magnets. The QD0 magnet would also have a
compensating solenoid winding around the quadrupole and dipole
windings.}
\vspace*{2mm}
\centering
\begin{tabular}{lcccl}
\hline
\hline
                      & \multicolumn{1}{p{1.8cm}}{ \centering Length (m)    }
                      & \multicolumn{1}{p{1.8cm}}{ \centering Starts at (m) }
                      & \multicolumn{1}{p{2.5cm}}{ \centering Strength }
                      & \multicolumn{1}{p{3.5cm}}{ \centering Comments } \\\hline
$\mathrm{L}^\star$    &   0.3    &   0.0    &                      & Drift \\
         QD0          &   0.46   & $\pm 0.3$  & $-820.6$\kGauss/m  & Shared quadrupole    \\
        QD0H          &   0.29   & $\pm 0.76$ & $-820.6$\kGauss/m  & HER quadrupole       \\
        B00L          &   0.4    &  $-1.05$   &   $-2.2$\kGauss    & Dipole inc. LER only \\
        BOOH          &   0.4    &   1.05     &    1.5 \kGauss     & Dipole inc. HER only \\
         QF1          &   0.4    & $\pm 1.45$ & 292.3 \kGauss/m    & Quadrupole LER only  \\
         QF1          &   0.4    & $\pm 1.45$ & 589.1 \kGauss/m    & Quadrupole HER only  \\
         B0L          &   2.0    &   2.05     &    0.3 \kGauss     & Dipole LER only      \\
         B0H          &   2.0    &   2.05     &   0.526 \kGauss    & Dipole HER only      \\
                      &          &           &                   & \\
     QD0 $x$ offset   & $+6.0$\mm &          &                     & Incoming HER side    \\
     QD0 $x$ offset   & $+7.5$\mm &          &                     & Incoming LER side    \\
\hline
\end{tabular}
\end{table}

\subsection{QD0H Magnet Design}
\begin{figure}[htb]
\centering
\includegraphics[width=0.5\textwidth,height=0.49\textwidth]{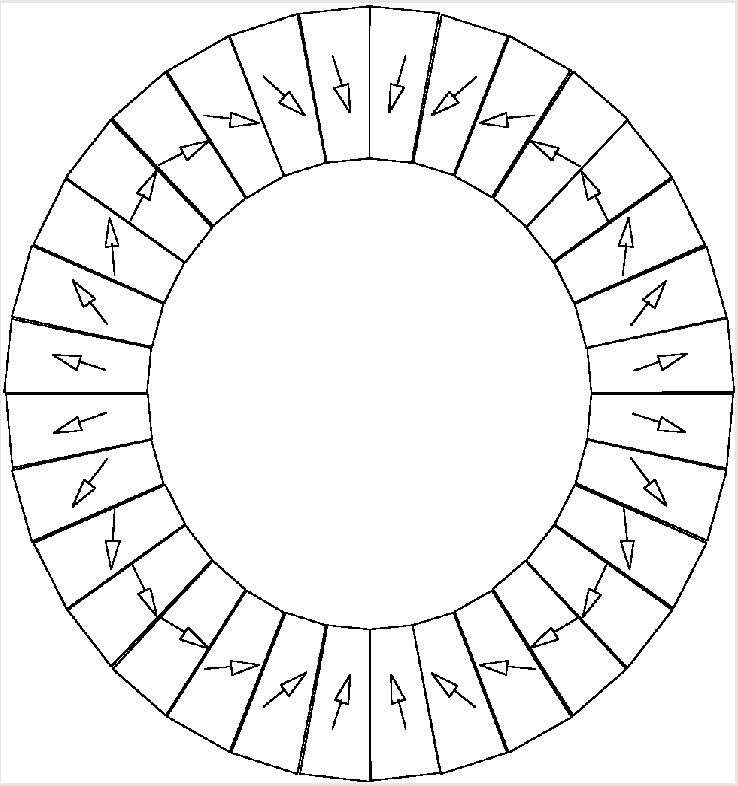}
  \caption{A picture of the magnetic block position
and magnetic field orientation for each block in a Halbach-style
permanent magnet. This is a 32 block design, which allows for the
correction of the first 16 higher harmonics of the magnet.}
\label{fig:IR_Fig_3}
\end{figure}

We now look at some more details of the QD0H magnet. We rely on the experience
of the \pepii accelerator design, which used permanent magnet (PM) material to
construct the final shared quadrupole. The design uses the Halbach 
method~\cite{bib:ir2} for placing magnetized blocks in a cylindrical
geometry to achieve the desired magnetic field. Figure~\ref{fig:IR_Fig_3} shows
the magnetic block layout. The quality of the magnetic field is determined by
the accuracy with which one can position the magnetized blocks and then
maintain these block positions.

 \begin{figure}[htb]
 \centering
 \includegraphics[width=0.45\textwidth]{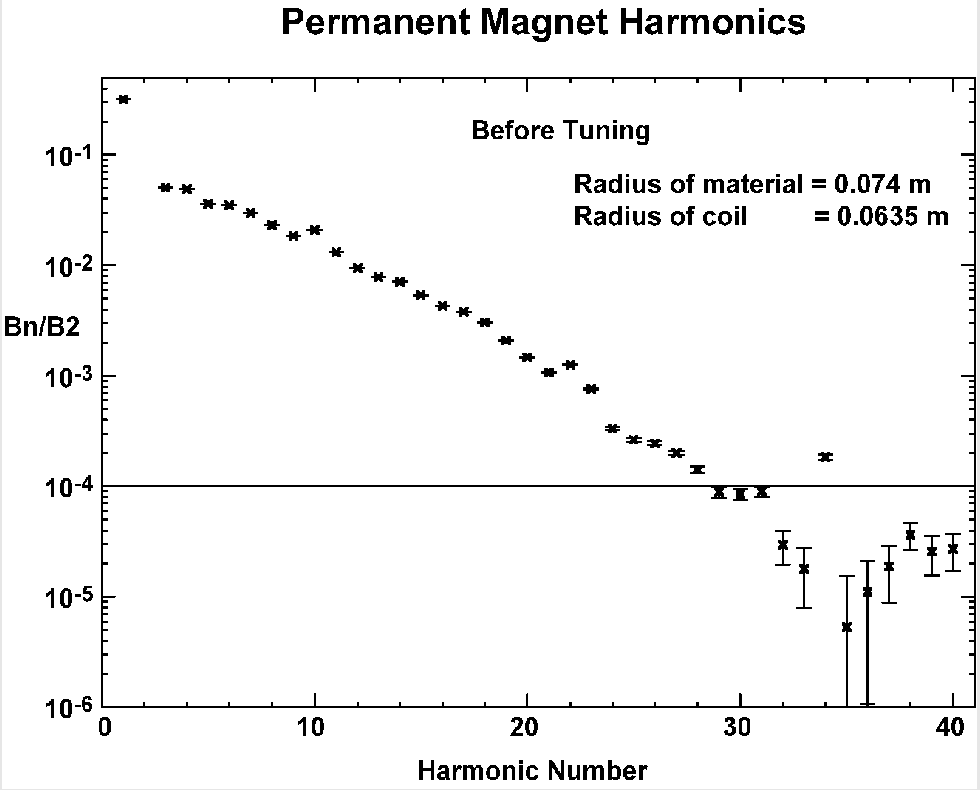}
\includegraphics[width=0.45\textwidth]{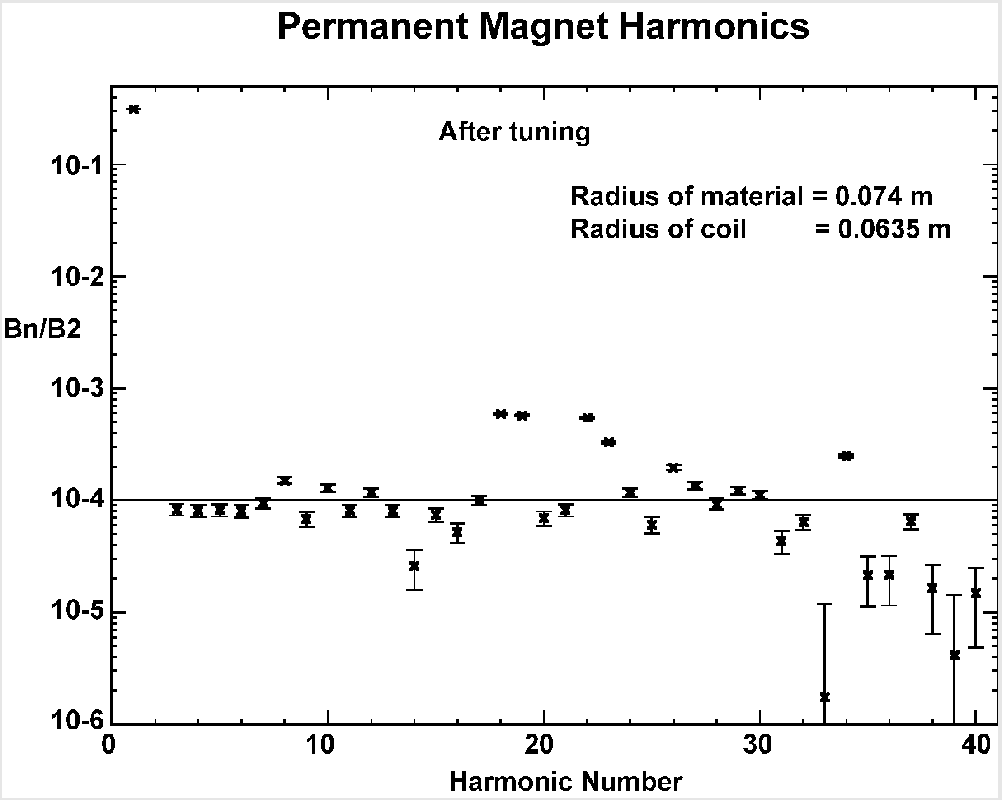}
\vspace*{5mm}
\caption{Measured harmonic content of a Halbach-style 
 magnet as shown in Fig.~\ref{fig:IR_Fig_3} prior to correcting 
 the higher harmonics (left) and after
 adjusting the block positions (right). The accuracy of the block position was
 about 50--75\mum\ with the inner radius of the material at 74\mm. 
The harmonics shown are measured very close to the material (at
 63.5\mm). This means that the field quality if this magnet is as good
 as or better than the plot shows over 85\% of the magnet aperture.}
\label{fig:IR_Fig_4a}
\end{figure}

The \pepii team developed a technique for correcting the higher harmonics of
these magnets during assembly~\cite{bib:ir3}. The technique can correct half as
many harmonic numbers as there are magnetic blocks used to make up a slice of
the magnet. The magnetic slices used to build the \pepii quadrupoles had an
inner radius of $57\mm$, while the \superb\ design has an inner radius for 
magnetic material of $12\mm$. The block position control in the \pepii magnet was
about $50\mum$. Figure~\ref{fig:IR_Fig_4a} shows the result of
correcting the block positions on the higher harmonics of a prototype magnet
slice. Because the inner radius of the magnet material in the present design
is about five times smaller than for the \pepii design we would have to control
the magnetic block positions to about $10\mum$. Although this is somewhat
more challenging, we do not consider it overly difficult. In the \pepii
design, the block assembly was epoxied together to maintain position stability and
the temperature was stabilized by water cooling.

\subsection{Transverse Beam Profile}
\begin{figure}[htb]
 \centering
 \includegraphics[width=0.6\textwidth]{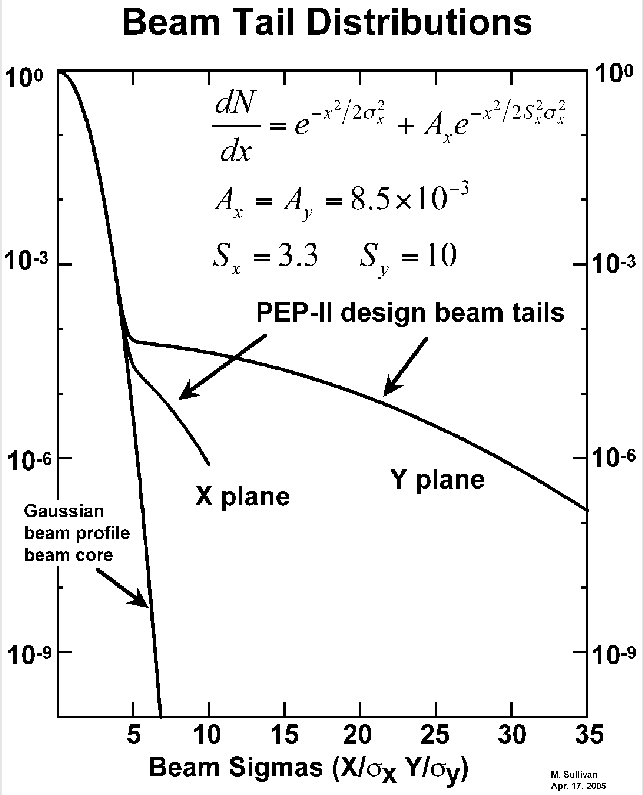}
\vspace*{5mm}
\caption{Plot of the beam tail distributions used in the SR
 simulation. The plot is in beam $\sigma$ along the bottom axis so as
 to be able to plot both the $x$ and $y$ plane at the same time. The actual
 $y$ beam size is much smaller than the actual $x$ beam size.  The $x$ beam
 tail distribution is traced out to 10 beam \sx while the $y$ tail
 distribution is traced out to 35 beam \sy.}
\label{fig:IR_Fig_5}
\end{figure}

In order to study detector backgrounds from SR we use a two-gaussian
transverse beam profile, as employed in the \pepii design. The primary gaussian
distribution matches the sigma $x$ and sigma $y$ given by the nominal beam
emittances and beta functions. To this we add a small fraction of a wider
gaussian that simulates the non-gaussian or
``tail'' distribution of the beams. These tail distributions are produced from
several effects; beam-beam forces from the collision, intra-beam scattering,
Touschek scattering, {\it etc}. Consequently, it is difficult to quantify the exact
nature of the tail distribution. We have conservatively chosen a relatively high tail
distribution since, with collimation of the beam at $10\sx$ for the $x$
plane and $35\sy$ for the $y$ plane, we estimate a beam lifetime of about an
hour. Figure~\ref{fig:IR_Fig_5} shows the transverse beam distributions, together
with the expressions used.

\begin{figure}[htb]
 \centering
\includegraphics[width=0.75\textwidth]{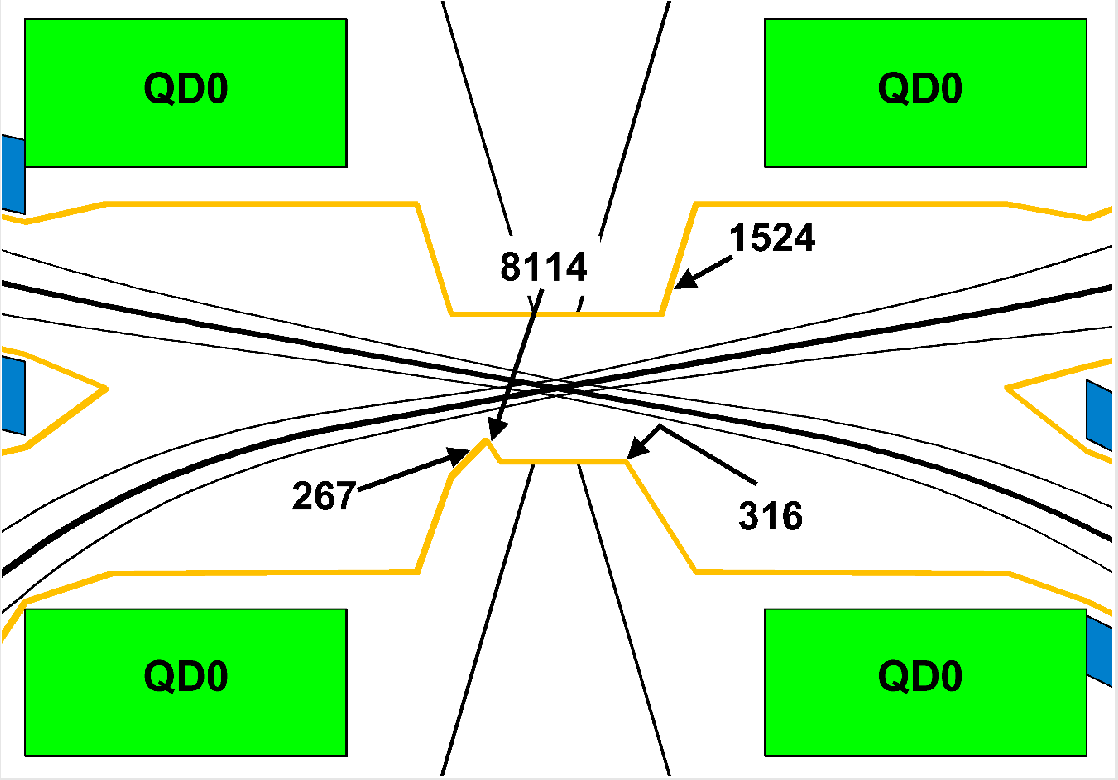}
\caption{Close-up view of the IR with predictions for the photon
 rate from the incoming beams that strike surfaces near the detector
 beam pipe. The rates are for photons per beam crossing having energy
 greater than 10 keV. 
 Table~\ref{table:IR_Table_3} provides more detailed information
 about the photon rates. Most surfaces will be sloped, so that 
 scattered photons have no solid angle acceptance
 to enter the detector beam pipe. The surfaces that do have a
 non-zero back-scatter probability are the inside
 mask surface of the HEB mask (the surface near the IP with 8114
 $\gamma$/crossing) and the downstream surface of the central beam pipe of
 the HEB (the surface with the 316 $\gamma$/crossing value).}
\label{fig:IR_Fig_6}
\end{figure}

We model a beryllium detector beam pipe with a $1\cm$ radius and
a $\pm 4\cm$ long physics window, which accommodates a $\pm 300 \mrad$
detector acceptance. We also assume a $1\cm$ radius beam pipe out to
$\pm 10\cm$. Since we must set the QD0 axis closer to the incoming
beam trajectory, the axis is consequently farther away from the outgoing beam
orbits. This generates more SR from the outgoing beams, which is not a direct
source of backgrounds, but can become a source from backscattered radiation. In
addition, the total amount of SR power produced in QD0 grows rapidly as the
offset moves closer to the trajectory of the incoming beams. With this in mind,
we move the QD0 axis only as close as needed to make sure the SR from QD0 does
not strike the detector beam pipe, and the photon rates on nearby
downstream mask surfaces are acceptably low.
We place a mask on the incoming HEB side to help shadow the detector beam
pipe from HEB radiation. This mask is located on the $-x$ side of the detector
beam pipe, and is modeled as an ellipse offset from the detector axis.
The mask tip is $3\mm$  in from the $10\mm$  radius pipe. Since this is the
only surface that is inside the $1\cm$ radius of the detector beam pipe, the
overall masking design is quite open, and there are no obvious cavities for trapping
higher-order-mode (HOM) power. Figure~\ref{fig:IR_Fig_6} shows the photon rates
on the surfaces that intercept SR from the incoming beams.

\begin{figure}[htb]
\centering
\includegraphics[width=0.8\textwidth]{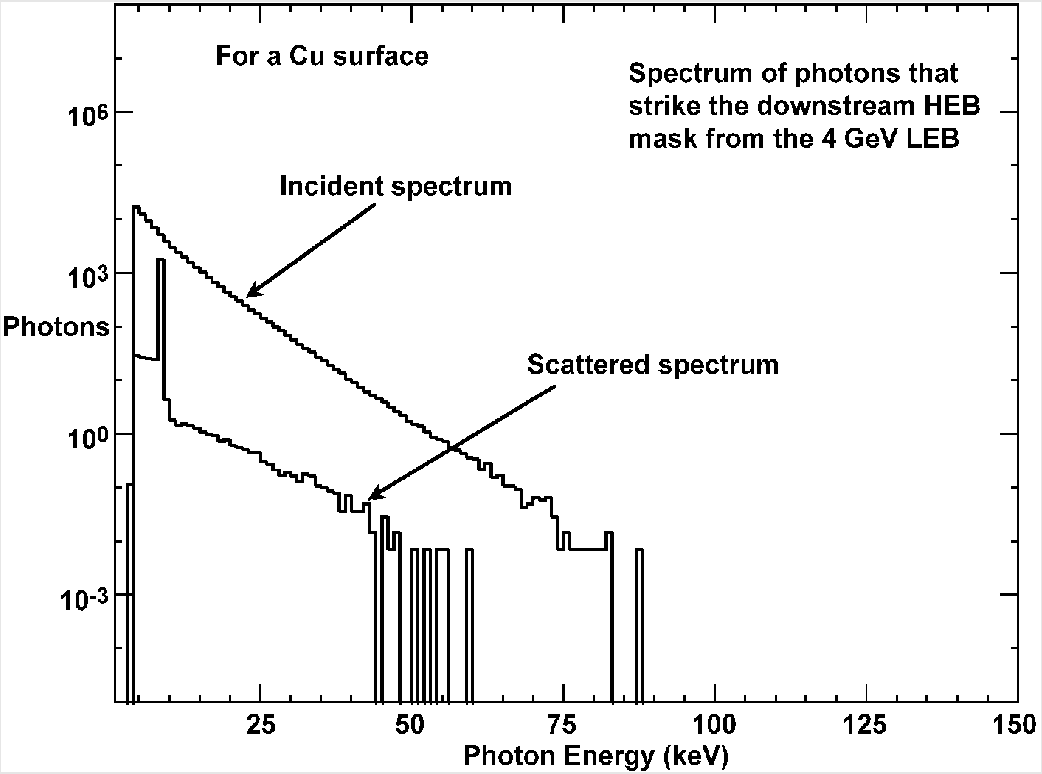}
\includegraphics[width=0.8\textwidth]{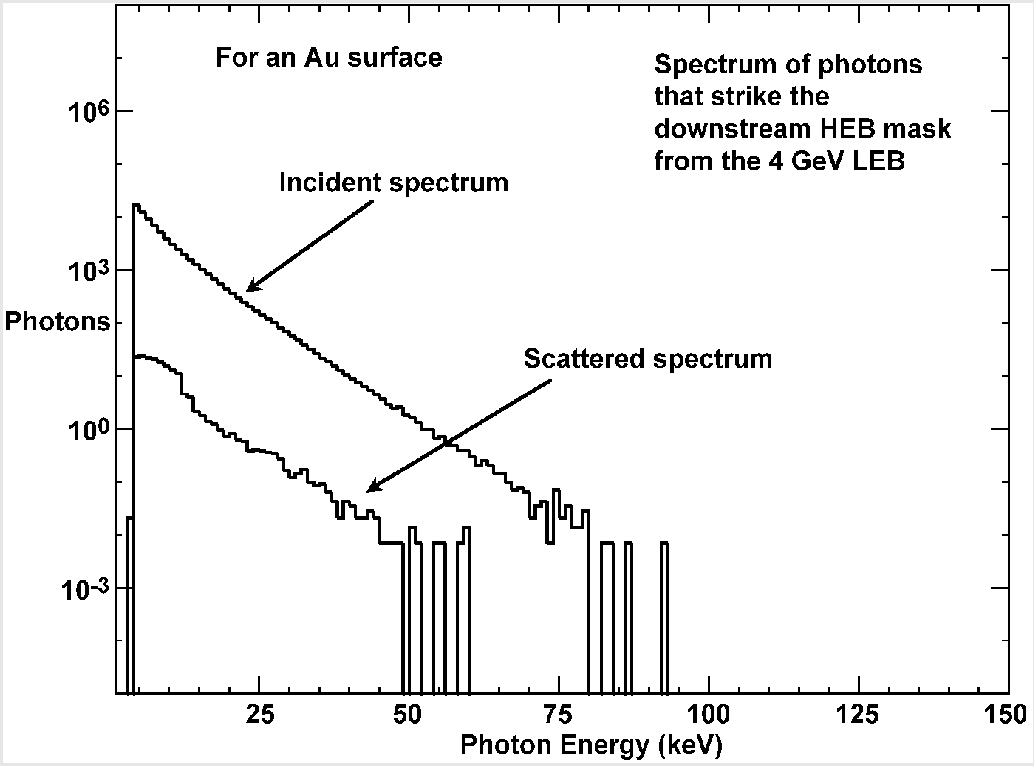}
\vspace*{3mm}
\caption{
Photon energy spectrum incident on the inside
 surface of the HEB mask from the incoming LEB. Both the incident
and backscattered spectra are shown for a Cu surface LEB mask (above) and 
 a Au surface (below).}
\label{fig:IR_Fig_7a}
\end{figure}

The SR photons that strike the surfaces near the detector beam pipe
originate from beam particles that are greater than five beam
sigma from the beam center; the background rates are thus sensitive to the particle
density of the tail distribution. The photons from the LEB that strike
the inside surface (detector beam pipe side) of the HEB mask have a
chance of backscattering from the mask surface and hitting the
detector beam pipe. This is also true of the photons from the HEB that
strike the beam pipe downstream of the detector physics
window. Figure~\ref{fig:IR_Fig_7a} shows the energy
spectrum of the photons that strike these surfaces, as well as the
energy spectrum of photons that backscatter from these surfaces. The
backscatter spectra are obtained using the incident spectra and a
program that uses an Electron-Gamma Shower (EGS) simulation~\cite{bib:ir4}.

\begin{figure}[t]
\centering
\includegraphics[width=0.8\textwidth]{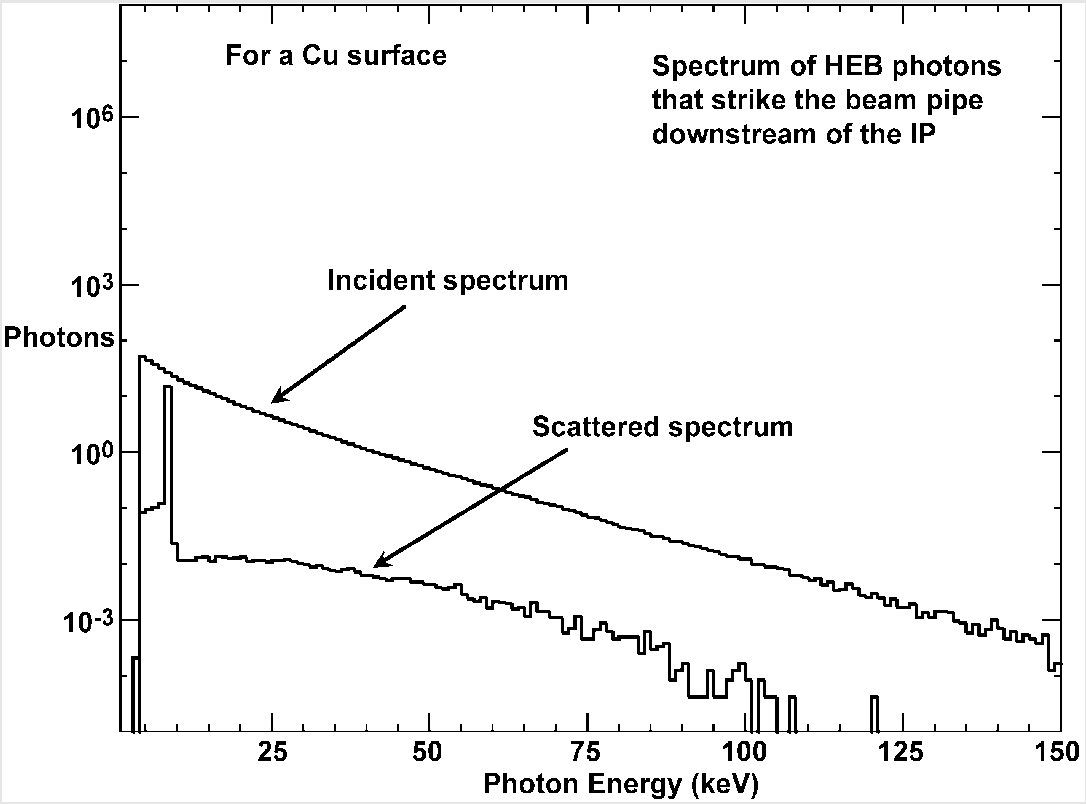}
\includegraphics[width=0.8\textwidth]{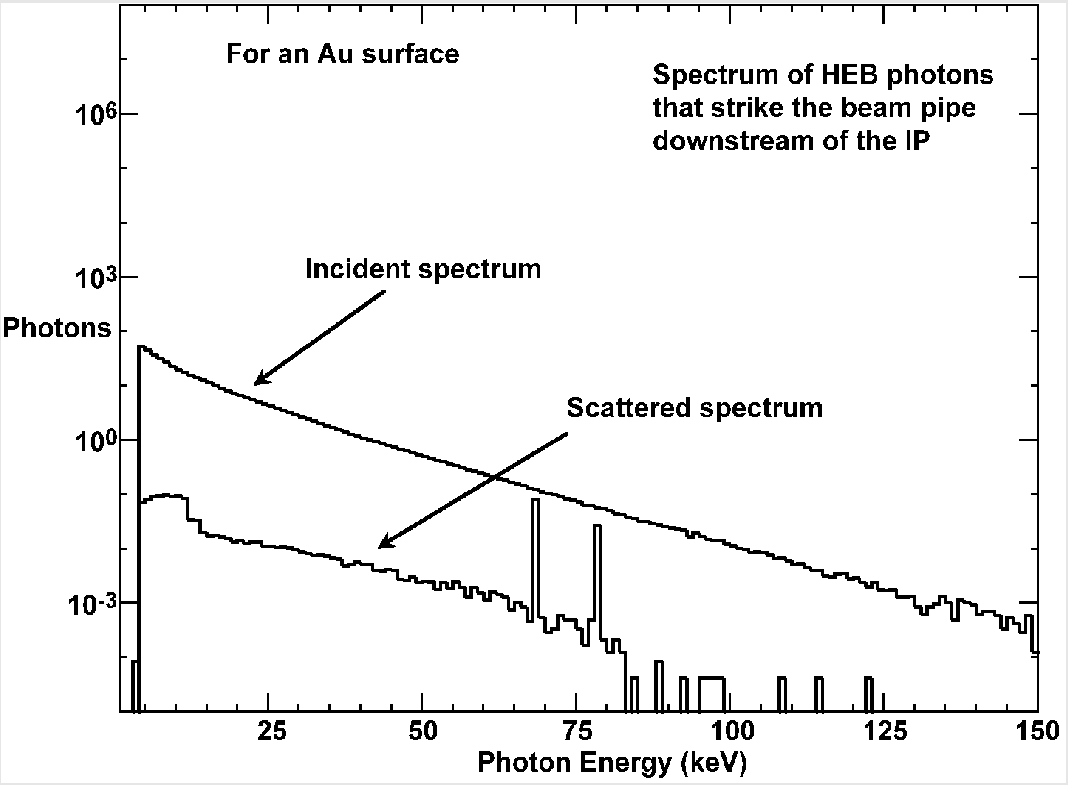}
\vspace*{3mm}
\caption{
Photon energy spectrum incident on the beam pipe
 surface downstream of the IP from the incoming HEB. Both the incident
and backscattered spectra are shown for a Cu surface HEB mask (above) and 
 a Au surface (below).  
 Notice the
 incident HEB photon energy spectrum does not fall off as quickly as
 the LEB photon energy spectrum. Both the incident and backscattered
spectra are shown for an Au surface inbound mask.}
\label{fig:IR_Fig_8a}
\end{figure}

We study two cases: a \Cu surface and an \Au surface. The backscattering
rates are for normal incident photons. Based on the backscatter rates
for these nearby surfaces, we can estimate the rate for photons
incident on the detector beam pipe. Calculating the solid angle
acceptance of the detector beam pipe from these nearby surfaces that
are struck by photons, we find an acceptance fraction of about
$1.1\%$. Applying this fraction to the total backscattered photon rate from
these nearby surfaces, we can estimate the rate of photons striking the
detector beam pipe. The results are summarized in Table~\ref{table:IR_Table_3}.
We find the rate of photons incident on the detector
beam pipe is lower if the backscattered photons are produced from
a \Cu rather than Au surface. However, this conclusion is drawn by integrating the
spectra above $10\kev$, which is also above the \Cu K-shell photo-emission line
at $8.9\kev$. The 8.9\kev emission line is evident in the Cu
backscattered spectrum shown in Fig.~\ref{fig:IR_Fig_8a}. Although these photons
are low energy, we may prefer the \Au coating
in order to suppress this potential background source. The rate of
photons on the detector beam pipe is sufficiently low in either case.
These predicted rates are significantly lower than the 10 per beam crossing for 
the \pepii IR design.

\subsection{Outgoing SR Fans}

\begin{figure}
\centering
\includegraphics[width=0.8\textwidth]{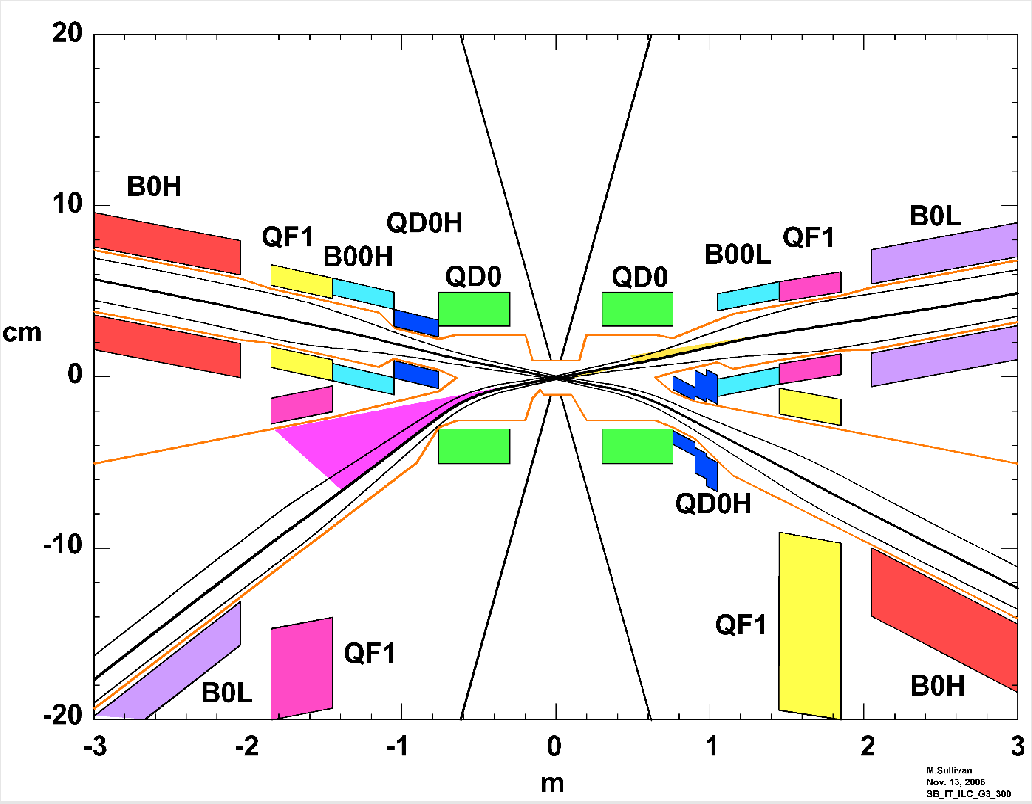}
\includegraphics[width=0.8\textwidth]{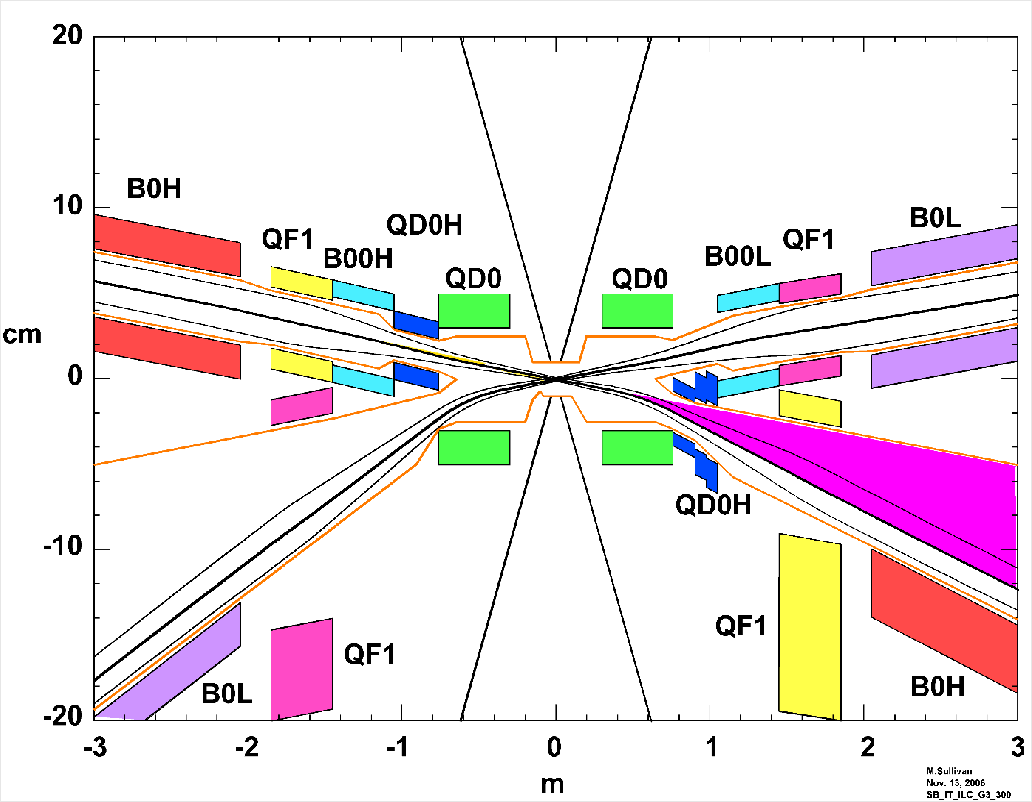}
\caption{
Layout of the IR with the LEB (above) and HEB (below) synchrotron radiation
 fans shown. The highest power fan comes from the beam as it goes
 through QD0 after the collision. Note the beam pipe for
 both beams flares out in order to avoid the out-going SR fans from these
 beams.}
\label{fig:IR_Fig_9a}
\end{figure}

As discussed above the exiting beams
are strongly bent as they travel through the QD0 magnets. The outgoing
LEB generates $88\kW$ of SR power in QD0 and the outgoing HEB generates
$141\kW$.  The beam pipe for the outgoing beams is designed so that
these high power fans do not strike any nearby surfaces and are
absorbed on beam pipe surfaces that are meters away from the collision
point. Figure~\ref{fig:IR_Fig_9a} shows the SR fans generated by
the beams and Table~\ref{table:IR_Table_4} lists the SR power produced by the beam
for each IR magnet.

{ \setlength{\tabcolsep}{1pt}  
\begin{table}[p]
\centering
\begin{threeparttable}
\caption{\label{table:IR_Table_3}
Photon rates from nearby beam pipe (b.p.) surfaces and
calculations of photon rates incident on the detector beam pipe for
the two cases of a Cu and a Au surface. See Fig.~\ref{fig:IR_Fig_6} for an
illustration of the surfaces studied.}
\vspace*{2mm}
\setlength{\extrarowheight}{0pt}
\centering
\begin{tabular}{lccccc}
\hline
\hline
 Surface  && LER beam                 &    HER mask       & HER mask                  & b.p. 10 cm \\
                    &&  incoming surf.\tnote{a} \ \ &    inside surf.    & outside surf.\tnote{a} \ \    & from the IP \\

\hline
  Distance to IP   &&          15--20 cm           &          10 cm           &          10--15 cm           & 10 cm \\
    Source(s)      &&           LER QF1           &    LER QD0,     &           HER QF1           & HER QF1 \\
                              &&                             &   B00L, QF1     &                             &         \\
 Fraction of solid angle &&    0.000\tnote{a}     &          0.011           &    0.000\tnote{a}     & 0.011 \\[-2mm]
         \ \ \ to det. b.p.           & \\
\hline
\multicolumn{6}{c}{Energy of $\gamma > 10 \kev$}\\
\hline

Incident photons    & N/xing &  1524        &    8114     &      267      & 316  \\
       & \Hz    &  3.63\EE{11} & 1.93\EE{12} &  6.36\EE{10} &7.52\EE{10} \\
Backscatter from Cu  & N/xing &   ---      &    8.85          &   ---      & 0.69  \\
       & \Hz    &   ---      &    2.11\EE{9}    &   ---      & 1.64\EE{8} \\
Inc. on b.p. from Cu  & N/xing &   ---      &    0.097         &   ---      &  0.0076 \\
       & \Hz    &   ---      &    2.32\EE{7}    &   ---      &  1.81\EE{6} \\
Backscatter from Au  & N/xing &   ---      &    24.2        &   ---      & 1.12  \\
       & \Hz    &   ---      &    5.75\EE{9}   &   ---      & 2.65\EE{8}   \\
Inc. on b.p. from Au  & N/xing &   ---      &    0.27        &   ---      & 6.34\EE{7}  \\
       & \Hz    &   ---      &   0.12    &   ---      &   2.93\EE{6} \\
\hline
\multicolumn{6}{c}{Energy of $\gamma > 20 \kev$}\\
\hline

Incident photons    & N/xing &   15.4       &    1072     &     102       &  120 \\
       & \Hz    &  3.67\EE{9} & 2.55\EE{11} &  2.43\EE{10} &  2.86\EE{10}\\
Backscatter from Cu  & N/xing &   ---      &    2.63         &   ---      & 0.49  \\
       & \Hz    &   ---      & 6.26\EE{8}       &   ---      & 1.16\EE{8} \\
Inc. on b.p. from Cu  & N/xing &   ---      &     0.029        &   ---      & 0.0054 \\
       & \Hz    &   ---      &  6.90\EE{6}       &   ---      &  1.28\EE{6}\\
Backscatter from Au  & N/xing &   ---      &       2.83       &   ---      & 0.57  \\
       & \Hz    &   ---      &    6.74\EE{8}         &   ---      & 1.36\EE{8} \\
Inc. on bp from Au  & N/xing &   ---      &      0.031       &   ---      &  0.0063\\
       & \Hz    &   ---      &     7.41\EE{6}        &   ---      & 1.49\EE{6} \\

\hline
\end{tabular}
\begin{tablenotes}
\item[a]These surfaces are sloped to eliminate the possibility of
backscattered photons hitting the Be beam pipe at the IP.
\end{tablenotes}
\end{threeparttable}
\end{table}
}

\begin{table}[htb]
\caption{\label{table:IR_Table_4}
Summary of the total SR power generated in the
 IR, for a LER beam current of $4\amp$ and a HER beam current of $2.2\amp$. 
 These are the upgrade parameters rather than the design values
 of $2.3\amp$ and $1.3\amp$ respectively.}
\vspace*{2mm}
\centering
\setlength{\extrarowheight}{3pt}
\begin{tabular}{cccccc} \hline\hline
          & \multicolumn{2}{c}{LER} & \multicolumn{2}{c}{HER} & Type \\\hline
 $z$ ctr & \ \ Magnet \ \ & \ \ Power \ \ & \ \ Magnet \ \ & \ \ Power \ \ & \\
(m)   &        & (W)  &        & (W)  & \\\hline
  -3.05   &  B0L   &       147       &  B0H   &       755       & Bend \\
  -1.65   &  QF1   &       29        &  QF1   &       114       & Quad    \\
  -1.25   &        &                 &  B00H  &      1228       & Bend \\
 -0.905   &        &                 &  QD0H  &       24        & Quad \\
  -0.53   &  QD0   &      87970      &  QD0   &      6592       & Quad+Bend \\
  0.53    &  QD0   &      1718       &  QD0   &     141270      & Quad+Bend \\
  0.905   &        &                 &  QD0H  &       24        & Quad \\
  1.25    &  B00L  &      1568       &        &       --        & Bend \\
  1.65    &  QF1   &       29        &  QF1   &       114       & Quad \\
  3.05    &  B0L   &       147       &  B0H   &       755       & Bend \\\hline
\end{tabular}
\end{table}

The larger beam pipes on the outgoing beam lines require that the
magnets on these lines have larger apertures. A possible design
for the outgoing QF1 magnet is one similar to the horizontally split
quadrupole magnets used in SPEAR III~\cite{bib:ir5}.  These magnets have no
material in the horizontal plane, which allows the horizontal fan to
pass through the magnet without striking the beam pipe. The outgoing
B0 magnets can be C-shaped bend magnets. Both of these designs employ
iron core magnets, which would need to be shielded from the
detector magnetic field. We would accomplish this by inserting a
compensating superconducting solenoid around both beam lines and into
the detector far enough so that at least the QF1 and B0 magnets can be
iron core magnets. Figure~\ref{fig:IR_Fig_2} shows a suggested layout of the
compensating solenoid.  

At some point the outgoing SR fans will
strike the beam pipe. We assume this occurs somewhere near $10\m$ from
the IP (similar to the HEB \pepii design), resulting in only a
small probability for backscattered photons to hit the
detector beam pipe. Although this has not been fully studied, steps can be
taken to minimize background from this source. The
backscatter rate can be made very small by keeping the source point as
far as possible from the detector beam pipe (minimizing the solid
angle acceptance of the detector beam pipe). The beam pipe surface can
be coated with a layer of high-$Z$ metal such as gold. The angle of the
beam pipe surface can, in some cases, be such as to eliminate the
possibility of backscattering into the detector beam pipe. Another
approach is to add a small ($\sim 1\mm$) mask near the detector beam pipe
that completely shadows the Be beam pipe from this source.
We believe that any background from the
strong outgoing SR fans striking the beam pipe far from the detector
can be made negligible at the IP, as
it is in the \pepii design.

\subsection{Beam-Gas Bremsstrahlung (BGB)}
Another source of backgrounds in the detector comes from 
beam-gas bremsstrahlung (BGB), which
results from a beam particle colliding with a gas molecule in the vacuum
chamber, resulting in the production of a gamma and an off-energy beam particle. Both
of these particles can cause backgrounds if they strike the beam pipe near or
inside the detector. In order to minimize this background source,
the residual pressure must be as low as possible just
upstream of the detector for both beams.
The \pepii experience indicates that we can achieve a dynamic pressure
of 1--2\nTorr in these regions with sufficient pumping. Bending magnets further
outboard of the detector ($\sim 10$--20\m) can help minimize this background
source. Suitable arrangement of the last bend magnet on the inbound beam line 
can be used to direct the stream of gammas to a known location away from the 
detector. Likewise this will overbend low-energy beam particles out the beam 
pipe, again well away from the detector.

\subsection{Radiative Bhabhas}
\begin{figure}[htbp]
\centering
\includegraphics[width=0.8\textwidth]{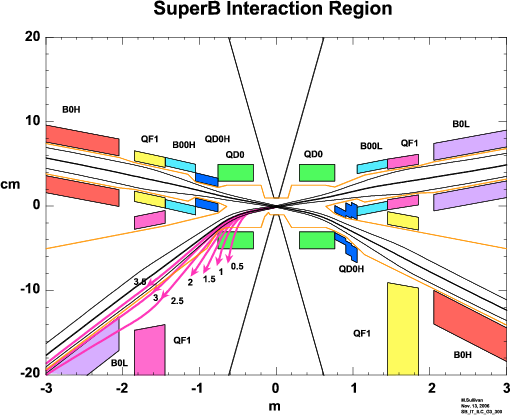}
\includegraphics[width=0.8\textwidth]{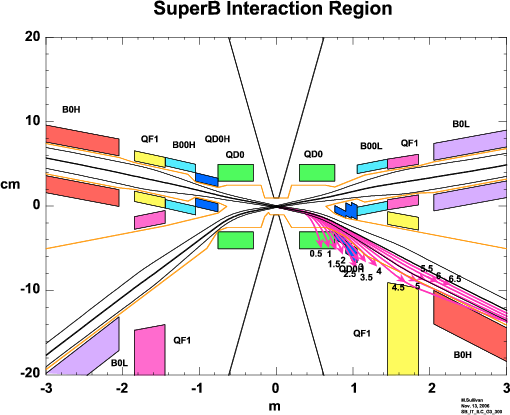}
\caption{ Representative trajectories for off-energy
particles due to radiative Bhabha events from the LEB (above) and HEB (below).} \label{fig:IR_Fig_10}
\end{figure}

The final-state particles produced in radiative Bhabha scattering are a source
of detector background. The gamma produced by this
reaction and the reduced energy beam particle can both generate backgrounds in
the detector if these particles strike material close enough to the detector.
The gammas are generally produced in the direction of the beam, and hence exit
the IP along the crossing angle trajectories. In our design, this means
the outgoing gammas travel along the edge of the strong outgoing SR fans and hence
do not intersect the beam pipe until several meters away from the
collision point. However, the outgoing off-energy beam particles from
this reaction can be a source of background for the detector. Figure~\ref{fig:IR_Fig_10} 
illustrates the orbits of these
off-energy particles for Bhabha scattering from the LEB and HEB. A Monte Carlo simulation is needed
to fully study this background source, and ways of shielding the detector from
it. Section~\ref{sec:det:Backgrounds} on detector backgrounds has further details on this
subject.

\subsection{Luminosity Monitor}
The \pepii accelerator uses a zero-angle luminosity
monitor that detects the gammas from the radiative Bhabha 
reaction~\cite{bib:ir7}. The monitor detector is located next to the HER beam pipe
where it intercepts the gammas from LEB radiative Bhabha events through
 a relatively thin Al window. This method works quite well, and is an invaluable tool
for tuning up the accelerator. The good signal-to-noise ratio
allows measurement of the luminosity contribution from individual bunches. The
beam pipe design for \superb\ allows a
luminosity monitor similar to the \pepii type to be installed
7--10\m from the IP. In the \superb\ case, a detector could be
installed on either side of the IP, one to intercept the radiative
Bhabha signal from the LEB and another to intercept the radiative
Bhabha signal from the HEB. The main background signal for this
detector is BGB generated by the incoming beam. This background
signal is integrated over the length of the beam trajectory near the
collision point following the last inbound bend. In the \pepii case, this is $42\cm$. 
The \superb\ design has bending magnets for the incoming beams
between the QD0 and QF1 magnets, as well as just outboard of the QF1
magnets. This would make the integrated distance for the BGB signal
about $1\m$, very similar to \pepii.


\clearpage

\section{Magnet Lattice and Optics}
\label{section:MagnetLatticeAndOptics}
\subsection{HER Lattice}
\label{sec:lattice}

The high-energy ring (HER) has 6 arcs, with 12 cells each. Each arc is
approximately 250\m long. Six straight sections connect the arcs; one
contains the final focus (FF), while in the opposite straight section
a magnetic chicane is used to keep the
beams separated. The other four straight sections will house the
wiggler magnets, needed to control emittances and damping times, and the
RF cavities.
In the initial configuration, the ring will have a horizontal emittance of $1.6\nm$-rad
with no wiggler magnets installed; in the second phase, which has
$0.8\nm$-rad emittance, two wigglers with a $0.83\Tesla$ field will be
used. In this way we can control both emittance and damping time.

The HER lattice was inspired by the OCS lattice used as the
baseline design of the \ilc Damping Rings (DR), given the similarities
of the \ilc and \superb\ designs. The beam energy was scaled to the
$4\times 7 \gev$ required by \superb, and the circumference was
shortened from the original $6.1 \km$ to the final design value
of $2.25 \km$, providing shorter
(cheaper) rings with the required design emittance and damping time.

The OCS lattice achieves a very small emittance at $5\gev$, by employing a TME
(theoretical minimum emittance) $45\m$ long cell \cite{bib:mlo_1_1}
 with two dipoles and two
quadrupole families (QF and QD). The $\beta$ functions are well
separated and high at the sextupoles locations, allowing for easy
chromaticity correction.

\begin{figure}[htb]
 \centering
 \hspace{-10mm}
\includegraphics[width=0.9\textwidth]{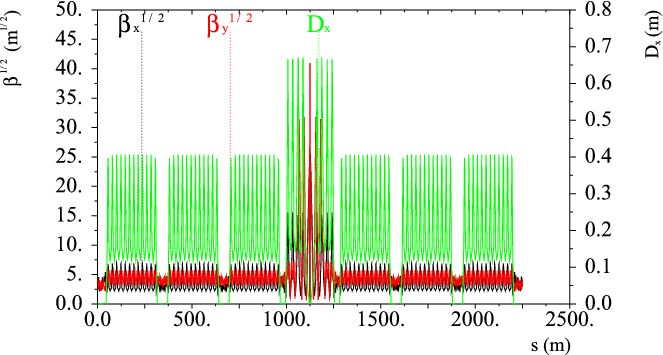}
    \caption{Optical functions in the HER.}
    \label{fig:HER}
\end{figure}

%

\begin{figure}[thbp]
 \centering
\includegraphics[width=0.65\textwidth,angle=-90]{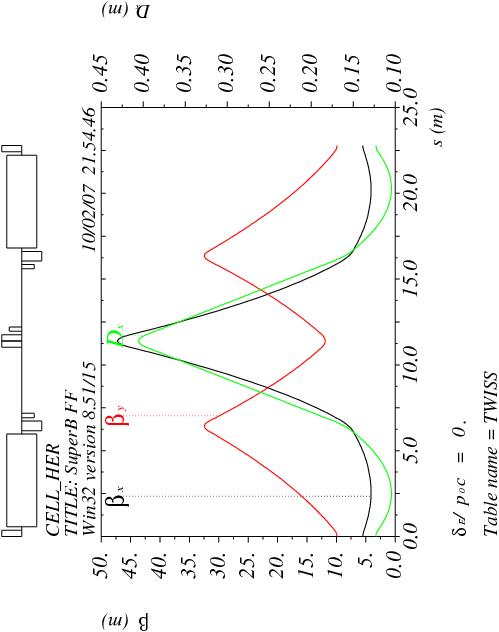}
\includegraphics[width=0.65\textwidth,angle=-90]{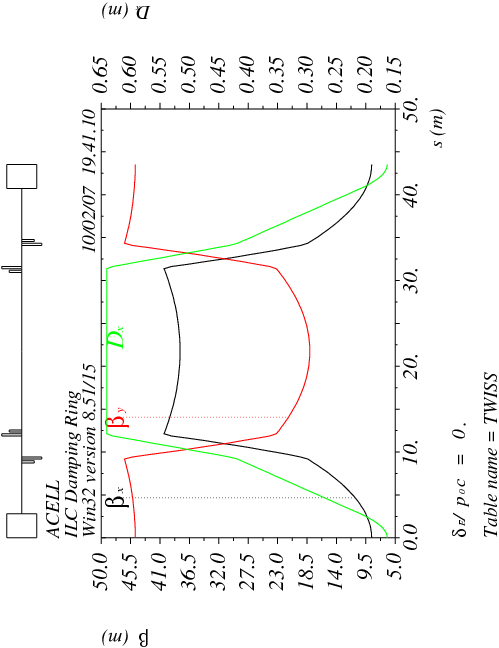}
    \caption{   \label{fig:HERcell}Optical functions for the HER cell (top)
    and OCS cell (bottom).}
\end{figure}

We were able to construct a similar cell using the \pepii HER bending
magnets, which are $5.4\m$ long. However, to shorten the ring
circumference, we chose a cell with a phase advance of
$\pi$ in the horizontal plane and $0.4\times\pi$ in the vertical
plane, with even smaller intrinsic emittance. With the shorter arcs we
gained other benefits: smaller natural horizontal chromaticity,
which drops from $-80$ to $-55$, and fewer elements. The number of
sextupole families was also reduced from three to two per arc, requiring
higher strength magnets. 
The optical $\beta$ functions in the ring are shown in Fig.~\ref{fig:HER}.
Figure~\ref{fig:HERcell} compares the
optical functions for the new cell \vs\ the
OCS design.

\begin{figure}[thbp]
 \centering
\includegraphics[width=0.75\textwidth]{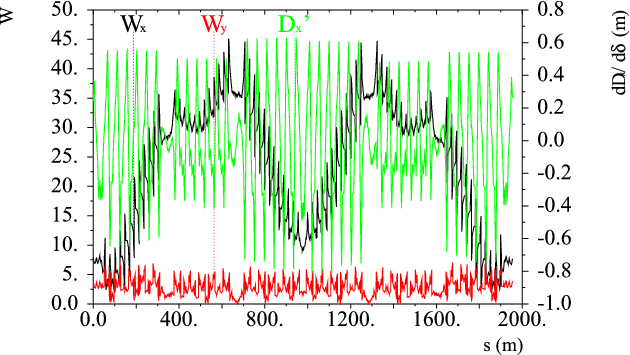}
\includegraphics[width=0.75\textwidth]{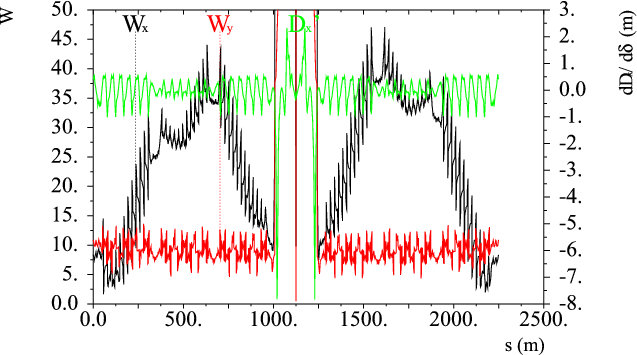}
 \caption{
HER $W$ chromatic function and second-order dispersion function with sextupoles on,
for the ring with (bottom) and without (top) a final-focus insertion.}
\label{fig:HERW}
\end{figure}

\begin{figure}[!hbp]
 \centering
\includegraphics[width=0.75\textwidth]{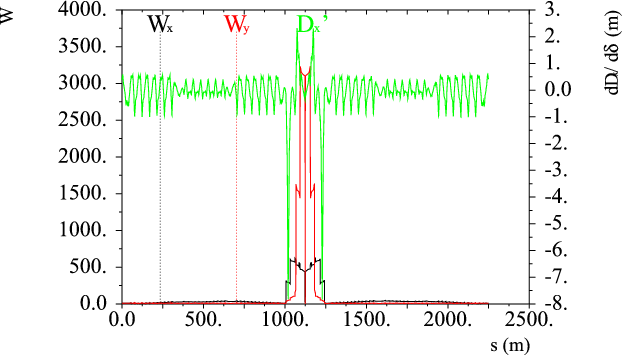}
  \caption{\label{fig:HERWFF}
HER $W$ chromatic function and second-order dispersion function with
 the insertion of FF.}
\end{figure}

\begin{figure}[!hbp]
 \centering
\includegraphics[width=0.65\textwidth,angle=-90]{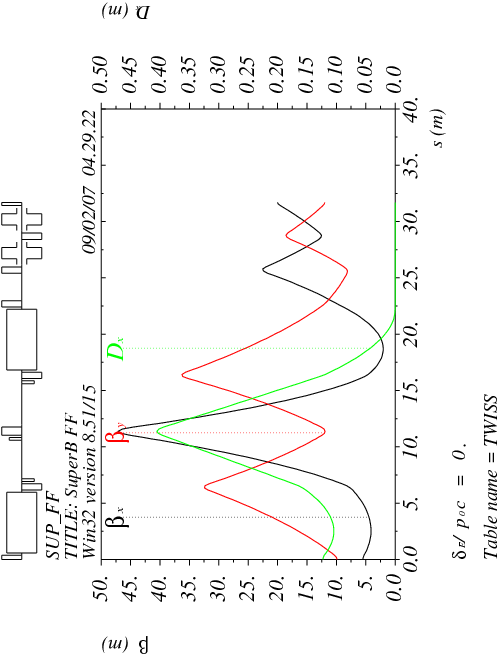}
\includegraphics[width=0.65\textwidth,angle=-90]{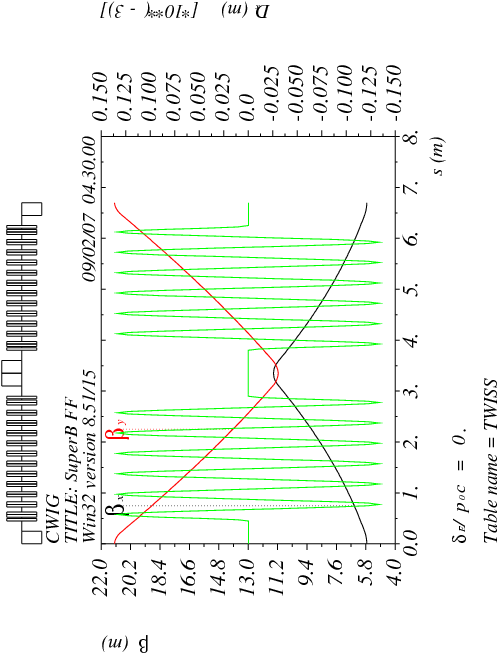}
    \caption{\label{fig:HERdisps}
    HER dispersion suppressor region (top) and wiggler cell (bottom).}
\end{figure}


\begin{figure}[!hbp]
\centering
\includegraphics[width=0.75\textwidth]{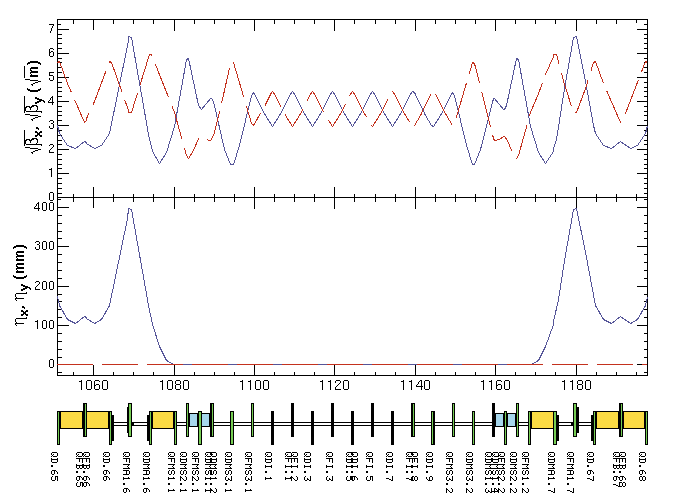}
\caption{\label{fig:straight-her}
Lattice design of the straight section opposite to the IP in HER.
A matching section to adjust the betatron tunes is included.}
\end{figure}

The phase advance between the arcs has been also optimized in order
to minimize chromatic terms. Chromaticity in the arcs is
corrected with three families of sextupoles, two SD and one SF. The
behavior of the chromatic $W$ functions and second-order dispersion
function for the ring without the insertion of the FF, and with
sextupoles to correct chromaticity to zero, is shown in
Fig.~\ref{fig:HERW}. The vertical $W$ function is always very small;
however, by adjusting the horizontal phase advance, the horizontal
$W$ function can be made periodic and goes to zero in the two main
straight sections. This behavior is slightly perturbed by the
insertion of the FF, with its high gradient sextupoles, as seen in
the right-hand plot of Fig.~\ref{fig:HERW}, where a blowup allows a
better appreciation of the $W$ behavior in the arcs. Its value in the FF is, of
course, quite different, as can be seen in Fig.~\ref{fig:HERWFF} in
full scale, due to the large chromaticity induced by the high
$\beta$ values in the IR quadrupoles.

Each arc has a dispersion suppressor (an example is shown in
Fig.~\ref{fig:HERdisps}) so that the dispersion is zero in all the
straights connecting the arcs, where the RF cavities and wigglers
will be installed. The straight section where we can install the
wigglers has an optical function behavior similar to the ILC OCS design
(see Fig.~\ref{fig:HERdisps}, righthand plot). Of course similar $\beta$
functions can be obtained in wiggler-free sections.  Tuning of the phase
advances can easily be performed in the straight section opposite the FF
(see Fig.~\ref{fig:straight-her}), without perturbing the chromatic
characteristics of the ring.

All the \pepii HER magnets (originally built for an $18\gev$ machine)
have been used in the ring. We will need to build some additional
sextupoles and quadrupoles.

\subsection{LER Lattice}

\begin{figure}[htbp]
 \centering
\includegraphics[width=0.5\textwidth,angle=-90]{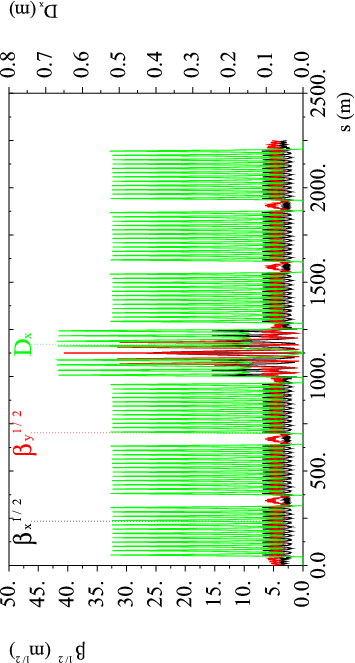}
    \caption{Optical functions in the LER.}
    \label{fig:LER}
\end{figure}

\begin{figure}[htbp]
 \centering
\includegraphics[width=0.65\textwidth,angle=-90]{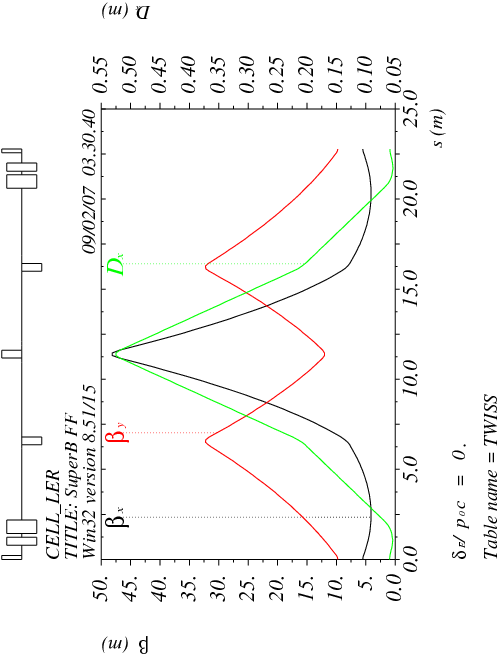}
\includegraphics[width=0.65\textwidth,angle=-90]{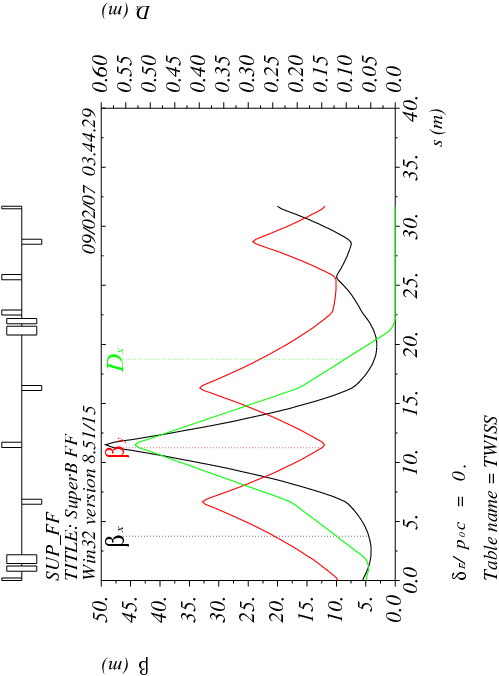}
    \caption{   \label{fig:LERcell} LER cell (top) and dispersion
    suppressor (bottom).}
\end{figure}

The Low Energy Ring lattice is very similar to the HER design, as can be
seen in Fig.~\ref{fig:LER} where the optical $\beta$ functions are
shown.
However, since the \pepii LER dipoles are very short ($0.45 \m$), in order to
have the same emittance as the HER, the cell has been modified to incorporate four
bending magnets, two of the \pepii LER type and two of a new type
($0.75 \m$). The LER cell is very similar to the
HER cell. Fig.~\ref{fig:LERcell} shows the optical functions for the new
cell. The dispersion is somewhat lower in this cell then in the
HER cell. The dispersion-suppressor straight sections are very similar 
(righthand plot of Fig.~\ref{fig:LERcell}).

\begin{figure}[htbp]
 \centering
\includegraphics[width=0.65\textwidth,angle=-90]{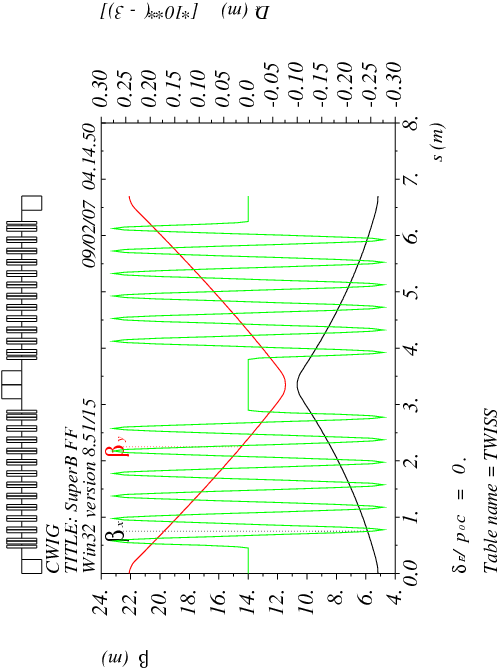}
\includegraphics[width=0.65\textwidth,angle=-90]{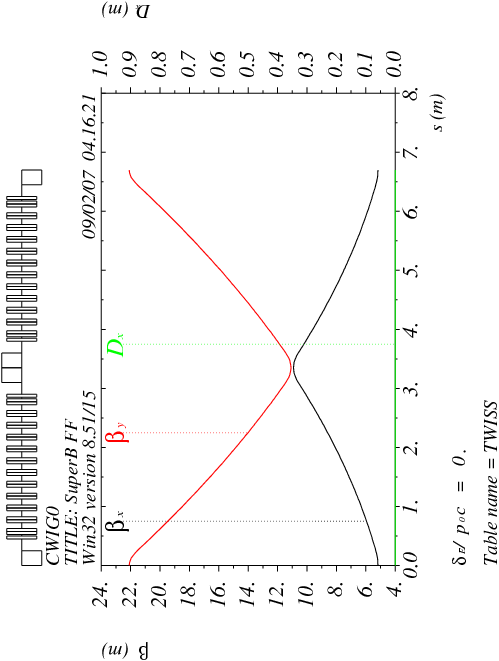}
    \caption{   \label{fig:LERwig} LER wiggler straight with
    wiggler on (top) and off (bottom).}
\end{figure}

\begin{figure}[htb]
\centering
\includegraphics[width=0.75\textwidth]{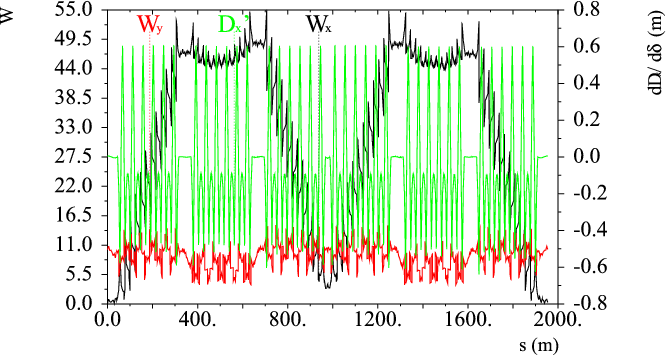}
\includegraphics[width=0.75\textwidth]{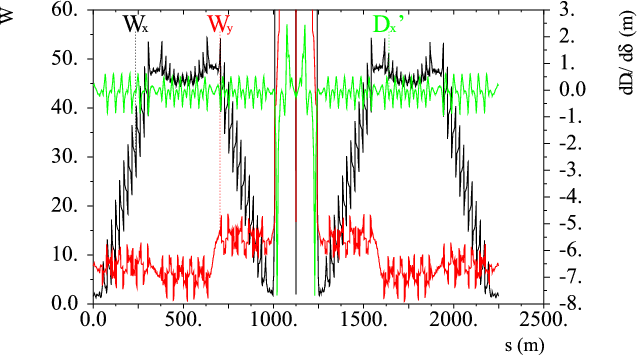}
\caption{
LER $W$ chromatic function and second-order dispersion function for the ring
with (bottom) and without (top) a final focus insertion.}
\label{fig:LERWFF}
\end{figure}

The $1.6\nm$-rad emittance in the LER is obtained using two
$1\Tesla$ wiggler magnets of the same design as the wigglers in the
\ilc OCS lattice. The magnet technology will be the subject of further
study; this is an example of an issue common to \superb\ and
the \ilc DR. With four wigglers, we will
be able to reach $0.8\nm$-rad emittance and the same damping time as in the
HER ring. The wiggler section is very
similar to that for the HER, as can be seen from
the lefthand plot in Fig.~\ref{fig:LERwig}. A cell with the same optical characteristics
but without the wiggler field is shown in the righthand plot of Fig.~\ref{fig:LERwig}.
The chromatic behavior is also the same since the same constraints
on the phase advances have been chosen (see Fig.~\ref{fig:LERWFF}).
Finally, all \pepii LER magnets
(built for a $3.1\gev$ machine) have been
used in the new ring design; we also need to build some
new sextupoles and quadrupoles.

%

In conclusion, we have been able to develop a lattice design that is very similar for both
rings, \ie, with the same design values for parameters such as emittance and damping
times, in spite of their different energies and the constraint of re-using all the \pepii components.

\subsection{Final Focus}

\paragraph{Conceptual Design.}
The design of the \superb\ final focus section must satisfy
several different constraints:
\begin{itemize}
\item Very small $\beta$ functions at the interaction point (IP);
\item Small geometric aberrations, with non-interleaved sextupoles; and
\item All bends with the same sign, to avoid chicanes, keep the
  geometry simple and reduce the arc length.
\end{itemize}

Producing low-$\beta$ functions at the IP requires the design
to locate the final focus quadrupole doublets as close as possible
to the IP.
The \superb\ final focus (FF) has been designed with this
principle, using the experience gained in designing the FFTB/NLC final
focus \cite{bib:Nlc}. The design parameters for the $\beta$ function
at the IP are $20\mm$ in the horizontal plane and $200\mum$ in the vertical
plane, with a minimum distance ($L^\star$) between the IP and the first FF
quadrupole
magnet of $30\cm$. The FF quadrupole magnets must have strong
magnetic gradients, in order to realize the extremely low-$\beta$
functions at the IP.
We have chosen to design the FF as a section of a regular arc, so
that it is naturally embedded in the lattice without breaking the
6--fold symmetry. The interaction region is then geometrically, but not
optically, equivalent to all the other straight sections.

The horizontal crossing angle is taken to be $34\mrad$; it will be adjusted
with bending magnets in the dispersion suppressor of the arc section
connected to IR.  The solenoidal field of the detector will be
compensated with compensation solenoids on each side of the detector.
An example of compensation scheme is described in
Sec.~\ref{sec:comp}.

\begin{figure}[htb]
\centering
\includegraphics[width=0.75\textwidth]{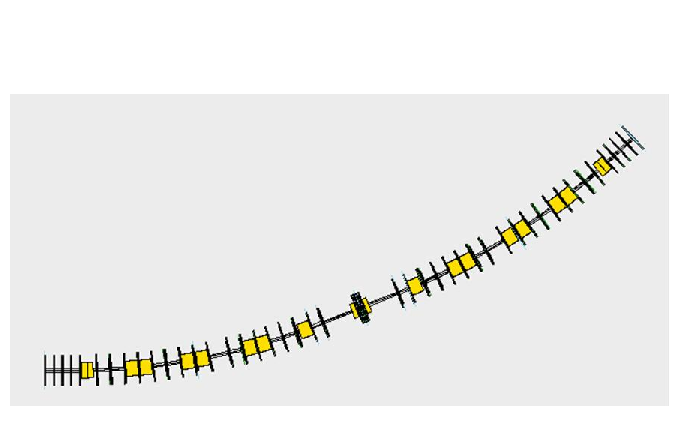}
\caption{
Lattice design for the final focus.
}
\label{fig:ff-lay}
\end{figure}

\begin{figure}[htb]
\centering
\includegraphics[width=0.8\textwidth]{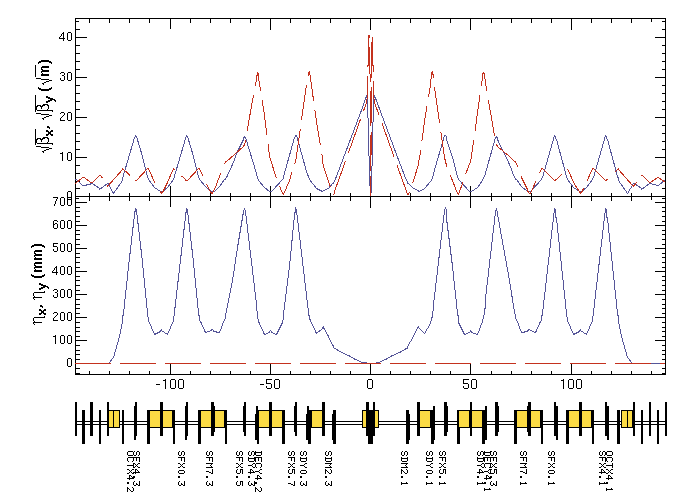}
\caption{
$\beta$ functions in the interaction region.
}
\label{fig:ff-layout}
\end{figure}

The lattice optical functions of the FF are shown in
Fig.~\ref{fig:ff-layout}. A sketch of the magnetic structure is shown
in Fig.~\ref{fig:ff-lay}.

Due to the high $\beta$ values in the first
FF doublets, the FF quadrupole magnets generate large chromaticity,
which should be corrected as locally as
possible. This is accomplished in the \superb\ FF design by using the scheme
studied for the FFTB/NLC at SLAC: the local chromaticity correction
is accomplished by two sextupole magnets in each plane (SDY0 and SDY4,
SFX0 and SFX4), each pair being connected with a $-I$ transformation matrix:

\begin{eqnarray}
\left(\begin{array}{cccc}
  -1   & 0  &   0    & 0 \\
m_{21} & -1 &   0    & 0 \\
  0    & 0  &   -1   & 0 \\
  0    & 0  & m_{43} & -1
\end{array}\right).
\end{eqnarray}

The sextupole pairs should reduce nonlinearities for each other
in correcting the chromaticity, while also increasing the FF
bandwidth or momentum acceptance. The dispersion
at the sextupoles is created with several bending magnets and
matching quadrupole magnets to make the dispersion
($\eta_x$,$\eta'_x$) zero at the IP and localized in the IR. The
layout of the IR is geometrically symmetric and the sign of the
bending angle is the same for all bending magnets. These bending
magnets help to reduce arc length; 32 \pepii HER dipoles will be
used for both LER and HER IRs. The HER and LER branches
have different quadrupole strength requirements; the first
quadrupole, QD0, will therefore be split into two pieces so that the LER beam
sees just the first segment. Additional sextupoles are also used to
correct the third-order chromaticity. These do not reduce the dynamic
aperture, since the $\beta$ function at the sextupole magnets is
small.

\begin{figure}[htb]
 \centering
\includegraphics[width=0.8\textwidth]{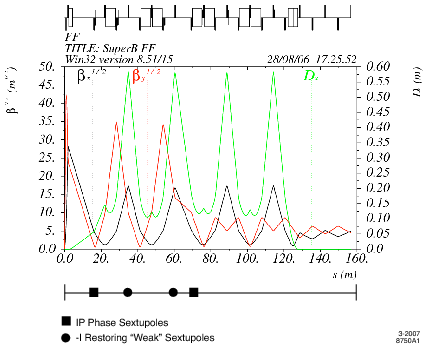}
 \caption{
    $\beta$ functions in one half the IR. The positions of the IP in-phase
    sextupoles and the two $-I$ restoring sextupoles are indicated
    on the scale below.}
\label{fig:ff-beta}
\end{figure}

\begin{figure}[htbp]
\centering
\includegraphics[width=0.55\textwidth]{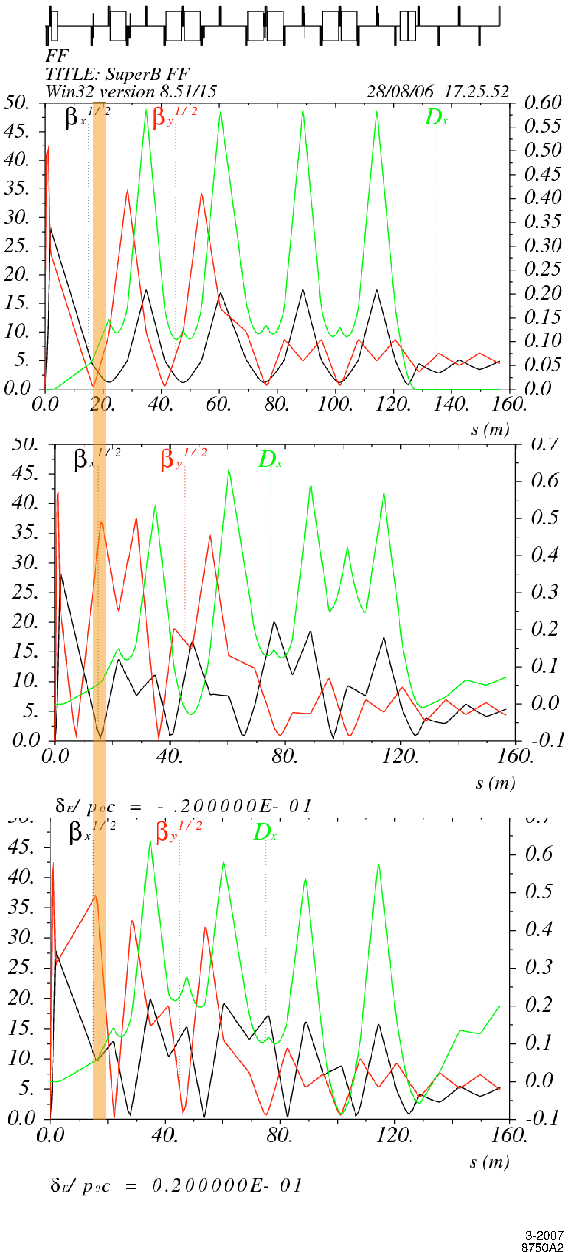}
 \caption{
Optical functions ($\beta$, $\eta$) for on (top plot) and off-momentum
($-2\%$ middle plot, $+2\%$ bottom plot) particles. The vertical shaded
band indicates a location where minimum $\beta_y$
for on-momentum particles corresponds to a maximum $\beta_y$
for $\pm 2\%$ off-momentum particles.}
\label{fig:ff-off}
\end{figure}

Two identical weak sextupole magnets are interleaved with the main
local chromaticity correction sextupoles (with about $10\%$ of their
intensities) to improve the behavior of the off-momentum particles,
and thereby the momentum acceptance of the rings. In Fig~\ref{fig:ff-beta} the
$\beta$ functions for half the IR and the position of the sextupoles are
shown. As can be seen in Fig.~\ref{fig:ff-off}, they
are located at a minimum $\beta$ for on-momentum particles, but
off-momentum particles see a maximum $\beta$.

Two weak octupole magnets are used to correct the chromatic
effect for the off-momentum particles.  There is a $5\%$ emittance
growth in the HEB due to the insertion of the FF in the ring, while for
the LEB this effect is negligible.

\begin{figure}[!htbp]
\centering
\includegraphics[width=0.6\textwidth]{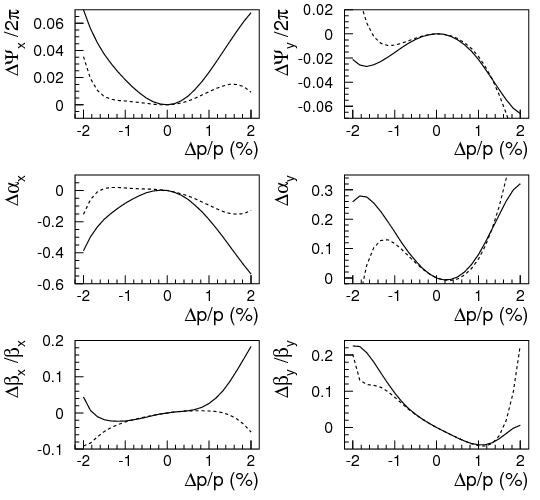}
\includegraphics[width=0.6\textwidth]{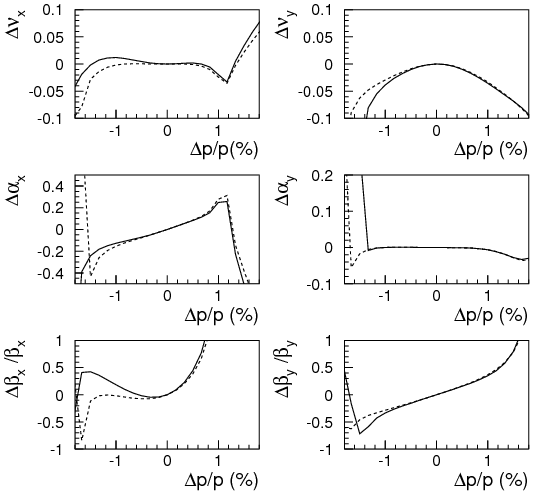}
\caption{Chromatic effects at the IP for half the IR (upper half) and for the full
HER (lower half): betatron phase advance (top plots), $\alpha$
(middle plots), and $\beta$ (bottom plots) functions as a function
of momentum deviation (between $-2\%$ and $+2\%$) in the horizontal
and the vertical plane. The dashed and solid lines show the cases in which the
octupole magnets are turned on and off, respectively. } \label{fig:chrom-ir}
\end{figure}

\paragraph{Chromaticity Correction.}
Figure~\ref{fig:chrom-ir} shows the betatron phase advance between the
end of the IR and the IP, as well as the $\alpha$ and $\beta$ function
at the IP for the off-momentum particles (bandwidth).
In this calculation, half of the lattice in the IR is
treated as a transport line.  Solid lines
show the case in which the octupole magnets are turned off. The dashed
lines show the impact of turning on the octupole magnets, which correct
the second-order chromaticity. The same distributions for the full HER
ring are shown at the bottom.

The betatron tunes for the HEB have been chosen to be $48.57$ in the
horizontal and $23.64$ in the vertical plane. The linear
chromaticity is adjusted to be close to zero using three families of the
sextupoles in the arc section and seven families in the IR.
Solid lines (dashed lines) show the case in which the octupole magnets are
turned off (turned on). If we increase number of sextupole families in the arc
section, we can correct not only the chromaticity for the betatron
tunes, but also the Twiss parameters at the IP.

%

\paragraph{Dynamic Aperture.}
The dynamic aperture is commonly defined by requiring stability in one transverse
damping time. The transverse damping time of the HER
is $32\msec$, which corresponds to 4200 turns. It is difficult to
apply either an analytic approach or perturbative methods to this problem, since the
sextupole magnets cause strong nonlinearities.
Therefore, the dynamic aperture is estimated by numerical tracking simulations.
A particle-tracking simulation was performed using SAD~\cite{bib:SAD},
an integrated code for optics design, particle tracking, machine
tuning, \etc, that has been successfully used for years at several machines
such as KEKB and KEK--ATF.
Six canonical variables, $x$, $p_x$, $y$, $p_y$, $z$, and $\delta$ are
used to describe the motion of a particle,
where $p_x$ and $p_y$ are transverse canonical momenta
normalized by the design momentum, $p_0$, and $\delta$ is the relative
momentum deviation from $p_0$.
The injected beam is round in the transverse direction.
Coherent oscillations due to injection kickers are assumed to be negligible.
Thus, initial conditions $y=x$, $p_{x0}=0$, $p_{y0}=0$,
and $z=0$ were set to evaluate the acceptance of the injected beam in these
sections. Tracking with off-momentum particles, within $\pm 2 \%$, was also
performed in order to check the momentum aperture.
As a criterion for defining stability, we required that
the maximum amplitude of the particle
orbit be within $10 \cm$ in the $x$ and $y$ coordinates during one
damping time.
The linear chromaticity was adjusted to be nearly zero by
optimizing the strength of the sextupole magnets.
Synchrotron radiation damping was turned on but quantum excitation
was turned off to avoid statistical fluctuations during tracking
simulations.

\begin{figure}[h!tb]
\centering
\includegraphics[width=0.8\textwidth]{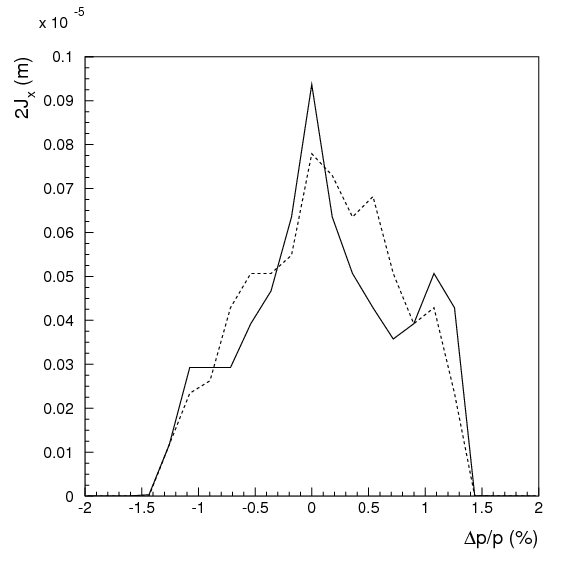}
\caption{
Dynamic aperture for HER.  Solid (dashed) lines show the case for the
octupole magnets are turned off (turned on).  The ratio of vertical to horizontal
Courant-Snyder invariant is fixed to be $J_y/J_x=0.25\%$.}

\label{fig:da-her}
\end{figure}

Figure~\ref{fig:da-her} shows the dynamic aperture obtained
from the tracking simulations of the HER, assuming the ideal lattice,
including nonlinear wigglers. No magnet errors were included in the lattice.
The transverse acceptance for the initial condition is given by
$J_{y0}/J_{x0}=0.25 \%$, where $2J_{x,y}$ is the Courant-Snyder invariant.

\subsection{Detector Solenoid Compensation}
\label{sec:comp}
One of the key issues in high luminosity colliders is the control of coupling
between horizontal and
vertical planes. With the extremely small emittances
required in the \superb\ design, the coupling correction is of primary importance.
The main source of betatron coupling is the detector solenoidal field. Hence
an efficient local correction scheme for the coupling arising from the detector
solenoid is mandatory. Other coupling sources,
such as quadrupole tilts and sextupoles misalignments, are much
smaller, and can be corrected by skew quadrupoles distributed along the ring.

A solenoid rotates the normal transverse modes of the beam by an angle defined
by the integral of the longitudinal field component along the beam orbit and
inversely proportional to the beam rigidity \br:
$$
        \thr = \frac{1}{2\; \br} \int B_z(s)\; ds\,.
$$
Compensation by two anti-solenoids placed on either side of the detector,
with opposite magnetic field to make the total integral of $B_z$ along the beam
trajectory vanish, is sufficient only if there are no quadrupoles between
the detector and the anti-solenoids. However, to achieve the very small IP beta
functions needed for high luminosity, the low-$\beta$ quadrupoles cannot be
installed too far from the IP, and correction by two anti-solenoids
is no longer sufficient.

At the Frascati \phifactory\daphne a coupling factor as low as $0.2\%$ has
been measured with single beams~\cite{bib:compensation_1}. The correction scheme
implemented at \daphne (the so called ``Rotating Frame Method'', RFM
\cite{bib:compensation_2}) is very efficient: at
$510\mev$ the effect of the \KLOE detector solenoid is a
rotation of $45^\circ $ of the normal modes. RFM, based on the general
properties of the solenoid matrix, allows the insertion of quadrupoles between the
detector solenoid (DS) and the anti-solenoids (AS) without affecting the
coupling correction. The principle is very simple: in order to cancel the
coupling created by the low-$\beta$ quadrupoles, each quadrupole immersed in
the DS magnetic field has to be tilted by the angle $\thr$ defined above,
where the integral is performed from the IP to the quadrupole location.
The exact application of the RFM implies that each quadrupole be continuously
tilted as a helix. This is of course not practical, since, apart
from technical difficulties, the rigidity of the scheme would require very
strict tolerances on the DS field and beam energy.
However it has been proven at
\daphne that small adjustments of the quadrupole tilts and the AS field
allow the measured coupling to be be corrected to a very small value in practice.

\begin{figure}[htb]
\begin{center}
\includegraphics[width=0.85\textwidth]{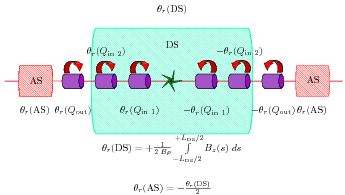}
\end{center}
\caption{\label{fig:compensation} Sketch of the RFM coupling correction scheme.}
\end{figure}

A sketch showing how the RFM scheme can be applied for the case where
there are quadrupoles inside and outside the DS is shown in
Fig.~\ref{fig:compensation}, where $\thr(DS)$ refers to the rotation
angle of the detector solenoid, $\thr(AS)$ to the rotation angle of
the anti-solenoid, $\thr(Q_i)$ to the rotation angle of the
quadrupoles inside the detector and $\thr(Q_o)$ to the rotation angle
of the quadrupoles outside the detector before the AS.

For \superb\ the first defocusing quadrupole will be common to both beams,
which means that it will be tilted by an angle \thr averaged over the values
needed for each beam. The residual coupling will be compensated by the
rotation of the following quadrupoles on the separate lines and of the nearby
skew quadrupoles, since we need just four variables to correct the coupling.
Table~\ref{table:IR_rotation_angle} provides a first estimate of the tilt
angles required for both beams, with the geometry described in
Sec.~\ref{section:InteractionRegion}.

\begin{table}[t!]
  \caption{
  \label{table:IR_rotation_angle}
  Estimate of the rotation angles inside the IR
}
\setlength{\extrarowheight}{2pt}
\vspace*{2mm}
  \centering
  \begin{tabular}{ccccccc}
    \hline
    \hline
    & \babar & Anti & QD0 & QF1L & QD0H & QF1H \\
    & ($1.5\Tesla \times 3.5 \m$) & Solenoid & centre & centre & centre & centre \\ \hline
LER & $11.3 ^\circ$ & $-5.6^\circ$ & $1.7^\circ$ & $4^\circ$ & & \\
HER & $6.4 ^\circ$ & $-3.2^\circ$ & $1^\circ$ & & $1.7^\circ$ & $2.3^\circ$ \\
\hline
\end{tabular}
\end{table}

\subsection{Dynamic Aperture}

\begin{figure}[htb]
\centering
\includegraphics[width=0.45\textwidth]{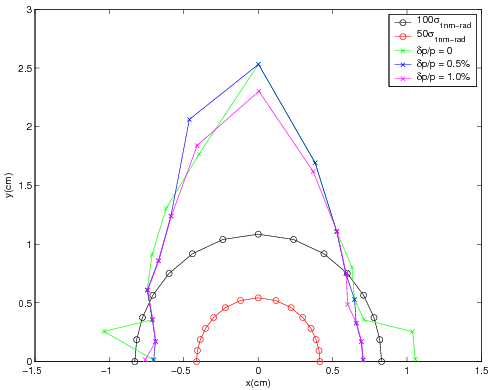}
\includegraphics[width=0.45\textwidth]{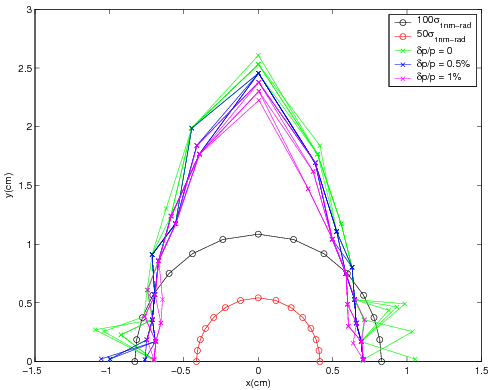}
\caption{\label{fig:Cai_1} Dynamic aperture of the ``ideal'' arc
lattice (left) and an arc lattice with five different seeds for magnetic
errors (right). The paraxial approximation is assumed. }
\end{figure}

Evaluation of dynamic aperture for the lattice with $\pi$ cells, but
without the insertion of the final focus, was first
carried out using LEGO~\cite{bib:Cai_Lego}. Particles were tracked for 1024
turns with synchrotron
oscillations, but no radiation damping or quantum excitation.
The dynamic aperture is defined as the boundary between surviving and lost
particles. 
The results of the simulation are shown in Fig.~\ref{fig:Cai_1} where
the lefthand plot is the case of the ``ideal'' lattice and the
righthand plot is the case in which lattice magnetic errors are included.
For the ideal lattice, without the insertion of the final focusing
optics, the dynamic aperture is $70 \sx$ (1\nm-rad) and 200 $\sigma_y$
(0.5\nm-rad) in the horizontal and vertical planes respectively. The aperture
does not degrade significantly either with off-momentum oscillations up to $1\%$
or with the measured multipole errors in dipole, quadrupole, and sextupole
magnets \cite{bib:Cai_mulerr}.
The magnet errors were based on those observed for the \pepii ring
magnets (see Table~\ref{table:errors_LER} for the LER and
Table~\ref{table:errors_HER} for the HER). They are parameterized
in terms of a multipole expansion:
\begin{equation}
(B_y+iB_x)/B_0(r) = \sum_{n=1} (b_n+ia_n)({x\over r}+i{y\over
r})^{n-1},
\end{equation}
where $r$ as the reference radius and $B_0$ is the main field of the magnets.

\begin{table}[!hbtp]
\caption{Multipole errors used in the study of dynamic aperture in the LER.}
\label{table:errors_LER}
\vspace*{2mm}
  \centering
\setlength{\extrarowheight}{2pt}
\begin{tabular}{ccccc} \hline\hline
\multicolumn{1}{p{5.1cm}}{\centering
\textrm{Multipole index(n)}} &
\multicolumn{1}{p{3cm}}{\centering
 \textrm{Systematic:} $b_n$} &
\multicolumn{1}{p{3cm}}{\centering
\textrm{Random:} $b_n$}\\
\hline
\multicolumn{1}{p{5cm}}{\centering
 dipole magnet: (r=0.03m) }  &                  & \\
             3               & -0.50 x  10$^{-4}$ & 1.00 x  10$^{-4}$ \\
             5               & 3.00 x  10$^{-4}$  & 1.00 x  10$^{-4}$ \\
             7               &        -         & 1.00  x  10$^{-5}$ \\
             9               &        -         & 1.00  x  10$^{-5}$ \\
\multicolumn{1}{p{5cm}}{\centering
  quadrupole: (r=0.05m) }    &                  & \\
             3               & 1.02 x  10$^{-4}$  & 4.63 x  10$^{-5}$ \\
             4               & 1.91 x  10$^{-4}$  & 8.09 x  10$^{-5}$ \\
             5               & 1.89 x  10$^{-5}$  & 8.86 x  10$^{-6}$ \\
             6               & 5.69 x  10$^{-4}$  & 2.80 x  10$^{-5}$ \\
             7               & 6.60 x  10$^{-6}$  & 3.45 x  10$^{-6}$ \\
             8               & 9.60 x  10$^{-6}$  & 5.72 x  10$^{-6}$ \\
             9               & 7.14 x  10$^{-6}$  & 3.85 x  10$^{-6}$ \\
             10              & 3.37 x  10$^{-4}$  & 5.62 x  10$^{-6}$ \\
             11              & 6.08 x  10$^{-6}$  & 3.32 x  10$^{-6}$ \\
             12              & 5.34 x  10$^{-5}$  & 6.20 x  10$^{-6}$ \\
             13              & 1.10 x  10$^{-5}$  & 6.53 x  10$^{-6}$ \\
             14              & 6.65 x  10$^{-5}$  & 8.20 x  10$^{-6}$ \\
\multicolumn{1}{p{5cm}}{\centering
  sextupole: (r=0.05652m)}   &                  & \\
             5               &        -         & 2.20 x  10$^{-3}$ \\
             7               &        -         & 1.05 x  10$^{-3}$ \\
             9               & -1.45 x  10$^{-2}$ & - \\
             15              & -1.30 x  10$^{-2}$ & - \\
\hline
\end{tabular}
\end{table}

 \begin{table}[hbt]
\caption{Multipole errors used in the study of dynamic aperture in
the HER.}
\label{table:errors_HER}
\vspace*{2mm}
  \centering
\setlength{\extrarowheight}{2pt}
\begin{tabular}{ccccc}
\hline\hline
\multicolumn{1}{p{5.1cm}}{\centering
\textrm{Multipole index(n)}} &
\multicolumn{1}{p{3cm}}{\centering
\textrm{Systematic:} $b_n$}  &
\multicolumn{1}{p{3cm}}{\centering
\textrm{Random:} $b_n$}\\
\hline
\multicolumn{1}{p{5cm}}{\centering
 dipole magnet: (r=0.03m) }  &                  & \\
             3               & 1.00 x  10$^{-5}$  & 3.20 x 10$^{-5}$ \\
             4               &        -         & 3.20 x  10$^{-5}$ \\
             5               &        -         & 6.40 x  10$^{-5}$ \\
             6               &        -         & 8.20 x  10$^{-5}$ \\
\multicolumn{1}{p{5cm}}{\centering
 quadrupole: (r=0.0449m) }   &                  & \\
             3               & 1.03 x  10$^{-3}$  & 5.60 x  10$^{-4}$ \\
             4               & 5.60 x  10$^{-4}$  & 4.50 x  10$^{-4}$ \\
             5               & 4.80 x  10$^{-4}$  & 1.90 x  10$^{-4}$ \\
             6               & 2.37 x  10$^{-3}$  & 1.70 x  10$^{-4}$ \\
             10              & -3.10 x  10$^{-3}$ & 1.80 x  10$^{-4}$ \\
             14              & -2.63 x  10$^{-3}$ & 7.00 x  10$^{-5}$ \\
\multicolumn{1}{p{5cm}}{\centering
  sextupole: (r=0.05652m)}   &                  & \\
             5               &        -         & 1.70 x  10$^{-3}$ \\
             7               &        -         & 1.80 x  10$^{-3}$ \\
             9               & -1.45 x  10$^{-2}$ & - \\
             15              & -1.30 x  10$^{-2}$ & - \\
\hline
\end{tabular}
\end{table}

\begin{figure}[tb]
\centering
\includegraphics[width=0.7\textwidth]{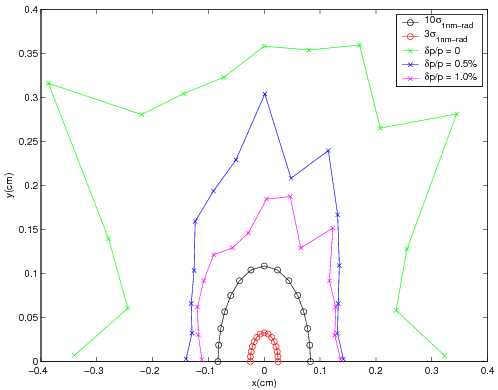}
\caption{\label{fig:Cai_3}
 Dynamic aperture of the ideal lattice, with Hamiltonian treatment up to the
 4th order.
}
\end{figure}

When the final focus is inserted in the lattice, its impact on
the dynamic aperture is rather significant, as shown in Fig.~\ref{fig:Cai_3}.
Even for the on-momentum particles, the dynamic
aperture for the ``ideal'' lattice is reduced to $30 \sigma$ in both
horizontal and vertical planes. For off-momentum particles, the
aperture reaches to nearly $15 \sigma$. From the tracking study, we
also found that the paraxial approximation is not accurate enough
for the quadrupole magnets in the final focusing system. As a
result, we used a better approximation that includes the
fourth-order momentum terms in the Hamiltonian \cite{bib:Cai_norma}.

\begin{figure}[tb]
\centering
\includegraphics[width=0.45\textwidth]{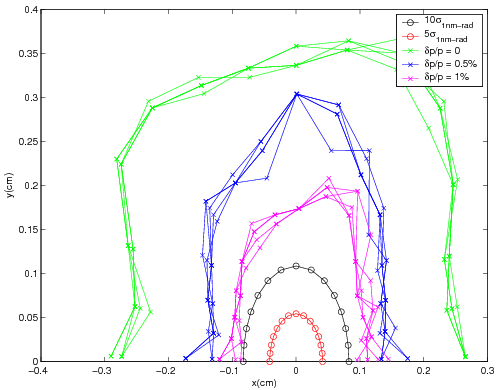}
\includegraphics[width=0.45\textwidth]{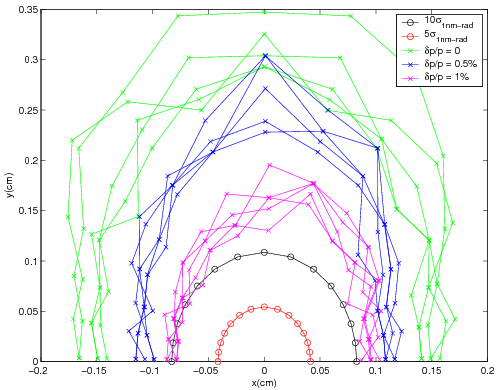}
\caption{ \label{fig:Cai_4}
 Dynamic aperture of the lattice with FF, including the magnetic
 errors in the arcs only (left) and in the arcs and FF (right). }
\end{figure}

In addition to the study of the ``ideal'' lattice, we also investigated the
effects of magnetic errors on the dynamic aperture for the FF. As
shown in the left plot of Fig.~\ref{fig:Cai_4}, the errors in the
arcs alone do not significantly reduce the dynamic aperture.
However, including the multipole errors in the FF
section (right plot), further reduces the dynamic aperture to
16$\sigma$ and  10$\sigma$ for on and off momentum particles
respectively.

A sixth-order Taylor map is extracted by tracking of truncated power
series through the lattice. The map is then canonically transformed
into the standard normal form, from which we can extract
information about tune dependencies on the amplitudes of the particles
as well as the high-order chromaticities. The result of the
normal-form analysis is shown in Table~\ref{table:Cai_1}.
It can easily be seen that some amplitude dependent terms are rather
large, for instance the crossing terms between the horizontal and vertical
planes. These large terms are largely a result of the interference among
the non-interlaced sextupoles and may well be the reason behind the
small dynamic aperture.

\begin{table}[b!]
\caption{\label{table:Cai_1}
 Coefficients of the term:
 $(2J_x)^{n_x}(2J_y)^{n_y}\delta^{n_z}$.
}
\vspace*{2mm}
\centering
\setlength{\extrarowheight}{3pt}
\begin{tabular}{c c ccc}
\hline
\hline
\multicolumn{3}{c}{index} \\
$n_x$ & $n_y$ & $n_z$ & $\nu_x$ & $\nu_y$\\
\hline
         0          &    0    & 0 &       0.57892        & 0.60492  \\
         0          &    0    & 1 &       -0.14E-2       & 0.56E-1\\
         1          &    0    & 0 &  -2347  $(m^{-1})$  & -55553 $(m^{-1})$\\
         0          &    1    & 0 &  -55553 $(m^{-1})$  & -533217 $\m^{-1})$\\
         0          &    0    & 2 &         -172         & -69\\
         1          &    0    & 1 & -583538 $(m^{-1})$  & -15155279 $(m^{-1})$\\
         0          &    1    & 1 & -15155279 $(m^{-1})$& -12292257 $(m^{-1})$\\
         0          &    0    & 3 &         3156         & -2753\\
\hline

\end{tabular}
\end{table}

\begin{figure}[htb]
\centering
\includegraphics[width=0.45\textwidth]{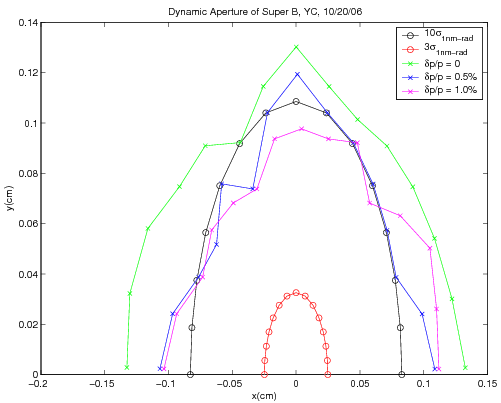}
\includegraphics[width=0.45\textwidth]{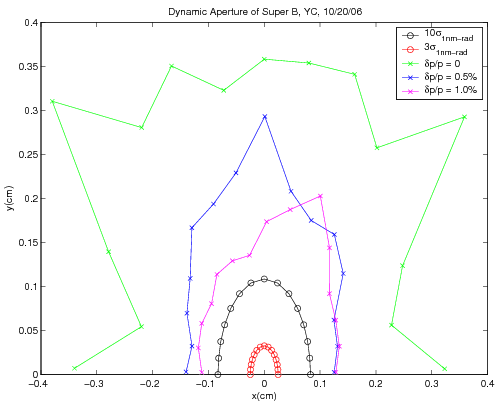}
\caption{ \label{fig:Cai_6} Dynamic aperture with the paraxial
approximation (left) and the exact Hamiltonian up to
fourth order (right), for the HER lattice.}
\end{figure}

For the ideal \superb\ lattice, including a strong focusing section, we have
carefully studied the validity of the paraxial approximation:
\begin{equation}
H = -\sqrt{(1+\delta)^2-p_x^2-p_y^2}+1+\delta
  \approx {1\over 2(1+\delta)} (p_x^2+p_y^2).
\end{equation}
As can be seen in Fig.~\ref{fig:Cai_6}, we find that the dynamic aperture obtained
using the paraxial approximation (left) is
artificially small compared to that obtained with the exact Hamiltonian (right).
The study was carried out by globally substituting the integrating
Hamiltonian for all elements including the quadrupoles. We also find
that the reduction in dynamic aperture could be mostly restored
by adding the fourth-order term
$(p_x^2+p_y^2)^2/[8(1+\delta)]$ into the paraxial approximation as previously
seen in Fig.~\ref{fig:Cai_3}. Clearly, the difference between
Fig.~\ref{fig:Cai_3} and Fig.~\ref{fig:Cai_4} is rather small.

\begin{figure}[hbt]
\centering
\includegraphics[width=0.7\textwidth]{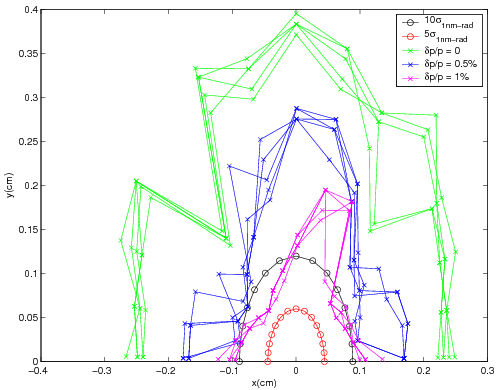}
\caption{\label{fig:Cai_8}
  Dynamic aperture including all magnetic
  errors in the LER lattice.
}
\end{figure}

Simulations have also been  performed for the LER, with similar results
to those for the HER. In particular, the dynamic
aperture for the design lattice with all the multipole errors
is shown in Fig.~\ref{fig:Cai_8}.
We conclude that the dynamic aperture for the LER is just slightly smaller
than that for the HER (Fig.~\ref{fig:Cai_4}).

The crab waist correction of the beam crossing angle at the IP requires the 
use of sextupole magnets. Since these sextupoles are near the IP, they can 
have a side effect of reducing the dynamic aperture for the rings because of their 
high gradients and nonlinearity. The dynamic aperture calculations discussed above 
have been performed without crab sextupoles in the final 
focus lattice. A study of the final focus design that minimizing these 
effects is ongoing; preliminary results suggest a 
reduction of dynamic aperture by less than a factor of two. Further work
on the optimization of locations and phase advacnes for the crab sextupoles
will be needed, together with optimization of the working point for the rings.
It should be noted that tune scans from beam-beam simulations 
without crab sextupoles show that the baseline luminosity of \tenTo36 is still 
achievable. However, the working point in tune space must be chosen closer to the half-integer and 
the beam footprint is reduced, leaving less freedom to change the working point. 

Based on our simulation results, it is quite clear that the dynamic
aperture is limited by the final focusing system in the
\superb\ lattices. Although the dynamic aperture is small, it is more
than adequate for the stored beam, which is extremely small.
The acceptance for a larger injected beam remains a problem to be
studied. We may need to optimize further the locations of octupole magnets
insert more higher-order multipole correctors near the final focusing
quadrupoles, or simply scan the betatron tunes to find a better
operating point.


 
\section{Imperfections and Errors}
\label{section:ImperfectionsAndErrors}


\subsection{Tolerances, Vibrations and Stability}

The movement of elements in the magnetic lattice of the \superb\ accelerator will affect 
the equilibrium emittance of the beam. The horizontal emittance, and particularly the 
vertical emittance, are quite small and will require special care to achieve. We will first discuss 
errors in the rings outside the interaction region. The roll stability of quadrupole 
magnets, and the horizontal and vertical offset stability of the quadrupole magnets, are the 
most important sources of  errors. There are several recent studies for the next generation 
low-emittance storage rings that have looked 
extensively at this stability issue. PETRA-III, NSLS-II, and the ILC Damping Rings all 
have lattice specifications that are similar to the arc and straight section magnets for 
\superb. The design reports of these accelerators discuss these tolerances~\cite{biblio:alignment}. 
The total effect is estimated by including magnet errors 
around the complete ring with the appropriate betatron and phase weighting. The 
amplitude of fast magnet motion due to normal ground motion has only
a small impact on the emittance. However, slow magnet motion can lead to an increased emmittance.
according to ``$ATL$'' models, which incorporate temporal and spatial correlations in 
reasonable agreement with observations. In the model, $<y^2> = ATL$, where $y$ is the transverse offset, $A$ is a constant 
about $4\times 10^{-6}\mum^2$/m/s, $T$ is the time and $L$ is the separation distance between points 
of interest, for example two adjacent quadrupoles. As a result, orbital steering corrections at the
5--10\mum level are required over timesclaes of a few minutes in order to 
keep the vertical emittance within specifications. BPM resolutions of 
order 1\mum are also needed. These studies show that in order to be acceptable, ring quadrupoles must
have three-sigma truncated {\it rms} 
misalignments of 100\mum and {\it rms} rolls of 100$\,\mu$radians. These 
alignment tolerances are possible but challenging. 

The final quadrupole doublets adjacent to the IP 
have strong fields and the beams have large beta functions. Vibration tolerances for 
these magnets are especially tight. Typically, there is a one-to-one correspondence
between the size and direction of vertical motion by final doublet quadrupole magnets 
and motion of the beam at the IP. 
The vertical beam size at the IR is 20--35\nm. Since we need 
keep the beam in collision with tolerances at the 0.1 sigma level or less, quadrupole magnets must be 
be kept stable to 2--4\nm. The vibration of large objects such as quadrupoles 
depends on the design of the mechanical supports and the local ground excitation. 
Typical motion is about 50\nm in the 50\Hz range. Since there are only a few of 
these magnets, active vibration controls in the mechanical supports can be employed to 
bring vibrations within specification. An active vibration suppression by a factor 
of 10--20 is within industry standards.

The bunches will collide in \superb\ at 476\MHz. To maintain luminosity, it 
is important to keep the bunches transversely centered on one another. Feedback systems using 
the position monitors and the luminosity signal will be required. Arc BPMs, with resolutions 
of about 1\mum, can be used to bring the beams close to 
collision. In the IP region, specially constructed BPMs with resolutions of
0.5\mum will be needed. The position monitors in the interaction region quadrupoles are located in a 
very high beta region, which also increases their sensitivity. These BPMs will be used to bring the 
beams very close to collision. A 476\MHz luminosity signal can be extracted from beamstrahlung
under these conditions. Very rapid horizontal and vertical 
position feedback systems $(\sim 100\Hz)$ based on this signal will keep the beams in collision. 
\pepii and KEKB both already use such a fast luminosity feedback very effectively to keep the beams in 
collision.

\subsection[Coupling and Dispersion Tuning]
{Coupling and Dispersion Tuning for Low Vertical Emittance Rings}
A variety of collective effects can increase the vertical beam
emittance at high currents; however, in the low-current limit, which we
consider in this section, three effects dominate contributions 
to the vertical emittance.  The non-zero vertical opening angle
of the synchrotron radiation in dipole magnetic fields excites
vertical betatron motion of particles as they ``recoil'' from photon
emission.  Vertical dispersion from steering errors generates vertical
emittance, in the same way that horizontal dispersion from the bending
magnets determines the horizontal emittance of the beam.  Betatron
coupling from skew quadrupole errors leads to a transfer of horizontal
betatron motion (and hence horizontal emittance) into the vertical
plane.  The first of these effects, the non-zero vertical opening
angle of the synchrotron radiation, places a fundamental lower limit
on the vertical emittance that can be achieved in any storage ring;
this can be calculated for a given lattice design.  In most rings,
including the \superb\ rings, the lower limit is a fraction of a
picometer, and is significantly smaller than the specified vertical
emittance.  The effects of vertical dispersion and betatron coupling,
which arise from magnet alignment and field errors, invariably dominate 
the vertical emittance in an operating storage ring;
reducing the vertical emittance in the \superb\ rings to the value
required to achieve the specified luminosity will require highly
precise initial alignment of the machine, followed by careful tuning
and error correction.

The lowest vertical emittance achieved in an operating storage ring
is $4.5\picom$ in the KEK Accelerator Test Facility (ATF)
\cite{biblio:impere_1}; the \superb\ rings are specified to operate
at $4\picom$ in the first stage, so the alignment and tuning issues
require some attention. We note however that the Damping Rings for
the \ilc are specified to operate at $2\picom$, and that
significant effort has already been devoted to understanding the alignment
and tuning requirements in these systems \cite{biblio:impere_2}.
While an experimental demonstration is still required, theoretical
and simulation studies suggest that $2\picom$ vertical emittance is
a realistic goal for the \ilc damping rings.  The question we then
must consider is how the likely difficulty of achieving $2\picom$
in the \superb\ rings (and the more stringent constrint for the 
\superb\ upgrade) compares with the
difficulty of achieving $2\picom$ in the \ilc damping rings.

Broadly speaking, we may characterize the behavior of the
vertical emittance in a given lattice by calculating the vertical
emittance generated by a variety of magnet alignment errors.  The
principal errors to consider, in this context, are vertical sextupole
misalignments and rotations or tilts of quadrupoles around the beam
axis, both of which generate unwanted skew quadrupole components.
Also relevant is the closed orbit distortion generated by vertical
misalignments of the quadrupoles, which results in vertical beam
offsets in the sextupoles with the same consequences as vertical
misalignments of the sextupoles themselves.  Estimates of the
sensitivity of a lattice to these errors can be made using analytical
formulae \cite{biblio:impere_2} involving the magnet strengths and lattice functions; it
is usually found that simulations support the results of these
analytical calculations.

Table~\ref{table:emittance_comparison} shows the results of
analytical estimates of the sensitivity of the \superb\ rings to
various alignment errors, compared to the ATF and the baseline
design for the \ilc damping rings.  We emphasize that the
results given in Table~\vref{table:emittance_comparison} are statistical,
in that they represent the mean over many different sets of random
errors: the spread in the response of a lattice to a given set of
alignment errors is large, usually $100\%$ of the mean.

\begin{table}[h]
\caption{\label{table:emittance_comparison}
Specified vertical emittance in the \superb\ rings, the ATF, and the \ilc
 Damping Rings, with sensitivity indicators.
 }
\vspace*{2mm}
\centering
\setlength{\extrarowheight}{2pt}
\begin{tabular}{lcccc}
\hline
\hline
 & \superb\ & \superb\ & \ilc\ & KEK \\[-2mm]
 & LER & HER & DRs & ATF \\ \hline
Vertical emittance (pm)                  & 4   & 4   & 2   & 4.5 \\
Orbit amplification factor               & 46  & 44  & 32  & 21  \\
Quadrupole jitter sensitivity (nm)       & 209 & 217 & 221 & 227 \\
Sextupole alignment sensitivity ($\mu$m) & 95  & 87  & 70  & 50  \\
Quadrupole tilt sensitivity ($\mu$rad)   & 166 & 183 & 79  & 800 \\
\hline
\end{tabular}
\end{table}

The sensitivity indicators given in Table~\vref{table:emittance_comparison}
should be interpreted as follows:
\begin{itemize}
\item The orbit amplification factor is the mean {\it rms} vertical orbit
distortion divided by the {\it rms} vertical quadrupole misalignment;
\item The quadrupole jitter sensitivity is the mean {\it rms} quadrupole
misalignment required to generate an {\it rms} closed orbit distortion equal
to the vertical beam size at the specified vertical emittance;
\item The sextupole alignment sensitivity is the mean {\it rms} sextupole
vertical misalignment required, in an otherwise perfect lattice, to
generate the specified vertical emittance; and
\item The quadrupole tilt sensitivity is the mean {\it rms} quadrupole tilt
error required, in an otherwise perfect lattice, to generate the
specified vertical emittance.  Smaller values therefore indicate a
\emph{greater} sensitivity to quadrupole tilts, and larger values
are more desirable.
\end{itemize}
It is important to note that these sensitivity
indicators should not be taken as alignment tolerances: they simply
indicate the mean response of the beam to errors of a given magnitude.
Generally, alignment of the magnets will be significantly worse than
the indicated sensitivities, but coupling correction and tuning
techniques can then be used to achieve the specified vertical
emittance.  The sensitivity values that we calculate may be taken to
indicate the difficulty of implementing the tuning successfully, and
the frequency with which tuning might be required to maintain the
specified emittance.

With the exception of the quadrupole tilts, the values given in
Table~\vref{table:emittance_comparison} indicate that tuning \superb\
to achieve $4\picom$ and $2\picom$ in the upgrade stage should not
be significantly more difficult than tuning ATF to achieve the
already-demonstrated emittance of $4.5\picom$, or tuning the \ilc
damping rings to achieve the specified $2\picom$ vertical
emittance. However, it is important to note that for \superb, the
strong sextupoles and quadrupoles in the final focus region were
omitted from the calculations: the beam orbit and emittance tend to
be particularly sensitive to motion of these elements, which will
therefore need special consideration.

Finally, we comment that a range of tuning techniques and algorithms
have been tested in simulation and experiment on the ATF and on other
electron storage rings, including those in colliders and
third-generation synchrotron light sources.  One procedure applied to
the ATF is described in the references \cite{biblio:impere_1,
biblio:impere_2,biblio:impere_3}; studies to determine
an optimum tuning procedure for storage rings required to operate
routinely with emittance of around $2\picom$ are in progress
\cite{biblio:impere_4}.  For
\superb, it is expected that further development of the lattices could
reduce the sensitivity to alignment errors.  Detailed studies,
including simulations, are needed to characterize fully the
sensitivities and determine specifications for the magnet support
scheme, survey and alignment tolerances, and diagnostics and
instrumentation performance.

\subsection{Final Focus Tuning}

The final focus ``tuning knobs'' are adjustments of magnet field and
alignment to compensate the linear and non-linear beam aberrations
and beam size growth at the IP caused by ``slow'' field or tilt errors
in the FF quadrupoles. Sextupoles, octupoles and decapoles can be
used in the tuning knobs. Alternatively, the normal and skew
quadrupole correcting coils can be considered, which have the
advantage of not creating second-order orbit distortions. This method has been
studied for the FF systems of the NLC, ILC, ATF2 (see for example
ref.~\cite{bib:yuri}) since all these machines employ the same design principles.
A short summary is provided here.

\begin{figure}[h!tb]
\centering
\includegraphics[width=\textwidth]{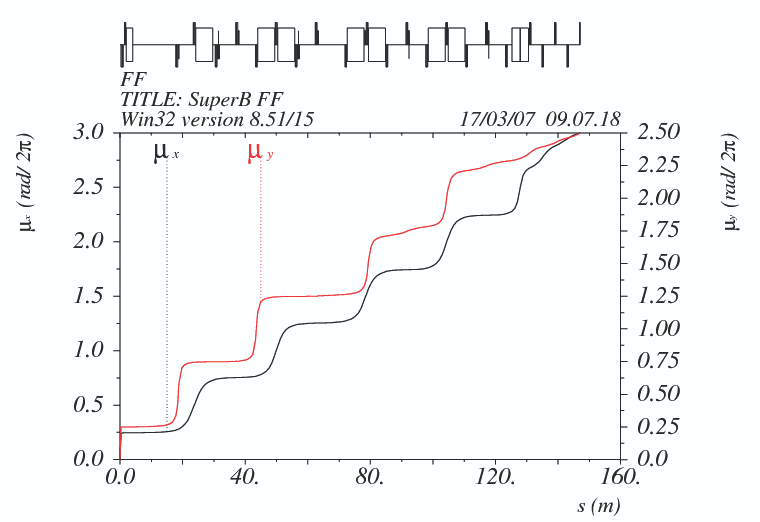}
\caption{\label{fig:FF_phase} Phase advances in the final focus (IP
is at $s=0$).}

\end{figure}

Very large peaks produce a characteristic $90^\circ$-to-IP phase advance
at most of the FF magnets (see Fig.~\ref{fig:FF_phase}). This
$90^\circ$ phase advance reduces the number of efficient
tuning knobs, but also helps in correcting the FF errors, since the FF
correctors are effectively at the same phase as the FF errors.
However, this assumes that
the out-of-$90^\circ$ phase aberrations propagating to the
IP from the upstream optics can be corrected prior to the FF.

A number of linear and non-linear tuning knobs can be
implemented.
Examples of orthogonal linear knobs are:
\begin{itemize}
  \item Horizontal offset in a sextupole to correct the horizontal dispersion
at IP and the longitudinal offset of $\betxs$, $\betys$ waists (3
knobs); and
  \item Vertical offset in a sextupole to correct the vertical IP
dispersion and the dominant $(x,y)$ coupling term ($R_{32}$) at the IP (2
knobs).
\end{itemize}

Adjustment of field and tilt angles of the FF sextupoles can be used
to correct second-order optical aberrations at the IP, which is
needed as well. Additionally, adjustment of the octupole and
decapole fields can be used for the third- and fourth-order corrections.
These magnets create many high-order terms;
``absolute'' orthogonality between different terms is therefore typically not
possible to achieve using a limited number of correctors. Hence the
goal is to create approximately orthogonal knobs that excite one
dominant term per knob, while keeping the other terms small. The sextupole
knobs can be calculated with second-order matrix optimization using
MAD code \cite{bib:MAD}. A simple octupole knob can correct the
octupole field error, and two decapole knobs can correct the
decapole field error and the field difference between the two
decapoles. The fixed $90^\circ$ phase to the IP limits the number of
matrix terms (knobs) which can be created. To improve the
orthogonality of knobs based on sextupole fields, sextupoles can
also be added to the lattice. The effect of the knob is equivalent
to exciting the corresponding matrix term at the IP, for example:
\begin{align}
\Delta x^{*} &= T_{162} \cdot x^{*\prime} \cdot \delta,\\
\Delta x^{*} &= T_{166} \cdot \delta^{2},\\
\Delta y^{*} &= T_{342} \cdot x^{*\prime} \cdot y^{*\prime},\\
\Delta y^{*} &= U_{3422}\cdot (x^{*\prime}){^2} \cdot y^{*\prime}\,.
\end{align}
The actual effect of a matrix term depends, of course, on the IP beam
parameters.

\begin{figure}[h!tb]
\centering
\includegraphics[width=\textwidth]{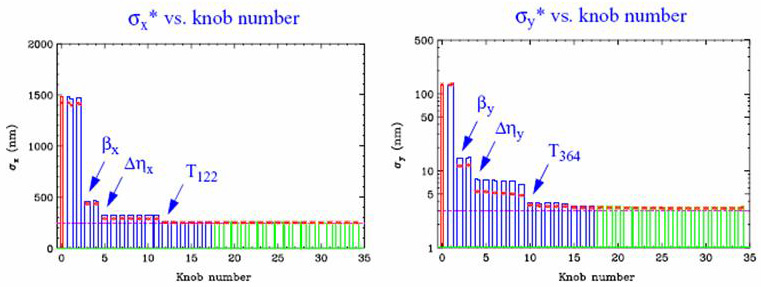}
\caption{\label{fig:FF_sim} Phase advances in the final focus (IP is
at s=0).}
\end{figure}

The effectiveness of these knobs depends on the set of the random
machine errors, which cause the IP aberrations. Tracking of many
sets of errors would show which aberrations are the largest at the
IP, and therefore which correcting knobs are most important. An
example of the iterative procedure for FF tuning can be found in
ref.~\cite{bib:yuri}. An ideal initial beam distribution is first
generated with a large number of particles, and tracking is done without
magnet errors, thereby characterizing the ideal beam at the IP.
Random field and alignment errors are then assigned to magnets and
BPMs, and tracking with the errors before any correction and
measurement of the beam at IP is performed. The initial orbit is
corrected using the corrector quadrupole $x$, $y$ offsets, and the
a response matrix between the correctors and BPMs, and
tracking performed again. The IP tuning correction obtained by
applying the tuning knobs one-by-one with the orbit correction after
each knob is determined, followed by tracking and measuring again.
In the tuning loop, the linear knobs are applied first, then the
second-order vertical and horizontal knobs. Finally, octupole and
decapole knobs can be applied. This procedure can be iterated as
needed, and various combinations of {\it rms} errors must be
studied. An example of the efficiency of this method for the NLC
final focus tuning simulation is shown in Fig.~\ref{fig:FF_sim}.

A beam-based alignment (BBA) procedure is required to minimize
misalignments and improve the effectiveness of the tuning. The orbit
correction in the simulation may need improvement. Tracking with
various levels of misalignment will demonstrate the level of
residual alignment error required for good tuning.

\clearpage

\section{Intensity Dependent Effects}
\label{section:IntensityDependenteEfx}
\subsection{Beam-Beam Interactions }
Beam-beam interactions are the most important limitation to
luminosity performance.
They depend on a number of different beam parameters and running
conditions; their impact on collider performance can only be calculated using
computer simulations, which are also used to choose optimum operating
conditions.

Beam-beam simulations for \superb\ started with a beam
parameters set similar to that of the \ilc Damping Rings (see
Table~\ref{tab:BBParam}).  The simulations have been carried out with two
weak-strong codes, BBC~\cite{bib:bb_BBC} and LIFETRAC~\cite{bib:bb_Lifetrac},
that have been successfully used for beam-beam
collision studies for the KEK $B$-Factory~\cite{bib:bb_KEKB} and \daphne
\cite{bib:bb_Dafne}. In the following we will summarize the steps
that, starting from a very different parameter set, have led to the
final choices.

\begin{table}[t!bh]
\caption{IP Parameters for early \ilc-like design and current \superb\ concept. For 
the \superb\ design, the first entry is for the LER and the bracketed number of for the HER.}
\label{tab:BBParam}
\vspace*{2mm}
\centering
\setlength{\extrarowheight}{2pt}
\begin{tabular}{lcc} \hline\hline
       Parameter                         & Early \ilc-like     & \superb\ \\ \hline
     Horizontal emittance \epsx (nm-rad) & 0.8                 & 1.6  \\
      Vertical emittance \epsy (pm-rad)  & 2                   & 4    \\
        IP horizontal \betx (mm)         & 9                   & 20   \\
         IP vertical \bety (mm)          & 0.08                & 0.30 \\
      Horizontal beam size \sx ($\mu$m)  & 2.67                & 5.66 \\
       Vertical beam size \sy (nm)       & 12.6                & 35   \\
          Bunch length \sz (mm)          & 6                   & 6    \\
    Momentum spread $\sigma_{e}$ ($\times 10^{-4}$ & 10        & 8.4 (9.0) \\
        Crossing angle \thx (mrad)       & $2\times 25$        & $2\times 17$\\
   No. particles/bunch $N_{part}$ ($\times 10^{10}$) & 2.5      & 6.2 (3.5) \\
     No. bunches $N{_{bunch}}$           & 6000                & 1733 \\
           Circumference (m)             & 3000                & 2250 \\
Longitudinal damping time $\tau{_s}$ (ms)& 10                  & 16   \\
          RF frequency (MHz)             & 600                 & 476  \\

\hline
\end{tabular}
\end{table}

\begin{figure}[htb]
\centering
\includegraphics[width=\textwidth]{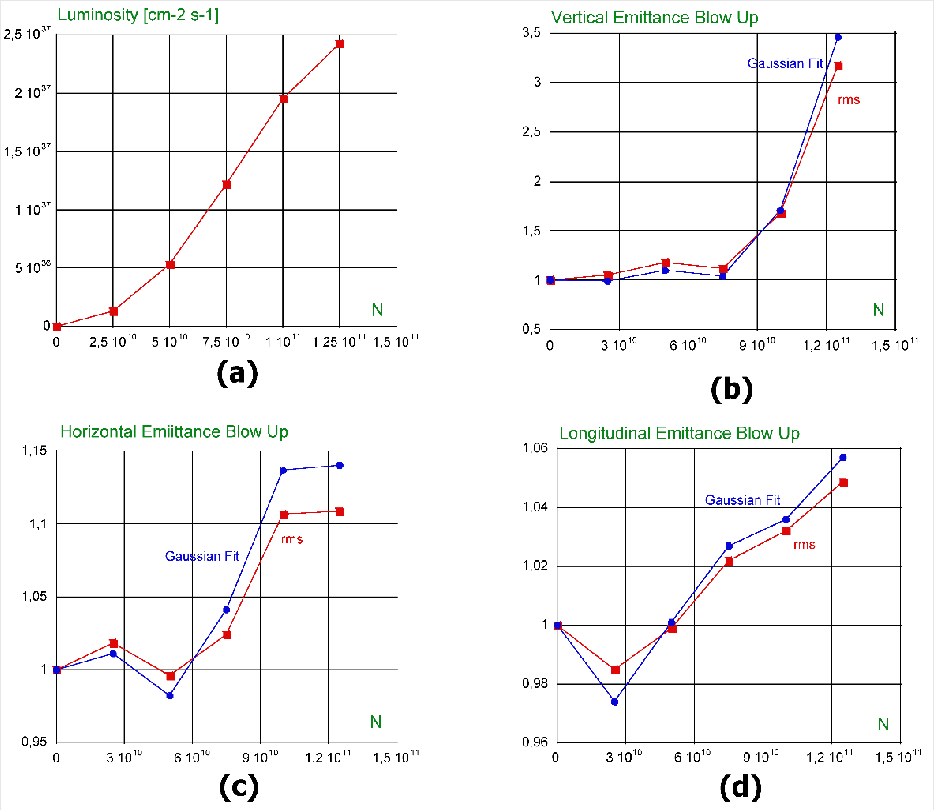}
\caption{\label{fig:BB_Fig_1}
Simulation results for luminosity (a), and blowup of vertical (b), horizontal (c) and longitudinal (d) emittance
as a function of the single bunch intensity.}
\end{figure}

Results from simulations
with these parameters are summarized in
Fig.~\ref{fig:BB_Fig_1}, where the luminosity (a) and
the blowups of vertical emittance (b), horizontal emittance (c) and
longitudinal emittance (d) are shown as a function of the number of
particles per bunch. The luminosity has been calculated assuming 6000
colliding bunches. As can be seen, the luminosity grows quadratically
with the bunch current, exceeding a luminosity of \tenTo37 with no
blowup for single bunch populations up to $7.5 \times 10^{10}$. This is
possible due to the crabbed waist scheme, which allows
for a decrease in the vertical beta function \betys at the IP and
an increase in the vertical tune shift \xiy by a factor of 2--3 with respect
to that seen in ordinary head-on collisions.

The design luminosity of \tenTo36 is achieved with only 2--$2.5 \times
10^{10}$ particles per bunch. This corresponds to an average
beam current of $2.4\amp$, which is a value close to the best results
obtained so far at particle factories. We consider $2.5 \times 10^{10}$
particles/bunch to be a conservative choice for the nominal design value.
According to numerical simulations, the design beam-beam tune shift is well
below the maximum achievable value. We have used this safety
margin to significantly relax and optimize many critical parameters, including
damping time, crossing angle, number of bunches, bunch length,
bunch current, emittances, beta functions and coupling, while maintaining
a design luminosity of \tenTo36.  We should stress that
 the condition imposed by the crabbed waist
scheme has been always satisfied during the optimization process.

In order to explain our
optimization strategy we first discuss how beam-beam
interactions are affected by the different parameters.

\begin{figure}[htb]
\centering
\includegraphics[width=0.9\textwidth]{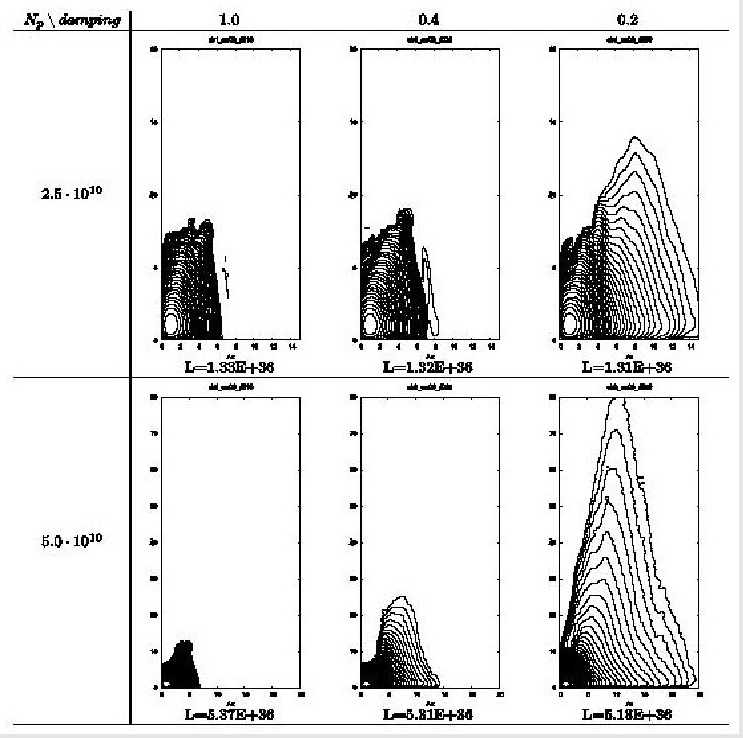}
\caption{\label{fig:BB_Fig_2}
 Beam-beam non-gaussian tails in the space of normalized betatron amplitudes for
 three values of the damping time: initial 10 msec, 25 msec and 50 msec
 (see columns 1, 2 and 3, respectively). }
\end{figure}

\paragraph{Damping time.} Damping time and quantum noise fluctuations
play important roles in beam dynamics. They affect the instability
thresholds for high current operations and the maximum achievable
beam-beam tune shift parameter, the resulting luminosity and beam
lifetime. In the \superb\ rings, the damping time is shortened by
means of wiggler magnets. Since the wigglers are a
non-negligible contribution to the overall machine cost, we have investigated
with numerical simulations the degree to which an
increase in the damping time via a reduction in the number of
wigglers affects the luminosity and beam-beam induced tails.
Figure~\ref{fig:BB_Fig_2} shows the beam-beam non-gaussian tails in
the space of normalized betatron amplitudes for three values of the
damping time: $10$, $25$ and $50\ms$ (columns 1, 2 and 3,
respectively). The simulations have been carried out for $2.5 \times
10^{10}$ (upper plots) and $5.0 \times 10^{10}$ (lower plots)
particles/bunch. As can be seen, a damping time increase by a factor
of 2.5 does not lead to any substantial luminosity degradation.
However, in order to be conservative, we have chosen a longitudinal
damping time of $16\ms$, similar to that for the \pepii $B$-Factory.

\begin{figure}[htb]
\centering
\includegraphics[width=\textwidth]{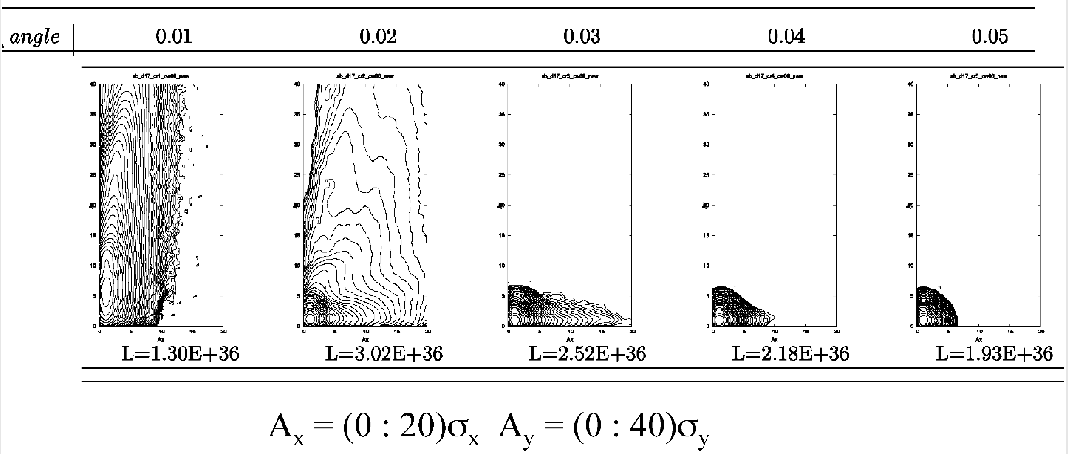}
\caption{\label{fig:BB_Fig_3}
Distribution contour plots as a function of the crossing angle, in radians,
and \bety.}
\end{figure}

\begin{figure}[htb]
\centering
\includegraphics[width=0.7\textwidth]{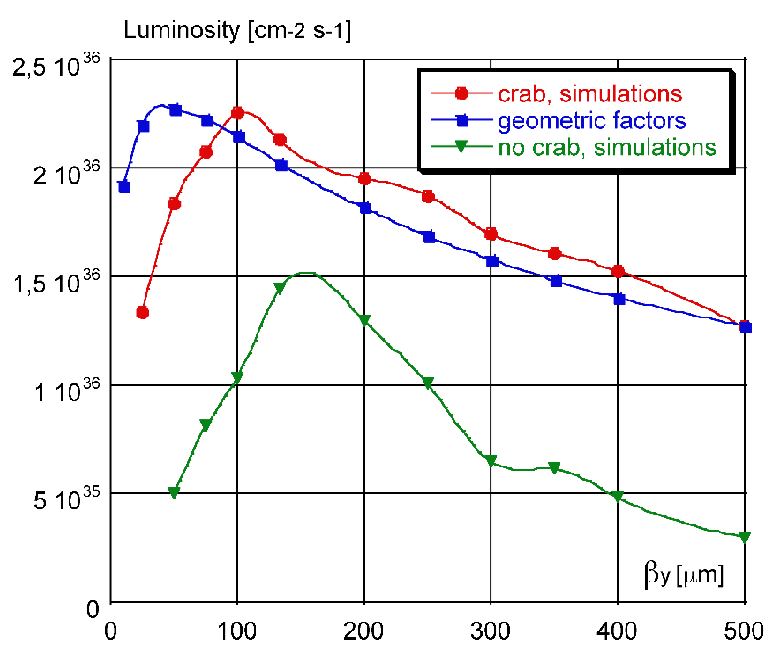}
\caption{\label{fig:BB_Fig_4}
Luminosity as a function of \bety, with \betx fixed at 20\mm.}
\end{figure}

\paragraph{Crossing angle and vertical beta function.} Having designed
a safety margin in
the vertical tune shift, it is possible to increase the \bety
function at the IP. This makes the IR design easier, by reducing the collider
chromaticity and simplifying the dynamic aperture optimization. However, the crossing angle must be proportionally reduced in order
to keep the \bety function comparable to the overlap area of the
colliding bunches: clearly an optimum should exist.
The tune shift and the luminosity grow with increasing \bety and decreasing
horizontal crossing angle \thx. However, at some point, the tune shift reaches
its limit, and beam blowup and tail growth occur. In the opposite direction, with
lower \bety and higher \thx, the luminosity drops due to geometric factors,
without beam blowup.
Such a situation can be seen in Fig.~\ref{fig:BB_Fig_3} where
contour plots are shown as a function of the crossing angle and \bety,
respectively. The optimum is at about $2\thx = 30 \mrad$ and
$\bety = 133 \mum$. Due to IR design requirements, the final value of the crossing
angle has been chosen to be only slightly different, $2\thx = 34 \mrad$.
Nevertheless, the \bety function can be further increased, at the
expense of a slight luminosity reduction. Indeed, as can be seen in
Fig.~\ref{fig:BB_Fig_4}, an increase from 133 to $200\mum$ for \bety
would lead to just a 10\% luminosity reduction.

\begin{figure}[htbp]
\vspace*{5mm}
\centering
\includegraphics[width=0.7\textwidth]{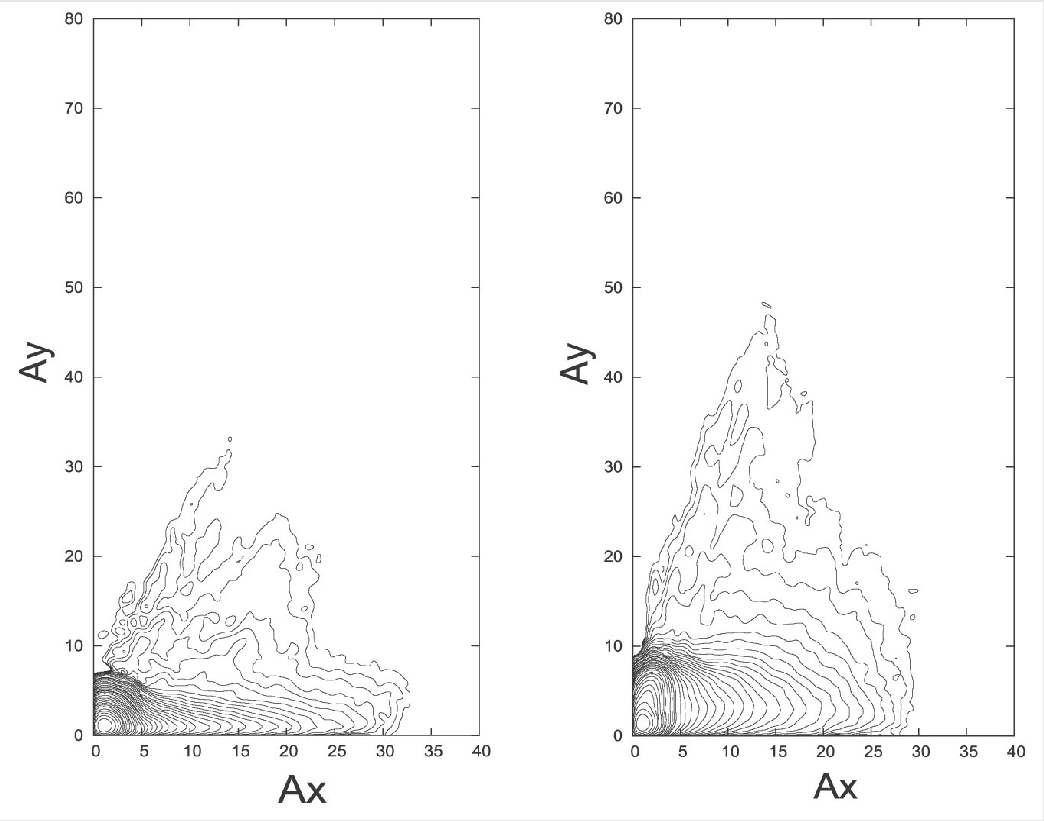}
\caption{\label{fig:BB_Fig_5}
Growth of beam tails for relaxed coupling factor (left), and
relaxed emittance and \bety (right).}
\end{figure}

\paragraph{Betatron coupling.} Betatron coupling can also be relaxed. This can be
very important at the initial stage of the collider operation. If we
assume that we can relax the coupling factor, by as much as a factor
of four, for example, then the luminosity can be recovered by using
half as many bunches with twice the single bunch charge. In this
case, the beam current will remain the same and we can obtain the
design luminosity by exploiting the fact that the luminosity grows
quadratically with the bunch intensity before the tune shift limit
is reached. The beam tails are also limited, as can be seen in
Fig.~\ref{fig:BB_Fig_5} (top plot).

\paragraph{Emittances.} The same strategy can be used for the emittances. By doubling
the bunch intensity and reducing the number of bunches by a factor of two, the
same luminosity can be obtained with emittances two-times higher and a \bety
function increased by a factor of $\sqrt{2}$. As shown in Fig.~\ref{fig:BB_Fig_5}
(bottom plot) the bunch tails are limited in this scenario as well.

\begin{figure}[htpb]
\centering
\includegraphics[width=0.9\textwidth]{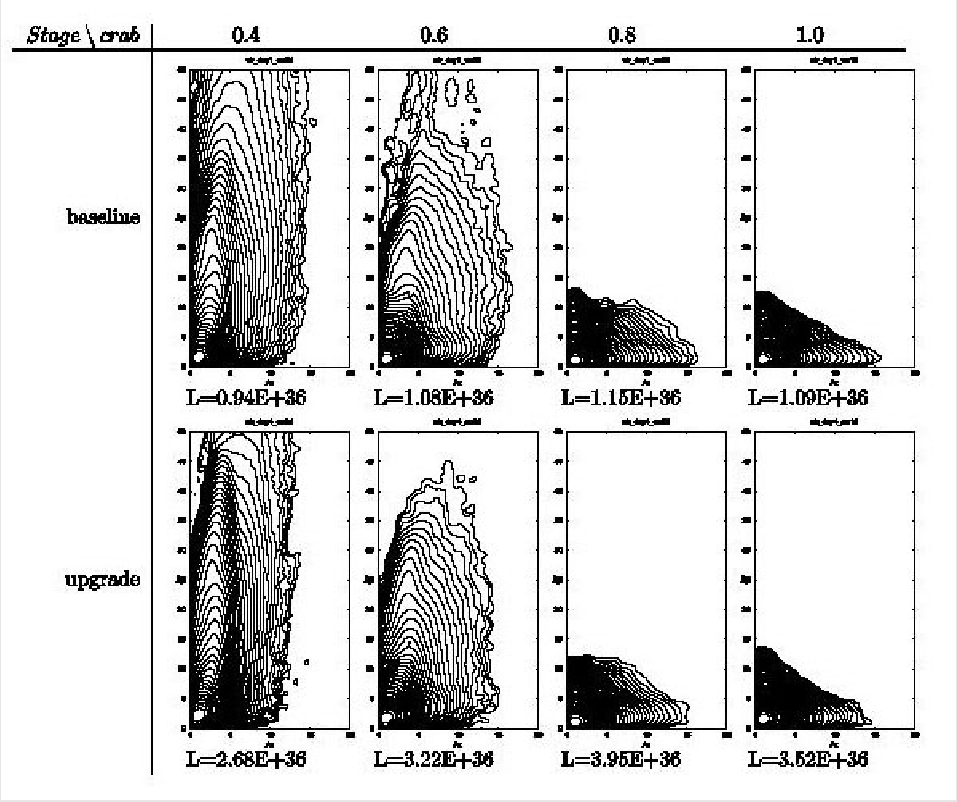}
\caption{\label{fig:BB_Fig_6}
Bunch distribution contour plots for the ``Baseline''
and ``Upgrade'' parameters for different strengths of the crabbing sextupole. }
\end{figure}

After several iterations, beam parameters for the ``baseline'' and
``upgrade'' stages have been chosen that give a luminosity of
\tenTo36 and $3.4 \times \tenTo36$ respectively. The corresponding
parameter sets can be found in Table~\ref{tab:params}.
Figure~\ref{fig:BB_Fig_6} shows bunch distribution contour plots for
the baseline and upgrade parameters for different strengths
of the crabbing sextupoles. Based on simulations, no lifetime
problems due to electromagnetic beam-beam interaction are expected
for the two design sets. The maximum luminosity and shortest tails
are provided by operating the sextupoles at 80\% of their nominal
crabbing strength.

\begin{figure}[htb]
\centering
\includegraphics[width=0.9\textwidth]{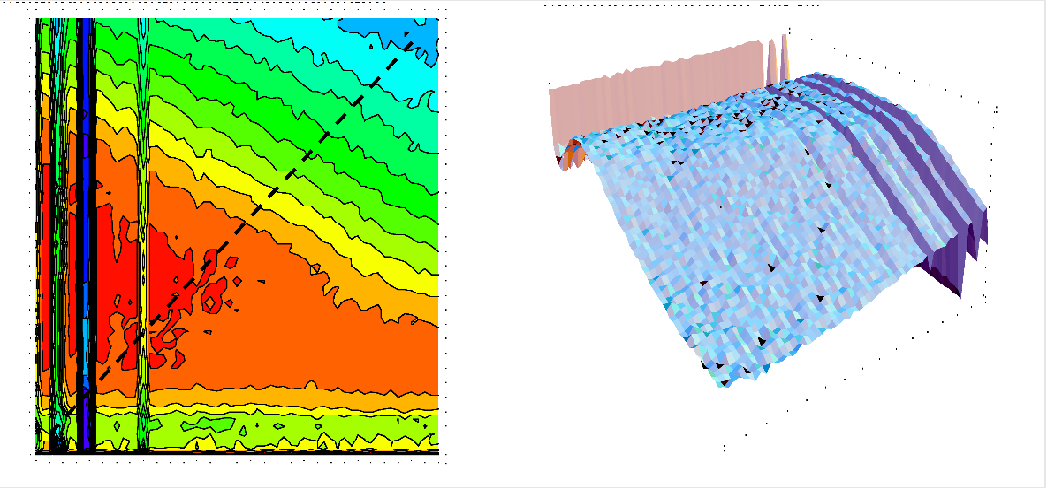}
\caption{\label{fig:BB_Fig_7}
2D and 3D surface luminosity plots. The red color on the contour plot corresponds
to the highest luminosity while the blue is the lowest.
Each contour line corresponds to a 10\% luminosity reduction.
Here $\lum_{\rm min} = 2.25 \times \tenTo34$, $\lum_{\rm max} = 1.21 \times \tenTo36$.}
\end{figure}

All the simulations have been performed for one of the best working points.
In order to define how large the ``safe'' tune area is, a luminosity
tune scan has been performed for tunes above the half integer, which is typical for
the operating $B$-Factories. The 2D and 3D surface plots for the scans are
shown in Fig.~\ref{fig:BB_Fig_7}, where red corresponds to the
highest luminosity, and blue the lowest.
Individual contours differ by a 10\% in luminosity.
The maximum luminosity found inside the scanned area is
$\lum_{\rm max} = 1.21 \times \tenTo36$, while $\lum_{\rm min} = 2.25 \times \tenTo34$.
We conclude that the design luminosity can be achieved over
a wide tune area. However, for the final choice of the operational
working point, one also needs to take into account the main coupling resonance
(dashed line), which can affect the luminosity performance; the working point
should be chosen quite far from this resonance.

\afterpage{\clearpage}
\subsection{Lifetimes and Backgrounds}

\subsubsection{Luminosity lifetime}
An important contribution to beam lifetime is the loss of particles
due to scattering at the interaction point at a rate proportional
to the machine luminosity.

In the following, we consider the loss of particles due to QED processes
$\epem \to \epem \gamma$
(radiative Bhabha) and $\epem \to \epem$ (elastic Bhabha) that scatter
beam particles outside the ring acceptance.
The loss rate for the ring $i$ depends on luminosity \lum and on cross
section $\sigma = \sigma_{\mathrm{rad.}} + \sigma_{\mathrm{el.}}$ according to
$$
\frac{dN_i}{dt} = - \sigma_i\, \lum \,.
$$
Assuming \lum constant, the following approximation holds:
$$
N_i(t) \approx N_i e^{- \Delta t \frac{  \sigma_i\, \lum}{N_i}}\,,
$$
where $N_i$ is the mean number of particles in the ring $i$ and
$\Delta t$ is the time elapsed since injection.
The beam luminosity lifetime $\tau_i$ quoted in Table~\ref{tab:params}
is defined as:
$$
\tau_i = \frac{N_i}{\sigma_i\, \lum}\,.
$$

An excellent approximation for the cross section to lose a particle
from beam $i$ due to radiative Bhabha process is~\cite{bib:lumilife_lumiXsec}:
\begin{equation} \label{eq:lumilife_RadXsec}
\sigma_{\mathrm{rad.}} \approx
\frac{16\, \alpha\, r_e^2}{3} \left[
\left( \ln \frac{E_{\mathrm{c.m.}}^2}{m_e^2} -\frac{1}{2} \right)
\left( \ln \frac{E_i}{k_{\mathrm{min}\, i}} -\frac{5}{8} \right)
+\frac{1}{2} \left( \ln \frac{E_i}{k_{\mathrm{min}\, i}}\right) ^2 -
\frac{3}{8} - \frac{\pi^2}{6}
\right]\,,
\end{equation}
where $k_{\mathrm{min}\, i}$ is the minimum energy of a radiated photon that cause
the loss of a particle from beam $i$; thus $k_{\mathrm{min}}\, i/E_i$
can be taken as the fractional energy aperture for the ring $i$.
Note that this expression depends only logarithmically on the energy
acceptance of the ring.

\begin{figure}[t]
\begin{center}
\includegraphics[width=0.8\textwidth]{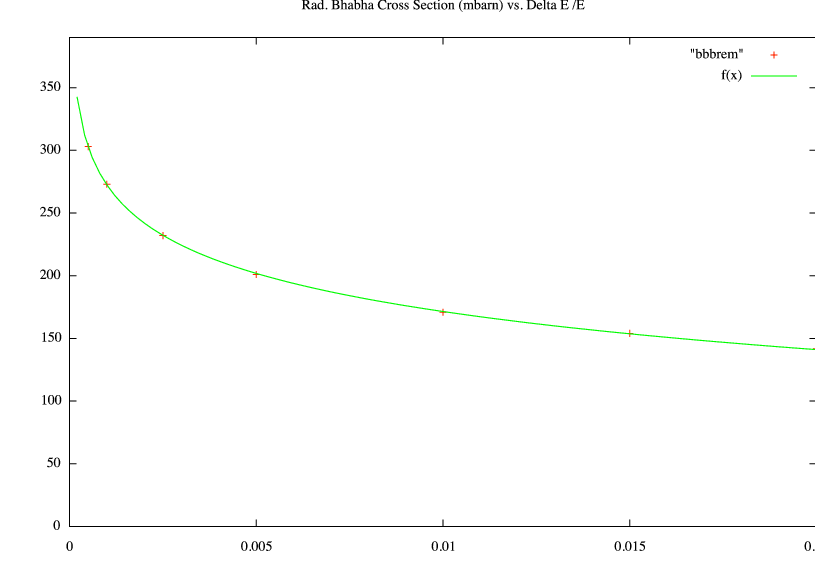}
\end{center}
\caption{ \label{fig:lumilife_1} Radiative Bhabha cross section in mbarn
for loss of a particle as a function of the ring energy
acceptance: crosses are the BBBrem predictions, continuous line is
a phenomenological fit. }
\end{figure}

Actual measurements of this cross section~\cite{bib:lumilife_3} find
a value smaller than the prediction; this reduction
can be ascribed to the effect of finite bunch density.
To correctly model this effect, the BBBrem Monte Carlo generator
\cite{bib:lumilife_Kleiss} was used. The predicted cross section as
a function of the energy acceptance is shown in Fig.~\ref{fig:lumilife_1}
together with the best fitting function:
$$
\sigma_i  = \ln \frac{E_i}{ 2 {k_{\mathrm{min}\, i}}} \times 43.9 \mbarn \,.
$$

The cross section predicted by BBBrem for a ring energy acceptance of
1\% is $170 \mbarn$, corresponding to a beam lifetime of $10.4\min$
for the LER and $5.9\min$ for the HER. This is to be compared to
values of $265 \mbarn$, $6.7\min$ and $3.8\min$ respectively that would be
obtained with Eq.~\ref{eq:lumilife_RadXsec}.

Another loss mechanism, typically not as important as the
bremsstrahlung contribution considered so far, is the loss of particles due
to elastic Bhabha ($\epem \to \epem$) scattering at sufficiently
large angles to escape the acceptance of the ring. A tree-level
approximate formula for this cross section to lose a particle from
the beam $i$ is:
$$
\sigma_{\mathrm{el.}} \approx \frac{8\,\pi\,\left( \hbar\, c\, \alpha\right)^2}{ E_{\mathrm{c.m.}}^2}
\frac{E_j}{E_i}\left(\frac{1}{\vartheta_{\mathrm{min. x}}^2} +
\frac{1}{\vartheta_{\mathrm{min. y}}^2 }\right)\,,
$$
where $\vartheta_{\mathrm{min. x,y}}$ is the minimum horizontal/vertical
scattering angle in the laboratory frame leading to particle loss.
The particle loss cross sections are $3.0\mbarn$ for the HER and
$9.0\mbarn$ for the LER under the usual assumption of a 10$\sigma$ limiting
aperture, calculated using the uncoupled horizontal and the fully
coupled vertical beam sizes.

The total particle cross sections and lifetime for these processes are shown in
Table~\ref{tab:xsectlife}.

\begin{table}[hbt]
\caption{Total particles cross sections and lifetime.}
\label{tab:xsectlife}
\vspace*{2mm}
\centering
\setlength{\extrarowheight}{1pt}
\begin{tabular}{p{3cm}cp{5mm}c}
\hline
\hline
                         & LER & & HER \\
\hline
$\sigma_{\mathrm{rad.}}$ & 170 $\mbarn$ & & 170 $\mbarn$ \\
$\sigma_{\mathrm{el.}}$  &  3 $\mbarn$  & & 9 $\mbarn$ \\
$\sigma_{\mathrm{tot.}}$ & 173 $\mbarn$ & & 179 $\mbarn$\\
      Lifetime           &10.3 $\min$   & & 5.7 $\min$ \\
\hline
\end{tabular}
\end{table}

\subsubsection{Touschek lifetime }
\label{subsubsec:accel_Touschek_tau}

The Touschek beam lifetime in \superb\ is expected to be small, particularly
for the LER, because of the extremely small beam emittance. In
order to estimate the Touschek beam lifetime we use the formula for
the particle loss per unit time given by Le~Duff~\cite{bib:CERN_rpt}:
\begin{equation}
\frac{1}{N}\frac{dN}{dt} = \frac{1}{\tau}
=\frac{Nr_0^2c}{8\pi\sigma_x\sigma_y\sigma_s}\frac{\lambda^3}{\gamma^2}
D(\xi),
\end{equation}

where $\lambda$ is the momentum acceptance $\sigma_i$ the beam size
in the three planes, $\gamma$ is the Lorentz factor, and
\begin{equation}
\xi = \left(\frac{\Delta
E/E}{\gamma}\right)^2\frac{\beta_x}{\varepsilon_x}\,.
\end{equation}

 For the function $D(\xi)$, we employ Bruck's
approximation~\cite{bib:Bruck}, valid for $\xi<0.01$:
\begin{equation}
 D(\xi) =
\sqrt{\xi}\left(\ln\left(\frac{1}{1.78\xi}\right)
-\frac{3}{2}\right)\,.\end{equation}

The total machine acceptance in $\Delta p/p$ is the lesser of the RF
acceptance and the lattice acceptance,

\begin{equation}
\lambda_{mach} = Min(\lambda_{rf},\lambda_{latt})\,.
\end{equation}

This approach was used successfully to describe experimental data from the
\pepii LER~\cite{bib:Wie_Touschek}.

The RF acceptance in turn is given by:

 \begin{equation}
 \lambda_{rf} = \sqrt{\frac{V_0}{\pi|\eta|hE_0} F(\frac{eV_rf}{V_0})},\\
 F(q) = 2\left(\sqrt{q^2-1}-\cos^{-1}(\frac{1}{q})\right)\,.
\end{equation}

The RF acceptance for \superb\ is quite large, about 2.5\%~$\Delta p/p$;
however, it would difficult to maintain such an acceptance with the chosen
lattice designs. We will therefore assume a reasonable value of 1\%
for the acceptance due to this limitation. The \superb\ parameters relevant
for Touschek beam lifetime are summarized in
Table~\ref{tab:Touschek_parms}.

 \begin{figure}[tb]
 \centering
 \includegraphics[width=0.75\textwidth]{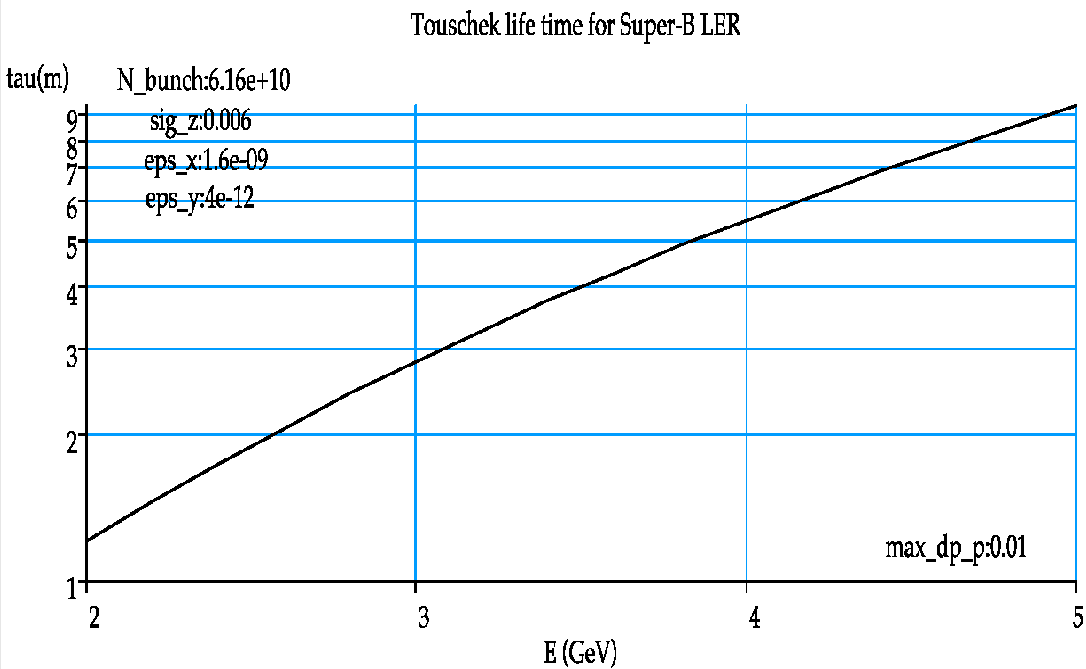}
 \caption{Touschek beam lifetime {\it vs.} beam energy for the LER.}
 \label{fig:Touschek_LER}
 \end{figure}

\begin{table}[!hb]
\caption{Nominal \superb\ beam parameters.}
\label{tab:Touschek_parms}
\vspace*{2mm}
\setlength{\extrarowheight}{1pt}
\centering
\begin{tabular}{lcc}
\hline\hline
                                    & HER & LER \\
\hline
Beam Energy (GeV)                   & 7   & 4   \\
Bunch length (mm)                   & 6   & 6   \\
Energy spread (\%)                  & 0.1 & 0.1 \\
Horizontal emittance (nm)           & 1.6 & 1.6 \\
Vertical emittance (pm)             & 4   & 4   \\
Energy acceptance (\% $\Delta p/p$) & 1   & 1   \\
$\beta_x$ avgerage (m)              & 10  & 10  \\
$\beta_y$ avgerage (m)              & 22  & 22  \\
Particles/bunch & $3.52\times10^{10}$ & $6.16\times10^{10}$ \\
\hline
\end{tabular}
\end{table}

\begin{figure}[htb]
\centering
\includegraphics[width=0.75\textwidth]{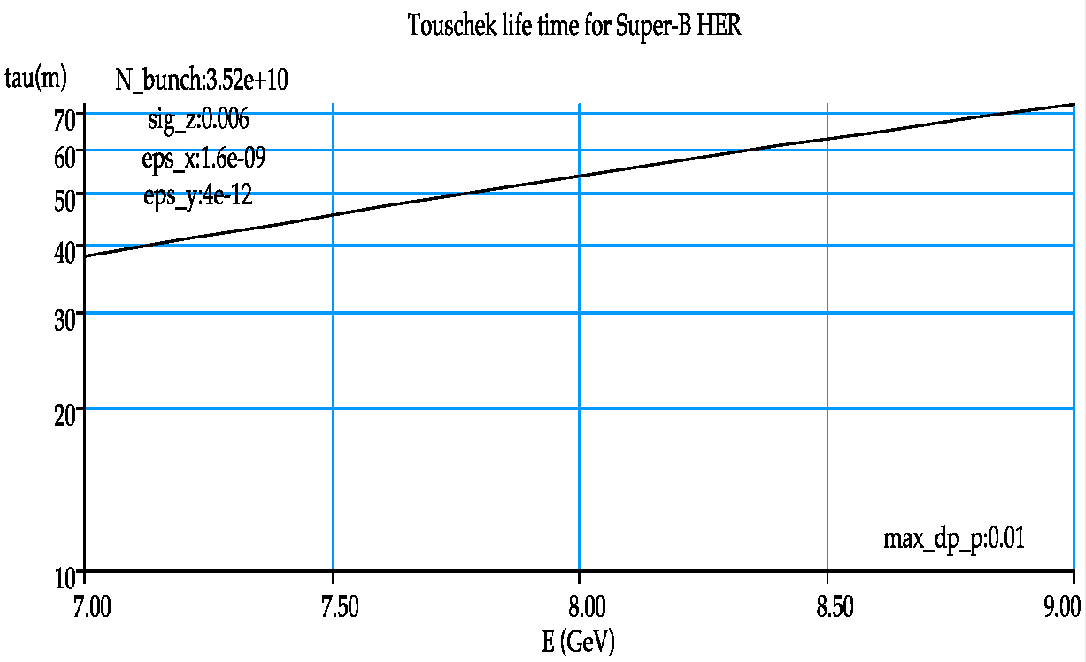}
\caption{Touschek beam lifetime {\it vs.} beam energy for the HER.}
\label{fig:Touschek_HER}
\end{figure}

In Fig.~\ref{fig:Touschek_LER}, we show the Touschek beam lifetime for the
LER as a function of beam energy. At 4\gev, it is slightly above $5\min$.
The penalty paid in terms of beam lifetime for an increased
energy asymmetry is evident. For the \superb\ HER, the corresponding
result is shown in Fig.~\ref{fig:Touschek_HER}. Beam lifetime is close
to $40\min$ at 7\gev. In collision, however, since the luminosity
beam lifetime will be much lower for the HER than the LER, due to the
smaller number of particles present in an
HER bunch, the actual beam lifetimes are expected to be similar;
a few minutes for each ring.

In Table~\ref{tab:Touschek_Summary} we summarize the Touschek beam
lifetimes for both rings at their design energy, for both the nominal
and the upgrade parameter set for \superb.

\begin{table}[!htb]
 \caption{Touschek beam lifetime summary.}
 \label{tab:Touschek_Summary}
\vspace*{2mm}
\setlength{\extrarowheight}{2pt}
\centering
\begin{tabular}{lccc}\hline\hline
{Parameter set} &  {Luminosity}   & {Lifetime HEB} & {Lifetime LEB} \\
                    & (m$^{-2}$s$^{-1}$)  &       (min)        & (min) \\
\hline
      Nominal       & $1.0\times10^{36}$  &         38         & 5.5 \\
      Upgrade       & $2.44\times10^{36}$ &         19         & 3 \\
\hline
\end{tabular}
 \end{table}

\subsubsection{Touschek backgrounds}
\label{subsubsec:accel_Touschek_bkg}

Simulation studies of background from Touschek scattering
\cite{bib:Tou_1} have been performed for the LER, using a program developed for
\daphne; preliminary results are presented here. The reliability
of the calculation has been tested with KLOE data,
showing good agreement \cite{bib:Tou_2,bib:Tou_3}. Further checks
for the \superb\ LER case are under way. Touschek scattering is a
source of background due to the off-energy particles arising from
the elastic scattering of particles within a bunch. Such scattering
results in two particles with energy errors $+\Delta p/p$ and
$-\Delta p/p$ that follow betatron trajectories around the
off-energy closed orbit. In the simulation Touschek particles are
taken within one transversely Gaussian bunch with the proper energy.
Particles are tracked over many turns or until they are lost. In
this way, an estimate is obtained for the Touschek losses around the
entire ring and for the IR alone. Essentially all losses at the IR
arise from particles that are Touschek scattered in dispersive regions.
Touschek-scattered particles have a betatron oscillation
proportional to the dispersion $D$, to the invariant $H$ and to the
momentum spread $\Delta p/p$:

$$x = ( |D|+\sqrt{H\; \beta}) \Delta p/p\,.$$

The parameter $H$ is defined by the following relation:

$$ H = \gamma_x {D_x}^2  + 2\alpha_x D_x D'_x + \beta_x {D'_x}^2\,. $$

Given an energy spectrum $P(E)$, one can either throw the single
scattered particle energy shift accordingly or use a uniform
distribution and weight particles contributions with $P(E)$. We use
the latter approach, which allows us to cope with the tails of both
the Touschek probability density function and the probability of
beam loss {\it vs.} energy deviation. For a lower energy shift, the
Touschek scattering probability increases while the probability of
loss decreases and vice versa. The Touschek density function is
mostly related to beam parameters such as bunch volume, emittance,
momentum deviation and bunch current. On the other hand, particle
losses are related mostly to machine parameters and optics, such as
the physical aperture, the phase advance between dispersive regions
and collimators, and between dispersive regions and the IR.

The calculation
of the energy spectra starts from the formula~\cite{bib:Bruck}:

$$
\frac{1}{\tau} =
\frac{\sqrt{\pi} r_e^2 \; c \; N}
 {\gamma^3 (4\pi)^{3/2}\; V \; \sigma_x' \varepsilon^2} C(u_{\mathrm{min}})\,,
$$

where:
\begin{alignat*}{1}
  \varepsilon &= \frac{\Delta E}{E}, \\
  u_{min}   &= \left( \frac{\varepsilon} {\gamma\; \sigma_x'}\right)^2, \\
        V  &= \sigma_x \sigma_y \sigma_z , \\
\sigma_x'  &= \sqrt{\frac{\varepsilon_x}{\beta_x} +
              \sigma_p^2 \left({D'}_x + D_x \frac{\alpha_x}{\beta_x}\right)^2 }\,.
\end{alignat*}

Here $C(u_{\mathrm{min}})$ accounts for the \Moller cross-section and momentum
distribution.  For a chosen machine section the Touschek probability
is evaluated in small steps (30 per element) to account for the beam
parameter evolution within the element. Each element is sampled 100 times.
The density function for the chosen section
is obtained by interpolating between the results
using the Touschek
scaling law $A_1\varepsilon^{-A_2}$.  Further details on the
simulation can be found in ref.~\cite{bib:Tou_1}.

\begin{figure}[htb]
\centering
\hspace{-1cm}
\includegraphics[width=0.75\textwidth]{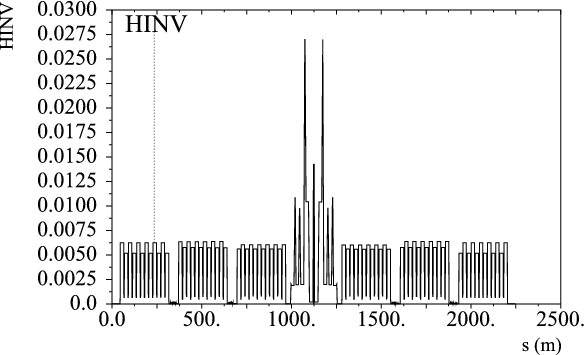}
\caption{\label{fig:Tou_1H_LER} H function for the LER.}
\end{figure}

Figure~\ref{fig:Tou_1H_LER} shows the behavior of the $H$ function
along the ring, with the IP located at $s = 1124\m$. It appears that
the value of $H$ is almost constant at $6.\EE{-3}$ for most of the
ring, except for the IR. For this reason, Touschek particles are
generated continuously along the whole ring in three element steps.
The Touschek-scattered particles undergo large betatron oscillations
in the regions where $H$ and $D$ are high, with very similar energy
spectra, but different phase advance, leading to different loss
probability.

The beam parameters used for our simulations are shown in
Table~\ref{tab:Toubacks}. Full tracking has been performed for one
machine turn, and only particles with a relative energy deviation
between $0.003$ and $0.02$ have been simulated. Particles with
higher energy deviations are lost locally, and do not contribute to
backgrounds in the experiment, while particles with relative energy
deviation $< 0.003$ almost always remain inside the beam pipe. A
beam pipe with a 2\cm\ radius was assumed for the entire ring,
outside of the IR.

\begin{table}[htb]
\caption{Relevant beam parameters used for Touschek background simulations.}
\label{tab:Toubacks}
\vspace*{2mm}
\setlength{\extrarowheight}{2pt}
\centering
\begin{tabular}{cccccc}
\hline
\hline
$N_{\mathrm part}/\mathrm{bunch}$ & $I_{\mathrm bunch} (mA)$
& $\varepsilon_x$ (nm-rad) & Coupling (\%)  & $\sigma_z (mm)$ &  $\thx (mrad)$ \\
\hline

$6.2\EE{10}$ &    $1.3$ & $0.8$  &   $0.25$  &  $6.$  &  $17$ \\
\hline
\end{tabular}
\end{table}



\begin{figure}[hbt]
  \centering
\includegraphics[width=0.8\textwidth]{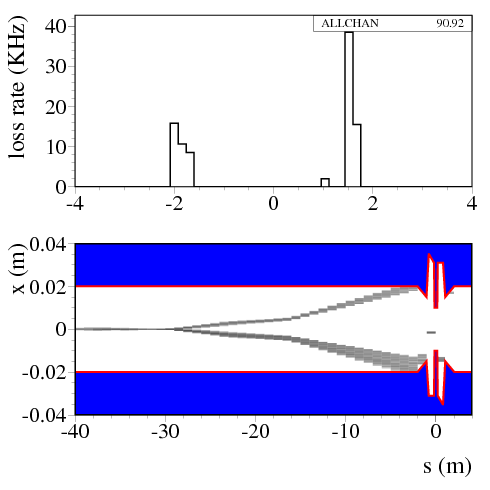}
\caption{\label{fig:Tou_4_TouTraj2}
Touschek particle trajectories generated at $1.5 \pi$
phase advance from the IP are all lost at the IR.
Collimators placed further away from the IP do not remove these particles.}
\end{figure}

Almost all IR losses are due to Touschek scattering occurring far
away the IP; they can, therefore, be very effectively reduced with a
suitable arrangement of collimators. The most effective location for
collimators would be at longitudinal positions corresponding to
large radial oscillation of scattered particles. A detailed study on
the optimal position of collimators is ongoing. However, a
preliminary set of locations has been identified, giving a loss rate of
about $90 \kHz$ within the IR ($-4 < s < 4\m$) for a $1.3 \mA$
single bunch current. The upper plot of
Fig.~\ref{fig:Tou_4_TouTraj2} shows the distribution of IR Touschek
particle losses, while the lower plot shows trajectories of
scattered particles that are eventually lost at the IR. This preliminary
collimation scheme
appears to be effective for all particles generated along the ring and
eventually lost at the IR, except for those scattered at $s \sim -30
\m$. At this location, the phase advance from the IP is about $1.5
\pi$, resulting in large radial oscillations just at the IR. This
residual source of IR background is difficult to remove by inserting
additional collimators very close to the IR. However, the phase
advance between positions where Touschek scattering results in
significant IR losses and the IP can be adjusted in the final
design.


Preliminary studies show that particle losses due to the Touschek
effect are expected to be quite high in the LER, consistent with the
Touschek lifetime calculations. Since Touschek particles
are close in energy to the beam and have small divergence, they
can be an important source of background for the detector. Further
studies are in progress to optimize the position of
collimators, and to improve the machine model, taking into account
non-linear terms in the final focus quadrupoles, which can be a
relevant issue when tracking Touschek particles.

\afterpage{\clearpage}
\subsection{Intrabeam Scattering }
Intrabeam scattering \cite{bib:IBS_1, bib:IBS_2} is associated with
the Touschek effect; while single large-angle scattering events
between particles in a bunch leads to loss of particles (Touschek
lifetime), multiple small-angle scattering events lead to emittance
growth, an effect that is well known in hadron colliders and
referred to as intrabeam scattering (IBS).  In most electron storage
rings, the growth rates arising from IBS are usually very much
longer than synchrotron radiation damping times, and the effect
is not observable.  However, IBS growth rates increase with
increasing bunch charge density, and for machines that operate with
high bunch charges and very low vertical emittance, the IBS growth
rates can be large enough that significant emittance increase can be
observed.  Qualitative observations of IBS have been made in the
LBNL Advanced Light Source \cite{bib:IBS_3}, and measurements in the
KEK Accelerator Test Facility (ATF) \cite{bib:IBS_4} have been shown
to be in good agreement with IBS theory. IBS is expected to increase
the horizontal emittance in the ILC damping rings by roughly $30\%$
\cite{bib:IBS_5}; the \superb\ rings will operate with comparable
bunch sizes and beam energy, and with somewhat larger bunch charge,
so we may expect similar emittance growth from IBS in \superb\ to
that in the ILC damping rings.  There is a strong scaling with
energy, with IBS growth rates decreasing rapidly with increasing
energy. Therefore, we expect significantly larger IBS emittance
growth in the \superb\ low energy ring than in the high energy ring.

Several formalisms have been developed for calculating IBS growth
rates in storage rings, notably those by Piwinski~\cite{bib:IBS_1} and
by Bjorken and Mtingwa~\cite{bib:IBS_2}. IBS growth rates depend on
the bunch sizes, which vary with the lattice functions around the
ring; to calculate
accurately the overall growth rates, one should therefore calculate
the growth rates at each point in the lattice, and average over the
circumference.  Furthermore, since IBS results in an increase in
emittance, which dilutes the bunch charge density and affects the IBS
growth rates, it is necessary to iterate the calculation to find the
equilibrium, including radiation damping, quantum excitation and IBS
emittance growth.  The full IBS formulae include complicated integrals
that must be evaluated numerically, and can take significant
computation time; however, methods have been developed
 \cite{bib:IBS_5,bib:IBS_6} to allow
reasonably rapid computation of the equilibrium emittances, including
averaging around the circumference and iteration.

For calculation of the IBS emittance growth in the \superb\ rings, we
use the formulae of Kubo \etal~\cite{bib:IBS_6}, which are based on
an approximation to the Bjorken-Mtingwa formalism~\cite{bib:IBS_2}.
This approximation has been shown to be in good agreement with data
on IBS emittance growth collected at the ATF~\cite{bib:IBS_4,bib:IBS_6}.
In our calculations, the average growth
rates are found from the growth rates at each point in the lattice,
by integrating over the circumference; we use iteration to find
the equilibrium emittances in the presence of radiation and IBS.

IBS effects tend to be most significant in the horizontal plane.
This is due to the effect of dispersion, which has consequences
for the horizontal emittance similar to those in the case of quantum
excitation from synchrotron radiation.  When two particles scatter,
there tends to be a transfer of horizontal to longitudinal momentum;
this changes the energy deviations of the particles, which, if the
scattering takes place at a location with large dispersion, leads to
an increase in horizontal emittance.  The principal difference in
this respect between synchrotron radiation and IBS is that
synchrotron radiation is only significant in the bending magnets,
where the dispersion is low by design; IBS occurs throughout the
lattice, including regions with relatively large dispersion.

\begin{figure}[!htb]
\centering
\includegraphics[width=0.8\textwidth]{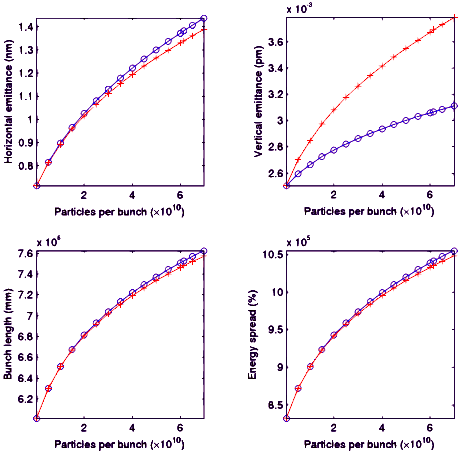}
\caption{\label{fig:IBS_emit_LER}
 Transverse emittance growth, and growth in bunch length and energy spread
 in the \superb\ LER, as functions of the bunch charge.
 The blue line with data points marked as circles shows the case that
 betatron coupling makes a $10\%$ contribution to the vertical emittance,
 with vertical dispersion contributing $50\%$.  The red line with data points
 marked as crosses shows the case that betatron coupling and vertical
 dispersion make equal contributions to the vertical emittance.
}
\end{figure}

Figure~\ref{fig:IBS_emit_LER} shows
the equilibrium transverse emittances, bunch
length and energy spread in the \superb\ LER as functions of the
bunch charge.

At the nominal bunch charge of $6.16\times 10^{10}$, the horizontal
emittance is nearly doubled.  However, with the design natural
emittance of $0.7 \nm$, the final emittance is still below the
specified operating horizontal emittance of $1.6 \nm$.  There is
also an increase in the vertical emittance, of between $20\%$ and
$50\%$, depending on whether the vertical emittance is generated
predominantly by vertical dispersion, or by roughly equal
contributions from vertical dispersion and betatron coupling.  The
increase in vertical emittance is significant, but there are
possibilities for reducing the impact. If it is felt undesirable to
reduce the specification on the vertical emittance below the nominal
$4 \picom$ (at low bunch charge), then the synchrotron radiation
damping time may be reduced by increasing the length of wigglers.

As indicated in our results, the increase in vertical emittance from
IBS depends on the relative contributions of betatron coupling and
vertical dispersion to the vertical emittance.  If betatron coupling
dominates, then the proportional increase in vertical emittance from
IBS will be equal to the proportional increase in horizontal
emittance.  If we assume roughly equal contributions to the vertical
emittance from betatron coupling and from vertical dispersion, then
the relative increase in the vertical emittance ($50\%$) is half the
relative increase in the horizontal emittance ($100\%$).  However,
if the betatron coupling contributes only $10\%$ to the vertical
emittance, then the proportional increase in the vertical emittance
is reduced to approximately $20\%$ at the nominal bunch charge.  It
is difficult to know at this early stage the likely relative
contributions of betatron coupling and vertical dispersion to the
vertical emittance, and this requires further study.  A residual {\it rms}
vertical dispersion of $4 \mm$ will generate about $25\%$ of the $4
\picom$ vertical emittance in either the LER or the HER lattice; if
the {\it rms} vertical dispersion is increased to $5.5 \mm$, then this
will generate about $50\%$ of the $4 \picom$ vertical emittance.

\begin{figure}[htb]
\centering
\includegraphics[width=0.8\textwidth]{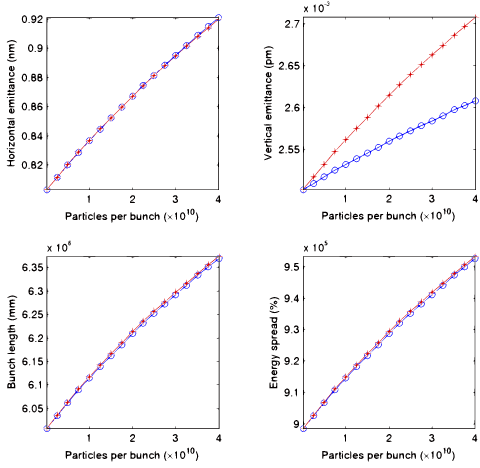}
\caption{\label{fig:IBS_emit_HER} Transverse emittance growth, and
growth in bunch length and energy spread in the \superb\ HER, as
functions of the bunch charge. The blue line with data points marked
as circles shows the case that betatron coupling makes a $10\%$
contribution to the vertical emittance, with vertical dispersion
contributing $50\%$.  The red line with data points marked as
crosses shows the case that betatron coupling and vertical
dispersion make equal contributions to the vertical emittance.  }
\end{figure}

The strong scaling of IBS growth rates with energy means that in the
HER the emittance growth from IBS is much less than in the low
energy ring; the effects of IBS are further mitigated by the lower
bunch charge in the high energy ring. Fig. \ref{fig:IBS_emit_HER}
shows the transverse emittances, bunch length and energy spread in
the \superb\ HER as functions of the bunch charge.

There is a $14\%$ increase in horizontal emittance at the nominal
bunch charge of $3.52\times10^{10}$ particles, and an increase in
vertical emittance of between $4\%$ and $8\%$, depending on whether
the betatron coupling makes a contribution of $10\%$ or $50\%$ to
the vertical emittance (with the remaining contribution coming from
vertical dispersion). We again assumed that betatron coupling and
vertical dispersion make roughly equal contributions to the vertical
emittance.

\afterpage{\clearpage}
\subsection {Space Charge Effects in the LER}

Space charge effects in the LER have been studied using a
weak-strong model of dynamics,
as implemented in the code Marylie/Impact (MLI).  The impact of
space charge is noticeable, but our results suggest the
existence of workable regions of the tune space in which the design
emittance is minimally affected.  However, additional studies are
recommended to fully substantiate this conclusion.

\begin{figure}[!hbt]
\centering
\includegraphics[width=0.9\textwidth]{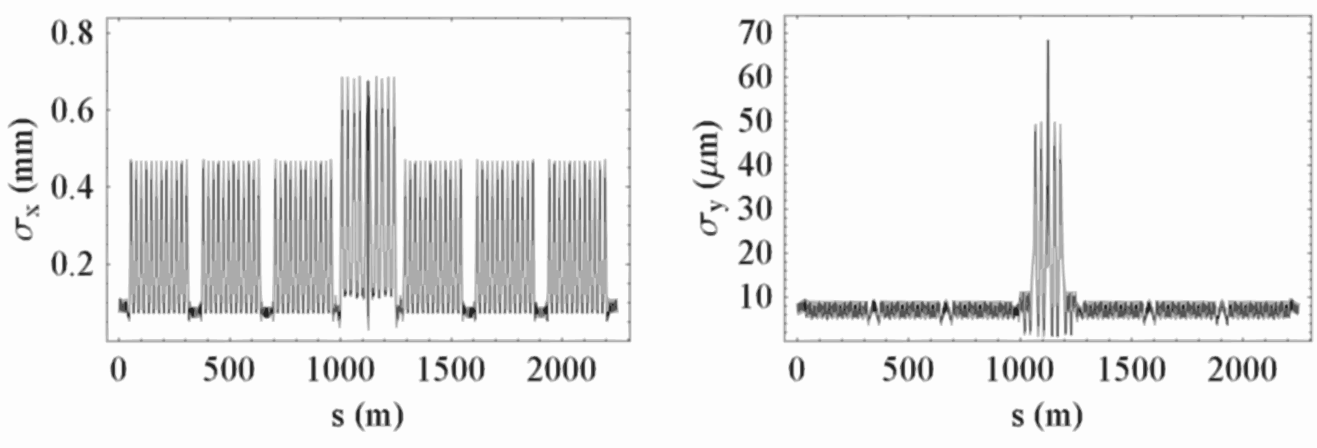}
\caption{\label{fig:sigmas} {\it rms} transverse beam sizes along the
LER lattice at equilibrium. The horizontal size (left picture)
includes the effect of a finite energy spread in the dispersive
regions.}
\end{figure}

The large bunch population and small beam sizes result in
appreciable space charge tune shifts in the \superb\ rings, and in
particular in the LER, as space charge effects scale inversely with
the beam energy. For the LER at the design equilibrium and bunch
population ($N=6.16\times10^{10}$) linear theory ($i=x,y$):
 \begin {equation}
 \Delta \nu_i  = - \frac{1}{4\pi} \frac{2 r_e}{\beta^2 \gamma^3} \int\limits_0^C
\frac{\lambda \beta_i}{\sigma_i(\sigma_x + \sigma_y)}\; ds\,,
  \end {equation}
yields the following horizontal and vertical space charge tune
shifts: $\Delta{\nu_x} = -0.004$, $\Delta{\nu_y} = -0.179$. This
equation, in which $\beta$ and $\gamma$ are the relativistic
factors, $\beta_x$, $\beta_y$ are the lattice functions, $\sigma_x$,
$\sigma_y$ the horizontal and vertical {\it rms} beam sizes,
$\lambda=N/\sqrt{2\pi}\sigma_z$ the longitudinal peak density
($\sigma_z$ is the {\it rms} longitudinal bunch length), applies to
particles undergoing infinitesimally small betatron and synchrotron
oscillations about the center of a gaussian bunch. Plots of the
transverse beam sizes for the LER at equilibrium, as determined using
the design emittances $\varepsilon_x=0.71\nm$-rad, $\varepsilon_y=2.5\,{\rm pm}$-rad,
and $\varepsilon_z=5$\mum-rad are shown in Fig.~\ref{fig:sigmas}.

While space charge should have little effect on injection
efficiency, since its effects become noticeable only after several
damping times, it could cause particle beam losses at later times,
if the working point in tune-space is sufficiently close to an
unstable lattice resonance. Proximity to stable resonances would
be less damaging, but could also be detrimental, and could lead to
unacceptable emittance degradation. Far from resonances, space
charge may still compromise the target vertical
equilibrium emittance, when its impact is considered in combination
with radiation and linear coupling in a non-ideal lattice. The
latter effect, however, should be small~\cite{bib:spa_cha_1}, and was
neglected here, as we limited our attention to an error-free
lattice in the absence of any radiation effects.

Our study was conducted using a weak-strong model for space
charge, as implemented in the numerical tools recently developed
to study similar effects in the ILC damping rings \cite{bib:spa_cha_2,
bib:spa_cha_3}. In the weak-strong model the space charge force is
calculated as if it were produced by a 6D gaussian bunch matched to the
ideal linear lattice and with {\it rms} emittances equal to those
expected at equilibrium for a realistic lattice with some residual
linear coupling. A collection of macroparticles, initially
distributed according to a bunch density at equilibrium, is then
tracked element-by-element, with the inclusion of the lattice
nonlinearities and treating the space charge force with the kick
approximation. Because of its non-self consistent nature, this
model will likely overestimate the effect of any detected
emittance growth, and should be used mainly as a tool to search the
tune space for regions of minimal emittance growth. An accurate
characterization of emittance growth would require more detailed,
and considerably more computationally intensive, models of beam
dynamics.

In our study we used an augmented version of the Marylie/Impact (MLI)
code \cite{bib:spa_cha_4}. The code was validated during the ILC damping
ring studies  by calculations carried out independently using SAD
\cite{bib:SAD, bib:spa_cha_2}. For more details on the implementation of the
weak-strong space charge model in MLI we refer to \cite{bib:spa_cha_2}.

\begin{figure}[t!bh]
\centering
\includegraphics[width=\textwidth]{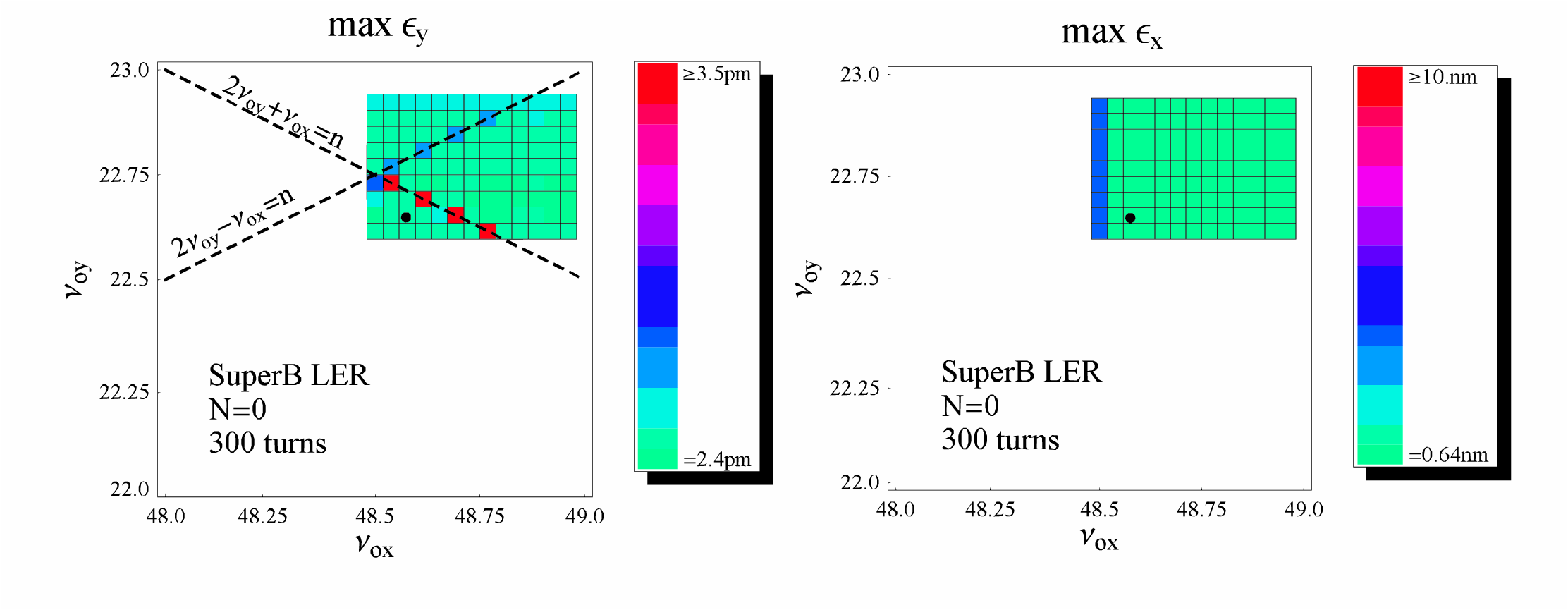}
\caption{\label{fig:scan2}  Tune scan of horizontal and vertical
maximum {\it rms} emittance growth over 300 machine turns, not including
the effects of space charge. The color coding shows the vertical (left)
and horizontal (right) emittance on a linear scale from minimum to maximum.
The design working point
$\nu_{0x} = 48.575$, $\nu_{0y} = 48.647$ is shown as a black dot.}
\end{figure}

\begin{figure}[htb]
\centering
\includegraphics[width=\textwidth]{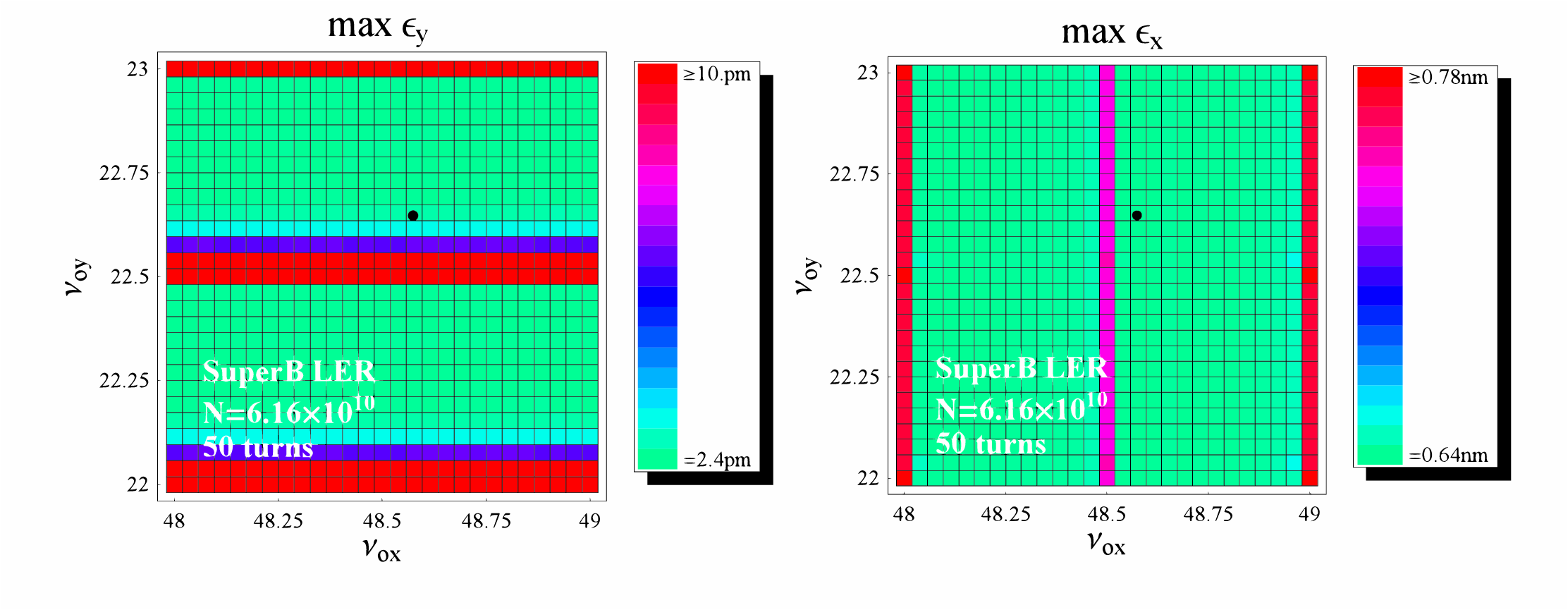}
\caption{\label{fig:scan1}  Tune scan of horizontal and vertical
maximum {\it rms} emittance growth over 50 machine turns, including the effect
of space charge.
The color coding shows the vertical (left) and horizontal (right) emittance
on a linear scale from minimum to maximum. The design working point
$\nu_{0x} = 48.575$, $\nu_{0y} = 48.647$ is shown as a black dot.}
\end{figure}

\begin{figure}[htb]
\centering
\includegraphics[width=\textwidth]{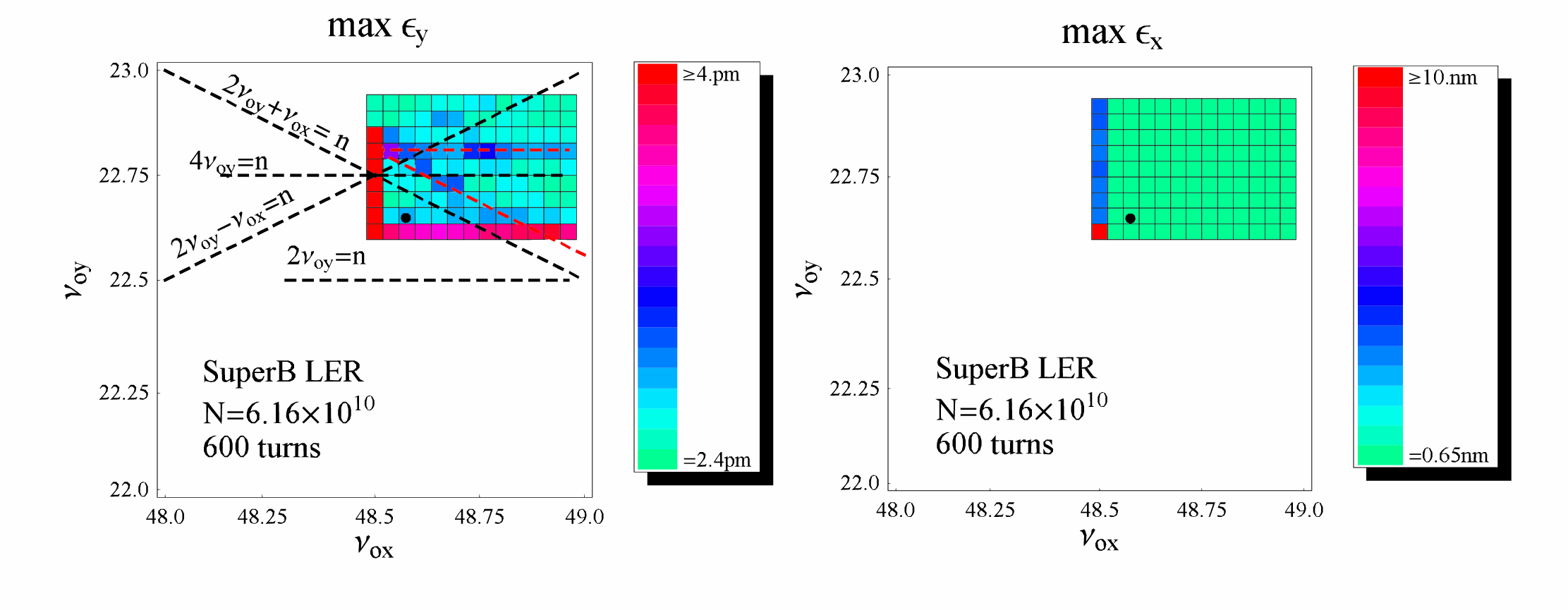}
\caption{\label{fig:scan3}  Tune scan of horizontal and vertical
maximum {\it rms} emittance growth over 600 machine turns, including the
the effects of space charge. The color coding shows the vertical (left)
and horizontal (right) emittance on a linear scale from minimum to maximum.
The design working point
$\nu_{0x} = 48.575$, $\nu_{0y} = 48.647$ is shown as a black dot.}
\end{figure}

We assessed the space charge effects in the LER lattice by
producing tune space scans and looking for the {\it rms} emittance
changes in the transverse plane. The results of our
investigations are reported as color-density plots  showing the
maximum value of the {\it rms} emittance experienced by the
macroparticle beam within the indicated duration of tracking, see
Figs.~\ref{fig:scan1}, \ref{fig:scan2}, and \ref{fig:scan3}.

We tracked a bunch population of 200 macroparticles. Because of the
weak-strong nature of the model, evaluation of the space charge
kick is independent of the number of macroparticles used; a
modest number should therefore be sufficient to provide an acceptable
sampling of the phase space available to the beam. The detuning of
the lattice was done by inserting pure phase rotations at the end
of the one-turn lattice with proper matching, so as not to perturb
the value of the lattice functions. This amounts to a linear kick
causing a small discontinuity in both the particle transverse
position and momentum.
A short term tracking (50 machine turns) tune scan of the
$[0,1]\times[0,1]$ region around the design working point exhibited
strong half-integer resonances (Fig.~\ref{fig:scan1}). These
resonances are already present in the bare lattice and the effect
of space charge is to noticeably enlarge their width. This is seen,
for the vertical plane, in the lefthand plot of Fig.~\ref{fig:scan1}, where
the apparent width of the resonance lines is about $\Delta \nu_y
\simeq 0.1$.

Notice, however, that this value is somewhat smaller than the tune
shift predicted by linear theory. This should be interpreted as a
consequence of the highly nonlinear nature of the space charge
force, which causes the tune shift experienced by particles to
decrease quickly from the value predicted by linear theory with
increasing amplitude of the betatron (or synchrotron) oscillations.
In the horizontal plane (lefthand plot in Fig.~\ref{fig:scan1}) the
space charge tune shift is too small to have any detectable effect
on the scale of resolution used for these tune scans.
The minimum emittance reported in the figures,
$\varepsilon_y =2.4\,{\rm pm}$-rad and $\varepsilon_x =0.65\nm$-rad (regions with
dark green shading), representing the {\it rms} emittances of
the simulated macroparticle distributions,
differ slightly from the nominal equilibrium values
$\varepsilon_y =2.5\,{\rm pm}$-rad and $\varepsilon_x =0.71\nm$-rad, because of
statistical fluctuations associated with the use of a limited number of
macroparticles.

For practical reasons, in order to study space charge effects on a longer
time scale (up to 600 machine/turns), we restricted our
investigation to a smaller area of tune-space. The case with space charge
(Fig.~\ref{fig:scan2}) is to be compared with the case without
space charge (Fig.~\ref{fig:scan3}). In the absence of space
charge, the vertical emittance tune scan (lefthand plot in
Fig~\ref{fig:scan2}) shows evidence of two third-order resonances
at $2 \nu_{0y} + \nu_{0x}=n$ and $2 \nu_{0y} - \nu_{0x}=n$, with
the first being considerably stronger, and resulting in about
$100\%$ emittance growth over 300 machine turns. The other
resonance resulted in a smaller $\sim 10\%$ growth over the same
tracking time. Outside these narrow resonances, the vertical {\it rms}
emittance appears to remain largely unchanged.

Inclusion of space charge causes some additional degradation of
the {\it rms} vertical emittance that is not apparent in short term-tracking
(lefthand plot in Fig.~\ref{fig:scan3}). Not unexpectedly, the
largest growth occurs along the half-integer $\nu_{0x}=48.5$ line.
This resonance is already present in a bare lattice, but with
visible consequences only on the horizontal motion. Its impact on
the vertical motion is fostered by the $x/y$ coupling introduced
by space charge. The emittance growth detected along this line was
very large, and for some choices of the vertical tune was found to
lead to particle losses. Outside this resonance line and the upper
part of the region affected by the $\nu_{0y}=22.5$ resonance we
observe some smaller, but clearly noticeable, emittance growth up to
about $30\%$ over 600 turns (region with bluish shading). The
emittance growth appears to be concentrated mostly along two
lines--we highlighted these strips of growth in the lefthand plot of
Fig.~\ref{fig:scan1} with two red dashed lines. These lines are not
traceable to some obvious resonances and identification of the
exact growth mechanism will require some more work. At this time
our tentative interpretation is that these regions of growth are
related  to the resonance lines $2 \nu_{0y} + \nu_{0x}=n$ and
$4\nu_{0y}=n$; they appear shifted upward for reasons that remain to be
understood. Notice that the magnitude of this shift is
well within the value of space charge detuning -- an argument in
favor of this interpretation. The resonance $2 \nu_{0y} +
\nu_{0x}=n$ is a lattice resonance already present in the bare
lattice whereas the fourth order $4\nu_{0y}=n$ could be ascribed
to a space charge induced resonance caused by the space charge
force nonlinearity and the breathing of the the beam envelopes.
Recall that the latter is among the lowest-order
nonlinear resonances that can be driven by space charge.

In conclusion, this preliminary study indicates that space
charge effects are noticeable in the low energy ring. One clear
consequence is the enlargement of strong half-integer structural
lattice resonances present in the bare lattice, causing fast
emittance growth and possibly, particle losses. This alone poses a
significant limitation to the choice of the working point because of
the sizeable space charge  vertical tune shift. On a longer time
scale, we encountered some areas of moderate, but clearly detectable,
emittance growth. Encouragingly, however, our calculations also show
the existence of regions in the tune space that appear little
affected by emittance growth. Further studies are needed to
insure that motion stability persists on a longer timescale, up
to a few damping times, and in the presence of lattice errors.

 \afterpage{\clearpage}
\subsection{HOM Heating and Intensity-Dependent Effects}

In this section we discuss the effects that must be considered in the
design of the vacuum chamber for a very high luminosity
\epem factory. We investigate the influence of the various
intensity-dependent effects on the actual performance of the
accelerator. The analysis is based on the parameters
listed in Table~\ref{table:HOM_1}.

\begin{table}[htb]
\caption{\label{table:HOM_1}
Main parameters of \superb\ used in the HOM analysis.}
\vspace*{2mm}
\centering
\setlength{\extrarowheight}{3pt}
\begin{tabular}{lcc}
\hline
\hline
                           &    LER         &         HER   \\ \hline
Energy (GeV)              &    $ 4   $     & $    7   $    \\
Beam current (A)        &    $ 2.3 $     & $    1.3 $    \\
Bunch length (mm)         &    $ 4.7 $     & $    5.0 $    \\
Beam energy spread         & $ 8.8\EE{-4}$  & $ 9.0\EE{-4}$ \\
SR energy loss (MeV/turn) &    $ 1.9 $     & $    3.3 $    \\
Long. damping time (ms)   &   $ 16   $ & $16   $           \\
Momentum compaction        & $ 1.8\EE{-4}$ & $ 3.0\EE{-4}$ \\
RF voltage (MV)           &    $ 6   $ & $ 18 $    \\
Synchrotron tune           &    $ 0.011 $ & $ 0.02 $ \\
Number of cavities         &    $ 14   $ & $   22 $    \\
RF frequency (MHz)        &  $ 476   $ & $  476   $ \\
Circumference (m)         & $ 2250   $ & $ 2250   $    \\
Revolution frequency (kHz)&  $ 130   $ & $  130   $    \\
Harmonic number            & $ 3570   $ & $ 3570   $    \\[5mm]
Number of bunches          &
\parbox{3cm}{\centering \vspace{3pt}
  $ 1733   $ \\ every second bucket \\$+5\%$ gap \vspace{3pt}}
                           & \parbox{3cm}{\centering \vspace{3pt}
                             $ 1733   $ \\ every second bucket\\$ +5\%$ gap
                             \vspace{3pt}}\\
\hline
\end{tabular}
\end{table}

In terms of collective effects, the dominant issue is the relatively
high beam current that must be supported in each ring. A beam circulating
in a storage ring interacts with its surroundings electromagnetically
by inducing image currents in the walls of the vacuum chamber and
exciting higher-order-modes (HOMs) in the chamber elements, such as
RF cavities, vacuum valves, collimators, bellows, beam position monitor
(BPM) electrodes, kickers, {\it etc}. This interaction leads, in turn, to
a temperature rise of the chamber elements, and may cause beam
instabilities. In the worst case, the HOM electric fields may be large
enough to produce sparking, or even breakdowns that may lead to beam
aborts due to bad vacuum conditions.

These issues fall into the
broad categories of single-bunch and multi-bunch phenomena and
higher-order-mode (HOM) heating. The main concern is
coupled-bunch instabilities, where different bunches ``communicate''
through the narrow-band ring impedances, \ie, wakefields
deposited in various resonant objects can influence the motion of
following bunches, and can cause the motion to become unstable if the
beam currents are too high. To avoid coupling to the bunch motion at
the bunch spacing resonance, HOMs must have a damping time
(loaded filling time) $\tau_l = 2 Q/\omega$ smaller than the bunch spacing
$\tau$.

The multi-bunch instabilities are mainly driven by the total beam
current, with little regard to how it is distributed in the ring,
\ie, once the bunch separation is small enough for bunches to see
fully the wakefields left by proceeding bunches, the growth rates become
independent of the details of the bunch pattern. Thus, if high beam
current is required, coupled-bunch instabilities become almost
unavoidable.
\begin{table}[htb]
\caption{\label{table:HOM_2}
Impedance and $Q$ values for monopole modes estimated from calculations and
measurements.  Shunt impedance is defined as $R =V^2/2P$.
}
\vspace*{2mm}
\setlength{\extrarowheight}{5pt}
\centering
\begin{tabular} {r  r@{.}l r@{} l r@{}l r r r}
\hline
\hline
$f_{\mathrm meas} $    &
\multicolumn {2}{c}{$ R/Q_{\mathrm meas} $  }&
\multicolumn {2}{c}{$ Q_{\mathrm meas} $  }&
\multicolumn {2}{c}{$ R_{\mathrm meas} $  }&
$ f_{\mathrm calc} $  &
$ R_{\mathrm calc} $  &
$ Q_{\mathrm calc}        $   \\

$(MHz)$    &
\multicolumn {2}{c}{$ (\Omega) $ }&
\multicolumn {2}{c}{$          $ }&
\multicolumn {2}{c}{$ (\Omega) $ }&
$ (MHz)      $  &
$ (\Omega)   $  &
$            $  \\
\hline

 $  476  $&$ 117$&$3\PM{0.00}{18.5} $&\multicolumn{2}{c}{$32469$}
 &\multicolumn{2}{c}{$ 3.809\EE{6} $}&$  476 $& &\\
 $  758  $&$ 44$&$6\pm{13.4}        $&$   18 $&\PM{0.0}{4.0} &$  809 $
 &\PM{241}{362} &$  758 $&$  879 $&$   15  $  \\
 $ 1009  $&$ 0$&$43\PM{0.00}{0.05}  $&$  128 $&\PM{0.0}{3.0} &$   55 $
 &\PM{0.0}{7.0} &$ 1010 $&$   35 $&$  100  $  \\
 $ 1283  $&$ 6$&$70\PM{6.4}{0.00}   $&$  259 $&\PM{47}{92}   &$ 1736 $
 &\PM{2272}{617}&$ 1291 $&$ 1013 $&$   88  $  \\
 $ 1295  $&$ 10$&$3\pm{2.1}         $&$  222 $&\PM{0.0}{88}  &$ 2287 $
 &\PM{455}{1184}&$ 1307 $&$ 1831 $&$  203  $  \\
 $ 1595  $&$ 2$&$43\PM{0.00}{2.14}  $&$  300 $&\PM{0.0}{170} &$  729 $
 &\PM{0.0}{691} &$ 1596 $&$  214 $&$   52  $  \\
 $ 1710  $&$ 0$&$44\pm{0.11}        $&$  320 $&\PM{125}{0.0} &$  141 $
 &\PM{104}{35}  &$ 1721 $&$  476 $&$   54  $  \\
 $ 1820  $&$ 0$&$13\pm{0.013}       $&$  543 $&\PM{0.0}{120} &$   70 $
 &\PM{7.0}{21}  &$      $&$      $&$       $  \\
 $ 1898  $&$ 0$&$17\pm{0.043}       $&$ 2588 $&\PM{0.0}{1693}&$  442 $
 &\PM{111}{328} &$ 1906 $&$  715 $&$  685  $  \\
 $ 2121  $&$ 1$&$82\pm{0.18}        $&$  338 $&\PM{69}{100}  &$  616 $
 &\PM{199}{226} &$ 2113 $&$ 1346 $&$  163  $  \\
 $ 2160  $&$ 0$&$053\pm{0.011}      $&$  119 $&\PM{10}{35}   &$    6 $
 &\PM{2.0}{3.0} &$ 2153 $&$  293 $&$  300  $  \\
 $ 2265  $&$ 0$&$064\pm{0.016}      $&$ 1975 $&\PM{0.0}{1314}&$  126 $
 &\PM{32}{95}   &$ 2263 $&$  450 $&$  306  $  \\
 $ 2344  $&\multicolumn{2}{c}{}      &$  693 $&\PM{0.0}{511} &$      $
 &              &$      $&$      $&$       $  \\
\hline
\end{tabular}
\end{table}

\begin{figure}[htb]
\centering
\includegraphics[width=0.9\textwidth]{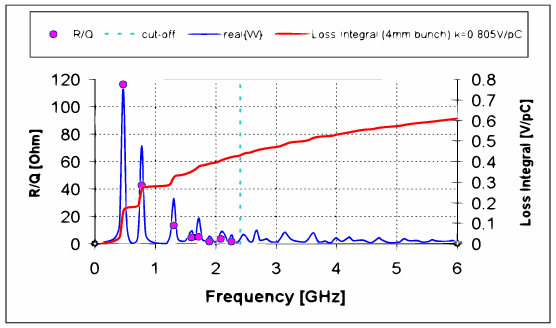}
\caption{\label{fig:HOM_1} \pepii cavity spectrum, R/Q and loss integral.}
\end{figure}

\subsubsection{RF cavities}

The main contribution to the narrow-band impedance comes from the
RF cavities. This means that HOMs trapped in the cavity must be very
well damped, as was done with the \pepii cavities. Measured and
calculated frequencies and Q~values of monopole higher-order modes
for the \pepii cavities are shown in Table~\ref{table:HOM_2}
(from ref~\cite{bib:HOM_1}).
The \pepii spectrum, calculated from the wake potential of a $4\mm$
bunch \cite{bib:HOM_5}, is shown in Fig.~\ref{fig:HOM_1}.

\paragraph{HOM power below cut-off frequencies.}

Power loss into the $n^{\mathrm{th}}$ mode in a cavity, according
to \cite{bib:HOM_2,bib:HOM_3}, is given by:

\begin{align}
\label{eq:HOM_1}
P_n &=I^2\times \tau\times k_n\\
P_n &=I^2\times \tau \times \frac{\omega_n}{2}\left(\frac{R}{Q}\right)
\frac
{1-\exp\left(-2 \frac{\tau}{\tau_{l,n}}\right)}
{1-2\;\exp\left(\frac{\tau}{\tau_{l,n}}\right) \cos \omega_n\tau +
\exp\left(-2 \frac{\tau}{\tau_l,n}\right)} \\
\tau_{l,n} &= 2 \frac{Q_{l,n}}{\omega_n}\,,
\end{align}
where $k_n$ is the loss factor, $I$ is the beam current, $\tau$ is the
bunch spacing and $Q_{l,n}$ is a loaded $Q$.  Table~\ref{table:HOM_3}
shows the HOM power into each bellow cut-off mode, and the
total loss of a \pepii cavity for the beam current of $1\amp$.

\begin{table}[tb]
\caption{\label{table:HOM_3}
HOM power in a \pepii cavity for modes below the cut-off for a current
of $1\amp$.}
\vspace*{2mm}
\setlength{\extrarowheight}{3pt}
\begin{tabular}{lcrrrrrrr} \hline\hline
\multicolumn{1}{c}{\parbox{2cm}{
\centering{
Mode frequency\\ (GHz)}}} &

\multicolumn{1}{c}{\parbox{1.14cm}{ \vspace{3pt}
\centering{ R/Q\\ (${\rm \Omega}$) }\vspace{3pt}}} &

\multicolumn{1}{c}{\parbox{1.14cm}{ \vspace{3pt}
\centering{ $Q_{\mathrm{load}}$}}} &

\multicolumn{1}{c}{\parbox{1.14cm}{ \vspace{3pt}
\centering{ Loss \\ factor \\ (V/\pC)}}}  &

\multicolumn{1}{c}{\parbox{1.14cm}{ \vspace{3pt} \centering{
Filling \\ time \\ (ms)}}}  &

\multicolumn{1}{c}{\parbox{1.14cm}{ \vspace{3pt}
\centering{$\cos()$ }}}&

\multicolumn{1}{c}{\parbox{1.14cm}{ \vspace{3pt}
\centering{ $\exp()$}}} &

\multicolumn{1}{c}{\parbox{1.14cm}{ \vspace{3pt}
\centering{Bunch \\ spacing \\ (ns)}}}   &

\multicolumn{1}{c}{\parbox{1.14cm}{ \vspace{3pt}
\centering{Power \\loss \\ (kW)}}} \\
\hline

0.475997 & 117.3  & 8000  & 0.1754  & 2.675   &  1.000   & 0.9969
& 4.202  & 0      \\
0.758    & 44.6   & 18    & 0.1062  & 0.004   &  0.398   & 0.1082
& 4.202  & 0.4701 \\
1.009    & 0.43   & 128   & 0.0014  & 0.020   &  0.066   & 0.6595
& 4.202  & 0.0013 \\
1.283    & 6.7    & 259   & 0.0270  & 0.032   & -0.774   & 0.7699
& 4.202  & 0.0083 \\
1.295    & 10.3   & 222   & 0.0419  & 0.027   & -0.933   & 0.7349
& 4.202  & 0.0140 \\
1.595    & 2.43   & 300   & 0.0122  & 0.030   & -0.299   & 0.7552
& 4.202  & 0.0055 \\
1.71     & 0.44   & 320   & 0.0024  & 0.030   &  0.398   & 0.7542
& 4.202  & 0.0023 \\
1.82     & 0.13   & 543   & 0.0007  & 0.047   & -0.602   & 0.8378
& 4.202  & 0.0002 \\
1.898    & 0.17   & 2588  & 0.0010  & 0.217   &  0.988   & 0.9620
& 4.202  & 0.0065 \\
2.121    & 1.82   & 338   & 0.0121  & 0.025   &  0.850   & 0.7180
& 4.202  & 0.0519 \\
2.16     & 0.053  & 119   & 0.0004  & 0.009   &  0.889   & 0.3835
& 4.202  & 0.0033 \\
2.265    & 0.064  & 1975  & 0.0005  & 0.138   & -0.994   & 0.9412
& 4.202  & 0.0000 \\[5mm]
Total    & 184.437 &       & 0.3811  &         &          &
&        & 0.5635 \\
\hline
\end{tabular}
\end{table}

\paragraph{HOM power for above cut-off frequencies.}

\begin{figure}[htb]
\centering
\includegraphics[width=0.8\textwidth]{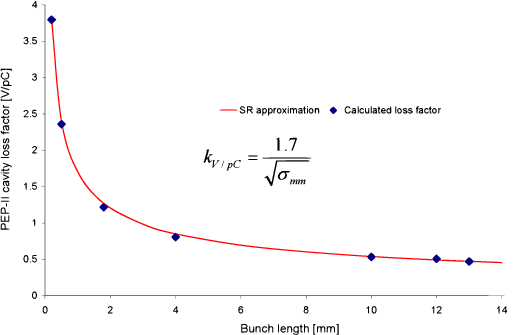}
\caption{ \label{fig:HOM_2}
Wakefield loss factor.
}
\end{figure}

The calculated loss factor~\cite{bib:HOM_4,bib:HOM_5}
for different bunch lengths is shown in Fig.~\ref{fig:HOM_2}.
HOM power above the cut-off frequency for a PEP-II cavity~\cite{bib:HOM_5} is given by
$$
P_{\kW} = \left( \frac{1.7}{\sqrt{\sigma_{\mm}}} - 0.3811 \right) \times 4.2
\times I_{\amp }{^2}\,.
$$
HOM losses below and above the cut-off frequencies are shown in Table~\ref{table:HOM_4}.
The total HOM power for all cavity losses is also shown for the \pepii and \superb\ parameters.

\begin{table}[htb]
\caption{ \label{table:HOM_4}
Total HOM power for all cavities.}
\vspace*{2mm}
\setlength{\extrarowheight}{1pt}
\centering
\begin{tabular}{lcccc}
\hline
\hline
&\multicolumn{2}{c}{LER}&\multicolumn{2}{c}{HER}\\
& \pepii & \superb\  & \pepii & \superb\  \\
\hline
Beam current (A)             & 2.9    & 2.3   & 1.8   & 1.3   \\
Bunch length (mm)            & 13     &  6    & 11    &  6    \\
Number of cavities           &  8     & 12    & 26    & 22    \\
Power below cut-off (kW)     & 37.91  & 23.85 & 47.47 & 22.86 \\
Power above cut-off (kW)     & 25.54  & 55.62 & 46.51 & 53.31 \\
Total HOM cavity power (kW)  & 63.46  & 79.4  & 93.98 & 76.16 \\
\hline
\end{tabular}
\end{table}

\subsubsection{Resistive-wall wakefields}
The resistive wall loss factor from~\cite{bib:HOM_6} is given by
$$
s_0 = \left(2 a^2 \frac{\rho}{Z_0} \right)^{\frac{1}{3}}
\quad \mathrm{ when } \quad
\frac{s_0}{\sz} \ll 1
$$
$$
k_{RW} \approx 0.2 \times \frac{Z_0 c}{4 \pi a^2}\times
\left( \frac{1}{\sz} \right)^{\frac{3}{2}}\times \sqrt{2\frac{\rho}{Z_0}}
\times F(a,b)\,.
$$
The calculated resistive wall losses for the LER and HER rings of \pepii and
\superb\ are shown in Table~\ref{table:HOM_5}.

\begin{table}[htb]
\setlength{\extrarowheight}{1pt} \caption{ \label{table:HOM_5} LER
and HER resistive wall losses. Losses were calculated under the assumption
that the \superb\ and \pepii chambers have the same material composition.
The \pepii LER chambers consist of 10\% copper, 50\% aluminum, and
40\% stainless steel, while the HER chambers are 60\% copper and 40\%
stainless steel.}
\centering \vspace*{2mm}
\begin{tabular}{lcccc}
\hline
\hline
&\multicolumn{2}{c}{LER}&\multicolumn{2}{c}{HER}\\
& \pepii & \superb\  & \pepii & \superb\  \\
\hline
Bunch length (mm)       & 13    & 6      & 11    & 6     \\
Bunch spacing (nsec)    & 4.2   & 4.2    & 4.2   & 4.2   \\
Beam current (A)        & 2.9   & 2.3    & 1.8   & 1.3   \\
Power (kW)              & 71.74 & 143.92 & 36.15 & 46.81 \\
\hline
\end{tabular}
\end{table}

\subsubsection{Beam chamber elements}

\paragraph{\pepii collimator wakefield.}

\begin{figure}[htb]
\centering
\includegraphics[width=\textwidth]{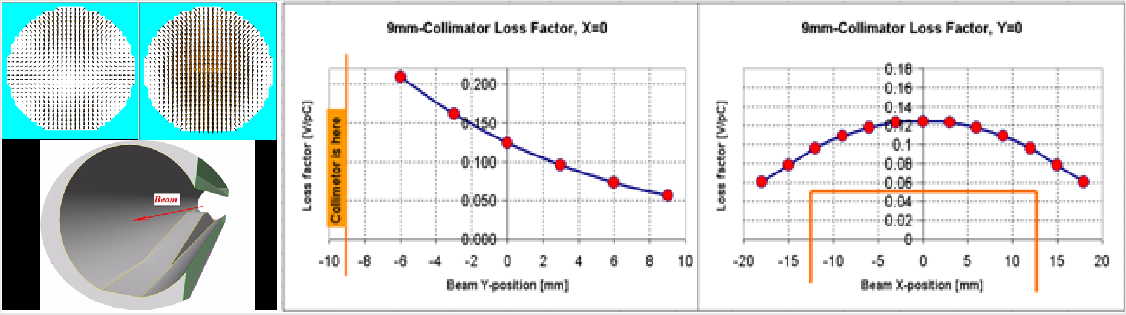}
\caption{ \label{fig:HOM_3}
\pepii collimator (left) and calculated loss factor (right) for different
beam positions relative to the collimator.}
\end{figure}

\begin{figure}[htb]
\centering
\includegraphics[width=0.7\textwidth]{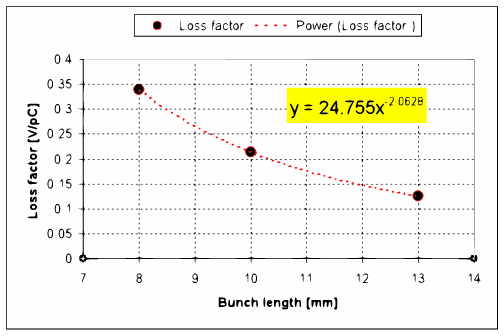}
\caption{ \label{fig:HOM_4}
Bunch length dependence of the collimator loss factor.
}
\end{figure}

The geometry of a \pepii collimator is shown in Fig.~\ref{fig:HOM_3}.
The loss factor for a $13 \mm$
bunch is also shown for different positions of the beam
collimator~\cite{bib:HOM_7}.
The bunch length dependence is shown in Fig.~\ref{fig:HOM_4}.
Wakefield collimator losses for \pepii and \superb\ parameters are
shown in Table~\ref{table:HOM_7}.

\begin{table}[htb]
\caption{ \label{table:HOM_7}
Wakefield power from \pepii collimators for \pepii and \superb\ parameters.
}
\vspace*{2mm}
\setlength{\extrarowheight}{1pt}
\centering
\begin{tabular}{lcccc}
\hline
\hline
                         &\multicolumn{2}{c}{LER}&\multicolumn{2}{c}{HER}\\
                         & \pepii & \superb\ & \pepii & \superb\ \\
\hline
Beam current (A)         &  2.9  &  2.3  &  1.8 &  1.3 \\
Bunch length (mm)        & 13    &  6    & 11   &  6   \\
Number of collimators    &  7    &  7    &  6   &  6   \\
Wakefield power (kW)    & 18.1  & 53.5  & 16.7 & 29.3 \\
\hline
\end{tabular}
\end{table}

\paragraph{HOM power in injection and abort kickers.}

\begin{figure}[htb]
\centering
\includegraphics[width=0.8\textwidth]{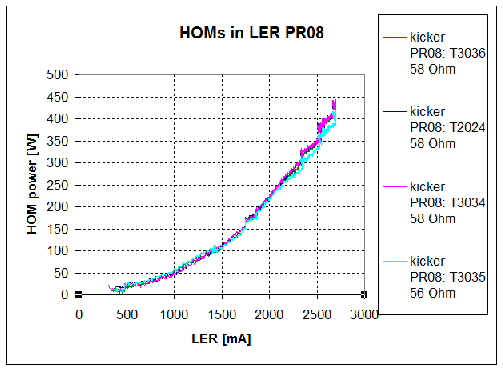}
\caption{\label{fig:HOM_5}
Dissipated power in injection and abort kickers of the \pepii LER.
}
\end{figure}

Figure~\ref{fig:HOM_5} shows the beam current dependence of the power
dissipated in the injection and abort kickers of the \pepii LER.
At a beam current of $3\amp$, the power in these four LER kickers reaches
$2\kW$ for \pepii parameters. If we assume that the bunch length dependence
is the same as that for the
resistive-wall wake-field losses, $\sigma^{-3/2}$, then the \superb\ LER
will have $3.8\kW$ dissipated for a beam current of $2.3 \amp$.

\paragraph{Loss factor and wake-field power of longitudinal kickers.}

\begin{figure}[htb]
\centering
\includegraphics[width=0.9\textwidth]{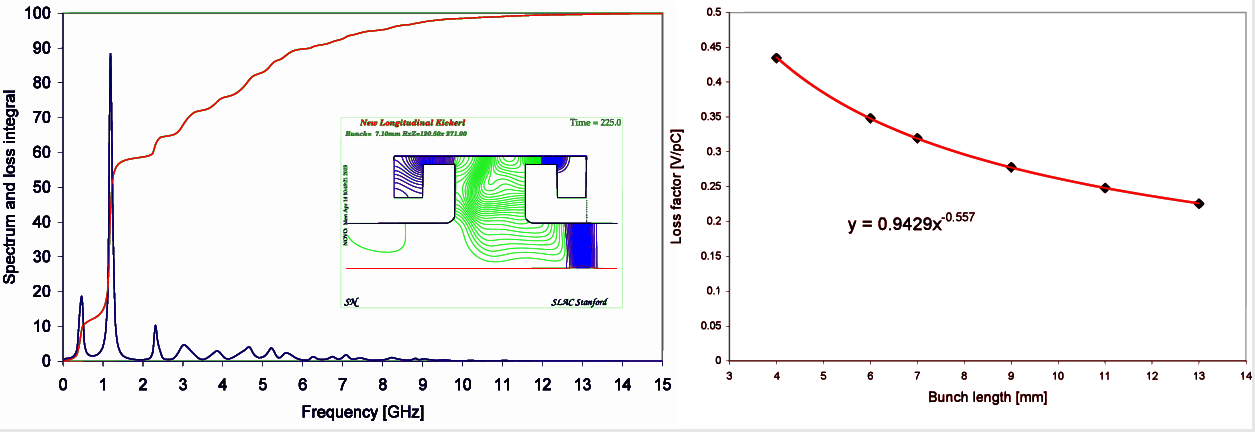}
\caption{\label{fig:HOM_6}
  Longitudinal kicker spectrum (left) and loss factor as a
  function of bunch length (right).
}
\end{figure}

\begin{figure}[htb]
\centering
\includegraphics[width=0.8\textwidth]{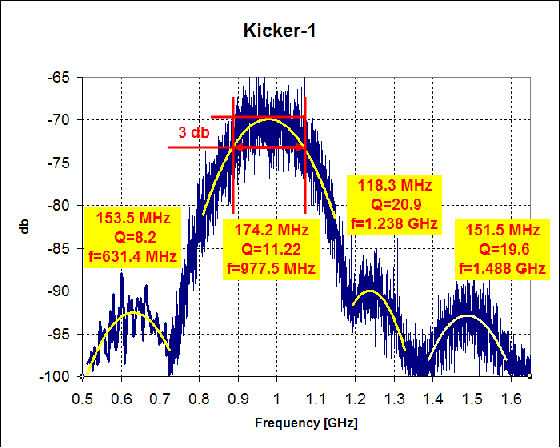}
\caption{\label{fig:HOM_7}
Measured single bunch spectrum for a longitudinal kicker.
}
\end{figure}

The longitudinal kicker spectrum and the loss factor as a function of bunch
length from an azimuthally symmetric model are shown in Fig.~\ref{fig:HOM_6}.
The measured single bunch spectrum for a longitudinal kicker is shown in
Fig.~\ref{fig:HOM_7}.
The wake-field power in two longitudinal kickers is shown in Table~\ref{table:HOM_8}
for the \pepii and \superb\ LER and HER parameters.

\begin{table}[htb]
\caption{ \label{table:HOM_8}
Wakefield power from two longitudinal kickers for
\pepii and \superb\ parameters.
}
\vspace*{2mm}
\setlength{\extrarowheight}{1pt}
\centering
\begin{tabular}{lcccc}
\hline
\hline
&\multicolumn{2}{c}{LER}&\multicolumn{2}{c}{HER}\\
& \pepii & \superb\  & \pepii & \superb\  \\
\hline
Beam current (A)               & 2.9  & 2.3  & 1.8  & 1.3  \\
Bunch length (mm)              & 13   & 6    & 11   & 6    \\
Number of longitudinal kickers & 2    & 2    & 2    & 2    \\
Wakefield power (kW)          & 5.93 & 7.74 & 2.68 & 2.47 \\
\hline
\end{tabular}
\end{table}

\paragraph{Loss factor and wake-field power for transverse kickers.}

\begin{figure}[htb]
\centering
\includegraphics[width=0.8\textwidth]{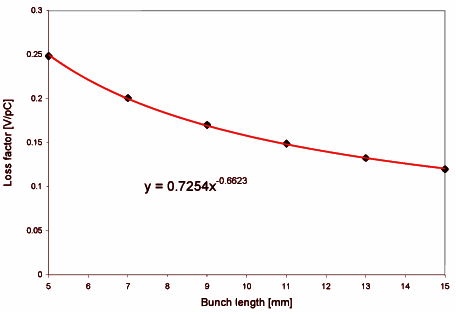}
\caption{\label{fig:HOM_8}
Transverse kicker loss factor as a function of bunch length.
}
\end{figure}

The transverse kicker loss factor as a function of bunch length from an
azimuthally symmetric model is shown in Fig.~\ref{fig:HOM_8}.
Wake-field power in two transverse kickers is shown in Table~\ref{table:HOM_9}
for \pepii and \superb\ LER and HER parameters.

\begin{table}
\caption{ \label{table:HOM_9}
Wake-field power from two transverse kickers for \pepii and
\superb\ parameters.
}
\vspace*{2mm}
\setlength{\extrarowheight}{1pt}
\centering
\begin{tabular}{lcccc}
\hline
\hline
&\multicolumn{2}{c}{LER}&\multicolumn{2}{c}{HER}\\
& \pepii & \superb\  & \pepii & \superb\  \\
\hline
Beam current (A)             & 2.9  & 2.3  & 1.8  & 1.3  \\
Bunch length  (mm)           & 13   & 6    & 11   & 6    \\
Number of transverse kickers & 2    & 2    & 2    & 2    \\
Wake-field power (kW)        & 9.37 & 7.84 & 2.68 & 2.47 \\
\hline
\end{tabular}
\end{table}

\paragraph{Distributed Pumps.}

The loss factor per unit length is estimated from~\cite{bib:HOM_8}:
$$
k = \frac{5 Z_0 c}{144 \pi^{\frac{7}{2}}} \frac {r^6}{a^2\sigma^5} g
$$
where $g \approx 10$ (coherence effect).
Results for the coupled power in distributed pumps are shown
in Table~\ref{table:HOM_10} for \pepii and \superb\ parameters.

\begin{table}[htb]
\caption{ \label{table:HOM_10}
Coupled power in distributed pumps.
}
\vspace*{2mm}
\setlength{\extrarowheight}{1pt}
\centering
\begin{tabular}{lcccc}
\hline
\hline
&\multicolumn{2}{c}{LER}&\multicolumn{2}{c}{HER}\\
& \pepii & \superb\  & \pepii & \superb\  \\
\hline
Beam current (A)      & 2.9  & 2.3   & 1.8  & 1.3   \\
Bunch length (mm)     & 13   & 6     & 11   & 6     \\
Wake-field power (kW) & 1.24 & 37.21 & 5.50 & 59.44 \\
\hline
\end{tabular}
\end{table}

\subsubsection{IP wakefields}

\begin{figure}[htb]
\centering
\includegraphics[width=0.7\textwidth]{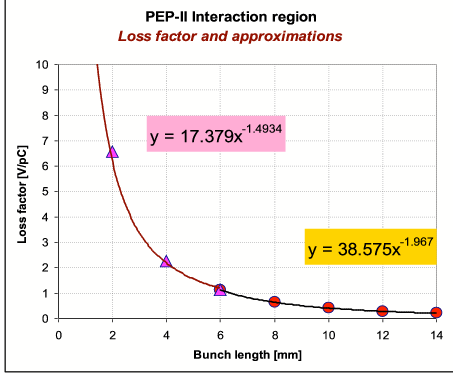}
\caption{\label{fig:HOM_9}
IP geometrical loss factor as a function of bunch length.
}
\end{figure}

The calculated geometrical loss factor is shown in Fig.~\ref{fig:HOM_9} 
as a function of bunch length. The IP ``geometrical'' HOM power is given in
Table~\ref{table:HOM_11} for \pepii and \superb\ parameters.

\begin{figure}[htb]
\centering
\includegraphics[width=0.8\textwidth]{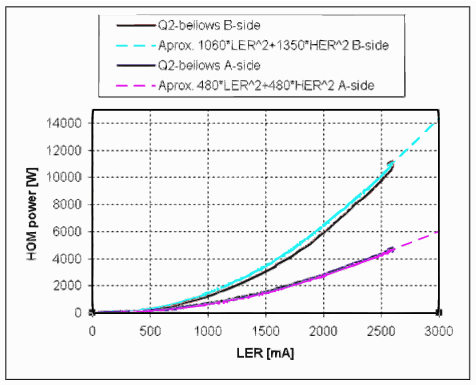}
\caption{\label{fig:HOM_10}
Measured power loss in the \pepii Q2 bellows. The ``A'' side is at the rear end
of \babar, while the ``B'' side is at the forward end.
}
\end{figure}

Additional substantial HOM power loss will likely occur in the transition
bellows from the single beam pipe in the IR to the separate HER and
LER beam pipes in the rest of the ring. In PEP-II, these
forward and backward Q2 transition bellows contain ``open'' ceramic tiles
that generate power loss through Cherenkov radiation~\cite{bib:HOM_9}. The
loss factors are given by:
\begin{alignat*}{3}
\mathrm{when } \; \sigma & >  s = \frac{a \sqrt{\varepsilon-1}}
  {2 \varepsilon} &\quad \mathrm{ loss\ factor } \quad
k & =  \frac{c z_0 L}{ 2 \pi a^2} \times \frac{s}{\sqrt{\pi} \sigma} \\
\mathrm{when } \; \sigma & <  s
   &\quad \mathrm{ loss\ factor} \quad
k & =  \frac{c z_0 L}{ 2 \pi a^2}
\end{alignat*}
Power loss due to Cherenkov radiation in the forward and backwards
Q2 bellows in \pepii, and
extrapolations for equivalent components with \superb\ parameters, is shown
in Table~\ref{table:HOM_12}. The measured
power in the Q2 bellows is shown in Fig.~\ref{fig:HOM_10}.
Additional numerical and experimental study is needed to make a more
precise prediction for \superb.

\clearpage
\begin{table}[htb]
\caption{\label{table:HOM_11}
IP ``geometrical'' HOM power.
}
\vspace*{2mm}
\setlength{\extrarowheight}{1pt}
\centering
\begin{tabular}{lcc} \hline\hline
Parameters            & \pepii & \superb\ \\ \hline
Bunch length (mm)     & 13    & 6     \\
Loss factor (V/pC)    & 0.248 & 1.137 \\
LER current (A)       & 2.9   & 2.3   \\
HER current (A)       & 1.8   & 1.3   \\
Bunch spacing (ns)    & 4.2   & 4.2   \\
LER power loss (kW)   & 8.77  & 25.26 \\
HER power loss (kW)   & 3.38  & 8.07  \\
Total power loss (kW) & 12.15 & 33.33 \\
\hline
\end{tabular}
\end{table}

\begin{table}[htb]
\caption{\label{table:HOM_12}
Power loss in the forward and backward Q2 transition bellows for \pepii and
extrapolations to \superb\ parameters.
}
\vspace*{2mm}
\setlength{\extrarowheight}{1pt}
\centering
\begin{tabular}{lcc} \hline\hline
Parameters                & \pepii & \superb \\ \hline
$\varepsilon$ (mm-rad)    & 30.00 & 30.00   \\
L (mm)                    & 59.20 & 59.20   \\
Bunch length (mm)         & 13.00 &  6.00   \\
s/$\sigma$                &  0.41 &  0.90   \\
Loss factor (V/pC)        &  0.07 &  0.30   \\
LER current (A)           &  2.9  &  2.3    \\
HER current (A)           &  1.8  &  1.3    \\
Bunch pattern by 2 (ns)   &  4.20 &  4.20   \\
LER power loss (kW)       &  4.89 & 13.15   \\
HER power loss (kW)       &  1.88 &  4.20   \\
Total power loss (kW)     &  6.77 & 17.36   \\
\hline
\end{tabular}
\end{table}

\subsubsection{Total HOM power}

The calculated total HOM power for \pepii and \superb\ is summarized in
Table~\ref{table:HOM_13} for the LER and HER.

\begin{table}[htb]
\caption{\label{table:HOM_13}
Calculated total HOM power.}
\vspace*{2mm}
\setlength{\extrarowheight}{1pt}
\centering
\begin{tabular}{lcccc} \hline\hline
 & \multicolumn{2}{c}{\centering LER}
 & \multicolumn{2}{c}{\centering HER} \\
 & \pepii & \superb & \pepii & \superb \\ \hline
RF cavities      &  63.46 &  79.47 &  93.98 &  76.16\\
Resistive wall   &  71.74 & 143.02 &  36.15 &  46.81\\
Collimators      &  18.11 &  53.47 &  16.7  &  29.29\\
Kickers          &  17.3  &  21.38 &   6.08 &   6.14\\
Screens          &   1.24 &  37.21 &   5.5  &  59.44\\
IP wakes         &  13.66 &  38.41 &   5.26 &  12.27\\
Total power (kW) & 185.51 & 372.96 & 163.67 & 230.11\\
\hline
\end{tabular}
\end{table}

\subsubsection{Total RF power}

Table~\ref{table:HOM_14} shows the measured~\cite{bib:HOM_10}
total RF power for the \pepii LER and HER. 
The measured losses, particuarly for the \pepii HER, are greater 
than the calculated losses listed in Table~\ref{table:HOM_13}.
The source of this discrepancy remains to be understood, although
it should be noted that the determination of the total measured 
consumption depends on a knowledge of klysteron and modulator
efficiencies. Rescaling the power losses observed in \pepii to
the \superb\ design gives the revised predictions also
shown in Table~\ref{table:HOM_14}. 

\begin{table}[htb]
\caption{\label{table:HOM_14} Measured RF power distribution in the HER and LER
of \pepii with extrapolation to \superb. }
\vspace*{2mm}
\setlength{\extrarowheight}{1pt}
\centering
\begin{tabular}{lcccc} \hline\hline
 & \multicolumn{2}{c}{\centering LER}
 & \multicolumn{2}{c}{\centering HER} \\
                       & \pepii & \superb & \pepii & \superb \\ \hline
Beam current (A)       & 2.9   & 2.3     & 1.8   & 1.3   \\
Bunch length (mm)      & 13    & 6       & 11    & 6     \\
S.R. loss (MeV/turn)   & 0.56  & 1.9     & 3.16  & 3.3   \\
RF voltage (MV)        & 4.05  & 10      & 15.4  & 15    \\
Number of cavities     & 8     & 14      & 26    & 22    \\
Number of klystrons    & 4     & 7       & 9     & 8     \\
Reflection coefficient & 0.142 & 0.202   & 0.083 & 0.010 \\
S.R. power (kW)        & 1624  & 4370    & 5688  & 4290  \\
Cavity loss (kW)       &  250  &  549    & 1187  & 1622  \\
Refl. power (kW)       &  345  & 1355    &  652  &   64  \\
HOM power (kW)         &  210  &  422    &  292  &  378  \\
Total RF power (kW)    & 2429  & 6695    & 7819  & 6354  \\
\hline
\end{tabular}
\end{table}

\clearpage

\afterpage{\clearpage}
\subsection{Single Bunch Impedance Effects}

We have made an estimate of single-bunch instabilities for
\superb. We use the parameters for the 12 cell design in
ref.~\cite{bib:agoh_1}, which are listed in
Table~\ref{table:bunch_instability} along with \kekb parameters.

\begin{table}[htb]
\caption{\label{table:bunch_instability}
\superb\ parameters for the 12 cell design and \kekb parameters for comparison.
}
\vspace*{2mm}
\setlength{\extrarowheight}{1pt}
\centering
\begin{tabular}{lcccc} \hline\hline
 & \multicolumn{2}{c}{\centering \superb}
 & \multicolumn{2}{c}{\centering \KEKB} \\
                                 & LER & HER & LER & HER \\ \hline
Energy (GeV)                           & 4    & 7      & 3.5   & 8   \\
Circumference (m) & \multicolumn{2}{c}{2250}& \multicolumn{2}{ c}{3016}\\
Bunch length (mm)                      & 4.7  & 5.0    & 6     & 6   \\
Energy spread $(\times 10^{-3})$       & 1.0  & 1.0    & 0.73  & 6.9 \\
Momentum compaction $(\times 10^{-4})$ & 1.8  & 3.0    & 3.4   & 3.4 \\
Particles per bunch $(\times10^{10})$  & 6.16 & 3.52   & 7     & 5.2 \\ \hline
\end{tabular}
\end{table}

\subsubsection{Longitudinal impedance}

The instability threshold in longitudinal impedance is given by the
Keil-Schnell-Boussard criterion \cite{bib:agoh_2, bib:agoh_3}

\begin{equation}
\label{eq:agoh}
\frac{Z}{n} = Z_0\sqrt{\frac{\pi}{2}}\frac{\gamma\,\alpha_p\,
            \sigma_\delta^2\, \sz}{N\, r_e}.
\end{equation}

Using this equation, we estimated the thresholds $Z/n$ for \superb\ and \kekb
listed in Table~\ref{table:agoh}.

\begin{table}[hbt]
\caption{ \label{table:agoh}  Instability threshold in longitudinal impedance
$Z/n$ for \superb\ and \KEKB along with the measured values in \KEKB \cite{bib:agoh_4}. }
\vspace*{2mm}
\setlength{\extrarowheight}{1pt}
\centering
\begin{tabular}{lcccc} \hline\hline
 & \multicolumn{2}{c}{\centering \superb}
 & \multicolumn{2}{c}{\centering \kekb} \\
 & LER & HER & LER & HER \\ \hline
Threshold $Z/n$ by Eq.~\ref{eq:agoh}
       & $ 18 \mOhm $ & $ 98 \mOhm $ & $ 17 \mOhm $ & $ 49 \mOhm $ \\
Measured impedance $Z/n$ & & & $ 72 \mOhm$ & $ 76 \mOhm$ \\ \hline
\end{tabular}
\end{table}

By comparing with the measured impedance in the \KEKB HER~\cite{bib:agoh_4},
the \superb\ HER seems
to be safe from longitudinal instability, although the safety margin may
be small. On the other hand, the threshold impedance is very small in the
\superb\ LER. Since the momentum compaction factor is small in the present
\superb\ LER design with $\alpha_p=1.8 \times 10^{-4}$, it takes time to
damp the microbunching, \ie, structures inside the bunch can last a long time
and even grow in the LER.
As a result of the potential well distortion from longitudinal
impedances, bunches in the LER will be lengthened by roughly $1 \mm$, to an
equilibrium value around $6 \mm$. Although this effect relaxes somewhat the
instability threshold, due to the 20\% reduction in peak current, the
longitudinal instability in the \superb\ LER remains severe.

However, it is instructive to note that the \KEKB LER operates
without a longitudinal microwave instability.
In this case, Eq.~\ref{eq:agoh} predicts $ Z/n=17 \mOhm$ for the KEK LER,
similar to the value calculated for the \superb\ LER, while the actual
measured longitudinal impedance is $Z/n=72 \mOhm$.
This experience suggests that we may be able to operate the \superb\ LER
using the present design, although the momentum compaction factor may be too
small. It has been noted in ref.~\cite{bib:agoh_3,bib:agoh_5} that
our argument using the simple criterion is not sufficiently reliable, because
the instability threshold strongly depends on the impedance characteristics.
Therefore, more detailed studies are
necessary to address the longitudinal impedance problem.
We should at the same time make efforts
to reduce the longitudinal impedance, especially in LER.

\subsubsection{Transverse impedance}
The threshold impedance for transverse single-bunch instability is given by
\begin{equation}
\label{eq:agoh_2}
Z_\perp \beta_\perp = Z_0 \frac{4 \sqrt{2}}{3}
\frac {\gamma\, \alpha_p \, \sigma_\delta \,C}{N\, r_e}.
\end{equation}

The threshold impedances for \superb\ are listed in Table~\ref{table:agoh_2}.
\begin{table}[hbt]
\caption{\label{table:agoh_2}
Instability threshold in transverse impedance for
\superb\ and measured impedance for \KEKB.}
\vspace*{2mm}
\setlength{\extrarowheight}{1pt}
\centering
\begin{tabular}{lccc} \hline\hline
 & \multicolumn{2}{c}{\centering \superb}
 &  \kekb \\
 & LER & HER & LER \\ \hline
Threshold $Z_\perp \beta_\perp$ by Eq.~\ref{eq:agoh_2}
   & $ 13  \MOhm $ & $ 66  \MOhm $ & $ 18  \MOhm $ \\
Measured impedance $Z_\perp \beta_\perp$
& & & $1.2 \sim 2.1 \MOhm$ \\ \hline
\end{tabular}
\end{table}

The transverse impedance in the \KEKB LER was measured
to be $Z=80\sim139 \kOhm/\m$
\cite{bib:agoh_4}. The average beta function in \KEKB is $\beta\sim 15\m$;
$Z\beta$ is estimated to be between $1.2$ to $2.1 \MOhm$. Compared with the
values of the
\KEKB impedance, we find that both the \superb\ LER and HER will operate well
below threshold, and should have no problem with transverse single bunch
instability. However, a more detailed investigation should be made of both the
transverse and the longitudinal impedance. In particular,
as noted in ref.~\cite{bib:agoh_4}, collimators will dominate the
transverse impedance
in the ring, and so special care will be needed in their design.

It may also be that the transverse mode coupling (TMC) instability will
limit beam intensity in
\superb. We made a rough estimation for the threshold of the TMC
instability using the following equation:
\begin{equation}
\label{eq:agoh_3}
Z_\perp \beta_\perp =Z_0\frac{4 \, \gamma\, \nu_s\, b}{\pi\, N\, r_e}\,.
\end{equation}
In the \superb\ design, \pepii magnets will be reused to reduce costs, so we
use the \pepii vacuum chamber size of $b = 25\mm$
(full height $h=50 \mm$ in the \pepii LER bends).
The threshold for the \superb\ LER is $Z \beta = 6.5 \MOhm$,
which corresponds to
$Z/n \sim Z_\perp\, \pi\,b^2\,/C \sim 280 \mOhm$, where we are assuming that
the average beta function is $\beta=20 \m$. It appears that this threshold
is high enough to avoid transverse mode coupling instability.

\afterpage{\clearpage}
\subsection{Coherent Synchrotron Radiation}

With a very short bunch length, coherent synchrotron radiation (CSR)
emission can drive microwave instability. In particular, the low
energy beam could be unstable if the bunch charge is large, because of the
smaller energy. We will discuss possible single bunch instabilities due to
CSR in the \superb\ LER, and estimate the margin to the instability threshold.
If the low energy beam is stable for CSR-source microwave instability,
the HER should be as well. For the HER, the higher energy results in greater rigidity
and stronger damping, while CSR does not depend on the particle energy in this energy regime. It is therefore sufficient to investigate CSR effects for LER
alone.

The bunch length is designed to be $5 \mm$ in the
\superb\ LER~\cite{bib:agoh_4}.
The ring will have three types of bending magnets in the present design: short,
middle bends for arc-sections and long bends for final focus section, to obtain
a small emittance and to reduce the cost by re-using \pepii magnets. The dipole
parameters are listed in Table~\ref{table:csr1}.

\begin{table}[h!tb]
\caption{ \label{table:csr1}  Dipole parameters for the \superb\ LER.}
\vspace*{2mm}
\setlength{\extrarowheight}{1pt}
\centering
\begin{tabular}{cccc}
\hline\hline
\multicolumn{1}{c}{\parbox{4.5cm}{\centering Name   (length)}}
  & Bending radius     & Number of bends &\\\hline
Short  ($0.45\m$) & $22.5 \m$  & $144$ & \pepii \\
Middle ($0.75\m$) & $28.4 \m$  & $144$ & new \\
Long   ($5.4 \m$) for FF
                   & $116 \m$ (average) & $16$  & \pepii \\ \hline
\end{tabular}
\end{table}

Since the vacuum chambers currently used in the \pepii LER bends have
an inside dimension of $90 \mm$ in width and $50 \mm$ in height
\cite{bib:agoh_12},
we use a $w \times h = 90 \times 50 \mm$ rectangular copper pipe
in our analysis.

The current \superb\ LER design has four wiggler sections for damping.
However, we do not consider CSR emitted in the wigglers in our
analysis, because the bending magnets in the arc sections will
dominate. The wigglers are, however, considered in calculating the synchrotron
radiation loss of the beam. In this section we consider CSR generated
in the arc sections and the resistive wall wakefield for a $2 \km$ long
ring. We note that the bunches may be lengthened by 0.5--$1\mm$
because of other wakefields, \eg, due to vacuum components and
cavities. As a result, our analysis may give a somewhat overly
severe threshold for the CSR-source microwave instability.

\begin{figure}[htb]
\centering
\includegraphics[width=0.8\textwidth]{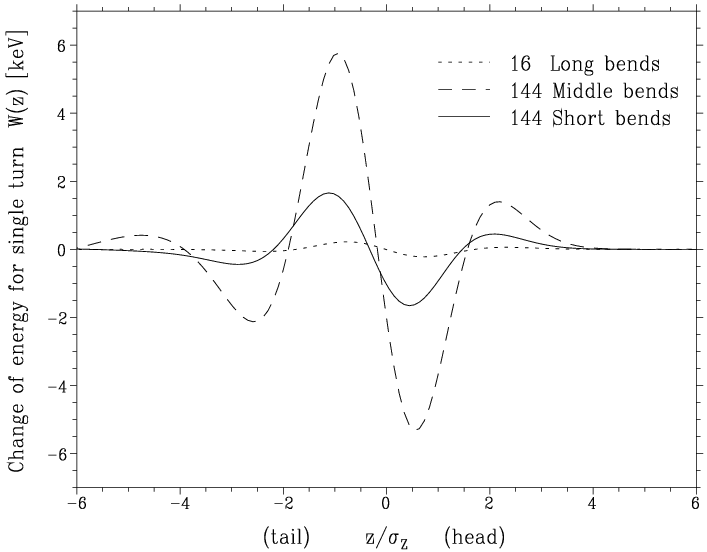}
\caption{\label{fig:csr_1}
CSR longitudinal wakefield for single turn. The three
lines show the CSR wakefield for three types of bend magnets:
dotted line for the 16 long bends, dashed line for the 144 middle bends, and
solid line for the 144 short bends.}
\end{figure}

The CSR wakefield for a single turn is shown in Fig.~\ref{fig:csr_1}, where
we assume a perfectly conducting pipe, in order to remove the resistive
wall wakefield. The middle bending magnets give the largest
contribution, but the magnitude of the longitudinal wakefield is just
$\pm 6 \kev$ for a single turn.

\begin{figure}[htb]
\centering
\includegraphics[width=0.8\textwidth]{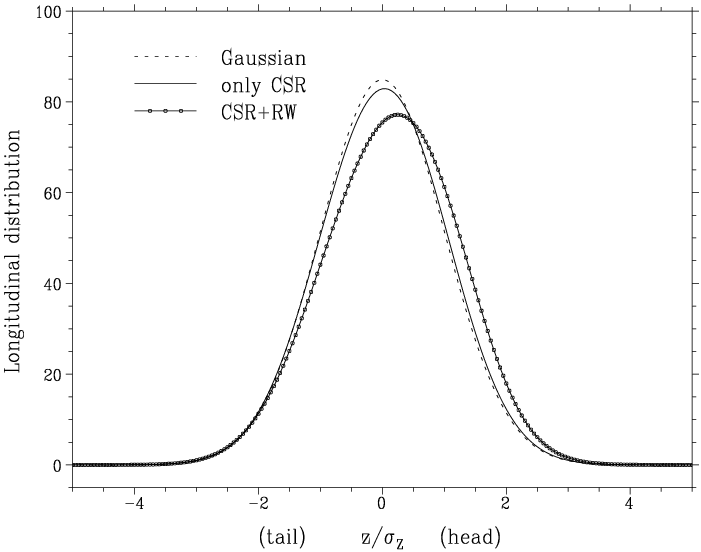}
\caption{\label{fig:csr_2}
Longitudinal distribution with CSR and resistive wall
(RW) wakefields.  The dotted line is a gaussian distribution (with no
wakefield), the solid curve considers only the CSR contribution,
and the solid curve with
circles shows the combined CSR and resistive-wall wakefields for a $2\km$
length.
}
\end{figure}

We have studied the effect of CSR and resistive
wall wakefields on the longitudinal dynamics of the machine design by
solving the Fokker-Planck equation. For the design bunch
charge, $N_e = 9.87 \nC$  ($N = 6.16 \times 10^{10}$),
the resulting longitudinal bunch distribution is shown in Fig.~\ref{fig:csr_2}
for a $2\km$ drift length.
In order to isolate the CSR contribution alone, we also show
the distribution after removing the resistive wall wakefield contribution.
We observe that the distortion of the longitudinal
bunch distribution is mainly determined by the resistive wall
wakefield.

\begin{figure}[htb]
\centering
\includegraphics[width=0.8\textwidth]{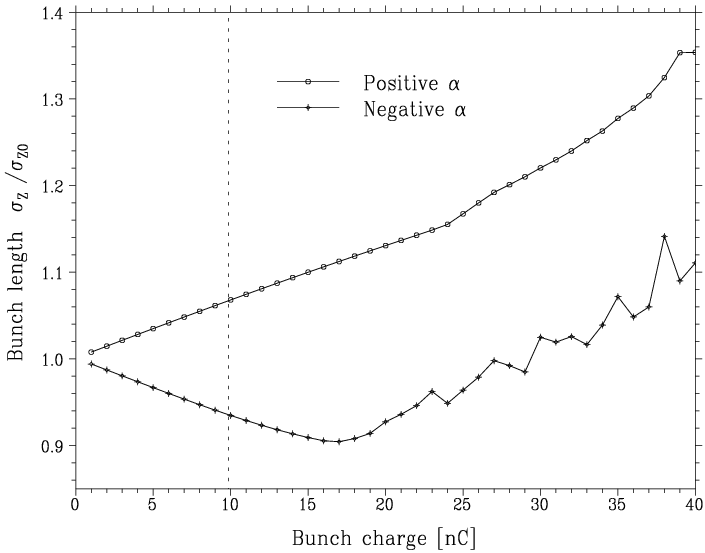}
\caption{\label{fig:csr_3}
Bunch charge dependence of the {\it rms} bunch length.
The circles ($\circ$) and crosses ($+$)
correspond to positive and negative momentum compaction factors. The
dotted vertical line is the design bunch charge of $9.87 \nC$.
}\end{figure}

\begin{figure}[tb]
\centering
\includegraphics[width=0.8\textwidth]{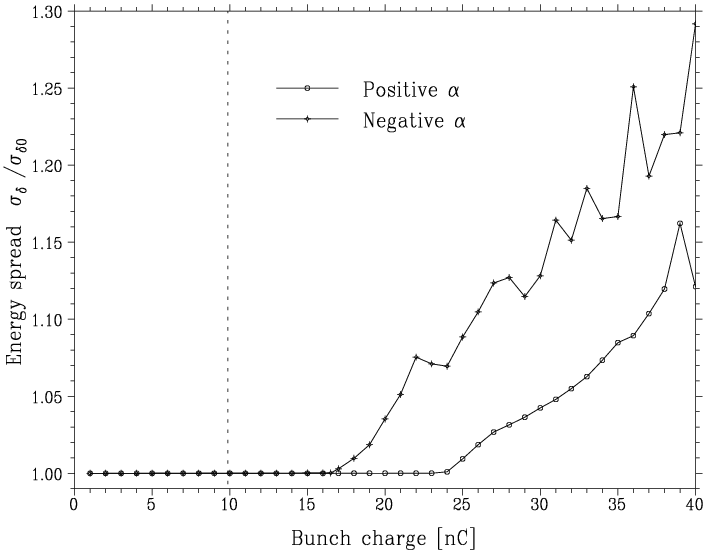}
\caption{\label{fig:csr_4}
Bunch charge dependence of the {\it rms} energy spread.
The circles ($\circ$) and crosses ($+$)
correspond to positive and negative momentum compaction factors. The
dotted vertical line is the design bunch charge of $9.87 \nC$.
}
\end{figure}

With increasing bunch charge, CSR will induce microwave
instabilities. The bunch length and energy spread dependence on bunch
charge are shown in Fig.~\ref{fig:csr_3} and
\ref{fig:csr_4}. The energy spread starts increasing at
$24 \nC$ in bunch charge ($N=1.5\times 10^{11}$ particles/bunch).
The bunch is no longer stable above this threshold, with both the bunch
length and the energy spread oscillating with a
saw-tooth shape in the time domain. Since the design bunch charge
is $9.87 \nC$ ($N=6.16\times 10^{10}$ particles/bunch) the safety
margin is about $140\%$.

We made the same stability analysis for a negative momentum
compaction factor, where we assume that the absolute value is the
same as the positive case. The resulting bunch length and energy spread are
shown in Figs.~\ref{fig:csr_3} and \ref{fig:csr_4}
with the crosses ($+$). With the negative momentum compaction factor, the
threshold charge is $16.5 \nC$, \ie, smaller than the positive
case. Although the bunch length will be a little shorter than the
design value below the threshold, this is due, not to the CSR, but
to the resistive wall wakefield instead.

Based on these studies, we conclude that the bunches in
the \superb\ LER will not be affected by CSR. The instability threshold for
the bunch charge is $24 \nC$, which is about $2.4$ times larger than the
design value; the margin is enough to avoid the microwave
instability. For the design bunch charge, bunch lengthening due to CSR
is small; CSR will not be a concern in the present LER design.

With negative momentum compaction, the threshold charge will
be smaller than in the positive case, although it still has some margin
with respect to the threshold. Negative momentum compaction seems to be
interesting, since the bunch will be somewhat shorter than in the positive
case. However, since the \superb\ Factory does not require short bunches
because of the proposed crabbed waist collision, a design with
negative momentum compaction may not be a practical scheme.

\afterpage{\clearpage}
\subsection{Transverse Multibunch Stability}

\begin{figure}[!htb]
\centering
\includegraphics[width=0.8\textwidth]{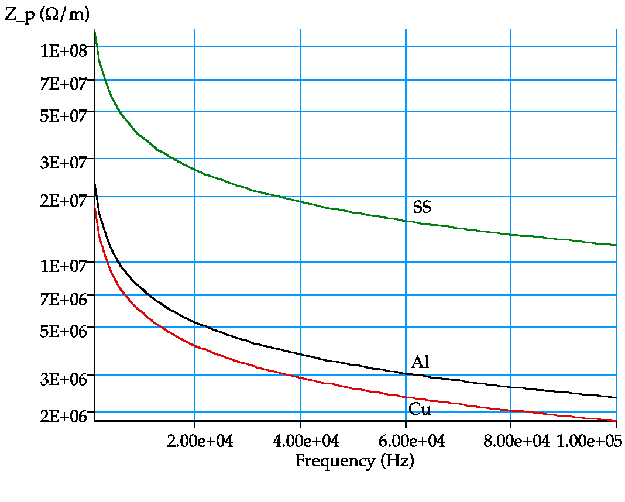}
\caption{Resistive wall impedance {\it vs.} frequency for copper,
aluminum and stainless steel.}\label{fig:Res_wall}
\end{figure}

In
the transverse plane, the coherent multibunch growth rate is dominated by
the resistive wall impedance (we do not concern ourselves in this
section with two-stream-type instabilities). The transverse HOMs in
the RF cavity are sufficiently well damped that their growth rates are
significantly smaller.
The resistive wall impedance may be estimated by
the following formula:
\begin{equation}
Z_\perp\left(\omega\right) =\frac{RZ_0}{b^3}\delta_s\left(\omega\right),
\end{equation}
where $\delta_s$ is the skin depth, which is proportional to the square root
of the conductivity of the chamber wall.
Figure~\ref{fig:Res_wall} shows the frequency dependence of $Z_\perp$
for copper, aluminum and stainless steel, the most common vacuum
chamber materials, for an aperture radius $b$ of 4.5\cm. As can be
seen, copper and aluminum have similar impedance whereas stainless
steel is significantly worse. We will therefore avoid using
stainless steel for the vacuum chambers.
In principle, a larger aperture radius could be chosen.
However, since magnet (especially
quadrupole) costs scale proportional to the aperture and the
\pepii magnets are to be reused in \superb, the dimensions used here are
reasonable.

Experience at
\pepii indicates that the actual growth rate for transverse instabilities
is close to 1/ms,
several times larger than originally estimated. However, this can
be damped by the transverse
feedback system. Practically all of the growth occurs in low-lying
modes. At \superb, avoiding stainless steel chambers will help
keep the impedance lower than that of the \pepii vacuum system.
However, we use the \pepii experience to estimate parameters for the
\superb\ transverse feedback systems. To control a maximum transient
of 1\mm, we require about 1.5\kV integrated gradient, or about
50\W amplifier power for the 0.63\m long, 4.5\cm aperture-radius
stripline kickers used at \pepii, consistent with the 150\W
amplifiers per electrode. These kickers and amplifiers are suitable
for the 238\MHz bunch frequency of the nominal \superb\ parameter
set.

For the upgrade parameters, the bunch frequency doubles to
476\MHz, requiring kickers of 0.315\m length. Everything else remains
the same, except the voltage and bandwidth need to double and the amplifier
power quadruple. The resulting hundreds of watts in a bandwidth of
almost 500\MHz are not straightforward to obtain in commercially produced
amplifiers. Ways to reduce the requirements include higher $\beta$
functions at the kicker and smaller kicker apertures, \eg,
increasing $\beta$ from 25 to 75\m and reducing the kicker
aperture radius to 3.75\cm would just about compensate for the
shorter length, thus maintaining the power requirement of about 100\W.
The increased bandwidth requirement is not affected by these
changes. However, growth rates at the high frequencies will
need to be evaluated, as they may be much smaller, and therefore
require less power to control.

An issue requiring further study is the level of noise
that can be tolerated in the transverse feedback systems. Fundamentally, the
amplitude growth due to noise should be significantly less than the
damping. Given the extremely small beam size, especially in the
vertical plane, frontend noise and quantization noise from the
digital portions of the system can spoil the equilibrium emittance of
the beam. This issue is complex and requires further study to
quantify these effects and understand the ramifications for the transverse
feedback system for \superb.

\afterpage{\clearpage}
\subsection{Electron Cloud Instability }

\subsubsection{Analytic approach}
A single-bunch instability is caused by a short range
transverse wakefield induced by the electron cloud \cite{bib:Ohmi_OZP}.
The wakefield is analytically estimated by a simple model:
\ie,
the beam and an electron cloud having the same transverse size interact with
each other.
We will focus on the vertical instability with this treatment.
The wakefield is represented by a resonator model.
The resonator frequency ($\omega_e$) corresponds to oscillation
frequency of electrons in the beam field,
\begin{equation}
\omega_{e,y}=\sqrt{\frac{ \lambda_+ r_e c^2}{\sy (\sx+\sy)}},
\label{eq:omegae}
\end{equation}
where $\lambda_+$ and $\sigma_{x(y)}$ are the beam line density in a bunch
and the transverse beam sizes, respectively, $r_e$ is the classical electron 
radius and $c$ is the speed of light.
The frequency in the horizontal plane for a flat beam is low.

The wakefield is expressed by
\begin{equation}
W_1(z) [{\rm m}^{-2}] = c\; \frac{R_S}{Q}
  \exp \left(\frac{\omega_e z}{2\; Q\; c}\right)
 \sin \left(\frac{\omega_e}{c}z \right) ,
\label{eq:coastwake}
\end{equation}
where
\begin{equation}
c \frac{R_S}{Q}=
K\; \frac{\lambda_e}{\lambda_+}\; \frac{L}{\sigma_y (\sigma_x+\sigma_y)}
  \frac{\omega_e}{c}.\label{eq:RQ}
\end{equation}
The density of the electron cloud $\lambda_e$, which is the local line density
near the beam, is related to the electron volume density
$\rho_{e}$ via $\lambda_e=2\pi \rho_e \sigma_x\sigma_y$, where
$K$ is the enhancement factor due to the cloud size.
The wake force can be calculated by a numerical method \cite{bib:Ohmi_OZP}.
$K$ is 2--3 for a sufficiently large cloud compared to the beam size.
$Q$ characterizes damping of electron coherent motion due to
the nonlinear interaction with the beam: it is estimated to be 5--10 for
a coasting beam by the numerical method.
$Q$ is reduced by other effects, such as
variations of beam charge density as a function of $z$ and beam size
as a function of $s$, which induce a frequency spread of $\omega_e$
and make it difficult to estimate an accurate value.

The electron phase advance in the bunch, $\omega_e\sigma_z/c$, is
an important parameter for the instability characteristics.
A large phase advance helps Landau damping, but induces
a strong cloud pile up and pinching near the beam,
with the result that $K$ increases.
The wake force, with a range characterized by $Q$,
is efficient only inside the bunch with a length,
$\omega\sigma_z/c > Q$; \ie, the effective $Q$ value is the minimum
of the true $Q$ value and $\omega\sigma_z/c$.

The Keil-Schnell-Boussard criteria for the transverse wake force,
which is based on coasting beam model, gives the threshold of
the fast head-tail instability.
The threshold cloud density for a given bunch intensity is expressed by
\begin{equation}
\rho_{e,th}=\frac{2 \gamma \nu_s \omega_{e,y} \sigma_z/c}
      {\sqrt{3} K Q r_e \beta L}, \label{eq:anath}
\end{equation}
where $\beta$ and $\nu_s$ are average \bety function and
synchrotron tune, respectively.
For finite chromaticity ($\xi$), $\omega_e$ is replaced by
$\omega_e+\omega_0\xi/\eta$.

The threshold value of the electron cloud density is estimated from
Eq.~\ref{eq:anath} to be,
\begin{equation}
\omega_e \sigma_z/c=15,\; \rho_{e,th}=1.3\times 10^{12} \mbox{m}^{-3}
\end{equation}
for \superb, where $\nu_s=0.025$.
The threshold density is given for $K\times Q=3\times 5=15$ and
$\beta=30$\m.

\subsubsection{Numerical simulation}
Although the wakefield approximated by the resonator model permits us
to study the instability with a simple analytic formula,
the estimation of the threshold includes factors, such
as $K$ and $Q$, where there is some uncertainty in determining an appropriate
value.
Since $K$ is related to pinching, one could choose $K\sim \omega_e \sigma_z/c$.
A value of $Q$ larger than $\omega_e \sigma_z/c$ will not help.
These uncertainties are removed using tracking simulations
\cite{bib:Ohmi_OZ,bib:Ohmi_rumolo,bib:Ohmi_ohmiPIC}.

We report on simulation results using a strong-strong code, named PEHTS
\cite{bib:Ohmi_ohmiPIC}.
A bunch and an electron cloud are represented by macro-particles,
and the interactions between them are determined by solving a two-dimensional
Poisson equation using
the particle-in-a-cell method.

\begin{figure}[htb]
\centering
\includegraphics [width=0.8\textwidth]{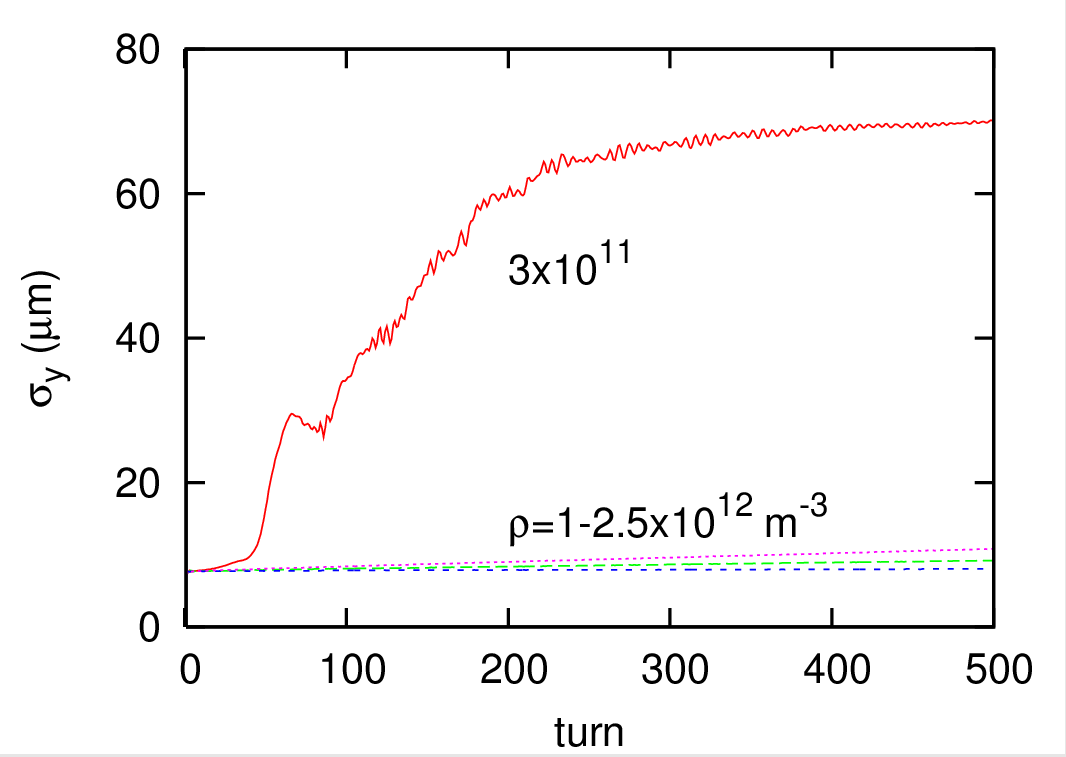} 
\caption{\label{fig:gr_M}
Emittance growth due to the fast head-tail instability caused by the
electron cloud effect.
}
\end{figure}

The interaction between beam and cloud is evaluated at eight positions
around the ring, where the $\beta$ function is uniform.
Figure~\ref{fig:gr_M} shows emittance growth due to the fast head-tail instability
caused by the electron cloud effect in the \superb\ ring.
Each line shows an emittance growth for various cloud densities.
The threshold density is determined by the density at which the growth
starts. From this numerical simulation, we determine that
the instability starts between $\rho_e=2.5$ and
$3\times 10^{11}\m^{-3}$.

\begin{figure}[htb]
\centering
\includegraphics [width=0.8\textwidth]{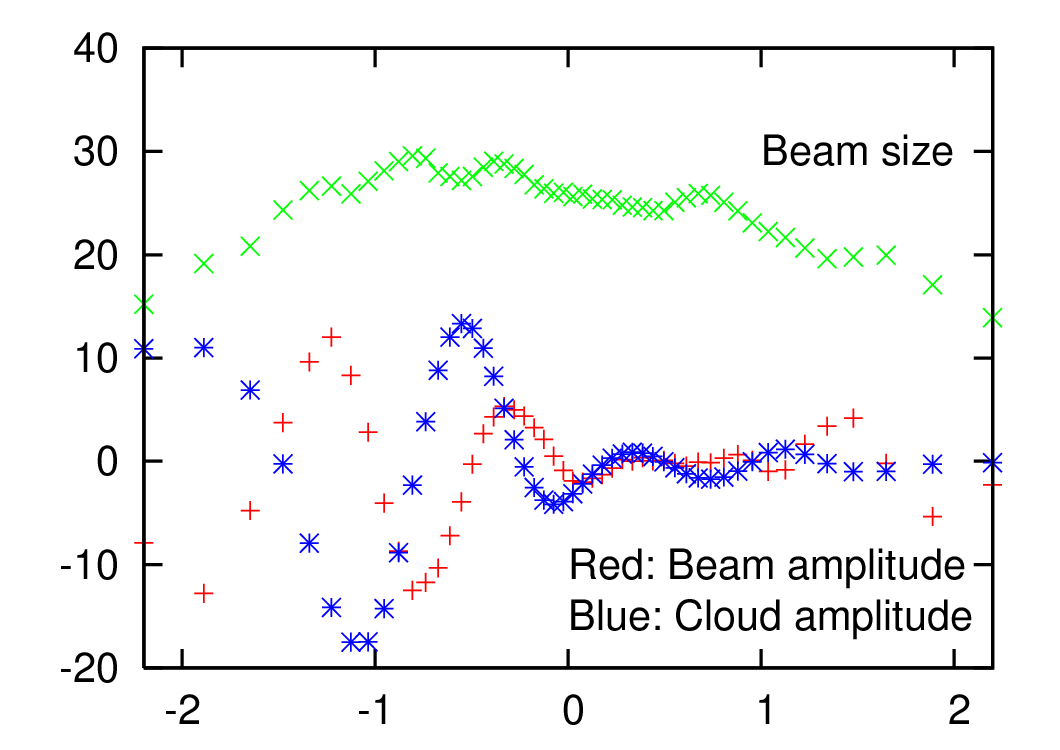} 
\caption{\label{fig:examplebcamp}
Variation of the bunch size, and the centroid location of the bunch and
cloud, with longitudinal position along the bunch, after 60 simulated turns
around the ring.
}
\end{figure}

Figure~\ref{fig:examplebcamp} shows a snapshot of the beam and cloud centroid
oscillation amplitude, as well as
the vertical beam size, as a function of the bunch length, from a 60-turn
simulation at an above threshold density $\rho_e=3\times 10^{11}\m^{-3}$.
The amplitude of beam-cloud coherent motion is
similar to the increase in vertical beam size.
We conclude that the fast head-tail instability is dominant at these bunch
densities. The coherent motion is smeared, due to the nonlinear beam-electron
cloud interactions, and the amplitude is reduced after 300 turns.
This behavior has been observed in experiments at KEKB \cite{bib:Ohmi_JFl},
where a sawtooth coherent instability arises.

The analytical estimate can be compared with the threshold value
given by simulation,
\begin{equation}
\rho_{e,sim}=0.3\times 10^{-12}\m^{-3}.
\end{equation}
The density given by the simulation is systematically lower than the analytic
estimate
for a very low emittance ring, as already seen for the \ilc Damping Ring
\cite{bib:Ohmi_ohmi:snowmass05}.
We believe that the lower threshold density in the simulation
is caused by pileup and pinching
of electrons due to the attractive force of beam.
The characteristic constant for the attractive force is
the electron phase advance in the beam, $\omega_e \sigma_z/c$.
In the \KEKB and \pepii $B$ Factories, where
the phase advance is far lower than for the \ilc Damping Rings,
the analytical estimate is in good agreement with the simulation.

\subsubsection{Electron cloud density}
Positrons create photons with a line density $0.15$/meter for the
\superb\ ring. We assume that 99\% of these photons are absorbed in the
antechamber slot, and the remaining 1\% create photoelectrons with
a quantum efficiency of 10\%. With these assumptions, positrons create
photoelectrons with a line production density $\lambda_e=0.15\times
10^{-3}$ per positron per meter. The photoelectrons stay in the chamber
for 20--40\ns or 10--20 bunch passages.
Such an estimate can be obtained, for example, from an average electron velocity
of $10^6$--$10^7$\m/s and a chamber size of 0.05--0.1\m. This
is the duration of the electron build-up time, when multipactoring is
not dominant. The amplification factor ($A_e$), defined as
the number of stored electrons normalized to the produced number
of electrons, is the ratio of the dwell time (build-up time) and the bunch
spacing, \ie, $A_e\approx 10$--20.

The electron cloud density can be estimated as
\begin{equation}
\frac{\lambda_e N_p A_e}{D^2}=
\frac{0.15\times 10^{-3}\times 3.3\times 10^{10}\times 10}{0.05^2}
=2\times 10^{10}\m^{-1}.
\end{equation}
If the photon absorption in the antechamber is less than 80--90\%,
electrons
created by photoemission will exceed the threshold density for
the single bunch instability.

\begin{figure}[htb]
\centering
\includegraphics [width=0.8\textwidth]{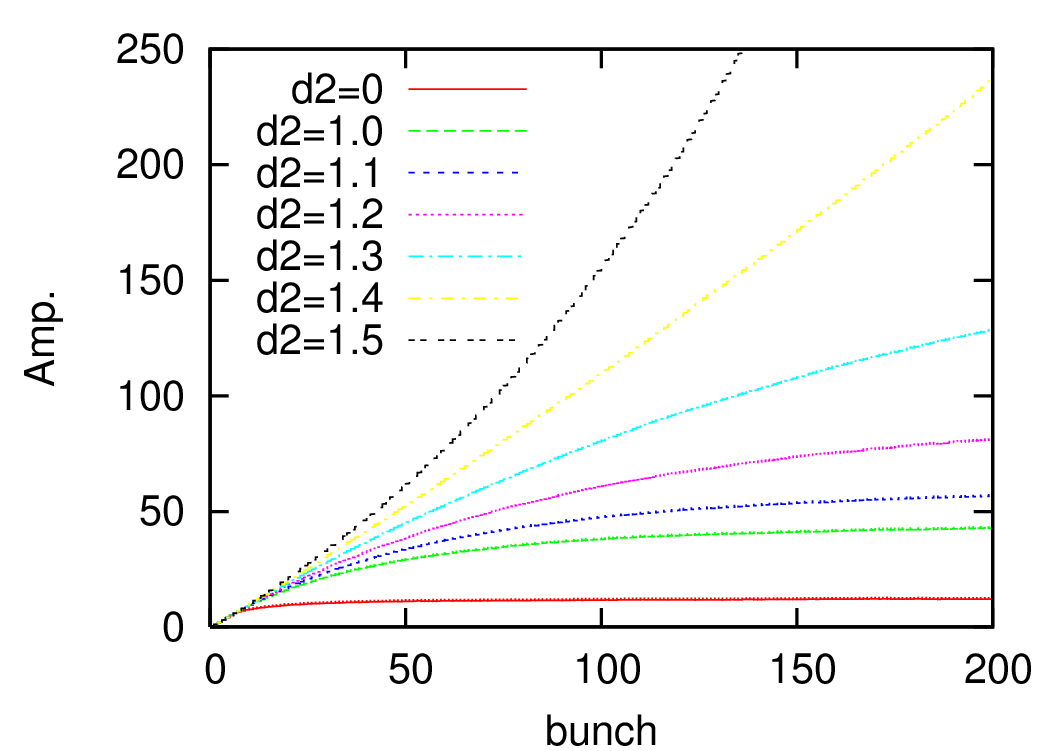} 
\caption{\label{fig:MulP} Amplification of electrons as a function
of the secondary emission coefficient $D^2$ (labeled d2 here). }
\end{figure}

The amplification factor increases due to multipactoring of
electrons. Figure~\ref{fig:MulP} shows the amplification factor obtained
from a simulation of electron cloud build up. The limit of the
amplification is $\pi R^2 \rho_{e,th}/\lambda_e\approx 200$. For no
secondary electrons, an amplification of 13 is predicted.
Increasing the secondary emission rate, the amplification factor
increases. From this simulation, the secondary emission rate should be
less than  $\delta_{2,max}=1.3$.

\afterpage{\clearpage}
\subsection{Electron Cloud Remediation Techniques}

Recent simulation results for
electron cloud build-up in the \superb\ positron ring are discussed here,
assuming beam parameters similar to those of the
\ilc Damping Ring, but adopting a shorter bunch
spacing~\cite{bib:ref2}.
Possible remedies
for the electron cloud formation considered recently include
clearing electrodes and vacuum chamber grooves \cite{bib:ref4,bib:ref5}.
Our simulations show that the insertion of
clearing electrodes in the vacuum chamber is indeed a
extremely powerful way to suppress electron cloud formation.
We will describe the effect of clearing electrodes in the dipole magnetic field
regions and the chamber layout.

\subsubsection{Electron cloud build-up and clearing electrode effect}

We have used the simulation code POSINST to evaluate the
contribution to the electron cloud build-up in the arc bends of
\superb.

The \kekb and \pepii $B$ Factories have adopted external solenoid fields to
mitigate the electron cloud effect in field-free regions, which constitute
a large fraction of the rings \cite{bib:ref6,bib:ref7}. The \superb\
rings typically do not have long field free regions. Over much of the
ring, the beam pipe is surrounded by magnets, such as wigglers and
dipoles, where large electron cloud densities may develop. In
magnetic field regions, external solenoid fields are not effective
in suppressing the build-up of the electron cloud. Thus, we have
focused our simulations on the build-up of an electron cloud
in the arc bend regions.

\begin{figure}[htb]
\centering
\includegraphics[width=\textwidth]{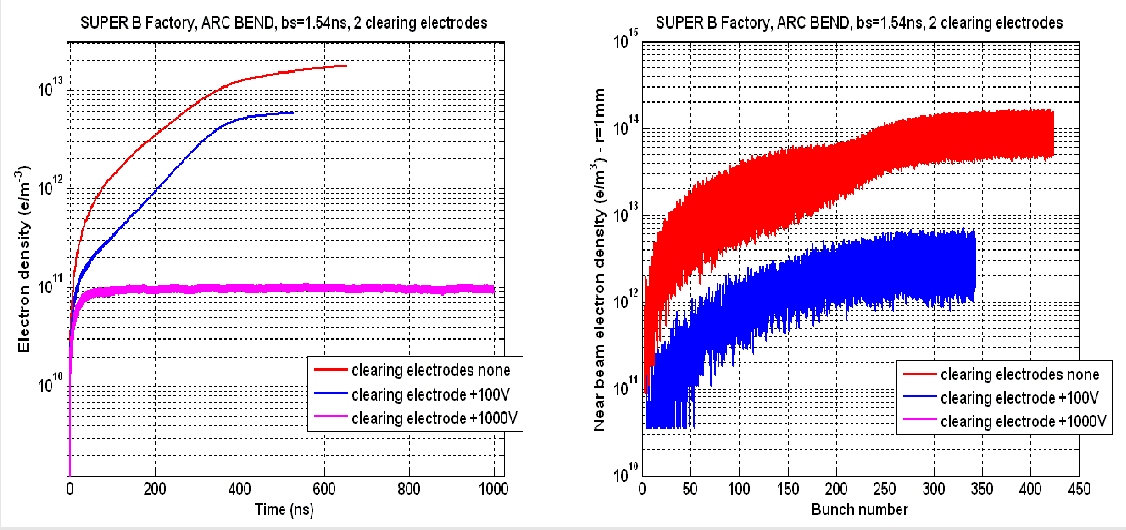}
\caption{Simulation of electron cloud build-up in \superb,
using two clearing electrodes. Average (left) and
central (right) electron density, with and without clearing electrodes
are illustrated.  Note: we have used up to $1\times 10^{6}$
macroparticles to represent the electrons.} \label{fig:Pivi_due}
\end{figure}

\begin{figure}[htb]
\centering
\includegraphics[width=\textwidth]{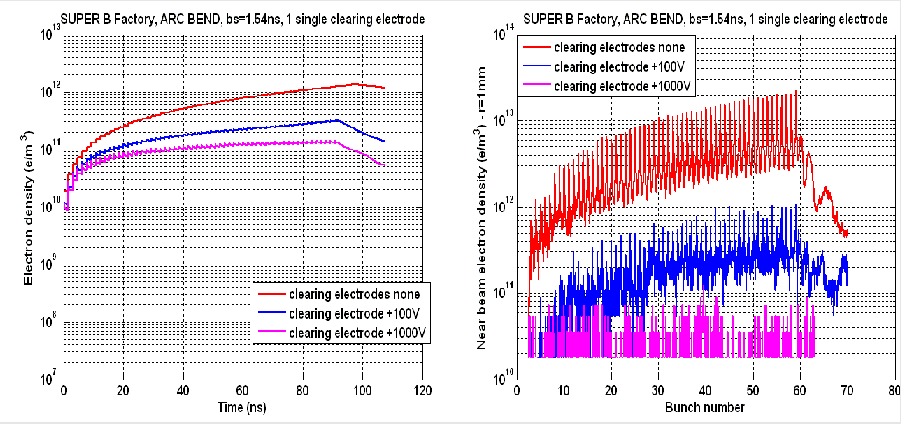}
\caption{Simulation of electron cloud build-up in \superb,
using one single clearing electrode during the passage of the beam
followed by a gap. Average (left) and central (right) electron
densities, with and without clearing electrodes, are illustrated.
The clearing effect is already visible after
applying just $+100$\V and becomes stronger at $+1000$\V.}
\label{fig:Pivi_uno}
\end{figure}

To remove most of the synchrotron radiation emitted in the arc
sections, we have assumed a vacuum chamber with an antechamber
design.
For these preliminary simulations, we have assumed the same bunch
population of $2\times 10^{10}$ particles per bunch but a reduced
bunch spacing of 1.5\nsec in comparison with the ILC DR (6.154\ns).
Results for the electron cloud build-up are shown in
Fig.~\ref{fig:Pivi_due} and \ref{fig:Pivi_uno}.

\begin{figure}[!htb]
\centering
\includegraphics[width=0.5\textwidth]{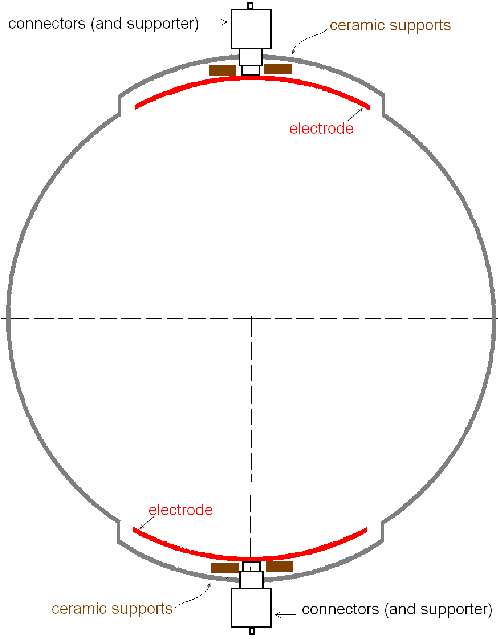}
\caption{Sketch of the simulated electrode arrangement for the
vacuum chambers in the \superb\ bend magnets.}
 \label{fig:Pivi_tre}
\end{figure}

To mitigate the formation of an electron cloud, we have also
simulated the effect of clearing electrodes installed in the bend
vacuum chamber, and extending along the longitudinal direction of the
magnet. The electrodes are biased with a positive potential. A
sketch of a possible clearing electrode configuration is shown in
Fig.~\ref{fig:Pivi_tre}. In a bend or wiggler magnet, the electrodes can be
arranged along the top and bottom, since the electron cloud forms
mostly along stripes directed along the vertical magnetic field lines
\cite{bib:ref1,bib:ref8}. The effect of the electrodes is to compensate, on
average, for the electric field from the positron bunch, which tends
to attract the electrons to the center of the chamber. The electrons
at the wall are first accelerated to the center by the bunch, and
then accelerated back to the surface by the electrodes, during the
time interval between bunches.

The effect of the two clearing electrodes is shown in
Fig.~\ref{fig:Pivi_due}. The average cloud chamber density
and the central cloud density are plotted on the left
and right side of the figure, respectively, for different electrode bias
potentials. A bias voltage of 1\kV is sufficient to suppress electron
cloud formation and drastically reduce the central cloud density
near the beam. The effect of applying a potential to only one
clearing electrode, rather than two, is shown in
Fig.~\ref{fig:Pivi_uno}. A single electrode is also very effective in
suppressing electron cloud formation.

These preliminary simulations show the effect of the
clearing electrode suppression in \superb, although with beam
parameters (bunch population and bunch spacing) that differ from
the \superb\ configuration. Future simulations will be performed with updated
parameters. Clearing electrodes have been proposed for the \ilc DR
and LHC magnetic field regions. An extensive R\&D program is ongoing to
test their effect with operating accelerators~\cite{bib:ref9,bib:ref10}.

\afterpage{\clearpage}
\subsection{Fast Ion Instability}

\subsubsection{Model}

We consider CO$^+$ ions as the instability source, because
the major components of residual gas in vacuum systems are CO and H$_2$, and
the ionization cross-section of CO is 5 times higher than that of H$_2$.
The ionization cross-section is $1.9\times 10^{-22}$~m$^{-2}$ for CO
at the electron beam energy, $E=7 \gev$.
We assume that the partial pressure of CO gas is $P=3\times 10^{-8}\Pa$.
The number of ions created by the electron beam with a population $N_e$
is expressed by
\begin{equation}
n_i[\mbox{m}^{-1}]=0.046 N_e P[\Pa].
\end{equation}
In our case $n_i=27\m^{-1}$
for $N_e=1.9\times 10^{10}$ and $P=3\times 10^{-8}\Pa$.

\begin{figure}[htb]
\centering
\vspace*{5mm}
\includegraphics [width=0.8\textwidth] {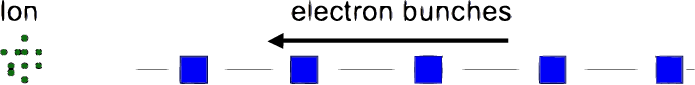} 
\caption{\label{fig:IonModel} Model of beam-ion interaction.
}
\end{figure}

We investigate ion instabilities for various bunch filling pattern
in \superb.
A simulation method based on the model shown in Fig.~\ref{fig:IonModel}
is used.
Ions are represented by macro-particles, and each bunch
is represented by a rigid transverse gaussian macro-particle.
The beam size of the bunch
is fixed, as determined by the emittance and $\beta$ function,
and only dipole motion is considered.

Beam-ion interaction is expressed by the Bassetti-Erskine
formula~\cite{bib:Ohmi_Bassetti}
for a beam with gaussian distribution in the transverse plane.
The equations of motion for electrons and ions are expressed as
\begin{eqnarray}
\frac{d^2\mbox{\boldmath $x$}_{e,a}}{ds^2}&+&K(s)
  \mbox{\boldmath $x$}_{e,a}
   =\frac{2 r_e}{\gamma} \sum_{j=1}^{N_{i}}
    \mbox{\boldmath $F$}
    (\mbox{\boldmath $x$}_{e,a}-\mbox{\boldmath $x$}_{i,j}),
      \label{beq0} \\
\frac{d^2\mbox{\boldmath $x$}_{i,j}}{dt^2}
   &=& \frac{2 r_e c^2}{M_i/m_e}
    \sum_{a=1}^{N_{e}}
    \mbox{\boldmath $F$}
   (\mbox{\boldmath $x$}_{i,j}-\mbox{\boldmath $x$}_{e,a}),
   \label{ieq0}
\end{eqnarray}
where the suffixes $i$ and $e$ denote the ion and electron,
respectively. $M_i$ and $m_e$ are masses, and
$N_i$ and $N_e$ are their number. $\gamma$ and $r_e$ are the
Lorentz factor of the beam and the classical electron radius,
respectively.
$\mbox{\boldmath $F$}(\mbox{\boldmath $x$})$ is the Coulomb force
expressed by the Bassetti-Erskine formula.
These consist of $N_e+N_i$ differential equations, where
each electron couples to the motion of all ions, and
each ion couples to the motion of all electrons.

It is easy to solve the equations simultaneously
with a numerical method \cite{bib:Ohmi_ohmiion}.
The structure of the bunch train and $\beta$ function variation are also
taken into account with this approach.
The effect of a bunch-by-bunch feedback system is included in the simulation.
The feedback system has a damping time of 50 turns and fluctuation of
$0.02 \sigma_y$.
This gain is rather conservative with present technology.

\subsubsection{Simulation of ion instability}

\begin{figure}[htb]
\centering
\includegraphics [width=0.9\textwidth]{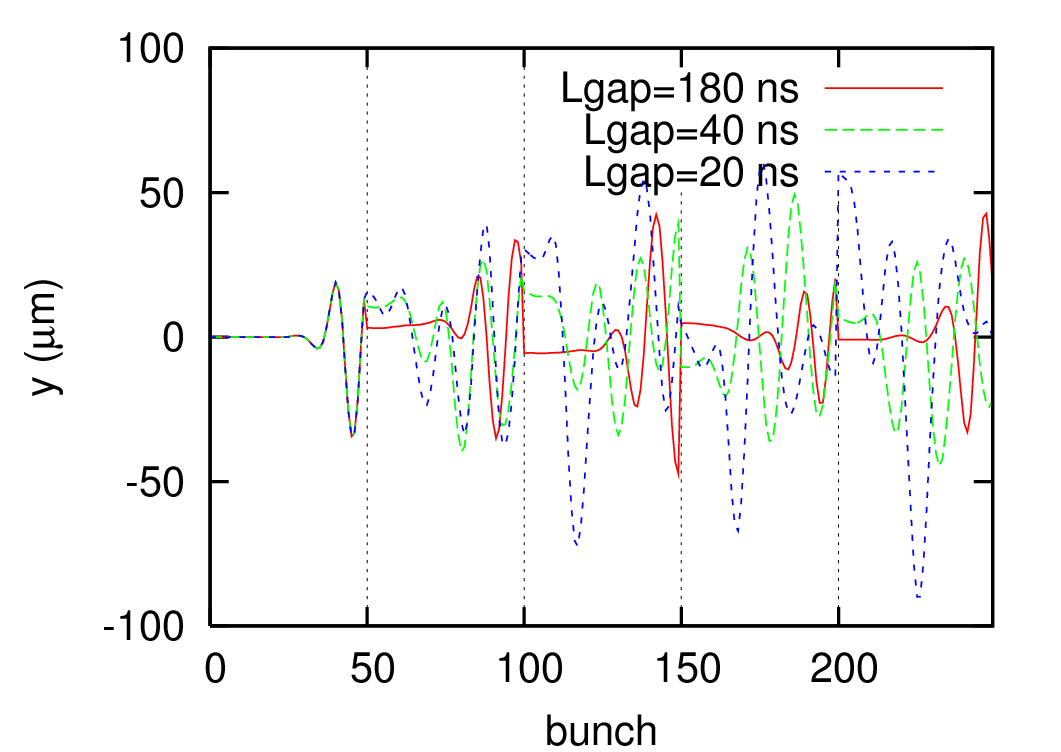} 
\caption{\label{fig:bpos}
Vertical position of all bunches after 1000 turns for various
train gap lengths.}
\end{figure}

The simulation gives the position and momenta of every bunch, turn by turn.
Figure~\ref{fig:bpos} shows the vertical position of every bunch after 1000
turns. We use as filling parameters the bunch population
($N_e=1.9\times 10^{10}$), the
bunch spacing ($L_{sp}=2$\ns), the number of bunches in a train ($N_{b}=50$),
the number of trains ($N_{tr}=5$).
Gaps between trains are simulated for three cases,
$L_{gap}=10, 20$ and $90\times 2 \ns$.
In the figure, the gap is removed: \ie,
$y$ at 1--50, 51--100 \etc\ are the vertical bunch positions of the first, second
\etc\ trains, respectively.
The amplitude of the head of the first train is exactly zero, because there is
no ion effect, and the amplitudes of the first 50 bunches do not depend on
the gap length.
Those of the second, third \etc\ trains are not zero, and depend on the gap length.
Some ions
remaining after the passage of previous trains affect the head part
of the subsequent trains.
The maximum amplitude is saturated for all trains at
$L_{gap}\le 40\ns$. This means that the gap length is efficient for clearing the ions.
On the other hand, the maximum amplitudes increase along trains for
$L_{gap}\le 20 \ns$; \ie, the gap length is not sufficient, and ions
are built up.

\begin{figure}[htbp]
\centering
\includegraphics [width=0.6\textwidth]{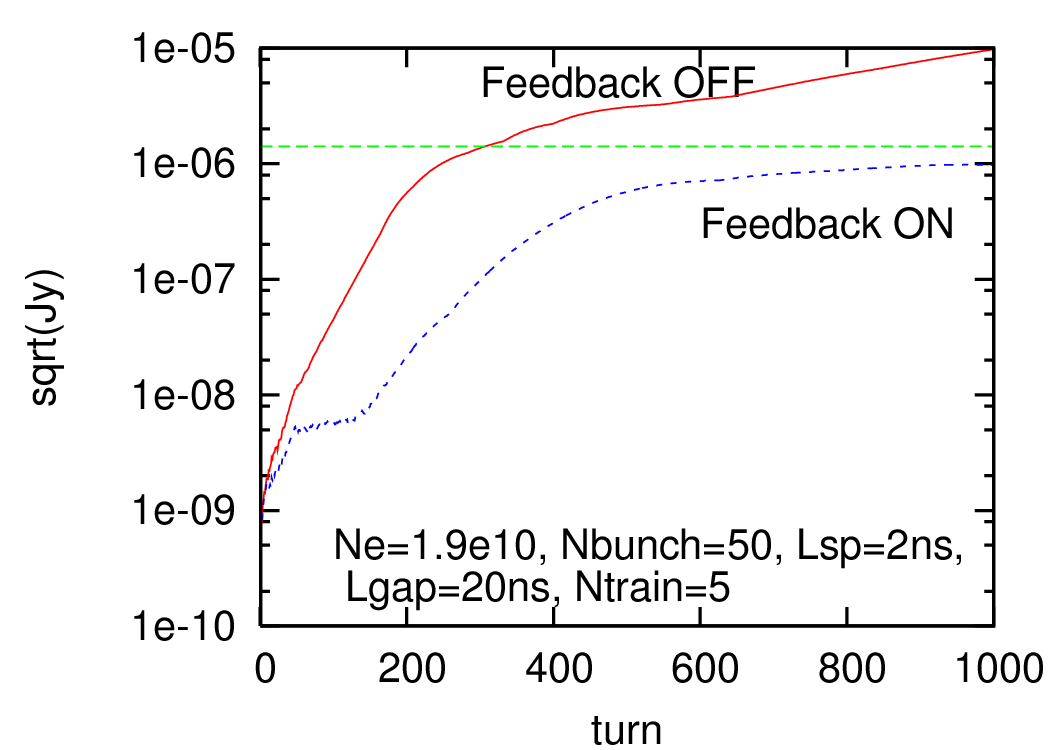} 
\includegraphics [width=0.6\textwidth]{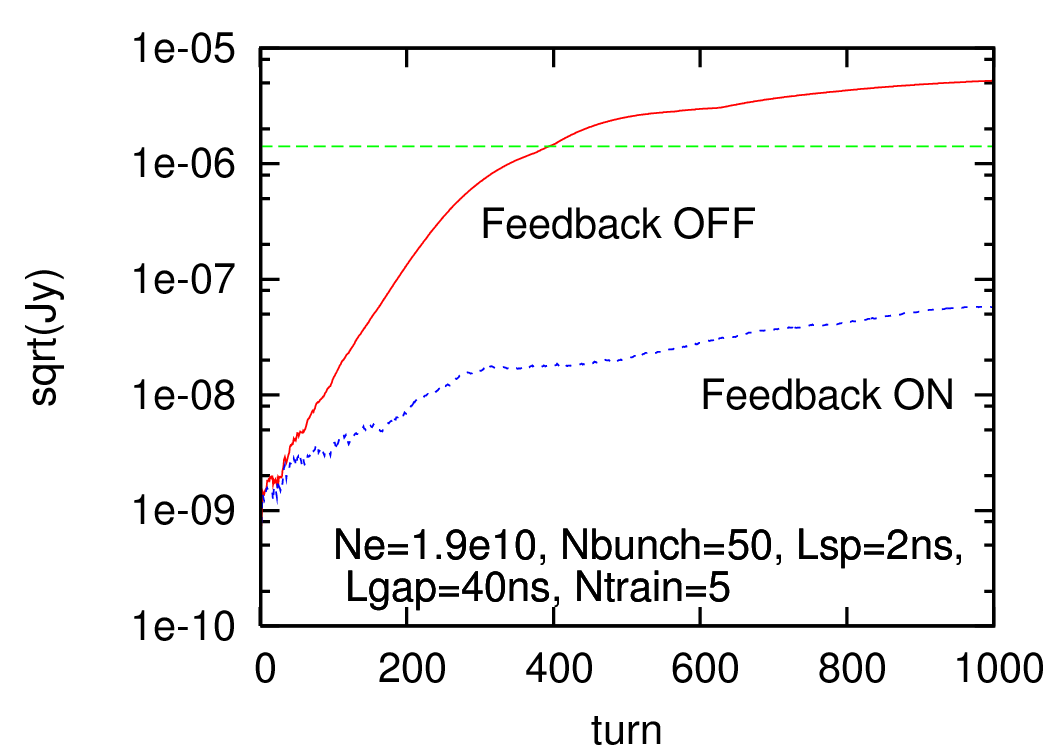} 
\includegraphics [width=0.6\textwidth]{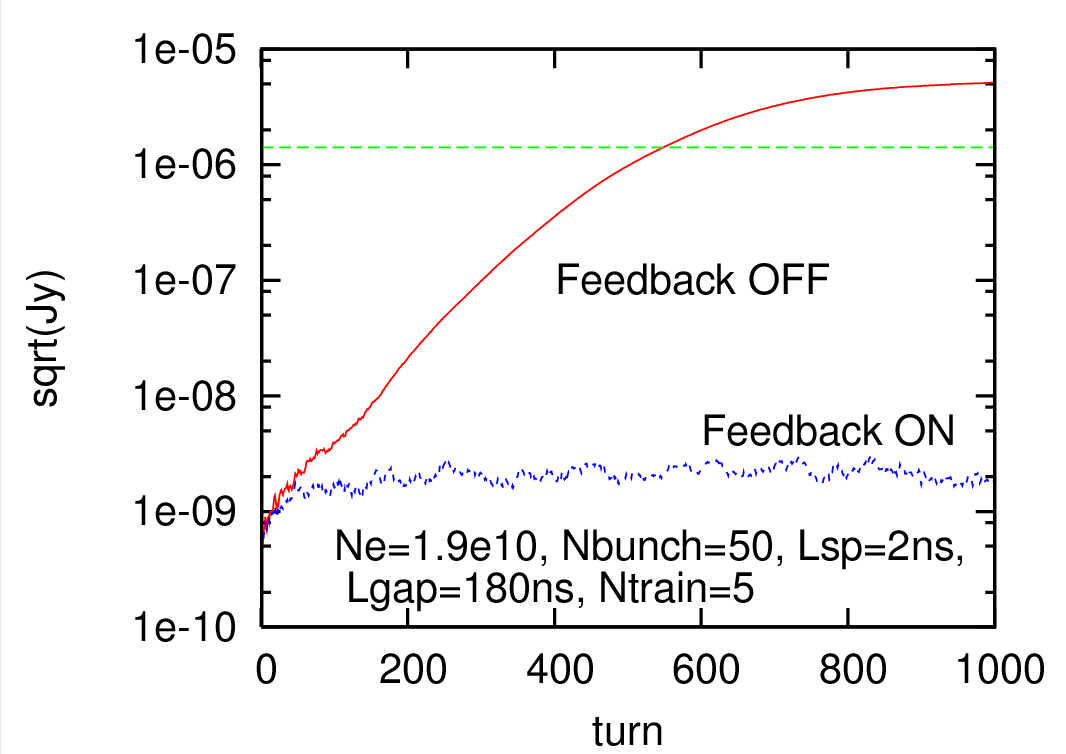} 
\caption{\label{fig:gr1}
Evolution of the maximum amplitude ($\sqrt{J_y}$)
for train gap lengths (top to bottom)
$L_{gap}=20$, 40 to 180\ns.}
\end{figure}

\begin{figure}[htbp]
\centering
\includegraphics [width=0.6\textwidth]{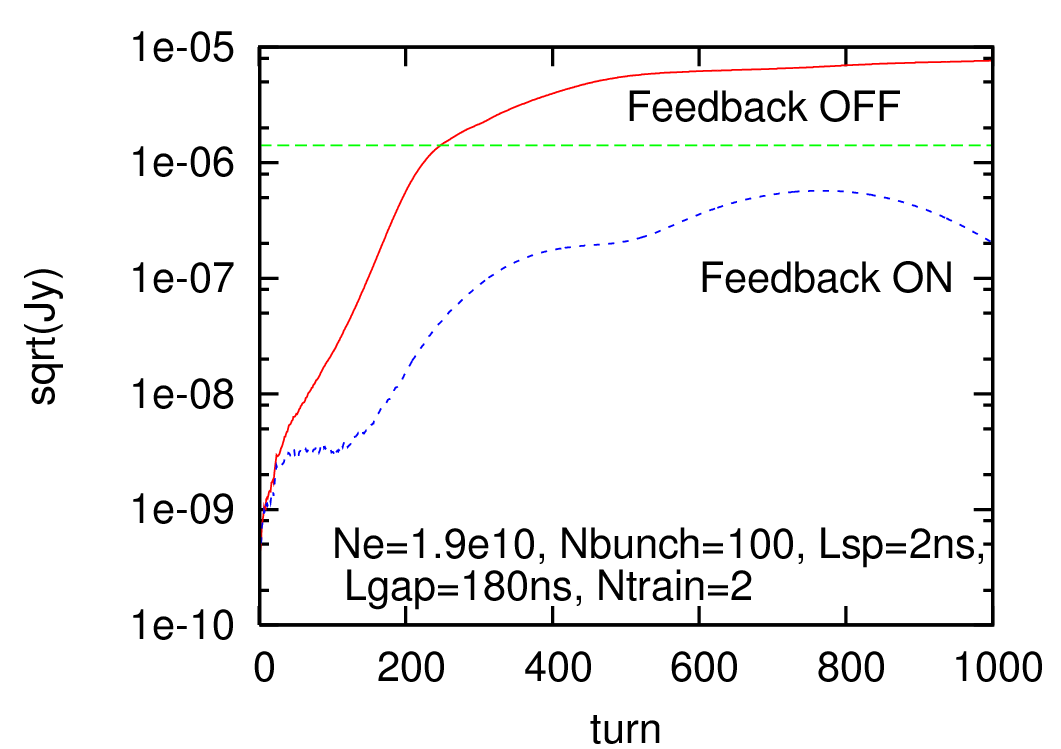} 
\includegraphics [width=0.6\textwidth]{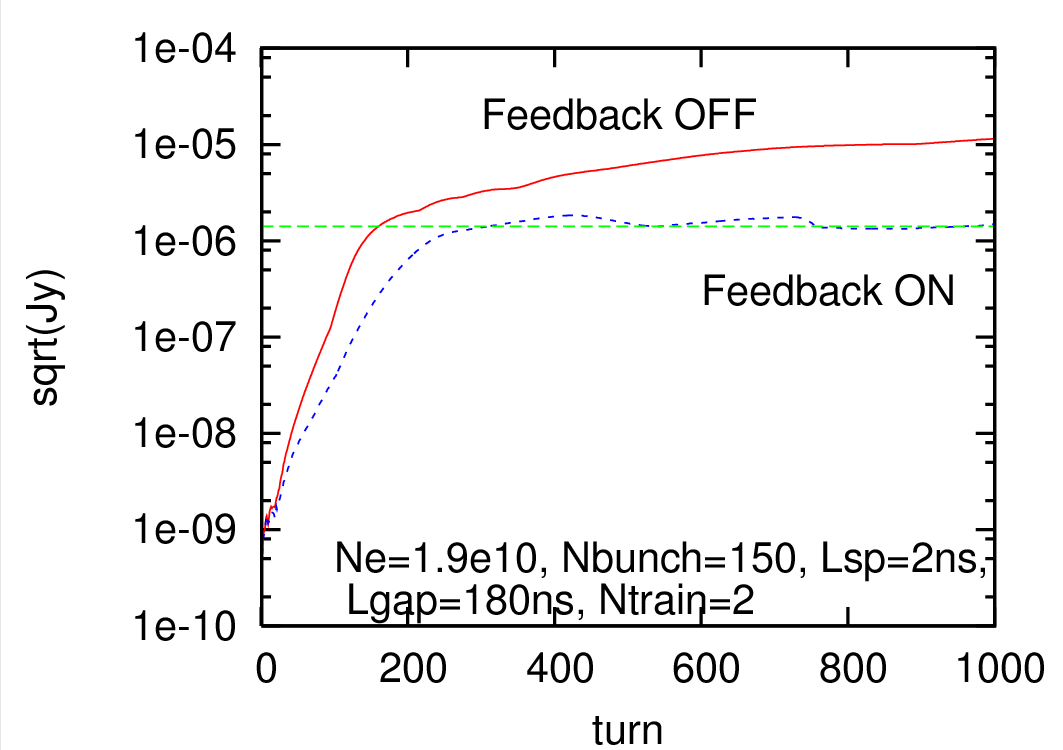} 
\includegraphics [width=0.6\textwidth]{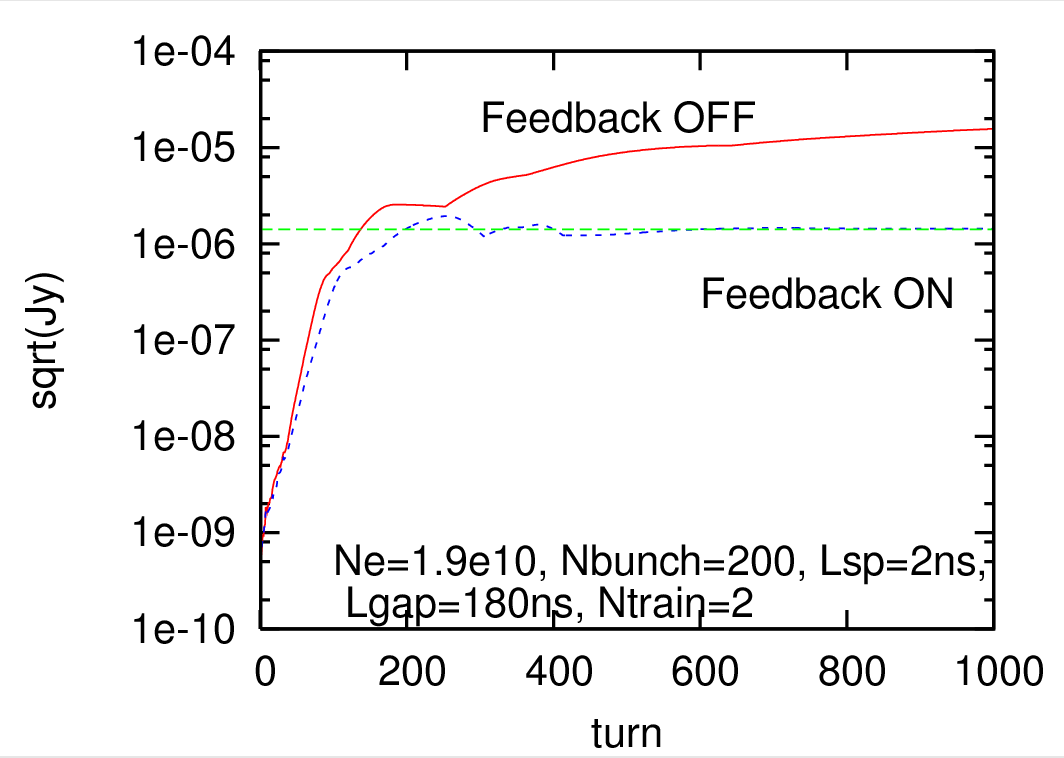} 
\caption{\label{fig:gr2}
Evolution of the maximum amplitude ($\sqrt{J_y}$)
for various train lengths (top to bottom) $N_b=100, 150, 200$.}
\end{figure}


The maximum amplitude for all bunches, $\sqrt{J_y}$, is obtained
turn-by-turn from the simulation.
Figure~\ref{fig:gr1} shows the evolution of $\sqrt{J_y}$ with turn number.
The red and blue lines show the evolution with and without
the bunch-by-bunch feedback system, respectively. From top to bottom,
the amplitude growth is shown for the three
gap lengths, $L_{gap}=20$, 40 and 180\ns.
Beam oscillations are suppressed by the feedback system for
$L_{gap}\le 40 \ns$, while
considerable residual oscillation remains for $L_{gap}\le 20 \ns$.

Figure~\ref{fig:gr2} shows the variation in amplitude growth with the number
of bunches in a train ($N_b=100$, 150, 200), where $L_{gap}=180\ns$.
The instability for $N_b=100$ is suppressed by the feedback system,
but it is not suppressed for longer trains, $N_b\ge 150$.

In summary, the bunch filling pattern for \superb\ is
$N_b=100$ and $L_{gap}=40 \ns$ (20 buckets).
In this filling pattern, secondary efficiency should be less than the
$\delta_{2,max}=1.3$ required to keep below electron cloud instability.
These results depend on various conditions, such as
chamber geometry, magnetic field, and vacuum pressure.
Further studies should be done as the design is updated.



\afterpage{\clearpage}

\section{Magnet Systems}
\label{section:MagnetSystems}
\subsection {Introduction}
The \superb\ rings will be built with room-temperature magnets. The
lattice has been designed to take maximum advantage of the potential
availability of the \pepii ring magnets without compromising performance.
This is possible,
since the energies and the circumference of the \pepii rings are quite
comparable to those of \superb. The \superb\ HER magnet requirements
are summarized in Table~\ref{tab:HER_magnet_Needs}, while the LER
requirements are summarized in Table~\ref{tab:LER_magnet_Needs}.

\begin{table}[htbp]
\caption{Magnet parameters for the \superb\ High Energy Ring.}
\label{tab:HER_magnet_Needs}
\vspace*{2mm}
\setlength{\extrarowheight}{1pt}
\centering
\begin{tabular}{lcccc}
\hline\hline
{Dipoles} & {Length} & {Max. Field} & {Min. $\rho$} & {Quantity}
 \\(Location) & (m) & (T) & (m) & \\
\hline
        Arc         &     5.4      &      0.168       &        139        & 144 \\
    Final focus     &     5.4      &      0.212       &        110        & 16 \\
    Soft bends      &      2       &      0.053       &        444        & 2 \\
 & & & & \\
\hline
{Quadrupoles} & {Length} & {Max. Gradient} & {Int. Strength}
& {Quantity} \\
      (Location)        &     (m)      &        (T/m)        &         (T)         & \\
   \hline   Wiggler     &     0.5      &        21.7         &        10.9         & 34 \\
          Arc           &     0.56     &        17.0         &         9.5         & 213 \\
     Arc, final focus.  &     0.56     &        20.0         &        11.2         & 30 \\
      Final focus       &     0.56     &        32.4         &        18.1         & 2 \\
     Straight section.  &     0.73     &        16.6         &        12.1         & 138 \\
 & & & & \\
\hline
{Sextupoles} & {Length} & {Max. Strength} & {Int. Strength} & {Quantity} \\
   (Location)    &     (m)      &      (T/m$^2$)      &        (T/m)        & \\
\hline
      Arc        &     0.25     &         100         &        25.0         & 216 \\
 High gradient   &     0.25     &         350         &        87.5         & 10 \\
  Final focus    &     0.6      &         320         &        192.0        & 4 \\

\hline
\end{tabular}
\end{table}

\begin{table}[htb]
\caption{Magnet parameters for the \superb\ Low Energy Ring.}
\label{tab:LER_magnet_Needs}
\vspace*{2mm}
\setlength{\extrarowheight}{1pt}
\centering
\begin{tabular}{lcccc}
\hline \hline
{Dipoles}      & {Length} &  {Max. Field}   &  {Min. $\rho$}  & {Quantity} \\
(Location)          &     (m)      &         (T)         &         (m)         & \\ \hline
Arc              &     0.45     &        0.592        &        22.5         & 144 \\
Arc              &     0.75     &        0.469        &        28.4         & 144  \\
Final focus          &     5.4      &        0.121        &         110         & 16 \\
Soft bend           &      2       &        0.03         &         444         & 2  \\
 & & & & \\
\hline
{Quadrupoles}       & {Length} & {Max. Gradient} & {Int. Strength} & {Quantity} \\
(Location)          &     (m)      &        (T/m)        &         (T)         & \\
\hline
All                 &     0.43     &        9.74         &       4.1882        & 341  \\
Wiggler             &     0.5      &        6.59         &        3.295        & 36 \\
Final focus         &     0.56     &        10.36        &       5.8016        & 42 \\
 & & & & \\
\hline
{Sextupoles} & {Length} & {Max. Strength} & {Int. Strength} & {Quantity}  \\
(Location)          &     (m)      &      (T/m$^2$)      &        (T/m)        & \\
\hline
Arc              &     0.25     &         55          &        13.75        & 218  \\
Final focus          &     0.25     &         192         &         48          & 8  \\
Final focus          &     0.6      &         180         &         108         & 4  \\
\hline
\end{tabular}
\end{table}

In order for \pepii magnets to be suitable for \superb, the magnet apertures must
 be sufficient, but not too much larger than needed
to avoid excessive power consumption. At \superb, with its small beam
sizes, the apertures will be dominated by impedance and vacuum
conductance considerations, rather than the size of the beams, and the
apertures required will be similar to those of \pepii. Thus, there
is a good match between the size of the \pepii magnet apertures
and the anticipated \superb\ requirements.

Table~\ref{tab:HERMagnets} lists the magnet inventory of
the\pepii HER. In many cases, these
magnets are capable of higher field strengths than used
operationally at \pepii, 
since they were originally designed for the 18\gev\ PEP-I
rings. This has been taken into account in
Table~\ref{tab:HERMagnets}.

{ \setlength{\tabcolsep}{5pt}  
\begin{table}[hbtp]
\caption{\pepii High Energy Ring magnets.}
\label{tab:HERMagnets}
\vspace*{2mm}
\setlength{\extrarowheight}{1pt}
\centering
\begin{tabular}{lcccccc}
\hline\hline
{Dipoles}   & {Length} & {Aperture} &  {Field}   & {Int. Strength} & {Current} & {Quantity} \\
   (Location)     &     (m)      &      (mm)      &      (T)       &        (Tm)         &      (A)      & \\
\hline
       Arc        &     5.4      &       60       &      0.27      &        1.45         &      950      & 194 \\
  IR Soft bends   &      2       & $150\times100$ &     0.092      &        0.184        &      170      & 6 \\
& \\
\hline
{Quadrupoles} & {Length} & {Aperture} & {Gradient} & {Int. Strength} & {Current} & {Quantity} \\(Location) & (m) & (mm) & (T/m) & (T) & (A) & \\
\hline
       Arc        &     0.56     &      R 50      &     16.96      &         9.5         &      350      & 202 \\
   Inj. sect.     &     0.45     &      R 50      &     11.11      &          5          &      200      & 4 \\
  Straight sect.  &     0.73     &      R 50      &     17.53      &        12.8         &      350      & 81 \\
       IR         &     1.5      &                &      6.67      &         10          &      650      & 2 \\
       IR         &     1.5      &                &       10       &         15          &     1150      & 2 \\
   Global skew    &     0.3      &      R 90      &      2.33      &         0.7         &      250      & 4 \\
     IR skew      &     0.2      &      R 50      &      0.32      &        0.064        &      50       & 4 \\
     IR skew      &     0.3      &      R 50      &      1.33      &         0.4         &      12       & 4 \\
& \\
\hline
{Sextupoles}  & {Length} & {Aperture} & {Strength} & {Int. Strength} & {Current} & {Quantity} \\(Location) & (m) & (mm) & (T/m$^2$) & (T/m) & (A) & \\\hline
       Arc        &     0.3      &      R 60      &      210       &         63          &      400      & 104 \\
& \\
\hline
{Correctors}  & {Length} & {Aperture} &  {Field}   & {Int. Strength} & {Current} & {Quantity} \\
   (Location)     &     (m)      &      (mm)      &      (T)       &        (Tm)         &      (A)      & \\
\hline
      Arc X       &     0.3      &  $90\times50$  &     0.018      &       0.0054        &      12       & 96 \\
      Arc Y       &     0.3      &  $90\times50$  &      0.01      &        0.003        &      12       & 96 \\
    Straight      &     0.3      &      R 50      &     0.012      &       0.0036        &      12       & 91 \\\hline
\end{tabular}
\end{table}
}
\clearpage
The \pepii LER magnet inventory is listed in
Table~\ref{tab:LERMagnets}. These magnets were built specifically
for the \pepii LER. In most cases, the maximum field was
specified such that the \pepii LER can reach 3.5\gev\ in
energy.

{ \setlength{\tabcolsep}{3pt}  
\begin{table}[tbhp]
\caption{\pepii Low Energy Ring magnets.}
\label{tab:LERMagnets}
\vspace*{2mm}
\setlength{\extrarowheight}{1pt}
\centering
\begin{tabular}{lcccccc}
\hline\hline
{Dipoles}         & {Length} & {Aperture} &  {Field}   & {Int. Strength} & {Current} & {Quantity} \\
(Location)          &     (m)      &      (mm)      &      (T)       &        (Tm)         &      (A)      & \\
\hline
Arc dipole          &     0.45     &      63.5      &      0.93      &        0.42         &      750      & 192 \\
Straight BB+/-        &     0.45     &                &                &                     &               & 10 \\
Straight BM..., BV...     &     0.5      &                &      0.56      &        0.28         &      850      & 10 \\
Straight, BC..        &     1.5      &                &      0.37      &        0.562        &      175      & 10 \\
& \\

\hline
{Quadrupoles}       & {Length} & {Aperture} & {Gradient} & {Int. Strength} & {Current} & {Quantity} \\
(Location)          &     (m)      &      (mm)      &     (T/m)      &         (T)         &      (A)      & \\
\hline
Arc, Q58Al4          &     0.43     &      R 5       &      9.5       &         4.1         &      160      & 196 \\
Straight Q58Al4        &     0.43     &      R 5       &      9.5       &         4.1         &      160      & 127 \\
IR 2 Q58Cu4          &     0.43     &      R 5       &      11.9      &         5.1         &      200      & 30 \\
Insertion QF2         &     0.5      &                &      13.6      &         6.8         &     1200      & 2 \\
Insertion QD1         &     1.2      &                &                &                     &     (pm)      & 2 \\
Insertion SK1         &     0.2      &                &                &                     &     (pm)      & 2 \\
Skew quad           &     0.2      &                &      2.6       &        0.52         &      12       & 15 \\
& \\

\hline
{Sextupoles} & {Length} & {Aperture} & {Strength} & {Int. Strength} & {Current} & {Quantity} \\
(Location) & (m) & (mm) & (T/m$^2$) & (T/m) & (A) & \\\hline
Arc SF, SD1          &     0.25     &      R 60      &      192       &        48.1         &      310      & 76 \\
Arc SD2            &     0.35     &      R 60      &      245       &        85.6         &      500      & 8 \\
IR 2             &     0.25     &      R 60      &       0        &                     &               & 7 \\
& \\

\hline
{Orbit correctors}    & {Length} & {Aperture} &  {Field}   & {Int. Strength} & {Current} & {Quantity} \\
(Location)          &     (m)      &      (mm)      &      (T)       &        (Tm)         &      (A)      & \\
\hline
Arc X             &    0.233     & $130\times90$  &     0.0365     &       0.0085        &      12       & 96 \\
Arc Y             &    0.312     & $250\times90$  &     0.0212     &       0.0066        &      12       & 92 \\
Arc X wide          &              &                &                &        0.012        &      12       & 4 \\
Straight           &     0.3      &                &     0.0252     &       0.00755       &      12       & 104 \\
\hline
\end{tabular}
\end{table}
}

\subsection{Dipoles}

\begin{figure}[htb]
 \centering
 \includegraphics[width=0.9\textwidth]{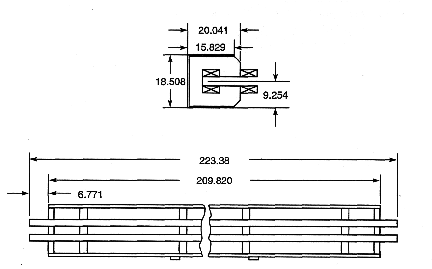}
\caption{Cross section of a \pepii HER main dipole. All dimensions are in inches.}
\label{fig:HER_Dipole}
\end{figure}

The \pepii HER dipoles have {\sf C}-shaped yokes and
2.2\cm sagitta based on their design 165\m bending radius. For
\superb, the bending radius for the main arc dipoles is 139\m and the
sagitta will increase to 2.6\cm, which is close enough to the original
value. For some of the dipoles in the final-focus region the bending
radius is reduced to 110\m, but the increase in sagitta to 3.3\cm will
be tolerable given the more than 5\cm total width for the good-field region
and the fact that the magnets can always be centered on the average
beam orbit. Figure~\ref{fig:HER_Dipole} shows a sketch of a \pepii
HER dipole.

While the \pepii HER dipoles are the original PEP dipoles, the
magnets were completely overhauled and refurbished during
construction of \pepii, serialized, and mechanically and magnetically
measured. The measurement data--$\int{Bdl}$, field harmonics at 
0.9$\,$Tm and gap height \vs\ $s$--are available in the archives of the
Magnetic Measurement Group at SLAC~\cite{bib:SLAC-MMG}.
They have been in constant use since \pepii commissioning began. 
Despite the high beam current, the
radiation environment in the \pepii arcs is actually quite benign,
and no evidence for significant radiation damage to the magnet coils
has been seen. We therefore, at present, see no need to re-measure or
refurbish the dipole magnets, although each magnet coil will be
carefully inspected for signs of aging. Comparing Tables
\ref{tab:HER_magnet_Needs} and \ref{tab:HERMagnets} shows that the
\superb\ HER dipole magnet requirements, including the soft
bends, are well satisfied, including a significant number of spare 
\pepii HER dipole coils.

\begin{figure}[htb]
\centering
\includegraphics[width=0.8\textwidth]{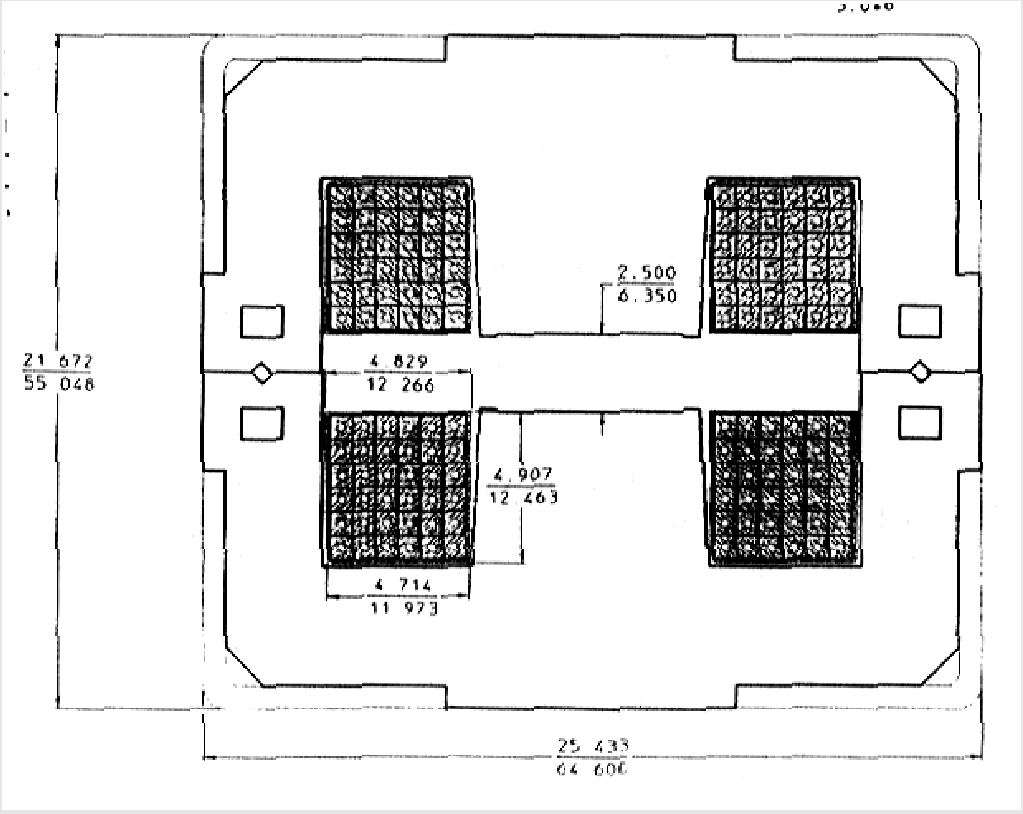}
\caption{Cross section of a \pepii LER main arc dipole.
Dimensions are given in inches (upper numbers) and cm (lower numbers).}
\label{fig:LER_Dipole}
\end{figure}

The 0.45\m-long \pepii LER dipoles are box-type magnets.
Because of their short length, there is no issue with the different sagitta
at any reasonable bending angle. In \superb\ the angle will be less
than at \pepii because of the added 0.75\m dipole magnets. These long
dipoles will be newly built, likely using the
laminations cut for the existing \pepii LER arc dipoles. It is also
conceivable to rebuild some of the excess short dipoles,
combining two into a 0.75\m long unit; a similar conversion was done
with PEP-I quadrupoles to create additional magnets for the \pepii
HER. The sixteen 5.4\m-long dipoles needed for the \superb\ LER will be
covered by left-over \pepii HER dipoles. The \pepii LER dipole
magnets were built new at the time of \pepii construction. They were
measured at the factory at that time, however, an individualized set
of measurements does not exist for each magnet. There is also a
certain variation of field shape with excitation in these magnets.
We therefore anticipate re-measuring each of the dipoles at the
operating field for \superb, before installation in the \superb\ LER.
As in case of the HER dipoles, however, there is no need to
refurbish the \pepii LER dipole magnets; a careful inspection should
suffice. Figure~\ref{fig:LER_Dipole} shows the cross
section of the LER dipoles. There is
significant space available in the horizontal plane to accommodate
an antechamber for the vacuum system.

\subsection{Quadrupoles}

It is anticipated that most \pepii quadrupoles will find use in
\superb. All of the 0.56\m quadrupoles of the \pepii HER will be used
in the \superb\ HER; in fact, more magnets are needed than
exist. Some of the new quadrupoles will be of the same design as the
exciting \pepii HER quadrupoles, having similar field gradients.
However, there are about 30 quadrupoles with gradients exceeding the
specification for the \pepii HER 0.56\m-long quadrupoles. These will
be built using new designs optimized for the higher field
requirement. Since these are DC magnets, matching of new and old
magnets is straightforward. In the same way, an additional sixty 0.73\m-long
quadrupoles have to be built.

\begin{figure}[htb]
\centering
\includegraphics[width=\textwidth]{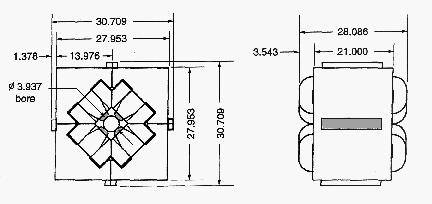}
\caption{Cross section and side view of a \pepii HER main arc
quadrupole.  All dimensions are in inches.}
\label{fig:HER_Quadrupole}
\end{figure}

\begin{figure}[htb]
\centering
\includegraphics[width=0.7\textwidth]{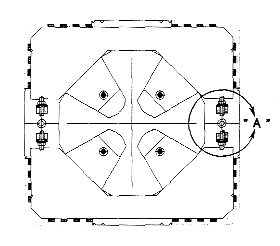}
\caption{Cross section of a \pepii quadrupole.}
\label{fig:LER_Quadrupole}
\end{figure}

For the 0.43\m-long \superb\ LER quadrupoles the needs are covered
by the existing \pepii LER quadrupoles. The latter come in
three different coil configurations with somewhat different maximum
excitation, so care will be taken in matching the coil type to the
requirements. In addition, over forty new 0.56\m-long quadrupoles are
needed. While these can use the \pepii HER lamination design, a
modified coil design may be necessary to accommodate an antechamber
for the vacuum system. Again, a complete audit trail exists for the
measurements of the \pepii HER quadrupoles, while for the \pepii LER
quadrupoles only a sparse data set is available. As a result, we will
need to re-measure the \pepii LER quadrupoles as well. Careful inspection of
all coils will detect any sign of aging, and there is a significant
number of spare coils available in case it is decided to replace
some of the coils. There may, however, be cases of quadrupoles in
\superb\ being excited at higher current than in \pepii. In these
cases we will change the cooling circuits to connect all coils in
parallel, thus minimizing the total temperature increase during
operation. Figure~\ref{fig:HER_Quadrupole} shows a cross sectional and
side view of a \pepii HER quadrupole, while Fig.~\ref{fig:LER_Quadrupole}
shows a \pepii LER quadrupole.

\subsection{Sextupoles}

Altogether, the two \superb\ rings will use over 500 sextupoles, a
little less than half of which exist in \pepii. Additional
magnets will be built using the original \pepii lamination die. Since the
sextupoles in the \superb\ LER have less strength, we may use two
different coil configurations optimized for the application. The
0.6\m-long sextupoles in the final focus will be new.

\subsection{Correction Magnets}

The basic orbit-corrector dipoles exist in three types for each ring:
horizontal arc-type, vertical arc-type and straight-section type, which
are mounted either as horizontal or vertical correctors around the
chamber of circular cross section. Where prudent these magnets will be
reused. However, the \superb\ vacuum system will differ from
that of \pepii, which may prevent reuse of some of these magnets. Since
the cost for orbit correctors is fairly modest, it appears
prudent to avoid compromising the vacuum chamber geometry in order to
reuse existing orbit corrector magnets. The same principle applies to
other correction magnets, such as skew quadrupoles.

\subsection{Field Quality}

\begin{figure}[htb]
\centering
\includegraphics[width=0.7\textwidth]{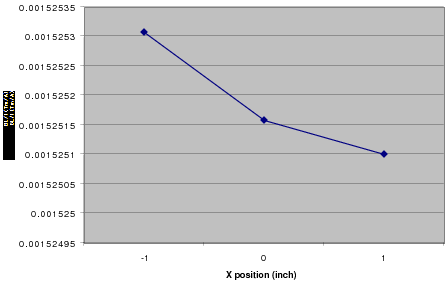}
\caption{Field uniformity of a sample of \pepii HER 5.4-m dipole magnets.}
\label{fig:Dip_Sum}
\end{figure}

\begin{figure}[htb]
\centering
\includegraphics[width=0.7\textwidth]{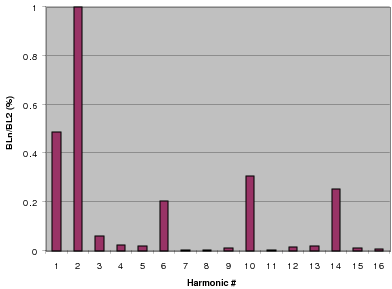}
\caption{Harmonic spectrum of a sample of \pepii HER 0.56\m  quadrupoles.
Harmonic \# 2 is the gradient.}
\label{fig:Quad_harm}
\end{figure}

Field uniformity requirements for \superb\ magnets will be determined
following more detailed tracking studies. However, since the beam
sizes are small and orbit excursions will have to be tightly
controlled in order to preserve the small emittances, the beams do not
sample field regions far from the nominal center line. We
therefore expect the field uniformity tolerances of the \pepii magnets
to be sufficient for \superb\ applications. The field uniformity of the
\pepii HER dipoles is shown in Fig.~\ref{fig:Dip_Sum}; the field
harmonics of the \pepii HER 0.56\m quadrupoles magnets are shown in
Fig.~\ref{fig:Quad_harm}. Since we will have individual measurement
data for each magnet, sorting algorithms will be employed as necessary to
mitigate the effect of field differences between the magnets in a
family, as was done for \pepii.

\subsection{Power Conversion}

The DC power supplies for the magnets represent a significant share
of the overall cost of the magnet system. The approach taken in
\pepii is to power long strings of identical magnets with 500\V,
400\amp chopper units fed from a bulk power supply (one for each
ring), which in turn is fed by a 480\V ac line. Shorter strings (or
``families'') are fed from smaller individual supplies operating on
208 or 480\V ac feeds. All of these supplies are of relatively
modern switching type and therefore, in principle, capable of being
operated at the 50\Hz ac frequency used in Europe (as opposed to
the 60\Hz used in North America) including the large bulk power
supplies for the choppers~\cite{bib:Bellomo}. While the details must
worked out, reuse of components of the \pepii magnet power
system appears to be feasible.

\subsection{Summary of Regular Lattice Magnets}

\superb\ magnet requirements are well within the performance envelope
of the \pepii magnets, and almost all \pepii magnets, with the
possible exception of specialty magnets such as the insertion
quadrupoles, will be reused. Additional magnets will be built to
existing designs wherever feasible. Only a limited number of \superb\
magnet designs have no \pepii counterpart and will be of a new
design. Table~\ref{tab:Magnet_Summary} summarizes the magnet types
and building needs.

\begin{table}[htbp]
\caption{\superb\ magnet summary.}
\label{tab:Magnet_Summary}
\vspace*{2mm}
\setlength{\extrarowheight}{1pt}
\centering
\begin{tabular}{lcp{2cm}p{2cm}cl}
\hline\hline
{Type} & {Length} & {Required} & {Extant} & {Build} & {Design} \\
       & {(m)} & {for \superb} & {at \pepii} & {new} & \\
\hline
  Dipole   &       0.45       &            144            &          194           &        0        & \\
  Dipole   &       0.75       &            144            &           0            &       144       & PEP-II (lamin.) \\
  Dipole   &       5.4        &            176            &          194           &        0        & \\
  Dipole   &        2         &             4             &           6            &        0        & soft bends \\[5mm]
Quadrupole &       0.43       &            341            &          353           &        0        & \\
Quadrupole &       0.5        &            70             &           0            &       70        & PEP-II or new \\
Quadrupole &       0.56       &            255            &          202           &       53        & PEP-II, new coil \\
Quadrupole &       0.56       &            32             &           0            &       32        & new (high field) \\
Quadrupole &       0.73       &            138            &           81           &       57        & PEP-II \\[5mm]
Sextupole  &       0.25       &            452            &          188           &       264       & PEP-II (2 coil configs.) \\
Sextupole  &       0.6        &             8             &           0            &        8        & new \\
\hline
\end{tabular}
\end{table}

\begin{figure}
\centering
\includegraphics[width=\textwidth]{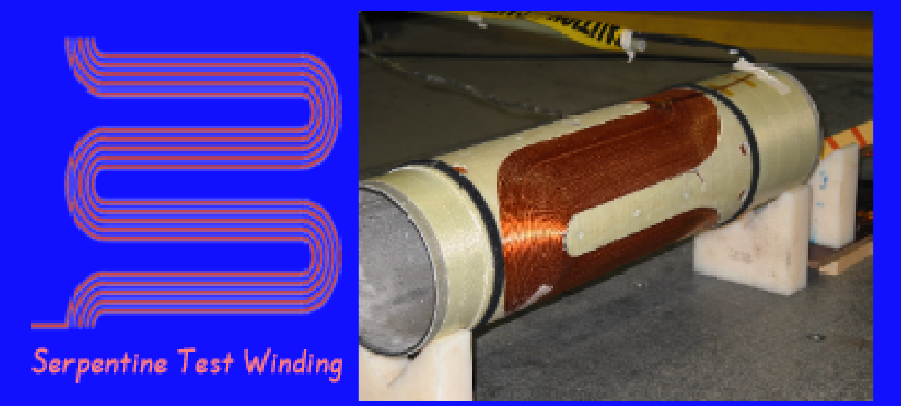}
\caption{IP SC quadrupole during manufacture.}
\label{fig:Parker2}
\end{figure}

\subsection{IR Quadrupoles}

\begin{figure}[hbt]
\centering
\includegraphics[width=0.6\textwidth]{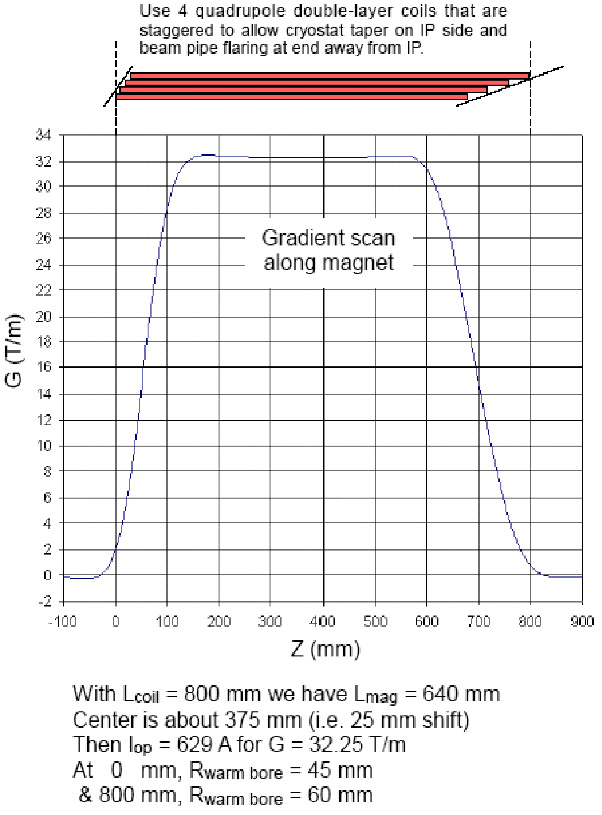}
\caption{Field profile of a superconducting quadrupole using thin SC windings.}
\label{fig:Parker1}
\end{figure}

The IR QD0 magnet can be a permanent magnet as described in
Section~\ref{sec:QD0}. The next IR quadrupoles must be
compact, since we need to separate into two beamlines as soon as possible
in order to reduce IR backgrounds. This suggests 
the use of
superconducting quadrupoles, which can be
constructed in a fashion similar to the IR quadrupoles built at
Brookhaven Laboratory for HERA and BEPC-II~\cite{bib:Parker}. This
is also the choice for the ILC final focus. One possible version of the
\superb\ IR quadrupole design is shown in Fig.~\ref{fig:Parker2}
and \ref{fig:Parker1}. This specific superconducting
quadrupole design is 0.8\m long and has a gradient of 25.8$\,$T/m. The inner
warm bore of the cryostat in this example is about 6\cm radius
and the outer warm shell diameter about 20\cm.
Thus, the design has about 14\cm in radius, including the
cryostat, to provide main quadrupole windings as well as solenoidal,
skew-quadrupole, and dipole windings in a combined design.
Other designs
are also possible, and a study of alternatives is in progress.

\subsection{Permanent Magnet Damping Wiggler}

The srong damping required for the \superb\ rings 
will be provided by wiggler magnets. These will be constructed as
permanent magnets in order to minimize energy consumption.
In addition, we will use a wedge-shape pole-tip design
to concentrate magnetic flux lines, thereby minimizing the volume of
magnetic material. The field amplitude for each individual pole
will be tunable at the level of about 400\Gauss, using a design that
provides for independent adjustment of individual poles.
The wiggler parameters are shown in Table~\ref{tab:wig}.

\begin{table}[htb]
\caption{Wiggler parameters.}
\label{tab:wig}
\vspace*{2mm}
\setlength{\extrarowheight}{2pt}
\centering
\begin{tabular}{cccccc}
\hline
\hline
Field & Pole gap & Period length & Transverse homogeneity \\
(T) & (mm) & (mm) & ($@\pm 2\cm$)  \\
\hline
          $1$           &        $30$         &     $400$     & $7\EE{-4}$\\
\hline
\end{tabular}
\end{table}

\begin{figure}[htb]
\centering
\includegraphics[width=\textwidth]{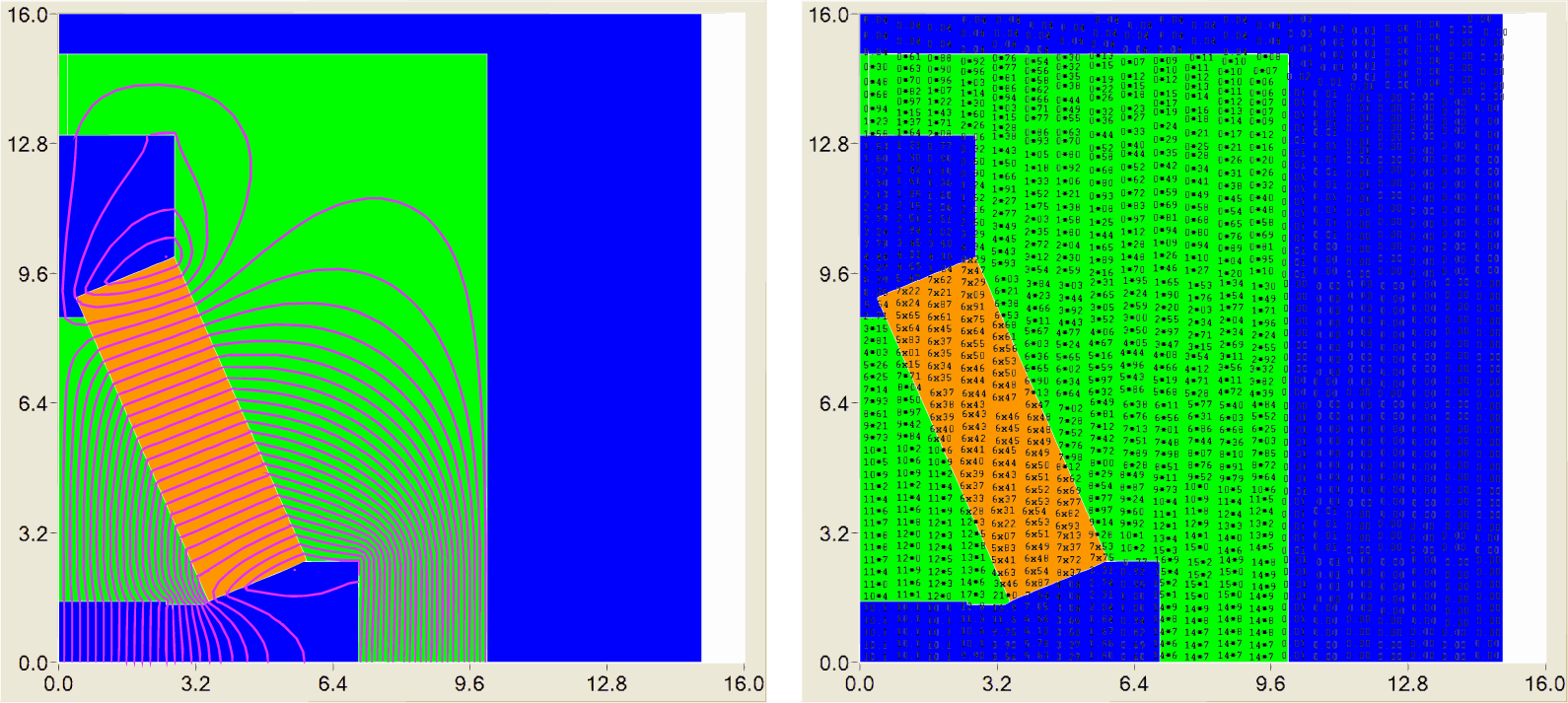}
\caption{Magnetic flux distribution for the transverse cross-section
of the wiggler quarter (left).
Color palette: green--iron yoke, orange--permanent magnet, blue--air.
Field distribution in the wiggler transverse cross-section in kG (right).}
\label{fig:Wig1}
\end{figure}
\begin{figure}[htb]
\centering
\includegraphics[width=0.6\textwidth]{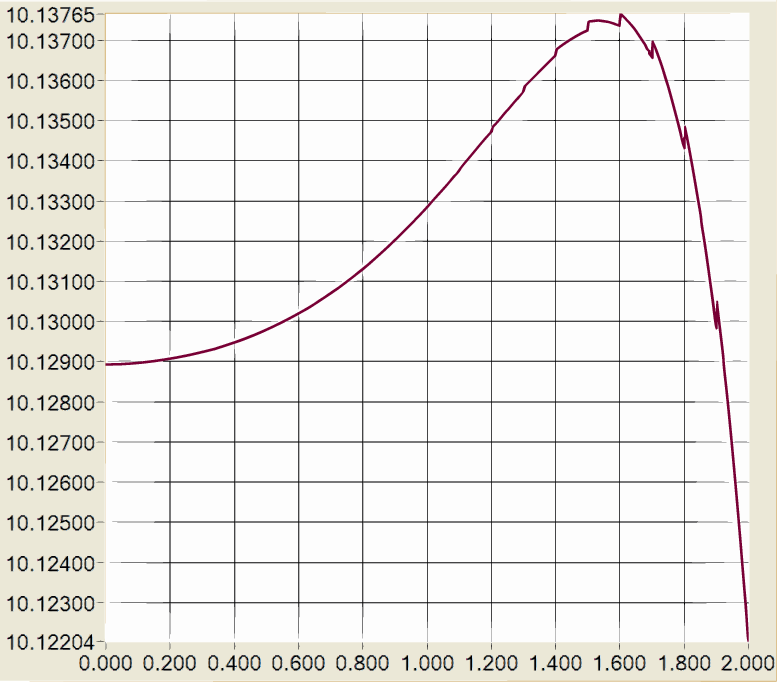}
\caption{Transverse distribution of the wiggler magnetic field (kG).}
\label{fig:Wig3}
\end{figure}

Figure~\ref{fig:Wig1} shows the magnetic flux lines and
the maximum field in the relevant parts
of the wiggler. The maximum field in the iron yoke reaches 1.2--1.4\Tesla,
which is far from saturation. Figure~\ref{fig:Wig3}
shows the transverse homogeneity of the wiggler field.

\begin{figure}[htb]
\centering
\includegraphics[width=\textwidth]{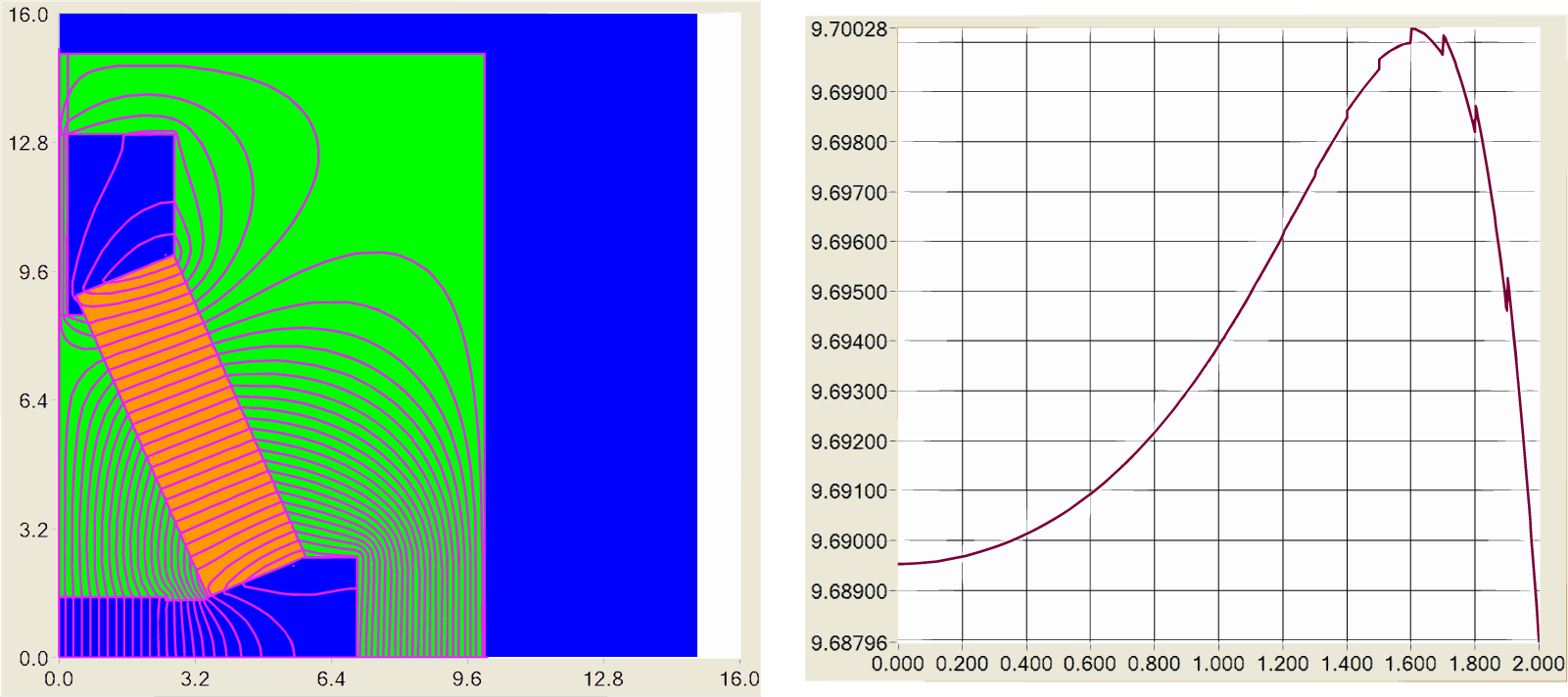}
\caption{Flux line distribution (left) and
transverse distribution of the wiggler field with the magnetic
shunt inserted (right).}
\label{fig:Wig4}
\end{figure}

To achieve a uniform field profile in the longitudinal direction, a
special magnetic shunt adjustment is foreseen between the pole and the
upper part of the iron yoke (see lefthand plot of Fig.~\ref{fig:Wig4}). 
This shunt expels some fraction of magnetic flux, allowing the field in the
wiggler gap to be tuned (righthand plot).

\begin{figure}[htb]
\centering
\includegraphics[width=\textwidth]{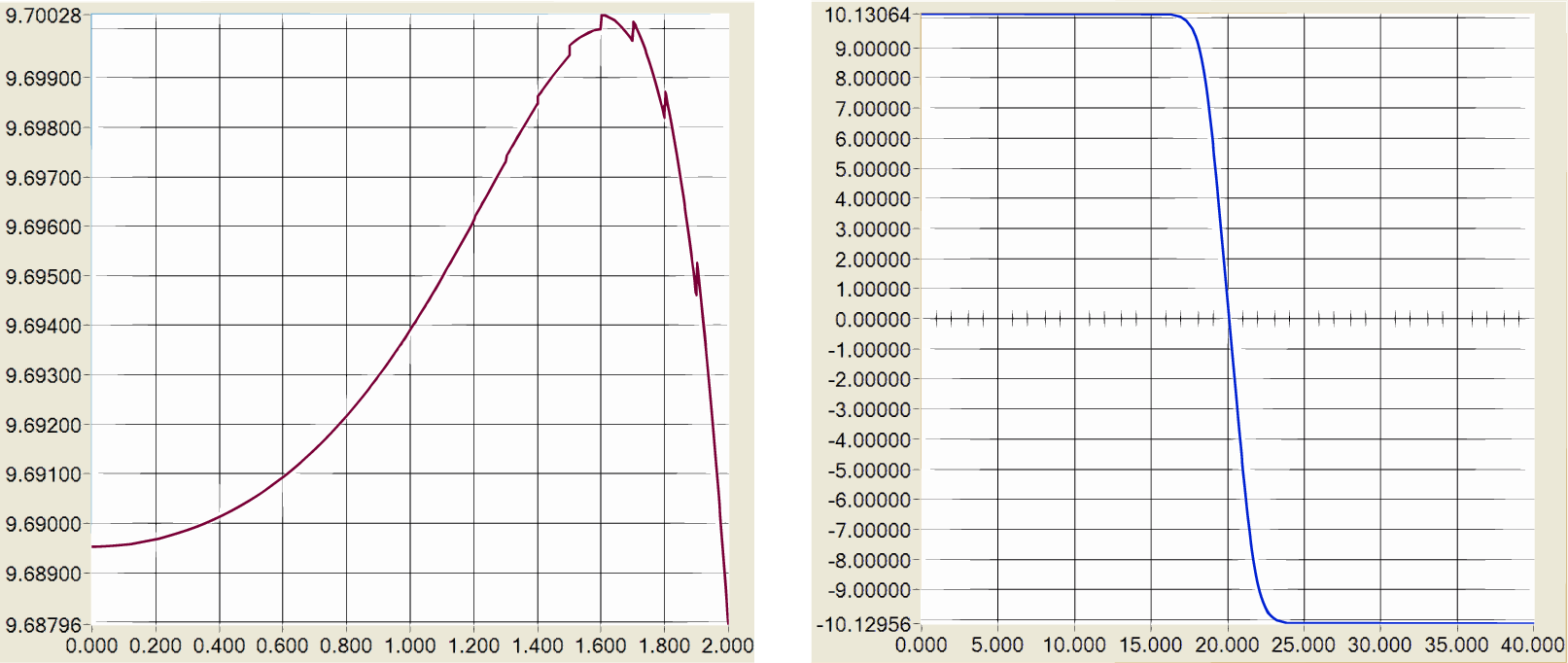}
\caption{Longitudinal magnetic flux distribution (left) and longitudinal
distribution of the wiggler field (right).  }
\label{fig:Wig6}
\end{figure}

Fig.~\ref{fig:Wig6} shows the longitudinal distribution of the
magnetic flux and the longitudinal profile
of the wiggler field. The permanent magnets, which are shown in
Fig.~\ref{fig:Wig6} shield the magnetic field of adjacent poles,
providing easy wiggler assemble and adjustment.
\clearpage

\afterpage{\clearpage}

\section{RF Systems}
\label{section:RF}
\subsection{Parameters}
The RF system provides the energy to the beam necessary to make up for
synchrotron-radiation losses and---to a lesser extent---losses due to
the loss factor of the RF system and the cavities.
In addition to replenishing the beam energy, the RF system also
provides longitudinal focusing, \ie, the RF voltage, together with
the momentum compaction $\alpha_p$ of the magnet lattice, controls the
bunch length. For \superb, the bunch length is
about 6\mm in each ring at the operating current; 
in order to achieve this, the
zero-current bunch lengths should be less by about one mm. One of the most
challenging aspects of a high-current \epem\ storage ring,
the RF system for \superb\ draws heavily on the experience gained at
\pepii, and in fact it is planned to reuse the \pepii RF hardware.

The RF voltage, however, cannot be chosen solely on the basis of
bunch length. In order to control the cavity at high beam current a
minimum amount of stored energy, \ie, voltage on the cavity,
is required. In addition, the maximum power transmitted to the beam
by each cavity is limited by the coupling hardware and by the RF
window necessary to separate the cavity vacuum from the waveguide,
thus setting a lower limit on the number of cavities necessary to provide a
certain amount of power to the beam. The power limit for each cavity
window of the \pepii RF system is $0.5 \MW$~\cite{bib:PEP_rf_paper}.

The fundamental operating parameters for the system are given in
Table~\ref{tab:RF_parms}.

\begin{table}[htb]
\centering
\begin{threeparttable}
\caption{RF parameters for the nominal \superb\ beam currents. \label{tab:RF_parms} }
\vspace*{2mm}
\setlength{\extrarowheight}{1pt}
\centering
\begin{tabular}{lcc}
\hline\hline
           Parameter               &    HER     & LER \\
\hline
      RF frequency f$_{rf}$        (MHz) &    476     & 476 \\
         Harmonic no. h                  &    3572    & 3572 \\
 Beam current I$_{\mathrm{beam}}$  (A)   &    1.3     & 2.3 \\
     Energy loss/turn  U$_0$       (MeV) &    3.3     & 1.9 \\
      Power loss/turn  P$_0$       (MW)  &    4.3     & 4.4 \\
 Momentum compaction  $\alpha_c$         & 2.3\EE{-4} & 3.3\EE{-4}\\
Total RF voltage V$_{\mathrm{RF}}$ (MV)  &     15     & 10\\
          No. cavities                   &     22     & 14\\
          Volts/cavity             (MV)  &   0.682    & 0.714\\
 No. klystrons with 2 cavities           &     5      & 7 \\
 No. klystrons with 4 cavities           &     3      & -- \\
     Forward power/klystron        (MW)  &    0.85    & 0.76\\
 Coupling factor
       $\beta_{\mathrm{opt}}$            & 2.5/5.5$\tnote{a}$  & 8.3\\
       Detuning frequency          (kHz) &    -100    & -220\\
      Synchrotron frequency        (kHz) &    2.2     & 1.9\\
    Bunch length (0-current)       (mm)  &    4.4     & 5\\
\hline
\end{tabular}
\begin{tablenotes}
\item[a]$\beta_{opt}$ for 2/4 cavities per klystron.
\end{tablenotes}
\end{threeparttable}
\end{table}

At the planned beam currents, $4.3 \MW$
radiation power have to be provided to the beam in the HER and $4.4\MW$
to the LER beam. These values are comparable to the corresponding values
for the \pepii rings. An RF frequency of
$476\MHz$ has been chosen to allow reuse of the \pepii RF components. 
Our design requires 34
of an available total of 36
cavities to provide the power to both beams. A total of
fifteen $1.2\MW$ klystrons provide the power. This somewhat exceeds the
minimum required and provides an operating margin allowing the system
to operate reliably with a minimum number of trips, while at the same
time increasing the beam-current reach of the rings by
about $20\%$. In order to compensate the beam loading, the cavities
are detuned from resonance, where the detuning frequencies shown in
Table~\ref{tab:RF_parms} are less than those routinely
achieved at \pepii ($\approx300\kHz$.)

The klystrons required, plus several spares, exist
at SLAC, although more klystrons may eventually
be built to replenish the supply as tubes age.

\subsection{RF System Description}

\begin{figure}[htb]
\centering
\includegraphics[width=0.9\textwidth]{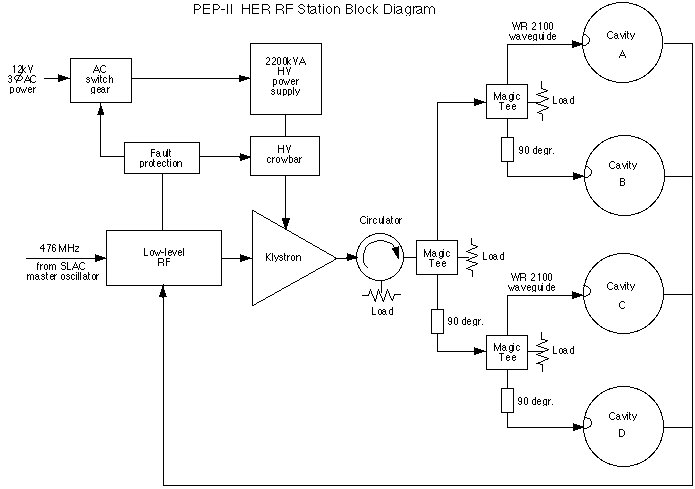}
\caption{Block diagram of an RF station.}
\label{fig:RF_A}
\end{figure}

Each station consists of a DC power supply, klystron amplifier and a
high-power circulator and waveguide distribution system at
surface level feeding cavities down in the tunnel. A low-level RF
system provides control
and feedback for stable multi-bunch high current
operation.

\begin{figure}[htb]
\centering
\includegraphics[width=0.7\textwidth]{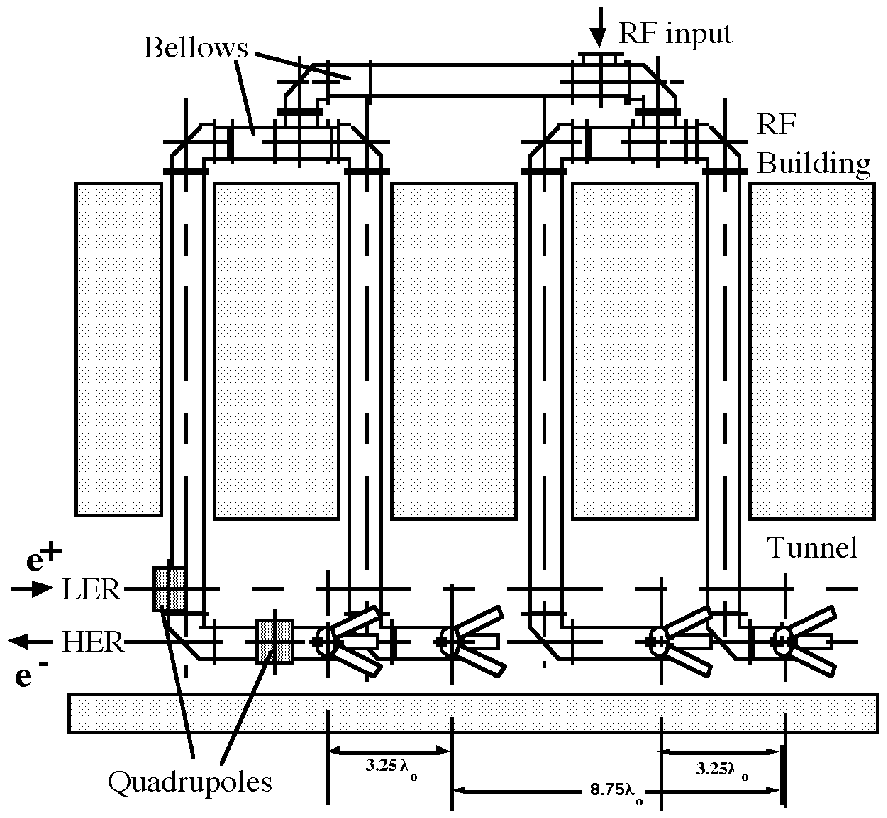}
\caption{Layout of the wave guides for a 4-cavity RF station.}
\label{fig:RF_B}
\end{figure}

\begin{figure}[htb]
\centering
\includegraphics[width=0.8\textwidth]{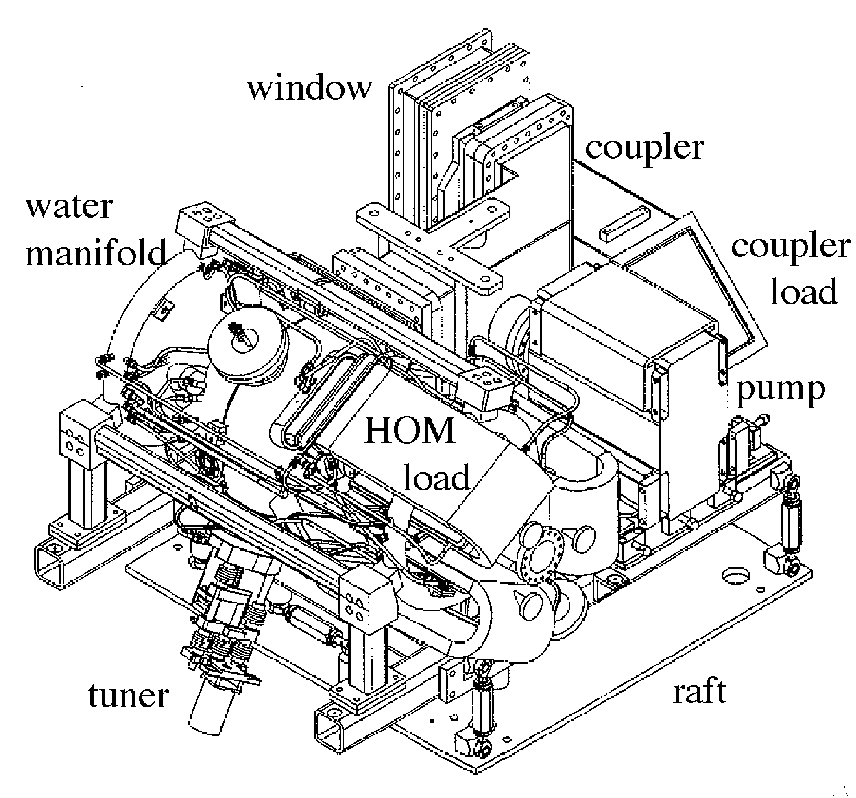}
\caption{Assembly of an RF cavity.}
\label{fig:RF_C}
\end{figure}

\clearpage

\begin{figure}[htb]
\centering
\includegraphics[width=\textwidth]{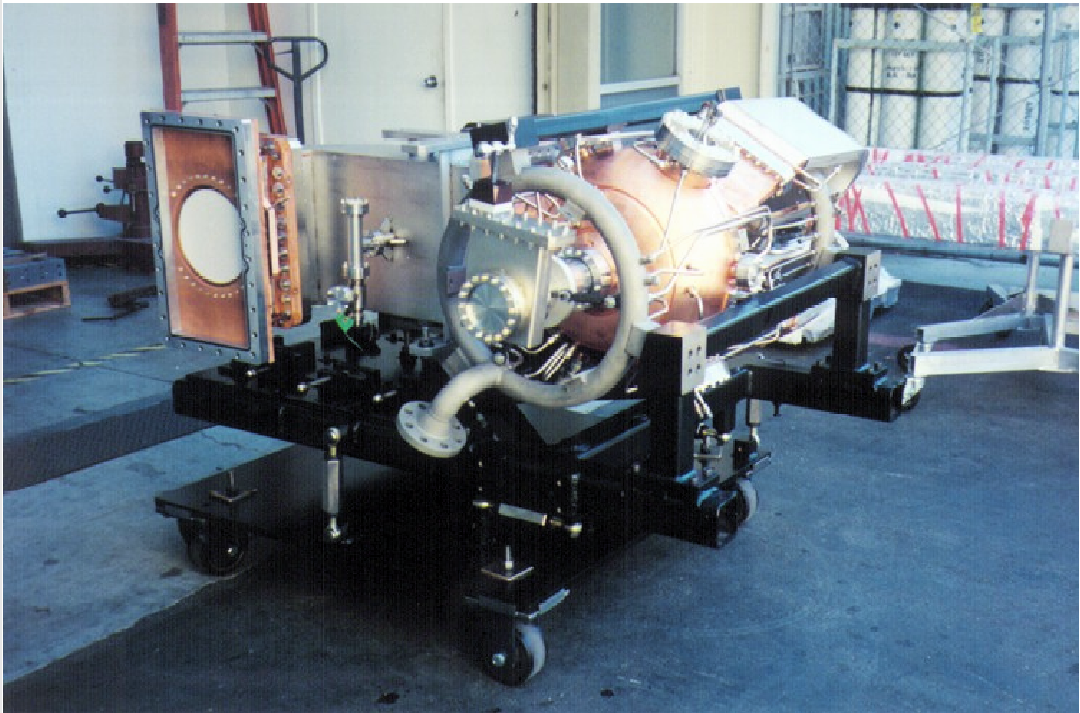}
\caption{Photo of an assembled RF cavity raft.}
\label{fig:RF_D}
\end{figure}

Figure~\ref{fig:RF_A} shows a block diagram of a typical RF station
layout. Figure~\ref{fig:RF_B} shows the waveguides running down
penetrations into the tunnel. Figure~\ref{fig:RF_C} shows a cavity
raft assembly with all ancillary equipment mounted, and
Figure~\ref{fig:RF_D} shows a production raft assembly.

The DC power
supply uses a modern SCR-switched design for both voltage
control and crowbar functions. The major components and the
enclosures were sourced from industry while the SCR switching
and crowbar stacks were fabricated in house. The klystrons
are required to provide a high average power with good
reliability, low group delay (to enable fast feedback), and
good efficiency. The RF distribution is via WR2100 waveguide, chosen
primarily for low group delay, and a circulator is used on the
output of each klystron.  The low-level RF system is a 
state-of-the-art implementation in the VXI environment and
incorporates fast RF electronics, programmable feedback loops,
tuner loops, a built-in network analyzer and other
diagnostics. High-level controls and the user interface are
implemented in EPICS.
The windows, tuners and HOM loads
have been specified for operation at up to a 3\amp beam
current to give safety margin and headroom for future upgrades
of the machine.

\subsection{RF Cavities}
\begin{figure}[htb]
\centering
\includegraphics[width=0.7\textwidth]{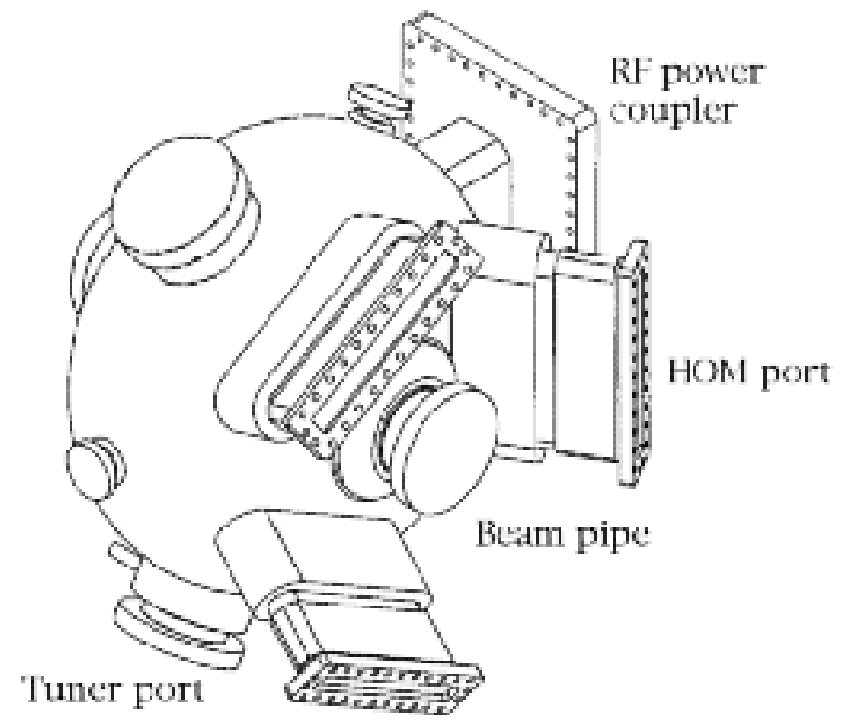}
\caption{RF Cavity.}
\label{fig:RF_G}
\end{figure}

The RF cavity is a reentrant design with the addition of
three HOM damping waveguides as shown in Fig.~\ref{fig:RF_G}. The design
parameters are listed in Table~\ref{tab:RF_T3}. The cavity also has ports for an aperture
coupler, a PEP-style plunger tuner, pick-up loop, {\it etc}. The location of
the HOM damping ports has been chosen to couple to all modes, but most
strongly the worst ones.
Measurements on a cold test model have
shown that the damping scheme reduces the impedance of the worst HOMs
by more than three orders of magnitude \cite{bib:rf_E}. Subsequent analysis,
measurement of the high-power cavity and observations with beam have
confirmed this \cite{bib:rf_F}. The coupling factor has been set to 3.6 for both rings
in order to match the transient response in each ring from the gap in
the beam.

\begin{table}[htb]
\caption{RF cavity parameters.}
\label{tab:RF_T3}
\vspace*{2mm}
\centering
\setlength{\extrarowheight}{1pt}
\begin{tabular}{p{6cm}c}
\hline\hline
       Parameter        & Value \\
\hline
     RF frequency       (MHz)  & 476 \\
 Shunt impedance $R_s$  (\MOhm)  & 3.5 \\
      Gap voltage       (MV)   & 1.02\\
 Accelerating gradient  (MV/\m) & 4.6\\
    Wall power loss     (MW)   & 150\\
Coupling factor $\beta$         & 3.6\\
      Unloaded Q                & 32000\\
\hline
\end{tabular}
\end{table}

In order to optimize the application of the cavity to \superb, we
note that the coupling factor $\beta$ should increase significantly
for the LER. This factor, which measures the coupling of the
generator to the cavity, is dependent on the details of the coupling
system, \ie, the physical size of the coupling slot.
Fortunately, there appears to be a way to affect this coupling with
a change of the size of the waveguide leading to the coupler slot:
by reducing its vertical (short-) side length by a moderate amount,
we can raise the coupling factor, since it scales inversely with the
square of this dimension. Thus $\beta$ up to 10 is achievable
without changes to the cavity itself. Since the quality of
the match varies only slowly with $\beta$, we plan to optimize for a
common coupling factor for all cavities.

\subsection{HOM Damping}
\begin{figure}[htbp]
\centering
\includegraphics[width=0.9\textwidth]{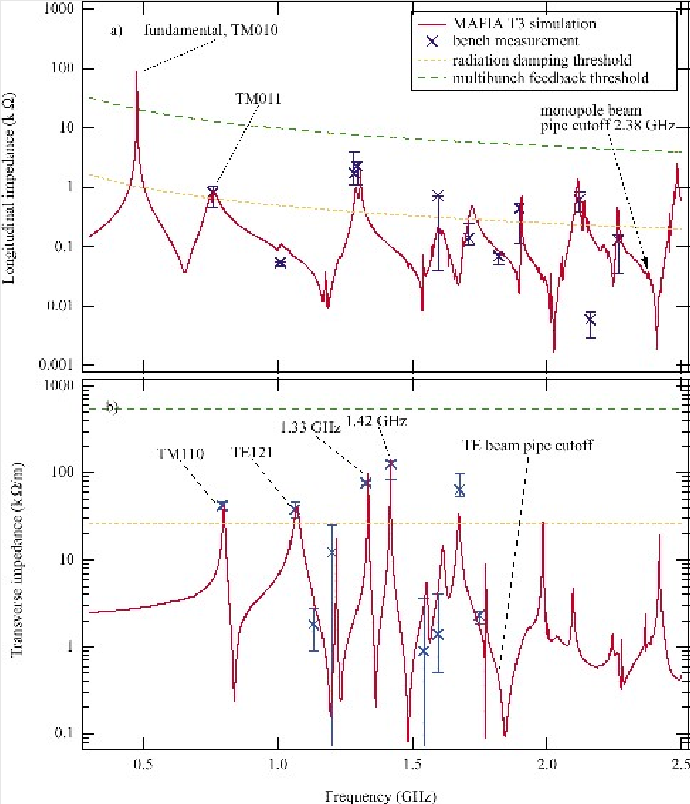}
\caption{Impedance spectrum of RF cavity.}
\label{fig:RF_K}
\end{figure}

\begin{figure}[htbp]
\centering
\includegraphics[width=0.9\textwidth]{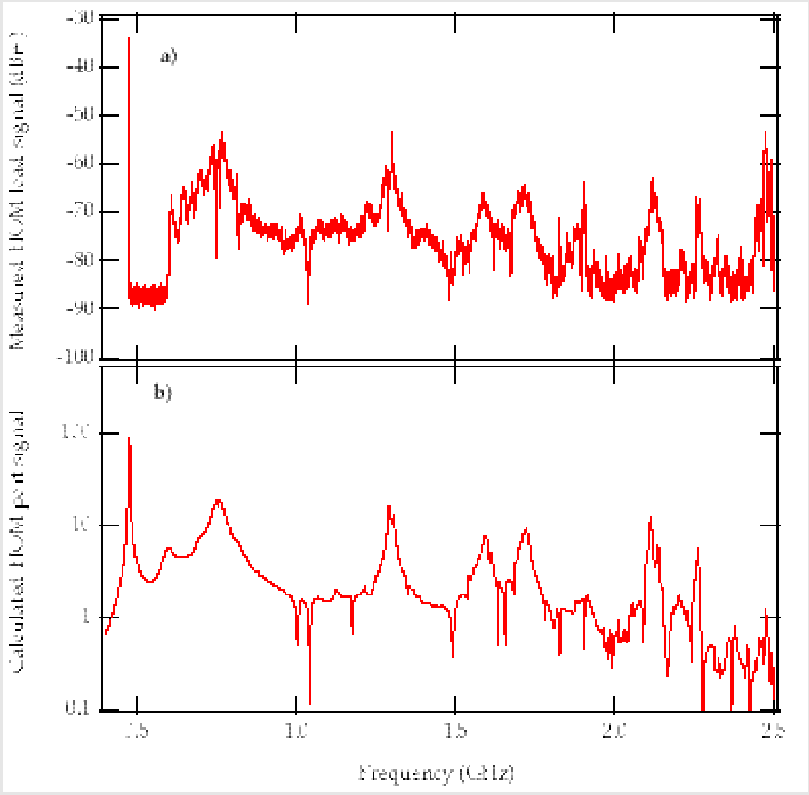}
\caption{Measured and calculated spectrum of a single bunch at the HOM load.}
\label{fig:RF_L}
\end{figure}

Figure~\ref{fig:RF_K} shows the longitudinal and transverse
impedance spectra of the cavity, calculated using a time
domain method \cite{bib:rf_F} along with the original lab measurements.
There is reasonably good agreement for the strongest HOMs. One cavity
is instrumented with a pick-up probe to monitor the signal reaching
the HOM load. This signal has been observed with a single bunch in the
machine, which excites a signal in the cavity---in
proportion to the impedance---at all harmonics of the revolution
frequency. Due to the short bunch length the spectrum extends to very
high frequencies and excites all modes up to the beam pipe cutoff and
beyond. Figure~\ref{fig:RF_L} shows the measured and calculated spectra reaching
the HOM load. There is reasonable agreement in the general
features of the two spectra. Initial operation of the feedback systems
indicated that true HOM-driven instability growth rates were
consistent with expectations \cite{bib:rf_J}.

The HOM loads have a design power rating of 10\kW each. In an updated
estimate based on the measured HOM spectrum of a cavity, and for a
bunch length of 1\cm, the total
beam induced HOM power for 4.2\ns bunch spacing is estimated to be
15\kW nominally, and 21\kW worst-case, if a
particular HOM happens to coincide in frequency with a beam
harmonic. For 2.1\ns spacing the numbers are 6\kW and 13\kW, respectively.
Scaling up to 4\amp at 2.1\ns, even the worst-case estimate is only about
23\kW, within the operating envelope of the existing HOM loads,
even when taking into account the shorter bunches at \superb.

\subsection{Beam Stability}
\begin{figure}[htb]
\centering
\includegraphics[width=0.8\textwidth]{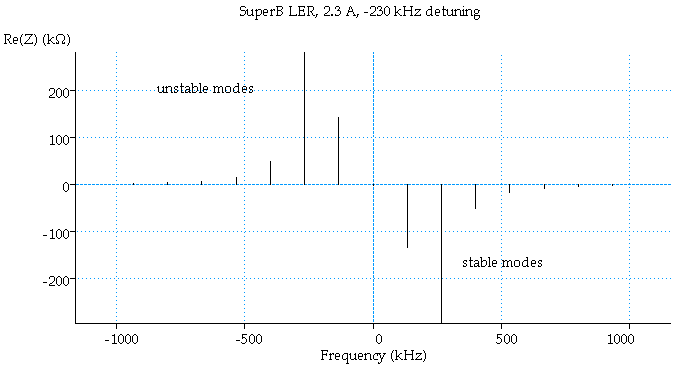}
\caption{Driving impedances for modes -7\ldots7.}
\label{fig:modal_impedance}
\end{figure}

The combination of high beam current and many bunches can lead to
longitudinal multibunch instability depending on the cavity impedance
presented to the beam. For each longitudinal mode there is an upper and a lower
synchrotron sideband. The real component of the impedance at the lower
sideband damps that particular mode while the component at the upper
sideband drives that mode unstable. Thus, the difference between these
two impedances determines the growth or damping rate. The particular
impedance depends on the amount of detuning. For the \superb\
RF system at nominal parameters, the resulting modal spectrum of the
impedance is shown in Fig.~\ref{fig:modal_impedance}.

The driving terms for $m=-1$ and $m=-2$ would cause strong instability
of these modes; fast and comb-filter feedback loops in the LLRF
are employed to reduce the driving impedance by a factor of around
200~\cite{bib:Corredoura_93}.
To get a sense of the significance of these numbers we compare to the
situation at \pepii. At \superb, there will be 14 cavities rather than
8, but the beam current will be 2.3\amp rather than an achieved 3\amp and
projected 4\amp at
\pepii. Since the growth rates depend on the product of impedance times
the number of
cavities, and on the beam current, we expect growth rates at \superb\ to
be some 30\% above the presently measured 2 to 3\ms$^{-1}$. This is
within the performance envelope of the low group delay woofer of the
longitudinal feedback system without further effort to reduce the
impedance of the cavities. For the upgraded
beam current, however, we will likely exceed the \pepii growth rate by a
significant amount. Possible measures to prevent beam instability under these
conditions include reduction of group delay through the system to
allow higher loop gain of the feedback system, and a more sophisticated
comb-filter loop, allowing one to individually taylor the gain at each
sideband, thus changing the difference of the impedances for each mode
in favor of the damping impedance~\cite{bib:Dmitry_private}.
The impedance for each of the HOMs does not depend on the specific
detuning conditions of the cavities. An increase in growth rate of
30\% compared to \pepii appears to be well within the capability of an
upgraded longitudinal feedback system. \pepii plans to reach
4\amp beam current in the LER before the end of 2008, providing
valuable additional data on high-beam-current running.

\subsection{Low Level RF System}

\begin{figure}[htb]
\centering
\includegraphics [width=0.7\textwidth]{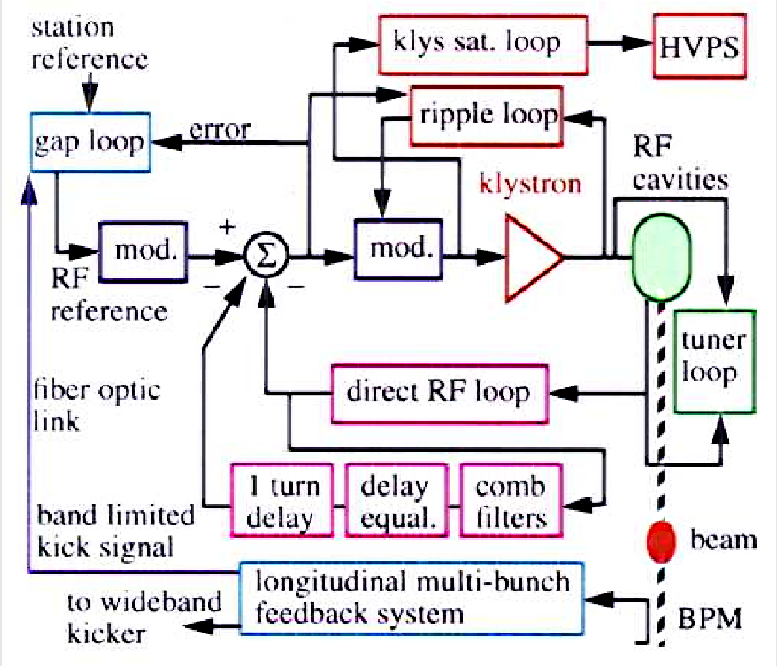}
\caption{Low level RF system diagram.}
\label{fig:PEP_LLRF}
\end{figure}

Figure~\ref{fig:PEP_LLRF} shows a block diagram of the low level RF
system, as implemented at \pepii. The fast, direct RF feedback
loop reduces the impedance presented to the beam by a factor in excess
of 10, with the comb-filter loop reducing the impedance at the modal
frequencies by another factor of 10. In order to maintain stable
operation, the gain and phase of these loops are adjusted to follow the
beam current. The ``Ripple loop'' stabilizes the phase through the
klystron. Another loop stabilizes the measured gap voltage by
adjusting the high voltage on the klystron. The gap feedforward system
is an adaptive loop that compensates the transients caused by the
kicker gap as the beam passes by, thus preventing saturation of the
klystron due to the gap transient. Finally, on the bottom of the
diagram the connections to the longitudinal feedback system
implementing the woofer
link are shown. In addition to these functions, diagnostics in the
system include a network analyser to record transfer functions
when tuning a station, and the ability to save important system
parameters in ``fault files'' for the last few tens of milliseconds 
before a beam
abort, thus allowing analysis of the cause of beam abort and RF
trips. The state of the system is maintained in ``bump-less reboot'' files
to ensure that a station comes up in its previous state after a reboot.
The LLRF is implemented in VME/VXI technology, and
controlled by an EPICS IOC. The system is described in numerous
references in the literature.

\subsection{Summary}

The \pepii RF stations are a very good match to the \superb\
requirements, providing sufficient voltage and power to exceed the
nominal requirements by a healthy margin. By reusing the \pepii RF
system, considerable savings in development and fabrication effort will
be realized. The HOM damping in the cavities reduces the impedance to
a level where the required damping rates for multibunch instabilities will be well
within the performance envelope of present-day technology. The
low-lying multibunch modes are effectively damped by the low-level RF
feedback loops, with the low-frequency beam-feedback system
(``woofer'') providing the balance of the damping required. It is
important to note that the parameters called for are a moderate
extrapolation of the well established performance
at \pepii, thus minimizing the risk in building such a system for
\superb. Some of the new technologies anticipated for \superb\ will
already be implemented at \pepii: the asymmetric comb filter and the Gboard-based
longitudinal feedback system~\cite{bib:Gboard}. For the upgraded beam
currents, the required additional power can be provided by adding a
moderate number of stations (2 or 3 in the LER, 1 station with two cavities in
the HER).



\afterpage{\clearpage}

\section{Vacuum System}
\label{section:vacuum}
\subsection{General Considerations}

Vacuum system parameters for \superb\ can be derived from two different
approaches. One is from the standpoint of the \pepii rings, which 
benefits from significant and lengthy experience gained in operating
a comparable vacuum system~\cite{bib:Wienands_PAC_PEPvac}. The other is
to take advantage of effort already spent on setting the \ilc Damping Ring 
vacuum system parameters~\cite{bib:Malyshev_ILCDR},
which takes into account the extremely
small beam emittance. For \superb\ we have adopted the pressure
specifications for the \ilc Damping Rings, which have beam emittances
similar to \superb, rather that the somewhat less stringent pressure
specifications for \pepii. The other vacuum parameters for the \superb\
rings (\eg\ apertures) in general do not deviate significantly from those
for \pepii. We summarize pressure requirements in
Table~\ref{tab:vacuum_parms}.

{ \setlength{\tabcolsep}{2pt}  
\begin{table}[htb]
\caption{\label{tab:vacuum_parms}Basic vacuum system parameters for \superb.}
\vspace*{2mm}
\setlength{\extrarowheight}{3pt}
\centering
\begin{tabular}{p{9.2cm}cccc}
\hline
\hline
 Parameter                 & $HER $    & $LER$     & $ HER$      & $ LER$       \\
                           &  Arc      &   Arc     &  Straight   &   Straight   \\\hline
 Beam Energy               (GeV) & 7         & 4         & 7           & 4      \\
 Beam $\gamma$                   & 13700     & 7829      & 13700       & 7829   \\
 Beam current              (A)   & 1.3       & 2.3       & 1.3         & 2.3    \\
 Chamber halfwidth         (m)   & 0.045     & 0.15      & 0.045       & 0.045  \\
 Bending angle             (rad) & 0.052     & 0.052     & --          & --     \\
 Bending radius            (m)   & 139       & 28        & --          & --     \\
 Length of radiation fan   (m)   & 3.1       & 2.9       & 3.6         & 3.6    \\
 E$_{loss}$/turn           (MeV) & 2.0       & 0.8       & --          & --     \\
 Total SR power            (MW)  & 2.6       & 1.9       & --          & --     \\
 Average pressure          (nTorr) & 0.5       & 0.5       & 0.1         & 0.1  \\
 Max. linear power density (W/\cm) & 39        & 75        & 39          & 35   \\
 Vertical height of s.r. fan (mm) & 0.45    & 0.74      & 0.45        & 0.74    \\
 No. of photons \EE{+21} (1/\s)  & 7.4       & 7.4&      &                     \\
 Critical photon energy    (keV) & 7.2       & 5.1       &                     \\
 \raggedright Photon desorption coeff. \EE{-7}  (molecules/$\gamma$) & 1   & 1   & 1   &  1 \\
 Photo desorption gas load \EE{-5}  (Torr/\s)    & 2.2 & 2.2 & 2.2 & 2.2        \\ 
 \raggedright Thermal desorption coeff. \EE{-11} (\Torr L/\s/\cma) & 1   & 1   & 1   & 1   \\\hline
\end{tabular}
\end{table}
}

The requirements are not beam-lifetime driven, but rather
have been determined in \ilc investigations to reduce the chance for
collective instabilities, such as the fast ion instability in the electron
ring, or excessive electron-cloud effects in the positron ring. Such
instabilities--even if controlled by feedback systems--can be
detrimental to the small emittance goals for \superb.

To avoid excessive impedance, high-conductivity materials, such as copper
or aluminum, are preferred. The dissapation of HOM energy
will be dealt with using localized HOM
absorbers~\cite{bib:Sasha_absorbers}. While this considerably increases 
the power density at the absorption point, it is believed to
reduce the resistive wall impedance, thus lowering the requirements
for the transverse feedback systems. The concern for \superb\ arises
from the small emittances: while growth rates are not excessive
compared to other facilities, preventing emittance growth from noise
in the feedback systems will be a concern.

NEG coating provides the bulk of the pumping speed for the proposed
\ilc Damping Rings. This technique,
developed at CERN for the LHC warm sections~\cite{bib:Benvenuti}, is
attractive, as it provides high pumping speed with low secondary
emission. It does require {\em in situ} baking of the
vacuum system to a relatively high temperature of about $200^\circ$C,
thus limiting the materials that can be used and incurring the added
expense of a suitable bake-out system. Operationally, {\em in situ}
baking adds the risk of opening vacuum leaks, especially if baking is done
at higher temperatures. For \superb, we will investigate this technique
carefully, and possibly apply it in regions that lend themselves to
efficient manufacture and deployment of such a system. On the other
hand, more conventional pumping schemes may provide similar pressures
at less overall cost. This optimization will require more investigation
before a final decision can be made. 

In the following we outline a
design approach that will likely achieve the required parameters,
given the present state of investigations.
Choosing a common technology and common cross sections for the HER
and LER vacuum systems would have benefits (economy of scale and
less design effort, to name just the obvious ones). However, the two
rings exhibit significant differences in layout due to the
relatively short dipoles in the LER compared to the HER. They
also have different technical requirements arising from the
susceptibility of the positron ring to electron-cloud effects, which
favors an antechamber design. At this point, for \superb\ we pursue
individually optimized vacuum designs for the arcs in each ring.
For the straight sections (excluding the interaction region), a
common design appears to be quite feasible.

\subsection{HER Arc Vacuum System}

\begin{figure}[htb]
\centering
\vspace*{5mm}
\includegraphics[width=\textwidth]{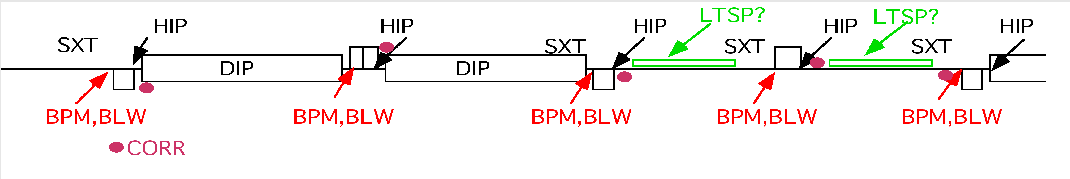}
\caption{Possible layout of an HER arc cell.}
\label{fig:HER_ArcCell}
\end{figure}

A possible layout of an arc cell of the HER is shown
in Fig~\ref{fig:HER_ArcCell}.
Discrete ion pumps will be placed at every quadrupole and distributed pumps will
be in the dipole as well as in the long drift tubes. The chambers will
be anchored to the quadrupoles at the location of the BPMs, with
the other end allowed to expand in the longitudinal
direction. Shielded bellows will be inserted next to the BPMs to avoid
mechanical stress that could cause the BPM pickups to move with respect to the
nearest quadrupole magnet.

\begin{figure}[htb]
\centering
\includegraphics[width=0.8\textwidth]{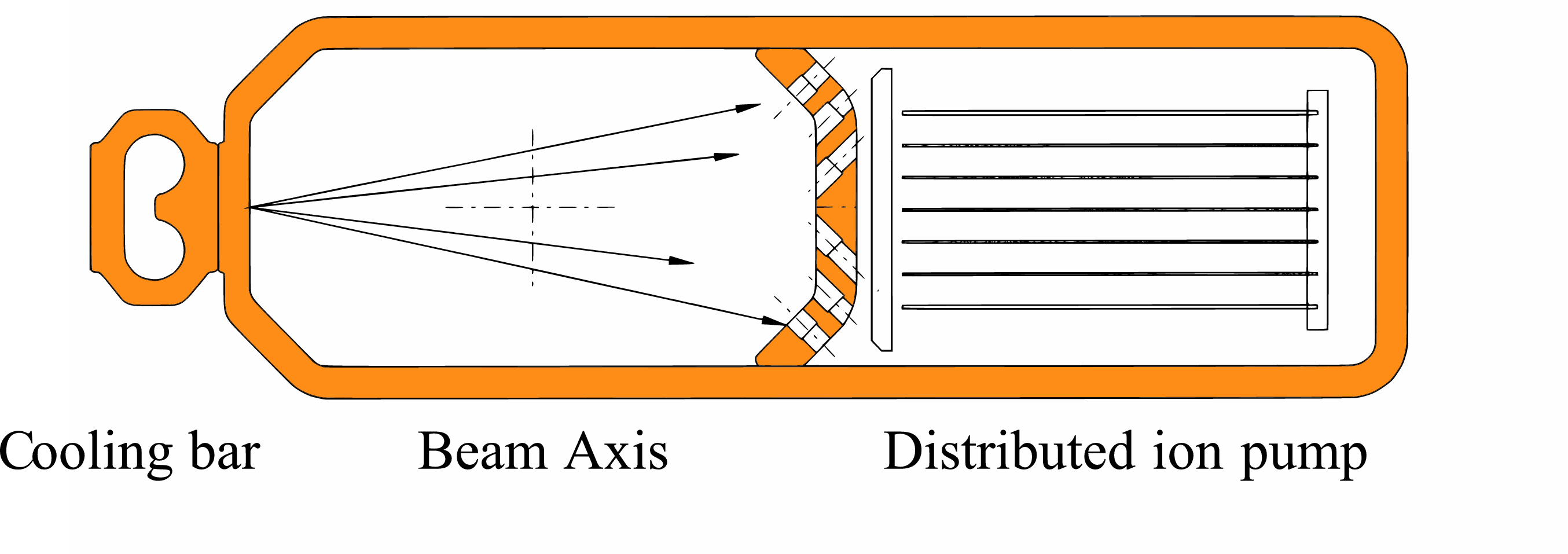}
\caption{Cross section of a \pepii HER main dipole vacuum
chamber.}
\label{fig:HER_dipole_xsection}
\end{figure}

The pumping requirement for the HER arcs to maintain 0.5\nTorr
pressure at 1.3\amp beam current is about 150$\,$L/s/m average pump
speed. The low photon-desorption coefficient ($\eta = 10^{-7}$) used
here has been achieved at
\pepii~\cite{bib:Wienands_PAC_PEPvac}. Since the conductance of any
reasonable aperture over the length of a half cell will be smaller
than necessary to achieve this pressure with lumped ion pumps,
distributed pumping is required. The \superb\ HER will use the
existing 5.4\m long dipoles of the \pepii HER. In \pepii, the dipole
vacuum chambers house distributed ion pumps (DIPs) of 120$\,$L/s/m
effective pumping speed (including the screen) that use the field
of the dipoles. The outside wall of the chambers constitutes the
absorbing surface for the synchrotron radiation. The power handling 
capability of
the \pepii HER chamber is 100\W/m, more than sufficient for \superb,
even for the upgrade option at 2.2\amp beam current. The aperture of
the dipole chambers ($5\times9\cm$ with an octagonal shape) is
sufficient for \superb\ even taking into account the extra 0.4\cm of
sagitta due to the smaller bending radius of the dipole in \superb.
Figure~\ref{fig:HER_dipole_xsection} shows a cross section of the HER
arc vacuum chamber.
Since the development of such vacuum chambers with DIPs is a
significant effort there would be considerable savings in reusing
the \pepii dipole vacuum chambers in the \superb\ HER. Only 144 of
the 192 dipoles will be used in \superb\ so there would be a
significant number of spare chambers in case failures occur. To
ensure reliability we would open a few of the \pepii chambers to
inspect the condition of the pump and assess the expected service
life.

It should be noted that there are 16 additional dipole magnets of the
same design at higher bending angle in the final focus section. The
design of the vacuum system for these magnets will be the subject of further
investigation.

Reusing the \pepii
HER dipole chambers naturally sets the cross section of the other arc
vacuum chambers as well. Because of the different magnet lattice we do
not plan to reuse the \pepii HER quadrupole chambers, but will instead
build new ones. These could be fabricated from aluminum extrusion for
cost savings, pending more detailed thermal analysis. Alternatively,
copper extrusions, like those used in \pepii, could be used for
the quadrupole chambers.

The drift tubes in the straight portions of the arcs will be required
to have some kind of pumping as well, as the power dissipated on the
walls is significant, although smaller than in the dipole chambers. A
straightforward solution would be to use a chamber similar to the
dipole chamber but with a getter pump in the pumping compartment of
the chamber. This could be a long NEG pump such as used in some
areas of \pepii or a long titanium sublimation pump (TSP) such as
one recently developed for the \pepii IR\cite{bib:Long_TSP}. The TSP
would have the advantage of being less prone to separating dust into
the beam channel, a significant problem in the early days of \pepii.
A potential issue with HOM heating of NEG pumps due to the high beam
current is avoided by ensuring that the screen is sufficiently dense
to minimize leakage of electromagnetic fields. An
alternative option for these chambers would be a NEG-coating scheme
similar to that proposed for the \ilc damping rings, but with an
octagonal vacuum-chamber cross section.
Table~\ref{tab:HER_vac_parms} summarizes the HER arc vacuum system
components.

{\setlength{\tabcolsep}{2pt}  
\begin{table}[htb]
\caption{HER vacuum system components.}
\label{tab:HER_vac_parms}
\vspace*{2mm}
\setlength{\extrarowheight}{2pt}
\centering
\begin{tabular}{lcccp{4.7cm}}
\hline
\hline
 Component           &  Pump speed   &  Length   &  Number &      Comment \\
                     &      (L/s)    &     (m)   &         &               \\
\hline
 Distrib. ion pumps  &       600     &      5    &   144   & in dipoles \\
 Holding pumps       &        60     &     0.3   &   216   & adjacent to quadrupoles \\
 Long TSP or NEG pump&       600     &      4    &    72   & drift sections, speed is screen-dominated \\
 Hot filament gauges &               &     0.1   &    24   & \\
 Pirani gauge        &               &     0.1   &    12   & to protect gate valves \\
 Roughing ports      &               &     0.1   &    24   & \\
 Bellows             &               &     0.2   &   216   & at each quadrupole \\
\hline
\end{tabular}
\end{table}
}

\subsection{LER Arc Vacuum System}

The average pumping requirement for the LER is similar to that for
the HER. However, the relatively short dipoles, coupled with the
desire for antechambers to reduce the number of photoelectrons near
the beam, favor a different vacuum chamber design. While NEG coating
is attractive, this approach would favor the use of stainless steel
chambers, which for \superb\ would likely have to be copper clad on
the inside. The substantial thickness necessary for the copper layer
raises concerns about the stability of the cladding, especially after
numerous heating cycles required to regenerate the NEG material. On the other
hand, copper extrusions of the required shape, while probably
obtainable, may be quite expensive.

At \pepii, a TiN-coated aluminum chamber with a
substantial antechamber and discrete photon stops was chosen for the
LER. The photon stops have a design limit in power handling of about
15\kW \cite{bib:PS_power}. For 144 arc dipole pairs, the \superb\ LER
synchrotron radiation power per dipole comes out to 13\kW for the nominal
2.3\amp beam, but rises to
22\kW for the 4\amp upgraded beam current. With one photon stop per dipole
pair, this approach would require a certain extrapolation of
the technology used in the \pepii photon stops. Moreover, the
clustering of four dipole magnets in close proximity in the lattice
creates a large opening angle for the radiation fans, making for a more
difficult photon-stop geometry.

\begin{figure}[htb]
\centering
\includegraphics[width=0.8\textwidth]{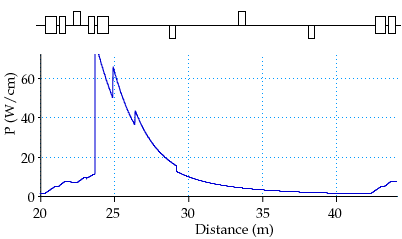}
\caption{Linear synchrotron radiation power density in a standard LER arc period.}
\label{fig:Power_dens}
\end{figure}

A variation of this approach may be feasible at \superb\ as well:
retaining the antechamber, but rather than using discrete photon
stops laying out the chamber geometry in such a way as to absorb the
photon power along most of the length of the chamber. Magnet
dimensions are such that an antechamber width of about 20\cm
can be accommodated. Due to the particular geometry of the lattice,
the peak power (75\W/cm) will hit inside of the dipole pair just
after the focusing quadrupole. If the absorbing surface were slanted
with a 1:10 ratio, the real power density would be reduced by a
factor of 10 to a manageable 100\W/cm$^2$ for 2.3\amp beam
current, and about twice that for the upgrade scenario.
Figure~\ref{fig:Power_dens} shows the anticipated
linear-power-density profile for a standard LER arc period.

\begin{figure}[htb]
\centering
\includegraphics[width=0.8\textwidth]{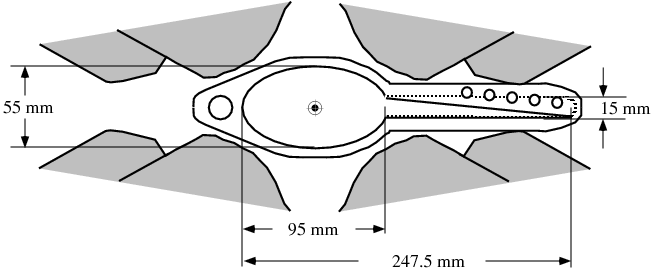}
\caption{LER
Arc quadrupole vacuum chamber concept. The shaded structures represent
the envelope of the magnet poles and coils.}
\label{fig:LER_arcvac}
\end{figure}

\begin{figure}[htb]
\centering
\includegraphics[width=0.8\textwidth]{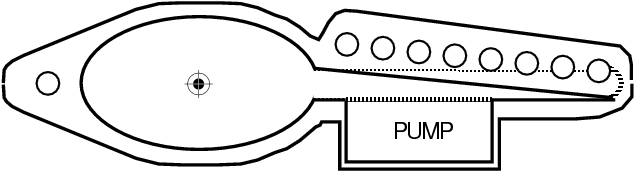}
\caption{LER
dipole vacuum chamber concept. Inside dimensions match those of
Fig.~\ref{fig:LER_arcvac}.}
\label{fig:LER_dipolevac}
\end{figure}

There is sufficient vertical space in the dipoles to provide for
the necessary cooling and pump channels. The more limited vertical
space in the quadrupole pockets restricts the amount of cooling
above and below the antechamber, but, given the more moderate power
densities (about 20\W/cm or 30\W/cm$^2$) at these locations, this
should be sufficient. This approach would make use of relatively
economical aluminum extrusion and proven technology. Due to the
temperature restriction of most aluminum alloys, it is unlikely that such extrusions
would be
suitable for incorporating a NEG coating design; distributed
pumping by either a long NEG or a long TSP is anticipated.
Figure~\ref{fig:LER_arcvac} shows a conceptual layout of the
quadrupole chamber, while Fig.~\ref{fig:LER_dipolevac} shows a dipole
chamber. The latter is also applicable to the drift sections, and
more room can be provided in the drift sections for the pumps, if
required.

While the actual chamber would be an extrusion, the channel for the pump
would be a separate welded-on section, to allow for cutting the screen
shielding above the pump.

All LER vacuum-chamber sections not located
within magnets would carry current windings to create a solenoidal
field of about 50\Gauss on the beam axis as a measure designed to prevent
electron-cloud built-up near the beam. If it were
determined to be of benefit,
longitudinal grooves can be provided in the beam channels of the
extrusions. All aluminum chambers would be TiN coated inside the beam
channel. Table~\ref{tab:LER_vac_parms} provides a summary of the LER arc
vacuum components.

{\setlength{\tabcolsep}{2pt}  
\begin{table}[htb]
\caption{LER arc vacuum system components.}
\label{tab:LER_vac_parms}
\vspace*{2mm}
\setlength{\extrarowheight}{2pt}
\centering
\begin{tabular}{lcccp{4.7cm}}
\hline
\hline
 Component            &  Pump speed &  Length    &  Number    &  Comment \\
                      &      (L/s)  &    (m)     &            &            \\\hline
 Long TSP or NEG      &        240  &        1.2 &        144 & in dipoles \\
 Holding pumps        &         60  &        0.3 &        300 & adjacent to quadrupoles \\
 Long TSP or NEG pump &        600  &        3   &        288 & drift sections \\
 Hot filament gauges  &             &        0.1 &         24 &            \\
 Pirani gauge         &             &        0.1 &         12 & to protect gate valves \\
 Roughing ports       &             &        0.1 &         24 &            \\
 Bellows              &             &        0.2 &        300 & at each quadrupole \\
\hline
\end{tabular}
\end{table}
}

\subsection{Straight Sections}

\begin{figure}[htb]
\centering
\includegraphics[width=0.8\textwidth]{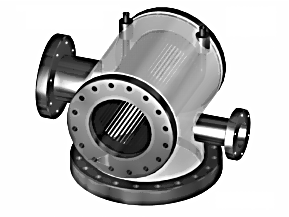}
\caption{Rendering of a \pepii straight-section pump tee.}
\label{fig:pump_tee}
\end{figure}

The straight sections in both rings will likely use the same chamber
materials and cross sections. For economy, a circular chamber
geometry is preferred, since (after a matching section) there will
not be significant synchrotron radiation left to absorb. In order to
keep the impedance low, aluminum or copper is preferred
over stainless steel. The gas load in most of the straight section
is dominated by thermal outgassing from the chamber wall. For the
specified 0.1\nTorr pressure, an average pumping speed of about
300$\,$L/s/m is required. This is a substantial pumping speed, which is
only practical with distributed pumping. In these sections, the
NEG-coating technology is a serious contender: the circular cross
section is ideally suited for this purpose and if copper is used, the
temperature of about $200^\circ$C needed to regenerate the
low-temperature NEG material would not weaken the
chamber. The NEG coating would be applied in the drift
sections between the quadrupoles. With NEG being primarily a
hydrogen pump, however, significant additional pumping is
needed to remove other common gases such as Ar, Co and CO$_2$ from the
system; this pumping will be provided by ion sputter pumps. Having
one 400$\,$L/s ion pump in each half cell should reduce the partial
pressure of these gases sufficiently. To achieve high conductance,
the pump tee uses a large screen and a plenum.
Figure~\ref{fig:pump_tee} shows a \pepii straight-section pump tee,
a design which could be adapted for \superb\ with little modification.

Experience with
\pepii has shown that the straight sections can be subject to
significant electron-cloud effect. While weak solenoidal fields (on the
order of 30--50\Gauss) on the beam axis have been shown to be very
effective in suppressing the electron-cloud effect in the present
$B$ Factories~\cite{bib:PEP_solenoids}, there is concern that the tight
emittance requirements of rings like \superb\ and the \ilc damping
rings necessitate stronger measures. At \pepii, there are presently
several test chambers installed with different surface preparation
to systematically study the efficacy of grooves in the chambers and of
different materials and coatings on the secondary
emission~\cite{bib:ILC_testchambers}. Results of these experiments are
expected to be available sufficiently soon to be fully taken into account
for the detailed design of the \superb\ vacuum system. A summary of
the straight-section components is provided in
Table~\ref{tab:Straight_vac_parms}.

{\setlength{\tabcolsep}{2pt}  
\begin{table}[htb]
\caption{Vacuum system components for the straight sections,
excluding the IR.} \label{tab:Straight_vac_parms}
\vspace*{2mm}
\setlength{\extrarowheight}{2pt}
\centering
\begin{tabular}{lcccp{4.7cm}}
\hline
\hline
Component           &  Pump speed &  Length &  Number  &  Comment \\
                    & (L/s)       & (m)     &          &          \\\hline
NEG-coated chamber  & $>300$/m &  $\approx 5$ & tbd &                                \\
Holding pumps       & 400      & 0.3         & 180 & adjacent to quadrupoles \\
Hot filament gauges &          & 0.1         & 10  &                                \\
Pirani gauge        &          & 0.1         & 10  & to protect gate valves\\
Roughing ports      &          & 0.1         & 10  &                       \\
Bellows             &          & 0.2         & 180 & at each quadrupole\\
\hline
\end{tabular}
\end{table}
}

\subsection{Expansion Bellows}
\begin{figure}[htb]
\centering
\includegraphics[width=0.9\textwidth]{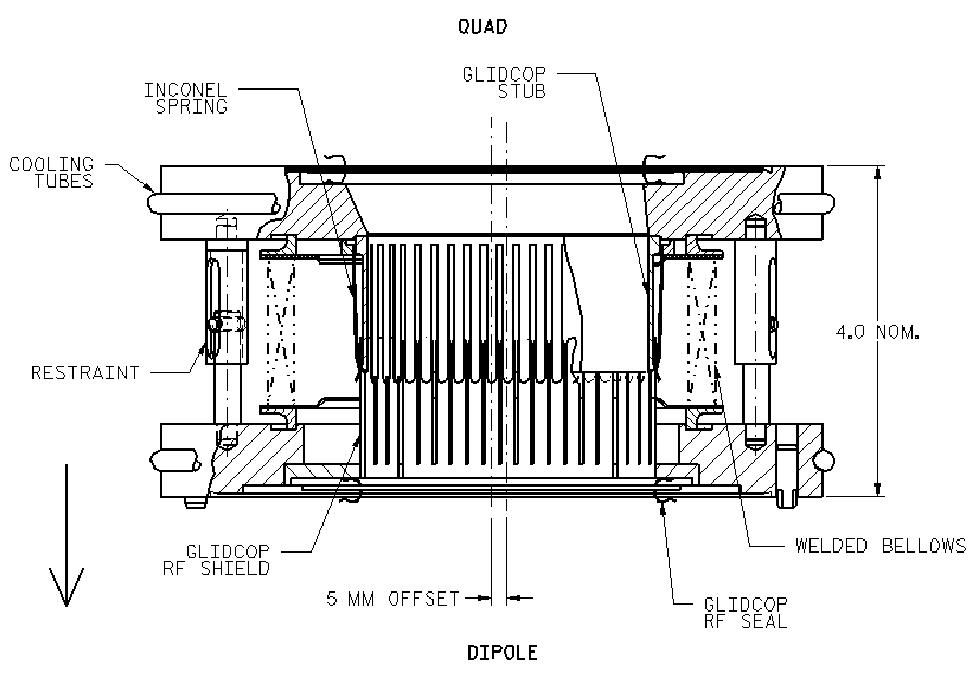}
\caption{Sketch
of a \pepii HER shielded arc
bellows.}
\label{fig:bellows}
\end{figure}

Shielded bellows designs have been used successfully at high beam
current, both at KEKB and at \pepii. Very similar designs will be
used at \superb, with incremental improvements mainly to reduce the
chance of damage to the shield during installation. The \pepii
bellows have been produced with circular, octagonal and antechamber cross section.
Figure~\ref{fig:bellows} shows a straight-section bellows design.

\subsection{HOM Absorbers}

\begin{figure}[htb]
\centering
\includegraphics[width=0.7\textwidth]{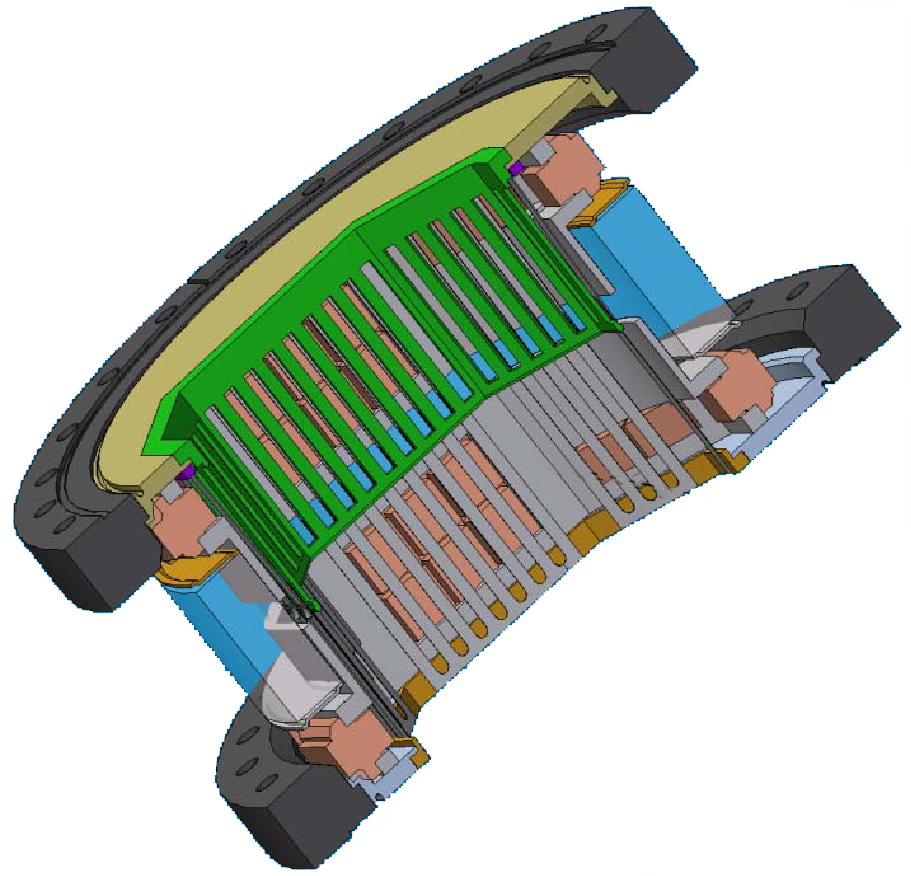}
\caption{HOM-absorbing
module integrated with a straight-section
bellows.}
\label{fig:HOM_absorber}
\end{figure}

With the vacuum system of the \superb\ rings made entirely from
material with high electrical conductivity, there is concern that
higher-order modes (HOMs) excited by the beam will not be quickly absorbed,
leading to excessive localized heating and other
difficulties. Using a less conductive material in some areas, \eg, the straight
sections, could improve the dissipation of HOM power into the vacuum
system. However, this would significantly raise the impedance
presented to the beam, especially the resistive wall component. At
\pepii, the threshold for resistive-wall instability has been found to be
significantly lower than expected. It is quite possible that the
stainless-steel vacuum system in the straight sections is one of the
culprits for the observed behavior. Dedicated HOM absorbers have
recently been developed to address the localized heating observed for
certain bellows units. These designs use SiC tiles with high electrical
permittivity and relatively high electrical loss (``$\tan\delta$'')
to absorb HOM power. The tiles are prevented from extracting power
directly from of the beam, while at the same time HOM power is let through,
by an array of longitudinal slots in front of the tiles.
Figure~\ref{fig:HOM_absorber} shows a rendering of such an absorber
module. This particular absorber is integrated with a bellows module
for reduced demand on space.

\subsection{Flanges}

\begin{figure}[htb]
\centering
\includegraphics[width=0.9\textwidth]{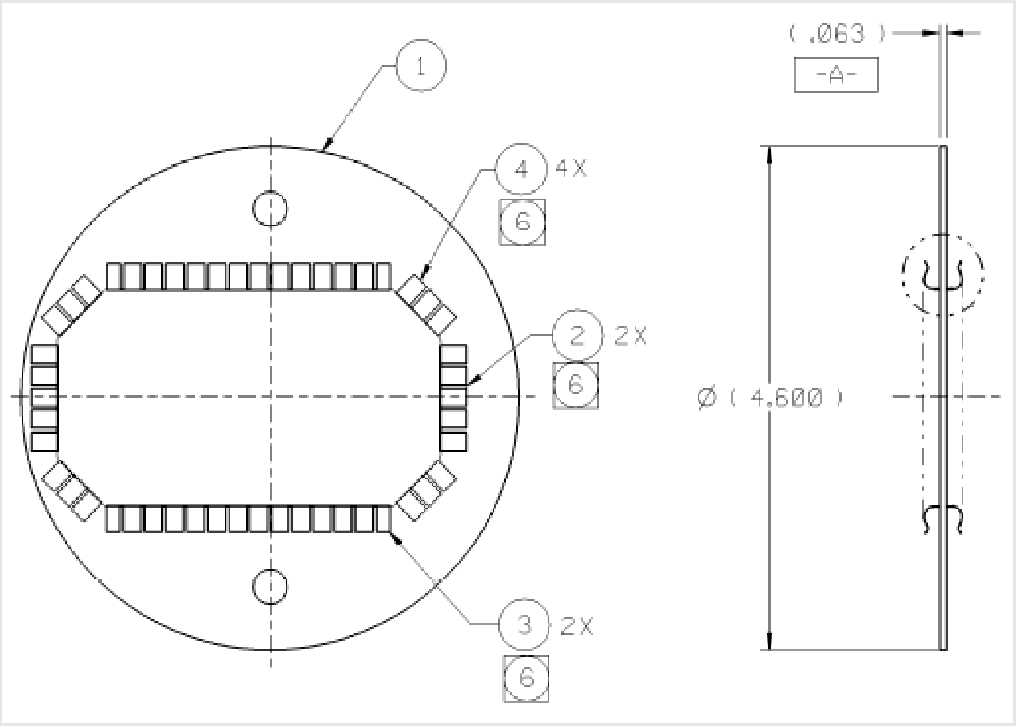}
\caption{Drawing of a \pepii HER ``$\Omega$'' RF seal.}
\label{fig:Omega_seal}
\end{figure}

Stainless steel flanges with ConFlat seals will be used throughout.
For aluminum chambers, flanges made of aluminum may be easier to
fabricate and attach to the chambers. However, operational
experience at \pepii has shown the vacuum joint using Compression
``C'' seals to be much more difficult to assemble and keep leak tight.
This presents a significant operational issue due to the large
number of flanges in a typical machine. ConFlat seals are robust and
straightforward to seal after a repair to the vacuum system. They do
present a gap to the beam aperture that must be bridged, either by
a ``gap ring'', usually made of copper, or by a more flexible
RF seal involving spring fingers. The latter design is preferred
where the joint can flex as the chambers heat up and cool down with changing
beam current. This can particularly be the case where chambers are bolted
together, not with a bellows, but using a ``flex flange'' to
allow some angular movement between the two chambers.
Figure~\ref{fig:Omega_seal} shows a flexible ``$\Omega$'' seal for the
\pepii HER. An upgraded version with better
flexibility of the spring fingers is presently being built; any design for \superb\ will
certainly incorporate the latest experience with such seals in
\pepii.

\subsection{Vacuum Monitoring}
The electrical current drawn by ion pumps is directly
related to the pressure and provides a convenient way to monitor the
vacuum pressure, as long at pressures are in the operable region of
ion pumps. At \superb, with the very low anticipated pressures, this
may not be always the case, especially in the straight sections, and wherever
significant NEG or TSP pumping is installed. For these cases, we
anticipate the installation of sufficient hot-filament gauges to allow
monitoring of the system.

Even in cases where the pump currents do provide meaningful pressure
readings, care has to be taken to avoid the pump currents being
affected by photo- and secondary electrons in the beam pipe. A good
pump screen is mandatory, but it may be necessary also to use
(relatively small) magnets around the pump ports to deflect any
electrons and thereby prevent them from reaching the active pump 
area. Gauges
will be mounted using right-angle adapters to prevent line-of-sight
into the vacuum system. The DIPs in the HER chambers are well
protected by the magnetic field of the dipole magnets and not
susceptible to stray electrons to any significant degree.

In the IR straight sections, in addition to these diagnostics, there
will also be a number of residual gas analyzers (RGAs) to
track down even very small leaks. The widespread deployment of RGAs
around the rings is prohibitively expensive; following \pepii
experience, it is not really necessary, as long as pump ports are
provided to quickly flange in an RGA where needed.

\subsection{Summary}

The vacuum system described in this section
fulfills the \superb\ requirements for pressure and synchrotron
radiation power handling. Some reuse of
\pepii components is anticipated, although this is not a driving
factor. A significant design and optimization effort will be undertaken
to ``flesh out'' the details of the design, and ensure an effective and
cost-efficient solution. This effort will include a
thorough evaluation of the NEG coating technology and other concepts
presently under consideration for the \ilc damping rings.


\afterpage{\clearpage}

\section{Instrumentation and Controls}
\label{section:InstrumentationAndControls}
\subsection{Beam Position Monitors}\label{sec:BPM}
\subsubsection{Requirements}
The plan for beam-position monitors (BPMs) for  \superb\ benefits from
the experiences of other rings. In particular, the growing number of
synchrotron light sources, with their demanding requirements for orbit
stability, has led to impressive commercial processors, while the high
beam currents in \pepii have exposed thermal issues not seen
elsewhere but which will be relevant for  \superb.

The BPMs must serve a range of conditions, from tracking the orbit of
a small injected charge on its first turn, with an accuracy of $100
\mum$, to measuring a stable orbit to $200 \nm$ in a full ring with over
$2 \amp$ of circulating beam. The measured orbit must be insensitive to the
fill pattern. Measurements such as the phase advance require
turn-by-turn beam positions for 1000 or more consecutive turns all
around the ring. The position history of the last 1000 or more turns
must be available after a beam abort for post-mortem
investigation. Data must be available on a speed compatible with orbit
feedback, which may be applied both globally and locally near the IP.

\subsubsection{Buttons}
\pepii uses $15 \mm$ diameter electrodes (``buttons'') mounted flush
with the chamber walls to measure beam position. Identical buttons
are used as pick-ups for other diagnostics, such as feedback and
bunch-current monitoring. The buttons are mounted at approximately
45 degrees to the horizontal and vertical axes (the exact location
depends on the cross sections of the different vacuum chambers) to
avoid direct hits from synchrotron radiation. The buttons are
stainless steel, mounted on molybdenum pins that pass through a
ceramic feedthrough to an SMA connector outside the chamber. When
the vacuum chambers are copper (LER near the IP, HER arcs) or
stainless steel (standard straights for both rings), these button
assemblies are electron-beam welded into place. Since the buttons
are not suitable for welding into aluminum chambers (LER arcs and
wiggler straights), buttons there are mounted on flanges.

\begin{figure}[htb]
\centering
\includegraphics[width=0.8\textwidth]{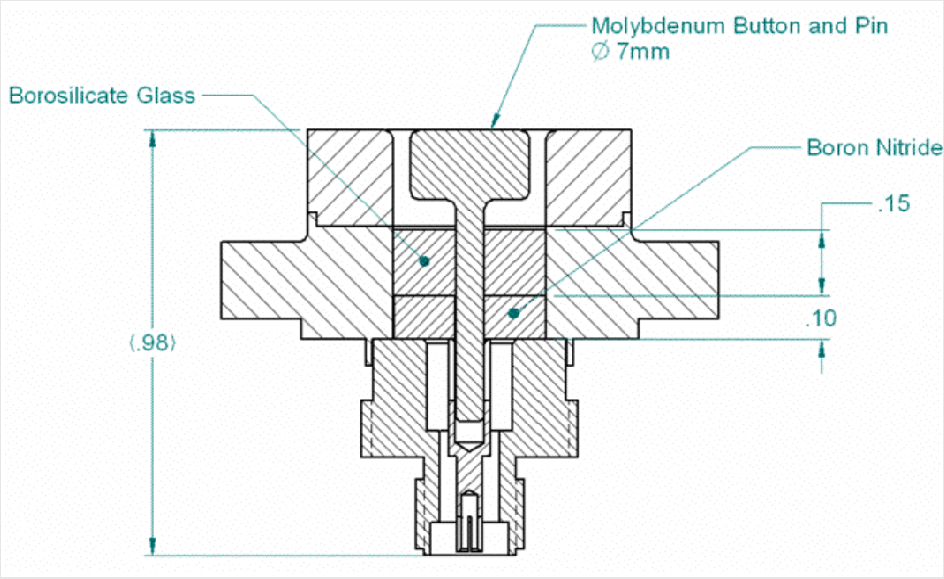}
\caption{\label{fig:Dia1}
 New \pepii BPM $7\mm$ button assembly, mounted in a vacuum
flange. Note the integral molybdenum button and pin.}
\end{figure}

In June 2005, with the LER current in \pepii at $2.4 \amp$, the RF voltage was
increased from $4.05$ to $5.4 \MV$ to shorten the bunch length. Within a
week, some buttons on the upper half of a few chambers dropped. The
end of the molybdenum pin is captured inside a socket on the back
surface of the stainless button with a press fit requiring some spring
force. This force appears to have weakened after years of thermal
cycling and with the increased high-order-mode power from the shorter
bunches.
The flanged buttons in \pepii are being replaced with $7\mm$
diameter buttons. These buttons and pins are made together from a
single piece of molybdenum. This is the design proposed for \superb,
shown in Fig.~\ref{fig:Dia1}. Given the difficulty in replacing welded
buttons, all the \superb\ buttons should be mounted on flanges.

\subsubsection{Processors}

The growing number of light sources around the world in recent years
has led to the introduction of various commercial BPM processors that
could satisfy the requirements for \superb. Electronics change
rapidly, and so it is too early to select a processor for this
project, but the performance available commercially is illustrated by
the Libera Electron processor \cite{bib:ins_1_dia_1} from
Instrumentation Technologies in Slovenia.

Each Libera Electron is a separate one-unit-high rack-mounted
chassis. The inputs are the four button cables, a ring-turn clock
($133.3 \kHz$ for \superb), an acquisition trigger, and a beam-abort
trigger. An internal Linux processor can run EPICS and so serve the
measurements to the control system over ethernet. 

Each button signal is filtered to a $10\MHz$ bandpass, and then the gain
is adjusted by a $62\dB$ automatic gain control, for a wide dynamic
range of over 80\dB. The signals are sampled at a frequency near $120
\MHz$ (adjusted for each ring's revolution frequency) and downconverted
digitally. The resulting beam position may then be read at various
rates: sample by sample, turn by turn, $10 \kHz$ for fast orbit feedback,
down to $10 \Hz$ for position monitoring. Depending on the requested
rate, digital filters further narrow the bandwidth to reduce noise and
remove dependence on the fill pattern. For data-rate modes of $10
\kHz$ and below, the buttons may be automatically cycled among the four
input channels to even out any gain differences.

In turn-by turn mode, the processor records data from up to hundreds
of thousands of consecutive turns, beginning with an acquisition
trigger synchronized either with stored beam or with an injection
fiducial. Similarly, the abort trigger saves a $16,000$-entry buffer of
turn-by-turn beam positions measured prior to the abort.

\subsection{Beam Size Monitors}

\begin{figure}[htb]
\centering
\includegraphics[width=0.8\textwidth]{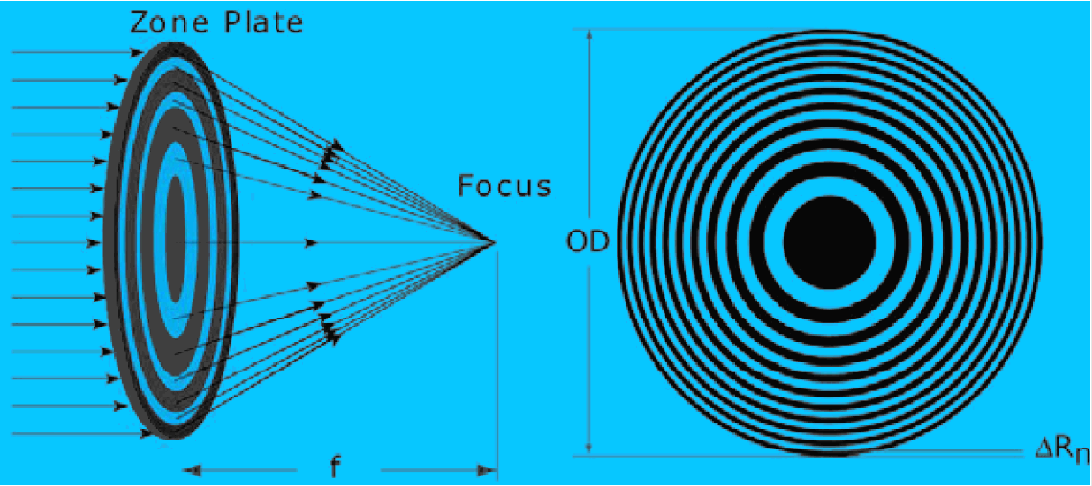}
\caption{\label{fig:Dia2}
A monochromatic x-ray beam focused by a zone plate.}
\end{figure}

In storage rings, synchrotron radiation from bend magnets provides the
standard measurement of beam size. Since the vertical size in
\superb\ will be very small--typically $20 \mum$--even at the defocusing
quadrupoles, and so the diffraction limit precludes measurements using visible
light. Instead, we turn to x-rays.  The simplest x-ray imaging
technique, a pinhole camera, is also not suitable, since the necessary
hole diameter would be impractically small and would pass very little
x-ray power. X-ray zone plates
\cite{bib:ins_2_dia_2,bib:ins_3_dia_3,bib:ins_4_dia_4,bib:ins_5_dia_5,
bib:ins_6_dia_6}, however,
provide an effective approach.  A zone plate is essentially an x-ray
lens of radius $r$ that focuses using diffraction rather than refraction
or reflection. An x-ray-opaque metal, typically gold, is electroplated
in a pattern of $N$ (typically hundreds) of narrow ($\sim 1 \mum$) circular
rings (Fig.~\ref{fig:Dia2}) onto a thin membrane of an x-ray-transparent
material, such as $\mathrm{Si}_3 \mathrm{N}_4$.
The thickness and separation of the rings
vary systematically so that, when illuminated by a collimated and
monochromatic x-ray beam, each ring forms a first-order diffraction
maximum that adds in phase at a focal point downstream, as shown in 
Fig.~\ref{fig:Dia2}. Zone plates are produced commercially by firms such
as Xradia \cite{bib:ins_7_dia_7} for use at synchrotron light sources.

Since the zone plate's focal length \cite{bib:ins_8_dia_8}:
        $$ f = \frac{r^2}{N\lambda}$$
depends on the wavelength $\lambda$, it is well defined only for a
monochromatic beam, but not for the broadband x-ray spectrum
of synchrotron emission from a dipole magnet. Narrowing the
bandwidth is also essential to reduce the power striking the zone
plate to a safe level for the delicate zone-plate structure. We therefore
precede the zone plate with a monochromator.

\begin{figure}[htb]
\centering
\includegraphics[width=0.9\textwidth]{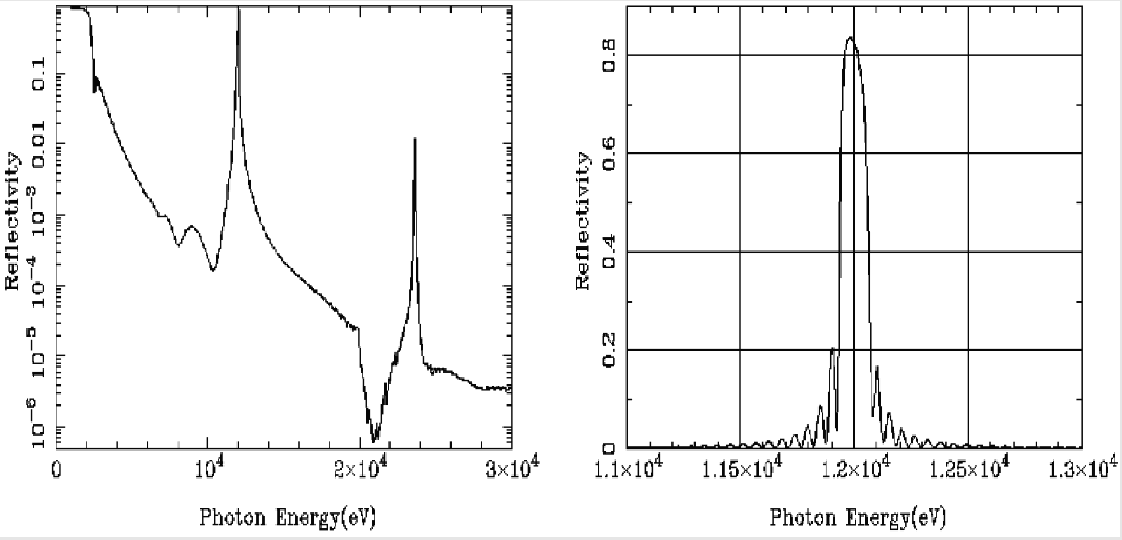}
\caption{\label{fig:Dia3}
Calculated reflectivity {\it vs.} energy for a single multilayer
mirror. P-polarized x-rays incident at $1.007^\circ$ to grazing on a mirror
with $200$ layers each of $2.1 \nm$ of $\mathrm{B}_4\mathrm{C}$
 and $0.9 \nm$ of $\mathrm{Mo}$, and with an
interdiffusion thickness of $0.5 \nm$, deposited on a silicon substrate.}
\end{figure}

An x-ray monochromator commonly uses Bragg diffraction from a single
crystal, with a typical bandpass $\DeltaE/E$ of $10^{-5}$. This is far
narrower than is needed for imaging and is costly in terms of
flux. Instead, a bandwidth of about 1\% can be obtained with a
grazing-incidence multilayer mirror, a substrate coated with
alternating thin layers of light and heavy materials. Here we consider
a mirror with layers of $\mathrm{B}_4\mathrm{C}$ and $\mathrm{Mo}$,
with the reflectivity~\cite{bib:ins_9_dia_9} shown in Fig.~\ref{fig:Dia3}. 
The center of the band
may be tuned by small variations in the angle of incidence. To
preserve the direction of the incident beam while tuning, such mirrors
are commonly used in pairs, with the outgoing beam parallel to the
incoming beam, but displaced slightly. Like zone plates, such mirrors
are available commercially for use at light sources, from firms such
as Rigaku/Osmic~\cite{bib:ins_10_dia_10}.

The large heat load now strikes the first multilayer mirror, rather than
the zone plate. Although the mirror is far more robust than the zone
plate, it is important to reduce the surface heating to maintain the
flatness and thickness of the layers. After the x rays enter a
separate beamline, the heat load is reduced in several steps that
follow the design of the x-ray beam-size monitor in the LER of 
\pepii~\cite{bib:ins_11_dia_11}.

First, the x rays pass through a water-cooled conical beampipe with a
$5\mm$ aperture at the downstream end. Grazing incidence on the inside
wall of this Glidcop~\cite{bib:ins_12_dia_12} cone allows for safe
absorption of photons that are not aimed at the zone plate.

\begin{figure}[htb]
\centering
\includegraphics[width=0.7\textwidth]{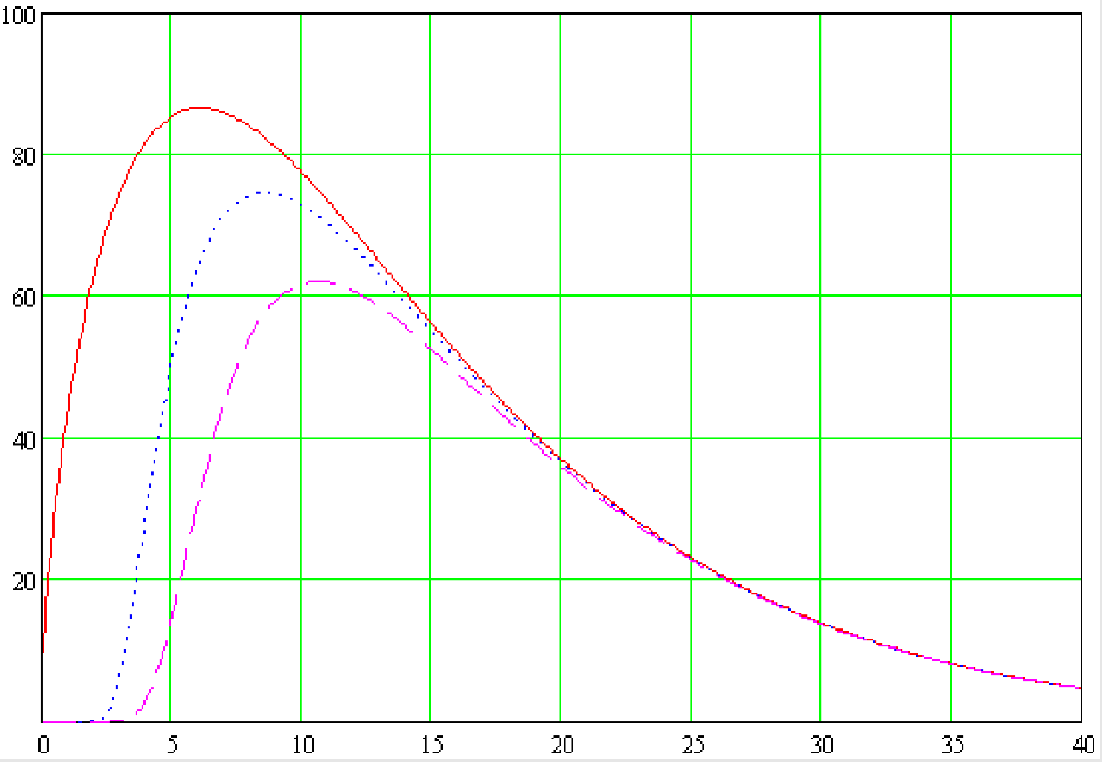}
 \caption{ \label{fig:Dia4}
 Spectrum of HER synchrotron radiation ($\W \cm^{-2} \kev^{-1}$)
 entering the filter (solid curve), after the graphite (dotted),
 and after both graphite and aluminum filters (dashed). }
\end{figure}

Next, a high-pass x-ray filter removes visible, ultraviolet, and
lower-energy x rays. The preliminary design considered here images
photons at $12 \kev$, somewhat above the critical energy in the dipoles
($5.5 \kev$ for LER and $7.4 \kev$ for HER). The filter's first stages use
five thin (5 to $50 \mum$) layers of pyrolytic graphite, which absorb
photons below about $4\kev$, and radiate the heat toward the walls of a
water-cooled vacuum chamber. Two thin sheets of aluminum foil then
remove energies below $6\kev$. The combination cuts the heat flux in
half, as seen in Fig.~\ref{fig:Dia4}.

Just before the first multilayer mirror, a $3\mm$ conical hole in a
Glidcop plate further narrows the pencil of x rays and also serves as
a photon BPM. A pair of photodiodes compares the intensity of x-ray
fluorescence from the upper and lower surfaces of the cone, so that
the vertical angle of the electron or positron orbit in the source
dipole can be adjusted to center the x-ray beam on the hole. A similar
pair for the horizontal direction ensures that the middle of the x-ray
fan is passed through to the next stage.

The beam then reaches the first multilayer mirror, where grazing
incidence at one degree spreads the remaining heat. The dimensions and the
water-cooling channels in the silicon substrate are carefully designed
to conduct this heat away with minimal distortion.

A magnification of ten would be helpful to measure the small beam
size, but, to get this magnification with a single zone plate, the
detector would have to be well over $100 \m$ away, given the long
distance from the source to the zone plate. Instead, we use a two-lens
system, demagnifying with the first zone plate and magnifying with a
second. A resolution (referred to the source plane) of $2 \mum$ is readily
achievable with such a system.

Just before the image plane, the x rays pass out of the vacuum through
a thin beryllium window. The basic detector then has a scintillator,
lens and video camera. Inspired by the wire scanners used to measure
beam profiles in the SLAC linac, \pepii is now preparing a more
elaborate system, designed for rapid measurement of the size of each
bunch \cite{bib:ins_11_dia_11}. An x-ray-opaque mask with three slots
at three orientations scans across the image plane. A fast ($1 \ns$)
scintillator \cite{bib:ins_13_dia_13} and photomultiplier after the
mask respond to each bunch as it passes by. Fast digitizing and
processing electronics sort the data into profiles for each bunch over
many turns as the mask moves, so that the major axis, minor axis and
tilt of each bunch's ellipse can be computed every few seconds. A
linear motion, like that of the wire scanner, could be used, but a
rotating device, similar to an optical chopper wheel, would be more
robust. With a radius of $100 \mm$, the change in slot orientation with
rotation would not cause difficulty.

\subsection{Longitudinal Feedback}

\begin{figure}[htb]
\centering
\includegraphics[width=0.9\textwidth]{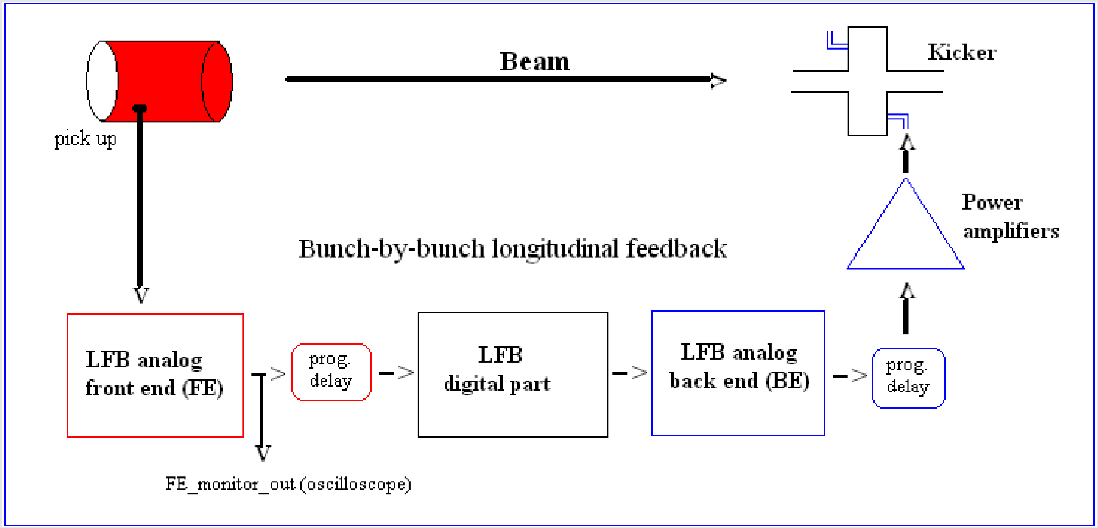}
\caption{\label{fig:Dia5}
 Bunch-by-bunch Longitudinal feedback block diagram.}
\end{figure}

The \pepii longitudinal feedback system, although designed in the first half of
the 1990s, still performs very well. Some parts of the system
would be redesigned for \superb, others can be used without changes.
The main components of the longitudinal feedback system are
(see Fig~\ref{fig:Dia5}):

\begin{itemize}
\item Pickup (buttons and the vacuum chamber in which they are located);
\item Frontend analog electronics;
\item Bunch-by-bunch digital processing system and feedback setup controls;
\item Backend analog electronics;
\item Power amplifiers, circulators and loads; and
\item Kickers
\end{itemize}

Other important components for the system are
SUN/Solaris computers servers with the high \& low level software,
including the source codes;
VME/VXI controllers, VxWorks software licenses for software development;
spare parts (very important for the obsolete components);
instrumentation needed for timing the system and diagnostics;
racks and crates; and
cables

If vacuum components and BPM buttons are reused from \pepii, the existing
pickups are perfectly adequate for the feedback systems.
The analog frontend and backend electronics still work very well,
and, in principle, can be reused. Their control is based on the
VXI bus, which is still in use, and on a now obsolete National Instruments
(VxWorks-based) controller module. As a consequence, there are a number
of options to be considered in an implementation for \superb. The
minimal cost solution would be to retain the analog front- and back-end
modules unchanged. A more realistic and robust
approach would be to update the software using more recent
versions of controllers, operating systems and EPICS tools. Alternatively,
a completely
modern approach would be to use the VME64 bus and redesign with setup
control based on FPGA (field programmable
gate array) technology.

The power amplifiers, circulators and power loads can be reused, 
saving a significant amount of money ($>500$ k\$).  Cavity kickers
(\daphne-style) can be reused or, eventually, redesigned, since these are
relatively inexpensive components.

The digital processing system, as well as the feedback setup controls,
still perform well and could be reused, although
a large number of the components, as well as the VME controllers, are now
obsolete. A new digital approach based on FPGA technology has been 
developed and tested on \pepii, \KEKB
and \daphne. The ILC-DR will use the new digital approach as well,
suggesting that, for
synergy with ILC project, it may be attractive to implement a new
FPGA-based design.

The minimal cost solution for the feedback system
would be to reuse all existing components
of the system, simply replacing the cables. 
However, it would be prudent to replace
obsolete parts of the system, while taking advantage of synergies 
with the ILC project. For example,
the FPGA technology approach for the digital processing should be
implemented, using new controllers and computer servers, refreshing
all the software controls and porting these to the last version of
operating systems and EPICS operator interface. 

\subsection{Transverse Feedback}

The transverse feedback systems, just like the LFB, were
designed in the 1990s. 
The main components of the transverse (horizontal and vertical)
feedback systems are:
\begin{itemize}
\item Pickups (buttons and the vacuum chamber in which are located);
\item Frontend analog electronics;
\item Digital processing system and feedback setup controls;
\item Backend analog electronics;
\item Power amplifiers, absorbers and loads; and
\item Stripline kickers
\end{itemize}

Other important components of the system are spare parts (particularly 
for obsolete components); instrumentation; racks and crates; and cables.

A minimal cost strategy would be to reuse all system components. 
In particular, re-using the power
amplifiers would save a significant amount of
money. The only new costs involved would be associated with new cabling.
However, to ensure future reliability,
as well as to engender synergy with the ILC-DR project, a preferable approach
is to implement a solution using 
FPGA technology for the digital processing modules, 
new controllers 
and servers, and porting all software
controls to the latest system. 

\subsection{Beam Abort System}

At design currents, the stored energy in both the HER and LER beams
of \superb\ is $68 \kJ$. A sudden beam loss that deposited this energy
into a small region could melt the beam pipe; we therefore require a
system that quickly detects faults and extracts the beam into a
dump. Several types of faults should trigger beam aborts, such as a
trip of a main magnet string, a fault in an RF station, a rapid loss
of beam current, or excessive background radiation in the detector.
Table~\ref{table:Dia1} provides a detailed list.

\begin{table}[htb]
\caption{\label{table:Dia1} Triggers for beam aborts.}
\vspace*{2mm}
\setlength{\extrarowheight}{3pt}
\centering
\begin{tabular}{p{6cm}l}
\hline\hline
Manual abort from control room & Faulty beam-abort trigger sys.     \\
Beam-stopper insertion         & HV on abort kicker $< 80\%$        \\
Vacuum-valve insertion         & Rapid drop in beam current         \\
Faulty dipole or quad. string  & Sudden large orbit excursion       \\
Faulty RF station              & High radiation level at experiment \\
Faulty LFB                     & Temp. over limit on a thermocouple \\
Faulty TFB                     & Trip of a klixon (thermal switch)  \\
\hline
\end{tabular}
\end{table}

The fastest of these mechanisms is a loss of RF, causing beam to
spiral inward and scrape within some tens of turns. A suitable
response speed can be attained only with a hard-wired system that
bypasses the latency inherent to the network of control-system
computers. Other fault processes are substantially slower. Magnet
trips are slowed by inductance, but the response time (milliseconds)
is still fast enough to hard-wire the trigger. Thermocouple trips are
still slower due to heat capacity, and so can be detected by the
control system, which then triggers the abort.

In a large machine, abort triggering is necessarily distributed, with
processing electronics at several stations around the ring. At \pepii,
these are connected together in a bidirectional loop for each
ring. Each station passes on a request for an abort to the next
station. For fail-safe operation, this abort-request line normally
propagates a fast clock (the "heartbeat") that is halted to initiate
an abort. The loop starts and ends at the controller for the abort
kicker, which monitors the heartbeat.

The triggering hardware must latch the source of the abort and pass
this information along to the control system. In this way, if an abort
is triggered by a momentary excursion, it will still be possible to
determine the source. Also, an abort often causes the firing of other
abort triggers. For example, RF stations will indicate high reflected
power after the beam is dumped. The automatic recording of precise
time stamps for each trigger is essential to determine the sequence of
events.

The dump itself need not be under vacuum. In \pepii, the beam exits
the vacuum through a thin aluminum window on a chamber downstream of
the kicker. It is then stopped by blocks of graphite, aluminum, and
finally, copper, that make up the meter-long dump. To ensure that the
beam has not burned a hole through the dump, there is a small pocket
of gas, at a pressure somewhat above ambient, trapped between the
second and third layers.

If the pressure in this burn-through monitor drops, then an interlock
halts all further injection. (The \pepii dumps have never shown such
damage.)  The abort kicker must dump the beam within one turn. Since a
bunch passing through the kicker magnet while its field is rising
would not get a sufficient kick to exit into the dump, but
would instead start a large orbit oscillation, the kicker must have a
fast risetime that is synchronized with a short gap in the fill
pattern. The field must then decay slowly over the course of one
turn, so that all bunches strike the dump, but each deposits its
energy at a different point, in order to avoid damage to the dump
window and to the dump itself.  The abort gap in the fill pattern must
be short, in part so that the total number of bunches available for
beam is not greatly reduced. Also, at high currents a long gap would
allow time for the RF cavities to gain energy while unloaded, leading
to a current-dependent slew in synchronous phase from bunches at the
head of the pattern to those at the tail.

\subsection{Control System}
\label{CS}

The control system outlined here takes advantage of the considerable
body of experience from other accelerator laboratories, while leaving
the flexibility to draw upon new technology. In particular, the global
EPICS collaboration provides a standard architecture, with a distributed
database and a large collection of software tools that are continually
developed, shared, supported and upgraded by the many participating labs.
The collaboration is large, mature, and invaluable, since it is no longer
necessary to write custom code for tasks that are common to many machines.
The architecture of the control system has three tiers of distributed computing.
At the front end, EPICS IOCs (input-output controllers) communicate with instruments,
process the measurements, and serve this data by way of gateway computers
and middleware to user applications at the top layer.

\subsubsection{Frontend designs}
Older instrumentation commonly employs modules in VME and VXI crates,
or in CAMAC crates for even older installations. Stand-alone instruments,
such as oscilloscopes, communicate through short-range GPIB connections to a
local computer or to a GPIB-to-ethernet interface, allowing control by a
distant machine.
This arrangement is substantially changed in new installations. CAMAC,
VXI and GPIB are no longer used, and the need for VME is greatly reduced.
Some devices interface to an IOC through PLCs (programmable logic controllers).
Newer instruments communicate directly over ethernet, and often include embedded
processors, arranged with one for each device or for a collection of like devices.
The EPICS collaboration has developed drivers for a wide range of hardware
and instruments, such as motors, video cameras, and oscilloscopes.
An oscilloscope is now essentially a computer hidden behind a front
panel with the usual oscilloscope knobs and display. EPICS communicates
with the scope through its ethernet port. It is interesting to note that
these instruments often allow remote control via a web browser, using a
web page served by the scope itself. While this method is of limited use
for our application, since it is not integrated with the control system,
the concept illustrates the evolution of instrument architecture.

Some devices, such as the BPM processors discussed in Section~\ref{sec:BPM},
 can run EPICS on their embedded processors, turning the device itself
 into part of the control system.
These also have the capability to save data from many ring turns and to
work jointly with other processors and higher-level applications to
implement fast orbit feedback.

Other diagnostics need special hardware for bunch-by-bunch data capture.
For example, transverse and longitudinal feedback, and bunch-current monitoring,
all begin with a task-specific analog front end that combines signals from beam
pick-ups, mixes the result with an appropriate harmonic of the ring's RF,
and outputs a signal suitable for digitizing at the RF rate or faster.
All bunch-by-bunch tasks can use identical digital hardware, starting with
a fast digitizer, followed by an FPGA (field-programmable gate array),
and finally a fast DAC (digital-to-analog converter) to drive the feedback
correction signal. A computer, either nearby or on an additional board in
the same box, loads the FPGA with firmware written for the specific job,
reads the data accumulated by the FPGA, and serves as an IOC to communicate
with the rest of the control system. The FPGA data includes both the
essential results (such as the charge in each bunch) and a considerable
body of supplemental beam-diagnostic information (such as the spectrum
of modes being corrected by feedback). All of this can be monitored by the
user over EPICS.

As always, video is needed in many places, such as at screens on the
injection line, or for measuring beam size with synchrotron light.
In older systems, analog cameras send signals over coaxial cable either
to modulators for a closed-circuit cable television system that brings
multiple channels to users in the control room, or to digitizers on
frame-grabber boards in computers outside the tunnel. Digital cameras
have also been available, but with interfaces that do not allow transmission
over the long distances typical of large rings or linacs.

Recently, a new camera standard has been introduced that replaces the
coaxial analog video output with a gigabit ethernet port. The output is
entirely digital, and can be transmitted over 100\m with no loss of resolution.
 Once on the network, the image can be displayed or analyzed by any computer.
 Many such images might overwhelm the capacity of the network, delaying
 communications with other instrumentation. One way to preserve network
 bandwidth is to set the cameras for a lower update rate for slowly changing images.
 A more thorough approach gives the cameras a separate gigabit network.

\subsubsection{High-level applications}

Many applications fundamental to running \superb\ are similar to
those at other accelerators, and are available from the EPICS collaboration,
with only modest modification. This category might include BPM orbit displays,
steering, orbit feedback, video and oscilloscope displays, a history buffer
(an archive that records all signals periodically, typically at 1\Hz), and an
error log (a recording of each change to a setting of an accelerator component,
such as a magnet, or state, such as an excessive temperature).
A high-level mathematical language such as Matlab is useful for writing applications,
but tools must be added to provide access to EPICS data, ideally in a manner
structured by physical devices to organize the many EPICS channel names. SNS,
for example, is using XAL, a Java class hierarchy providing a programming
framework based on the physical layout of the accelerator. The user interfaces
for broadly used applications should be
designed with input from operators and physicists. For less elaborate tasks,
the tools should allow the accelerator physicists themselves to write the necessary code.

\subsubsection{System management}

Several items must be organized at an early stage. For example, a relational
database of control items must be set up at the outset, along with a well-planned
naming convention that includes both an overall scheme and many examples.
Another early need is an environment for developing and testing code.
This provides a basis for code management and bug tracking, and for code
testing and release.

Also, the timing system should be carefully planned and started early.
Timing includes both a means of generating triggers and a means of distributing
pulse information to devices or processes which need that information.
This combination allows triggering and data acquisition linked to events
like the travel of an electron or positron bunch along the linac, to
the injection of a bunch into a ring, or to one or more turns of a stored
bunch in either ring.

\subsubsection{Safety and security}

The computers on the control-system network must be highly secure, but still
must allow remote users to connect and control the machine. These requirements
need secure firewalls and gateways restricting outside access, and also good
security even within the firewalls.

The control system is also responsible for safety, both for machine
protection and for personnel protection. These functions, and especially the latter,
must be kept distinct from the rest of the control system to ensure a fast
and reliable response.



\afterpage{\clearpage}

\section{Injection System}
\label{section:InjectionSystem}
\subsection{Requirements}

The injection system for \superb\  must provide electrons and 
positrons with a injection rate of about $10^{12}$ particles per second in order
to compensate for beam losses due to the short beam lifetimes. This requirement will
be particularly demanding for the positron source. The injected beams must also
have small emittance to fit into the resticted phase space volume allocated for the 
injection, given the limited dynamic aperture of the machine and the requirements for the stored beam. 
Two possible solutions are considered here: the first based on experience with the injector for
the \daphne \epem collider and the second uses a 6\gev linac and two 1\gev damping rings for
electrons and positrons. 

\subsection{Extrapolation from the \daphne Linac}
We first briefly describe
the \daphne linac and injection scheme, and then examine
the scaling of key parameters to the
\superb\ reuirements.
The \daphne injector is composed of a $\approx 60 \m$-long Linac and
an Accumulator~\cite{bib:inj_LCgen1,bib:inj_LCgen2,bib:inj_Accgen1,bib:inj_Accgen2},
a $\approx 33\m$-circumference ring used for longitudinal and transverse phase
space damping. 

\subsubsection{The linac}

The \daphne linac, built on the basis of a turn-key commercial contract,
accelerates both
positron and electron beams to the collider operational energy. 
The linac is an S-band accelerator ($2.865 \GHz$)
driven by four $45 \MW$ klystrons, each followed by a SLED peak power
doubling system. It delivers $10 \ns$ pulses at a repetition rate of
$50 \Hz$. A quadrupole FODO focusing system is distributed along the entire
structure~\cite{bib:inj_LCgen3}. The
relevant beam parameters for both electron and
positron beam operations are shown in Tables~\ref{tab:inj_linacpar}, 
\ref{tab:inj_HEele}, \ref{tab:inj_HCele} and \ref{tab:inj_linacpos}. 
The injection of the positron and electron
beams is not simultaneous; the switching time between the two modes
is about one minute.

\begin{table}[!hbt]
\caption{\label{tab:inj_linacpar} Linac parameters.}
\vspace*{2mm}
\centering
\setlength{\extrarowheight}{1pt}
\begin{tabular}{p{7cm}r}
\hline
\hline
RF frequency                            & 2856\MHz   \\
Klystron power                          & 45\MW      \\
No. of klystrons                        & 4          \\
No. of SLED peak power doublers         & 4          \\
No. of accelerator sections             & 16         \\
Repetition rate                         & 50\Hz      \\
Beam pulse width                        & 10\ns      \\
\hline
\end{tabular}
\end{table}

\begin{table}[htb]
\caption{\label{tab:inj_HEele}High-energy electron mode.}
\vspace*{2mm}
\centering
\setlength{\extrarowheight}{1pt}
\begin{tabular}{p{7cm}r}
\hline
\hline
Output Energy          & 800\mev            \\
Bunch Charge           & $9.4\times 10^{9}$ \\
Output emittance       & $\leq 1$\mm-mrad  \\
Energy spread          & 1\% FWHM     \\
\hline
\end{tabular}
\end{table}

\begin{table}[htb]
\caption{\label{tab:inj_HCele}High-current electron mode for positron target.}
\vspace*{2mm}
\centering
\setlength{\extrarowheight}{1pt}
\begin{tabular}{p{7cm}r}
\hline
\hline
No. of accelerator sections   & 5                              \\
Input charge from gun         & $\approx 4\EE{11}$ particles   \\
Input energy from gun         & 120\kev\\
Output current                & $\approx 3\EE{11}$ particles   \\
Output energy                 & 250\mev\\
Output emittance              & $> 1$\mm-mrad\\
Energy spread                 & 10\% FWHM\\
Beam spot radius              & 1\mm\ {\it rms}
\\ \hline
\end{tabular}
\end{table}

\begin{table}[b!t]
\caption{\label{tab:inj_linacpos}High-energy positron mode.}
\vspace*{2mm}
\centering
\setlength{\extrarowheight}{1pt}
\begin{tabular}{p{7cm}r}
\hline
\hline
 No. of accelerator sections  & 10                     \\
 Output energy                & 550\mev                \\
 Input energy                 & $2 < E < 14$\mev       \\
 Output bunch charge          & $\approx 3.7\EE{9}$ particles \\
 Output emittance             & $<5$\mm-mrad           \\
 Energy spread                & 2\% FWHM   \\
\hline
\end{tabular}
\end{table}

\subsubsection{The injector}
The injector subsystem includes a thermionic electron gun, a
prebuncher and a buncher. The gun, a triode with Pierce geometry, is
equipped with a $3 \cma$ dispenser cathode able to deliver up to $8
\amp$ peak current in a $10 \ns$ FWHM pulse, with a maximum repetition rate of
$50 \Hz$ at $120 \kev$. Typical operational values are $120 \kV$ and $7
\amp$ in the positron mode and $120 \kV$ and $0.5 \amp$ in the electron
mode.  The prebuncher is a RF cavity tuned at the fundamental frequency
of $2856 \MHz$ followed by a drift of $21.3 \cm$, and by a five cell
$2/3\pi$ traveling-wave constant-gradient buncher.

\subsubsection{The RF structures and modulators}
\begin{figure}[htb]
\begin{center}
\includegraphics[width=\textwidth]{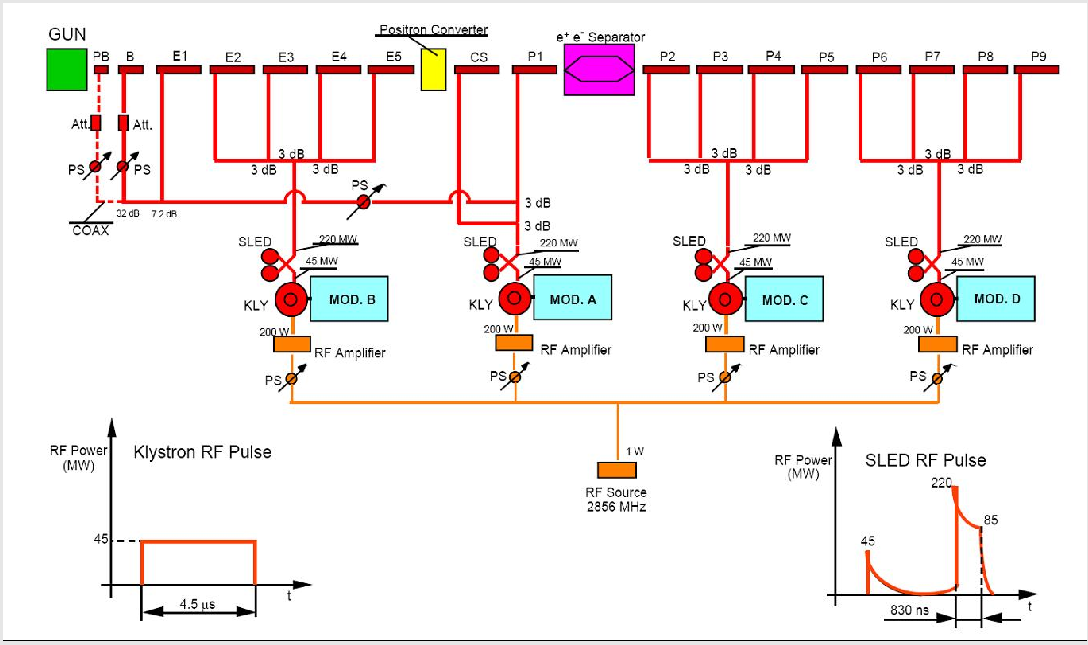}
\caption{Schematic layout of the RF system for the \daphne linac.}
\label{fig:linacRF}
\end{center}
\end{figure}

The accelerating sections (E1-E5, CS, P1-P9) are all of the same
type: $3\m$ long, $2/3 \pi$ traveling wave constant gradient SLAC
design structures.  In our configuration, with an output of $45 \MW$ from
the klystron, the nominal accelerating component of the electric
field is $27.7 \MV/\m$ in the CS and $19.6 \MV/\m$ in the remaining
accelerating sections.  The phase adjustments between sections
are done by means of low power $360^{\circ}$ phase shifters
upstream of the RF amplifiers of each klystron, and by a high power
$360^\circ$ phase shifter that uncouples the CS and E1 sections. The four
modulators are able to produce a video pulse of $6 \mus$ FWHM and
$4.5 \mum$ flat top, with a peak power of 100\MW at 50 pps.  A HV
power supply with a resonant charging circuit charges a pulse
forming network, composed by 8 LC cells, to 50\kV. The switching
thyratron is an EEV CX2168.
A schematic layout of the linac RF system is shown in Fig.~\ref{fig:linacRF}.

\subsubsection{The positron source}
The
positron beam is produced by a high-energy electron beam impinging
on a high $Z$ target~\cite{bib:BM2}. Due to the initial small transverse
size, but large divergence, of the beam emerging from the target,
a flux concentrator is placed between the
target and the capture section in order to trade a reduction in the divergence 
for increase in beam size (SLAC design \cite{bib:PCSlac}). The concentrator
consists of
a tapered field DC solenoid, with a peak
of $1.2 \Tesla$, plus a pulsed coil generating a solenoidal field that drops
adiabatically from a peak of $\sim 4 \Tesla$ to zero in about $12 \cm$.
A $7 \m$ length of uniform field $0.5\Tesla$ DC solenoid, wrapped around
accelerating sections CS and P1, completes the magnetic focusing of the
capture system.
The system allows the choice of 3 different targets, with thickness
varying around 2 radiation lengths, built with an alloy of $75\%$
tungsten and $25\%$ rhenium.  A remotely controlled actuator permits
the extraction of the target from the beam path during electron mode
operation.  The $28 \MV/\m$ accelerating gradient of the CS reduces the
large energy spread of the outcoming beam. Another efficient knob to
minimize this effect is the CS RF phase. The RF scheme for the linac
allows operation in both accelerating and decelerating modes.  The
nominal conversion efficiency is $0.9 \%$, which
is consistent with operating experience of $\sim 1\%$.


\subsubsection{The accumulator ring}
\begin{figure}[htb]
\begin{center}
\includegraphics[width=0.9\textwidth]{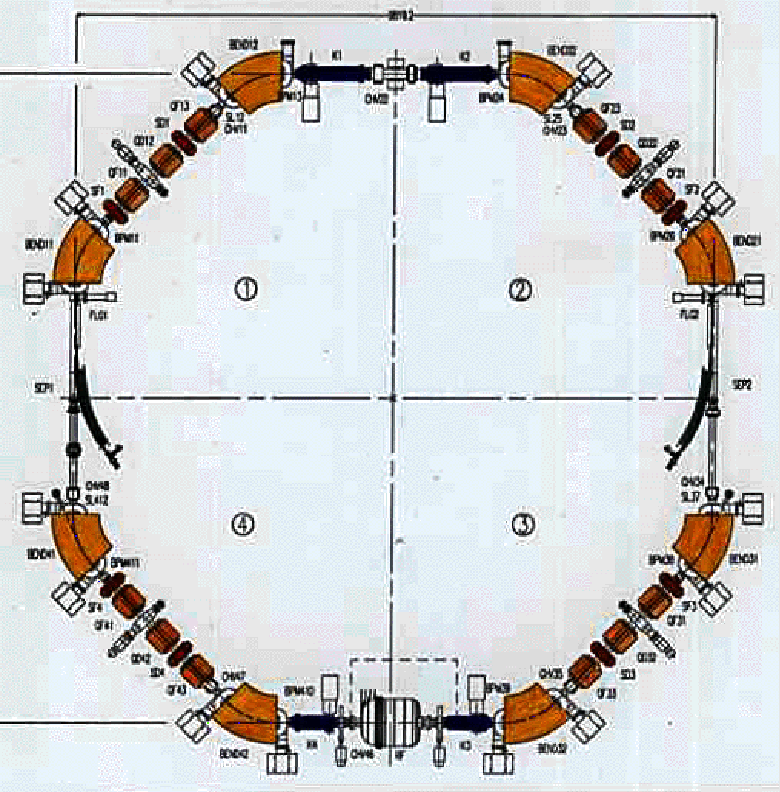}
\caption{Schematic layout of the \daphne accumulator ring.}
\label{fig:accumulator}
    \end{center}
\end{figure}

The accumulator is a quasi-octagonal ring with a total length of $32.5\m$
along the nominal trajectory. Its lattice is made of four almost
achromatic arcs, each consisting of two 45 degree full iron H-type
sector dipole magnets with a small gradient to optimize the damping
distribution, a quadrupole triplet and two sextupoles to correct the
ring chromaticity. All the dipoles are powered in series. The
quadrupoles are connected in three independent families, and the
sextupoles in two families. The electron beam from the linac
is injected into the ring by a system of two septum
magnets, the first bending the beam by 34 degrees and the second
performing the final deflection of 2 degrees into a special $3.5 \m$
vacuum vessel between two achromats. The stored beam is extracted by
a mirror-symmetric system placed in the opposite straight section.
The positron beam follows the opposite path. The remaining two
straight sections host the pulsed kicker magnets used to deflect the
beam at injection and extraction and the RF cavity. A system of 8
correctors and 10 position monitors allows a careful correction of
the closed orbit in the ring for the purpose of optimizing
injection efficiency. Two synchrotron light monitors and two stored
current monitors are also part of the diagnostic system. A
transverse feedback system is implemented on the ring: it consists
of a stripline pick-up and a stripline kicker. The vacuum chamber is
fully stainless steel and a pumping system consisting of 18 sputter
ion pumps is designed to reach an average dynamic pressure in the
ring of $5 \nTorr$. A parameter list for the accumulator is provided 
in Table~\ref{tab:accpar} and a schematic layout is shown in 
Fig.~\ref{fig:accumulator}.

\begin{table}
\caption{\label{tab:accpar}Accumulator ring parameter list.}
\vspace*{2mm}
\centering
\setlength{\extrarowheight}{1pt}
\begin{tabular}{p{7cm}r}
\hline
\hline
Energy                             & 510\mev        \\
Circumference                      & 32.56\m        \\
Emittance                          & 0.26\mrad-\mm  \\
Horizontal betatron tune           & 3.12           \\
Vertical betatron tune             & 1.14           \\
RF frequency                       & 73.65\MHz      \\
RF voltage                         & 200\kV         \\
Max storable current               & 100\mA         \\
Bunch length                       & 3.8\cm         \\
Synchrotron radiation loss per turn& 5.2\kev        \\
Horizontal betatron damping time   & 21.4\ms        \\
Vertical betatron damping time     & 21.4\ms        \\
Longitudinal damping time          & 10.7\ms        \\
\hline
\end{tabular}
\end{table}

\subsubsection{Injection schemes}
\label{inj}
The injection system for the \daphne main rings is designed to provide
full flexibility in the bunch patterns~\cite{bib:injscheme1,bib:injscheme2}. 
The circumference of the accumulator is $1/3$ of the
main ring, so that a single bunch stored in the booster can be
transferred on each turn into one out of three equidistant buckets of
the main ring. The accumulator RF harmonic number is 8, and
therefore 24 buckets of the main ring can be filled by shifting the
relative phase of the linac gun with respect to the accumulator
cavity. The main ring revolution frequency acts as a clock for the
whole system. The accumulator RF system, running at the 24th harmonic
of the clock is phase-locked to the main ring system, and its phase is
shifted with respect to the clock in steps of $2.72 \ns$ during the
damping time before extraction ($100 \mus$) when more than 24 bunches are
to be injected, or if the desired bunch pattern is not in coincidence
with one of the 8 accumulator buckets. Five steps are necessary to
fill all the main ring buckets. The trigger of the linac gun is locked
to the accumulator RF generator, with the capability of reaching any
of the 8 accumulator buckets.  The acceptance of the booster, due to
high cavity voltage and low energy of the beam, is almost one bucket
length in phase ($13.6 \ns$) and $\pm 2.3 \%$ in energy spread. A $10
\ns$, $\pm 1.5 \%$ bunch from the linac is accepted in longitudinal
phase space with more than $95 \%$ efficiency. 
The standard positron
injection cycle in \daphne involves positron accumulation
at $50\Hz$ in the accumulator, with extraction and injection into the 
main rings at $2\Hz$. The bunch charge injected in the MRP is about
$2\times 10^{10}$ $\ep$ per pulse at $2\Hz$ (15 pulses in the accumulator).

\subsubsection{Scaling for \superb}

In the positron converter scheme, where a primary electron beam
impinges on a metallic target, the number of pairs emerging from the target
depends on the target thickness $t$ (radiation length units) and the
primary electron beam energy $E_0(\mev)$ in a form that can be
approximately expressed by \cite{bib:BM2,bib:Barnett}:
\begin{equation}
N(t,E_0)\propto E_0b(E_0)\frac{[b(E_0)t]^{a(E_0)-1}e^{b(E_0)t}}{\Gamma[a(E_0)]}\,,
\end{equation}

where:
\begin{equation}
a(E_0)=1-\frac{1}{2}b(E_0)+b(E_0)ln{\frac{E_0}{E_c}}\,.
\end{equation}

The function $b(E_0)$ has a small dependence from $E_0$ and can be
approximated as $b(E_0)\approx b=0.5$.
The {\it critical energy} $E_c$ can be approximated by:

\begin{equation}
E_c=\frac{610(\mev)}{Z+1.24}\,,
\end{equation}
where $Z$ is the target atomic number.
The number of produced pairs is maximum when $t=t_{max}$:

\begin{equation}
t_{max}=-\frac{1}{2} + \ln{\frac{E_0}{E_c}}\,.
\end{equation}

In the case of \daphne, the tungsten $(Z = 74)$ target thickness
$t_{max}$ is optimized for $E_0=250 \mev$.
For other primary beam energies $E$ an estimate of the relative 
variation in the positron yield can be
obtained from:

\begin{equation}
\begin{split}
R(E)=\frac{N[t_{max}(E_0=250 \mev),E]}{N_{max}(E_0=250\mev)}=\\
\frac{E(\mev)}{250}
\frac{\Gamma[a(250\mev)]}{\Gamma[a(E)]}[a(250\mev)-1]^{a(E)-a(250\mev)}
\end{split}
\end{equation}

\begin{figure}[htb]
\begin{center}
  \includegraphics[width=0.8\textwidth]{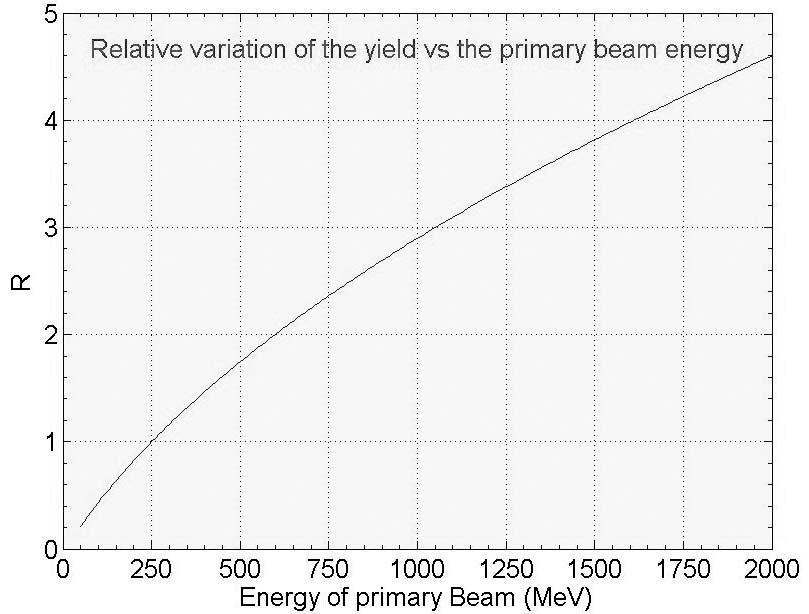}
  \caption{
    Relative variation of the positron yield {\it vs.} the energy of the primary
    electron beam, normalized to the yield at $250 \mev$.}
    \label{fig:RvsE}
    \end{center}
\end{figure}

\begin{figure}[htb]
\begin{center}
    \includegraphics[width=\textwidth]{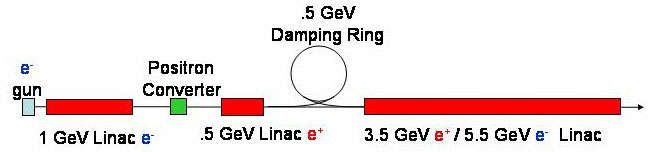}
    \caption{Schematic drawing of the \superb\ injection scheme. }
    \label{fig:Sbinj}
    \end{center}
\end{figure}

The variation in this relative yield $R(E)$ is shown in Fig.~\ref{fig:RvsE}
for beam energies up to 2\gev. In
recent years there has been intense effort directed towards developing high
intensity positron sources \cite{bib:crystal1,bib:crystal2,bib:crystal3}. These
studies show that, with crystalline targets, it is
possible to achieve additional enhancement factors of 2--3 in the normalized
positron yield over the corresponding amorphous material.
Starting from the yields of the \daphne positron
injection system (Section~\ref{inj}) of
about a $2\EE{10}$ $\ep/{\rm s}$ injection pulse, and assuming a $1 \gev$
primary electron beam and a crystalline target, we estimate that an 
eightfold enhancement factor could, in principle, be obtained providing a 
$1.6\EE{11}$ $\ep/{\rm s}$ injection rate at
$2\Hz$. Other gains could be realized through a higher value for the 
magnetic field in the adiabatic matching system downstream of the target, 
and through a more fully optimized design of the transport line 
optics~\cite{bib:CLIC}. In any case,
the overall improvement must be carefully evaluated with proper
particle tracking codes. A schematic drawing of
a possible injection scheme is shown in Fig.~\ref{fig:Sbinj}.

\subsection{Alternative Solution for the Injection System}

An alternative design for the injection system requires
a polarized electron gun, a 6\gev linac operating
at 2856\MHz, a 
positron converter at the 5\gev location, two damping rings operating at 1\gev, two beam 
transport lines (BTL) to the \superb\  LER and HER rings, and finally spin 
manipulation solenoids. Many 
components of the SLAC electron and positron sources could be reused in this scenario.
The SLAC polarized gun produces
two longitudinally polarized $e^-$ bunches of about $6\times 10^{10}$ particles at 120\Hz; 
we assume the linac will operate at 100\Hz. There are 
several possible injection schemes for the \superb\ accelerator using this system, but we describe
a solution where the 100\Hz repitition rate for the linac is divided into 25\Hz for 
electrons and 75\Hz for positrons.
The length for the linac tunnel, assuming the same average gradient as the SLAC 
linac (17\mev/m), is $(6\gev/0.017\gev/m) = 353\m$. Such
an injection linac, even allowing extra length for injection lines, would easily fit 
into the tunnel planned for the Italian SPARX FEL project, which has length of about 1\km.

\subsubsection{Electron injection} 

The electron gun produces two longitundinally polarized
bunches with $6\times 10^{10}$ $e^-$ at a rate of 25\Hz.  These bunches are accelerated
to 1\gev and injected into an electron damping ring. In the injection line for the 
damping ring, the bunches pass through a spin rotator to rotate the spin into the 
vertical direction in the damping ring. The transverse emittance of the electron bunches 
is damped over a time of 0.04\sec (1/25\sec). The two bunches are extracted, injected 
into the linac, and then accelerated the full 6\gev to a final energy of 7\gev,
before beam transport to the \superb\ HER. Spin manipulation solenoids 
are inserted into the damping ring-to-linac transport line to provide vertically polarized electrons for 
injection into the electron storage ring. There will be some losses 
from injection and extraction from the damping rings, and from injection into the 
\superb\ ring, which we estimate to be about 15\%. The number of electrons available 
for injection with this scheme
is $0.85$ (eff) $\times 25$ (Hz) $\times 2$ (bunches) $\times 6 \times 10^{10}$ (per bunch) $= 
2.6\times 10^{12}$ $e^-$ per 
second. 

\subsubsection{Positron injection} 

The electron gun produces two bunches with $6\times 10^{10}$
$e^-$ at a rate of 75\Hz.
The two bunches are accelerated to 5\gev before impinging on a positron converter 
target. The resulting positron bunches are subsequently captured by an accelerating section. We
assume that polarization is not required for the positrons. The SLAC SLC positron 
converter source produced two usable positrons for each incident electron at 30\gev. For the
\superb\ injector, with a 5\gev beam, the same system will yield about six times 
fewer positrons, since the yield is proportional to the  incident $e^-$ energy. After the 
positron target, the positrons are captured 
and accelerated to 1\gev in the remaining section of the linac. The positrons are then 
transported back to the beginning of the linac and injected into the positron damping ring 
at 1\gev. The positrons remain in the damping ring four injection cycles (three $e^+$ and 
one $e^-$) or 0.04\sec (1/25\sec) in order to achieve low emittance. The positron damping ring will 
have six bunches stored at any one time. After four storage cycles the oldest two positron 
bunches are extracted from the damping ring and accelerated to 3\gev,
before beam transport to the \superb\ LER. The positrons, with very 
large emittance from the target, will likely have a smaller injection efficiency into the 
damping ring than the electrons; we assume an efficiency of 67\%. A pulsed magnet chicane 
near the target will be needed to allow the $e^+$ bunch pair to miss the target and one $e^-$ 
bunch pair to strike the target on the same linac pulse. The number 
of positrons available for injection with this scheme
is $0.67$ (eff) $\times 2/6$ (conversion) $\times 75$ (Hz) $\times 2$ 
(bunches) $\times 6 \times 10^{10}$ (per bunch) $= 2.0\times 10^{12}$ positrons per sec. 

\subsection{A Polarized \ep Source as an Upgrade}
\label{sec:Posipol}

Polarized electron-positron pairs can be produced by
converting circular polarized-high energy gammas in a solid/liquid
target. The most efficient process for energy amplification in the
production of high energy photons is Compton scattering, where
the collision between a high energy
electron and a photon boosts the energy of the recoiling photon. In
this way, it is possible to produce gammas in the 10--100\mev
energy range by colliding electron beams of 1--2\gev
with a $1 \mum$ pulse from a solid state laser~\cite{bib:Posipol_1}. 
Higher electron energy is required if a \CO2
laser ($10 \mum$) is used, but in this case, the advantage is
the linear increase of the photon number per pulse energy. In all
cases, the final photon polarization depends only on the laser polarization,
and not on that of the electron beam. It is therefore possible 
to select the polarization of the gamma, and consequently
of the positron, by switching the laser polarization.

Compton production is very attractive from an energy
conversion efficiency point of view, but has the disadvantage of a very small
elastic cross section ($ 8 \pi r_e^2 / 3$, where $r_e$
the classical electron radius). In order to increase the rate, in the case
of a solid
state laser, the number of photons per pulse can be amplified by stacking
laser pulses into a Fabry-Perot optical resonator \cite{bib:Posipol_1}. Typical
gains of order $10^3$ for a pulsed laser have already been achieved
\cite{bib:Posipol_2}, but development aimed at
attaining gains of $10^4$ or $10^5$ \cite{bib:Posipol_1} are underway.

After production, the polarized gammas can be converted into linearly
polarized \epem\ pairs in a solid target. To maximize the production rate,
0.4--0.5 radiation length thick tunsgten or
titanium targets have been studied \cite{bib:Posipol_3}. To maximize the
collection efficiency for the produced positrons, the target is placed in a
``capture section'' composed of strong solenoidal fields and
accelerating cavities \cite{bib:Posipol_4}.
For $60\%$ polarized positron beams, an $ \ep
({\mathrm captured}) /\gamma$
conversion efficiency parameter $\eta \sim ~ 1 \mbox{--} 2 \%$ is 
observed \cite{bib:Posipol_5}. To optimize the efficiency for injection
and capture in the main ring, it is possible to stack different bunches
in an intermediate accumulator ring \cite{bib:Posipol_6}. After a few
milliseconds of stacking and cooling the beam can be transferred to
the main ring.

\begin{figure}[tb!]
\psfragscanon
\psfrag{em}{$e^-$}
\psfrag{ep}{$e^+$}
\psfrag{1}[lt][lt]{
\parbox[t]{5.5cm}{\footnotesize
 1 - Gamma production : Electron beam impinging on a
laser pulse. The electron beam can be stored in an
accumulator ring, or delivered by a high current linac
or an ERL. Laser pulses can be stored in an optical
resonator (solid state) or regenerative amplifier
cavities ($\mathrm{CO}_2$)}
}
\psfrag{2}[lt][lt]{
\parbox[t]{3.6cm}{\footnotesize
2 - Pair production: The polarized gammas impinge on a W or Ti
target, creating a linear polarized pair.} } \psfrag{3}[lt][lt]{
\parbox[t]{3.6cm}{\footnotesize
3 - Capture: The polarized positrons are focussed, and accelerated in
the capture section. The acceptance of this section has a strong
impact on the total efficiency of the source} }
\psfrag{Acc}[lt][lt]{}

\includegraphics[width=\textwidth]{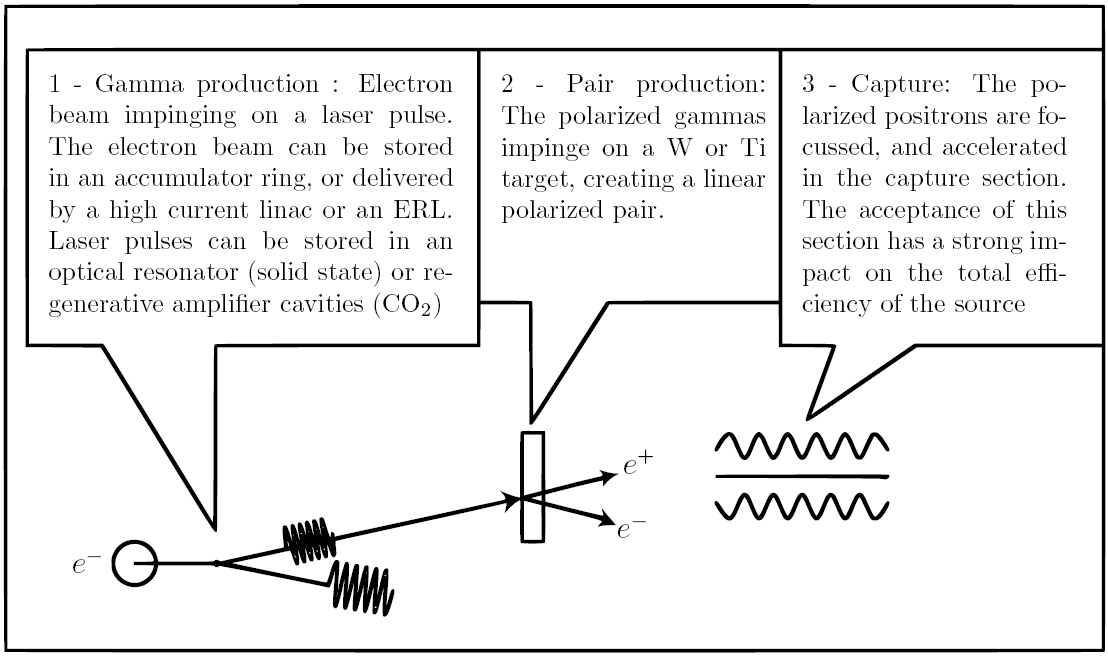}
\caption{\label{fig:Posipol_1} Schematic summary of the scheme for a
polarized positron source based on Compton backscattering. }
\end{figure}

A schematic summary of the full scheme for a polarized positron source based on
Compton scattering is shown in Fig.~\ref{fig:Posipol_1}, where the process
is divided into three parts: gamma production, pair
production and positron capture.

\subsubsection{Estimate of the positron rate and application to \superb}
\superb\ requires $2 \times 10^{12}$ polarized positrons
per second (60--70\% beam polarisation). The optimized
time structure for injection must still be defined, depending on the
injection scheme; we will therefore evaluate here only the
beam characteristics and the number of
Compton collisions needed to meet the \superb\ requirement.

The total rate $R_{\ep}$ of polarized positrons that can be
delivered to the main ring is defined by:
$$
R_{\ep} = 2 \times 10^{12}\; (\mbox{\superb\ case})
= \Ngacoll\, \vartheta\, \eta\,,
$$
where \Ngacoll is the number of backscattered gammas per collision,
$\vartheta$ is the number of collisions per second and $\eta$ is the
efficiency for production and capture of the
polarized positrons. If we assume $\eta = 2\%$, the result is:
$$
\Ngacoll\, \vartheta = 10^{14}\; (\mbox{\superb\ case})\,.
$$
To determine the necessary number of collisions needed to attain the 
required flux we must
estimate the number of backscattered gammas per collision.

\subsubsection{\ilc simulations and scaling to \superb\ parameters}
A number of simulations have already been carried out in connection
with the \ilc. \Ngacoll scales linearly as a function of the laser
pulse and electron bunch intensity up to the threshold of the
non-linear Compton regime. Given the small backscattered
gamma/electron ratio we are assuming, this should not be a factor for \superb.
Table~\ref{tab:Posipol_1} summarizes simulation results
obtained with the Monte Carlo code CAIN for a number of relevant
configurations. Two different solutions are illustrated
for the electron machine used for gamma generation 
to emphasize the range and
flexibility of solutions envisaged for the \superb\
case. These two solutions are under consideration for the
\ilc: an accumulation ring
in which the Compton cavities are inserted in the collision region, and
an energy recovery linac with the interaction region inserted in the
recirculation line.
In the Compton ring, because of the crossing angle and the long
electron bunch, the flux is reduced. The ``OPO'' (optimized path for
overlap) crab angle scheme strongly reduces this effect
\cite{bib:Posipol_7}, but has not yet been proven experimentally.

\begin{table}[htbp]
\caption{\label{tab:Posipol_1}
Results from gamma production simulations, for two different configurations for
the electron source.
}
\vspace*{2mm}
\centering
\setlength{\extrarowheight}{3pt}
\begin{tabular}{llll}
\hline
\hline
\multicolumn{4}{c}{\parbox{12cm}{\vspace{3pt}
\centering
Compton ring + laser + optical resonator.\\
One interaction point. (\ilc parameters)} }\\ \hline
Electron beam
& Laser pulse
& Angle collision
& \Ngacoll \\ \hline

Energy = $1.3\gev$
& Wavelength = $1 \mum$
& $0^\circ$
& $5.15 \times 10^8$ \\

Charge = $10 \nC$
& Pulse energy = $0.1 \J$
& No Crab
& \\

 $\sx =30\mum$
& Waist = $20 \mum$
& $8^\circ$
& $7.25 \times 10^7$ \\

 $\sy = 5 \mum$
& Pulse length = $ 240 \mum$
& No Crab
& \\

 $\sz = 6\mm$
&
& $8^\circ$
& $4.55 \times 10^8$ \\

&&  OPO Crab & \\ \hline

&\\

\multicolumn{4}{c}{\parbox{12cm}{\vspace{3pt} \centering

ERL + laser + optical resonator. \\ One interaction point. (ILC parameters)
} }\\ \hline
Electron beam &
Laser pulse &
Angle collision &
\Ngacoll \\ \hline

Energy = $1.3 \gev$ &
Wavelength = $1 \mum$ &
$5^\circ$ &
$1 \times 10^9$ \\

Charge = $1.5 \nC$ &
Pulse energy = $0.6 \J$ &
No Crab &
scaled to\\
 & & &  $0.1 \J$ laser:\\
 & & &  $1.6 \times 10^8$ \\

$\sx =15 \mum$ &
Waist = $15 \mum$ & & \\

$\sy = 15 \mum$ &
Pulse length = $240 \mum$
& & \\
 & (bunch compression) & & \\
$\sz =200 \mum$  & & & \\ \hline
\end{tabular}
\end{table}

Based on the \superb\ parameters, and assuming only one interaction point,
we must accumulate $\vartheta \sim 10^5 \mbox{--} 10^6$ pulses in one
second to satisfy the injection requirements ($10^{14}$ gammas/sec
to provide $2 \times 10^{12}$ positrons/sec). Continuous
injection at $0.1\mbox{--} 1 \MHz$ would be enough to satisfy this requirement.
Nevertheless, it should be emphasized that the simulations summarized above
were performed assuming very high performance parameters, for which
R\&D is still ongoing in connection with the \ilc. The \superb\
polarized positron source design will be the subject of dedicated
studies in collaboration with LAL-Orsay.



\afterpage{\clearpage}

\section{Polarization}
\label{section:SpinPolarization}
\subsection{Introduction}

The study of \CP and $T$ violation in the lepton sector, in particular, the
search for \CP or $T$ violation in the production and decay of $\tau$ lepton 
pairs, is an important objective of the \superb\ program. Thus, the provision
of polarized electrons and/or positrons~\cite{bib:Koop_1} is an
important design consideration.
The requirements for a polarization facility include:
\begin{itemize}
\item A stable longitudinal direction for spin at the IP;
\item A depolarization time longer than one beam lifetime;
\item Fast switching of the sign of the polarization, within or
      less one beam lifetime;
\item The ability to provide arbitrary filling patterns, \eg,
      it would be very useful to have opposite
      polarizations in neighboring RF buckets;
\item Polarization of the electron beam in the initial design, with
the possibility of positron polarization as an upgrade; and
\item High degree of polarization.
\end{itemize}

Existing laser gun technology can provide electrons with up to $90\%$
polarization at the required intensity and
repetition rate \cite{bib:Koop_2,bib:Koop_3}.
Direct acceleration of electrons in a linac up
to the full LER/HER energy completely eliminates all problems related
to crossing integer or intrinsic resonances during acceleration
in a booster ring. Such an injection scheme for
polarized electrons will be described in more detail below.

\begin{figure}
\centering
\includegraphics[width=0.68\textwidth]{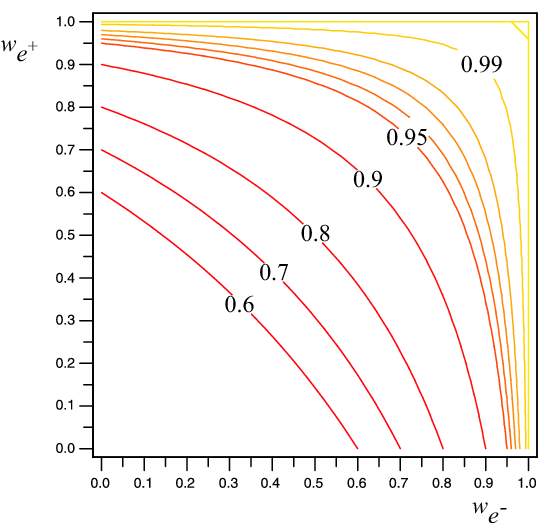}
\caption{\label{fig:tau_pol}
$\tau$ polarization in one-photon \epem annihilation as a function of the
polarization of the electron and positron beams.
}
\end{figure}

\begin{figure}
\centering
\includegraphics[width=0.68\textwidth]{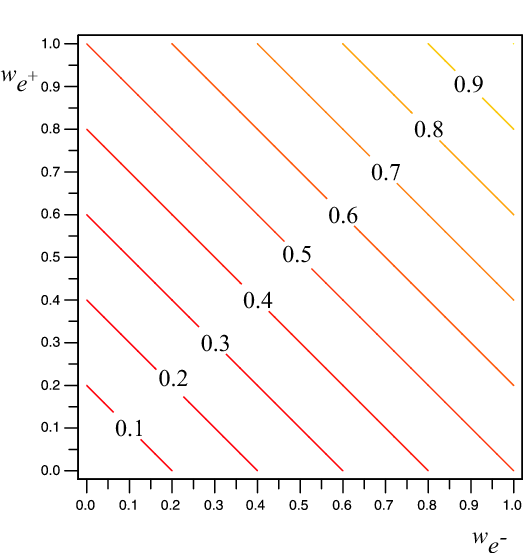}
\caption{\label{fig:tau_fm}
Luminosity enhancement factor for the one
photon exchange processes as a function of the
polarization of the electron and positron beams.
}
\end{figure}

The effective polarization
asymmetry $w$ for a collision:
$$
w = \frac {{w_{e^-}} + {w_{e^+}} }{ 1+{w_{e^-}}{w_{e^+}} }\,,
$$
where ${w_{e^-}}$ and ${w_{e^+}}$ are the linear polarizations of the electron and positron beams,
approaches unity if both beams are polarized in the
same direction with high individual polarizations, as shown in Fig~\ref{fig:tau_pol}.
For example, $w=0.995$ if ${w_{e^-}} = {w_{e^+}} = 0.9$,
while $w=0.90$, if ${w_{e^-}} = {w_{e^+}}=0.63$.
In addition, the effective luminosity enhancement for
one-photon annihilation processes due to polarization is:
$$
\frac{\lum}{\lum_0} = 1 + {w_{e^-}}{w_{e^+}}.
$$
This becomes a factor of two in the limit $w = {w_{e^-}} ={w_{e^+}}=1 $
as illustrated in Fig.~\ref{fig:tau_fm}.

Despite these advantages, the difficulty in obtaining polarized positrons
seem sufficiently large that we defer this the installation to a possible
upgrade of \superb.

Spin direction manipulations are most easily done at $100 \kev$ energies; this
is the approach  taken, for example, at NIKHEF's AmPS facility \cite{bib:Koop_4}, where a so-called $Z$-manipulator is used. This device employs a combination
of two electrostatic bends to rotate spin by $\pm 90^\circ$
around the vertical axis while solenoids between these bends provide
spin rotation around the longitudinal direction. The spin direction of
electrons injected into the HER can be changed very quickly with
this arrangement, perhaps with every source
laser pulse as the ultimate limit, simply by changing the sign of circular
polarization for the laser light.

\subsection{Siberian Snake Solution}
\begin{figure}
\centering
\includegraphics[width=0.8\textwidth]{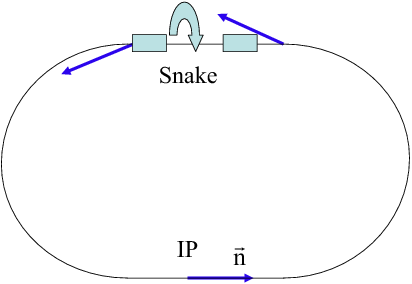}
\caption{\label{fig:Pol1}
Two $\pi/2$ solenoids of the Siberian Snake installed at a half
turn away from the interaction point rotate spin by $\pi$ around the
velocity direction. As a result, an equilibrium closed spin orbit
$\VE{n}(\theta)$ has a purely longitudinal spin direction
at the IP. In the arcs, the spin always lies in the horizontal plane.
}
\end{figure}

There are many ways to create a stable closed orbit with
a longitudinal direction for spin at the collision point. However, a Siberian
Snake is perhaps the simplest method~\cite{bib:Koop_5}.
The spin trajectory in a ring with one Siberian Snake installed a
half-turn from the IP is shown in the Fig.~\ref{fig:Pol1}. Two
$90^\circ$ solenoids, with intervening normal quads to
provide decoupling of the betatron oscillations, rotate
spin by $180^\circ$ around the longitudinal axis. This arrangement results
in the formation of a closed spin
orbit $\VE{n} (\theta)$ with a purely longitudinal equilibrium spin
direction at the IP. Everywhere along the arcs, the spin lies in the
horizontal plane, rotating around the vertical axis, which is
directed along the bending magnetic field of the ring. Perpendicular to
\VE{n}, spins make a half turn around \VE{n} each turn, and thus the
total spin tune equals $\nu=0.5$.

\begin{figure}
\centering
\includegraphics[width=0.9\textwidth]{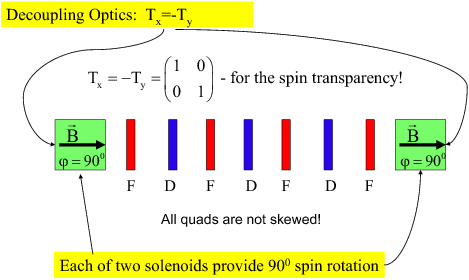}
\caption{\label{fig:Pol2}
A FODO lattice decouples horizontal and vertical motions in the $180^\circ$
spin rotator of the full Siberian Snake.
}
\end{figure}

The coupling induced by the two solenoids of a full Siberian Snake must
somehow be compensated in the ring optics. The simplest, and at the same time very
convenient way to do this, was suggested by Litvinenko and
Zholents in 1980~\cite{bib:Koop_6}. If matrices of the FODO lattice
inserted between solenoids satisfies the requirement:
$$
T_y = - T_x
$$
then the horizontal and vertical betatron oscillations became
fully decoupled, as shown in Fig.~\ref{fig:Pol2}. An additional requirement
comes from the spin transparency condition \cite{bib:Koop_7}:
$$
T_x = - T_y = \begin{pmatrix} 1 & 0 \\ 0 & 1\end{pmatrix}\,.
$$
In the case of the partial Siberian Snake, which rotates spin by the total angle
$\varphi \le \pi$, this expression becomes:
$$
T_x = - T_y = \begin{pmatrix} -\cos \varphi & -2 r \sin \varphi
\\ (2 r)^{-1} \sin \varphi & -\cos \varphi\end{pmatrix} ,\quad r = \frac{p c}{e B}\,.
$$

\subsection{Equilibrium Polarization and Depolarization Time}

The degree of polarization at equilibrium and the spin relaxation time are
described by the formulae of Derbenev and Kondratenko \cite{bib:Koop_8}:
\begin{alignat*}{3}
w_{\mathrm{rad}} &=&
- \frac{8}{5\sqrt{3}}
\frac
{ \left\langle \left| r \right|^{-3} \VE{b} \cdot \left( \VE{n} - \VE{d}\right)
\right\rangle}
{ \left \langle
    \left| r \right|^{-3}
    \left(
      1 - \frac{2}{9} \left( \VE{n} \cdot \VE{v} \right) ^2
      + \frac{11}{18}\; \VE{d}^2
    \right)
\right\rangle}\\
\tau_{\mathrm{p}}^{-1} &=&
\frac{5 \sqrt{3}}{8} \lambda_e r_e c \gamma^5
\left \langle
 \left| r \right|^3
 \left(
   1 - \frac{2}{9} \left( \VE{n} \cdot \VE{v} \right)^2 +
   \frac{11}{18}\; \VE{d}^2
 \right)
\right \rangle\,,
\end{alignat*}

where $\left| r \right|$ is the modulus of the radius of curvature, $\VE{b}
= \VE{B}/|\VE{B}|$ is a unit vector directed along the bending
field \VE{B},\VE{\nu} is the velocity vector (assuming $c=1$), and
$\VE{d}\equiv \gamma \left( \partial \VE{n}/\partial\gamma \right)$
is the so called spin-orbit coupling vector. 

In a sense, vector \VE{d} describes the chromaticity of the
equilibrium spin direction \VE{n}. Its value is proportional to the energy:
$$
 | \VE{d} | \propto \nu_0 \quad , \quad \nu_0 = \gamma a
\quad \left( \nu_0 = 4.54 \mbox{ for } E=2\gev \right)\,.
$$
In a normal ring with one spin-transparent Siberian Snake, $\VE{d}^2$, averaged over the circumference, can be evaluated exactly:
$$
\left\langle \VE{d}^2 \right\rangle = \frac{\pi^2}{3} \nu_0^2
\quad , \quad \left( \left\langle \VE{d}^2\right\rangle =68
  \mbox{ for } E=2\gev\right)\,.
$$

A general formula for the vector \VE{d} is:
$$
\VE{\Delta n}(\theta) = \Re \left( i\VE{\eta}(\theta)^\star
\int\limits_{-\infty}^{+\infty} \VE{w} \cdot \VE{\eta} \; d\theta'
\right) \quad , \quad
\VE{d}(\theta) \equiv \gamma \frac{\partial \VE{n}}{\partial \gamma}\,,
$$
where  it is assumed that a spin perturbation $\VE{w}(\theta)$
is adiabatically switched from zero to its final value over the
azimuthal interval $-\infty<\theta'<\theta$.
The three components of the spin perturbation are given by:
\begin{alignat*}{2}
w_x & \sim \nu_0 &
 \left(-K_x \frac{\Delta \gamma}{\gamma} -y'' \right)
\quad,&\quad \nu_0 =\gamma a\\
w_y & \sim \nu_0 &
 \left(-K_y \frac{\Delta \gamma}{\gamma} +x'' \right)
\quad&,\quad a \sim 1.16\EE{-3} \\
w_z & \sim &
 -(1+a) K_z \frac{\Delta \gamma}{\gamma}
\quad&,\quad K_{x,y,z} = B_{x,y,z} / \left\langle B_y \right\rangle\,.
\end{alignat*}
The complex vector $\VE{\eta} = \VE{\eta_1} -i\VE{\eta_2}$,
where \VE{\eta_1}, \VE{\eta_2}, and \VE{n} are three real unit vectors
representing the three orthogonal solutions of the spin motion
equation for the equilibrium particle, provides a convenient
description of a rotation of spin around the \VE{n}
direction by the angle $\varphi$. Writing
$\VE{\eta}(\theta) = \VE{\eta}(0) E^{i\varphi}$,
$\VE{\eta}(\theta)^\star = \VE{\eta}(0)^\star E^{-i\varphi}$, the
periodicity conditions are:
\begin{alignat*}{3}
\VE{n}(\theta+2 \pi) &=& \VE{n}(\theta) \\
\VE{\eta}(\theta+2 \pi) &=& \VE{\eta}(\theta) e^{i 2\pi\nu}\,.
\end{alignat*}

Thus, $\VE{n}(\theta)$ is periodic, while $\VE{n}(\theta)$ receives
a phase advance $\varphi=2 \pi \nu$ after one turn. As already noted,
in the case of a ring with a Siberian Snake the spin tune is $\nu=0.5$.

The equilibrium degree of self-polarization becomes nearly zero
$w_{\mathrm{rad}} \sim 0$ in the case of a Siberian Snake
and the spin depolarization time at $E=2\gev$ and $|r| = 20 m$
is approximately:
$\tau_{\mathrm{p}} = 4000\sec$ ($\tau_{\mathrm{p}} \propto \gamma^7$
for $|r| = \mbox{constant}$).

Mixing fresh polarized beam through continuous injection with an
old partly depolarized circulating beam, one finds an
equilibrium polarization:
$$
w_{\mathrm{average}} = \ze0 \frac{\tau_p}{\tau_0+\tau_p}+
w_p \frac{\tau_0}{\tau_0+\tau_p}+
$$
Here $w_0$ is the initial polarization of the injected particles,
$\tau_0$ is the beam life time, $w_p$, $\tau_p$  are the asymptotic
radiative self-polarization degree and the radiative polarization time,
respectively. From thisrelation we conclude that
$w_{\mathrm{average}} \sim w_0 $  if $\tau_p \gg \tau_0$.
The beam lifetime is determined mainly by the luminosity and by
Touschek losses, and is estimated to be approximately $\tau_0=200 \sec$.
Taking $w_0 = 0.90$, $w_p= 0.1$,and $\tau_p=4000\sec$
one obtains an average degree of polarization:
$$
w_{\mathrm{average}} = 0.86
$$

\subsection{Polarization Schemes for 4\gev and 7\gev}
The simplest scheme with one Siberian Snake in a ring works very well
below $2.5 \gev$. At higher energies, the depolarization time
becomes too short. Therefore, much more complicated schemes with
restoration of the vertical spin direction in the arcs must be considered for
the \superb\ implementation at $4$ and $7 \gev$.

\begin{figure}[tb]
\centering
\includegraphics[width=0.8\textwidth]{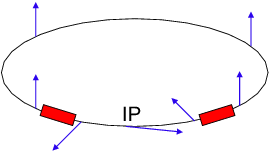}
\caption{\label{fig:Pol5}
A scheme with restoration of the vertical spin direction in the arcs by
solenoids.
}
\end{figure}

A possible layout is shown in Fig.~\ref{fig:Pol5}.
The first spin rotator (red box) rotates spin by $90^\circ$
around the longitudinal axis, then a $90^\circ$
rotation around the vertical axis brings the spin to a purely longitudinal
orientation at the IP. The second half of the full insertion is just
symmetric or anti-symmetric relative to the IP.
To be spin-transparent for the betatron oscillations, each spin rotator
is comprised of two $45^\circ$ solenoids, with the FODO lattice between.them.
However, in this case, the $2\times 2$ matrices of one spin rotator should satisfy
the condition \cite{bib:Koop_7}:
$$
T_x = - T_y = \begin{pmatrix} 0 & -2r \\ (2r)^{-1} & 0\end{pmatrix}\,.
$$
The field integral required for one $45^\circ$
solenoid at $4 \gev$ is equal to:
$$
\int B\;dl = 10.5 \Tesla \m\,.
$$
A symmetric layout does not work, in fact, because the \VE{n}-direction
is still
chromatic in the arcs, in other words $\VE{d}\ne 0$ and, as a result,
the beam will quickly depolarize.  The anti-symmetric option
meets this condition, but cannot be used because of of the final focus
design crierion that specifies only positive bends. Two symmetric insertions
could, in principle, compensate each others' spin-chromaticity, but such a
solution appears much too complicated.

\begin{figure}
\centering
\includegraphics[width=\textwidth]{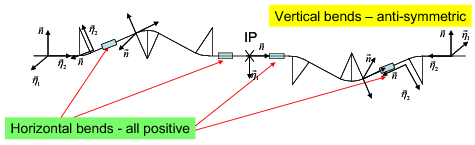}
\caption{\label{fig:Pol9} $7 \gev$ layout for achieving a
longitudinal spin direction at the IP by applying a sequence of
transverse bends. The scheme shown here restores the orbit plane
and the spin direction in the arcs simultaneously. Vector \VE{\eta}
makes a half turn around \VE{n}, hence the spin phase advance
equals $\pi$.}
\end{figure}

We therefore believe that a HERA-like solution is the only way
to obtain longitudinal polarization at high energies
\cite{bib:Koop_9}. The proposed scheme is shown in
Fig.~\ref{fig:Pol9}. A sequence of vertical and horizontal bends,
each rotating the spin by $90^\circ$, transforms an initial vertical direction for
\VE{n} into the longitudinal direction at the IP. Vertical bends are
anti-symmetric relative to
IP, while in contrast all horizontal bends are positive and symmetric. Every
vertical bend is made achromatic by being divided in two half-bends and
placing some optics in between. The two horizontal bends closest to the IP
belong to the FF-insertion.

Every $90^\circ$ spin rotation translates into $5.66^\circ$
of a velocity bend at $7 \gev\; (\nu=15.9)$.
After the two first bends, the $x$-plane is inclined by $97.5 \mrad$.
This could be rolled back by a weak solenoid with $Bl=0.455 \Tesla\m$.
The vertical orbit excursion will not exceed $3 \m$.

    The scheme discussed here could be made spin-achromatic for
the off-momentum particles, and simultaneously spin-transparent for
 particles with non-zero betatron amplitudes. The estimated
depolarization time at $7 \gev$ is:
$
\tau_p = 1800 \sec
$.
Taking a beam lifetime of $\tau_0=230/sec$
and a polarization $w_0 = 0.90$ for the injected beam:
$
w_{\mathrm{average}} = 0.81
$.

Some of the advantages of using transverse bend spin rotators include:
\begin{itemize}
\item Provision for a spin-achromatic and a spin-transparent solution;
\item Bending magnets are cheaper than solenoids; and
\item Positive horizontal bends match the final focus lattice.
\end{itemize}
Disadvantages include
\begin{itemize}
\item An orbit bump of several meters is needed in the vertical direction at $E=7\gev$; and
\item The implementation is a fixed energy solution, applicable only for $E\ge7\gev$.
\end{itemize}

\subsection{\superb\ Polarization Scenario}
There is no universal solution that can be applied at all intended
operating energies.
We therefore continue to develop two essentially different approaches.

The first, namely the Siberian Snake concept, works very well below
$2.5 \gev$. The Siberian Snake insertion is located in the straight section
opposite to the IP. A longitudinal polarization direction at the IP is
provided by two $90^\circ$ solenoids with a field integral:
$
B\,l=5.24 \Tesla \times m / \gev
$.
Non-skewed quadrupole lenses between the solenoids compensate the
betatron coupling. The solenoids could be switched off if experiment is
to be operated at higher energies, where the Siberian Snake could not be
used. Depolarization time with the snake exceeds one hour at $2 \gev$.

The second solution, based on HERA-like spin rotators with the
use of transverse fields, is applicable at $7 \gev$. At lower energies
the required vertical orbit bumps became too large. A bypass could be used
to switch off the spin perturbations generated by vertical
orbit bumps of these spin rotators in case of operation at low energies,
where the Siberian Snake option could be applied.
In both scenarios, the average polarization of the circulated beam
can reach $w_{\mathrm{average}} = 0.80$.



\afterpage{\clearpage}

\section{Site and Utilities}
\subsection{Tunnel}

The 2.3\km-circumference tunnel for the \superb\ Factory 
must have a diameter of 4
to 6 meters to accommodate the two accelerator rings, trays for the
power and control cables, cooling water pipes, an access path for equipment, 
and space for safety egress.
The two accelerator rings will be placed side-by-side in one
tunnel to keep both rings in the same plane. This reduces or
eliminates vertical bends that tend to increase the
vertical emittance. Since the rings as designed have six straight
sections, there will be six support buildings, one for each
straight. One of the straight sections will have the interaction
region housing the \superb\ detector, and will therefore be substantially
larger ($20\times 60$\m) than
the others. The magnet power supplies and controls, cooling water
conditioners, RF power supplies and controls, and diagnostic
controls will be housed in these straight section buildings. Other
than these buildings, the ring tunnel will be fully underground.
The tunnel for the \superb\ Factory could be constructed using several
methods, as appropriate: cut and cover, boring, blasting, \etc. 
In order to provide radiation
containment, the earth surrounding the tunnels must have a thickness
of about 3 to 6 meters, depending on detailed calculations. A floor
drainage system with sump pumps will be provided in the tunnel to
contain, collect and treat any free-running water in the tunnel.

\subsection{AC Power}
\label{sec:AC_power}
The accelerator requires power for electromagnets, RF
systems, diagnostics and controls, and air handling
systems. The largest power
contribution is from the RF system used to replace the energy lost by the beams
due to synchrotron radiation in the bending magnets and
wigglers.

The power requirements for \superb\ are shown
in Table~\vref{table:SU_1}. The table includes RF power
including inefficiencies of the klystrons and power supplies, magnet
power for the two rings, power for water distribution and cooling,
control power, injector power and the total estimated requirement.

The power required depends on the beam energies, since the beam current,
synchrotron radiation and injector energy change with the energy of
each ring. The needed power for three possible combinations of HER and LER
beam energies are shown. Within this range of configurations,
the minimum site power is about 34 MW and the
maximum is 43 MW. The minimum wall power requirement is achieved with
the design asymmetry of 4 on 7\gev.

\begin{table}[htbp]
\caption{\label{table:SU_1}
\superb\ Factory wall power requirements to achieve a luminosity of \tenTo36 
for three different beam energy configuations.}
\vspace*{2mm}
\setlength{\extrarowheight}{3pt}
\centering
\begin{tabular}{lccc}
\hline
\hline
Beam Energy (HER/LER)        (GeV)    & 7.0/4.0 & 8.0/3.5 & 9.0/3.1 \\
\hline
Beam current (HER/LER)       (A)      & 1.3/2.3 & 1.1/2.6 & 1.0/2.9 \\
HER RF power                 (MW)     &   8.6   &  12.8   & 18.2 \\
LER RF power                 (MW)     &   8.6   &   5.8   & 4.1 \\
HER magnet power             (MW)     &   4.0   &   5.2   & 6.6 \\
LER magnet power             (MW)     &   3.0   &   2.3   & 1.8 \\
Cooling system power         (MW)     &   2.4   &   2.6   & 3.1 \\
Control power                (MW)     &   0.5   &   0.5   & 0.5 \\
Injection system power       (MW)     &   4.0   &   4.6   & 5.1 \\
Lights and HVAC              (MW)     &   3.0   &   3.0   & 3.0 \\
Total Site Power             (MW)     &  34.1   &  36.8   & 42.4 \\
\hline
\end{tabular}
\end{table}

\subsection{Cooling System}

The electromagnets and RF systems require cooling water to operate
at a constant temperature. The cooling water must be pumped around the
ring with a supply line and a return line. Each subsystem will tap
into these lines. The cooling water will be chilled with cooling
towers, pumps, and heat exchangers outside the tunnel.  Approximately
80\% of the power listed in Section~\vref{sec:AC_power} must be
removed using a cooling tower. The remainder will be dissipated
external to the tunnel, mostly by the high-power high-voltage power
supplies for the RF system.

\subsection{Air Conditioning}

The water cooling system is needed to provide a steady temperature environment for
the six straight section buildings. All effort will be made to remove
excess heat from these building using the cooling water system, but the
remaining heat will removed using an air conditioning system, which will
need a capacity of about 2.5\MW to remove the power
unavoidably transferred to the building air via current-carrying cables \etc.


\label{section:Site_Util}
\clearpage


\afterpage{\clearpage}


%
\afterpage{\clearpage}

\graphicspath{{det/figures/}}
\chapter{The Detector}
\label{sec:Detector}
\section{Overview and overall design considerations}
\label{sec:det:Overview}

The \superb\ detector concept, 
is based on the current
\babar\ detector, with those modifications required to operate at a
luminosity of $10^{36}$ and with a reduced center-of mass boost.
optional configurations needed to cope with
higher beam-related backgrounds, as well as to improved detector
hermiticity are also discussed.  The necessary R\&D is to implement this upgrade is also discussed.  Cost estimates
and the schedule are described in Section~\ref{sec:CostSchedule}.

\begin{figure*}[thb]
  \begin{center}
  \includegraphics[angle=270.,width=0.9\textwidth]{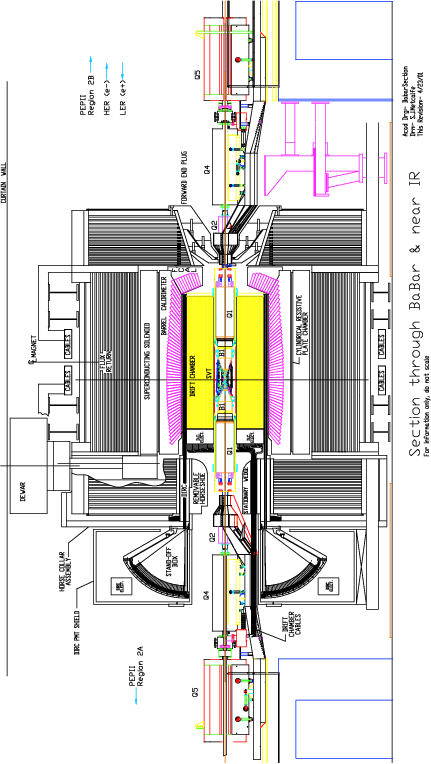}
  \vspace*{3mm}
  \caption{The current \babar\ detector.}
  \label{fig:det:babar}
  \end{center}
\end{figure*}

The current \babar\ detector is shown in
Fig.~\ref{fig:det:babar}.  \babar\ consists of a tracking system with
a 5 layer double-sided silicon strip vertex tracker (SVT) and a 40
layer drift chamber (DCH) inside a 1.5T magnetic field, a Cherenkov
detector with quartz bar radiators (DIRC), an electromagnetic
calorimeter (EMC) consisting of 6580 CsI(Tl) crystals and an
instrumented flux-return (IFR) comprised of both limited streamer tube (LST)
and resistive plate chamber (RPC) detectors for $\KL$ detection and $\mu$
identification.

The \superb\ detector concept reuses a number of components from \babar: the
flux-return steel, the superconducting coil, the barrel of the EMC and
the quartz bars of the DIRC. The flux-return will be augmented with
additional absorber to increase the number of interactions lengths for
muons to roughly $7\lambda$.  The DIRC readout will be replaced with
either faster PMTs in the current water tank or multi-channel plate
(MCP) photon detectors in a focussing configuration to reduce the
impact of beam related backgrounds and improve performance. The
forward EMC will feature cerium-doped LSO (lutetium orthosilicate) or LYSO (lutetium yttrium orthosilicate) crystals, hereafter referred to as L(Y)SO crystals, which have a much shorter
scintillation time constant, and lower Moli\'ere radius and better radiation hardness than the current
CsI(Tl) crystals, again for reduced sensitivity to beam backgrounds
and better position resolution.

\begin{figure*}[!h]
  \begin{center}
  \includegraphics[width=0.9\textwidth]{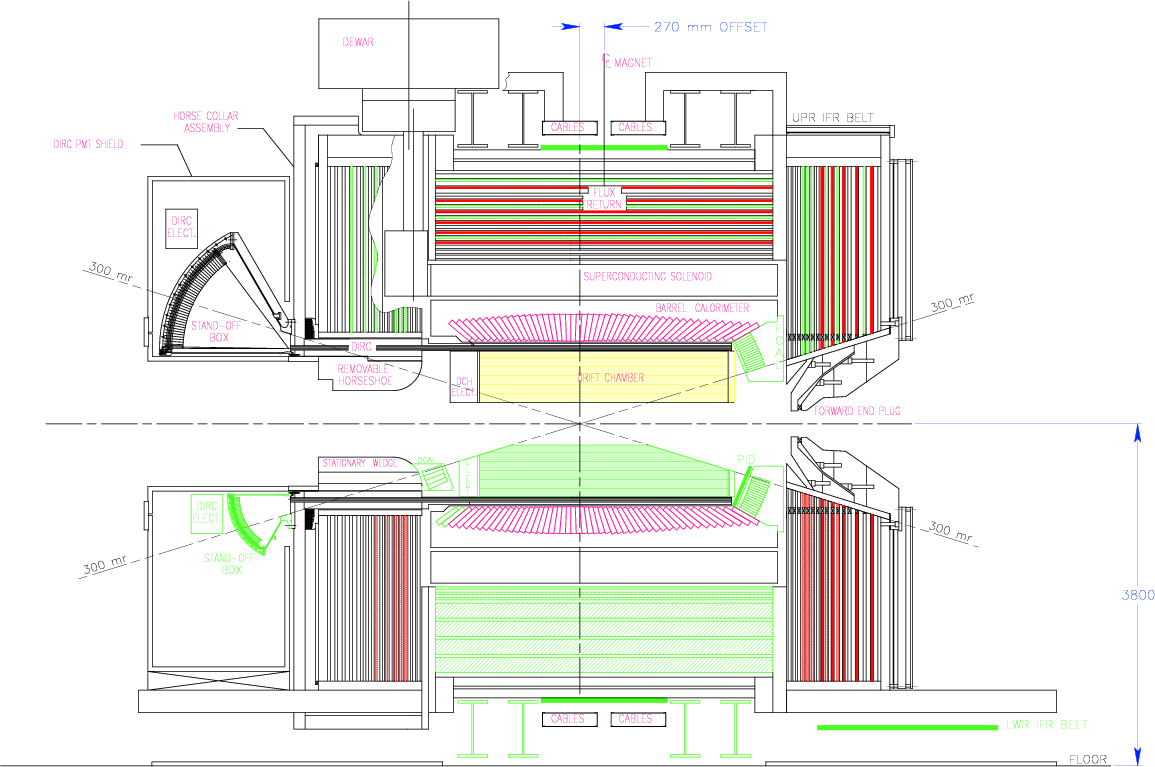}
  \vspace*{3mm}
  \caption{Concept for the \superb\ detector.  The upper half shows the
  baseline concept, and the bottom half adds a number of detector
  optional configurations.}
  \label{fig:det:superb}
  \end{center}
\end{figure*}

The tracking detectors for \superb\ will be new. The
current SVT cannot operate  at $\mathcal{L} = 10^{36}$, and the DCH
has reached the end of its design lifetime and must be replaced
at the end of \babar\ operation.  To maintain sufficient $\deltat$
resolution for time-dependent \CP violation measurements with the \superb\
boost of $\beta\gamma=0.28$, the vertex resolution will be improved
by reducing the radius of the beam pipe, placing the inner-most layer
of the SVT at a radius of roughly $1.2\cm$.  This innermost layer of
the SVT will be constructed of either silicon striplets or MAPS sensors,
depending on the estimated occupancy from beam-related backgrounds.
Likewise the cell size and geometry of the DCH will be driven by
occupancy considerations. To improve the hermeticity of the detector
\superb\ may also include a backwards EMC detector also consisting of L(Y)SO
crystals and forward and backward particle identification systems using either
a time-of-flight (TOF) or an Aerogel RICH (ARich) detector.

The \superb\ detector concept is shown in
Fig.~\ref{fig:det:superb}.  The top portion of this elevation view shows the
minimal set of new detector components, with the most reuse of
current \babar\ detector components;  the bottom half shows the
configuration of new components required to cope with higher beam backgrounds
and to achieve greater hermiticity.

\afterpage{\clearpage}

\section{Interaction Region}
\label{sec:det:IR}

\begin{figure}[htb]
\centering
\includegraphics[width=0.75\textwidth]{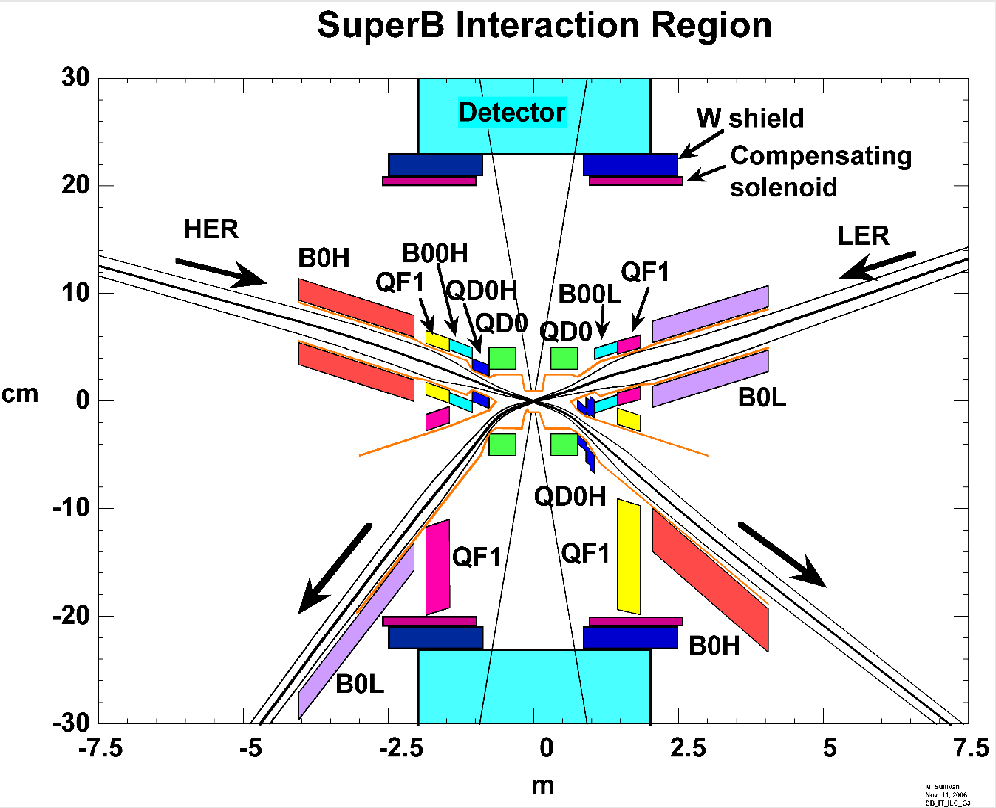}
\caption{
Layout of the interaction region. Note the asymmetric scales for the two axes.}
\label{fig:InteractionRegion}
\end{figure}

The interaction region design must satisfy the requirements imposed by the accelerator
design and the constraints determined by the detector geometry and sensitivity to
backgrounds.
The accelerator design based on small size of the beams at the interaction point
requires the vertical focusing magnets (QD0/QD0H) as close as possible to the IP
(cfr.~\vref{subsec:FF_geom}). This requirement constraints severely the detector
acceptance and stay clear; moreover the off-energy particles over-bent by the
final doublet are a major source of backgrounds in the detectors. 
Figure~\ref{fig:InteractionRegion} shows the layout of the \superb\ interaction region,
which is described in detail in Section~\ref{subsec:FF_geom}.
The first quadrupole magnet (QD0) starts $0.3\m$ away from the IP and its radial 
dimension and offset limits the detector acceptance to about 300\mrad in the forward 
and backward direction, which we assume as the baseline design acceptance.

The synchrotron radiation ``fans'' produced by the beam bending in the dipoles
and off-axis quadrupoles near the IP are kept away from the detector beam pipe
by a careful design of radiation masks and magnet mechanical apertures and position
(cfr.~\vref{subsec:sync_rad_fan}).

The off-energy particles produced by Touschek and inelastic Bhabha scattering 
are prevented from hitting the detector by a tungsten shielding enclosing the
beam lines up to $\pm 3$ meters from the IP.

The whole beam line and interaction region has been simulated with Geant4 and
the shielding design has been optimized to reduce as much as possible the
background in the detector satisfying the acceptance and stay clear constraints.

\afterpage{\clearpage}

\section{Backgrounds}
\label{sec:det:Backgrounds}

\label{sec:backgrounds}
\begin{figure}[thb]
  \centerline{
  \includegraphics[width=10 cm]{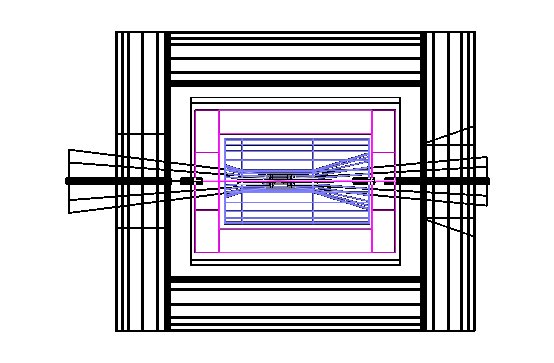}}
  \vspace*{0mm}
  \caption{Detector and beamline description in Geant4}
  \label{fig:G4Wired}
\end{figure}

Coping with machine-related backgrounds is one of the leading
challenges in designing the \superb\ detector.
Background considerations influence several aspects of the design:
readout segmentation, electronics shaping time, data transmission
rate, triggering and radiation hardness.
With the proposed collider design, the primary
sources of background are the beam-beam interaction,
radiative Bhabha production and Touschek scattering; photons from
synchrotron radiation  and lost beam particles
give smaller, though far from negligible, contributions.
These sources give rise to primary particles that can either hit
the detector directly, or generate secondary debris that
enters the apparatus. In addition, the heat load due to syncrotron
radiation photons striking masks must be
carefully evaluated.
We have simulated the different sources of background and
modeled them with detailed Geant4 detector and beamline
description to estimate their impact on the experiment.
The relevant magnetic and physical elements used in the
configuration are the two QD0 quadrupoles
surrounding the beampipe, the QD0H on the outgoing HER line,
the two QF1s and the two B0s elements on the downstream lines
(see Sec. \ref{section:InteractionRegion}).
Tungsten masks placed around the beamline protect the detector
on both sides of the interaction region. A view of the entire
interaction region and the detector generated with Geant4
is shown in Fig.~\ref{fig:G4Wired}.

\subsection{Beam-beam interaction}
The \superb\ design produces its high luminosity by employing a reduced
beam size having a high bunch charge density, as discussed in
\ref{sec:accel-overview}.
There is thus a strong beam-beam interaction and significant intrabeam scattering,
which in the \superb\ environment, are the dominant background sources, larger than
single-beam bremsstrahlung or Coulomb scattering.
The beam-beam interaction is studied by using the Guinea~Pig package
\cite{bib:GPig}. Photon emission is parameterized as an interaction
between $e^\pm$ and the collective beam-beam field, plus a component
due to the regime of photon emission in the collision of individual
particles \cite{bib:GPig2}. The $e^+e^-$ pair creation is
determined not only by low-energy electrons and positrons co-moving
with the beams and strongly deflected by the beam-beam field, but
also through the second-order QED process of pair creation during
the collision. This effect, at the \superb\ energy, is essentially a
mixture of an incoherent amplitude given by the interaction of
individual particles
\cite{bib:zolotarev} and a coherent process in which the emitted
photon interacts with the collective field of the oncoming
bunch\cite{bib:chen}. These processes are generated
in Gunea Pig and the photons and charged particles produced in the
interaction are then passed to a Geant4 simulation to model the detector response.
The expected hit rates from this source are expected to be small
in all subdetectors except the silicon tracker.

\subsection{Radiative Bhabhas}
\label{sec:radbhabha}
\begin{figure}[thb]
  \centerline{
  \includegraphics[width=6.5 cm]{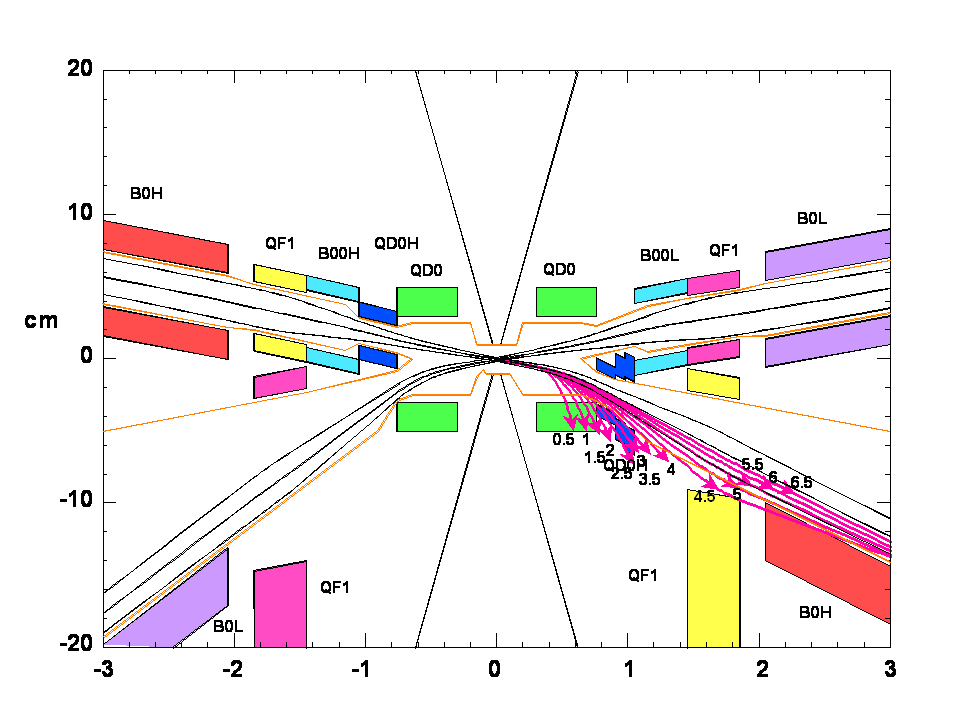} \ \ \
  \includegraphics[width=6.5 cm]{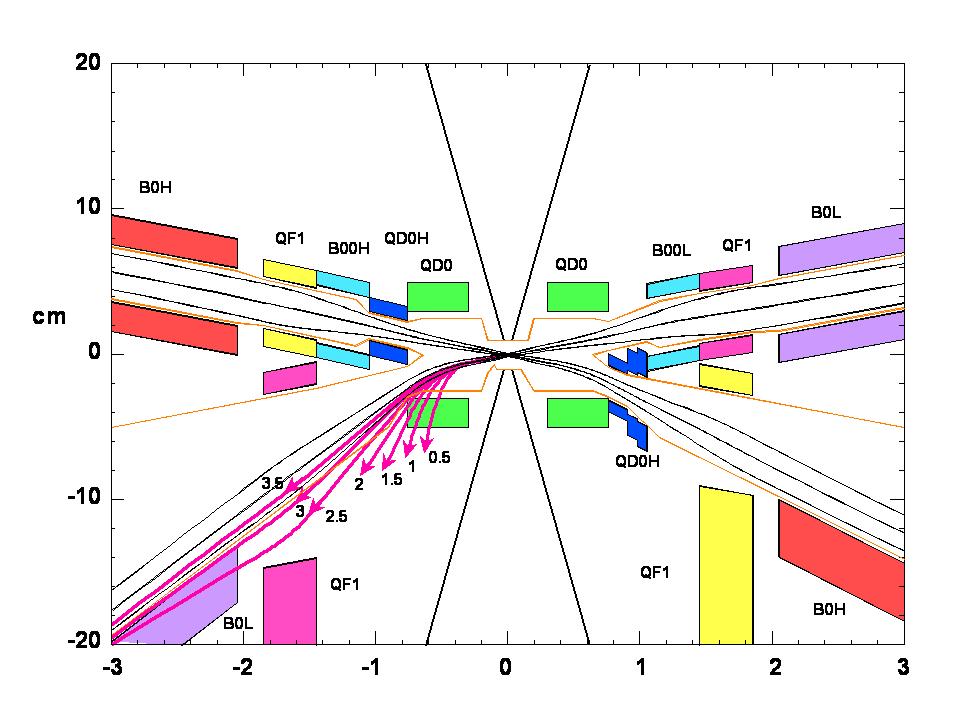}}
  \vspace*{0mm}
  \caption{Trajectories of primary electrons (left plot) and positrons
(right plot)
with different energies after the radiative Bhabha interaction}
  \label{fig:her-brem}
\end{figure}

The effect of particles scattered in radiative Bhabha processes is
studied with the BBBREM generator ~\cite{bib:bbbrem}, a Monte
Carlo program which simulates single bremsstrahlung in Bhabha
scattering, {\it i.e. }, $e^+e^- \rightarrow e^+e^-\gamma$ in the very forward
direction. An experimental cut is imposed on the energy loss of
the primary lepton, the secondaries are propagated
through the Geant4 detector simulation, and the hit rates are
studied.  Due to the dynamics of
the process, the impact of this source of background extends
to detector systems other than the tracker. As shown in Fig.~\ref{fig:her-brem},
a large fraction of off-energy electrons and
positrons hit the downstream  beamline elements,
producing electromagnetic showers. Low energy particles from
these showers enter in all detector subsystems.

\subsection{Touschek Scattering}
The Touschek scattering rate scales with the bunch charge
density (Sec.~\vref{subsubsec:accel_Touschek_tau}) hence
is expected to be way higher than in present B factories.

Simulation studies for background due to Touschek scattering
have been performed for the LER
(Sec.~\vref{subsubsec:accel_Touschek_bkg}). The expected loss rate
without collimators nor any other adjustment of the ring mechanical
apertures is $2.3\MHz$ within 4\m from the IP for a single LER
bunch ($I=1.3 \mA$.)

Touschek losses inside the detector are reduced by a factor 25
inserting three horizontal collimators far away from
the detector at $z=-117\m$, $z =-65\m$, and $z=-40\m$.

The remaining Touschek loss rate inside the detector is $90\kHz$
per single LER bunch ($156\MHz$ for 1733 LER bunches).
These particles are mainly scattered at $z\approx-31\m$ eventually
hitting the beam pipe near the IP and producing high multiplicity
electromagnetic showers.

The reduction of this dangerous source of background will be
achieved after further studies and optimizations of several
final focus parameters:
the phase advance between positions where Touschek
scattering occurs and the IP, the mechanical aperture of the
vacuum chambers upstream the IP, the insertion of additional
collimators and masks near the final doublet.

\subsection{Other sources of background}
Lost beam particles are a source of background proportional to
single beam currents.
Electrons or positrons circulating in the beam pipe lose momentum through Coulomb or bremsstrahlung interactions with
residual gas molecules. These interactions are more serious in regions of
the ring far from the interaction point, where the pressure is higher; these particles can
reach the interaction region, where they can be overbent by
the final focus elements hit the
detector.
An estimate of these lost particles backgrounds has been extrapolated
from studies made at \babar\cite{bib:backg-taskforce},
where this is one of the dominant sources of background.
For PEP-II, the rate of lost beam particles hitting the detector
was estimated to be less than 1\MHz/\cm$^2$ for currents of 1.2 on 2.8 A.
Scaling this value to the \superb\ currents
(see Sec.\ref{sec:accel-overview}),
leads to an expected rate of particle hits of the order of 1-2 MHz,
increasing to 2-5 MHz in the high luminosity regime.
The extrapolation from PEP-II is likely to be a pessimistic estimate,
since the permanent dipoles in the
final focus of PEP-II are eliminated in the \superb\ design. Nevertheless, the rate found
for \superb\ is negligible with respect to the other sources
of background.
Synchrotron radiation is another source of background proportional to single beam
currents. Despite the fact that
the interaction region has been designed to reduce as much as
possible the bending of beam trajectories in the incoming
beamline, some photons can still hit the
beampipe. This would result in additional background in the
detector and can give rise to outgassing due to
heating, thereby degrading the local
vacuum. The rate of photons above 10 keV range
hitting the pipe in the proximity of the beryllium beam pipe section
is expected to be about 1500 $\gamma$ per bunch crossing.
This is the same order of magnitude as
in PEP-II and the impact on the \superb\ detector is expected
to be negligible; a preliminary design of an adequate system of masks already exists.

\afterpage{\clearpage}

\section{Vertex Detector}
\label{sec:det:SVT}


\subsection*{Introduction}

The vertex detector provides precise information on
both the position and direction of charged particle trajectories as close as possible to the interaction point.
For very low momentum particles, the track parameters must be completely determined within the
vertex detector.
Precise vertex separation is fundamental to all time-dependent analyses,
which form the basis of the \superb\ scientific program, as is does for the existing asymmetric $B$ Factories.

Analytical calculation and Monte Carlo simulations have shown that the reduction in the precision
measurement on the \CP\ asymmetries is less than 10\% as long as the distance between the two \B\ vertices
is reconstructed with a resolution equal to half of the mean separation~\cite{ref:BaBar_LOI}.
For the boost value of PEP-II $B$ Factory ($\beta\gamma \simeq 0.55$),
the average separation along the $z$ coordinate between the vertices of
$<\Delta z>$ is  $\simeq  \beta\gamma c \tau = 250 \mum$, which implies a required precision on
 $\Delta z$ distance of the order of \hbox{$<\Delta z>/2 = 125 \mum$,} where
$\Delta z \simeq \beta \gamma c \Delta t$ and $\Delta t$ is the proper time difference between the two
\B\ decays.
This requirement has been met in the present \babar\ silicon vertex tracker (SVT) with a five layer
double-sided silicon detector~\cite{ref:NIMbabar}.
The precision of the vertex measurement also determines the ability to distinguish between signal and background.

For the proposed value for the center-of-mass boost of \superb,
$\beta\gamma = 0.28$ (a 7 \gev\ HER beam colliding with a 4 \gev LER beam), the average \B\ vertex separation
in the $z$ direction, $\langle \Delta z\rangle \simeq \beta\gamma c \tau_\B = 125 \mum$, is reduced by
nearly a factor of two with respect to the \babar\ experiment.
Vertexing performance must therefore be able to achieve a resolution on $\langle \Delta z\rangle$ of order
$60~\mum$ for optimal \CP\ time-dependent measurements.
%
%
The conceptual design of a \superb\ vertex detector,
based on the \babar\ SVT layout, with an additional innermost $\textrm{Layer~0}$, will be discussed herein.
For a vertex detector with a $\textrm{Layer~0}$  of $1.2-1.5~\cm$ radius,
fast simulation studies have shown that
 it is possible to achieve resolution  suitable for time-dependent analyses,
 as long as the total radial material before the first hit measurement is at the level of $\sim 1~\%~X_0$,
as discussed in section~\ref{sec:svt_performance}.
%
%
In the following, we will demonstrate that the performance requirements can be met with a design
based on a six layer silicon tracker detector, with a $\textrm{Layer~0}$ based on striplet or thin pixel technology
and $\textrm{Layer~1}$ to $\textrm{Layer~5}$ made of double-sided silicon microstrip sensors. $
\textrm{Layer~0}$ will be very close to, or even mounted on, the beampipe support.
The impact of machine background on the design for $\textrm{Layer~0}$ for the striplet and
the pixel option will be discussed in Sections~\ref{sec:svt_baseline} and~\ref{sec:svt_MAPS},
along with mechanical issues and electronic readout requirements.

\subsection{Detector concept}
\label{sec:concept}
\begin{figure*}[htb]
\includegraphics[width=1.0\textwidth]{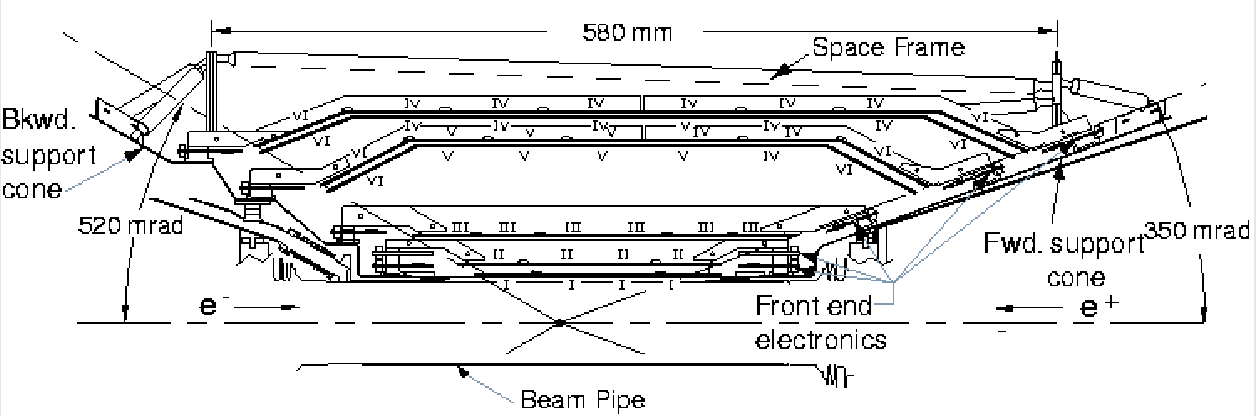}
\vspace*{1mm}
\caption{Longitudinal section of the SVT. The roman
numerals label the six different layers of sensors.}
\label{fig:svt_sideview}
\end{figure*}

%
The \superb\ interaction region configuration is based on a small beampipe radius ($\simeq 1\cm$)and new vertex detector layer ($\textrm{Layer~0}$) at a radius between $1.2-1.5~\cm$. This
allows measurement of the first hit of the tracks as close as possible to the production
vertex. There are five additional tracking layers, in a layout similar to
the \babar\ SVT system, at radii between 3 and 14 cm.
A longitudinal section of the \babar\ SVT detector, with Layer~0 added, is shown in
Fig.~\ref{fig:svt_sideview}.
The angular acceptance, constrained by the interaction region design, will be 300~mrad in both the forward and backward directions, corresponding to a solid angle coverage of $\sim$95\% in the
center-of-mass frame.
This six-layer vertex detector solution will significantly improve track parameter determination,
matching the more demanding requirements on vertex resolution, and will improve the standalone tracking
capabilities for low momentum particles as well.

Two options are viable for the $\textrm{Layer~0}$ sensors:
 striplets~\cite{bib:striplet} and CMOS monolithic active pixel sensors
(MAPS)~\cite{ref:rizzo06}-\cite{ref:bettarini}.
The choice must take into account the physics requirements on the vertex resolution, which depends on the
pitch and the amount of material of the sensors.
The intrinsic detector spatial resolution should not limit the track parameter
determination, even for the highest momentum tracks, for which multiple scattering effects are at a minimum.
For the highest momentum tracks, multiple scattering contributions to the uncertainty on the $z$ position
and $\tan \lambda$ are of the order of $10~\mum$ and 0.001, respectively.
In addition, to assure optimal performance for track reconstruction, occupancy must be maintained
under a few percent, imposing further requirements on the sensor segmentation and on the frontend
electronics.
Radiation hardness is also an important consideration, although it is not expected to be a particularly
demanding requirement compared to LHC detector specifications.
Background studies have been made to determine the hit rate in the detector region, and  particularly
in $\textrm{Layer~0}$, as discussed in Section ~\ref{sec:svt_bkg}.

Two technology solutions have considered for $\textrm{Layer~0}$ detectors. These are described in more detail in
section~\ref{sec:svt_baseline} and ~\ref{sec:svt_MAPS}.

A viable solution, which we take as the baseline,
is high resistivity silicon sensors with short strips
 (striplet detectors)~\cite{bib:striplet}.
Small standard double-sided high resistivity silicon detectors with short strip length
($ \sim 1~\cm$) having a pitch of $50~\mum$,
reduce the occupancy by geometrically reducing the area of a single channel.
This solution offers a reasonably low sensor material budget
($\simeq 200-300 \mum$ silicon thickness, $0.2-0.3~\%~X_0$) and a hit resolution of about $10~\mum$.
Detector occupancy is an important issue, as will be discussed
in sections ~\ref{sec:svt_baseline} and ~\ref{sec:svt_bkg},
but this solution does not require significant R\&D.

A second option for $\textrm{Layer~0}$ sensors is CMOS MAPS detectors,
as discussed in section ~\ref{sec:svt_MAPS},
with full in-pixel signal processing implemented at the pixel level.
In this case, the high segmentation of the detector ($\simeq 50 \times 50 \mum^2$ pixel area)
and the small amount of sensor material
($\simeq 50 \mum$ thick silicon sensor, $0.05~\%~X_0$),
provide optimal performance, both in terms of occupancy and multiple scattering.
A hit resolution of the order of $10-15 \mum$ is expected with MAPS detectors.
This solution, however, requires additional R\&D, as discussed in section~\ref{sec:svt_R&D};
important progress has already been made~\cite{ref:rizzo06}-\cite{ref:bettarini}.

\subsection{Physics Performance}
\label{sec:svt_performance}

Precise determination of the position of decay vertices is fundamental for the physics program of \superb.
Benchmark analyses sensitive to New Physics, such as $\phi \KS$, $\eta \KS$, $\eta' \KS$ require
time-dependent measurements. Improved vertexing performances will also increase our ability to separate signal events from background.
It may also be possible to consider the adoption of new tagging algorithm based on topological variables related to the separation
between the \B\ and $D$ vertices.

The approximation $\Delta z \simeq \beta\gamma c \Delta t$, where $\Delta t$ is
the proper-time difference between the \B decay vertices, still holds at the reduced boost
In order to account for the \B\ small flight length in the transverse plane ($\simeq 25~\mum$),
 we will, using fast simulation studies, evaluate the proper-time difference resolution
 by reconstructing the full 3-dimensional \B\ vertex separation.
The benchmark is to reach a resolution $\sigma(\Delta t) \simeq 0.6 \ps$,
 equivalent to that of \babar.

In order to simulate the resolution on the \B\ decay vertices and on $\Delta t$,
we have used the TRACKERR simulation progra~\cite{bib:trackerr},
 which employs analytic parametrizations to simulate the detector response.
We have reconstructed the $\B \to \pip\pim$ exclusive decay mode,
  evaluating the other \B\ decay vertex using the charged tracks of the rest of the event,
 after rejecting long-lived particles and tracks not compatible with the candidate vertex.

\begin{figure}[thb]
  \begin{center}
  \includegraphics[width=0.7\textwidth]{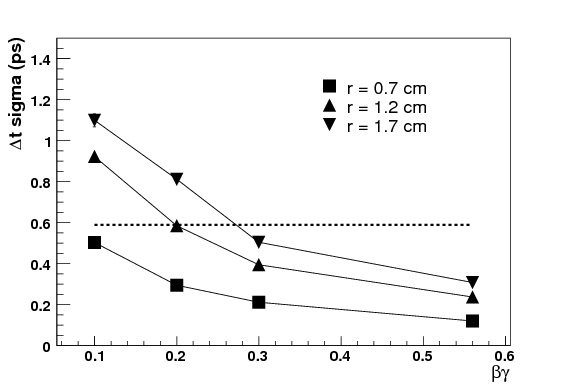}
  \caption{$\Delta t$ resolution as a function
of the $\beta\gamma$ boost value of the center-of-mass rest frame
adding a MAPS $\textrm{Layer~0}$  at different radii:
$0.7~\cm$ ($\blacksquare$), $1.2~\cm$ ($\blacktriangle$) and $1.7~\cm$ ($\blacktriangledown$).
 The resolution on the single hit ($z$ and $\phi$) was assumed to be
$5 \mum$ for $\textrm{Layer~0}$ in the $0.7~\cm$ radius configuration
and $10 \mum$ in the other cases.
The dashed line represents the \babar\ reference value according to the fast simulation.}
  \label{fig:DeltaT}
  \end{center}
\end{figure}

\begin{figure}[thb]
  \begin{center}
  \includegraphics[width=0.7\textwidth]{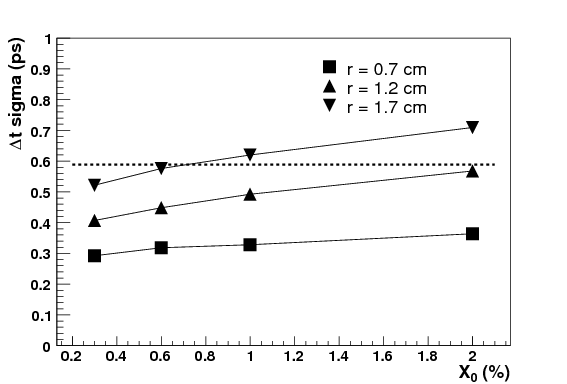}
  \caption{$\Delta t$ resolution for the nominal center-of-mass boost of $\beta\gamma=0.28$ as a function of the
 amount of radial material (in $X_0~\%$) before the first hit measurement, for different $\textrm{Layer~0}$ radii:
$0.7~\cm$ ($\blacksquare$), $1.2~\cm$ ($\blacktriangle$) and $1.7~\cm$ ($\blacktriangledown$).
The resolution on the single hit ($z$ and $\phi$) was assumed to be $5 \mum$ for the
$\textrm{Layer~0}$ in the $0.7~\cm$ radius
configuration and $10 \mum$ in the other cases.
The dashed line represents the \babar\ reference value according to the fast simulation.}
  \label{fig:DeltaT_X0}
  \end{center}
\end{figure}

Figure~\ref{fig:DeltaT} shows the resolution on $\Delta t$ for different
 $\textrm{Layer~0}$ radii, as a function of the $\beta\gamma$ value of the center-of-mass boost. The dashed line represents the \babar\ reference value ($0.6~\ps$).
We consider three $\textrm{Layer~0}$ radii: $0.7~\cm$ ($\square$),
$1.2~\cm$ ($\bigtriangleup$) and $1.7~\cm$ ($\bigtriangledown$). The amount of material of the
sensor is consistent with the MAPS solution ($\simeq 0.05~\%~X_0$),
while the beam-pipe radiation length is varied in the range $0.4-0.6~\%~X_0$ for the
three different configurations to
account for the potential variation of the required beam-pipe thickness with radius.
Figure~\ref{fig:DeltaT_X0} shows the resolution on $\Delta t$ as a function of the
amount of material before the first hit measurement in $\textrm{Layer~0}$.

These studies allow us to evaluate the feasibility of different solutions
for $\textrm{Layer~0}$,
 in terms of radial distance and the amount of material before the first hit measurement of the tracks.
For the proposed boost $\beta\gamma=0.28$, and for a $\textrm{Layer~0}$ radius of $1.2~\cm$,
the total radial material before the first hit measurement has to be kept below $2~\%~X_0$.
This constraint can be met with both MAPS and striplet detectors, as discussed in
 section~\ref{sec:svt_baseline}.

For the version usings a striplet $\textrm{Layer~0}$
 detector at $1.5~\cm$ radius, we obtain a resolution for $\Delta t$
of $0.55~\ps$.
In this case we considered a $1~\cm$ radius beampipe ($0.42\%~X_0$ radial material),
$200~\mum$  silicon wafer thickness, with $150~\mum$ equivalent silicon thickness for the three fanout and
$100~\mum$  equivalent silicon thickness for the support structure ($0.48\%~X_0$ radial material),
for a total amount of $0.90\%~X_0$ radial material, including $\textrm{Layer~0}$.

The good precision of the decay vertex determination
benefits several aspects of $B$ meson reconstruction. Reducing
the energy asymmetry without affecting the
proper time resolution enlarges the acceptance of the detector
to 95\% in the center-of-mass system, improving the reconstruction of
decay modes with neutrinos ($B \to \tau \nu$, $B \to D^{(*)} \tau
\nu$, $\tau$ decays, etc.).  The ability to separate the $B$ from the
$D$ decay vertex will help reject $q\bar{q}$ events ($q = u,
d, s$ quarks) and allow the adoption of analysis techniques for $B$ flavour
tagging based on topological algorithms~\cite{bib:topTag}.  The
resolution on the $B-D$ vertex separation in exclusively reconstructed
modes, with a $\textrm{Layer~0}$ radius of $1.2~\cm$, is
$\sim 40~\mum$.  This has to be compared with the average
separation of the $B-D$ decay vertices, which depends on the specific
reconstructed decay mode. As an example, for the decay $\B^0 \to D^-
\pi^+$, this separation is about $400~\mum$, corresponding approximately to 10 times
the resolution on the flight length distance.  An analogous study has
been made using an inclusive reconstruction technique, employing all possible
vertex combinations with tracks not associated with the
\B\ exclusively reconstructed candidate (the rest of the event).  The \B\ and
the $D$ decay vertex candidates are selected according to the most
probable two-vertex combination, based on a geometrical $\chi^{2}$
algorithm. In this case, the resolution on the flight length distance
is $\sim50~\mum$. The information on the $D$ flight length could be
used together with event shape variables in a general Fisher
discriminant function to distinguish \BB\ events from $q\bar{q}$ ($q =
u, d, s$ quarks) background events.  In addition a topological
algorithm based on the $D$ flight length information, and on a function
of the charge of the $D$ daughters could be used to tag the flavour of
the \B\ mesons.
%
%
%

\subsection{The Impact of Background on Performance}
\label{sec:svt_bkg}
\begin{figure}[thb]
  \centerline{
  \includegraphics[width=0.7\textwidth]{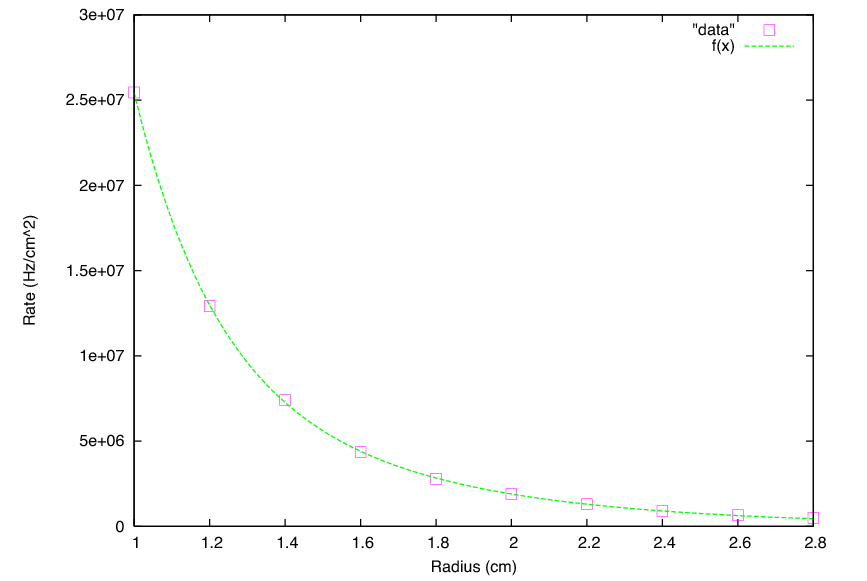}}
  \vspace*{0mm}
  \caption{Expected rate of charged tracks per $\cm^2$ from beam-beam
  background as a function of the Layer~0 radius}
  \label{fig:bkg_bb-ratevsr}
\end{figure}

\begin{figure}[thb]
  \centerline{
  \includegraphics[width=0.7\textwidth]{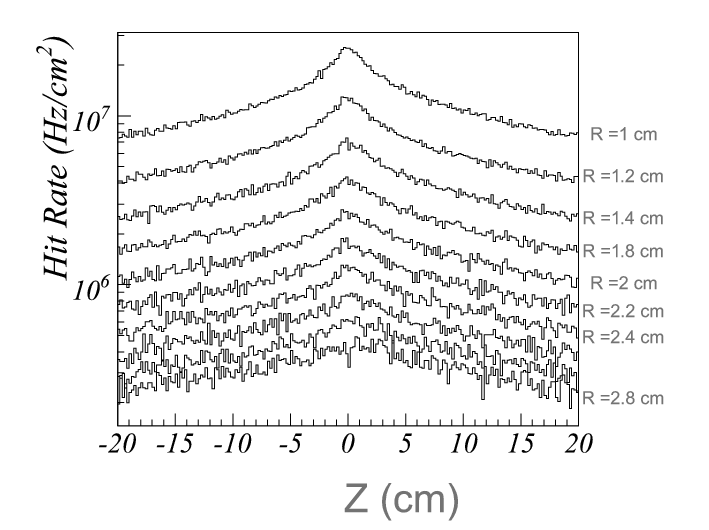}}
  \vspace*{0mm}
  \caption{Expected z distribution of hits from beam-beam
  background at different radii.}
  \label{fig:bkg_bb-z}
\end{figure}

Background considerations influence several aspects of the Silicon Vertex
Tracker design: readout segmentation, electronics shaping time,
data transmission rate and radiation hardness.
The main luminosity backgrounds considered in the simulations, as
described in Sect.\ref{sec:det:Backgrounds}, are
the beam-beam interaction and the secondary particles from radiative
Bhabha processes. Single-beam effects are mainly related to
bremsstrahlung in the quadrupole fields of the final focus elements,
and do not contribute substantially to the tracker backgrounds.
The beam-beam interaction, studied with the Guinea Pig generator
~\cite{bib:GPig}, turns out to be the most important source of background
for the Silicon Vertex Tracker. The expected hit rates in the tracker layers
have been studied, together with the azimuthal and polar dependence.
The average rate dependence as a function of the radius is shown
in Fig.~\ref{fig:bkg_bb-ratevsr}, while the dependence on $z$
for different radii is shown in Fig.~\ref{fig:bkg_bb-z}.

A typical hit rate of the order of 15MHz/$\cm^2$ is found at a 1.2\cm
radius, decreasing to about 5MHz/$\cm^2$ at a 1.5\cm radius.

Particles scattered in radiative Bhabha processes
also contribute to the background in the Silicon Vertex
Tracker. The process, as described in Sect.\ref{sec:det:Backgrounds},
is simulated with the BBBREM generator ~\cite{bib:bbbrem}.

\begin{figure}[htbp]
  \begin{center}
  \includegraphics[width=0.6\textwidth]{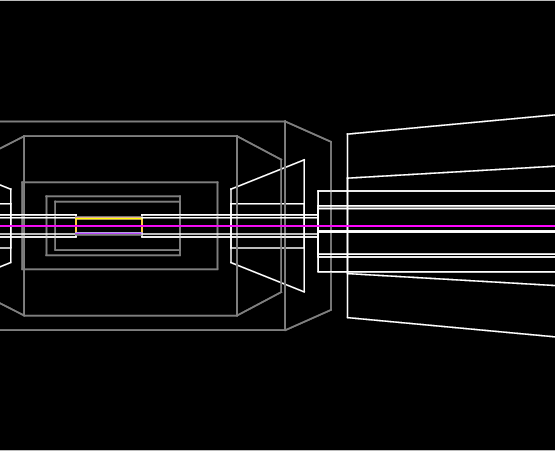}
  \vspace*{3mm}
  \caption{Detail of one the tungsten cones used to provide
  addtional shielding to SVT Layer~3.}
  \label{fig:bkg_shielding}
  \end{center}
\end{figure}

This background is mainly caused by particles hitting
the magnetic elements of the final focus in downstream regions
of the beampipe and being backscattered towards the detector.
The tungsten masks in the
IP region are of crucial importance for shielding this background source.
The geometry used in our studies meets the collider stayclear requirements, and
provides adequate protection for the Silicon Vertex Tracker. In the initial version of the shielding
geometry, the SVT layer which was actually least protected was the
Layer~3, which is at a radial position close to the inner
radius of the tungsten cones, see Fig.~\ref{fig:bkg_shielding}.
For this reason, we have provided additional tungsten discs at the
sides of the SVT for increased protection, as can be seen in
the fiure.

\begin{figure}[thb]
  \centerline{
  \includegraphics[width=7.0 cm]{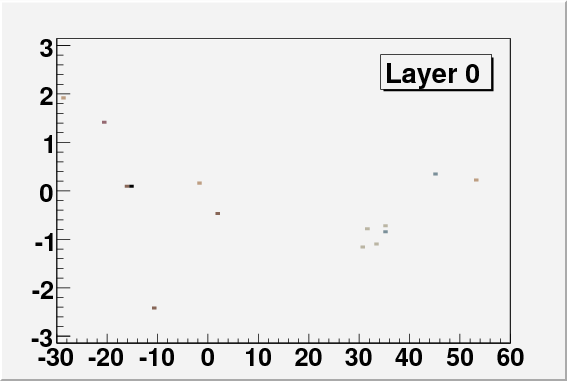} \ \ \
  \includegraphics[width=7.0 cm]{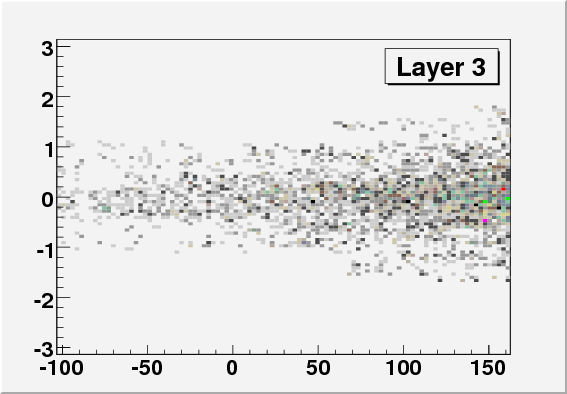}
}
  \vspace*{3mm}
  \caption{Expected hit rate distribution due to radiative Bhabha
  background process in Layer~0 (left plot) and Layer~3
  (right plot) of the
  Silicon Vertex Tracker.}
  \label{fig:bkg_rb-thetaphi}
\end{figure}

The azimuthal and polar dependence of the hits in Layer~0 and Layer~3
are shown in Fig.~\ref{fig:bkg_rb-thetaphi}.

A typical hit rate is of the order of 0.1 MHz/$\cm^2$ at 1.2 cm radius
and of 0.16 MHz/$\cm^2$ at 1.5 cm radius.

With these background rates, the electromagnetic component of the
expected integrated radiation dose produces peaks of the order of 6 Mrad/year,
for a total peak dose of about 30 Mrad over the experiment lifetime.


The decrease in the tracking performance with increased
occupancy is heavily dependent on the reconstruction algorithms employed.
Studies done for \babar~\cite{bib:SVT-LTTF},
provide figures of merit
for the expected changes in the tracking properties for benchmark
decay modes. Data taken during bad machine vacuum condition show a
significant deterioration of the tracking efficiency
when channel occupancy exceeds 10 to 15\%. With the
occupancy expected in \superb, even after the inclusion of the safety
factor of five, the reduction in performance due to the background occupancy
is not expected to be a serious issue.

\subsection{Layer~0 baseline design: Striplets}
%
\label{sec:svt_baseline}

The physics requirements imposed on the SVT Layer~0 design
(radius $\sim$1.5 cm, hit resolution of $\sim 10 \mum$,
reduced material budget $1~\%~X_0$) could, in principle, easily be met
using
double-sided silicon strips detectors (DSSD), $200 \mum$ thick, with
$50 \mum$ readout pitch and the provision of pulse height information to
improve the spatial resolution.

\begin{figure}[htbp]
  \begin{center}
  \includegraphics[width=0.6\textwidth]{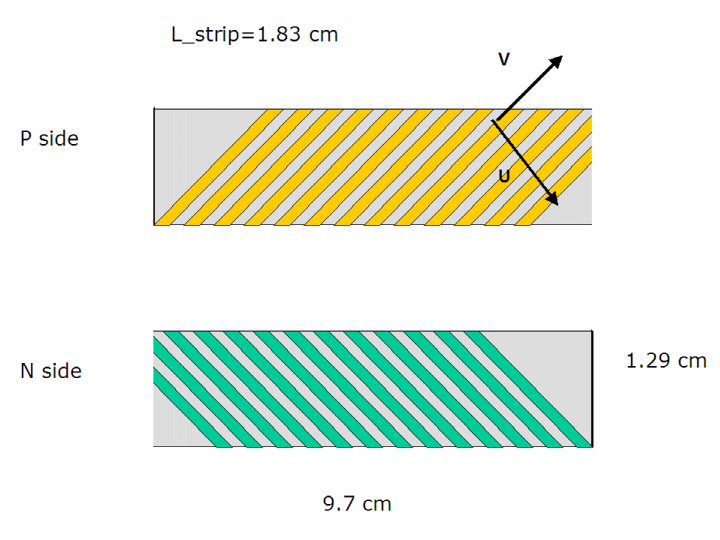}
  \caption{Schematic design of a 45 degree striplet sensor.}
  \label{fig:striplets}
  \end{center}
\end{figure}

Machine background at this small radius does, however, impose
a severe limit on area of each strip, if we are to maintain the occupancy below a few
percent to preserve the efficiency of track reconstruction.
The design objective can be met by using short strips, striplets, placed at an angle of 45 degrees with
respect to the detector edges. This is shown schematically in
Fig.~\ref{fig:striplets}; the strips on the two sides of the sensor
are orthogonal and are the same length.

Very high strip segmentation can, in principle, be achieved
by reducing the dimension of the short side of the sensor. This, however,
would increase the number of readout channels, as well as the
complexity of the module assembly and the amount of material in the active area.
A flexible printed circuit, glued on top of each silicon detector,
is required to bring the strip signals outside the active region
to the frontend electronics. Given the poor aspect ratio in the
Layer~0 SVT modules and the large number of channels, and the minimum achievable pitch for flex circuits
(assumed conservatively to be around $50 \mum$), the design must contend with the potentially large number of
layers of flex circuit required for each detector.

The proposed Layer~0 design meets all these criteria, as well as the constraints of adequate radiation hardness.
The baseline design is octagonal, with
the eight modules organized into a barrel-type structure,
placed at $r=1.5 cm$ from the interaction point.
The detector consists of double-sided silicon strip sensors,
$200 \mum$ thick, with $50 \mum$ readout pitch.
The readout electrodes are striplets placed at 45 degrees with
respect to the detector edges, and are  orthogonal on the two sides of the sensor
(junction and ohmic side).

Each module consists of:
\begin{itemize}
\item one silicon sensor,
\item a multilayer printed flex circuit, used
to bring the signals to the readout electronics located outside the fiducial region,
\item  two double-sided hybrid circuits, containing the frontend chips,
independently reading the two halves (forward/backward) of the silicon sensor.
\end{itemize}

\begin{figure}[htbp]
  \begin{center}
  \includegraphics[width=0.8\textwidth]{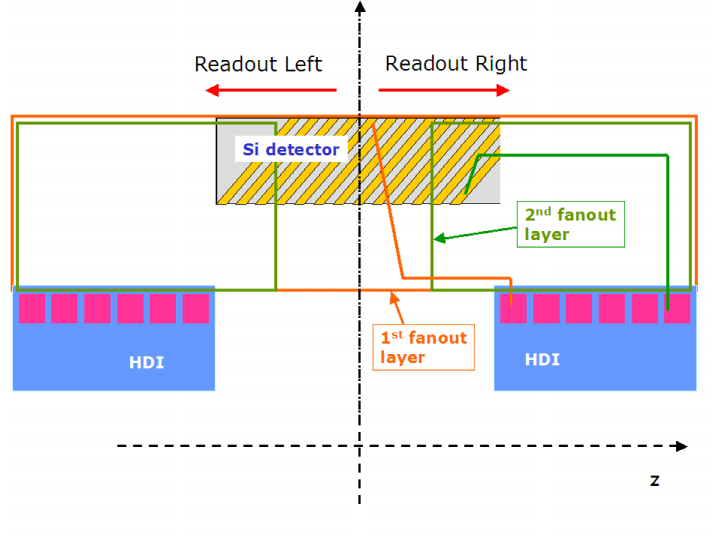}
  \caption{Schematic view of a Layer~0 striplet module.}
  \label{fig:striplets_module}
  \end{center}
\end{figure}

A Layer~0 module is shown
schematically in Fig.~\ref{fig:striplets_module}.

\subsubsection{Silicon Sensor}

The silicon detector fabrication technology will be similar to that
used for the present \babar\ SVT sensors, which have proved to be sufficiently
radiation hard~\cite{svt:bettarini:2005}
for the dose expected in the \superb\ environment.
%
The strips are biased from a ring surrounding the active region with
poly-silicon resistors,
and are AC-coupled to the readout electronics by means of capacitors
integrated on the sensor.
The silicon sensor active area will be $1.29 \times 9.7 \cm^2$, providing
an overlap region (~4\% of the sensor area) between adjacent modules,
which is useful for alignment of the sensors with reconstructed tracks.

\subsubsection{Connection of silicon sensors to readout electronics}

In the 45 degree striplet design with the chosen readout pitch,
each DSSD will have $\sim$3050 readout channels.
Each of the two double-sided hybrid circuits (Fig.~\ref{fig:striplets_module})
will mount six frontend chips per side, having 128 channels each.
To connect the silicon strips to the readout chips, a multilayer
flexible circuit will be glued to
the sensor and microbonded to the detector strips.
To increase the aspect ratio on the flex circuit and
reduce the number of flex layers/module needed, the printed circuit short side
is about two times larger than the short side of the silicon sensor
(Fig.~\ref{fig:striplets_module}).
This configuration can be realized using a pinwheel
assembly of the modules on the
support structure, as explained below.
With this lateral extension available for the trace routing (the flex circuit
is about 2.6 cm wide), with about 770 redout channels/side,
and a trace pitch on the flex of about $50 \mum$,
only 1.5 flex layers/module/side are required.
The flex circuit contribution to the total material budget(3 layers/module)
will be about $0.14 \% X_0$, assuming the use of the technology adopted for the
present SVT, based on a Upilex substrate with copper traces.
A different technology (which has been adopted for the ALICE silicon strip detectors)
is aslo being explored. It would
reduce the material budget and simplify the module assembly.
Kapton/aluminum microcables and Tape Automated
Bonding soldering techniques (ref) have been used in the ALICE design
for the connection between the silicon sensor and the readout electronics.
If this technology is available at a $50 \mum$ readout pitch, the material
budget for the connections to the frontend could be reduced by more than a
factor 4 ( to ~$0.03 \% X_0$).
This technology also simplifies the design of the multilayer connections, thereby simplifying module assembly,
since there are no fragile
wirebonds present on the surface of the module.

\subsubsection{Readout chip}

The choice of the frontend chip for the Layer~0 striplet detector is driven
by the expected machine background hit rate.
Simulation results presented in section \ref{sec:det:Backgrounds} indicate that at the Layer~0
location at a radius of 1.5\cm,
one can expect a peak hit rate of about $5\MHz/\cm^2$.
The design of Layer~0 and the choice of the frontend chip have been
optimized to safely handle a background rate
of $50\MHz/\cm^2$ (which will be used in the following discussion),
considering on average 2 strips/hit and increasing the expected background hit rate
by a factor five as a safety margin.
With this background figure, and a strip area of $50 \mum \times 1.83 \cm$,
the expected hit rate per strip will be about 450 kHz.

The present SVT frontend, the Atom chip, cannot sustain this
rate: with a readout window of $1 \mu s$, presently adopted the \babar\ SVT,
the Layer~0 strip occupancy would be about 45\% .
Even assuming a reduction of a factor two in the readout window, that is,
operating the chip with 100 ns shaping time instead of the
current 200 ns, this would ot provide acceptable occupancy.
Studies performed on SVT data during noisy machine conditions has
confirmed that there is a significant deterioration of reconstruction efficiency
above 15\% occupancy.

A different approach in the frontend is thus required to handle this
level of background. We propose to adopt the FSSR2 readout chip,
designed and optimized for the Forward Silicon Tracker of the BTeV experiment.
This chip implements a fast data driven readout architecture,
with no local pipeline, but enough bandwidth to ensure that no data are lost
due to readout deadtime.

The FSSR2 chip, described in detail in~\cite{svt:re:2006}
is a good match to the
Layer~0 striplet design. It has 128 analog channels, with a sparsified
digital output, and with address, timestamp and pulse height information for all
hits. It has a selectable shaper peaking time (65 ns is possible).
The chip has been realized in a $0.25 \mum$ CMOS technology for high radiation
tolerance.
The readout architecture has been designed to operate with 132 ns
machine bunch crossing, using a BCO clock with the same frequency,
that will define the timestamp granularity and the readout window.
A faster readout clock (70 MHz) is used in the chip, with a token pass logic,
to scan for the presence of
hits in the digital section, and to transmit them off-chip,
using a selectable number of output data lines.
With six output lines, the chip can achieve an output data rate of 840 Mb/s.

During the design of the FSSR2 the efficiency of the architecture
was investigated with detailed Verilog simulation~\cite{svt:hoff:btev}.
In BTeV, with nominal machine operation (132 ns bunch crossing and
2 interaction/bunch), an FSSR2 chip occupancy at the level of 2\% was expected,
with a 132 ns readout window.
Operated with a nominal BCO clock of 132 ns, the FSSR2 chip could
handle the nominal 2\% occupancy with an efficiency $> 99 \%$.
Even with higher occupancy (6 \% has been investigated), the FSSR2 could
read data with an efficiency above 98.5 \% , by operating the chip with
a BCO clock  of four times the nominal bunch crossing frequency.

In \superb\ the FSSR2 chip for Layer~0 striplets, would have an
occupancy of 6\%, including the five times safety factor,
using the nominal 132 ns readout window.
The minimum efficiency figure ($> 98.5\%$)
indicates the performance of the FSSR2 chip is adequate for our target application.

\subsubsection{Signal-to-noise performance with FSSR2 chip}

Noise performance of the FSSR2 has been carefully
measured~\cite{svt:re:2006}
using the second release of the chip.
Those results have been used to calculate the signal-to-noise ratio
for the baseline design of Layer~0.

The total capacitance load to the preamplifier input has been evaluated,
including the interstrip capacitance and the capacitance to
the back plane. The first term scales approximately with the ratio of the
width of the strip over the pitch, while the second term depends on the ratio
of the pitch over the thickness of the sensor.
Parameters for the calculation have been extracted from the design of those
existing SVT sensors that have the same pitch ($50 \mum$).
The total detector capacitance for the short Layer~0 striplets is $\sim$4 pF.
A similar contribution to the preamplifier load capacitance, about 5 pF,
comes from the traces on the flex circuit (both interstrip and back plane
contributions have been included).
\begin{figure}[htbp]
  \begin{center}
  \includegraphics[width=0.95\textwidth]{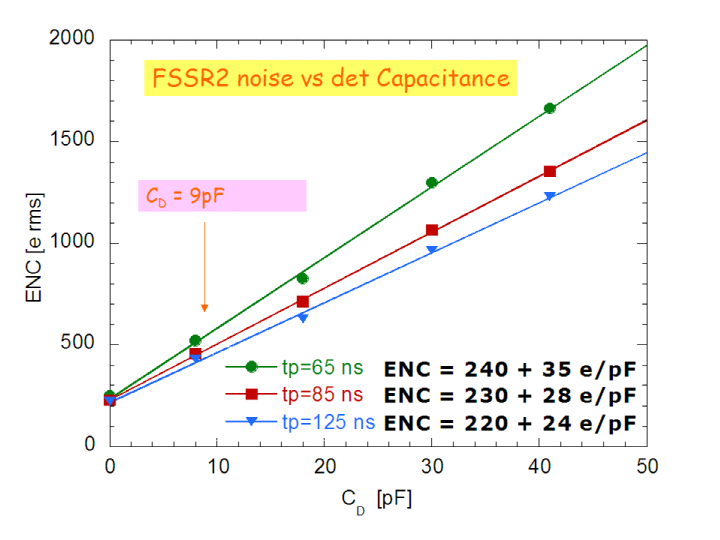}
  \caption{FSSR2 noise performance as a function of the load capacitance.}
  \label{fig:noise_fssr2}
  \end{center}
\end{figure}

With a total load of about 9 pF the Equivalent Noise Charge (ENC) measured in
the FSSR2 chip is about 500  $e^-$ $rms$, as shown in Fig.~\ref{fig:noise_fssr2}.
adding in the noise contribution from the $55\Ohm$ series resistance of
the strip, which includes the sensor and the flex circuit strips,
a total ENC of about 600 $e^-$ $rms$ is expected.
For a $200 \mum$ silicon thickness, the signal-to-noise ratio for MIPs
will be about 26. Even taking into account
the threshold dispersion measured for FSSR2 (about 300 $e^-$ $rms$),
this S/N figure will provide good performance.

\subsubsection{Radiation damage}

The background particle fluence at the 1.5 cm radius of the Layer~0 silicon sensor
will be about $2.5\times 10^{14}$ particle/cm$^2$/yr
includig the safety factor,
mainly due to electrons and positrons in the MeV energy range, corresponding
to a dose of about 6.5 Mrad/yr.

Considering that the non-ionizing energy loss (NIEL) for electrons in the
MeV range
is 40 times less than for 1 MeV neutrons~\cite{svt:lindstrom:2001}, the \superb\
silicon sensors will receive an equivalent fluence of about
$6\times 10^{12}$ $n_{eq}$/cm$^2$/yr, well below the limits explored for high
resistivity silicon sensor in the LHC experiments.
The expected radiation damage will cause an increase in the depletion voltage,
as well as S/N deterioration due to increased leakage current and a
reduction in charge collection efficiency.

The overall radiation damage
has been evaluated with the NIEL scaling hypothesis~\cite{svt:lindstrom:2001},
which is likely to be a conservative assumption for electrons in the MeV
energy range,
considering a total equivalent fluence of $3\times 10^{13}$
$n_{eq}$/cm$^2$
over the a five year experiment lifetime.

The expected change in the sensor depletion voltage would be marginal.
The increase in strip leakage current
($\Delta I_{leak}=\alpha \cdot fluence \cdot Volume $)
would be about 600 nA, calculated
assuming conservatively the current-related damage rate
$\alpha = 8\times 10^{-17} A/cm$, valid for 1 MeV neutrons.
Recent experimental results~\cite{svt:bosisio:tns2003} measured $\alpha$ of the same order of magnitude, for
high energy electrons (900 MeV), while older data for electrons of
1.8 MeV indicate much lower values for $\alpha$.
With this increase in the leakage current the corresponding noise,
after five years of operation,
will increase up to about 960 $e^-$ $rms$ (65 ns peaking time).
A charge collection efficiency drop of about 10\% would then be
expected~\cite{svt:bettarini:2005}.
The overall Layer~0 performance, after five years of operation, would
be still an acceptable S/N of about 15.

\subsubsection{Support structure}
\label{supporting structure}
\begin{figure}[htbp]
  \begin{center}
  \includegraphics[width=0.82\textwidth]{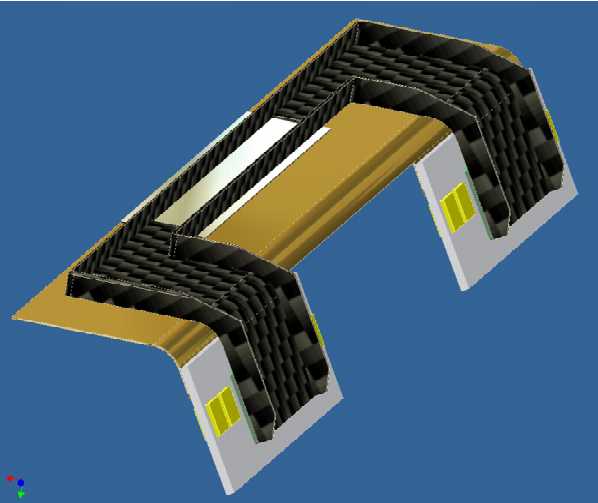}
  \caption{Design of a module of Layer~0.}
  \label{fig:L0 module 3D}
  \end{center}
\end{figure}

\begin{figure}[htbp]
  \begin{center}
  \includegraphics[width=0.85\textwidth]{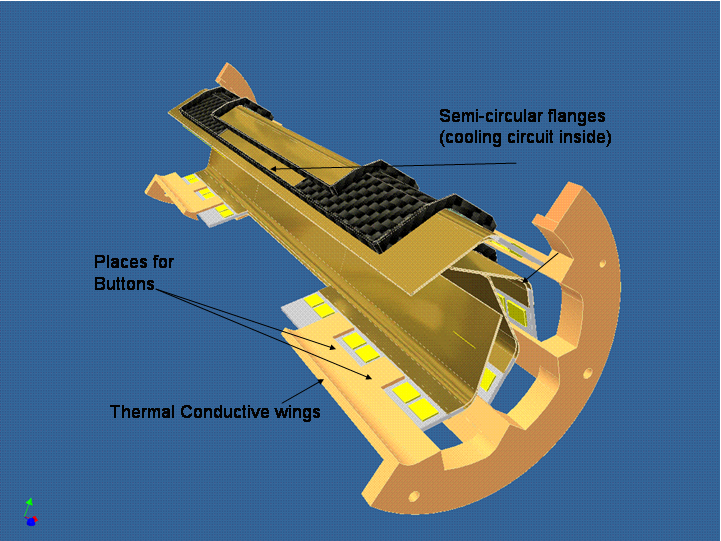}
  \vspace*{2mm}
  \caption{Positioning of a module of Layer~0 on a semi circular flange.}
  \label{fig:L0 module on a semi-flange}
  \end{center}
\end{figure}

\begin{figure}[htbp]
  \begin{center}
  \includegraphics[width=0.65\textwidth,angle=90]{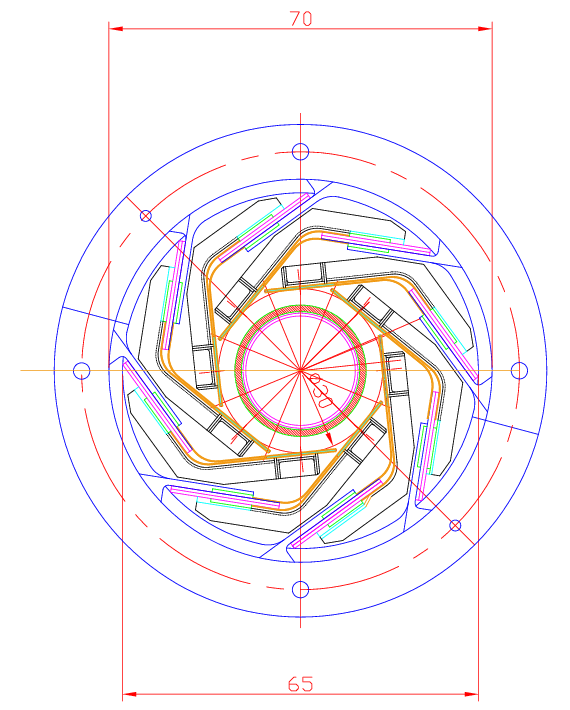}
  \caption{r-$\phi$ view of the layout of Layer~0.}
  \label{fig:L0}
  \end{center}
\end{figure}

\begin{figure}[htbp]
  \begin{center}
  \includegraphics[width=0.65\textwidth,angle=90]{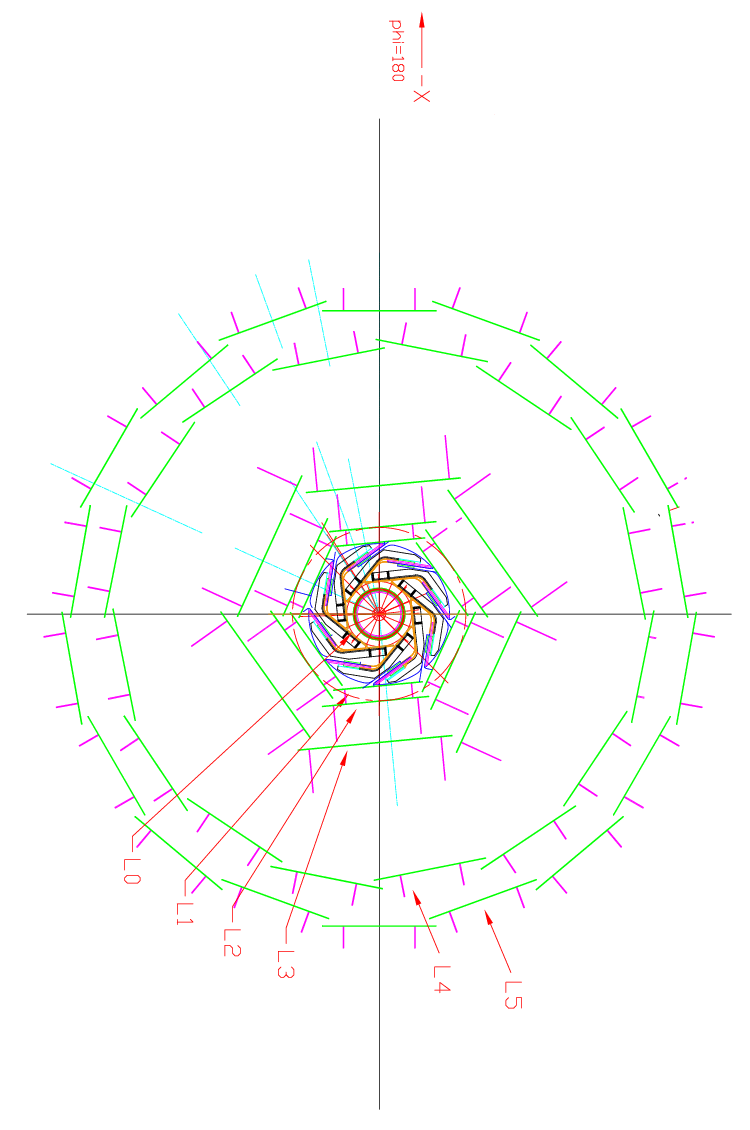}
  \caption{Layer~0 placed inside the five layer \babar\ SVT (r-$\phi$ cross section).}
  \label{fig:L0L5svt}
  \end{center}
\end{figure}

The mechanical details related to the Layer~0 modules, together with the procedures for the module assembling, up to
the final mounting on the flanges on the beam pipe have been worked out in some detail.
A 3D drawing of a module is shown in Fig.~\ref{fig:L0 module 3D}.
Each Layer~0 module,
having a lateral width of 54.5\mm and a length of 28.4\mm, is first assembled and microbonded while flat, using a planar chuck.
Both hybrids are then rotated around the border of the fanout extensions using a bending mask.
This rotation is necessary to allow the entire module to be positioned inside the radius of Layer~1 of the
SVT (see Fig.~\ref{fig:L0}).
Laminated carbon-kevlar ribs with carbon fiber end-pieces are glued to fix the relative position
of the hybrids and the sensor and to provide the needed rigidity.
The Layer~0 cooling circuit is placed inside the annular region of the Layer~0 flanges and
the hybrids are then mechanically and thermally
coupled to the wings of the flanges by two buttons.
Four modules are first mounted on each semi-circular flange (see Fig.~\ref{fig:L0 module on a semi-flange}), and
the two halves are coupled to wrap the Layer~0 structure over the beampipe cylinder.
Figure~\ref{fig:L0L5svt} reports the $r-\phi$ cross section of the whole SVT, with the Layer~0 positioned
inside the current \babar\ SVT.

\subsubsection{Material budget}

For the baseline design with striplets, the Layer~0 material
budget will be about $0.46 \% X_0$ for perpendicular tracks, assuming a
silicon sensor thickness of $200 \mum$, a light module
support structure (~$100 \mum$ Silicon equivalent),
similar to the one used for the \babar\ SVT modules, and the multilayer flex
contribution (3 flex layers/module, ~$45 \mum$ Silicon equivalent/layer).
A reduction in the material budget to about $0.35 \% X_0$ is possible
if kapton/aluminum microcable technology can be employed with a trace
pitch of about $50 \mum$.

\subsection{Layer 0 CMOS MAPS Option}
\label{sec:svt_MAPS}

The optimal technical choice for Layer~0 would be the
a pixel sensor. With a $50\times50 \mum ^2$ pixel area, the segmentation
needed to reach the target vertex resolution, the occupancy from machine
background is no longer an issue.
The expected Layer~0 pixel occupancy
is about 0.1\% per pixel in $1 \mu s$ time window with
the five times safety factor included.

Pixel technologies adopted in previous experiment are not adequate for this
application: hybrid pixels are too thick, while charge
coupled devices are too slow and are not sufficiently radiation hard.

New CMOS Monolithic Active Pixel Sensors (MAPS) are a
promising candidate for Layer~0: they incorporate the readout electronics and a very thin sensor
on the
same substrate, thereby reducing the detector material budget to
$\simeq 0.05~\%~X_0$ assuming a $50 \mum$ thick silicon chip.
The MAPS device uses an n-well/p-epitaxial diode to collect, through thermal
diffusion, the charge generated by a particle passing through the thin
epitaxial layer underneath the readout electronics.

CMOS MAPS matrices have been developed by several groups over the last few years.
These designs follow the very simple readout scheme already adopted for imaging
applications, based on the use of three transistors on the pixel cell (3T), with
a sequential readout.
Although these prototypes have shown excellent tracking performance,
their readout speed, limited by  the sequential processing, is a
major limitation for applications in environments having high
data throughput: a large area detector (1 Mpixel) might reach a frame
readout rate of about 1kHz, much smaller than required in our application.

A different approach to the design of high readout speed MAPS has recently been
proposed~\cite{ref:rizzo06}.
By exploiting the triple well option available in the CMOS commercial process,
a full signal processing chain
(charge preamplifier, shaper, discriminator and some elementary logic
functionality) has been implemented at the pixel level,
creating a monolithic pixel with a readout scheme easily compatible with data
sparsification.

Several prototype chips (APSEL series)
have been realized in the STMicroelectronics
0.13 $\mu$m triple well technology, including single pixel cells and
a small pixel matrix with a simple sequential readout.
The results of the tests performed that proved the new approach proposed
works as expected, and gave very encouraging results.
A single pixel signal of about 1250 $e^-$ has been measured for MIPs from
a radiative source. With further optimization of the frontend, a single
pixel noise of about 50 $e^-$ $rms$ has been achieved, giving a signal-to-noise ratio
of about 25 for MIPs.

Based on this new MAPS design, a dedicated readout architecture to
perform on-chip data sparsification is currently under development, to
incorporate in the same detector the advantages of the
thin CMOS sensors and functionalities similar to those in hybrid pixels.
In particular, a first prototype chip with a small pixel matrix and the first
block of a data
driven architecture was submitted in November 2006.
The readout of the final MAPS chip will be similar to that developed for
the FSSR2 chip that will be used for the SVT outer layers, to ensure
homogeneity in the peripheral electronics.

\subsubsection{MAPS module design}

A Layer~0 design based on MAPS sensor has been realized using, as in the
baseline striplet option, an octagonal module structure.
Each module will be composed of several MAPS chips
glued onto a support structure, providing the required the mechanical stability and hosting the
metal traces that connect the power, command and data lines to the
two hybrid circuits mounted at each end of the module, outside the fiducial
region.
For each MAPS chip some of the readout electronics will be located at
the chip periphery, outside the active part of the sensor: to fill in the
cracks
between the chips (in $z$ and $r-\phi$) a double layer of MAPS is used.
The two MAPS layers will be placed
on the same mechanical support, forming a module.
Each side of the module consist of 8 chips, each comprising $256 \times 256 $ pixels,
$50 \times 50 \mum^2$ pitch.

Power dissipation is one of the main issues for the MAPS module design:
since the sensor and electronics are integrated on the same chip,
a considerable amount of heat must be dissipated in the active area, while
keeping the material of the cooling system to a minimum.
In the present version of the APSEL chip, the power consumption
($60 \mu $W/pixel) is dominated by the analog part of the frontend, although
$10 \mu $W/pixel is easily achievable with minor modifications.
A new design of the analog part of the frontend has been explored, tests
aimed at reducing the power consumption to about
$5 \mu $W/pixel
are under way.
Different designs for the mechanical support of the
Layer~0 modules with MAPS are under study, depending on the minimum power
consumption achievable.

With a power budget below about $2 \mu $W/pixel,
operation without cooling in the active region would be possible,
using the module itself as thermal bridge to conduct the heat from the
active area to the ends of the module. Each
module will be mechanically and thermally coupled to the water-cooled end
flanges located outside the fiducial region.
For this option, with a Layer~0 material budget of about
$0.28 \% X_0$, including two MAPS layers (50 $\mu m$ Si each)
and the module support structure made of BeO (600 $\mu m$ thick),
a FEA simulation indicates a maximum operating temperature
of about 30$\deg$ C.

Above $2 \mu $W/pixel, water cooling in the active region is necessary.
In this case one solution investigated is to include in the module
support structure a water-filled microchannel.
Results from FEA simulation indicate a maximum operation temperature of a
few degrees celsius above the water temperature. For this option the total
material budget for Layer~0 would be about $0.41 \% X_0$, including 2 MAPS
layers (50 $\mu m$ Si each) and the the support structure made of AlN
(680 $\mu m$ thick, incorporating the water microchannel).

In the high power dissipation scenario, an alternative solution, currently
under study, is to use the external cylinder of the beam pipe both as a
mechanical support and as cooling source to evacuate the Layer~0 heat.
The beam pipe design already foresees a cooling system capable of
dissipating about 1 kW.
This approach could further reduce the total material budget for the
Layer~0 design with MAPS.

\subsubsection{MAPS radiation tolerance}

The radiation hardness of the MAPS sensors is an important issue that requires
further investigation, though preliminary tests~\cite{svt:winter:2006}
indicate that this technology can be applied, with
modest performance deterioration, in the \superb\ environment.
The triple well MAPS sensor (APSEL) is expected to be
even more radiation-tolerant than the standard MAPS
design~\cite{ref:rizzo06}, though the APSEL MAPS radiation resistance still
remains to be investigated.

The readout electronics for CMOS MAPS,
realized with modern deep submicron technology, and special
layout rules, can withstand the expected radiation levels.
The signal-to-noise performance of the
device could deteriorate in two ways due to radiation
damage to the active sensor:
a reduction in charge collection efficiency due to trapping (bulk damage,
from non ionizing radiation effects),
and an increase in the leakage current of the collecting diode
(surface damage, due to ionizing radiation),
causing a higher shot noise contribution.

Both effects have been partially investigated with irradiation tests
on several standard MAPS prototypes (3T readout).
First results from irradiation with neutrons and
protons~\cite{svt:deveaux:2003,svt:faps:2005} ,
indicates that fluences of  $\sim 10^{12}$
n$_{eq}$/\cm$^2$ can be tolerated, with only a modest (about 5\%) reduction in the collected signal,
Higher fluences might be allowable, by operating
the detector at low temperature to compensate the signal reduction
with a lower shot noise contribution.

Preliminary studies on ionizing radiation effects~\cite{svt:barbero:2005}
indicate that the noise increase for a standard pixel design could be
kept under control up
to a dose of 20 Mrad from a$^{60}$Co source.
The key requirements are to operate
the detector with a short integration time ($<$ 100 $\mu$s), or at low
temperature ($<$0$\deg$C), noting the integration time
dependence of the shot noise:
$V^2_n(t_{int})=qI_{leak}t_{int}/C^2_D$.

An alternative approach to reduce the ionizing radiation effects is to modify
the pixel design to avoid placing a thick oxide layer close to the
n-well/ p-epi junction. The leakage current contribution for irradiated
sensors is dominated by surface defects present at the interface between the
poor quality thick oxide and the silicon. The first structure realized with
a hardened pixel design has shown encouraging results~\cite{svt:winter:2006}.

\subsection{R\&D issues}
\label{sec:svt_R&D}

The technology for the Layer~0 baseline striplet design is well-established.
The multilayer flexible circuit, to connect the sensor to the frontend, may
benefit from some R\&D to reduce the material budget: either
reduce the minimum pitch on the Upilex circuit, or adopt
kapton/aluminum microcables and Tape Automated Bonding soldering techniques
with a $50 \mu m$ pitch.

The FSSR2 chip, proposed for the readout of the striplets and the outer layer
strip sensors, would haveto be produced with some minor modifications.

The CMOS MAPS technology is very promising for an alternative design of
the Layer~0, but extensive R\&D is still needed to meet all the requirements.
Key aspects to be addressed are: the readout speed, power consumption, radiation tolerance and
the development of a thin mechanical support structure to allow us to realize the benefits of the very thin
MAPS sensor. A detailed R\&D program will be pursued within the SLIM
collaboration (Silicon detectors with Low Interaction
with Material), funded by the Istituto Nazionale di Fisica Nucleare and the
Italian Ministry for Education, University and Research.
Among the final goals of this R\&D
project is the development of a MAPS matrix device, with sparsified readout
and timestamp information, suitable for use in a trigger system
based on associative memories {\bf(ref SLIM5 project)}.
A test beam run is foreseen for 2008 with a small prototype
tracker demonstrator consisting of a few planes of thin striplet
sensors read out by FSSR2 chips, and two planes of MAPS matrices.


\afterpage{\clearpage}

\section{Drift Chamber}
\label{sec:det:DCH}

\subsection{Introduction}

The drift chamber (DCH) is the main tracking and momentum-measuring system.
It provides precision momentum measurements, as well as good
particle identification for low momentum tracks (those below DIRC threshold) and
for tracks in the forward direction, outside the DIRC acceptance.  The
DCH design is based on the
\babar\ drift chamber, described in detail in the original \babar\
detector publication\cite{ref:NIMbabar}, which is summarized below.

\subsection{\babar\ DCH}

\begin{figure}[htbp]
  \begin{center}
  \includegraphics[width=0.9\textwidth]{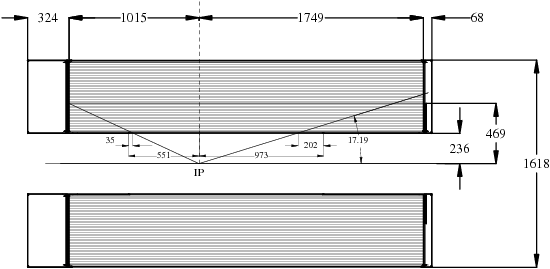}
  \caption{Side view of \babar\ drift chamber.  All dimensions in mm.}
  \label{dch:side}
  \end{center}
\end{figure}

The drift chamber is a conventional cylindrical design, 2.8~m long, with
flat aluminum endplates to hold plastic and metal feedthroughs for the
wires (see Fig.~\ref{dch:side}).  The inner cylinder is composed of three parts: a central
beryllium section, 1.5~m long and 1~mm thick, surrounding the interaction
point and the covering the full acceptance for \FourS\ decays; and forward and backward
aluminum sections, 5~mm think, with flanges for attaching the endplates.
The outer cylinder is a 5~mm think composite of Nomex honeycomb wrapped in
carbon fiber.

The forward ($+z$) endplate is machined with a step:  for $r<469$~mm, the
plate is 25~mm thick; outside that radius it is 12~mm thick.  This
provides sufficient strength to support the load of the wire tension while
minimizing material for tracks in the forward direction.

\begin{figure}[tbph]
  \begin{center}
  \includegraphics[width=0.39\textwidth]{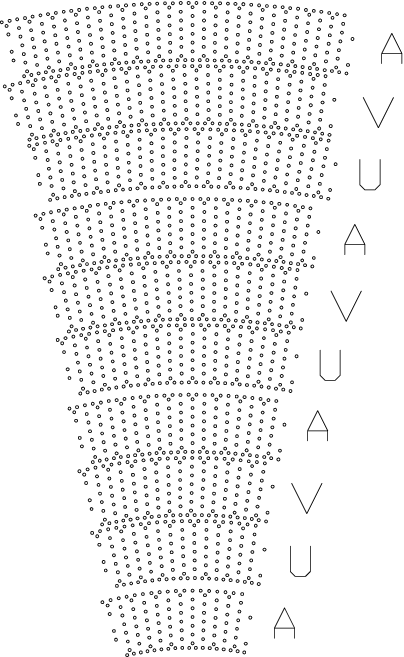}
  \vspace*{2mm}
  \caption{Drift chamber wire arrangement.  One 1/16th sector of the full
    chamber is shown, with axial (A) and small-angle stereo (U and V)
    superlayers indicated.}
  \label{dch:wires}
  \end{center}
\end{figure}

The drift system of the chamber consists of 40 layers of close-packed
hexagonal cells, each with a single sense wire surrounded by field-shaping
wires.  Each hexagonal cell is approximately 1~cm in radius.  The individual
layers are arranged in ten ``superlayers,'' as shown in
Fig.~\ref{dch:wires}.  To provide three-dimensional track reconstruction,
the superlayers alternate between axial (wires parallel to the $z$ axis),
and small-angle stereo (wire endpoints offset by 7 to 12 cells, in alternate
directions).

Three types of wires, at different voltages, define the cell electric field distribution.
The sense wires, at the center of each hexagonal cell, are
20~\mum diameter gold plated tungsten-rhenium alloy operated at
1930~V.
Each sense wire is surrounded by six grounded 120~\mum
gold-plated aluminum field-shaping wires.  Along the boundary of each
superlayer, the ground wires are replaced by a pair of 80~\mum
gold-plated aluminum wires at 335~V.  Adjacent to the inner and outer
cylinders of the chamber, a set of three wires are used with each cell, the
middle wire grounded, and the other two at 850~V.  These intermediate
voltage wires serve two purposes: they provide more uniform and symmetric
electric field configurations within the superlayer-edge cells, and they
assist in clearing residual ionization from the gaps between superlayers,
and between the wire field and the cylindrical boundaries.

The drift chamber is operated with a gas mixture of 80\% helium and 20\%
isobutane, passed through a bubbler to introduce 3500~ppm of water vapor.  The
bubbler also introduces approximately 100~ppm of oxygen into the gas mixture,
which has a small effect on the avalanche gain.  The chamber is maintained
at 4~mbar over atmospheric pressure with a recirculating pump;
freshly mixed gas in introduced as necessary to account for losses.

The performance of the \babar\ drift chamber in six years of operation has
been excellent.  The momentum resolution is
determined by reconstructing through-going cosmic ray events as two separate
``tracks,'' and taking the difference in the fitted transverse momentum
(inverse curvature in the $x-y$ plane) at the center of the chamber as the
resolution.  The result is
$$\sigma(p_T)/p_T = (0.13\pm 0.01)\% \cdot p_t + (0.45\pm 0.03)\%\ .$$

\begin{figure}[!t]
  \begin{center}
  \includegraphics[width=0.7\textwidth]{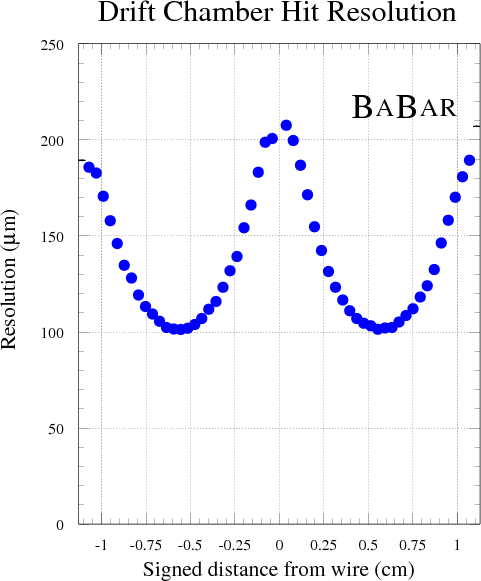}
  \vspace*{2mm}
  \caption{\babar\ single-hit resolution {\it vs.} distance from sense wire.
    Resolution is computed from the residual of the hit position compared with
    the fitted track excluding that hit.  The sign of the distance is positive
    (negative) for tracks passing to the right (left) of the radial vector
    to the sense wire.}
  \label{dch:resid}
  \end{center}
\end{figure}

The single-hit position resolution is determined for all tracks by
comparing the fitted trajectory excluding each measured hit with the
position of the hit determined from the readout timing and the calibration
time-to-distance relation for that cell.  The result
(Fig.~\ref{dch:resid}) is a weighted-average resolution of 125~\mum over
all cells; in the region of each cell with the most uniform electric field, the
resolution is 100~\mum.

\begin{figure}[htbp]
\centering
  \includegraphics[width=0.65\textwidth]{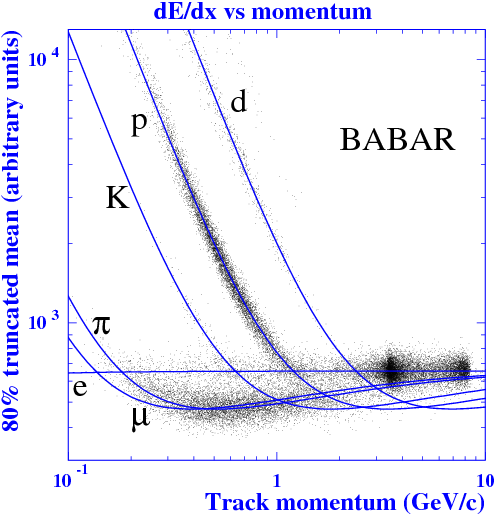}
  \caption{\babar\ relative $dE/dx$ {\it vs.} momentum for inclusive tracks.}
  \label{dch:dedx}
\end{figure}

The drift chamber readout system includes both timing, with 1~ns precision, and
integrated charge information.  The detector is calibrated for the
electronics gain of each channel, normalized for the charge deposition and
avalanche gain as a function of track trajectory in each layer of the
chamber.  With these calibrations, the integrated charge from each hit may
be used to compute a relative energy loss; the $dE/dx$ for each track is
computed from an 80\% truncated mean of the hits assigned to a track, as
shown in Fig.~\ref{dch:dedx}.  For electrons from radiative Bhabha events,
we obtain $\sigma(dE/dx)/(dE/dx) \lsim 7.5\%$.

\subsection{DCH Gas}

The occupancy rate in the drift chamber at \superb\ is a potential limiting
factor. It would be beneficial to find a gas with a shorter
collection time for the ions in a cell, so there is less chance
of having overlap hits from unrelated events. We have defined
a figure-of-merit for the collection times that can be used to
compare various gas mixtures with the \babar\ DCH cell geometry and compared several gas
mixtures to the mixture used in \babar -
Helium:Isobutane (80:20).

The drift chamber simulation program GARFIELD was used for this study.
A single hex cell was defined having one sense wire and 6 field wires
surrounded by another 6 bias wires at an appropriate voltage to
provide the same fields on the sense and field wire surfaces as in a
\babar\ cell, namely ~250 kV/cm on the sense wire and ~20 kV/cm on the
field wires. The same magnetic field (1.5 T) was also used in the
simulation.

Tracks were generated by GARFIELD, randomly positioned in the region from
0 to 1 cm from the sense wire with a vertical orientation in the cell
(emulating a radial straight track in a \babar\ cell). For each
track, ions were populated along the track, and GARFIELD made a
histogram of the arrival times of all ions that reached the sense
wire. The histograms generally have a broad maximum for a collection
time less than 200 ns, with a falling count rate at larger collection
times due to shortened track-segments near the boundary of the cell.

A figure-of-merit is established by noting the time ($\tau_{50}$)
at which 50\% of the charge is collected, and a
time ($\tau_{90}$) at which 90\% of the charge is collected. These
values of $\tau_{50}$ and $\tau_{90}$ were calculated for a variety of gas
mixtures to find the fastest gases.

Table~\ref{tab:dchgas} shows the results of the study. The table lists
the properties of the individual gases, such as density and radiation
length. The columns marked $\tau_{50}$ ns and $\tau_{90}$ ns give the
collection times for each mixture. The \babar\ gas, He(80)Isobutane(20)
has reference values of 400, 560 ns for $\tau_{50}$, $\tau_{90}$
respectively.

\begin{table}[htb]
\caption{\label{tab:dchgas} DCH Gas properties.}
\begin{center}
\begin{tabular}{ccccccccc}
\hline
\hline
\multicolumn{7}{c}{Gas Mixture (\%)} & \multicolumn{2}{c}{Drift Time} \\
He & Ne & Ar & CH4 &  C4H10 & CO2 & O2  & $\tau_{50}$ & $\tau_{90}$ \\\hline

80 &    &    &     &  20    &     &     &     400  &    560 \\

   &    &    & 100 &        &     &     &     560  &    700 \\
50 &    &    &  50 &        &     &     &     370  &    500 \\
50 &    &    &  45 &        & 5   &     &     365  &    495 \\
40 &    &    &  54 &        & 6   &     &     350  &    480 \\
40 &    &    &  57 &        & 3   &     &     320  &    480 \\
40 &    &    &  57 &   3    &     &     &     321  &    496 \\

   &    &    &  97 &        & 3   &     &     400  &    580 \\
   &    &    &  95 &        & 5   &     &     360  &    530 \\
   &    &    &  90 &        & 10  &     &     360  &    540 \\
   &    &    &  95 &   5    &     &     &     480  &    660 \\
   &    &    & 99.9&        &     &0.1  &     560  &    695 \\

   & 40 &    &  55 &        & 5   &     &     355  &    520 \\
   &    & 89 &   1 &        & 10  &     &     360  &    520 \\
\hline
\end{tabular}
\end{center}
\end{table}

Methane ($\rm{CH}_{4}$) is known to have a high drift velocity, but
this does not mean it has a shorter collection time, because the large
Lorentz angle for this gas makes the electrons spiral around the sense
wire. In fact 100\% $\rm{CH}_{4}$ has a longer collection time (560
and 700~ns) than the \babar\ gas. Various mixtures of $\rm{CH}_{4}$
with Isobutane ($\rm{C}_4\rm{H}_{10}$) or $\rm{CO}_{2}$ are shown. The
Helium:Methane:$\rm{CO}_{2}$ (40:57:3) mixture has the shortest
collection times of 320, 480 ns, approximately 20\% less than the
\babar\ gas. Replacing helium with neon or argon does not improve the figure-of-merit and
greatly increases the number of radiation lengths.

In summary, a gas mixture with methane could reduce the collection
time, but only by ~20\%, compared to \babar\ gas.

\subsection{Cell Geometry}

With the increased luminosity and beam-related backgrounds at \superb,
drift chamber cell size and occupancy become
important factors in determining the cell configuration.  An
increase in the number of cells, while decreasing the per cell
occupancy, increases the amount of material in the drift chamber, and
could have an effect on resolution.

We have studied the effect of cell size on tracking resolution in
simulation using the \babar\ detector as a starting point, and
varying size of the drift chamber cells.  The \babar\
drift chamber has hexagonal cells of ~2\cm diameter in each
dimension.  For comparison, we simulated drift chambers with
individual cell diameters
of 1.5\cm and 1.0\cm , keeping consistent the chamber inner and
outer radius, the amount of material in the inner part of the
detector, and the gas mixture.  We used a center-of-mass boost of $\beta\gamma=0.28$. We also included a level zero silicon vertex detector, but omitted the \babar-style support
tube.

\begin{table}[!b]
\caption{\label{tbl:DchSimWidths}
 Simulated resolutions of composite particle mass and $\Delta E$
  distributions for $B$ meson decays in simulated drift chambers with
  2\cm, 1.5\cm and 1\cm cell diameters.  There is a small worsening of
  the resolution for smaller cell sizes.}
\begin{center}
\begin{tabular}{lccc}
\hline
\hline
  & \multicolumn{3}{c}{Resolution (MeV)}\\
                        &2 \cm &1.5 \cm & 1 \cm\\\hline
\multicolumn{4}{c}{\Bz\to\jpsi\KS}\\
$m(\jpsi\to\mumu,\epem)$& 11.4 & 11.8 & 12.3 \\
$m(\KS\to\pip\pim)    $ & 1.64 & 1.67 & 1.74\\
$\Delta E $             & 15.9 & 16.6 & 17.3\\[1mm]
\multicolumn{4}{c}{$\Bz\to\phi\KS$}\\
$m(\KS\to\pip\pim)  $   & 1.73 & 1.78 & 1.84\\
$\Delta$ E              & 11.5 & 12.1 & 13.1\\[1mm]
\multicolumn{4}{c}{\Bz\to\pip\pim}\\
$m(\Bz\to\pip\pim)  $   & 20.1 & 20.6 & 21.3 \\
$\Delta E $             & 20.5 & 20.9 & 21.6 \\[1mm]
\multicolumn{4}{c}{$\Bz\to D^{*+}D^{*-}$}\\
$m(\Dz\to\Kpm\pimp) $   & 4.90 & 5.28 & 5.77 \\
$\Delta E $             & 11.8 & 12.4 & 13.4 \\
\hline
\end{tabular}
\end{center}
\end{table}

Because the cells are in a regular
hexagonal configuration, the number of cells in each layer
for the drift chambers with 1.5 and 1.0\cm diameter cells increased from 40
layers to 53 and 80,
respectively.  The
maximum and minimum stereo angles are kept the same as in the \babar\
drift chamber. The per cell resolutions for the 2\cm, 1.5\cm, and the
1\cm are taken to be 140, 157, and 178 \mum, respectively.  The first
is typical of \babar, while the last
two are obtained by assuming the increase in resolution in a smaller cell
caused by near-wire effects are the same for all cell sizes.

\begin{figure}[t]
\centering
  \includegraphics[width=0.95\textwidth]{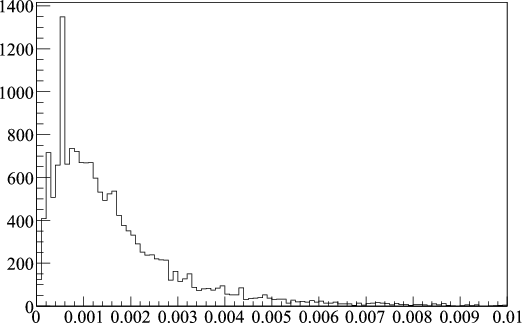}
  \vspace*{2mm}
  \caption{\label{dch:Ephotonbck} Energy of background photons entering the DCH (MeV)}
\end{figure}


For each of these drift chamber configurations, we simulated 50,000
events of each
of the following $B$ meson decay modes using the PRAVDA and TRACKERR
simulation programs: $\Bz\to\jpsi\KS$,
$\Bz\to\phi\KS$, $\Bz\to\pip\pim$, and $\Bz\to D^{*+}D^{*-}$.
Table ~\ref{tbl:DchSimWidths} shows the resolution from
gaussian fits to composite particle mass and $\Delta E$ distributions.
While in every case, the resolution of these quantities increases as
the drift chamber cells get smaller, the increase in per cell
resolution is largely offset by the addition of extra layers.

\subsection{Backgrounds}

The dominant source of background in the \superb\ DCH is expected to
be radiative Bhabhas, as discussed in
section~\ref{sec:radbhabha}.  Off-energy electrons and positrons
shower in the beamline elements, and the tails of those showers may
penetrate the passive shielding and reach the DCH. Most of the
particles reaching the DCH are photons; their energy spectrum is shown
in Fig~\ref{dch:Ephotonbck}. There is also a small component of
electrons.  At these $\mev$ energies, the photon
cross-section is dominated by Compton scattering.  To model the DCH
material, we take the mix of gas and wires from the \babar\ DCH,
corresponding to a density of $\rho = 8.4 \times 10^{-4} g/\cm^{3}$,
to estimate the probabilty of a photon interaction in the DCH.  At
$1~\mev$ the Compton cross-section is $\sigma = 0.066 \cm^{2}/g$,
corresponding to a path length of $1.8\times10^{4}~\cm$.  Thus most
photons do not interact in the gas volume, but those that do produce
$\mev$ Compton electrons that spiral along field lines in the $1.5~T$
magnetic field.  To estimate the occupancy from radiative Bhabhas we
assume that each electron produces a signal in three DCH cells.

\begin{figure}[!t]
\centering
  \includegraphics[width=1.0\textwidth]{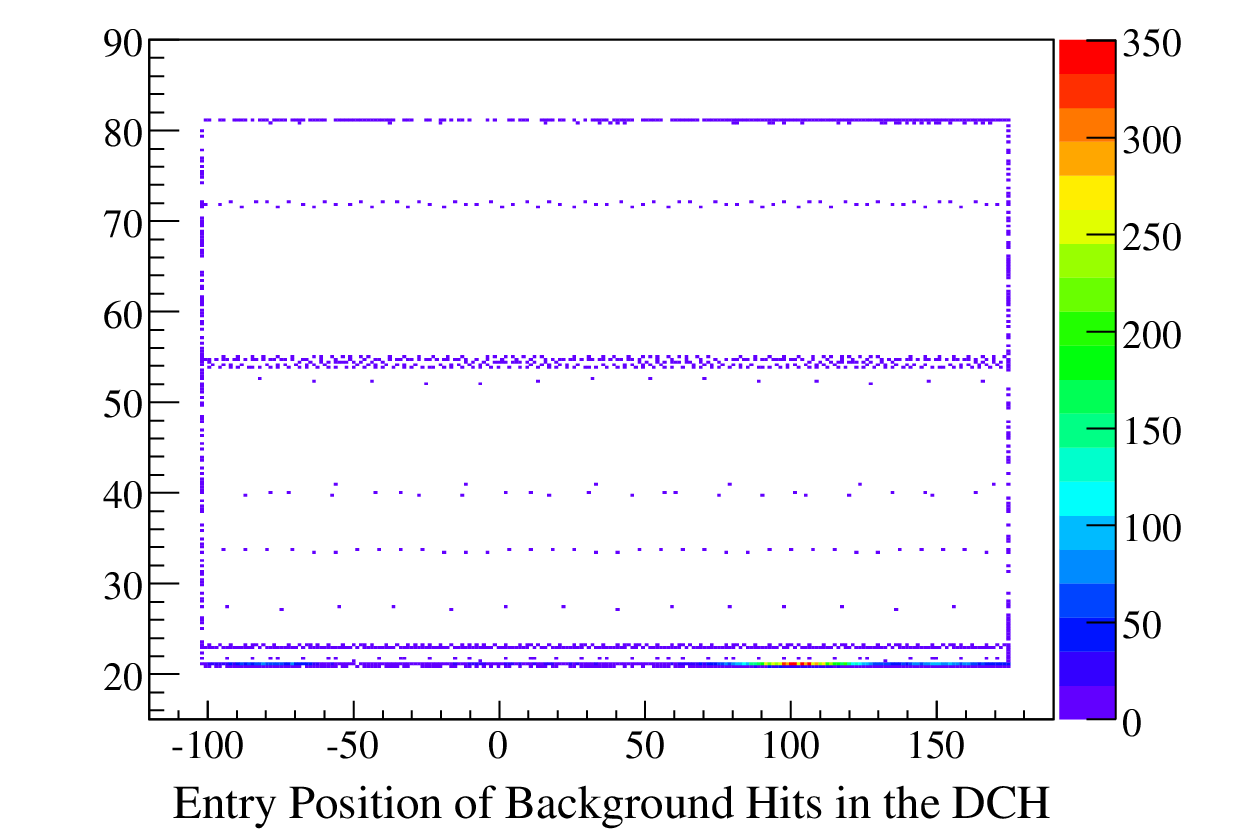}
  \caption{\label{dch:Rphibck} Position, $r\ vs.\ z$, of background electrons and photons
  entering the DCH.  Only the entry point of the photons is shown,
  while the electrons traverse the DCH by spiralling along the magnetic field lines.}
\end{figure}

Since the radiative Bhabhas originate along the
beam-line roughly $0.5 - 2$m from the IP, the number of background
hits will be highly dependent on the amount of passive shielding
inside the DCH.  Two different shielding schemes have been simulated,
to bracket the range of possible IP designs.  The entry point for
photons, along with the trajectory of a few entering electrons, is
shown in Fig.~\ref{dch:Rphibck}.  In this case, the DCH would have
an occupancy of roughly 7\%; with additional shielding the occupancy can
be reduced to $\sim$1.5\%.

\afterpage{\clearpage}

\section{Particle Identification}
\label{sec:det:PID}

\subsection{Introduction}

Excellent particle identification (PID) for hadrons and leptons over the
full momentum range for particles coming from $B$ decays is essential
to achieving the physics objectives of the \superb\ experiment.
In particular, precision measurements of \CP\ violation require full
particle identification to reconstruct exclusive final states,
to discriminate against backgrounds, and tag the quark flavour
of the other $B$ in the event.
In addition, studies of inclusive and exclusive decays of charm and
$\tau$ physics  benefit directly from high-momentum hadron
identification.
In general, these channels require the separation of pions and kaons
at considerably higher momenta than does $B$ physics -- indeed if the
full acceptance is to be used for these channels, particle separation
up to about 6 \gevc\ is needed in the forward direction.

Leptonic identification at \superb\ is provided by the EMC and IFR
detectors, while charge deposition (\dedx) in the central trackers can
be used to identify low-momentum hadrons.
However, these techniques are insufficient to distinguish pions and
kaons with momenta greater than approximately 0.7 \gevc\ or protons
above 1.3 \gevc, as is required to obtain efficient tagging and
$B$ event reconstruction.
To cover this momentum range, a dedicated hadronic PID system is needed.
The existing \babar\ \dirc\ ring imaging Cherenkov system, described
below, provides excellent
performance over the entire momentum range for $B$ physics, but
geometrically covers only the barrel portion of the detector.
A modest upgrade of this device would provide adequate performance
at \superb.
Further enhancements to the technology can improve the performance but
require significant R\&D and will be significantly more expensive.

The baseline detector described below contains a PID upgrade for
the forward end cap to improve the PID hermiticity.
However, as there are potential losses in performance as well from
including such a system, especially for photon detection, a cost/benefit
analysis is currently underway to ascertain the tradeoffs
associated with such a PID system in the end cap region.

\subsection{Baseline barrel PID for \superb\ --
\babar\ \dirc}

\subsubsection{Purpose and Design Requirements}

The particle identification system should be thin and uniform in
terms of radiation lengths (to minimize degradation of the calorimeter
energy resolution) and thin in the radial dimension to reduce the volume,
and hence, the cost of the calorimeter.
For operation at high luminosity, the PID system needs fast
signal response, and must be able to tolerate high backgrounds.

The PID system used in \babar\ is a new kind of ring-imaging Cherenkov
detector called the \dirc\ (the acronym \dirc\ stands for Detection of
Internally Reflected Cherenkov light).
It has be proven to provide pion/kaon separation of more
than 2.5 $\sigma$, for all tracks from $B$ meson decay from the pion Cherenkov
threshold up to 4.2 \gevc.
Particle identification below 700 \mevc\ relies primarily on
\dedx\ measurements in the DCH and SVT.

\subsubsection{\dirc\ Concept}

\begin{figure}[htb]
\centerline{\includegraphics[width=8cm]{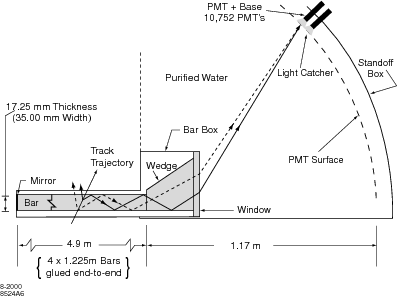}}
\vspace*{1mm}
\caption{Schematic of the \dirc\ fused silica radiator bar and
imaging region.
}
\label{fig:pid:princip}
\end{figure}

The \dirc\ is based on the principle that the magnitude of the Cherenkov
angles are maintained upon reflection from a flat surface.
Figure~\ref{fig:pid:princip} shows a schematic of the \dirc\ geometry
that illustrates the principles of light production, transport,
and imaging.
The \dirc\ radiator is synthetic fused silica in the
form of long, thin bars with rectangular cross section.
These bars are also light pipes for that portion
of the Cherenkov light trapped in the radiator by total internal reflection.
The Cherenkov photons are detected by an array of densely packed photomultiplier
tubes (PMTs), each surrounded by reflecting light-catcher cones to
capture light which would otherwise miss the active area of the PMT.
The PMTs are placed at a distance of about 1.2 m from the bar end.
The expected Cherenkov light pattern at this surface is essentially
a conic section, where the cone opening-angle is the Cherenkov
production angle modified by refraction at the exit from the fused
silica window.
The \dirc\ is intrinsically a three-dimensional imaging device, using
the position and arrival time of the PMT signals.
Photons generated in a  bar are focused onto the phototube detection
surface via a ``pinhole" 
defined by the exit aperture of the bar.
In order to associate the photon signals with a track traversing a bar,
the vector pointing from the center of the bar end to the center of
each PMT is taken as a measure of the photon propagation angles.
Since the track position and angles are known from the tracking system,
the three angles can be used to determine the two components of the
Cherenkov angles.

The arrival time of the signal provides an independent
measurement of the propagation time of the photon, and can be related to the
propagation angles.
This over-constraint on the angles and the knowledge of the timing of the signal are particularly
useful in dealing with ambiguities in the signal association,
especially in high background situations.

\subsubsection{\babar\ \dirc\ Design}

\begin{figure}[bhtp]
\begin{center}
\includegraphics[width=10 cm]{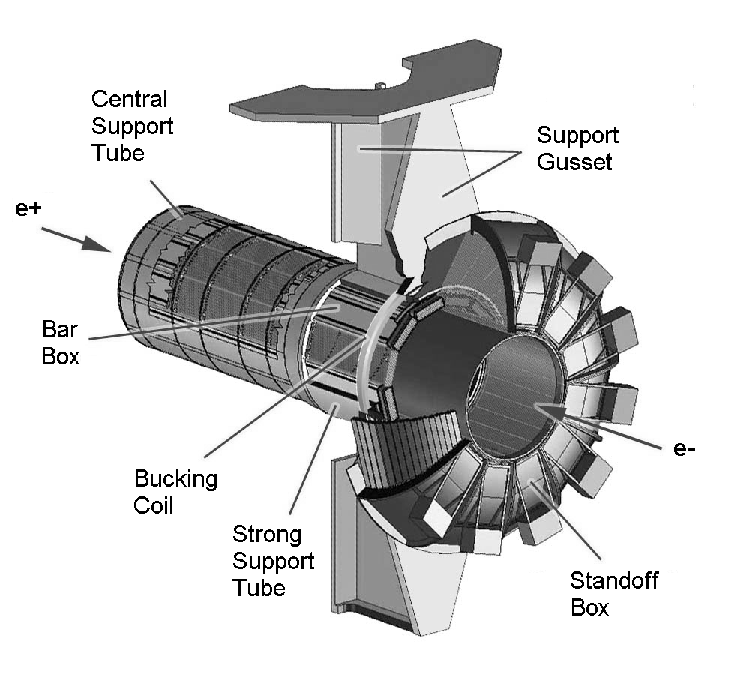}
\includegraphics[width=12 cm]{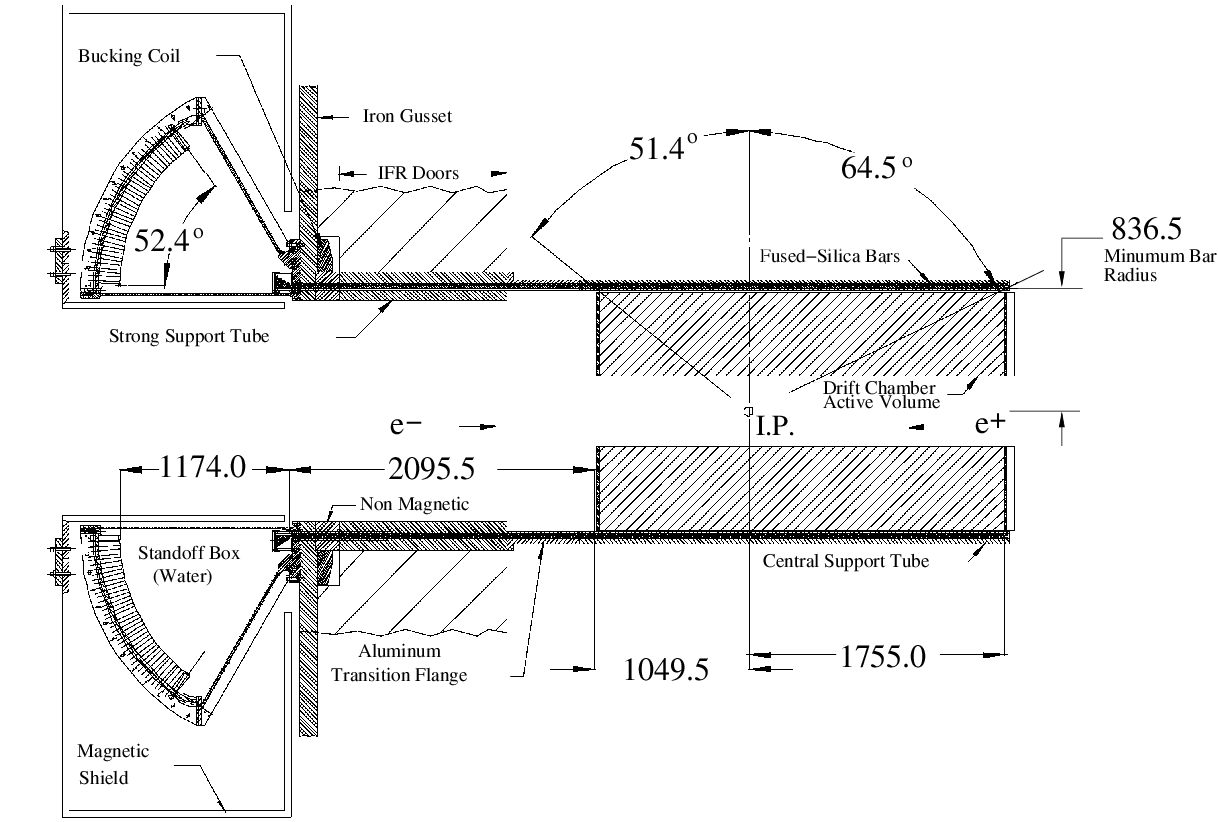}
\end{center}
\caption{Schematic view (upper) of the principal components
of the \dirc\ mechanical support structure.
The magnetic shield of the standoff box is not shown.
Elevation view (lower) of the nominal \dirc\ system geometry.
For clarity, the end plug is not shown. All dimensions are
in millimeters.}
 \label{fig:pid:components}
\end{figure}

The \dirc\ design and construction are described in detail
in ref.~\cite{ref:pid:dircnim}.
The principal components of the \dirc\ are shown schematically in
Fig.~\ref{fig:pid:components}.
The radiator bars are grouped into 12 hermetically sealed
containers, called bar boxes, made of very thin
aluminum-Hexcell panels.
Each bar box contains 12 bars, for a total of 144 bars.
Within a bar box the 12 bars are optically isolated by
an $\sim$150~$\mu$m air gap between neighboring bars,
enforced by custom shims made from aluminum foil.
The bars have nominal dimensions of 17.25\ mm thick (in the radial direction),
35\ mm wide (azimuthally), and 4.9\ m long. Each bar is assembled
from four 1.225~m pieces glued end-to-end (a 1.225~m
length bar was the longest obtainable with high
quality~\cite{ref:pid:quartzpaper}).
The standoff box (SOB), made of stainless steel, consists of a cone, a
cylinder, and 12 sectors of PMTs.
It contains about 6,000 liters of purified water.
Water is used to fill this region because it is inexpensive and has
an average index of refraction ($n \approx 1.346$) reasonably close
to that of fused silica, thus minimizing total internal reflection
at the silica-water interface.
Furthermore, its chromaticity index is a close match to that of fused
silica, effectively eliminating dispersion at the silica-water interface.
Iron gussets support the standoff box.
An iron shield, supplemented by a bucking coil, surrounds the standoff
box to reduce the magnetic field in the PMT region to below 1 Gauss.
The PMTs at the rear of the standoff box lie on a surface that is
approximately toroidal.
Each of the 12 PMT sectors contains 896 PMTs (ETL model 9125B)
with 29\ mm-diameter, in a closely packed array inside the water volume.
A double O-ring water seal is made between the PMTs and the vessel wall.
The PMTs are installed from the inside of the standoff box and
connected via a feed-through to a base mounted outside.
A hexagonal light catcher cone is mounted in front of the photocathode
of each PMT, improving the effective active surface area light
collection fraction to about 90\%.
The \dirc\ occupies 80\ mm of radial space in the central detector
volume, including supports and
construction tolerances, with a total thickness of about 19\%\ $X+0$
at normal incidence. The radiator bars subtend a solid
angle corresponding to about 94\%\
of the azimuth and 83\%\ of the center-of-mass polar angle.

The \dirc\ frontend electronics (FEE) is designed to measure the
arrival time of each Cherenkov photon detected by the PMT array
to an accuracy that is limited by the intrinsic 1.5\ ns transit
time spread of the PMTs.
The design contains a pipeline to deal with the Level~1 trigger latency of
12~$\mu{\rm s}$, and can handle random background rates of up to
2.5~MHz/PMT with less than 1~\% dead time.
The \dirc\ FEE are mounted on the outside of the SOB; the FEE are
highly integrated in order to minimize cable lengths and to retain
the required single photoelectron sensitivity.
The photon arrival time is measured by the time-to-digital-converter
(TDC) chip, a self-calibrating 16-channel microchip which performs
three major functions: digitization of the input signal time with
520~ps binning (250~ps resolution rms); and pulse pair separation
of 33.6~ns; and simultaneous handling of input and output data.

The \dirc\ uses two independent approaches for calibration of the
unknown PMT time response, and the delays introduced by the FEE and
the fast control system.
The first is a conventional pulser calibration using a light pulser
system to generate precisely timed 1~ns duration light pulses from
a blue LED.
The second uses reconstructed tracks from collision data.
The data stream and online pulser calibrations yield fully consistent
results.
The time delay values per channel are typically stable to an $rms$ of
less than 0.1\ ns over more than one year of daily calibrations.

\subsubsection{\babar\ \dirc\ Performance}

\begin{figure}[hbtp]
\centerline{
\includegraphics[width=6.5 cm]{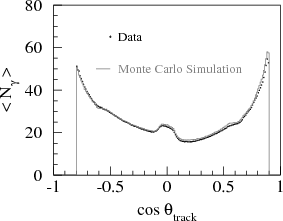} \ \ \
\includegraphics[width=6.5 cm]{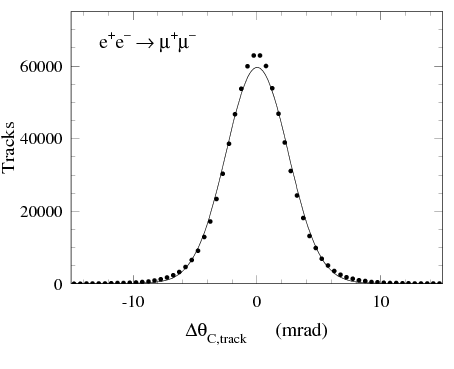}}
\vspace*{0mm}
\caption{
Average number of detected photons {\it vs.}
track polar angle (left) for reconstructed tracks in dimuon events
compared with Monte Carlo simulation.
Resolution of the reconstructed Cherenkov polar angle per track for
dimuon events.
The curve shows the result of a gaussian fit with a resolution of 2.5\mrad.}

\label{fig:pid:nphot}
\end{figure}

During seven years of running the \dirc\ has performed efficiently and
reliably.
The \dirc\ plays a significant role in most \babar\
physics analyses.
Details of the \dirc\ operational experience and the use of
\dirc\ in \babar\ physics analyses can be found in
ref.~\cite{ref:pid:dircnim}.

Some deterioration of the PMT front glass windows (made of B53
Borosilicate glass) that are immersed in the ultra-pure water
of the standoff box has been observed since 2000.
With water in the standoff box, these features are not very
noticeable, as water provides good optical coupling even to
corroded glass.
For most of the tubes, the observable effect is typically a slight
cloudiness, but for about 50 tubes, it is a much more pronounced
crazing.
Extensive studies have shown that this effect is associated with
a loss of sodium and boron from the surface of the glass.
Chemical analysis of the water has shown that the leaching rate is a
few microns per year, and is expected to be acceptable for the full
projected ten year lifetime of the experiment.
Loss of photon detection efficiency can arise from the corrosion
of the PMT front glass windows, as well as from photocathode aging,
dynode aging, and possible deterioration of the water transparency or
pollution of bar or window surfaces.
Direct measurements of the number of Cherenkov photons observed in
dimuon events as a function of time can be used to determine any
degradation of the photon yield.
An analysis using dimuon events from October 1999 through June 2006
shows a stable photon loss rate of 1-2\%/year.
There is no significant dependence of the loss rate on the radiator
bar number, the position of track long the bar length, or the location
of the Cherenkov ring in the PMT plane.
If the photon loss rate continues at this rate, the impact on the
particle identification power of the \dirc\ is negligible over the
lifetime of the experiment.

The background in the \dirc\ is dominated by low energy photons from \pepii  hitting the water-filled standoff box.
Our experience has been that attention to shielding in the
\dirc\ standoff box region is required
to reduce the sensitivity to beam-induced backgrounds and keep the
background rate under a limit of 300\ kHz/tube.
%

The single photon Cherenkov angle resolution has been measured
to be about 9.6 mrad, dominated by a geometric term that is due
to the sizes of bars, PMTs and the expansion region, and a
chromatic term from the photon production.
The measured time resolution is 1.7\ ns, close to the intrinsic
1.5\ ns transit time spread of the PMT's.

The average value of the number of detected photoelectrons,
$N_{\gamma}$, shown in Fig.~\ref{fig:pid:nphot}, varies between
about 17 for tracks with nearly perpendicular incidence to nearly
60 for polar angles towards the forward and backward regions. The
increase in the number of photons for tracks in the forward
direction compensates for the reduced average separation in the
Cherenkov angle for different particle hypotheses due to the
increased track momenta in this region.

The Cherenkov angle resolution, $\sigma_{C,track}$, for tracks
from dimuon events, $e^+e^- \to \mu^+\mu^-$, is shown in,
in Fig.~\ref{fig:pid:nphot}.
The width parameterized by a single Gaussian distribution, is 2.5~\mrad.
The resolution is $14\%$ larger than the design goal of 2.2~\mrad,
which was estimated from the extensive study of a variety of
prototypes, including a beam test.

\begin{figure}[bthp]
\centerline{
\includegraphics[width=6.5 cm]{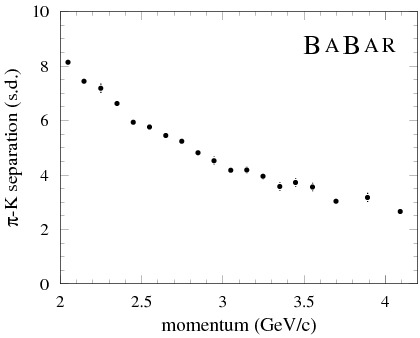} \ \ \ \ \
\includegraphics[width=6.5 cm]{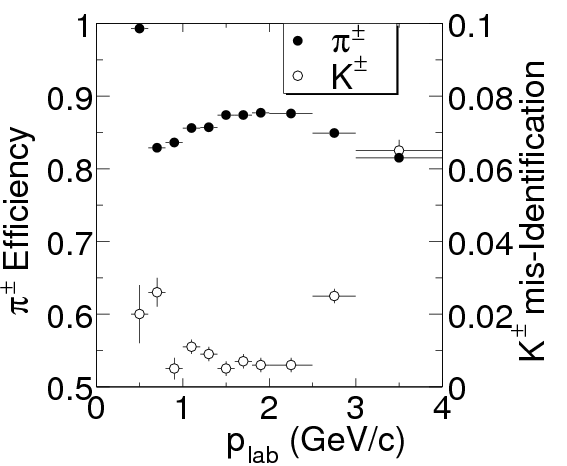}}
\vspace*{0mm}
\caption{
\dirc\ $\pi$-K separation {\it vs.} track momentum (left)
measured in $D^0 \rightarrow K^-\pi^+$ decays selected
kinematically from inclusive $D^{*}$ production.
The pion efficiency and kaon misidentification rate (right), as
a function of momentum in the laboratory frame, for the charged pion
selection used in the search for $B\rightarrow\rho\gamma$ and
$B\rightarrow\omega\gamma$.}

\label{fig:pid:pik}
\end{figure}

The $D^{*+}\to \pi^+ (D^0\to K^-\pi^+)$ decay
chain\footnote{Unless explicitly stated, charge conjugate decay
modes are assumed throughout this section.} is well suited to
probe the pion and kaon identification capabilities of the \dirc.
It is kinematically well-constrained and the momentum spectrum of
the charged pions and kaons covers the range accessible by $B$
meson decay products in \babar.

The pion-kaon separation power is defined as the difference of the
mean Cherenkov angles for pions and kaons assuming a gaussian
distribution, divided by the measured track Cherenkov angle
resolution.
As shown in Fig.~\ref{fig:pid:pik}, the separation between kaons and
pions is about $4~\sigma$ at 3 \gevc\ declining to about 2.5~$\sigma$
at 4.2 \gevc.

The efficiency for correctly identifying a charged kaon
that traverses a radiator bar and the probability to wrongly identify
a pion as a kaon, determined from the inclusive $D^{*}$ sample
are shown as a function of the track
momentum in Fig.~\ref{fig:pid:pik} for a particular choice of
particle selection criteria.
The kaon selection efficiency and pion misidentification, integrated
over the
$K$ and $\pi$ momentum spectra of the $D^{*}$ control sample,
are $97.97 \pm 0.07$\% (stat. only) and $1.83 \pm 0.06$\% (stat. only),
respectively.

\subsubsection{Summary of the barrel PID system}

\begin{figure}[htbp]
\centerline{\includegraphics[width=12 cm]{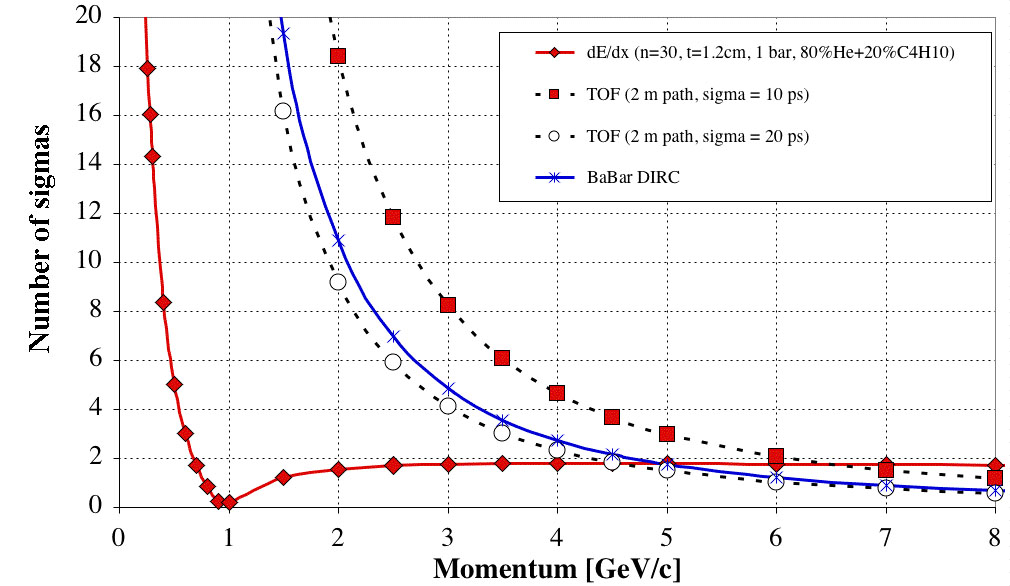}}
\vspace*{0mm}
\caption{
Expected PID performance as a function of momentum for
the barrel \babar\ \dirc\ (the Focusing \dirc\ option would be similar),
the forward end cap TOF option, and the \dedx\ method in the drift chamber.
A TOF resolution at a level of $\sigma$$\sim$20~ps with a path length
of $\sim$2 meters yields a performance equivalent to present
\babar\ \dirc, and is far superior to the \dedx\ method.}
\label{fig:pid:pid_perf}
\end{figure}

The barrel \dirc\ is a novel ring-imaging Cherenkov detector
well-matched to the hadronic PID requirements of \babar.
The detector performance achieved is excellent and close to that
predicted by the Monte Carlo simulations.
The \dirc\ has been robust and stable, and, indeed, serves also as a
background detector for \pepii tuning.
In combination with \dedx\ measurements in the DCH and SVT this
system will provide pion/kaon separation of more than 2.5 $\sigma$,
for all tracks from $B$ meson decays from the pion Cherenkov
threshold up to 4.2 \gevc, as shown in Fig.~\ref{fig:pid:pid_perf}

\subsection{PID Options}

\subsubsection{Baseline Barrel Solution}

Since the existing \babar\ \dirc\ photon detectors are aging, at
a minimum they need to be replaced with modern conventional phototubes.
To gain headroom with respect to background, these PMTs should be
faster than the present devices, such as the Hamamatsu R6427 with
0.5\ ns (FWHM) transit time spread.
This will increase the background rejection capability by a
factor 8 to 10.
A small extrapolation of the present ``non-focusing" design is
costed that replaces the large conventional PMTs with modern
pixelated PMTs, such as the Hamamatsu H-8500 MaPMT
($6\times6$ mm$^2$ pixels and spread with a transit time spread of less
than 0.15\ ns).
This allows a much smaller SOB to be used with fused silica coupling
and is expected to improve the background rejection capability by
another factor of 5-10.
With these modifications, the PID performance of the barrel
should be essentially identical to that of the present
\babar\ \dirc\ system described above, with good performance within
the expected background environment.

Further upgrades of the barrel detector system are possible that
could provide still further significant headroom in a high background
environment, and improve the PID performance, as described below.

\subsubsection{Barrel focusing DIRC option}

A new photon detection region will be placed within the SOB
(magnetic field shielded) volume.
This will consist of 12 modular focusing blocks, attached to each
bar box.
The light emitted from each bar will be focused onto a plane of fast
pixelated photodetectors, such as 64-channel Burle/Photonis microchannel
plate PMTs or 64/256-channel Hamamatsu multi-anode PMTs.
The time resolution of the PMTs is sufficient to provide significant
performance gain for measuring the Cherenkov angle, through correction
of the chromatic effects~\cite{ref:pid:fdirc}.
The shorter time resolution improves the background suppression
by more than an order of magnitude compared to the present DIRC, and,
in addition, the mass of standoff material
({\it i.e.} water) is reduced by more than an order of
magnitude, thus reducing the probability of secondary interactions
in the material, which is the major current source of background in
the \babar\ \dirc.

\subsubsection{Endcap upgrade}

It is understood that there are a number of cases where \babar\ physics
would have benefited if the PID system had covered the forward endcap
as well as the barrel.
Even though the endcap covers only a modest portion of the geometrical
acceptance, a more hermetic detector than \babar\ would allow higher
efficiency for exclusive $B$ physics channels, and could be especially
valuable for some of the types of studies expected to be important at
\superb.
For example, recoil physics studies benefit significantly (much faster
than linearly) from the highest possible acceptance.
Monte Carlo studies of the physics gains expected in multi-body $B$
decays are underway to quantify the physics gains.

Thus, the base detector presented here includes a forward TOF PID system.
However, as there are potential losses in performance as well from
including such a system, especially for photon detection, a
cost/benefit analysis, which is currently underway, is
needed to ascertain the tradeoffs  associated with such a
PID system in the end cap region.

Two candidate endcap detector designs are being considered - one utilizing
an aerogel radiator RICH to provide separation to the highest momenta
seen in the forward direction, and another using a fused silica
radiator to obtain very good time-of-flight (TOF) performance covering
the region up to about 4 \gevc.
Space considerations, PID performance needs, and the amount of total
material in front of the electromagnetic calorimeter, which degrades
its performance, led us to pursue further studies of the TOF system only.

\subsubsection{Status of time-of-flight system R\&D}

There has been progress in TOF capability recently, making it
worthwhile to consider such a system for \superb.
The progress derives from the introduction of new, very fast vacuum-based
photon detectors with a transit time distribution of
$\sigma_{TTS}$$\sim$30-50~ps, and the use of Cherenkov light
rather than the scintillation light for the TOF measurement.

Figure \ref{fig:pid:pid_perf} shows thata  time resolution of about
$\sigma$$\sim$10~ps is required to make the TOF method competitive with,
for example, a Forward Aerogel RICH.
A resolution of
$\sigma$$\sim$15-20~ps is likely to prove a more
realistic goal in practice.
The main advantage of the TOF system is its cost and simplicity
compared to RICH techniques.

The proposed endcap TOF system consists of a sheet of fused silica
radiator, 10 mm thick, coupled to a matrix of Burle/Photonis MCP-PMTs,
each $\sim$5x5cm$^{2}$ in size and each with 3mm thick MCP
windows.
The large radiator sheet is glued together from smaller segments and
it is slightly tilted to minimize the time spread of the Cherenkov
photons arriving at the photocathode.
The continuous radiator plate reduces the effect of gaps between
the MCP-PMT detectors.
It might, however, be useful to segment the radiator into smaller units to facilitate
maintenance; each segment would then carry several MCP-PMTs and be
removable as a unit.
An open question is whether we should allow photon reflections from the front surface of the
radiator.
If further analysis shows that thus is not desirable, we may choose
to provide a photon trap on the front surface, for example, by pressing
a soft rubber sheet onto it.)
The Burle/Photonis MCP-PMT can have either a  bialkali or an S20 multialkali
photocathode; the expected number of photoelectrons is
at least 30 photoelectrons per track.
Each MCP-PMT detector has two microchannel plates with
10~$\mu{}$m diameter holes in the chevron, and a hole
diameter-to-thickness ratio of $\sim$1:60.
In its nominal basic configuration, the MCP-PMT anode plane is segmented
to into 256 micro-pixels, each $\sim$3mm$^{2}$ in size.
The individual pixels are connected to a single timing point via
equal-time traces on a PC board.
This arrangement ensures that the time spread across the anode is
limited to
$\sigma_{Anode}$$\sim$(10/$\sqrt{12}$) $\sim$3~ps.
To maximize the signal risetime, the MCP anode structure must
be properly bypassed to ground.
The operating parameters, such as the MCP internal voltages and its
geometry remain to be optimized.

\begin{figure}[htbp]
\centerline{\includegraphics[width=12 cm]{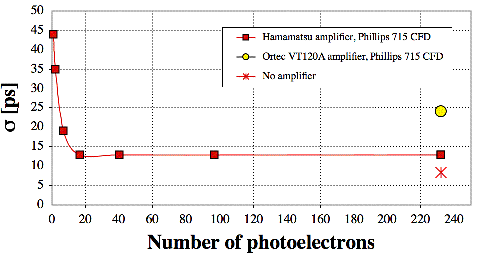}}
\vspace*{0mm}
\caption{
Multi-photoelectron timing resolution \cite{ref:pid:vavra}
as a function of number of photoelectrons using (a) a method,
employing a TDC with 25~ps/count, for the Hamamatsu C5594-44 and the
Ortec VT120A amplifiers, and (b) a method with no amplifier and a CFD
(see the method described in the text), all with the Burle
64-pixel MCP-PMT 85012-501 with 10~$\mu$m hole
diameter, and B=0 T. The data were obtained with a PiLas red laser
diode providing  diameter beam spot smaller than 1~mm.
}
\label{fig:pid:tof}
\end{figure}

What has been achieved until this point? The Nagoya group test beam
results \cite{ref:pid:nagoya}
indicate $\sigma$$\sim$6.2~ps with a small
($\sim$11mm diameter) Hamamatsu MCP-PMT R3809U-50-11X
with 3~mm thick MCP window, multialkali photocathode and
6~$\mu$m diameter MCP holes.
The tube was coupled to a 10~mm thick quartz radiator, yielding
$\sim$40 photoelectrons.
They used SPC-134 CFD/TAC/ADC electronics, with
$\sigma_{Electronics}$$\sim$4~ps.
We have done timing tests with laser diodes.
Our overall best result~\cite{ref:pid:vavra} so far, was obtained with
a 64-pad MCP-PMT with 10~$\mu$m diameter holes, operating without an
amplifier, and with no CFD, {\it i.e.}, the MCP-PMT pulses were directly viewed on
an oscilloscope.
The laser-induced spot size of less than 1~mm$^{2}$ diameter produced
a point-response timing resolution of
$\sigma$$\sim$8-9~ps.
This result was obtained using the ``histogram option" 
on Tektronix TDS-5104 digital oscilloscope after subtracting
trigger jitter contributions, determined in a separate
run with a precision pulser.
Results with a 25~ps/count TDC and with an amplifier and CFD, shown
in Fig.~\ref{fig:pid:tof}, yielded slightly worse results.

The detailed timing strategy must be simulated to find an optimum
scheme.
We are leaning towards an option either no amplifier
at all, or amplification by a factor of only  5 to 10.
In all tests thus far, the use of amplifier signals yielded worse
results compared to MCP direct signals.
However, realizing that there will be a variation in the leading edge
slope as a function of the number of photoelectrons and due to gain
fluctuations, we are considering either a double-threshold, or a
time-over-threshold timing strategy, located at each detector.
To achieve our timing resolution goal, we require  a TDC resolution
of at least 12~ps/count.
This requires a new development.
We may consider, for example, some simplified version of a new
ASIC-based TAC being developed at the University of Chicago, which aims
for a 1~ps resolution~\cite{ref:pid:chicago}.
It is important to have a second event-processing capability within
$\sim$100ns of the recovery of the MCP-PMT from saturation,
Tas well as a digital pipline.

Many possible degradation factors influence a high-resolution
TOF counter in a typical collider environment: (a) detector
design, (b) start time resolution, (c) detector signal timing strategy,
(d) cross-talk and charge sharing between anode pixels, (e) TDC
resolution, (f) tracking errors, (g) magnetic field effects on the gain
of the detector, (h) thermal drifts, (i) hermeticity of the radiator
and detectors, (j) variation of the number of photoelectrons near edges,
{\it etc.}
For example, the start time is a crucial issue in the proposed TOF
system.
We propose to use the accelerator RF pulse, which will be distributed
to the start TAC located on each MCP-PMT.
This start time is then corrected by tracking and vertexing information.
The calibration of the TOF system is crucial to its performance.
It must be performed often to keep track of thermal drifts.
We plan to use two methods, one based on Bhabha events and the other
on an ultra-precision light pulser.
At full luminosity, we expect a Bhabha rate of $\sim$1kHz.
This will allow a calibration precision to a few ps every 10-20 minutes.
In addition, we plan to use a low jitter pulser, producing a start pulse and a
sequence of randomly sequenced stop pulses with precise time difference
intervals of 5ns.
These stop pulses will drive a laser diode, with light directed through a fiber to selected
MCP-PMTs.
We have one such pulser, and we have verified that it is
capable of measuring a time interval between start and stop to
$\sigma$$\sim$7.5~ps.
With about 1000 successive calibration pulses covering 10 sequential
delays, we can track changes in the mean to the
few picosecond level.
To do that, however, pulsers have to be in thermal equilibrium, {\it i.e.},
they must run into a
dummy load for at least 30 minutes prior to calibration.

The magnetic field causes a loss of MCP gain, that depends on the firld sgtrength and the angle between the normal to
the MCP face and the field axis~\cite{ref:pid:vavra}.
The loss of gain must be compensated by adjusting the high voltage.
Our tests indicate that the present Burle tube with
10~$\mu$m diameter holes cannot be rotated by more
than $\sim$10-15~$\deg$.

MCP-PMT aging is an important issue.
The main effect of aging is damage of the photocathode by the ion backflow;
gain loss is thought to be much less significant.
The ions are created from residual hydrogen gas contamination left
in the MCP glass by the reduction process.
There are three methods presently considered to reduce photocathode
damage: (a) a thin film placed on top of the MCP surface, which blocks ions
but allows the electron transfer, (b) adding three MCP plates
instead of the usual two, or (c) better vacuum scrubbing of the residual
hydrogen.
 Recent tests by Burle/Photonis indicate that the vacuum scrubbing
technology has improved enough that only a 5\% degradation of the
photocathode QE has been observed even after a total charge dose limit
of $\sim$16 C/tube with a simple double MCP structure without
the protecting film.
This represents a total single photoelectron dose of
$\sim$10$^{13}$ per cm$^{2}$ (assuming a total gain of
$\sim$3x10$^{5}$), which is 5-10 years of the \superb\
expected total dose.
If the Burle tests can be reproduced in our lab, we may decide to
use the simple double MCP structure, which would reduce the cost
and improve the MCP efficiency, compared to the thin film technology.

\subsection{Summary of Requirements}

Experience with existing $B$ Factories at CESR; \pepii, and KEKB have
demonstrated the value of high quality PID.
High efficiencies for ``wanted" 
particles and low misidentification rates for ``unwanted" 
particles are \textit{both} crucial.
However, the capability to reach low misidentification rates is especially important
when low rate ``rare processes" 
are under study, as will be the case at \superb.
A positive signal for all particles (such as provided by imaging
Cherenkov counters, coupled with \dedx\ over the entire range of momentum)
is crucial for reaching the required efficiencies and levels of misidentification.
Modest upgrades to the barrel \babar\ \dirc\ and \dedx\ systems,
to cope with the higher data taking rates, and to provide more
headroom against backgrounds, appear to meet the minimal requirements
for the barrel region.
Further upgrades to the \dirc\ improve performance at higher cost.
A forward endcap is included in the proposed detector to improve
PID hermiticity, which should be especially beneficial for recoil
physics.

\afterpage{\clearpage}

\section{Electromagnetic Calorimeter}
\label{sec:det:EMC}



\subsection{Introduction}

\begin{figure}[!h]
  \begin{center}
  \includegraphics[width=0.6\textwidth]{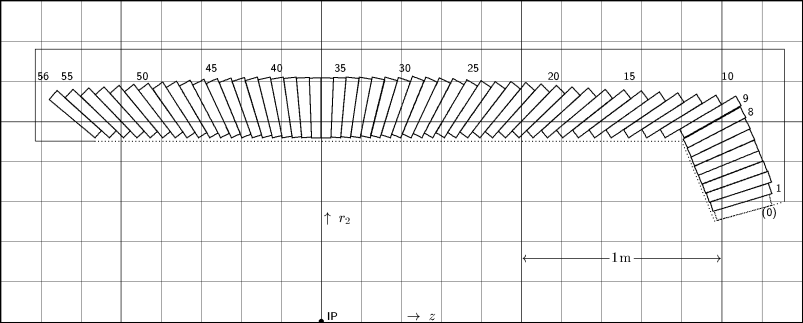}
  \includegraphics[width=0.242\textwidth]{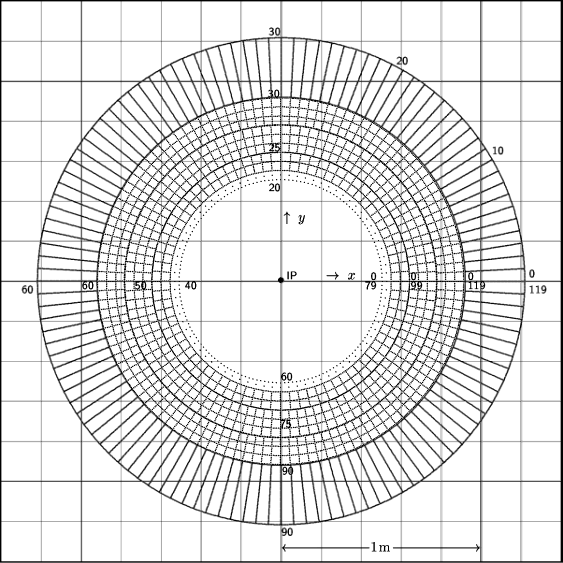}
  \caption{Schematic representation of the CsI(Tl) crystal layout of the \babar\ EMC}
  \label{fig:emc-babar}
  \end{center}
\end{figure}

The \superb\ electromagnetic calorimeter (EMC) is a cylindrically symmetric array of scintillating
crystals that measure the energy and direction of
electrons and photons, the direction of
neutral hadrons such as \KL, and discriminate between electrons and
charged hadrons.  The \superb\ EMC uses the barrel portion of the
\babar\ EMC, consisting of 5760 CsI(Tl) crystals, shown in
Fig.~\ref{fig:emc-babar}.  The forward endcap will be rebuilt using
a faster and more radiation resistant scintillating crystal, such as
L(Y)SO.  An optional backwards endcap calorimeter is also discussed.

Many general purpose $4\pi$ detectors have over the last three decade incorporated high
quality crystal calorimeters. Table~\ref{table:calorimeters} compares the parameters of a selection of these calorimeters. Several different types of crystals, having different light output, radiation length,scintillation decay time and radiation hardness,
have been used in these detectors. Table~\ref{table:crystals} compares some of the salient characteristics of the most widely used crystals.

{\setlength{\tabcolsep}{3pt}  
\begin{table}[!t]
\caption{\label{table:calorimeters}
Comparison between large crystal calorimeters.}
\setlength{\extrarowheight}{1pt}
\begin{center}
\begin{tabular}{lcccccc}
\hline
\hline
Experiment                    & L3 & CLEO II & \babar\ & Belle & KTeV & CMS \\
\hline
Crystal Type & BGO & CsI(Tl) & CsI(Tl) & CsI(Tl) & Pure CsI & PbWO$_4$ \\
Inner Radius (m) & 0.55 & 1.0 & 1.0 & 1.25 & N/A & 1.29 \\
\# of Crystals & 11400 & 7800 & 6580 & 8800 & 3300 & 76000 \\
Crystal Length ($X_0$) & 22 & 16 & 16-17.5 & 16.2 & 27 & 25 \\
Photosensor & Si PD & Si PD & Si PD & Si PD & PMT & APD \\
Energy Resolution (1GeV) & 2-3\% & 2-3\% & 2.7\% & 2.5\% & 2\% & 12\% \\
Noise (MeV)   & 0.8 & 0.5 & 0.15 & 0.2 & 1 & 40 \\
Dynamic Range & $10^5$ & $10^4$ & $10^4$ & $10^4$ & $10^4$ & $10^5$ \\
\hline
\end{tabular}
\end{center}
\end{table}
}

{\setlength{\tabcolsep}{3pt}  
\begin{table}[!h]
\caption{\label{table:crystals}
Properties of different crystals. }
\begin{center}
\setlength{\extrarowheight}{1pt}
\begin{tabular}{lcccccc}
\hline
\hline
Crystal Type & NaI & BGO & CsI(Tl) & Pure CsI & PbWO$_4$ & LSO \\ \hline
Density (g/cm$^3$) & 3.67 & 7.13 & 4.51 & 4.51 & 8.3 & 7.40 \\
Radiation Length (cm) & 2.59 & 1.12 & 1.86 & 1.86 & 0.89 & 1.14 \\
Moliere Radius (cm) & 4.13 & 2.23 & 3.57 & 3.57 & 2.00 & 2.07 \\
Interaction Length (cm) & 42.9 & 22.8 & 39.3 & 39.3 & 20.7 & 20.9 \\
Hygroscopic & Yes & No & Slight & Slight & No & No \\
Peak Luminescence (nm) & 410 & 480 & 550 & 420/310 & 425/420 & 402 \\
Decay Time (ns) & 230 & 300 & 1250 & 35/6 & 30/10 & 40 \\
Light Yield (\%) & 100 & 21 & 165 & 3.6/1.1 & 0.3/0.08 & 83 \\
d(LY)/dT (\%/degC) & 0 & -1.6 & 0.3 & -0.6 & -1.9 & 0 \\
Radiation Damage & Yes & 20\%/krad & 10\%/krad & 2\%/krad & Small & Small \\
Thermal annealing & Yes & Yes & Slow & Slow & Yes & Yes? \\ \hline
\end{tabular}
\end{center}
\end{table}
}

\subsection{Performance}

\begin{figure}[!h]
  \includegraphics[width=0.52\textwidth]{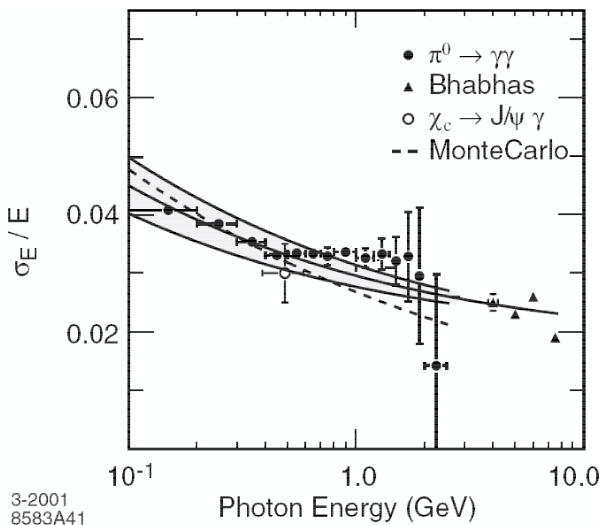}
  \includegraphics[width=0.49\textwidth]{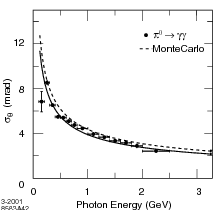}
  \vspace*{2mm}
  \caption{Photon energy resolution (left) and angular resolution (right)
  of the \babar\ electromagnetic calorimeter}
  \label{fig:emc_resolution}
\end{figure}
\begin{figure}[!h]
  \begin{center}
  \includegraphics[width=0.7\textwidth]{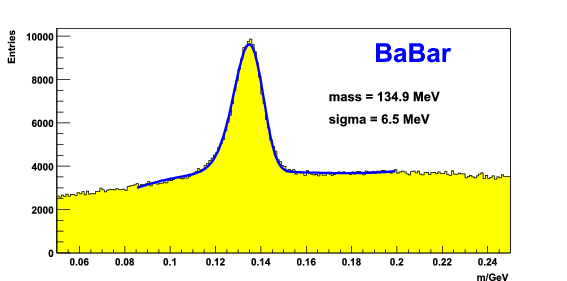}
  \vspace*{3mm}
  \includegraphics[width=0.6\textwidth]{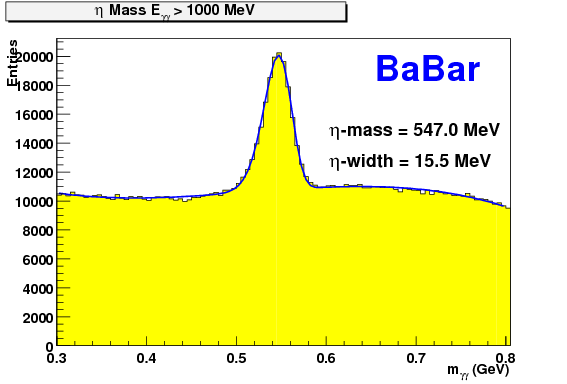}
  \caption{Mass peaks for $\pi^0$ (top) and $\eta$ mesons (bottom)}
  \label{fig:emc_pi0}
  \end{center}
\end{figure}

A summary of the performance of the \babar\ calorimeter can be
found in reference~\cite{ref:Kocian}.
The energy and position resolution for photons is
shown in Fig.~\ref{fig:emc_resolution},
and they are parametrised by:
\begin{displaymath}
{{\sigma_E}\over{E}} = {{2.30\%}\over{\sqrt[4]{E(GeV)}}} \oplus 1.35\% \hspace{1cm}
\sigma_{\theta}={{4mrad}\over{\sqrt{E(GeV)}}}
\end{displaymath}
The $\pi^0$ mass resolution is 6.5~MeV for $\pi^0$ energies above 300~MeV.
The mass peaks for $\pi^0$ and $\eta$ mesons are shown in Fig.~\ref{fig:emc_pi0}.
The resolution functions for the $\pi^0$ mass and energy have tails on the low side
which are well-described by a Crystal Ball function~\cite{ref:CBshape}.

The Belle calorimeter has a marginally better performance than \babar, with a $\pi^0$
mass resolution of 5-6~MeV in the energy range 100MeV-1GeV~\cite{ref:BelleEcal}.
The main differences between the two calorimeters are that the Belle CsI(Tl) crystals
are further from the interaction point, that they have a longer shaping time,
and do not use waveform digitization for the readout electronics.

The electromagnetic calorimeter, the key detector for electron identification,
uses a combination of $E/p$, track-shower matching, and shower shape information
to achieve an identification efficiency of 94\% for electrons above 700~MeV, with
misidentification rates of
order 0.1\%. There is also useful information for muon identification from the minimum
ionizing energy deposit and the shower shape. About half of all $K_L$ interact
in the CsI(Tl) crystals yielding a position measurement, but little energy information.

Combinatorial backgrounds are large for low energy $\pi^0$s; many analyses therefore
limit themselves to $\pi^0$ energies above 300MeV.  In \babar, this background arises
mostly from the rest of the $B\bar{B}$ event, with only a small component
from beam backgrounds. It is difficult to see how to improve this background
at \superb, but it is important not to make it significantly worse.

The angular coverage for reconstructing good photons in \babar\ is 82\%, with most
of the loss in the forward direction. After applying shower quality cuts and defining
a $\pi^0$ mass window, the reconstruction efficiency for $\pi^0$s is about 60\%.
Reconstruction of $B$ decays with one $\pi^0$ is quite efficient, but the
efficiency falls off rapidly as more $\pi^0$s are added.
The efficiency for $B$ reconstruction, which is
critical for $B$-tagged analyses, can be improved by increasing the solid angle
coverage for photons detection. Part of this improvement will come from the reduction in the
boost of the \FourS; further improvement is possible if good photon
identification can be added in the backward endcap region.

Improvements in EMC solid angle coverage benefit other analyses as well. In particular,
 better hermeticity improves the quality of neutrino reconstruction, and the
ability to veto background events on the the through the detection of excess
neutral energy is crucial to studies of $B^+\rightarrow\tau^+\nu_{\tau}$.

\subsection{Effect of dead material}


The \babar\ barrel crystals are separated from each other by 350$\mu m$ of wrapping material
(Tyvek, Al foil and mylar), and by up to 750$\mu m$ of carbon-fiber mechanical support
structure. This dead material between the crystals consumes $\sim$1\% of the solid angle
in $\phi$. It has less effect in $\theta$, because the crystal boundaries do not point
towards the interaction region. The dead material leads to a tail in the shower energy response
which has been studied with $\mu\mu\gamma$ events.
For 500 MeV photons an energy loss of 2\% is observed near the edges
of the crystals, consistent with the expectation from Monte Carlo simulation.
When this effect is calibrated out the photon energy
resolution improves by 10\%.

Since the solid angle loss is small, and the edge effects can be calibrated
out, there does not seem to be a strong argument for reducing the dead material
between the crystals. This would require the disassembly and reassembly of the barrel
modules down to individual crystals, which would be a significant amount of work.

The DIRC bar thickness is 19\% of a radiation length at normal incidence, and
30\% of a radiation length at the forward and backward ends of the barrel.
The effect of photon conversions in the DIRC has been studied by identifying
Cherenkov photons in the DIRC associated with neutral energy deposits in the calorimeter.
For 10\% of all 500 MeV photons more than 15 Cherenkov photons are found, with a smooth
distribution extending up to a maximum of 100 Cherenkov photons. Below 15 photons
it becomes difficult to identify the conversions. For the identified conversions,
an energy loss of 0.5 MeV/Cherenkov photon is observed; this is the leading factor in the tails of
the  photon energy response function. No satisfactory method has been devised to correct
for this tail, but it is possible to veto identified conversions, reducing
the tail of the response function significantly, at the expense of a 10\% loss in efficiency.

In the forward region, the 12 mm aluminium endplate of the current \babar\ drift chamber
places an average of 15\% of a radiation length in front of the forward endcap.
In the backward region this increases to 30\%, and the effect of the drift chamber readout
electronics and cables should also be included. If a backward endcap calorimeter
were to be added, it would be necessary to redistribute the dead material more
equally between the ends of the drift chamber, and to reduce the thickness in
 radiation lengths of the readout electronics as far as possible.

An option to add particle identification in the forward and backward regions
is being considered. This will add dead material on top of that from the drift chamber;
the amount depends on the choice of technology and the position of the
photodetectors. In some cases it may be possible to use these PID devices as active
preshower detectors in the same way as the DIRC.  The balance between improving
PID coverage and reducing the performance of the endcap calorimetry requires
further detailed physics studies.

\subsection{Backgrounds}

Calorimeter backgrounds arising from accelerator beam- and luminosity-related
effects impact calorimeter performance in several ways.  The most direct
impact is the deposition of energy in calorimeter crystals which exceeds the
effective threshold for reconstruction of a cluster, or for inclusion of the
crystal energy into an adjacent cluster arising from physics sources.  In the
first case, the result is the production of spurious neutral clusters, which
degrade the resolution of ``inclusive'' energy reconstruction (such as for
$\nu$/missing energy measurements) and increase the combinatorial background
in $\pi^0$ reconstruction.  In the second case, the resulting increased
crystal occupancy degrades cluster energy resolution and can negatively impact
cluster reconstruction performance.  Less direct, but no less important,
impacts are performance degradation due to cumulative radiation damage,
and data acquisition issues due to high calorimeter occupancy.

In the \babar\ and Belle experiments, the main contributions to
calorimeter backgrounds arise from single-beam lost-particle sources and
from small angle radiative Bhabha events. In both cases, a high-energy
primary $e^\pm$ or $\gamma$ strikes a beamline element within a few meters
of the IP and shower secondaries with energies ranging from sub-MeV to several
tens of MeV reach the calorimeter.  Calorimeter backgrounds resulting from
MeV-energy neutron production via the giant dipole resonance has also been
seen in \babar, in both simulation and data.  Neutron production appears
to be mainly associated with luminosity or single-beam sources in which the
primary electron or positron strikes the upstream or downstream septum chambers
in the vicinity of the Q2 magnets (about 2.5m from the IP).  Sub-MeV neutrons
then propagate into the detector and produce a delayed response with a rate that
scales as $\sim 1/r^2$ where $r$ is the distance from the production source.
Touschek scattering backgrounds have also been demonstrated by Belle, but not \babar, and
are expected to be potentially significant in at \superb, due
to the short Touschek lifetime of the beams.  These backgrounds have
not as yet been simulated, since they depend on the details of the accelerator
lattice design.  They are, however, expected to behave in a manner similar to
lost-particle bremsstrahlung events.  The impact of all of these sources can be
minimized by appropriate design of the accelerator final-focus region and IR apertures.

The primary concern for the calorimeter is the rate of small angle radiative Bhabha
events, which already dominates at \babar, and since it obviously scales with
luminosity, is potentially much worse at \superb.  The absence of dipole
magnets near the IP in \superb\  is expected to dramatically
improve the situation compared with naive scaling of \babar\ background rates.
The initial lattice design has not, however, yet been optimized to reduce this
background source.  Simulations using both simple ray-tracing (Magbends)
and full GEANT4 simulation of magnetic fields and particle interactions indicate
that the current IR layout results in a larger fraction of radiative Bhabha events
depositing energy near the IP than the present \babar/PEP-II layout.  Full GEANT4
simulation of radiative Bhabha events indicates that, in the absence of shielding,
the calorimeter background
would be dominated by $\sim 1$ MeV EM shower debris with an energy flux rate peaking at
approximately 20 MeV/$\mu$s per (CsI(Tl)) crystal in the forward barrel region
of the calorimeter.  A relatively weak $\phi$ dependence of the background rates
is observed, with the highest flux rate observed in the horizontal plane in the
positive $x$ direction.  Lower energy fluxes of at most $\sim 5$ - $10$ MeV/$\mu$s per crystal
are predicted in the forward and backward endcap regions. The resulting occupancy
rates and radiation doses are not anticipated to be problematic for L(Y)SO or pure
CsI crystals, which have considerably faster response time, are more radiation hard
and, in the case of L(Y)SO, permit finer segmentation than CsI(Tl).
Background rates
in the forward barrel are, however, potentially problematic.  With a decay time (for the slow
 component) of $\sim 1250$ns, CsI(Tl) crystal performance would be significantly degraded
 in the presence of a flux rate of much more than $\sim 1$ MeV/$\mu$s.  With additional shielding
added to the simulation, energy flux rates are significantly reduced, with rates
peaking at $\sim 1.5$MeV/$\mu$s per crystal.  It should
be noted that, since optimization of the IR magnets, apertures and shielding to minimize backgrounds
 has not yet been completed, these estimates are believed to be quite conservative.
  It will, however, be imperative to perform additional simulation studies as part of the
ongoing design effort, using an optimized IR and detector layout (including luminosity, Touschek
and single-beam lost-particle sources), in order to verify that calorimeter background
 rates are acceptable.

\begin{figure}[htbp]
  \begin{center}
  \includegraphics[width=0.8\textwidth]{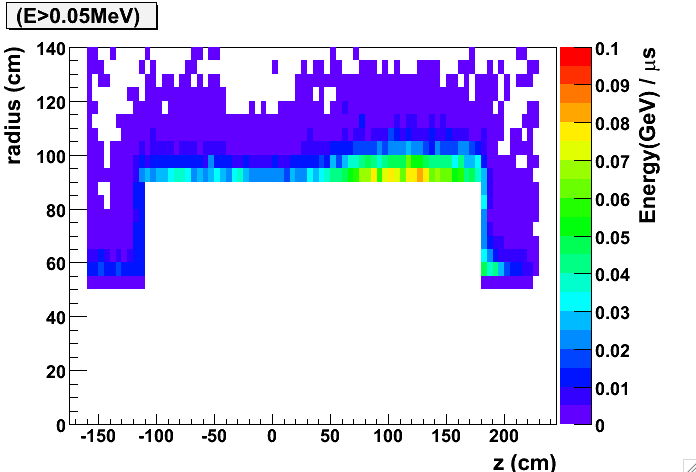}
  \caption{Energy flux rates from radiative Bhabha luminosity background simulation in the
EMC detector volume, with shielding in place.  Bins represent $5$cm $\times 5$cm regions in the
$r-z$ plane, integrated over all $\phi$.  The energy flux in a hypothetical CsI(Tl) crystal in the
forward barrel region would correspond to about 1 -1.5 MeV/$\mu$s. }
  \label{fig:emc_bg_shielding}
  \end{center}
\end{figure}

\subsection{Radiation damage}
\begin{figure}[htbp]
  \begin{center}
  \includegraphics[width=0.75\textwidth]{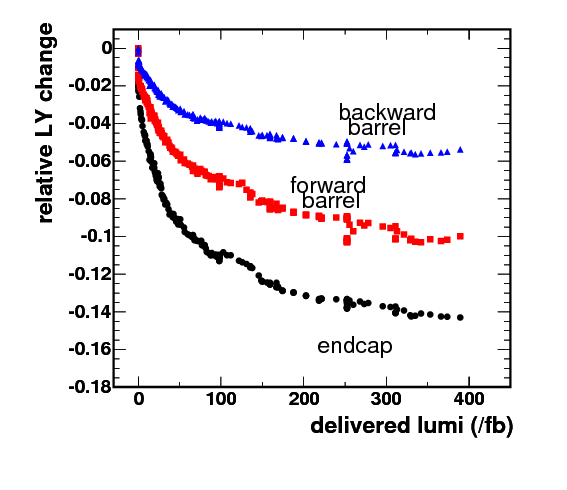}
  \caption{The light yield loss in the \babar\ CsI(Tl) crystals
due to radiation damage as a function of luminosity. The total dose received
after 300\invfb is 1.2krad in the endcap and 750rad in the barrel.}
  \label{fig:emc_raddam}
  \end{center}
\end{figure}

Radiation damage impacts CsI(Tl) through the creation of color centers in the crystals,
resulting in a degradation of response uniformity and light yield.  The nominal dose
budget for the \babar\ CsI(Tl) calorimeter is 10krad over the lifetime of the detector.
Pure CsI and L(Y)SO are considerably more radiation hard (see Table~\ref{table:crystals}).
The dominant contribution to the dose arises from luminosity and single-beam
background sources, and hence is due to MeV-level photons and (presumably) neutrons;
the integrated dose scales approximately linearly with integrated luminosity.  The
measured reduction of light yield due to radiation damage is shown as a function of
integrated luminosity in Fig.~\ref{fig:emc_raddam}.  To date, a total dose of
about 1.2krad has been received in the most heavily irradiated regions, resulting in a
 loss of about $\sim 15\%$ of the total light yield, but with no measurable impact on
physics performance.  It is notable that most of the observed light loss
occurred relatively early in \babar\ running, although radiation dose has been accummulating
 relatively steadily, and that crystals from different manufacturers have responded somewhat
 differently to irradiation.  It is anticipated that the CsI(Tl) barrel will have accumulated
 approximately 1.5krad in the most irradiated regions by the end of nominal \babar\ running
 in 2008.  In order for the barrel calorimeter to function in the \superb\ environment, beam
 background rates must be maintained at a level of approximately 1 MeV$/\mu$s or less per
CsI(Tl) crystal.  If this condition is achieved, then radiation dose rates are anticipated
 to be roughly comparable to current \babar\ levels.  A dose budget of well under 1 krad/year
 is expected to be achievable. At this a level, the CsI(Tl) barrel would survive for the
duration of \superb\ operations.  This assumption will, however, need to be verified by
detailed simulation.

\subsection{Barrel calorimeter}


The CsI(Tl) crystals used in the barrel calorimeters
of both \babar\ and Belle are the most expensive elements
of the two detectors. Based on the performance that has been achieved,
and the radiation damage that has been observed so far, both collaborations
have concluded that the re-use of the barrel crystals is possible at
a Super $B$ Factory.

The baseline assumption is that the geometry of the crystals is
unchanged from that of the current \babar\ detector. The one change that
should be made is to move the position of the interaction point
from -5cm to +5cm relative to the position of the crystal gap
normal to the beam axis. This adjustment retains the current
non-pointing geometry, but moves the barrel to a slightly more
symmetric position, in view of the reduced energy asymmetry.
The effect of the change in boost from $\gamma\beta=0.56$ to 0.28
and the shift of the IP is to increase the angular coverage of the
barrel from 79.5\% ($\cos\theta = -0.931$ to $+0.661$)
to 84.1\% ($\cos\theta = -0.883$ to $+0.798$).

If the crystal geometry is unchanged, it is possible to transport the
entire barrel calorimeter as one cylinder. Alternatively it
could be disassembled into its 280  individual modules, which would be transported
separately and reassembled on arrival. It would only be necessary
to disassemble the modules themselves if changes were being made to the
material between the crystals, or to the photodiode readout.
The costs of these alternatives are discussed in the chapter on the
re-use of existing \babar\ detector elements.

A possible change to improve the coverage in the backward region
would be to add one or two additional rings of crystals to the
last module ring in $\theta$, which currently only contains 6 rings of crystals.
However, this would lead to major changes in the mechanical support
structure and a redesign of the electronics readout, so it will not
be undertaken unless there is a significant gain from the extra ring(s).
Changes to the rear sectionof the barrel also clearly interact strongly with the
possible addition of a backward endcap calorimeter (see below).


\subsection{Forward endcap calorimeter}


In contrast to the barrel EMC, it is desirable to replace the EMC forward endcap, and possible
to do so at a comparatively modest cost.  In \babar, the innermost rings of the
forward endcap are subject to high radiation doses from both luminosity and single-beam background
sources, due to the proximity to the Q2 septum region which acts as a background source.  As such,
it is expected to have accumulated substantial radiation damage by the end of the nominal \babar\
program.  Redesign of the forward endcap region also permits the solid angle coverage to be optimized
for the SuperB machine and potentially permits space to be freed up, through the use of compact
L(Y)SO crystals, for a forward PID device.

The use of cerium-doped silicate crystals (GSO, LSO, LYSO) has
been developed for medical imaging. Large crystals of lutetium (yttrium) oxyorthosilicate
(L(Y)SO), have been obtained from several vendors and tested for properties important in their use
for high energy calorimetry by R.Y. Zhu at Caltech.
\cite{ref:RYZhu}. The combination of fast decay time (40ns), high light output
(50\% of CsI(Tl)), and negligible radiation damage, make them an attractive option.
 These crystals are also
non-hygroscopic and mechanically strong, simplifying the design of mounting structures.
In recent tests, readout of L(Y)SO crystals with APDs has been demonstrated
with a signal of 1500$p.e.$/MeV and readout noise $<$40keV.

The forward coverage is limited to 300 mrad by accelerator components in the IR.
The effect of the change in boost from $\gamma\beta=0.56$ to 0.28
and the shift of the IP by +10cm is to decrease the angular coverage of the
forward endcap from 9.4\% ($\cos\theta = 0.661$ to $0.849$)
to 6.2\% ($\cos\theta = 0.798$ to $0.921$), but also to decrease the loss of solid angle coverage
below 300 mrad by a factor of two.

Table~\ref{table:LSO} shows a possible layout for a forward endcap containing a
total of 2520 L(Y)SO crystals with dimensions of approximately 25mm$\times$ 25mm$\times$ 200mm.
The total volume of this design is 0.36m$^3$, giving a total crystal cost of
\$5.4M based on existing quotes for large LSO crystals of \$15/cc,
or \$3.6M if we assume a reduction to \$10/cc for a bulk order of a large number of crystals.
Note that the cost of the L(Y)SO crystals increases by 13\% if they are moved back from
the front corner of the barrel calorimeter by the maximum available space (125mm),
in order to make way for a forward particle identification system.

\begin{table}[htb]
\caption{\label{table:LSO}
A possible design for an L(Y)SO forward endcap.
All crystals are 200mm long, and the endcap is angled at 20$^{\circ}$ to the vertical
as in the current \babar\ detector.}
\begin{center}
\begin{tabular}{ccccc}
\hline
\hline
Ring in $\phi$ & Radius &  Crystal Face & Crystal Volume &  \# Crystals \\
  & (mm) &  (mm) & (cc) &  \\
\hline
1 & 597-620 & 24.4$\times$ 31.9 & 171 & 120 \\
2 & 620-643 & 24.4$\times$ 33.1 & 178 & 120 \\
3 & 643-666 & 24.4$\times$ 29.4 & 158 & 140 \\
4 & 666-689 & 24.4$\times$ 30.5 & 164 & 140 \\
5 & 689-712 & 24.4$\times$ 27.5 & 148 & 160 \\
6 & 712-735 & 24.4$\times$ 28.4 & 152 & 160 \\
7 & 735-758 & 24.4$\times$ 26.1 & 140 & 180 \\
8 & 758-781 & 24.4$\times$ 26.9 & 144 & 180 \\
9 & 781-804 & 24.4$\times$ 24.9 & 134 & 200 \\
10 & 804-827 & 24.4$\times$ 25.6 & 137 & 200 \\
11 & 827-850 & 24.4$\times$ 23.9 & 128 & 220 \\
12 & 850-873 & 24.4$\times$ 24.6 & 132 & 220 \\
13 & 873-896 & 24.4$\times$ 23.2 & 125 & 240 \\
14 & 896-919 & 24.4$\times$ 23.8 & 128 & 240 \\
\hline
\end{tabular}
\end{center}
\end{table}

An alternative choice for the forward endcap is pure CsI. This option has been
studied by the Belle collaboration.
In this case the geometry of the forward endcap can be kept the same
as the existing \babar\ endcap, with 820 crystals occupying a volume
of 0.71m$^3$ in 8 $\phi$ rings.
With an estimated cost of \$4/cc for pure CsI, the total crystal cost is \$2.7M,
which is less than the cost of L(Y)SO, but not by a large factor.


\begin{figure}[!b]
  \begin{center}
  \includegraphics[width=0.8\textwidth]{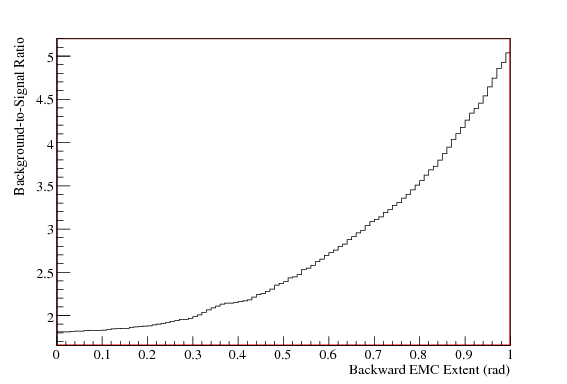}
  \caption[Background-to-signal ratio in the analysis of $B\to\tau\nu$ as a function
of the backward extent of the calorimeter]
{Background-to-signal ratio in the analysis of $B\to\tau\nu$ as a function
of the backward extent of the calorimeter. In this study, the energy resolution
of the calorimeter is severely degraded below 700 mrad to simulate performance
as a ``veto'' device.}
  \label{fig:emc_taunu_bwdacceptance}
  \end{center}
\end{figure}

\subsection{Backward region calorimetry}


The effect of the change in boost from $\gamma\beta=0.56$ to 0.28
and the shift of the IP by +10cm is to increase solid angle not covered by
the backward barrel from 3.5\% to 5.9\%, which makes it slightly larger
than the region below the forward endcap. Taking account of the 300 mrad
stay clear line, a maximum of 4.5\% of the total coverage could be
recovered by installing a backward endcap calorimeter. However, the
presence of the DIRC bars, and the material associated with the DCH
endplate and readout, make it very difficult to design an appropriate
calorimeter, and the actual gain in coverage is likely to be significantly
less than this.

One of the strengths of the \superb\ physics program is the ability to
study fully inclusive decays and decay modes containing missing energy
 ({\it e.g.} neutrinos) or single photons ({\it e.g.} radiative decays).  Hermeticity
 impacts these channels in two ways:  First, reduced hermeticity potentially
 degrades the resolution of inclusive measurements, such as the hadronic mass
 spectra in $b \to s \gamma$.  Second, it degrades the background rejection
 power in analyses that rely on missing energy or $\pi^0$ or $\eta$ vetos.
It is therefore important to maintain the most hermetic calorimetry possible.
The impact of a backward endcap calorimeter on such analyses was studied in
the context of \superb, using $B^+ \to \tau^+ \nu_{\tau}$ as a benchmark.
This analysis relies heavily on the detection
of soft neutrals to distinguish the low-multiplicity signal mode from
higher-multiplicity backgrounds. ``Irreducible'' backgrounds arise when one
or more particles pass outside of the detector acceptance.
Figure~\ref{fig:emc_taunu_bwdacceptance} shows the background-to-signal ratio in
a simulated analysis of $B^+ \to \tau^+ \nu_{\tau}$ as a function of the
acceptance in the backward direction. It is seen that the background can be
significantly reduced by extending the calorimeter in the backward direction.
  Since just the presence
of any significant energy in this region would indicate that an event
 is not signal, the energy resolution of the calorimeter is less critical
 than the angular coverage.  Consequently, the backward endcap option is
 being considered primarily as a ``veto'' device.  The baseline option
for this device is a series of L(Y)SO rings of design similar to the forward
 endcap which would be fitted behind the DCH endplate and electronics and
 inside the radius of the DIRC bars.  It would be desirable to avoid a gap
 in coverage between the back of the barrel and the backward endcap, but
 at this time it is not clear whether this can be achieved.  Space constraints
 dictate the use of L(Y)SO rather than pure CsI, due to the smaller radiation
 length and Moli\`ere radius.  Dead material associated with the DCH in
front of the endcap is expected to degrade the energy resolution of the
 endcap calorimeter, but not to significantly impact its operation as a veto
device.  The total amount of material in front, as well as the details of
 the layout and effective geometrical coverage will depend on the amount of space
 that can be gained by a redesign of the DCH readout.

\afterpage{\clearpage}

\section{Instrumented Flux return}
\label{sec:det:IFR}
%
%
%
%

\begin{figure}[htb]
\begin{center}
\includegraphics[width=0.5\linewidth]{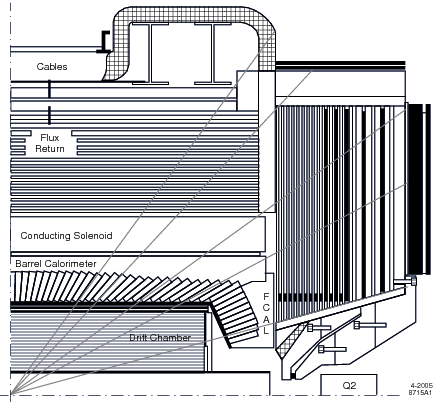}
\caption{ BaBar Instrumented Flux return}
\label{ifr:babar_ifr}
\end{center}
\end{figure}

The Instrumented Flux Return (IFR) is designed primarily to identify muons, and, in conjunction with the electromagnetic calorimeter, to identify neutral hadrons, such as $\KL$ and neutrons. This section describes the performance requirements and a baseline design for the IFR. The iron yoke of the detector magnet provides the large amount of material needed to absorb hadrons. The yoke is segmented in depth, with large area particle detectors inserted in the gaps between segments, allowing the depth of penetration to be measured. In Fig.~\ref{ifr:babar_ifr} we show a schematic view of the BaBar IFR.

\subsection{Physics Goals}

A muon identification system must have reasonably high efficiency for selecting penetrating particles such as muons, while at the same time rejecting charged hadrons (mostly pions and kaons).
Such a system is critical in separating signal events in $b \rightarrow s \ell^+\ell^-$  and $ \rightarrow d  \ell^+\ell^-$  processes from background events originating from random combinations of the much more copious hadrons.
  Positive identification of muons with high efficiency is also important in rare  $B$ decays as $B  \rightarrow \tau \nu_{\tau} (\gamma)$, $ B \rightarrow \mu \nu_{\mu} (\gamma)$  and  $B_d(B_s) \rightarrow \mu^+ \mu^-$ and in the search for lepton flavour-violating processes such as   $\tau \rightarrow \mu \gamma$.

\begin{figure}[htb]
\begin{center}
\includegraphics[width=0.45\linewidth]{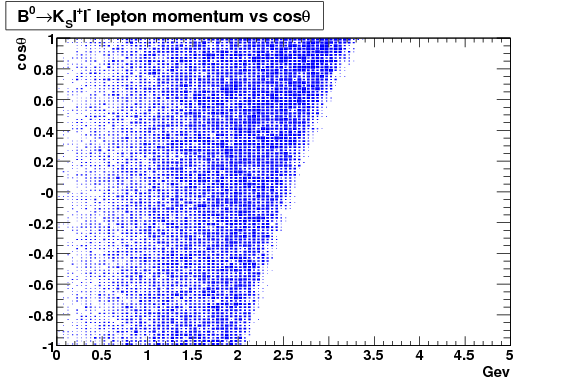}
\includegraphics[width=0.45\linewidth]{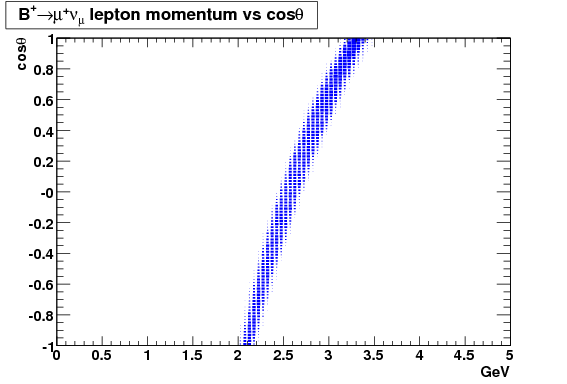}
\vspace*{2mm}
\caption{Scatter plot of $\cos{\theta}$ {\it vs.} momentum in the laboratory frame for muons from  $B^0  \rightarrow K_S \mu^+ \mu^-$ and $B^+ \rightarrow \mu^+ \nu_{\mu} $decays; $\beta \gamma = 0.28$ is assumed}
\label{ifr:benchmark_theta}
\end{center}
\end{figure}

    Momentum and polar angle distributions in the laboratory system for several of these channels are shown in Fig. \ref{ifr:benchmark_theta}.  The nominal boost of $\beta \gamma$ $=$ $0.28$ is assumed. Despite the boost, a muon detector system that is  symmetric around the interaction region is a suitable match to the physics goals.

    Background suppression in reconstruction of final states with missing energy carried by neutrinos (as in $ B \rightarrow \mu \nu_{\mu} (\gamma)$) can profit from vetoing the presence of energy carried by neutral hadrons.
     About 45\% of relatively high momentum $\KL$'s interact only in the \babar\ muon system. Some $\KL$ identification capability is therefore required.  On the other hand, having a muon system that is hermetic as possible down to small polar angles is problematic. since most of the background is concentrated in the region close to the beamline, thereby limiting the ultimate veto performance.


\subsection{Identification Technique}

Muons are identified by measuring their penetration depth in an absorber consisting of the iron of the return yoke of the solenoid magnet.
Hadrons shower in the iron, which has a hadronic interaction length   $\lambda_I = 16.5 \cm$~\cite{pdg2006}.
 The survival probability to a depth $d$ scales as $\exp^{-d/\lambda_I}$). Fluctuations in shower development and decay in flight of hadrons with muons in the final state are the main source of hadron misidentification as muons.   The penetration technique has a reduced efficiency for muons with momentum  below  1 $GeV$,  due to ranging out of the charged track in the absorber.  Moreover, only muons with a sufficiently high transverse momentum can penetrate the IFR to sufficient depth to be efficiently identified.

\begin{figure}[htb]
\begin{center}
\includegraphics[width=0.45\linewidth]{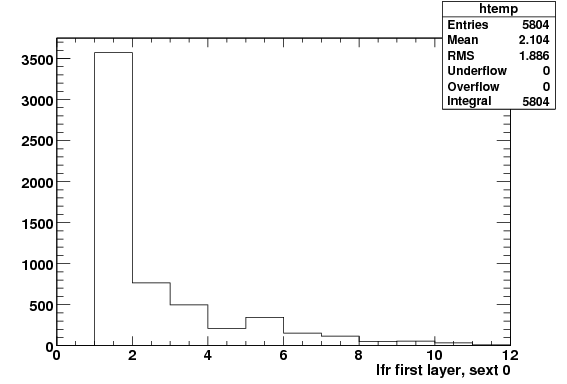}
\includegraphics[width=0.45\linewidth]{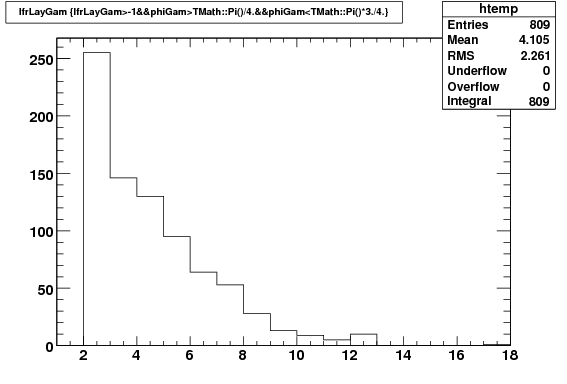}
\caption{\babar\ MC simulated events of $\KL$ in barrel sector of the \babar\ IFR, (left) distribution of the first layer hit by a $\KL$ and (right) distribution of number of hit layers in a $\KL$ shower}
\label{fig:ifr_KLhitsbarrel}
\end{center}
\end{figure}

 Neutral hadrons interact in the electromagnetic calorimeter as well as in the flux return.
A$\KL$ tends to interact in the inner section of the absorber. Using the \babar\ IFR simulation, we show the distribution of the first layer having signal due to a $\KL$ interaction in Fig.\ref{fig:ifr_KLhitsbarrel} for $\KL$'s impinging on the barrel sector with a momentum  in the range from  0.5  to 4 GeV$/c$. A $\KL$-initiated shower develops, on average, over about 4 layers (events with a single layer hit are not considered). See Fig.~\ref{fig:ifr_KLhitsbarrel}.

  $\KL$ identification capability is therefore mainly dependent on energy deposited in the inner part of the absorber.  Best performance can be obtained by combining the initial part of a shower in the electromagnetic calorimeter and with the rear part in the inner portion of the IFR. An active layer between the two subsystems, external to the solenoid, is therefore highly desirable.

\subsection{Baseline Segmentation Design}

\begin{figure}[htb]
\begin{center}
\includegraphics[width=0.5\textwidth]{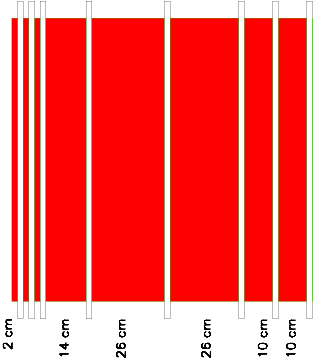}
\caption{Sketch of the longitudinal segmentation of the iron absorber (gray). Active detector positions are shown in white from the innermost (left) to the outermost (right) layers  }
\label{ifr:drawing}
\end{center}
\end{figure}

 The total amount of material in the \babar\ detector flux return (about 5 interaction length at normal incidence in the barrel region) is suboptimal for $\mu$ identification~\cite{Aubert:2001tu}.  Adding iron to the \babar\ flux return for the upgrade to the \superb\ detector can produce an increase in the pion rejection rate at a given muon identification efficiency.
  A possible longitudinal segmentation of the iron is showed in Fig.~\ref{ifr:drawing}.  The three inner detectors are most useful for $\KL$ identification; the coarser segmentation in the following layers preserves the efficiency for low momentum muons. The current \babar\ readout segmentation (strip with  3.7 cm pitch) will be retained.

\begin{figure}[htb]
\begin{center}
\includegraphics[width=0.7\textwidth]{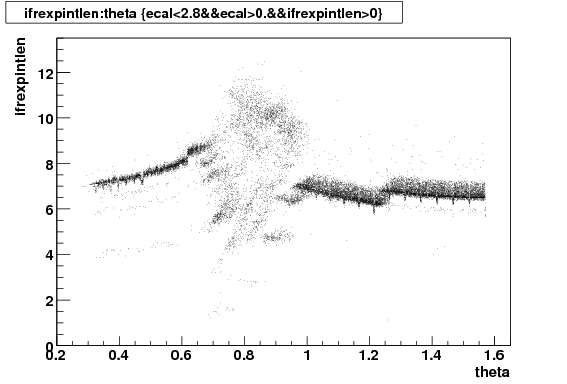}
\caption{ Number of interaction lengths {\it vs.} polar angle $\theta$ for muons of  5 \gev/$c$ momentum  in a baseline IFR  configuration for \superb\ detector.  This distribution was obtained with the current \babar\ simulation assuming maximal efficiency for active layers. The structure is due to the details of the arrangement of active layers.}
\label{fig:ifr_expintlenl}
\end{center}
\end{figure}

  Figure~\ref{fig:ifr_expintlenl}, shows the resulting number of interaction lengths as a function of the polar angle $\theta$ traversed by a muon of 5 \gev/$c$ momentum in the baseline \superb\ detector. Given the smaller boost, we adopt a nearly symmetric geometry around the interaction point; we thus show only the barrel and the region, $\theta < \pi /2$ ).

The efficiency and misidentification probability for muons and charged pions,
respectively, are shown in Table\ref{ifr:efftable} for several momenta.

\begin{table}
\caption{\label{ifr:efftable} Efficiency and misidentification probability for the baseline IFR for the \superb\ detector.}
\begin{center}
\begin{tabular}{lcccc} \hline \hline
   & \multicolumn{2}{c}  {\centering Loose selection}
   & \multicolumn{2}{c}  {\centering Very Tight selection}    \\
momentum &   $\mu$ eff.  &  $\pi$ misid.  &   $\mu$ eff.  &  $\pi$ misid. \\
\hline
0.8 \gevc          &  48        & 2.3     &  42   & 1.8 \\
1 \gevc             &   66       &  5.4    &  54  &    3.0 \\
2 \gevc             &   82       &  2.0     & 74    & 1.3 \\
5\gevc              &   84       &  1.9     &  79  &  1.2\\   \hline
\end{tabular}
\end{center}
\end{table}

\subsection{Technology Choice}

\subsubsection{The \babar\ Technologies: RPC and LST}

 The \babar\ detector uses two technologies in the IFR.

 The forward and backward endcaps use Resistive Plate Chambers (RPC's) with planar bakelite electrodes coated with linseed oil, and a mixture of freon, argon and isobutane gases.  The ionization produced along the path of the charged track is internally amplified and the electric signal induced on external copper strips is used to measure the position of the track impact point.
    The \babar\  RPC's are operated partly in streamer mode and partly in avalanche mode. Avalanche mode is preferred for the forward region around the beamline, where most of the background hits accumulate. To detect the avalanche signal, a preamplification stage is required.

       The barrel sector is instrumented with Limited Streamer Tubes (LST), made of  several square PVC cells coated with graphite and a center wire at high voltage.  The amplified ionization signal generated on the wires and induced on external readout strips is used to measure the position of the crossing track in a given layer.

       The \babar\ RPC operational experience was initially problematic~\cite{ifr_fwdupgrade}, but the forward endcap RPC's are still as efficient as they were when originally installed (single detector efficiency is, on average, in excess of 90 \%~\cite{ifr_fwdoper}).
         These chambers were manufactured to a higher standard than the initial chambers. A detailed quality assurance process, along with production improvements such as curing of the linseed oil, were similar to those adopted for the LHC and Opera detectors~\cite{ifr_fwdupgrade}.  These chambers have proven to be quite solid, and if operated in avalanche mode can sustain a rate up to several kHz/cm$^2$~\cite{ifr_atlasGIF}.
     The \babar\ LST operational experience, on the other hand,  has never been problematical, but these chambers cannot withstand rates higher than 100 Hz/cm\footnote{LST cells are about 1 cm wide, which translates into less than 100 Hz/cm$^2$} and are therefore usable only in regions in which relatively low background rates are expected.

      To reestablish the electric field after an avalanche has developed, dark current must flow through the RPC electrodes.  Dark currents are therefore a linear function of the rate of avalanches in the gas, each avalanche corresponding to charge of few tens of a pC. This current depletes the charge carriers  in the bakelite plates, which lowers the efficiency of the chambers \cite{ifr_vavraionic}.   Adding water to the gas mixture ameliorated this problem.  Nevertheless, it has been shown (\cite{ifr_atlasgraphite}) that the graphite layer used as an electrical contact to the cathode is damaged after an integrated charge of some hundreds of mC/cm$^2$ is accumulated. This sets the limit of the lifetime of such detectors, which is clearly dependent on the rate to which the chambers are exposed.
      LST ageing is characterized by the onset of a continuous discharge, caused by the Malter effect, which prevents further operation of the detectors.

\subsubsection{The \superb\ Environment}

 The operation of the RPC's in \babar\ in streamer mode was limited by the maximum rate the chambers can sustain. The LST's in the barrel have limited rate capability as well.
  High rates generate high dark currents, lowering the effective electric field across the gap and taking the counters out of their efficient operating regime.
  Avalanche mode operation for RPC  is preferred for the forward region at small polar angles where the background is highest. This background, due to particles from QED processes ({\it i.e.}, radiative Bhabhas) showering in beamline components, scales with luminosity, is only problematical in the endcap region.

\begin{figure}[htb]
\begin{center}
\includegraphics[width=0.7\linewidth]{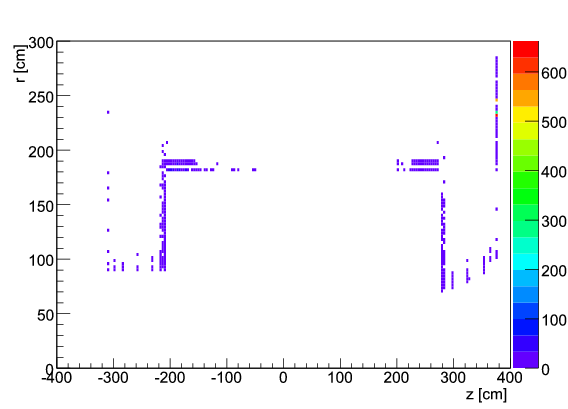}
\caption{ ($r,z$) distribution of Geant4  hits in IFR active layers according to the \superb\ interaction region simulation. The simulation corresponds to 1940 bunch collisions ( a 231 MHz collision rate).}
\label{ifr:hitdistr}
\end{center}
\end{figure}

    Other sources, such as scattering of beam particles on collimators far from the interaction region, were, in \babar, reduced to an acceptable level by external shielding of the outer layers of endcap RPC's. We have estimated these backgrounds in the \superb\ environment with the simulation described in Section \ref{sec:backgrounds}.  In Fig.~\ref{ifr:hitdistr} the ($r,z$) distribution of Geant4 hits  in IFR active layers is shown. Active layers are simulated as a 0.2 cm thick layer of RPC gas mixture (freon R134a 73.1\%,  Ar 22\%, isobutane 4.4 \% and $SF_6$ 0.5 \%). Any deposit of energy in these  gas layers is considered as a single hit. As expected, the critical regions are the small polar angles sections of the endcaps and the edges of the barrel internal layers, where we estimate that in the hottest regions the rate is a few $\times$ 100 Hz/cm$^2$.

These rates are too high for gaseous detectors. We have therefore chosen scintillator technology for the \superb\ baseline.  If more detailed background studies were to provide convincing evidence that the highest anticipated background rates were in the range of some 100 Hz/cm$^2$,  RPC's in avalanche mode could be considered as an alternative. LST's cannot sustain even these rates, and therefore the current \babar\ LST system cannot be reused.

 The effect of spurious hits on muon identification can be dealt with using offline pattern recognition techniques.  The precise timing information from RPC's and plastic scintillator detectors can be used to eliminate background hits, given suitable readout electronics.

\subsubsection{MINOS-type Scintillator Design}

 While the \babar\ experience with both RPC's~\cite{ifr_fwdoper} and LST's has been, in the end, positive, detectors with high rate characteristics are required in the high background regions of \superb. since a scintillator-based system provides much higher rate capability than the gaseous detectors; it can sustain machine backgrounds much higher than what is foreseen by the current simulation. For this reason, the baseline technology choice for the \superb\ detector is extruded plastic scintillator using WLS fiber read out with a pixelated APD. The basic scintillator bar design is similar to the active detector of the MINOS experiment~\cite{ifr_minos}).

              These detectors are straightforward to operate, as no high voltage or flammable gases are employed, do not have substantial ageing problems, and can sustain high rates. Production techniques, including characterization of scintillator bar and WLS fiber quality ,are well-established. Further optimization of the pixelated APD system and the associated data acquisition system is required.

 When the upgrade of the barrel IFR was under consideration in 2002, both LST and MINOS-type scintillator designs were developed.  The LST option was chosen for the upgrade; the scintillator design, however, is a close match to the requirements of the \superb\ environment. We will briefly describe this scintillator system~\cite{ifr_kim}.

\begin{figure}[htb]
\begin{center}
\includegraphics[width=0.5\textwidth]{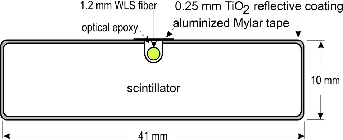}
\caption{Cross section of a MINOS scintillator bar.}
\label{ifr_scint}
\end{center}
\end{figure}

 The basic detector building block is a bar of inexpensive polystyrene scintillator, with PPO and POPOP doping, coextruded with a TiO$_2$ diffuse reflective coating, shown in Fig.~\ref{ifr_scint}.
 The bars, up to 4 meters in length, are read out by 1.2 mm diameter multiclad wavelength-shifting fibers (Kuraray Y11-175), viewed at both ends by a device such as a 64 channel RMD A6403 silicon avalanche photodiode (APD)~\cite{ifr_RMD}. Figure~\ref{ifr_scint} shows the MINOS bar dimensions and a single fiber readout. Designs with different bar dimensions and shapes, as well as multi-fiber readout, were explored in the context of the \babar\ upgrade; it is likely that a more conservative two fiber readout design would be the choice. The coordinate along the bar is provided by the time difference in the signals from both ends of the fibers, with a resolution of better than 15 cm, which is adequate.
 
\begin{figure}[htb]
\begin{center}
\includegraphics[width=0.6\textwidth]{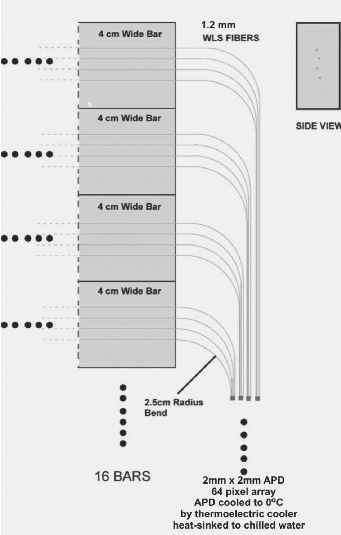}
\caption{Schematic layout of a sixteen bar module.}
\label{ifr_scint_box}
\end{center}
\end{figure}

 The bars will be assembled in groups of 16 by gluing into aluminum support structures, each read out by a single pixelated 64 channel APD device (see Fig.~\ref{ifr_scint_box}). These box structures are of uniform length in the barrel section; for the endcap region, a variety of different length bars and boxes must be produced. The noise performance of the APD's improves when they are cooled to temperatures around $0^\circ$C, which can be done with a Peltier effect cooler, heat-sinked to a cooling water loop. Dry nitrogen will be circulated within the support boxes to prevent condensation.

 The readout electronics must provide time and pulse height for each APD channel, which can be done in a straightforward way using electronics similar to that used in the DIRC system.


%

\afterpage{\clearpage}

\section{Electronics}
\label{sec:det:ELEX}

The conceptual design of the read-out electronics for the \superb
detector is based on the design used for \babar.  However, almost all
of the electronics will be new for the \superb detector, to accomodate
the greatly increased event rate.  The conceptual layout is as
follows.  Frontend analog-digital boards  shape, amplify and
digitize the detector signals. Some digital signals are ultimately
sent to the trigger system, described in Section~\ref{sec:det:TDAQ}.  All
digital information is pipelined to allow for the trigger decision
latency of roughly $12\mu sec$.  Upon receipt of a level one trigger,
data is written to a multi-event buffer to reduce dead-time.

Next data is transmitted serially to read-out interface boards, which
further sparsify the data. These boards will contain FPGAs with
integrated processors, allowing flexibility in the algorithms used.
Serial data is then sent to input/output boards for transmission over
fiber links to the DAQ's {\it cluster box}, also described in
Section~\ref{sec:det:TDAQ}.

\subsection{SVT Electronics}

The design of the SVT readout electronics is intimately connected with that of the silicon sensors, and is therefore discussed in Section~\ref{sec:det:SVT}.

\subsection{DCH Electronics}

\begin{figure}[bht]
\centerline{\includegraphics[width=5in]{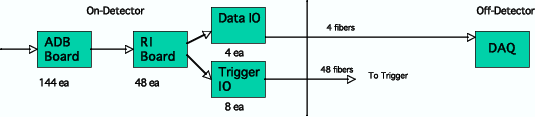}}
\caption{Schematic diagram of the \babar\ DCH electronics.}
\label{fig:elec:dch}
\end{figure}

A block diagram of the electronics system for the DCH is shown in
Fig.~\ref{fig:elec:dch}.  For the DCH, the analog-digital board
(ADB) consists of an amplifier ASIC and a combined TDC/ADC ASIC
(a replacement for the ELEPHANT chip in \babar).  If designed now, the
ASICs would use a $1/4$ micron TSMC process.  Individual ADB boards
could serve up to 64 channels per board, depending on the DCH
super-layer.  The output of the ADB board is sent to a read-out
interface board (RI), which uses an FPGA with integrated processor to
sparsify the data, and to sum the signals from the ADC for the
ionization from each DCH sense wire.  Data for the trigger is through
 dedicated trigger input/output cards, while the full readout
data is sent via input/output cards.  The input/output cards contain
the fiber links to the DAQ and Trigger systems.

\subsection{\dirc\ Electronics Upgrade}

\begin{figure}[bht]
\vspace*{8mm}
\centerline{\includegraphics[width=\textwidth]{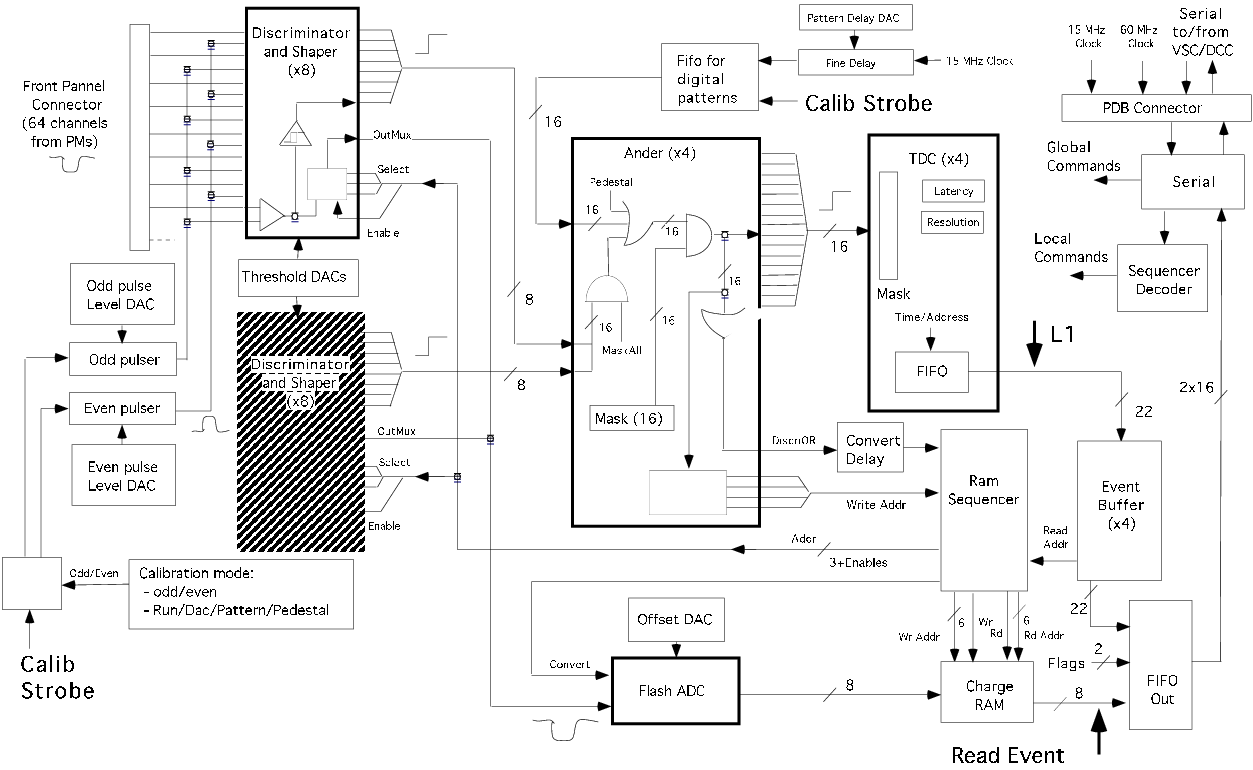}}
\caption{Schematic diagram of the \babar\ \dirc\ frontend board.
}
\label{fig:elec:dirc}
\end{figure}

The \babar\ \dirc\ frontend electronics (FEE) design is
briefly discussed in the PID portion of the detector section,
and is shown schematically in Fig.~\ref{fig:elec:dirc}.

The electronics upgrade assumed here will use similar design
principles, but is enhanced to cope with random background
rates of up to ~10\MHz/PMT with less than 1\% dead time,
while utilizing the timing resolution capabilities of
much faster PMTs (with $\sigma_{TTS}$ = 0.25-0.5\ns) that will
be employed.
The design will keep the same pipelined L1 trigger latency
of 12$\mu$s.
The frontend board's ICs and the PCB layout will
have to be modified to accommodate the higher throughput
rates, however, without major functional changes
in the architecture.
The DAQ maximum event rate will be upgraded from the
present 10 to 100\kHz, which means that the
1.2\GHz G-link will need to be upgraded.

Presently, we assume that we will be able to keep the same
VME crates, the CAEN high voltage power supplies and cables,
as well as the calibration system.
However a careful inspection of all components taken from
\babar\ will be required to verify that this is warranted.

\subsection{EMC Electronics}

\begin{figure}[bht]
\centerline{\includegraphics[width=5in]{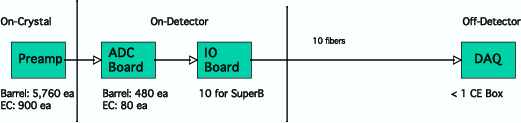}}
\caption{Schematic diagram of the \babar\ EMC electronics.}
\label{fig:elec:emc}
\end{figure}

Next a block diagram of the electronics system for the EMC is shown in
Fig.~\ref{fig:elec:emc}. Here preamplifier cards, located at each
crystal, shape and amplify the signals.  Analog signals are sent to an
analog-digital board (ADB) located on the detector.  The ADB board
contains an ASIC to auto-range the signal (a replacement for the CARE
chip in \babar) and an integrated 4~MHz ADC.  Again, if designed now
the ASICs would use $1/4$ micron TSMC process. The ADB boards will
contain the level one trigger latency pipeline, and a triggered data
buffer.  To reduce the data volume, the ADB boards will also use an
FPGA with integrated processor to determine the energy and time for
each crystal.  Thus only sparsified data will be sent to input/output
boards. From the input/output boards, serial data will be sent over
fiber links to the DAQ.

\subsection{IFR Electronics}

The MINOS-type scintillator system chosen as the \superb\ baseline for the IFR has no counterpart in the \babar\ detector. The APD readout is likely to be a 64 channel device, such as the RMD A6403~\cite{ifr_RMD}. Operated at a gain of 1000, this device would provide a signal of $\sim30,000$ electrons with two fibers per bar, each read out at both ends. After a simple preamplifier stage, the data acquisition system, which must provide amplitude and time information, can be very similar to the upgraded DIRC system discussed above, except that the trigger functionality is not required.

\afterpage{\clearpage}

\section{Trigger and DAQ}
\label{sec:det:TDAQ}
\subsection{Introduction}

The \babar\ and Belle experiments both chose to use ``open triggers'',
attempting to preserve nearly 100\% of \BB\ events of all topologies,
and a very large fraction of \tautau\ and \ccbar\ events.  This
choice has facilitated the very broad physics program of these experiments.
A cost of this approach is that it is quite difficult to separate
these events reliably from the \qqbar\ ($q=u,d,s$) continuum and from
higher-mass two-photon physics, resulting in a large cross-section of
events logged and reconstructed.

The physics program envisioned for the \superb\ experiment depends on
continuing this strategy, despite the resulting two order of magnitude increase
in data rate.
Very high efficiency for a wide variety of \BB\ events is of
great importance in carrying out recoil-based analyses and in looking
for the charmless \B\ decays that are central to the understanding
of $b\to s$ transitions and the angle $\alpha$.
Few classes of \B\ decays important to the physics program can provide
the kinds of clear signatures that would allow the construction of a specific
trigger for them.

The trigger and online designs presented here, therefore, focus on
the ability to deliver near-100\% efficiency and minimal deadtime.

The accelerator design, in which beam currents are comparable to those
in the current \B-factories while the luminosity is 100 times higher,
results in an event rate dominated by luminosity-driven processes.
A detector of the design envisioned here will have a cross section for
Bhabha events that produce detectable signals of approximately 50 nb.
This guarantees a minimum event rate in the detector of 50 kHz, due
solely to this process.  $s$-channel physics processes, including
\BB, will contribute approximately 5 kHz more.

It has not yet been possible to do a detailed background simulation
for the present design.  In addition, no satisfactory generic
two-photon simulation for these energies is presently available
(even for \babar\ itself).
This makes bottom-up estimates of interaction rates quite difficult.
In lieu of this, our rate estimates are based on simple scaling
arguments from the current \babar\ Level~1 trigger behavior.

\subsubsection{Event Size Estimation}

In recent \babar\ running, average event sizes are in the vicinity
of 35kB, a number which has grown only slightly since the beginning
of of the experiment.

Compared to \babar,
the \superb\ detector described herein will likely have similar channel
counts and similar amounts of data per hit in the barrel PID, EMC, and
muon detection system.
The use of smaller cells in the drift chamber could lead to an increase
in the number of wires.
The design of the innermost layer(s) of the silicon tracking system is not
yet final.
The use of either striplets or, {\it a fortiori}, MAPS pixel detectors
could greatly increase the number of channels.

In any event, a significant change in the portion of the event size
due to physics tracks is not foreseen.  The bulk of the \babar\ and
\superb\ event sizes are, of course, due to background occupancy, and
no detailed analysis of this has yet been possible.
As a result, we rely on naive estimates of these backgrounds.

For the purposes of this chapter, we assume an increase in the event
size to 75kB.
Should this turn out to be an underestimate, we believe that the
design presented below would scale linearly with an additional factor
of two in event size; beyond that might require the use of a hierarchical
event building network, and would significantly increase the cost.



\subsection{Trigger Algorithms and Implementation}

\subsubsection{Trigger Levels and Functional Requirements}

The trigger system consists of the following components:

\begin{itemize}

\item A ``Level~1'' trigger that receives a continuous data stream
  from the detector independently of the event readout,
  is fully pipelined to minimize dead time, and that can deliver
  readout decisions with the latency required by the finite depth of
  the frontend buffers. We expect this trigger to be a
  ``hardware-like'' system using FPGAs as specialized processors,
  operating with a reduced version of the data from the tracking
  and calorimeter systems.

\item A software ``Level~3'' trigger that runs on a commodity computer
  farm and can base its decision on a specialized fast reconstruction
  of complete events.

\item An offline ``Level~4'' trigger stage that is out of the ``deadtime
  loop'' may also be used to reduce the volume of permanently recorded
  data.  This is not part of the existing \babar\ architecture.
  (In \babar\, an offline filter is used to reduce the number of events
  fully reconstructed and skimmed for physics analysis.)

\end{itemize}

Note: While we do not explicitly foresee a ``Level~2'' trigger that acts on
partial event information in the data path, the data acquisition
system architecture would allow the addition of such a trigger stage at a later
time, hence the nomenclature.


\subsubsection{Level~1 Trigger architecture}

The \babar\ Level~1 trigger relies on data from the drift chamber
and calorimeter.
The trigger receives data at 4~MHz from the detector systems, and
evaluates ``trigger primitives'' -- essentially, tracks and clusters
meeting various sets of criteria -- independently in each time slice.
Interpolation in time allows the drift chamber primitives to be computed
at 8~MHz.

The primitives are transmitted to the global trigger logic, which
is able to combine them and form trigger decisions based on
specified predicates, including spatial correlations between
the drift chamber and calorimeter primitives.
Up to 24 parallel trigger decision candidates are evaluated in each
time slice, and an overall trigger decision is then generated by taking
into account the time resolution of each of the decision predicates
and their ``priority'', a measure of their reliability and importance
to achieving the desired trigger performance.

The current drift chamber trigger uses information from both axial and
stereo layers of the chamber, and performs a simple helix fit, using
coarse-grained timing to improve the fit over the use of hit-cell
information alone.

The drift trigger primitives represent observations of tracks with
a variety of constraints on their position and momentum.
The three-dimensional track fit thus provides a powerful means of
identifying, and ignoring in the trigger, those tracks which arise from
interactions away from the luminous region, such as beam particles
striking elements of the accelerator or detector structure.

The calorimeter trigger divides the barrel into 280 towers
(7 $\times$ 40 in $\theta \times \phi$), and the forward endcap
into 40 slices in $\phi$.
Trigger primitives are formed by applying several levels of minimum
energy requirements to the tower energies, ranging from one which
represents the energy deposition expected from a minimum-ionizing
particle, to primitives sensitive only to high-energy photons or
electrons from Bhabha scattering.
Calorimeter trigger primitives are computed from overlapping pairs of
towers in $\phi$ to avoid splitting clusters.

\paragraph{Level~1 Concept for \superb}

We believe that the existing conceptual architecture of the \babar\ trigger
will remain
suitable for use in \superb, with minor variations.

An increase in the sampling rate for the drift chamber signals
would provide an improvement in the position and momentum
resolution of the helix fit, and allow a narrower definition of
the luminous region, reducing the effects of detector background.

If the drift chamber cell size were to be reduced, it might
be necessary to increase the number of cells included in the
logic for each superlayer segment, in order to preserve the
track-finding efficiency at low transverse momenta.
(In \babar, eight
cells, centered around a single pivot, are used to construct each
segment).

It would also be desirable to increase the segmentation of the
calorimeter in the $\theta$ direction, reducing the impact of
overlapping energy deposits, allowing more sophisticated
topological triggers to be constructed, and improving the
ability of the trigger to associate tracks and clusters.
This will require logic that avoids cluster-splitting in
$\theta$ to be added to the design, in addition to the present
overlapping logic in $\phi$.

Should the forward endcap calorimeter be constructed from LSO crystals
with shorter decay times, it will be necessary to increase
the sampling rate of the calorimeter trigger accordingly.

The use of 2.5GBit/s fiber links to bring subsystem data to the
trigger (a technology foreseen for the DAQ system as well) and the
increased complexity of logic supported by current-generation FPGAs,
should allow the Level~1 trigger to be considerably more compact than
it was in \babar.  The ability to bring a larger fraction of the
subsystem data into a single device will also allow more complex
algorithms to be applied to the data without bandwidth limits from
inter-device connections.

\paragraph{Level~1 Rates}

The existing \babar\ Level~1 physics configuration produces a trigger
rate of approximately 3~kHz at a luminosity of $1 \times 10^{34}$.
Changes in background conditions can produce large variations in this
rate.
The present DAQ system performs well, with little deadtime, up to
rates of approximately 4.5~kHz, and continues to be upgraded.
This headroom is very useful in maintaining stable operation.
It may be possible to reach 7~kHz with improvements currently in
progress.

Since no detailed background simulation for \superb\ is yet available,
we are limited to fairly crude estimation of the capabilities of a
hardware trigger in this environment.

The present \babar\ offline physics filter's output corresponds to a
cross-section of approximately 20~\nb.  This filter includes a highly
efficient Bhabha veto.
We take this as an irreducible baseline for any open hardware trigger
design; in fact this is fairly optimistic since the offline filter
uses results from full event reconstruction.

The present \babar\ Level~1 trigger rate of 3 kHz, then, includes 200 Hz
from this source and a further 500 Hz of Bhabhas.
The remaining 2300 Hz of Level~1 rate arise primarily from
beam backgrounds, together with a small amount of low-mass two-photon
physics.

At the \superb\ luminosity, the combination of the 20 \nb
``irreducible'' cross-section and the 50 \nb of Bhabha would result
in an event rate of 70 kHz.
What is not known is how to scale the remaining portion of the
\babar\ Level~1 rate.
Since the beam currents in the envisioned accelerator design are
comparable to the present \pepii\ currents, and the IP design is
expected to produce lower backgrounds as a result of the removal of
the B1 dipoles, we cannot expect that this portion of the Level~1
rate will scale with luminosity.

Review of a variety of background scaling models used in \babar,
and the opinions of a number of \babar\ trigger experts, suggest
that scaling by 10\% of luminosity is a plausible goal.
This implies a contribution to the Level~1 rate from this source
of ~25 kHz, or a total rate from all sources of 100 kHz.
In order to maintain 50\% headroom, then, this suggests designing
the DAQ system to be capable of handling Level~1 Accept rates of
150 kHz.

This is a challenging goal (the LHC experiments are planning for
rates around 100 kHz) but would likely be achievable.
As presented below, we are confident that the event build and
downstream portions of the DAQ system could handle a 150 kHz rate.
Detailed analysis of this for the frontend systems has not been
done; this would be particularly important since there is no
existing HEP system that is designed for more than 100 kHz.

\paragraph{Bhabha Veto}

The trigger rate requirements could be reduced if a portion of
the Bhabha rate could be vetoed at Level~1.
The \babar\ Level~3 trigger is currently capable of
vetoing 90\% of Bhabhas; this, however, requires a fairly
high-precision reconstruction of tracks and clusters, so that
they can be matched reliably, especially in the edges of the
detector acceptance.
A less-challenging 50\% veto at Level~1 is very likely achievable,
and would reduce the rate estimates to 70 kHz, or 105 kHz
including headroom.

The design of the trigger primitives would have to be done with
the needs of a Bhabha veto taken into account.
This is a further reason why it may be desirable to run the
tracking trigger at higher rates, to improve tracking resolution,
and to use a finer-grained segmentation of the calorimeter in
the trigger.
The global trigger would have to be capable of performing
track-cluster matching.

Since it appears that the viability of an open trigger depends
on these conclusions, we recognize the need for additional R\&D
on the following topics:

\begin{itemize}

\item carrying out an analysis to confirm that
the estimated 10\% scaling of backgrounds is realistic;

\item evaluating the technical requirements and cost of
frontend systems capable of running at 150 kHz; and

\item defining the trigger primitive requirements of a
Level~1 Bhabha veto, and confirming that a 50\% or better
veto is achievable.

\end{itemize}

Either one of 150 kHz operation or 100 kHz operation with a
Bhabha veto appears to be an essential requirement.

In the discussion below we assume a 100 kHz requirement, and
identify the scaling issues associated with supporting
150 kHz instead.

\subsubsection{Level~3 Trigger architecture}

The \babar\ Level~3 trigger runs on a farm of Linux systems,
fed by the event builder with complete raw events.
It analyzes data from the Level~1 trigger, drift chamber, and
calorimeter, searching for charged tracks and calorimeter energy
clusters.
The primitive track segments from the Level~1 data are used to
seed the track-finding algorithm in the drift chamber, and
the actual drift chamber hits are then used to perform track fits.
The trigger selects charged particles coming from a limited
three-dimensional region around the interaction point (``IP tracks''),
to avoid contributions from machine backgrounds.

The physics triggers are based on a small number of straightforward
requirements on the detected tracks and clusters, \eg, a single IP
track of momentum greater than 600~MeV/c, or two IP tracks of any
momentum, or simple requirements on the total energy in the
calorimeter or its distribution.
These trigger selections are highly efficient for virtually all
processes of interest in \babar.
However, they do accept a substantial portion of the Bhabha
cross-section, and at the luminosity of \pepii, it has been
necessary to introduce a Bhabha veto.
The task of the veto is complicated by the fact that a large
part of the accepted cross-section consists of Bhabhas with
only one track detected and/or at least one incompletely
contained electron (positron) shower in the calorimeter.
The veto algorithms must therefore recognize these
degraded Bhabha signatures.
The current trigger's veto algorithms still pass
approximately 5~\nb of Bhabhas, roughly one-tenth of the initial
total.
A small additional rate of well-identified Bhabhas are
deliberately accepted to provide calibration samples for various
detector subsystems.

The current Level~3 configuration in \babar\ achieves a total
cross section of approximately 30\nb at
a luminosity of $1.2\times 10^{34}$.
This includes about 5\nb of events that are deliberately
accepted as calibration and injection monitoring samples,
and whose rate has been held constant as the \pepii\ luminosity has
increased, through the raising of prescale factors.

\paragraph{Level~3 Concept for \superb}
It is highly likely that the \babar\ Level~3 cross-section could be
further reduced, either by tightening the trigger requirements in such
a way that very high efficiency for \BB\ physics would be maintained,
but with some loss of efficiency for low-multiplicity physics such as
\tautau, or by the investment of additional CPU time to refine the
reconstruction and allow finer discrimination.

For the present purpose, however, we conservatively assume only that
the core 25~\nb cross-section of the present trigger can be maintained
at the \superb\ facility, and that no significant increase in the
calibration and monitoring event rates will be needed, compared to
\babar.
With this assumption, then, the computational requirements of the
Level~3 trigger should be similar per event as those for \babar,
except for possible increases arising from the increased wire
density of the drift chamber.
In the worst case, a doubling in the number of layers in the chamber,
we believe that the overall Level~3 processing time increase would
be less than 50\%.

In this conservative analysis, we estimate 25~kHz as the average event
rate to be expected after Level~3 in the \superb\ experiment.

\subsubsection{Level~4 Option}

In this document we imagine a ``trigger'' as a system that
irrevocably discards events not selected.
In \babar\ the final level of triggering in this sense is Level~3.
The raw data output of Level~3 (``XTC files'') is permanently
recorded to tape and archived at two sites.
Offline data processing (and reprocessing) in \babar\ begins with
these files.

Two further levels of filtering are used in \babar, however.
The reconstruction process for XTC files begins with a stage
that selects events for reconstruction, based on the event
analysis by Level~3.
The second stage of the process performs a basic reconstruction of the
event, only somewhat more detailed than that done by Level~3, and uses
this data to make a further selection before the event is subjected to
full reconstruction and written out in the \babar\ offline format.

At present \pepii\ luminosities, this ``physics filter'' selects
approximately 15~\nb from the event stream.
This is a very loose selection that is shared by all \babar\
analyses and would clearly be applicable to the \superb\ experiment.
This raises the question of whether a similar selection could
be performed as a Level~4 trigger in \superb--that is,
as an irrevocable selection preceding archival storage.

Experience in \babar\ has shown that this physics filter has
occasionally been modified as part of a reprocessing pass,
specifically in order to accept events from certain low-rate
low-multiplicity processes.
If this selection were applied as a Level~4, this flexibility
would be lost, and extensions to the filter could only be applied
to new luminosity.

The data volume and associated offline computing costs anticipated
for \superb\ are such, however, that this tradeoff may be found to be
worthwhile.

The \babar\ physics filter is evaluated after execution of a subset
of the full offline reconstruction.
It is unknown whether a similar selection with acceptable efficiency
for physics and a substantial rate of background rejection could be
constructed from the Level~3 quantities.
We recommend that this be studied, perhaps initially with an attempt
to implement the existing \babar\ physics filter selection ``as-is''
on Level~3 quantities.


\subsection{Online}


\begin{figure}[thb]
  \begin{center}
  \includegraphics[width=0.9\textwidth]{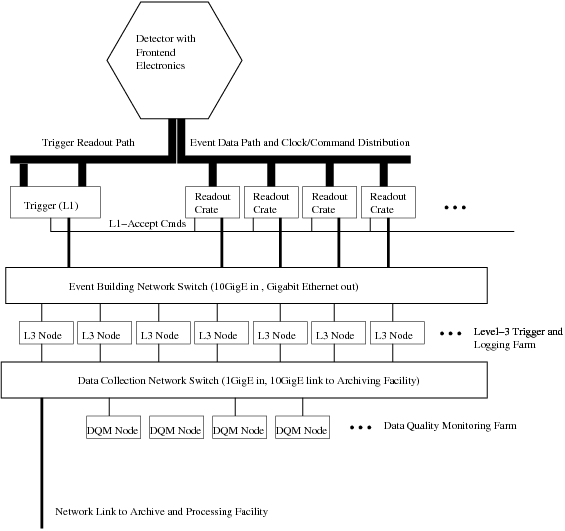}
  \caption{High-level view of the data acquisition system.}
  \label{fig:tdaq_dataflow}
  \end{center}
\end{figure}

\subsubsection{Data Acquisition}


In this section, we discuss the design of the data acquisition (DAQ) system
for the \superb\ detector, as well as the event data path and the fast
control and timing system.
The \superb\ DAQ system design will be similar to the current \babar\ data
acquisition system, but with faster fiber links to the frontend
electronics (2.5GBit/s) and much higher integration within the readout
modules.
It could be built from components of a modular data acquisition system
that is currently being developed at SLAC.

Cluster Element Modules (CEM) provide an FPGA incorporating a general-purpose
(PowerPC) processor, channels of generic high-speed serial I/O, and 10GBit
and 100MBit Ethernet commodity network connectivity (a "System-On-A-Chip").
Up to 32 CEMs can be housed in a crate (CE-box), interconnected by
Fast/Slow Cluster Interconnect Modules (fCIM/sCIM), providing managed
10GBit and 100MBit switched Ethernet connections within the crate, as well
as up to 8 10GBit-Ethernet external network connections per crate.

\begin{figure}[thb]
  \begin{center}
  \includegraphics[width=0.9\textwidth]{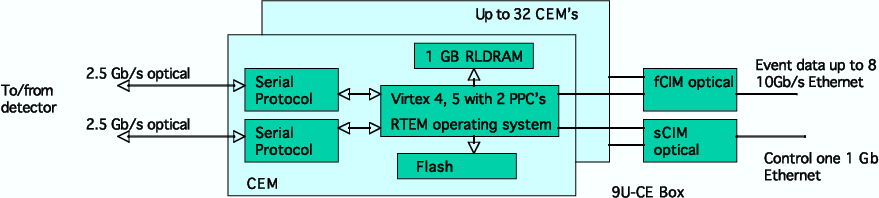}
  \caption{Schematic diagram of a CEM.}
  \label{fig:rom-electronics}
  \end{center}
\end{figure}

Each CEM can have two bidirectional 2.5GBit/s fiberoptic links to detector
frontends. These links are used to transmit a global clock and configuration
and readout commands to the frontend electronics and to receive event
data from the frontend electronics. In addition to the usual FPGA-style
processing capabilities, including DSPs, each CEM has two PowerPC processors
capable of running a real-time operating system.

As a baseline design, we foresee providing at least one crate per subdetector.

We anticipate that all detector systems will use a triggered readout scheme,
eliminating the untriggered readout used in the \babar\ EMC.

Because of the very high event rates anticipated, it will be important
to push as much of the task of ``feature extraction'' as possible into
intelligence in the frontend electronics.  The recent \babar\ drift
chamber upgrade provides a useful model, in which the new frontend
electronics include FPGAs that execute the feature extraction
algorithm previously hosted in the downstream general-purpose CPU,
and are still reprogrammable through the DAQ system, should a change
to the algorithm be needed.

In particular, we expect that the EMC feature extraction will have to
be migrated upstream in this way to handle the 20--30 times higher Level-1
Accept rates foreseen.

\babar\ experience suggests the possibility of single-event upsets (SEUs)
or other radiation damage affecting the operation of the frontends.
The design must take this into account, and provide features that
mitigate disruption to data acquisition or frontend configuration
from these sources.  That may require the incorporation of
redundant processing and/or frequent validation of configurations.

With the envisioned luminosity, one concern is the possibility of
overlap between physics events within the integration time of one of
the detector systems, notably the EMC barrel.  The probability of
a Bhabha event having occurred just before any given Level-1 Accept
is substantial, and it would be desirable to be able to disentangle
its energy deposition from that of the succeeding event.
In order to preserve event independence, and to allow for Bhabha vetoes
at Level~1, addressing this problem requires adding additional
information to each event.
This may require the feature extraction algorithms in the frontends
to record in each event a history of recent detector activity
sufficient to perform this analysis downstream.

A Fast Control and Timing System (FCTS) is responsible for providing
a common time base and clock synchronization for all components of the
data acquisition system and for the control of the frontend readout
electronics through configuration and readout commands.
The frontend readout electronics respond to a readout command by
extracting the data that correspond to the requested time window from
the sampling buffers and sending them up the event data path.

In contrast to the \babar\ design, it will be important to ensure that
there is no mandatory deadtime per Level-1 Accept.  This can be accomplished
either in a deterministic design in which the Level-1 Accept command can
be transmitted and the resulting data received in less time than the
minimum interval between possible triggers, or with a design in which
Level-1 Accepts can be queued in transit, with time tags that permit
retrieving data by address from frontend buffers.

The \babar\ design choice of synchronizing the event timebase to
the PEP-II revolution period proved to have unanticipated benefits,
as it permitted the straightforward implementation of the fast
event filtering required by trickle injection.
Because the present \superb\ accelerator concept requires trickle injection,
this capability must be preserved.
In order to avoid downstream requirements of preserving event
time ordering, the time relationship of events to the most recently
injected pulses should be determined and recorded in-band in
the event data stream at the CEM level.

In addition, experience from the integration of event and
ambient data in \babar\ indicates the desirability of
placing the recording of slow control data on the same timebase.

\subsubsection{Event Builder}


After the feature extraction algorithms have been applied to the event
contributions, whether in the frontends or the CEMs, they are combined
into complete events in a multi-stage event builder.
While not strictly
necessary, a crate-wide event build where all event contributions from a
crate are combined into a small number of CEMs may be desirable, in order
to have additional buffering and better control over the downstream
network traffic.

The final stage of event building is done in a network event builder that
combines the event contributions from multiple data acquisition crates
into complete events while fanning them out to the Level~3 trigger farm.
The \babar\ approach of a deterministic event distribution will be used,
avoiding the need for round-trips of ``worker available'' messages.

The event builder has to be able to handle the full Level~1 trigger
accept rate and the corresponding data volume.
At 100\kHz and 75\kByte event size, the data throughput would be
7.5\GByte/s or 60\GBit/s.
The physical network medium is 10\GBit Ethernet on the upstream side,
matching the interfaces provided on the CE-boxes, and 1\GBit Ethernet
on the Level~3 farm side in order to take advantage of cheap on-board
1-Gigabit electrical Ethernet interfaces of the farm machines, and
preserve the wiring flexibility of the longer cable runs thus allowed.

Network switches that have the required internal bandwidth
and can accommodate the necessary port counts are commercially available
today.
Since most (if not all) currently available network switches cannot
propagate flow control information across the backplane, the per-port
average output rate should be limited to no more than 50\% of the port
speed, in order to take advantage of the per-port output buffers.

The event builder should be connectionless, ideally using the UDP/IP
protocol with a simple flow control and retransmit mechanism.
This places much less stringent requirements on the network stack
processing time on the CEMs than would the use of TCP/IP.
The addition of retransmit, which was not used in \babar, reduces the
need for highly device-specific network tuning.

\subsubsection{Software Trigger Farm(s)}





The Level~3 trigger farm needs to provide sufficient aggregate network
bandwidth and CPU resources to handle the full Level~1 trigger rate on
its input side.
At 60\GBit/s total rate, and a maximum allowed average
rate of 0.5\GBit/s per output port, this requires a minimum of 120 1-GBit
connections to the event building switch.
While other configurations
are possible, we assume a baseline configuration of 120 Level~3 farm
nodes. At 100\kHz Level-1 Accept rate this translates into one event
per 1.2\ms per node.
Extrapolating from current technology, we expect the farm
nodes to have 8--16 CPU cores with clock speeds that are not
significantly higher than at present, requiring the parallel processing of
multiple events on a single node. With typical
SMP overheads and the CPU power needed to do the event building and
event data I/O, an 8-core node would allow 5--6 CPU-ms of Level~3
trigger processing time. If more CPU is required for the Level~3
trigger, the number of Level~3 trigger nodes can be increased.

The Level~3 trigger algorithms should be able to operate and log
data entirely free of event time-ordering constraints.
This greatly simplifies the use of multiple CPU cores, and
facilitates the provision of an efficient packet retransmission
protocol in the event builder.
With the exception of the highly \babar-specific retrofit of
trickle injection filtering, this was the already the case for
the \babar\ system and event independence is a
requirement for all downstream physics processing anyway.

\subsubsection{Data Logging}



As in the \babar\ data acquisition system, the output of the Level~3
trigger is logged to disk storage local to the farm nodes.
We assume 2--4\TByte of usable space per node, constructed from low-cost
disk in a mirrored (RAID-1) configuration.

At the estimated Level~3 trigger output rate of 210\Hz/node, and an
event size of 75\kByte, a single farm node needs to be able to write
16\MByte/s of Level~3 trigger output and to read at least twice that
rate for transfer to tape storage (about 16\MByte/s for keeping up with data
taking + 16\MByte/s contingency to drain buffered data).
Current-generation disks are capable of handling these rates.

A switched Gigabit Ethernet network separate from the event build is
used to transfer the data asynchronously from the farm machine disks
to archival storage and/or near-online farms for further processing.

In contrast to the \babar\ system, we do
not foresee an additional run-building stage that combines the
individual files from the farm nodes into per-run files that contain
all events from a run.
Instead, individual files corresponding to parts of a run will be
maintained in the downstream system.
The bookkeeping system and data handling procedures need to be
designed to handle non-monotonic runs and missing run contribution files.

At a total Level~3 trigger output rate of 25\kHz (1.9\GByte/s at
75\kByte event size), the aggregate farm capacity of 240-480\TByte
corresponds to approximately 1.5--3 days of data taking.
The size of this buffer is set by the number of nodes and the
availability of suitable internal disks.
We have assumed 1\TByte devices, which are just now coming to market;
somewhat larger ones will certainly be available on the \superb\
project time scale.

We assume that the site will provide the facilities for permanent
archival storage of the raw data, and network connectivity to the
experimental area with sufficient capacity for moving the Level~3
output.
It would be prudent to provide for double the normal output rate
of Level~3, in order to allow for timely recovery from temporary
outages.
This corresponds to a bandwidth requirement of about 4\GByte/s.
This could be provided as several 10\GBit/s fibers, or by using (costly)
higher-rate optical links; 40\GBit/s optical links are currently
available but are not commodity items.

Providing significantly more buffering as part of the online computing
system incurs large additional costs, most likely in the form of
a server farm of ${\cal O}(100)$ nodes, and quite possibly exceeding the cost
of the Level~3 farm itself.
This is probably not cost-effective to provide merely in order
to mitigate low-probability loss of connectivity accidents.

\subsubsection{Dead Time and Buffer Queue Depths}

As noted above, the goal of this design is to avoid mandatory
deadtime for each Level-1 Accept.  In \babar, there is a 2.6\mus
minimum interval between commands transmitted on the FCTS network.
This means that triggers which arrive within that interval after
a Level-1 Accept cannot be serviced and are lost.  The resulting
deadtime at the current peak \babar\ rates of ~5\kHz is 1.3\%.
At 100\kHz this would be 26\% and obviously unacceptable.

There is a minimum interval between possible triggers set by the
response time constants of the detector systems, such as the DCH drift
time and the EMC integration time.  It is possible to avoid mandatory
deadtime if the system is designed to be able to transmit Level-1
Accepts and read out the corresponding data in less than this time,
which is likely to be of order 1\mus.

Alternatively, if the FCTS is capable of queueing triggers in transit,
and if the frontend buffer memories are time-addressable, this
requirement can be relaxed, as long as there is sufficient buffering.
Model studies are required in order to assess the number of buffers
needed; it is likely that the answer will be 10--20.


Because of the high luminosity, overlapping events will be a reality
for this experiment, particularly Bhabhas overlapping with events of
all types.  This raises questions both of extending readout intervals
to allow disentangling final-state particles from closely-spaced
interactions, and of handling overlapping triggers.
The buffer queue design must take this into account, \eg, by allowing
consecutive Level-1 Accepts to cause the repeated readout of time
slices shared between them.

\subsubsection{Event Data Quality Monitoring and Display}

Event data quality monitoring is based on quantities calculated by the
Level~3 trigger, as well as quantities calculated by a fast
reconstruction pass. Depending on the CPU availability on the Level~3
nodes and the requirements of the data quality monitoring, the fast
reconstruction pass can either run on the Level~3 farm (low-overhead
machine virtualization can be used to isolate the Level~3 trigger from
the data quality monitoring CPU load), or on a separate data
quality monitoring farm. We will henceforth assume that a separate data
quality monitoring farm of 40 nodes is required. In both scenarios,
the fast reconstruction obtains event samples from the Level~3
processes. A distributed histogramming system (similar to the \babar\ DHP system)
collects the monitoring output histograms from all sources and makes
them available to automatic monitoring processes and operator GUIs.

\subsubsection{Other Components}

\paragraph{Experiment Control and Operator Console}

The Experiment Control system provides the operator interface for the
data acquisition system, configuring and controlling all trigger and
data acquisition components. It provides global sequencing of setup,
teardown, calibration, and error recovery, and interacts with the Slow
Control system to automate the data-taking operation as much as
possible. Most basic components of the \babar\ Run Control system
could be re-used with the overall control logic revised for the
operational needs of the facility.
The operator and expert user interfaces should be
fully integrated with the Slow Control user interfaces.


\paragraph{Slow Control}


The Slow Control system is responsible for controlling the detector
and detector support systems and for monitoring and recording
detector and environment conditions.
It will require a toolkit that provides the interface to whatever
industrial buses, sensors, and actuators may be used to monitor
and control the detector.
It must provide
a graphical user interface for the operator, have facilities to
generate alerts automatically, and have an archiving system to record
the relevant detector information. It must also provide software
interfaces for programmatic control of the detector.
To allow easy correlation of Slow Control information with event data, a
common timebase between the data acquisition system and the slow
control system is needed.

\paragraph{Electronic Logbook}

A web-based electronic logbook allows operators to keep an ongoing log
of the experiment status, activities, and changes and should be
considered an integral part of the experiment's configuration control
and management. In order to allow reliable bookkeeping, the electronic
logbook should be integrated with the Experiment and Slow Control
system and automate the recording of relevant information as much as
possible, and be readily extensible as directed by experience.

\paragraph{Configuration Management}


For the data acquisition and trigger systems, a strict configuration
management system is required. All hardware and software configuration
parameters must be defined in a configuration database. In addition a
strict software release management and tracking system needs to be in
place, so that it can always be determined what software version,
including any patches, was running at any given time, in any part of
the data acquisition system.
FPGA firmware is considered software in this respect
and must be included in the configuration management scheme.

\paragraph{Online Computing Infrastructure}


The online computing infrastructure consists of network components
file and applications servers. It must be designed to provide
high availability (where affordable) and be self-contained and
sufficiently isolated and firewalled to minimize the online system
downtime and dependencies on external resources (even the
downstream computing on the site). This assumes similar attention
to self-sufficiency in the essential operational computing of the
accelerator.
A controlled computing environment must be provided to facilitate
subdetector-specific computing tasks.

\subsection{Reuse of \babar\ Components}

The useful lifetimes of commodity computing components are such that
it is unlikely that, on the
time scales envisioned for this project,
any benefits could be obtained from attempting
to reuse components of the existing \babar\ online system.
The sole plausible exception
might be the reuse of some of the VME crates.

The existing \babar\ online system has been remarkably successful.
\babar\ records data with a net efficiency among the highest ever
attained in the field in the long term.
Much of this achievement was due to the great attention paid to
optimization and control system automation with regard to
highly \babar- and \pepii-specific details that cannot be expected
to reappear in the \superb\ experiment.
In addition, much of the data acquisition software is either tightly tied to
the hardware and cannot easily be reused,
or was designed according to computing and networking
constraints of the late 1990s.

The benefits of reusing existing \babar\ online system software are
therefore rather limited.
Some of the software frameworks and much of the basic design may
be applicable, but one must be careful, because their reuse might
prevent the clean incorporation of ``lessons learned'' and
limit the scope for performance optimizations over the lifetime
of the \superb\ experiment.

We recommend a clean redesign of the details of the online software
systems, taking advantage of existing knowledge, toolkits and
individual software packages only where appropriate.
As a result, only very modest savings in software engineering
effort are likely to be realized as a result of this reuse.

\subsection{Conclusions, Next Steps and Estimation Considerations}

On the whole, it appears that the trigger and online computing
systems for the \superb\ experiment can be constructed using
established techniques and commercial components,
with a design quite similar to that used in \babar.

The high Level-1 Accept rate in itself appears tractable, but some
questions remain and should be the subject of further R\&D.
Remaining uncertainties in the detector design also must be
resolved before a final design and costing of the trigger and
online systems can be completed.

\subsubsection{Questions for further research}

The following additional studies should be undertaken:

\begin{itemize}

\item A thorough evaluation of the channel counts and per-channel
data acquisition requirements, including the bit depth and
sampling rate requirements of each detector subsystem;

\item Estimates of the detector occupancy and the consequent
contribution to event size arising from backgrounds (note that
the event size for several \babar\ subsystems is dominated by
background occupancy);

\item Evaluation of the improvements in Level~1 trigger charged
particle tracking attainable by increasing the sampling rate of
the data supplied to the trigger, and the consequences for the
types of algorithm that could be implemented at Level~1;

\item A detailed estimate of the attainable performance of a
Bhabha veto at Level~1, and an evaluation of whether its use would
materially change the requirements and the design of the data
acquisition system, the overall system cost, or the ultimate
accelerator upgrade luminosity that could be handled;

\item Assessment of the cost and design implications of extending
Level-1 Accept capacity from 100\kHz to 150\kHz, to provide headroom at
the start, in case of the absence of a Bhabha veto, or to deal with a
later luminosity upgrade;

\item A more detailed investigation of the consequences of
overlapping events and/or overlapping triggers, and the implications
for the FCTS and other elements of the design; and

\item Queue modeling for the various components of the system,
including both its end-to-end behavior and a specific focus on
the behavior of the frontends and feature extraction,
to determine the buffering requirements in the frontends
and other design parameters.

\item Investigation of the adaptation of the existing
\babar\ physics filter to the use of Level~3 quantities.

\end{itemize}

Each of these studies can be carried out with a
few FTE-weeks of effort.

\subsubsection{Cost Estimation Considerations}

\paragraph{Online Farms and Network}

\babar\ experience with two major online system upgrades indicates that
the per-box cost for rack-mountable farm machines and network switches
stays roughly constant over time, while the per-unit performance increases.%

We assume that performance will continue to increase according to Moore's
Law, with the gain in CPU power coming primarily from the provision of
additional cores.

Networking hardware is assumed to improve somewhat more
slowly, but as noted above even the 2009 generation of equipment should be
entirely adequate for this proposal.

For the farm nodes we assume typical SLAC prices after a combination of
volume and educational discounts. Similar discounts should be achievable
by other institutions if farm machines are bought in sufficiently large
quantities. Since the quantities of network gear required for \superb\
online will most likely not be sufficient to qualify for volume discounts,
we base our estimate for the network cost on BaBar equipment list prices
without applying any discounts.

The cost for application servers, file servers and general network
infrastructure includes the infrastructure components needed by other
online subsystems, like run control or the slow control system.

\paragraph{Slow Control System}

Labor for Design and Implementation of online system. The table
contains only items that cannot directly be attributed to
specific electronics engineering

\paragraph{Labor requirements}

We base our estimate of labor required to design, implement and
commission the online system on the effort that was needed for
\babar.
The overall complexity of the \superb\ online system is
comparable to the \babar\ system, with the new challenges of higher
data rates and processing requirements being offset by the
availability of knowledge and ``lessons learned'' from the \babar\
experiment.

We anticipate that the overall labor investment in the \superb\
online system could end up comparable to that for \babar, but
with several caveats.

Most importantly, the subsystem-specific portions of the front-end
data acquisition (e.g., feature extraction) will likely require
more specialized expertise than was available for the original
\babar\ system development.
The design capacity of the \babar\ front-end DAQ was only approached
recently, after considerable revision of the original
subsystem-specific code by members of the core group with great
expertise in optimization of embedded systems.
The 100 kHz (or more) Level~1 Accept requirement for \superb\
would require this level of engineering \emph{ab initio},
particularly if feature extraction in the front-end FPGAs is needed,
as we expect.

We believe that these tasks, as well as the provision of
subsystem-specific detector control software, would be best served
by being provided by a core development group rather than by
contributions from the detector subsystem groups.

For the purposes of labor estimates, we include all aspects of
the core, detector-system-independent portion of the online,
as well as detector-system-specific software and firmware
running at any level of the data acquisition system or in
the detector control system.
We do not include front-end electronic engineering or any
computing system development downstream of the logging subsystem
described above.
We include in the labor estimate some work that was
deferred or descoped from the original \babar\ online system
as part of the ``triage'' required to deliver an initial
system with the personnel available, and which proved later
to be essential to providing the highly reliable system we
now have.  

Within this scope, we estimate, from \babar\ experience, that
approximately 80-90 FTE-years of effort would be required,
with the ramp-up of the group capable of being spread out
between 4 and 2.5 years in advance of the first physics running.
The total requirement could be slightly reduced if key developers
from \babar\ could be found to work on the \superb\ project,
simply because of the easier application of ``lessons learned''.
The number would likely be higher, or the initial system less
successful, if a new group started completely from scratch.

\afterpage{\clearpage}

\section{Computing}
\label{sec:det:COMP}

\subsection*{Introduction}

A luminosity of $10^{36}$ cm$^{-2}$s$^{-1}$ is nearly a
two orders of magnitude increase over current $B$~Factory experience,
requiring substantial growth in computing requirements.
Compared with other detector systems, the computing system
has the additional feature that ``construction'' is never
complete -- as more data is accumulated, the
computing resources must continue to grow. Mitigating these
aspects is the ``Moore's Law'' scaling of the computing
industry: The cost per unit of computing decreases rapidly
with time, whether it be CPU, storage, or networking.

In this section, we describe the computing requirements for
\superb. The model is based on the existing $B$~Factories. We discuss here the ``offline'' computing;
the computing requirements
associated with data acquisition are discussed in the Trigger and Data Acquisition section. Offline computing includes
event reconstruction, data handling, simulation, physics analysis,
and auxiliary tasks, such as high-level calibration and validation.

\subsection*{Event Rates}

The physics rates for a ${\cal L}= 10^{36}\,\rm{cm}^{-2}\rm{s}^{-1}$
$e^+e^-$ collider at the $\FourS$ resonance are high, with
a rate from $B\bar B$ production alone of approximately 1100 Hz.
It should be noted that a large fraction of the total cross section,
other than QED and two-photon processes, is useful in the physics
program.
We list various relevant rates in Table~\ref{tab:rates}.

\begin{table}[htb]
\caption{\label{tab:rates} Physics rates in $e^+e^-$ collisons at
the $\Upsilon(4S)$
 resonance.}
\smallskip
\begin{center}
\begin{tabular}{lc}
\hline
\hline
Process & Rate at ${\cal L}= 10^{36}\,\rm{cm}^{-2}\rm{s}^{-1}$\\
 & (kHz) \\
\hline
$\FourS\ to B\bar B$ & 1.1\\
$udsc$ continuum & 3.4 \\
$\tau^+\tau^-$ & 0.94 \\
$\mu^+\mu^-$ & 1.16 \\
$e^+e^-$ for $|\cos\theta_{\rm Lab}|<0.95$ & 30 \\
\hline
\end{tabular}
\end{center}
\end{table}

It is neither necessary nor desirable to record the entire QED ({\it e.g.}, Bhabha) rate.
Bhabha scattering is useful for detector calibrations, but the statistics available is
far greater than required. \babar\ records (with pre-scaling) only a few hertz of Bhabha scattering; this
absolute rate will also be sufficient, with perhaps a small increase, for \superb\ as well. It is also worth noting that
Bhabha and background events are both considerably smaller in size, and take less time to
process, than the $B\bar B$ events;  to be conservative in our estimates, we take no credit for the
slower than linear scaling requirements for these categories.

\subsection*{Requirements}

The considerable experience with the \babar~\cite{cmp:compBaBar} and
BELLE~\cite{cmp:compBelle} experiments
can be used to reliably estimate the computing requirements for the \superb\
detector. This experience is in the ${\cal L}= 10^{34}\,\rm{cm}^{-2}\rm{s}^{-1}$
regime; scaling by about two orders of magnitude is required.
Fortunately, much of this scaling exercise is quite straightforward.
We use here the \babar\ computing experience as our
basis for estimating the \superb\ computing requirements.

The computing model may be summarized as follows: The ``raw data'' from the
detector is permanently stored, and also run through a ``prompt calibration''
pass to determine various calibration constants. Once the constants are
derived, a full ``event reconstruction'' pass is performed, and reconstructed
data is then also permanently stored. Data quality is monitored at all steps in the process.

Once constants are known, and random trigger background ``frames'' are readied,
Monte Carlo simulated data, incorporating the constants and background
on a run-by-run basis, is prepared.

Data is made available to physics analysis in a convenient form through the process of
``skimming''. This involves the production of selected subsets of the data designed
for different areas of analysis.

From time to time, as improvements in constants, reconstruction code, or
simulation are implemented, the data may be ``reprocessed'' or new
simulated data generated. This is the reason, for example, why the CPU requirements
for data processing increase even when the luminosity is constant.

As a baseline, we simply scale all rates linearly with
luminosity.
In Table~\ref{tab:storassume} we list the assumptions used in computing the
disk and tape storage requirements, and in Table~\ref{tab:compassume} we
list the assumptions behind the CPU requirements.
We define some of the terms used in the tables as follows:
\begin{itemize}
 \item Mini: refers to an event data storage format containing
detector information as well as reconstructed tracks, {\it etc.}, but is relatively
compact, through noise suppression and efficient packing of data.
\item Micro: refers to data collections that contain
only information essential for physics analysis.
 \item Skim: a subset of the collection of all events. There may
be many skims, designed to be used for different analysis topics.
 \item Skim expansion factor: the total storage occupied by the skims,
divided by the storage required for one copy of the micro dataset.
\end{itemize}

\begin{table}[htb]
\caption{\label{tab:storassume} Assumptions for estimating computing storage requirements.}
\setlength{\extrarowheight}{2pt}
\begin{center}
\begin{tabular}{p{8cm}l}
\hline
\hline
Raw data size (TB/ab$^{-1}$) & 875 \\
Micro data size (TB/ab$^{-1}$) & 42  \\
Mini data size (TB/ab$^{-1}$) & 80 \\
Mini Monte Carlo data size (TB/ab$^{-1}$) & 78 \\
Copies of raw data & 1 \\
Copies of micro & 2 \\
Copies of mini & 1 \\
Skim expansion factor & 2 \\
Copies of skims & 1 \\
Fraction of micro data on disk & 1 \\
Fraction of mini data on disk & 0.1 \\
Fraction of micro Monte Carlo on disk & 1 \\
Fraction of mini Monte Carlo on disk & 0.1 \\
Fraction of skims on disk & 0.5 \\
\hline
\end{tabular}
\end{center}
\end{table}

\begin{table}[htb]
\caption{\label{tab:compassume} Assumptions for estimating computing CPU requirements.}
\begin{center}
\setlength{\extrarowheight}{2pt}
\begin{tabular}{p{11cm}l}
\hline
\hline
Physics analysis of data (MSpecInt2000/ab$^{-1}$) & 1.2 \\
Physics analysis of Monte Carlo  (MSpecInt2000/ab$^{-1}$) & 1.3  \\
Data reconstruction (MSpecInt2000/10$^{36}\,$cm$^2$s$^{-1}$) & 22.1  \\
Monte Carlo production (MSpecInt2000/10$^{36}\,$cm$^2$s$^{-1}$) & 70  \\
Skimming of data (MSpecInt2000/10$^{36}\,$cm$^2$s$^{-1}$) & 10.2   \\
Skimming of Monte Carlo (MSpecInt2000/10$^{36}\,$cm$^2$s$^{-1}$) & 10  \\
Duration of reprocessing (mo/yr) & 9  \\
Duration of reskimming (mo/yr) & 6  \\
Effective number of running days/month & 19.3 \\
Number of running months/year & 9 \\
\hline
\end{tabular}
\end{center}
\end{table}

The resulting CPU and storage requirements are shown in Table~\ref{tab:compreq}.
The numbers are year-over-year incremental requirements, with
year 1 showing the total initial computing complement.

\begin{table}[htb]
\caption{\label{tab:compreq} Summary of computing requirements for the first five years of \superb.
Year 1 numbers are the total required for the first year; subsequent years are
increments over the preceding year.}
\setlength{\extrarowheight}{2pt}
\begin{center}
\begin{tabular}{lccccc}
\hline
\hline
Parameter & Year 1 & Year 2 & Year 3 & Year 4 & Year 5\\
\hline
Luminosity (ab$^{-1}$) & 2 & 6 & 12 & 12 & 12 \\
Storage (PB) &&&&&\\
\quad Tape & 3.1 & 10.2 & 22.0 & 26.2 & 27.8 \\
\quad Disk & 0.83 & 3.35 & 7.55 & 10.2 & 10.2 \\
CPU (MSpecInt2000) &&&&&\\
\quad Data reconstruction & 3.0 & 8.8 & 14.7 & 8.8 & 0.0 \\
\quad Skimming & 2.7 & 9.4 & 16.1 & 12.1 & 0.0 \\
\quad Monte Carlo & 9.5 & 28.0 & 46.6 & 28.0 & 0.0\\
\quad Physics analysis & 5.1 & 15.0 & 30.0 & 30.0 & 30.0 \\
\quad Total & 20 & 61 & 107 & 79 & 30\\
\hline
\end{tabular}
\end{center}
\end{table}

\subsection*{Comparison with Requirements of LHC Experiments}

As shown in the previous section, the large increase in luminosity of \superb\ over PEP-II reuqires a substantial
increase in computing requirements.

We may put these requirements in perspective by comparing them with projections for the LHC.
Table~\ref{tab:compLHC} shows the projected requirements for the CMS experiment
and Atlas experiments in 2010,
summed over the different types of tiered computing centers. It is clear that
the computing requirements of the LHC experiments are significantly higher, even in 2010, than the
projections for the first year of \superb.

\begin{table}[htb]
\caption{\label{tab:compLHC} Summary of projected total
computing requirements for CMS~\cite{cmp:compCMS} and Atlas~\cite{cmp:compAtlas}
(Atlas numbers have been read from graphs) in 2010.}
\begin{center}
\begin{tabular}{lcc}
\hline
\hline
 & CMS & Atlas \\
\hline
Tape (PB) & 59.6 & 50 \\
Disk (PB) & 34.7 & 69 \\
CPU (MSpecInt2000) & 115.7 & 139 \\
\hline
\end{tabular}
\end{center}
\end{table}

The comparison with LHC provides a reality check. Nonetheless, the requirements are non-trivial.
Several considerations govern the feasibility of acquiring the
needed resources, as discussed in the following sections.

\subsection*{Industry Progress}

The first helpful fact is simply progress in the
computing industry. There is considerable uncertainty in projecting the
factors to be gained from this progress, but there is reason to believe
that many current trends will continue for some time to come.

There is much discussion of how far into the future the current ``Moore's Law''
rate for CPU performance will continue. However, there is confidence that the current rate
of a factor of ten in transistor count every five years will continue for
another decade~\cite{cmp:mooresLaw}.
CPU computing costs are thus expected to continue to decline by approximately a factor of
ten every five years.

Industry projections show tape densities doubling every two
years at least through 2010~\cite{cmp:tapeDen}.

Hard disk storage density has been following a similar trend as
Moore's Law for CPU's, with a factor of ten in drive density every five years~\cite{cmp:diskDen}.
In contrast with tape and CPU performance, it is not clear how this will
extrapolate into the future. There is thus some uncertainty in the relative size of the disk and tape
components of the computing model.

Historically, the improvement in wide-area network cost/performance has been
60\% per year~\cite{cmp:WANperf}. This remains an active area of industrial
development and the trend may be expected to continue for some time.

\subsection*{Software}

Another helpful factor is the expectation that we will be able to
build upon much of the \babar\ computing model, as well as on
Grid technology that has been developed by others. When \babar\
was constructed, we had a different situation, in which there was no existing
comprehensive software structure suitable for the
data-handling requirements. Nearly the entire software structure had to be built from scratch.

Fortunately, the present situation is rather different: the \babar\ code base, with the anticipated hardware
advances, provides a useful starting point for \superb. The Unix environment and C++ language
are reasonably stable; we don't anticipate the need for major development due to
operating system or language changes. Grid development is well-advanced, and
can be expected to form the basis for our distributed computing model, as detailed below.

Despite the major reduction in the required software effort expected due to the stability of
the model and the existing code base, substantial software
engineering effort will be required during both construction and
operation. To provide some perspective, after several years of operation, the service personnel requirement for
\babar\ is currently approximately 120 FTEs.
About half of this is for computing production and continuing physics software
tools development. Much of this is supplied by physicists, and is therefore not accounted in
construction or operational costs. It also excludes the substantial number of computing center personnel
required to maintain the computing facilities themselves.

\subsection*{Distributed Computing}

When \babar\ computing was developed, the tools for applying distributed computing,
still in a very early stage, tended to be purpose-built for the immediate
task at hand. Much has changed since then, including the deveoplment of
Grid technology. Coupled with the other advances, this
implies that we will be able to spread the computing
work over a much larger resource base than before.

The computing model for the \superb\ experiment assumes that computing
operations will be distributed over many sites on several
continents. This approach has the advantage that the disparate computing
capabilities at the participating institutes and in their countries can
be employed in an optimal way. The seamless integration of remote computing
centers will also ensure that participating groups of scientists will have
optimal access to \superb\ data, together with sufficient means to perform
data analysis at their home institutions.

\subsubsection*{Technology and Resources}

Though the precise state of development of computing technology at the startup
of the \superb\ facility cannot be predicted reliably, several
assumptions are relatively safe. The \babar\ experiment has already
used distributed computing to a large degree and obtained very good
experience with it. We are now in the ramp-up phase of a standardized Grid
infrastructure that will provide the backbone of the computing operations of
the LHC experiments. By the time of \superb\ startup, this style of
computing will have become routine, and an extended and well-commissioned
infrastructure will be in place. For this reason, the \superb\
computing model will be based on a Grid infrastructure, and it
is envisioned that a \superb\ facility will rely on a widely deployed
multi-purpose computing infrastructure. Whether these centers
will operate in a hierarchical tier-like structure, as in the present LHC
computing models, cannot safely be predicted; it is possible that in the not to distant
future, computing functionalities will have been
virtualized to such an extent that individual centers can assume varying
roles as needed, without a strict hierarchy. As long as tapes play a role, however,
there will be a distinction between sites that have powerful tertiary
storage facilities in place (``Level-1''), and sites that operate exclusively
with disk storage (``Level-2'').

\subsubsection*{Distributed Production}

Conservatism demands that we assume that the raw data produced by
the high level trigger will, at least initially, be stored at the experiment site
or in its vicinity.
Since fast turnaround of calibration and alignment processes is crucial, data
streams specialized to the corresponding information will be split off at
the reconstruction level and processed at a site with excellent network
and computing bandwidth to ensure fast updating of calibration constants. In order
to ensure small latency and fast data quality monitoring, the core site may also be the
location at which a significant part of the initial prompt reconstruction of
the fresh data is performed. Part of this initial reconstruction, however,
may also be performed at external sites, as is already done routinely by
\babar.

Since network technology will continue to develop at an even faster pace
than other computing resources, it is safe to assume
that data processing can be much more globalized than at present. Thus
data skimming, as well as reprocessing of data, can be performed at either local or
remote sites, depending on resources availability. As long as tapes are
involved, however, certain data-intensive tasks will be easiest to perform at Level-1 sites.
Event simulation, on the other hand, is far less demanding of
local storage, and many sites will be capable of
participating in simulation work, as it is already done for many
current experiments.

\subsubsection*{Data storage}
In the past, analysis models of experiments have been strongly shaped by
the limited bandwidth for access to data.  In the interest of smooth
analysis in a distributed model, substantial amounts of data replication
were necessary, which had a significant impact on storage requirements. The
ongoing advances in networking bandwidth, however, are likely to change
this paradigm; for \superb\ we will likely store events
for analysis only once within the collaborative
network, except for backup datasets. With this approach, certain classes of
reconstructed and skim data can be uniquely assigned to individual Grid
centers. Event-indexed data access patterns that collect individual events
from numerous sites for a specific analysis can be expected to play a
stronger role than at present; this allows reduction of the skim storage overhead,
since some portion of the skims can be virtualized. This approach is similar to \babar's
concept of ``Pointer skims''.

\subsubsection*{End-User Analysis}

Existing $B$~Factory projects such as \babar\ have already managed
to concentrate end user analysis to a high degree on a micro-DST event format, whose
contents are highly abstracted from the detector level and consist
mainlu of physics analysis objects. Such condensed event formats are crucial
with the large event samples of the \superb\ experiment. One of the
challenges will be to keep the latencies for end-user analysis, arising from the large number of events
that many studies will have to examine, at a
manageable level. It will therefore be very
attractive to introduce parallelism into the end-user analysis itself.
New approaches such as the PROOF (Parallel ROOT
Facility) project seem to be promising in this rspect. It is even conceivable that, with
abundantly available network bandwidth, such inherent parallelism will not be
restricted to one site, but can be run across the collaboration grid. This
would further reduce the need to replicate data due to access performance
considerations.

\subsection*{Conclusions}

Scaling \babar's computing requirements by almost two orders of magnitude na\"\i vely presents a
daunting picture for \superb\ computing. However, with anticipated industry advances
and the foundation of the existing technology base, scaling by this amount is feasible.  Costs are somewhat uncertain at this point, due
to the rapidly advancing computing industry. The substantial progress that has been made, and can be anticipated to continue, in distributed computing does, however, provide flexibility in the design of the system, and the ability to optimize the contribution of individual computing installations.

\afterpage{\clearpage}

\section{Reusability of existing hardware}
\label{sec:det:Reuse}


\subsection{Introduction}

The high energy physics community has made a substantial investment
over the past couple of decades in the design and construction of three
detectors, \babar, Belle and CLEO-II, specialized for \epem physics
measurement in the $\Upsilon$ system. \babar\ and Belle have been
optimized for asymmetric colliders. Experience gained with
these detectors has been fed into the design of  
the \superb\ detector. As the programs for this current generation of detectors
wind down, components from these detectors may become available. 
Use of these components could to reduce the cost and
construction time for the \superb\ detector. All three detectors have superconducting coils
placed within a steel flux return instrumented with a muon
identification system. All also have very costly CsI(Tl) calorimeters
which remain largely undamaged by the radiation exposure to which they have thus far been subjected. Both these types of systems
typically require long lead times for procurement and assembly.

\subsection{Component Suitability}

We will focus on the reuse of components of \babar, the detector
with which the majority of the community preparing this report is
familiar. Each of the detector systems will be considered in turn,
proceeding outward from the interaction point, with an eye toward
those elements with significant reuse value.

The \babar\ silicon vertex detector provides a template for design of
the outer layers of the \superb\ vertex detector. However, it does not
provide a viable source for detector components. Some of the Si sensor
modules will have radiation damage at the end of the \babar\
program. Many of the on-detector readout chips, designed and
fabricated before the advances in radiation-hard design and fabrication
realized in the LHC effort, have already received significant
radiation damage. The electronics bridging between newly designed
on-detector circuits and readout modules is not likely to be
useful.

The drift chamber will be adequate for the lifetime of the \babar\
program. However, the chamber already shows ageing effects from the
total charge integrated on the wires. The frontend electronics would require re-work for
higher rates. The DCH and its ancillary electronics are, consequently,
not candidates for reuse.

Many of the components of the \dirc\ (the \babar\ Cherenkov particle ID
system) are candidates for reuse, including the quartz bars, the 
bar boxes in which the quartz radiators are mounted, the ``strong support tube'' and the ``horse collar'' steel structure,
which bolts onto the barrel flux return steel, cantilevering
these elements into the \babar\ detector. The production of the
existing quartz bars was a serious challenge to the manufacturers, who
found it difficult to produce well-finished bars in a timely way;
replacement, because of the dispersal of the industrial team, would
present a potential schedule risk. The mechanical structure for the
support of the photomultiplier tubes, the standoff box, can also be reused if
the faster conventional phototube option for the \dirc\ is chosen. The \dirc\ magnetic
shield structure would be retained, since it also acts as the counterweight
for the backward end doors. The conventional compensating solenoid
coil that mounts to the horse collar could also be reused.

The electromagnetic calorimeter represents a large investment in
materials and effort (approx. \$25M). 
The system contains 6680 CsI(Tl) crystals, each weighing
approximately 4 kg. The majority of these (5760), located in
the barrel portion of the calorimeter, have minimal radiation damage
and are good candidates for reuse, with the caveat that detailed
background simulation, yet to be completed, must indicate acceptably
low background rates. The endcap crystals have been exposed,
especially in the regions closer to the beamline, to higher radiation
doses, and are consequently more severely damaged. This damage
manifests itself in reduced light yield, which leads to worsened energy
resolution. The high rates of luminosity-related background expected at \superb\ also make
the use of the endcap crystals problematic, since the scintillation
light decay time is slow: the principal fast component has a lifetime
of about 800 ns which will lead to unacceptable pulse pile-up. The barrel portion of the
calorimeter is a single environmentally sealed container, since
CsI(Tl) crystals are mildly hygroscopic. It consists of
two large support cylinders joined longitudinally close to the center.
This support structure contains 280 carbon fiber
honeycomb modules suspended inside from the rear. Each module
typically contains 21 crystals that are held into these honeycombs
with glue.

The superconducting coil, its cryostat and cryo-interface box and the helium compressor plant, are
prime candidates for reuse, providing substantial schedule
and financial benefits.

The flux return steel is segmented into layers, with detectors sandwiched
between the layers. In the barrel, the detectors are limited
streamer tubes (LST's), recently fabricated. They are
candidates for reuse. Resistive plate chambers, operated in streamer
mode (except for less than 10\% that have been operated in avalanche
mode starting this year), are used in the endcaps. The RPC's
located in the backward doors date from the initial construction of
the detector. A large fraction of these have failed; the
performance of the rest is decreasing with time. The RPC's in
the forward doors date from an upgrade performed in 2002. These
chambers were produced with improved quality
assurance measures. Experience gained with the first generation of RPC's has allowed the upgraded chambers to continue functioning with high
efficiency. There are, however, lregions the chambers that show
increased sensitivity to charge, which is a harbinger of chamber
failure. This is likely to limit the reusability of these components.

\subsubsection{Instrumented Flux Return}

The flux return steel is organized into five structures: the barrel
portion, and two sets of split end doors. Each of these is in turn
composed of multiple structures. Components were sized to match
the 50 ton load limit of the IR2 crane.

Each of the end doors is composed of eighteen steel plates organized
into two modules, joined together on a thick counterweighted steel platform
(Fig.~\ref{fig:det:babar}), which rests on four columns with
jacks and Hilman rollers.  There are 9 steel layers of 20mm thickness, 4 of 30mm
thickness, 4 of 50mm thickness, and 1 of 100mm thickness. During 2002,
five layers of brass absorber were installed in forward end door slots
in order to increase the number of interaction lengths traversed by muon
candidates. These doors are retained in the \superb\ baseline,
along with the five 25mm layers of brass installed in 2002, and 
the outer steel modules which house two double layers of RPC
detectors. Additional layers of brass (or steel) will be added,
following the specification of the baseline design in the instrumented
flux return section. Use of steel, though cheaper, would require
re-measurement of the magnetic field. A cost-benefit analysis will be
performed to choose between brass and steel.

The barrel structure consists of six cradles, each composed
of 18 layers of steel. The inner 16 layers have the same thicknesses
of the corresponding end door plates. The two outermost layers are
100mm each. The 18 layers are organized into two parts, the inner 16
layers, which are welded into a single unit along with the two side
plates, and the outer two layers, which are welded together and then
bolted into the cradle. The six cradles are, in turn, suspended from the
double I-beam ``belt'' that supports the detector. During the 2004/2006
barrel LST upgrade, layers of 22mm brass were installed, replacing 6
layers of detectors in the cradles. In the \superb\ baseline, all these
components will be retained, as well as all the additional flux return
steel attached to the barrel in the gap between the end doors and
barrel. As in the end doors, four additional layers of absorber will
be placed in gaps occupied in \babar\ by LSTs.

A modest upgrade to the baseline would improve the muon identification capability. The end door structures
and existing brass absorbers would be retained, as well as the gap
filling steel now located between the end doors and barrel. The
support belt and cradles would be redesigned with eight gaps to
accommodate the detectors. The outer two barrel layers, which form a
single module (steel layers 17 and 18), along with the flux bar that
covers the end of this module, would be retained to reduce cost. Some
of the thin plates can be cut out and reused to provide the structure
for the inner \KL identification gaps. The balance of the steel will
be 4 thick slabs for 4 layers of detector. In this way, tolerance gaps
are replaced with absorber. Redesign of the cradles also allows for
wider gaps between select steel plates, permitting detector redundancy
in select layers. As a cost-cutting measure, the barrel brass would be
retained for use as the additional endcap absorber. Remeasurement of the
magnetic field would be reuqired. The \babar\ support belt structure is
asymmetric: the I-beams under the bottom sextant are shorter than those of
the other sextants. This was driven by the existing beam line height
at IR2. In the redesign there would be a larger gap between the floor and base
plate of the end doors, allowing space for installation of belt
chambers under the end doors to complete the system of belt chambers
and absorber installed in the 2002 \babar\ upgrade.

\subsection{Component Extraction}

Extraction of components for reuse will require the disassembly of the
\babar\ detector. This begins with opening the end doors
and removing the \dirc\ plug. The support tube, which contains the
accelerator final beamline elements as well as the SVT, are then
removed. The brass absorber installed in the barrel and end doors
is removed next because the jacks which lift the detector do not
have adequate capacity to handle the full load. At the same time the
barrel LSTs are removed. The doors are then closed and the bolted up
detector moved off the beamline, freeing the detector of the space
restrictions imposed by the pedestals that hold the rafts that
cantilever accelerator components into the detector. The doors are
opened and moved aside for disassembly. The steel that is contained in
the gap between the doors and barrel is removed. The electronics hut
and services that connect it to the detector are removed. The EMC forward
endcap is removed. The SOB is removed. The \dirc\ is then removed from the
barrel. The DCH is removed from the DIRC. The barrel EMC is then
removed from the barrel steel. The solenoid is removed from the barrel
steel. A temporary structure is assembled inside the barrel hexagon to
support the upper cradles during disassembly. The upper half of the
support belt is removed. Because of the load limitations of the IR2
crane, the six cradles must be disassembled {\it in situ}. The outer
sections of the top cradles are removed, followed by the inner part of
each of the three cradles. The temporary support structure is
removed. The inner part of the lower cradles is removed, followed by
the outer portions. The balance of the structural belt is
disassembled.

\subsection{Component Transport}

The magnet steel components will be crated for transport in order to limit
damage to mating surfaces and edges. Most, if not all, of these
components can be shipped by sea. Some optimization of shipment 
involving air transport may be required, depending on the details of
the overall construction schedule. Detector components of the IFR that
are reused can be crated and shipped by air.

The \babar\ solenoid was shipped via special air transport from Italy
to SLAC. It is expected that this component can be returned to Italy
in the same fashion.

The \dirc\ and barrel calorimeter present transportation challenges. In
both cases transport without disassembly is preferred.

In the case of the \dirc, this would include the strong support tube
and bar boxes. Detailed engineering studies, which model accelerations
and vibrations during a flight, are needed to determine if the
\dirc\ can be safely transported. Special structures would have to be
designed to handle this object. If it proves impossible to ship the
detector as a unit, it is possible to disassemble the detector and
air-ship each bar box individually. The cost is time to disassemble
and reassemble the detector, with increased risk to damage these
larger components. A dry environment is required in all cases to avoid
condensation on the quartz bars. Disassembly of the bar
boxes exposes the valuable bars to damage; it is not considered a
viable option.

In the case of the EMC, there are two environmental constraints on
shipment of the device or its components. The glue joint that attached
the photodiode readout package to the back face of the crystal has
been tested, in mock-up, to be stable against temperature swings of
$\pm 5\degrees$C. During the assembly of the endcap calorimeter, due
to a failure of an air conditioning unit, the joints on one module
were exposed to double this temperature swing. Several glue joints
parted. The introduction of an air gap caused a light yield drop
of about 25\%. In order to avoid this reduction in performance,
temperature swings during transport must be kept small. Since the
crystals are mildly hygroscopic, it is best that they be transported
in a dry environment to avoid changes in the surface reflectivity, and
consequent modification in the longitudinal response of the
crystal. Individual endcap modules constructed in the UK were
successfully shipped to the US in specially constructed containers
that kept the temperature swings and humidity acceptably small.

Disassembly of the barrel calorimeter for shipment presents a
substantial challenge. Both the disassembly and assembly sites need to
be temperature and humidity controlled. The disassembly process
requires removal of the outer and inner cylindrical covers, removal of
cables that connect the crystals with the electronics crates at the
ends of the cylinder, splitting of the cylinder into its two component
parts and removal of the 280 modules for shipment. Though much of the
tooling exists, the environmentally conditioned buildings
used in calorimeter construction at SLAC no longer exist, though
alternative facilities could be fit out. The cooling and drying units
used in the module storage/calorimeter assembly building continue to
be available.

The clear preference is to ship the calorimeter as a single unit by
air. Including the tooling support stand and environmental conditioning
equipment, the load is likely to exceed 30 tons. It is anticipated
that such a load could be transported in the same way as the
superconducting coil and its cryostat, but verification is
needed. Detailed engineering studies, which model accelerations and
vibrations involved with flight that might cause the crystal-containing
carbon fiber modules to strike one another, are
needed to determine if the calorimeter can be safely
transported. Engineering studies of overall stability of the EMC
structure against the flight accelerations are also required.

\subsection{Detector Assembly}

Assembly of the \superb\ detector is the inverse of the disassembly of
the \babar\ detector. Ease of assembly will be influenced by available
facilities. In the case of \babar, the space limitations of the
IR2 hall led to engineering compromises in the design of the
detector. Assembly was made more complicated by the weight
restrictions posed by the 50 ton crane. Upgrades were made more
difficult because of limitations in movement imposed by the size of
the hall. A newly designed interaction region hall could ameliorate many of these problems.

\afterpage{\clearpage}

%

\afterpage{\clearpage}


%
\afterpage{\clearpage}

\chapter{Cost and Schedule}
\label{sec:CostSchedule}

The \superb\ project presented in this document relies heavily on the
experience of \pepii\ and \babar, with cost and schedule estimates
deriving directly from the \pepii\ and \babar budget and schedule.
Although the estimates
are deemed to be correct and are based on a bottoms-up evaluation using
a detailed work breakdown schedule, it should be emphasized that this
is a conceptual design report, and that therefore cost and
schedule have not received the level of close scrutiny and detailed
evaluation expected in a technical design report.

As is customary, cost estimates are presented in separate EDIA
(Engineering, Design, Inspection, Acceptance), Labor, and M\&S
(Materials and Services) categories. Manpower is always indicated in man-months,
since a monetary conversion is only possible after institutional
responsibilities are identified.
M\&S is estimated in 2007 Euros; no future escalation is applied,
since the project starting time cannot be defined at this time.
The total project cost can be calculated, once the responsibilities are
identified, by summing the monetary value of these three categories.

The reuse and refurbishing
of existing components has been assumed whenever technically possible
and financially advantageous. The replacement value of the reused components, \ie, how much would be required to build them from scratch,
has been obtained by escalating the corresponding cost (including
manpower) from the \pepii\ and \babar\ project from 1995 to 2007 using
the NASA technical inflation index~\cite{bib:NASA} and then converted
from US Dollars to Euros using the average conversion rate over the
1999--2006 period~\cite{bib:USDEURO}. The overall escalation factor
from 1995 dollars to 2007 Euro is thus $1.21 = 1.295*0.9354$. The same
escalation procedure has also been applied whenever the cost of new
components could be directly extrapolated from the original 1995
budget.

The replacement value ("Rep.Val.") of the reused components and the cost estimates to build
\superb\ are presented in separate columns of the cost tables.
One could be tempted to sum the two numbers to obtain an estimate of the cost of the
project if  built from scratch. This procedure does not yield a completely accurate result because
of the different treatment of the manpower (rolled up in the replacement
value; separated for the new cost estimate) and because the
refurbishing costs would then be added to the initial value, yielding
incorrect results.

Contingency is not included in the tables. Given the level of detail
of the cost estimates, a contingency of about 35\% would be appropriate.

\section{Accelerator}

The cost estimate for the \superb\ accelerator is shown in Table~\ref{tab:cost_accelerator},
broken down by major subsystem.
The costs are based on extrapolation from similar activities on other recent accelerator
projects, rather than a bottoms-up estimate produced by engineers. It is
anticipated that many \pepii components will be recycled, refurbished,
and moved to the new site. The last cost column contains the present replacement
value (2007 KEuro) of the accelerator components, as they exist in \pep2, that
are to be reused in \superb. The refurbishment, moving, and coordination costs
for the component transportation are included in the other columns.
The ``EDIA'' column lists the effort in man-months that will be needed for
engineering, design and inspection (EDIA).
The associated costs for this effort will reflect local labor rates and practices at the
laboratories and institutions worldwide that participate
in the construction of the \superb\ accelerator.
The ``Labor'' column contains an estimate of the man-months of technical labor for each subsystem,
again a laboratory-dependent cost. Technical labor includes local effort needed for R\&D support,
office service support, and assembly area support. The ``M\&S''  column lists the costs of
purchased parts and services from outside companies and vendors.

{ \setlength{\tabcolsep}{4pt}  
\begin{center}
{\footnotesize
\begin{longtable}{lp{6cm}p{1.5cm}rrrr}
\caption{\superb\ accelerator budget. }
\label{tab:cost_accelerator} \\

\hline\hline
\textbf{ } & \textbf{ } & \textbf{Number} & \textbf{EDIA} & \textbf{Labor} & \textbf{M\&S} & \textbf{Rep.Val.} \\
\textbf{WBS} & \textbf{Item} & \textbf{Units} & \textbf{mm} & \textbf{mm} & \textbf{kEuro} & \textbf{kEuro} \\
\hline
\endfirsthead

\multicolumn{7}{c}%
{\tablename\ \thetable{} -- continued from previous page} \\[3mm]
\hline\hline
\textbf{ } & \textbf{ } & \textbf{Number} & \textbf{EDIA} & \textbf{Labor} & \textbf{M\&S} & \textbf{Rep.Val.} \\
\textbf{WBS} & \textbf{Item} & \textbf{Units} & \textbf{mm} & \textbf{mm} & \textbf{kEuro} & \textbf{kEuro} \\
\hline
\endhead

 \\ \hline
  \multicolumn{7}{c}{Continued on next page} \\
\endfoot

\endlastfoot


   {\bf 1} & {\bf Accelerator} &     {\bf } & {\bf 5429} & {\bf 3497} & {\bf 191166} & {\bf 126330} \\[-1mm]
\hline
 {\bf 1.1} & {\bf Project management} &     {\bf } & {\bf 2112} &   {\bf 96} & {\bf 1800} &    {\bf 0} \\[-1mm]

    1.1.01 & Technical management &  15 people &        720 &         12 &        300 &          0 \\[-1mm]

    1.1.02 & Project physicists &  10 people &        480 &         12 &        300 &          0 \\[-1mm]

    1.1.03 & Accelerator physics &  10 people &        480 &         12 &        300 &          0 \\[-1mm]

    1.1.04 & Cost accounting and tracking &   5 people &        240 &         24 &        200 &          0 \\[-1mm]

    1.1.05 & Database and documentation &   3 people &        144 &         24 &        100 &          0 \\[-1mm]

    1.1.06 & Project travel & 100 trips/yr &         48 &         12 &        600 &          0 \\[-1mm]
\hline
 {\bf 1.2} & {\bf Magnet and support system} &     {\bf } &  {\bf 666} & {\bf 1199} & {\bf 28965} & {\bf 25380} \\[-1mm]

    1.2.01 & Engineering, design, and prototypes &  10 people &        360 &        120 &        500 &          0 \\[-1mm]

    1.2.02 & Dipole (0.5 m) removal and refurbish &        144 &          5 &         50 &        144 &       1440 \\[-1mm]

    1.2.03 & Dipole (0.5 m) shipping &        144 &          5 &          5 &        144 &          0 \\[-1mm]

    1.2.04 & Dipole (0.5 m) installation &        144 &          5 &         20 &        144 &          0 \\[-1mm]

    1.2.05 & Dipole (0.5 m) power supplies+cables &        144 &          2 &          4 &        450 &        432 \\[-1mm]

    1.2.06 & Dipole (0.5 m) water cooling connection &        144 &          2 &         12 &        144 &          0 \\[-1mm]

    1.2.07 & Dipole (0.5 m) supports/ship &        144 &          2 &          4 &        300 &        576 \\[-1mm]

    1.2.08 & Dipole (0.5 m) alignment &        144 &          3 &          6 &         72 &          0 \\[-1mm]

    1.2.09 & Dipole (0.75 m) construction &        144 &          3 &          3 &       1440 &          0 \\[-1mm]

    1.2.10 & Dipole (0.75 m) installation &        144 &          5 &         20 &        144 &          0 \\[-1mm]

    1.2.11 & Dipole (0.75 m) power supplies+cables &        144 &          2 &          4 &        600 &          0 \\[-1mm]

    1.2.12 & Dipole (0.75 m) water cooling connection &        144 &          2 &         12 &        144 &          0 \\[-1mm]

    1.2.13 & Dipole (0.75 m) supports &        144 &          2 &          4 &        450 &          0 \\[-1mm]

    1.2.14 & Dipole (0.75 m) alignment &        144 &          3 &          4 &         72 &          0 \\[-1mm]

    1.2.15 & Dipole (5.4 m) removal and refurbish &        176 &          6 &         60 &        176 &       2640 \\[-1mm]

    1.2.16 & Dipole (5.4 m) shipping &        176 &          5 &          5 &        352 &          0 \\[-1mm]

    1.2.17 & Dipole (5.4 m) installation &        176 &          5 &         20 &        352 &          0 \\[-1mm]

    1.2.18 & Dipole (5.4 m) power supplies+cables &        176 &          4 &          8 &        800 &        528 \\[-1mm]

    1.2.19 & Dipole (5.4 m) water cooling connection &        176 &          2 &         12 &        176 &          0 \\[-1mm]

    1.2.20 & Dipole (5.4 m) supports/ship &        176 &          3 &          6 &        500 &        352 \\[-1mm]

    1.2.21 & Dipole (5.4 m) alignment &        176 &          3 &          6 &         88 &          0 \\[-1mm]

    1.2.22 & Dipole (2 m) removal and refurbish &          4 &          1 &          1 &          4 &         40 \\[-1mm]

    1.2.23 & Dipole (2 m) shipping &          4 &          1 &          1 &          8 &          0 \\[-1mm]

    1.2.24 & Dipole (2 m) installation &          4 &          1 &          1 &          8 &          0 \\[-1mm]

    1.2.25 & Dipole (2 m) power supplies+cables &          4 &          1 &          2 &        100 &         25 \\[-1mm]

    1.2.26 & Dipole (2 m) water cooling connection &          4 &          1 &          1 &          4 &          0 \\[-1mm]

    1.2.27 & Dipole (2 m) supports/ship &          4 &          1 &          1 &         20 &          8 \\[-1mm]

    1.2.28 & Dipole (2 m) alignment &          4 &          1 &          1 &          2 &          0 \\[-1mm]

    1.2.29 & Quadrupole (0.43 m) removal \& refurbish &        341 &          5 &         50 &        341 &       3410 \\[-1mm]

    1.2.30 & Quadrupole (0.43 m) shipping &        341 &          5 &         10 &        500 &          0 \\[-1mm]

    1.2.31 & Quadrupole (0.43 m) installation &        341 &          5 &         20 &        682 &          0 \\[-1mm]

    1.2.32 & Quadrupole (0.43 m) PS+cables &        341 &          5 &         10 &       1000 &       1023 \\[-1mm]

    1.2.33 & Quadrupole (0.43 m) water cooling conn &        341 &          2 &         12 &        341 &          0 \\[-1mm]

    1.2.34 & Quadrupole (0.43 m) supports/ship &        341 &          4 &          8 &        700 &        288 \\[-1mm]

    1.2.35 & Quadrupole (0.43 m) alignment &        341 &          3 &          6 &        172 &          0 \\[-1mm]

    1.2.36 & Quadrupole (0.5 m) removal \& refurbish &         70 &          5 &         50 &         70 &        700 \\[-1mm]

    1.2.37 & Quadrupole (0.5 m) shipping &         70 &          2 &          4 &         70 &          0 \\[-1mm]

    1.2.38 & Quadrupole (0.5 m) installation &         70 &          6 &         40 &        140 &          0 \\[-1mm]

    1.2.39 & Quadrupole (0.5 m) PS+cables &         70 &          5 &         10 &        210 &        280 \\[-1mm]

    1.2.40 & Quadrupole (0.5 m) water cooling conn &         70 &          2 &          6 &         70 &          0 \\[-1mm]

    1.2.41 & Quadrupole (0.5 m) supports/ship &         70 &          4 &          8 &        150 &         70 \\[-1mm]

    1.2.42 & Quadrupole (0.5 m) alignment &         70 &          2 &          5 &         70 &          0 \\[-1mm]

    1.2.43 & Quadrupole (0.56 m) removal \& refurbish &        287 &          5 &         50 &        600 &       2870 \\[-1mm]

    1.2.44 & Quadrupole (0.56 m) shipping &        287 &          4 &          8 &        400 &          0 \\[-1mm]

    1.2.45 & Quadrupole (0.56 m) installation &        287 &          8 &         60 &        287 &          0 \\[-1mm]

    1.2.46 & Quadrupole (0.56 m) PS+cables &        287 &          5 &         10 &        800 &        861 \\[-1mm]

    1.2.47 & Quadrupole (0.56 m) water cooling conn &        287 &          4 &          8 &        287 &          0 \\[-1mm]

    1.2.48 & Quadrupole (0.56 m) supports/ship &        287 &          4 &          8 &        400 &        287 \\[-1mm]

    1.2.49 & Quadrupole (0.56 m) alignment &        287 &          4 &          8 &        150 &          0 \\[-1mm]

    1.2.50 & Quadrupole (0.73 m) removal \& refurbish &        138 &          5 &         40 &        200 &       1380 \\[-1mm]

    1.2.51 & Quadrupole (0.73 m) shipping &        138 &          3 &          6 &        250 &          0 \\[-1mm]

    1.2.52 & Quadrupole (0.73 m) installation &        138 &          6 &         40 &        138 &          0 \\[-1mm]

    1.2.53 & Quadrupole (0.73 m) PS+cables &        138 &          4 &          9 &        400 &        414 \\[-1mm]

    1.2.54 & Quadrupole (0.73 m) water cooling conn &        138 &          4 &          8 &        138 &          0 \\[-1mm]

    1.2.55 & Quadrupole (0.73 m) supports/ship &        138 &          4 &          8 &        300 &         81 \\[-1mm]

    1.2.56 & Quadrupole (0.73 m) alignment &        138 &          4 &          8 &        138 &          0 \\[-1mm]

    1.2.57 & Sextupole (0.25 m) removal \& refurbish &        452 &          5 &         40 &        376 &       3400 \\[-1mm]

    1.2.58 & Sextupole (0.25 m) shipping &        452 &          3 &          6 &        452 &          0 \\[-1mm]

    1.2.59 & Sextupole (0.25 m) installation &        452 &          6 &         40 &        452 &          0 \\[-1mm]

    1.2.60 & Sextupole (0.25 m) PS+cables &        452 &          8 &         16 &       1000 &        900 \\[-1mm]

    1.2.61 & Sextupole (0.25 m) water cooling conn &        452 &          4 &          7 &        350 &          0 \\[-1mm]

    1.2.62 & Sextupole (0.25 m) supports/ship &        452 &          4 &          7 &        500 &        188 \\[-1mm]

    1.2.63 & Sextupole (0.25 m) alignment &        452 &          4 &          7 &        226 &          0 \\[-1mm]

    1.2.64 & Sextupole (0.6 m) removal \& refurbish &          8 &          2 &          8 &         75 &         32 \\[-1mm]

    1.2.65 & Sextupole (0.6 m) shipping &          8 &          0 &          0 &          0 &          0 \\[-1mm]

    1.2.66 & Sextupole (0.6 m) installation &          8 &          2 &          3 &         16 &          0 \\[-1mm]

    1.2.67 & Sextupole (0.6 m) PS+cables &          8 &          2 &          4 &         50 &         35 \\[-1mm]

    1.2.68 & Sextupole (0.6 m) water cooling conn &          8 &          1 &          2 &          8 &          0 \\[-1mm]

    1.2.69 & Sextupole (0.6 m) supports/ship &          8 &          1 &          2 &         20 &          0 \\[-1mm]

    1.2.70 & Sextupole (0.6 m) alignment &          8 &          1 &          2 &          4 &          0 \\[-1mm]

    1.2.71 & Corrector (0.6 m) removal \& refurbish &        600 &          5 &         10 &        400 &        900 \\[-1mm]

    1.2.72 & Corrector (0.3m) construction &        236 &          4 &          8 &        100 &          0 \\[-1mm]

    1.2.73 & Corrector (0.6 m) shipping &        836 &          0 &          0 &        100 &          0 \\[-1mm]

    1.2.74 & Corrector (0.6 m) installation &        836 &          2 &          6 &        418 &          0 \\[-1mm]

    1.2.75 & Corrector (0.6 m) power supplies+cables &        836 &          2 &          6 &        700 &        500 \\[-1mm]

    1.2.76 & Corrector (0.6 m) water cooling conn &        836 &          2 &          6 &         10 &          0 \\[-1mm]

    1.2.77 & Corrector (0.6 m) supports/ship &        836 &          2 &          6 &        836 &        100 \\[-1mm]

    1.2.78 & Corrector (0.6 m) alignment &        836 &          2 &          6 &        200 &          0 \\[-1mm]

    1.2.79 & Corrector (0.3m) construction &        236 &          4 &          8 &        236 &          0 \\[-1mm]

    1.2.80 & Wiggler (1 m) construction &        200 &          6 &         12 &       3000 &          0 \\[-1mm]

    1.2.81 & Wiggler (1 m) installation &        200 &          2 &          6 &        200 &          0 \\[-1mm]

    1.2.82 & Wiggler (1 m) power supplies+cables &        200 &          3 &          6 &        600 &          0 \\[-1mm]

    1.2.83 & Wiggler (1 m) water cooling conn &        200 &          1 &          2 &        400 &          0 \\[-1mm]

    1.2.84 & Wiggler (1 m) supports/ship &        200 &          2 &          6 &        400 &          0 \\[-1mm]

    1.2.85 & Wiggler (1 m) alignment &        200 &          2 &          6 &        200 &          0 \\[-1mm]

    1.2.86 & Pol dipole (2 m) construction &          8 &          4 &          8 &        300 &          0 \\[-1mm]

    1.2.87 & Pol dipole (2 m) installation &          8 &          2 &          4 &         50 &          0 \\[-1mm]

    1.2.88 & Pol dipole (2 m) power supplies+cables &          8 &          4 &          8 &        150 &          0 \\[-1mm]

    1.2.89 & Pol dipole (2 m) water cooling connection &          8 &          1 &          2 &         20 &          0 \\[-1mm]

    1.2.90 & Pol dipole (2 m) supports &          8 &          2 &          4 &         64 &          0 \\[-1mm]

    1.2.91 & Pol dipole (2 m) alignment &          8 &          1 &          2 &         10 &          0 \\[-1mm]

    1.2.92 & IR skew quad  (0.5 m) refurb \& ship &         24 &          2 &          4 &         30 &        120 \\[-1mm]

    1.2.93 & IR skew quad (0.5 m) installation &         24 &          1 &          2 &         15 &          0 \\[-1mm]

    1.2.94 & IR skew quad (0.5 m) power supply/cable &         24 &          2 &          4 &         76 &          0 \\[-1mm]

    1.2.95 & IR skew quad (0.5 m) water cooling conn &         24 &          1 &          2 &         24 &          0 \\[-1mm]

    1.2.96 & IR skew quad (0.5 m) supports &         24 &          2 &          4 &         50 &          0 \\[-1mm]

    1.2.97 & IR skew quad (0.5 m) alignment &         24 &          1 &          2 &          5 &          0 \\[-1mm]

    1.2.98 & Injection kickers refurb \& installation &          4 &          4 &          8 &        400 &       1000 \\[-1mm]

    1.2.99 & Abort kicker refurb \&installation &          2 &          2 &          4 &        100 &        500 \\[-1mm]
\hline
 {\bf 1.3} & {\bf Vacuum system} &     {\bf } &  {\bf 620} &  {\bf 520} & {\bf 27600} & {\bf 14200} \\[-1mm]

    1.3.01 & Engineering, design, and prototypes &  15 people &        480 &        240 &        750 &          0 \\[-1mm]

    1.3.02 & Vacuum extrusion &     2600 m &          3 &          6 &       1300 &          0 \\[-1mm]

    1.3.03 & Vacuum chamber machining &     2600 m &         24 &         48 &       5200 &          0 \\[-1mm]

    1.3.04 & Vacuum chamber assembly &     2600 m &         24 &         48 &       5200 &          0 \\[-1mm]

    1.3.05 & Vacuum chamber distributed pumps &     2600 m &         12 &         24 &       4600 &          0 \\[-1mm]

    1.3.06 & Vacuum chamber lumped pumps &  700 units &         12 &         24 &       1400 &       2100 \\[-1mm]

    1.3.07 & Vacuum chamber controls &     2600 m &         12 &         24 &       1000 &       2600 \\[-1mm]

    1.3.08 & Vacuum chamber installation &     4600 m &         12 &         24 &       2300 &          0 \\[-1mm]

    1.3.09 & Vacuum chamber refurbish and remove &     2000 m &          5 &         10 &       1000 &       9500 \\[-1mm]

    1.3.10 & Vacuum chamber shipping &     2000 m &          5 &         10 &        600 &          0 \\[-1mm]

    1.3.11 & Polarization vacuum chambers &      100 m &          8 &         16 &       1000 &          0 \\[-1mm]

    1.3.12 & Beam abort chambers &    2 units &          5 &         10 &        750 &          0 \\[-1mm]

    1.3.13 & IR high power vacuum chambers misc &    8 units &          8 &         16 &       1500 &          0 \\[-1mm]

    1.3.14 & ECI electrodes and controls &  500 units &         10 &         20 &       1000 &          0 \\[-1mm]
\hline
 {\bf 1.4} & {\bf RF system} &     {\bf } &  {\bf 272} &  {\bf 304} & {\bf 22300} & {\bf 60000} \\[-1mm]

    1.4.01 & Engineering, design, and prototypes &  10 people &        240 &        240 &        500 &          0 \\[-1mm]

    1.4.02 & RF stations refurbish \& remove & 15 stations &         12 &         24 &        500 &      60000 \\[-1mm]

    1.4.03 & RF station shipping & 15 stations &          2 &          4 &       3000 &          0 \\[-1mm]

    1.4.04 & RF station install & 15 stations &         12 &         24 &      15000 &          0 \\[-1mm]

    1.4.05 & RF station new controls & 15 stations &          3 &          6 &       3000 &          0 \\[-1mm]

    1.4.06 & RF station HVPS pads & 15 stations &          3 &          6 &        300 &          0 \\[-1mm]
\hline
 {\bf 1.5} & {\bf Interaction region} &     {\bf } &  {\bf 370} &  {\bf 478} & {\bf 10950} &    {\bf 0} \\[-1mm]

    1.5.01 & Engineering, design, and prototypes &  10 people &        240 &        240 &        500 &          0 \\[-1mm]

    1.5.02 & IP Be vacuum chamber  &    2 units &          6 &         12 &        250 &          0 \\[-1mm]

    1.5.03 & IP Be vacuum bellows &    3 units &          2 &          4 &        200 &          0 \\[-1mm]

    1.5.04 & QD0H quadrupole (PM, 1.4 T, 0.46 m) &    2 units &          4 &          8 &        450 &          0 \\[-1mm]

    1.5.05 & QD0H quadrupole vacuum chamber &    2 units &          4 &          4 &        350 &          0 \\[-1mm]

    1.5.06 & QF1L quadrupole (SC, 0.4 m) &    2 units &          6 &         12 &        600 &          0 \\[-1mm]

    1.5.07 & QF1L quadrupole vacuum chamber &    2 units &          4 &          4 &        300 &          0 \\[-1mm]

    1.5.08 & Magic vacuum flange and remote act. &    2 units &          6 &         12 &        200 &          0 \\[-1mm]

    1.5.09 & Supports for QD0H and QF1 &    2 units &          3 &          6 &        250 &          0 \\[-1mm]

    1.5.10 & B00L magnet (0.4 m) &    2 units &          4 &          8 &        300 &          0 \\[-1mm]

    1.5.11 & QD0H (0.29 m) &    2 units &          3 &          6 &        400 &          0 \\[-1mm]

    1.5.12 & B00H (0.4 m) &    2 units &          3 &          6 &        250 &          0 \\[-1mm]

    1.5.13 & QF1H (0.4m) &    2 units &          3 &          6 &        400 &          0 \\[-1mm]

    1.5.14 &  B0L (2 m) &    2 units &          3 &          6 &        300 &          0 \\[-1mm]

    1.5.15 &  B0H (2 m) &    2 units &          3 &          6 &        300 &          0 \\[-1mm]

    1.5.16 & Forward support raft &     1 unit &          5 &          5 &        300 &          0 \\[-1mm]

    1.5.17 & Backward support raft &     1 unit &          5 &          5 &        300 &          0 \\[-1mm]

    1.5.18 & Cryogenic refrigerator &     1 unit &          3 &          6 &        700 &          0 \\[-1mm]

    1.5.19 & Croygenic distribution piping &     1 unit &          3 &          6 &        500 &          0 \\[-1mm]

    1.5.20 & Cryogenic controls &     1 unit &          3 &          4 &        300 &          0 \\[-1mm]

    1.5.21 & LN2 storage and distribution &     1 unit &          3 &          6 &        300 &          0 \\[-1mm]

    1.5.22 & H2O Cooling lines for IR components &      100 m &          3 &          6 &        250 &          0 \\[-1mm]

    1.5.23 & Water chillers for IR components &    4 units &          2 &          4 &        250 &          0 \\[-1mm]

    1.5.24 & Vacuum pump holding &   12 units &          2 &          3 &        200 &          0 \\[-1mm]

    1.5.25 & Vacuum holding controls and PS &   12 units &          2 &          3 &        100 &          0 \\[-1mm]

    1.5.26 & Vacuum pump TSPs &   20 units &          3 &          6 &        150 &          0 \\[-1mm]

    1.5.27 & Vacuum pump TSP controls and PS &   20 units &          3 &          6 &        150 &          0 \\[-1mm]

    1.5.28 & H2O temperature regulation &   10 units &          3 &          6 &        300 &          0 \\[-1mm]

    1.5.29 & Installation of IR components &   +/- 10 m &         12 &         36 &       1500 &          0 \\[-1mm]

    1.5.30 & Alignment of IR components &   +/- 10 m &         12 &         12 &        300 &          0 \\[-1mm]

    1.5.31 & Active vibration suppression for quads &    8 units &         12 &         24 &        300 &          0 \\[-1mm]
\hline
 {\bf 1.6} & {\bf Controls, Diagnostics, Feedback} &     {\bf } &  {\bf 963} &  {\bf 648} & {\bf 12951} & {\bf 8750} \\[-1mm]

    1.6.01 & Engineering, design, and prototypes &  15 people &        480 &        240 &        600 &          0 \\[-1mm]

    1.6.02 & Beam position monitor electronics & 1000 units &         24 &         48 &       2000 &          0 \\[-1mm]

    1.6.03 & Beam position monitor cables &    20000 m &         24 &         48 &       2000 &          0 \\[-1mm]

    1.6.04 & Master control computer &     2 each &          2 &          4 &        200 &          0 \\[-1mm]

    1.6.05 & Local computers &    14 each &          5 &         10 &        200 &          0 \\[-1mm]

    1.6.06 & Network and routers &    14 each &          5 &         10 &        400 &          0 \\[-1mm]

    1.6.07 & Control software &   6 people &        288 &         24 &        200 &          0 \\[-1mm]

    1.6.08 & Interface electronics &  500 units &         24 &         48 &        500 &          0 \\[-1mm]

    1.6.09 & Synchrotron light monitors (xray) &    2 units &          6 &         12 &       1500 &       1000 \\[-1mm]

    1.6.10 & Data storage &  1000 Gbit &          2 &          4 &         20 &          0 \\[-1mm]

    1.6.11 & Long fdbk kicker-electronics refurbish &    6 units &         12 &         24 &        500 &       2000 \\[-1mm]

    1.6.12 & Long fdbk shipping &    6 units &          1 &          1 &        100 &          0 \\[-1mm]

    1.6.13 & Trans fdbk kicker-electronic refurbish &    6 units &         12 &         24 &        500 &       2000 \\[-1mm]

    1.6.14 & Trans fdbk shipping &    6 units &          1 &          1 &        100 &          0 \\[-1mm]

    1.6.15 & Installation BxB feedback systems &    8 units &          6 &         12 &        150 &          0 \\[-1mm]

    1.6.16 & ECI monitors &    4 units &          6 &         12 &         50 &          0 \\[-1mm]

    1.6.17 & Bunch current monitors refurbish &    2 units &          2 &          4 &         20 &        300 \\[-1mm]

    1.6.18 & Bunch current monitors ship &    2 units &          1 &          1 &         20 &          0 \\[-1mm]

    1.6.19 & Bunch current monitors install &    2 units &          4 &          8 &        100 &          0 \\[-1mm]

    1.6.20 & Bunch length monitors refurbish &    2 units &          1 &          2 &         20 &        600 \\[-1mm]

    1.6.21 & Bunch length monitors ship &    2 units &          1 &          2 &         10 &          0 \\[-1mm]

    1.6.22 & Bunch length monitors install &    2 units &          3 &          6 &         80 &          0 \\[-1mm]

    1.6.23 & Luminosity monitor refurbish &     1 unit &          1 &          2 &         20 &        200 \\[-1mm]

    1.6.24 & Luminosity monitor ship &     1 unit &          1 &          2 &         10 &          0 \\[-1mm]

    1.6.25 & Luminosity monitor install &     1 unit &          3 &          6 &        100 &          0 \\[-1mm]

    1.6.26 & Polarization monitor &     1 unit &         12 &         24 &       1000 &          0 \\[-1mm]

    1.6.27 & RF master oscillator &     1 unit &          3 &          6 &         50 &          0 \\[-1mm]

    1.6.28 & Timing generator &     1 unit &          3 &          6 &         50 &          0 \\[-1mm]

    1.6.29 & Bunch injection controller &     1 unit &          3 &          6 &         50 &          0 \\[-1mm]

    1.6.30 & Loss monitors refurbish &  300 units &          1 &          2 &        100 &        900 \\[-1mm]

    1.6.31 & Loss monitors ship &  300 units &          1 &          1 &        150 &          0 \\[-1mm]

    1.6.32 & Loss monitors install &  300 units &          3 &          6 &        200 &          0 \\[-1mm]

    1.6.33 & IR background detectors refurbish &   30 units &          1 &          1 &         25 &        250 \\[-1mm]

    1.6.34 & IR background detectors ship &   30 units &          1 &          1 &         50 &          0 \\[-1mm]

    1.6.35 & IR background detectors install &   30 units &          3 &          6 &         80 &          0 \\[-1mm]

    1.6.36 & Fast IP position fdbk refurbish &    6 units &          1 &          2 &         10 &        200 \\[-1mm]

    1.6.37 & Fast IP position fdbk ship &    6 units &          1 &          2 &         20 &          0 \\[-1mm]

    1.6.38 & Fast IP position fdbk install &    6 units &          4 &          8 &         80 &          0 \\[-1mm]

    1.6.39 & Temp sensors for ring components & 2000 units &          5 &         10 &        666 &       1000 \\[-1mm]

    1.6.40 & Polarization controls &     1 unit &          5 &         10 &       1000 &        300 \\[-1mm]

    1.6.41 & IP HOM monitors &     1 unit &          1 &          2 &         20 &          0 \\[-1mm]
\hline
 {\bf 1.7} & {\bf Injection and transport systems} &     {\bf } &  {\bf 426} &  {\bf 252} & {\bf 86600} & {\bf 18000} \\[-1mm]

    1.7.01 & Engineering, design, and prototypes &  10 people &        360 &        120 &        600 &          0 \\[-1mm]

    1.7.02 & Injection e- gun  &            &         12 &         24 &       2000 &       2000 \\[-1mm]

    1.7.03 & Positron target and capture region &            &         12 &         24 &       2000 &       3000 \\[-1mm]

    1.7.04 & 6 GeV Linac &            &         12 &         24 &      50000 &       5000 \\[-1mm]

    1.7.05 & 2 x 1 GeV damping ring &            &         12 &         24 &      20000 &       3000 \\[-1mm]

    1.7.06 & Two injection transport lines &            &         12 &         24 &      10000 &       3000 \\[-1mm]

    1.7.07 & Polarization manipulation for injection &            &          6 &         12 &       2000 &       2000 \\[-1mm]

\hline
\end{longtable}
}
\end{center}
}

\subsection*{Site and Utilities}

The estimated cost of the site and utilities for the \superb\ project is presented in Table~\ref{tab:cost_site}.
This cost depends significantly on the site choice.

{ \setlength{\tabcolsep}{4pt}  
\begin{center}
{\footnotesize
\begin{longtable}{lp{6cm}p{1.5cm}rrrr}
\caption{\superb\ site and utilities budget. }
\label{tab:cost_site} \\

\hline\hline
\textbf{ } & \textbf{ } & \textbf{Number} & \textbf{EDIA} & \textbf{Labor} & \textbf{M\&S} & \textbf{Rep.Val.} \\
\textbf{WBS} & \textbf{Item} & \textbf{Units} & \textbf{mm} & \textbf{mm} & \textbf{kEuro} & \textbf{kEuro} \\
\hline
\endfirsthead

\multicolumn{7}{c}%
{\tablename\ \thetable{} -- continued from previous page} \\[3mm]
\hline\hline
\textbf{ } & \textbf{ } & \textbf{Number} & \textbf{EDIA} & \textbf{Labor} & \textbf{M\&S} & \textbf{Rep.Val.} \\
\textbf{WBS} & \textbf{Item} & \textbf{Units} & \textbf{mm} & \textbf{mm} & \textbf{kEuro} & \textbf{kEuro} \\
\hline
\endhead

 \\ \hline
  \multicolumn{7}{c}{Continued on next page} \\
\endfoot

\endlastfoot


 {\bf 2.0} & {\bf Site} &     {\bf } & {\bf 1424} & {\bf 1660} & {\bf 105700} &    {\bf 0} \\[-1mm]
\hline
 {\bf 2.1} & {\bf Site Utilities} &     {\bf } &  {\bf 820} & {\bf 1040} & {\bf 31700} &    {\bf 0} \\[-1mm]

    2.1.01 & Engineering, design, and prototypes &  10 people &        360 &        120 &        500 &          0 \\[-1mm]

    2.1.02 & Cooling towers & 3x10 MW ea &         30 &         60 &       6000 &          0 \\[-1mm]

    2.1.03 & Water pump pads &     3 each &         12 &         24 &       2000 &          0 \\[-1mm]

    2.1.04 & Water pumps &   20 pumps &         12 &         24 &       1000 &          0 \\[-1mm]

    2.1.05 & Water piping &    10000 m &        200 &        400 &      10000 &          0 \\[-1mm]

    2.1.06 & Power transformers & 3x10 MW ea &         50 &        100 &       3000 &          0 \\[-1mm]

    2.1.07 & Stepdown transformers & 12 at 500 KW ea &         50 &        100 &       1200 &          0 \\[-1mm]

    2.1.08 & Power AC wiring &      10 km &        100 &        200 &       5000 &          0 \\[-1mm]

    2.1.09 & Air conditioning for support halls & 6 buildings &          6 &         12 &       3000 &          0 \\[-1mm]
\hline
 {\bf 2.2} & {\bf Tunnel and Support Buildings} &     {\bf } &  {\bf 604} &  {\bf 620} & {\bf 74000} &    {\bf 0} \\[-1mm]

    2.2.01 & Engineering, design, and prototypes &  10 people &        360 &        120 &        500 &          0 \\[-1mm]

    2.2.02 & Ring tunnel boring &     2300 m &        150 &        300 &      35000 &          0 \\[-1mm]

    2.2.03 & Ring tunnel equipment alcoves & 6 ea mid-arc &         12 &         25 &       3000 &          0 \\[-1mm]

    2.2.04 & Ring cable and RF shafts &    12 each &         12 &         25 &       3000 &          0 \\[-1mm]

    2.2.05 & Straight equipment halls & 3x350 sq-m &         15 &         40 &      15000 &          0 \\[-1mm]

    2.2.06 & Interaction region hall &   500 sq-m &         25 &         50 &      10000 &          0 \\[-1mm]

    2.2.07 & Linac tunnel and beam transport lines &      500 m &         30 &         60 &       7500 &          0 \\[-1mm]

    2.2.08 & Linac technical support gallery &  2000 sq-m &         30 &         60 &       7000 &          0 \\[-1mm]

\hline
\end{longtable}
}
\end{center}
}

\section{Detector}

The \superb\ detector uses for the most part the same technology used
for \babar. The cost, detailed in Table~\ref{tab:cost_detector} broken down for detector subsystem, is therefore in general a direct extrapolation of
\babar\ costs. Where different or updated technology is used, the basis
of estimate is detailed in the following paragraphs.
As discussed in the Chapter~\ref{sec:Detector}, the \superb\ detector
is not completely defined: some components, such as the forward PID
and the backward calorimeter, have overall integration and performance
implications that need to be carefully studied before deciding to
install them; for some other components, such the SVT layer 0 or the
DIRC readout, new promising technologies exist that require additional
R\&D before they can be used in a full scale detector. The cost
estimate lists the different technologies separately, but in the
rolled-up value the technology that is considered most likely to be
used is included. The ones that are not included in the rolled-up value
 are shown in italics  in the table.


{ \setlength{\tabcolsep}{4pt}  
\begin{center}
{\footnotesize
\begin{longtable}{lp{6cm}rrrr}
\caption{\superb\ detector budget. }
\label{tab:cost_detector} \\

\hline\hline
\textbf{ } & \textbf{ }  & \textbf{EDIA} & \textbf{Labor} & \textbf{M\&S} & \textbf{Rep.Val.} \\
\textbf{WBS} & \textbf{Item}  & \textbf{mm} & \textbf{mm} & \textbf{kEuro} & \textbf{kEuro} \\
\hline
\endfirsthead

\multicolumn{6}{c}%
{\tablename\ \thetable{} -- continued from previous page} \\[3mm]
\hline\hline
\textbf{ } & \textbf{ }  & \textbf{EDIA} & \textbf{Labor} & \textbf{M\&S} & \textbf{Rep.Val.} \\
\textbf{WBS} & \textbf{Item}  & \textbf{mm} & \textbf{mm} & \textbf{kEuro} & \textbf{kEuro} \\
\hline
\endhead

 \\ \hline
  \multicolumn{6}{c}{Continued on next page} \\
\endfoot

\endlastfoot


   {\bf 1} & {\bf SuperB detector} & {\bf 3391} & {\bf 1873} & {\bf 40747} & {\bf 46471} \\[-1mm]
\hline
 {\bf 1.0} & {\bf Interaction region} &   {\bf 10} &    {\bf 4} &  {\bf 210} &    {\bf 0} \\[-1mm]

{\bf 1.0.1} & {\bf Be Beampipe} &   {\bf 10} &    {\bf 4} &  {\bf 210} &    {\bf 0} \\[-1mm]

   1.0.1.1 & Vertex chamber design &          4 &          0 &          0 &    {\bf 0} \\[-1mm]

   1.0.1.2 & Finalize Physics Req.mnts &          1 &          0 &          0 &    {\bf 0} \\[-1mm]

   1.0.1.3 & Fab method &          1 &          0 &          0 &    {\bf 0} \\[-1mm]

   1.0.1.4 & Design Review &          1 &          0 &          0 &          0 \\[-1mm]

   1.0.1.5 & Chamber detailing &          2 &          0 &          0 &          0 \\[-1mm]

   1.0.1.6 & Support procurement &          2 &          0 &          4 &          0 \\[-1mm]

   1.0.1.7 & Procure Beampipe Assembly &          0 &          0 &        197 &          0 \\[-1mm]

   1.0.1.8 & Procure Vx chamber Misc parts &          0 &          0 &          9 &          0 \\[-1mm]

   1.0.1.9 & Assemble Vx chamber, test, clean &          0 &          2 &          0 &          0 \\[-1mm]

   1.0.1.A & Assemble Vx chamber fixtures &          0 &          2 &          0 &          0 \\[-1mm]
\hline
 {\bf 1.1} & {\bf Tracker (SVT + L0 MAPS)} &  {\bf 248} &  {\bf 348} & {\bf 5615} &    {\bf 0} \\[-1mm]

{\bf 1.1.1} &  {\bf SVT} &  {\bf 142} &  {\bf 317} & {\bf 4380} &    {\bf 0} \\[-1mm]

   1.1.1.1 & Mechanical &         14 &        186 &        313 &          0 \\[-1mm]

   1.1.1.2 &    Cooling &          4 &          5 &         72 &          0 \\[-1mm]

   1.1.1.3 & Silicon Wafers and Fanout &         37 &        107 &       3240 &          0 \\[-1mm]

   1.1.1.4 & On-detector electronics &         69 &         11 &        672 &          0 \\[-1mm]

   1.1.1.5 & Detector monitoring &          4 &          4 &         73 &          0 \\[-1mm]

   1.1.1.6 & Detector assembly &          2 &          4 &          0 &          0 \\[-1mm]

   1.1.1.7 & System Engineering &         36 &          0 &         20 &          0 \\[-1mm]

\textit{\textbf{1.1.2}} & \textit{\textbf{L0 Striplet option}} & \textit{\textbf{23}} & \textit{\textbf{33}} & \textit{\textbf{324}} & \textit{\textbf{0}} \\[-1mm]

{\it 1.1.2.1} & {\it Mechanical} &    {\it 7} &   {\it 17} &    {\it 3} &    {\it 0} \\[-1mm]

{\it 1.1.2.2} & {\it Cooling } &    {\it 2} &    {\it 1} &    {\it 6} &    {\it 0} \\[-1mm]

{\it 1.1.2.3} & {\it Silicon Wafers and Fanout} &   {\it 10} &   {\it 15} &  {\it 257} &    {\it 0} \\[-1mm]

{\it 1.1.2.4} & {\it On-detector electronics} &    {\it 5} &    {\it 1} &   {\it 58} &    {\it 0} \\[-1mm]

{\bf 1.1.3} & {\bf L0 MAPS option} &  {\bf 106} &   {\bf 32} & {\bf 1235} &    {\bf 0} \\[-1mm]

   1.1.3.1 & Mechanical &         12 &         18 &         75 &          0 \\[-1mm]

   1.1.3.2 &    Cooling &          2 &          2 &         20 &          0 \\[-1mm]

   1.1.3.3 & MAPS Modules Components &         92 &         12 &       1140 &          0 \\[-1mm]
\hline
 {\bf 1.2} &  {\bf DCH} &  {\bf 113} &  {\bf 104} & {\bf 2862} &    {\bf 0} \\[-1mm]

     1.2.1 & System engineering &         24 &          0 &         50 &          0 \\[-1mm]

     1.2.2 &  Endplates &         14 &          0 &        550 &          0 \\[-1mm]

     1.2.3 & Inner cylinder &          4 &          0 &        157 &          0 \\[-1mm]

     1.2.4 & Outer cylinder &          4 &          0 &        100 &          0 \\[-1mm]

     1.2.5 &       Wire &          3 &          0 &        242 &          0 \\[-1mm]

     1.2.6 & Feedthroughs &          9 &         10 &        345 &          0 \\[-1mm]

     1.2.A & Gas System &          4 &          8 &         50 &          0 \\[-1mm]

     1.2.B &       Test &          3 &          6 &         40 &          0 \\[-1mm]
\hline
 {\bf 1.3} & {\bf PID (DIRC Pixelated PMTs + TOF)} &  {\bf 110} &  {\bf 222} & {\bf 7953} & {\bf 6728} \\[-1mm]

{\bf 1.3.1} & {\bf DIRC barrel - Pixelated PMTs} &   {\bf 78} &  {\bf 152} & {\bf 4527} & {\bf 6728} \\[-1mm]

   1.3.1.1 & Radiator Support Structure &          0 &          0 &          0 &       2372 \\[-1mm]

   1.3.1.2 & Radiator Bars and Assemblies &          3 &          5 &       2245 &       4256 \\[-1mm]

   1.3.1.3 & Standoff box &          4 &          8 &        655 &          0 \\[-1mm]

   1.3.1.4 &   Detector &         18 &         32 &       1590 &          0 \\[-1mm]

   1.3.1.5 & Calibration System &          1 &          3 &         19 &          0 \\[-1mm]

   1.3.1.6 & Mechanical Utilities &          4 &          8 &         19 &        100 \\[-1mm]

   1.3.1.7 & System Integration &         48 &         96 &          0 &          0 \\[-1mm]

\textit{\textbf{1.3.1}} & \textit{\textbf{DIRC barrel - Focusing DIRC}} & \textit{\textbf{92}} & \textit{\textbf{179}} & \textit{\textbf{6959}} & \textit{\textbf{6728}} \\[-1mm]

{\it 1.3.1.1} & {\it Radiator Support Structure} &    {\it 0} &    {\it 0} &    {\it 0} & {\it 2372} \\[-1mm]

{\it 1.3.1.2} & {\it Radiator Bars and Assemblies} &    {\it 5} &    {\it 8} & {\it 2806} & {\it 4256} \\[-1mm]

{\it 1.3.1.3} & {\it Standoff box} &    {\it 4} &    {\it 8} &  {\it 655} &    {\it 0} \\[-1mm]

{\it 1.3.1.4} & {\it Detector} &   {\it 18} &   {\it 32} & {\it 3461} &    {\it 0} \\[-1mm]

{\it 1.3.1.5} & {\it Calibration System} &    {\it 1} &    {\it 3} &   {\it 19} &    {\it 0} \\[-1mm]

{\it 1.3.1.6} & {\it Mechanical Utilities} &    {\it 4} &    {\it 8} &   {\it 19} &  {\it 100} \\[-1mm]

{\it 1.3.1.7} & {\it System Integration} &   {\it 60} &  {\it 120} &    {\it 0} &    {\it 0} \\[-1mm]

{\bf 1.3.2} & {\bf Forward TOF} &   {\bf 32} &   {\bf 70} & {\bf 3426} &    {\bf 0} \\[-1mm]

   1.3.2.1 & Support structure &          3 &          3 &         75 &          0 \\[-1mm]

   1.3.2.2 & Radiator/Detector box &          6 &         10 &        220 &          0 \\[-1mm]

   1.3.2.3 &   Detector &          5 &         18 &       3010 &          0 \\[-1mm]

   1.3.2.4 & Calibration System &          0 &          3 &         21 &          0 \\[-1mm]
\hline
 {\bf 1.4} &  {\bf EMC} &  {\bf 136} &  {\bf 222} & {\bf 10095} & {\bf 30120} \\[-1mm]

{\bf 1.4.1} & {\bf Barrel EMC} &   {\bf 20} &    {\bf 5} &  {\bf 171} & {\bf 30120} \\[-1mm]

   1.4.1.1 & Crystal Procurement &          0 &          0 &          0 &      20560 \\[-1mm]

   1.4.1.2 & Light Sensors \& Readout &          0 &          0 &          0 &       2570 \\[-1mm]

   1.4.1.3 & Crystal Support Modules &          0 &          0 &          0 &       2824 \\[-1mm]

   1.4.1.4 & Barrel Structure &          0 &          0 &          0 &       3306 \\[-1mm]

   1.4.1.5 & Calibration Systems &          0 &          0 &          0 &        625 \\[-1mm]

   1.4.1.6 & Project Management &          0 &          0 &          0 &        233 \\[-1mm]

   1.4.1.7 & Barrel Transport &         20 &          5 &        171 &          0 \\[-1mm]

{\bf 1.4.2} & {\bf Forward EMC} &         73 &        152 &       6828 &          0 \\[-1mm]

   1.4.2.1 & Crystal Procurement &         11 &         44 &       5542 &          0 \\[-1mm]

   1.4.2.2 & Light Sensors \& Readout &         23 &         31 &        460 &          0 \\[-1mm]

   1.4.2.3 & Crystal Support Modules &         19 &         59 &        316 &          0 \\[-1mm]

   1.4.2.4 & Endcap Structure &         15 &         17 &        357 &          0 \\[-1mm]

   1.4.2.5 & Calibration Systems &          2 &          1 &        101 &          0 \\[-1mm]

   1.4.2.6 & Project Management &          4 &          0 &         53 &          0 \\[-1mm]

{\bf 1.4.3} & {\bf Backward EMC} &         42 &         65 &       3096 &          0 \\[-1mm]

   1.4.3.1 & Crystal Procurement &          1 &          8 &       2446 &          0 \\[-1mm]

   1.4.3.2 & Light Sensors \& Readout &          4 &         17 &        199 &          0 \\[-1mm]

   1.4.3.3 & Crystal Support Modules &          8 &         31 &        226 &          0 \\[-1mm]

   1.4.3.4 & Endcap Structure LSO &          9 &          8 &        149 &          0 \\[-1mm]

   1.4.3.5 & Extended Barrel Structure CsI &         17 &         11 &        106 &          0 \\[-1mm]

   1.4.3.6 & Calibration Systems &          1 &          1 &         48 &          0 \\[-1mm]
\hline
 {\bf 1.5} & {\bf IFR (scintillator)} &   {\bf 56} &   {\bf 54} & {\bf 1268} &    {\bf 0} \\[-1mm]

     1.5.1 & System engineering &         24 &          0 &          0 &          0 \\[-1mm]

     1.5.2 & scintillator strips +  WLS fiber &          0 &          0 &        320 &          0 \\[-1mm]

     1.5.3 & Module factory retooling &          0 &          3 &         15 &          0 \\[-1mm]

     1.5.4 & Module fabrication &          0 &         27 &         68 &          0 \\[-1mm]

     1.5.5 & module installation &          0 &          0 &         14 &          0 \\[-1mm]

     1.5.6 & APD/preamp + cooling system &          5 &          0 &        663 &          0 \\[-1mm]

     1.5.7 &   Services &          3 &          6 &         14 &          0 \\[-1mm]

     1.5.8 & LED pulser system &          0 &          0 &         24 &          0 \\[-1mm]

     1.5.9 & Detector Assembly &         24 &         18 &        151 &          0 \\[-1mm]
\hline
 {\bf 1.6} & {\bf Magnet} &   {\bf 87} &   {\bf 47} & {\bf 1545} & {\bf 9623} \\[-1mm]

     1.6.0 & System Management &         36 &          0 &          0 &        577 \\[-1mm]

     1.6.1 & Superconducting solenoid &          0 &          0 &          0 &       2282 \\[-1mm]

     1.6.2 & Mag. Power/Protection &          0 &          0 &          0 &        171 \\[-1mm]

     1.6.3 & Cryogenics &         34 &         36 &       1377 &          0 \\[-1mm]

     1.6.4 & Cryo monitor/Control &         17 &         11 &        168 &          0 \\[-1mm]

     1.6.5 & Flux return &          0 &          0 &          0 &       6108 \\[-1mm]

     1.6.6 & Installation/test equipment &          0 &          0 &          0 &        485 \\[-1mm]
\hline
 {\bf 1.7} & {\bf Electronics} &  {\bf 286} &  {\bf 213} & {\bf 5565} &    {\bf 0} \\[-1mm]

     1.7.1 &        SVT &         11 &         21 &        343 &          0 \\[-1mm]

     1.7.2 &        DCH &         53 &         47 &       1203 &          0 \\[-1mm]

     1.7.3 &        DRC &         22 &          0 &       1491 &          0 \\[-1mm]

     1.7.4 & Forward PID &         43 &          0 &        702 &          0 \\[-1mm]

     1.7.5 &        EMC &         44 &         69 &        907 &          0 \\[-1mm]

     1.7.6 &        IFR &          0 &         64 &        141 &          0 \\[-1mm]

     1.7.7 & Infrastructure &          4 &         12 &        247 &          0 \\[-1mm]

     1.7.8 & Systems Engineering &         12 &          0 &          0 &          0 \\[-1mm]

     1.7.9 &    Trigger &         97 &          0 &        532 &          0 \\[-1mm]
\hline
 {\bf 1.8} & {\bf Online computing} & {\bf 1272} &   {\bf 34} & {\bf 1624} &    {\bf 0} \\[-1mm]

     1.8.1 &        DAQ &        420 &         22 &        163 &          0 \\[-1mm]

     1.8.2 & Event Flow &        258 &          0 &       1177 &          0 \\[-1mm]

     1.8.3 & Run Control / Slow Controls &        258 &          0 &         51 &          0 \\[-1mm]

     1.8.4 & Infrastructure &         48 &         12 &        234 &          0 \\[-1mm]

     1.8.5 & Software Triggers &        216 &          0 &          0 &          0 \\[-1mm]

     1.8.6 & Coordination and Commissioning &         72 &          0 &          0 &          0 \\[-1mm]
\hline
 {\bf 1.9} & {\bf Installation and integration} &  {\bf 353} &  {\bf 624} & {\bf 3830} &    {\bf 0} \\[-1mm]

     1.9.1 & Disassembly &         95 &        161 &        510 &          0 \\[-1mm]

     1.9.2 &   Assembly &        222 &        463 &       3320 &          0 \\[-1mm]

     1.9.3 & Structural analysis &         36 &          0 &          0 &          0 \\[-1mm]
\hline
 {\bf 1.A} & {\bf Project Management} &  {\bf 720} &    {\bf 0} &  {\bf 180} &    {\bf 0} \\[-1mm]

     1.A.1 & Project engineering &        300 &          0 &        100 &          0 \\[-1mm]

     1.A.2 & Budget, Schedule and Procurement &        300 &          0 &         40 &          0 \\[-1mm]

     1.A.3 &    ES \& H &        120 &          0 &         40 &          0 \\[-1mm]

\hline
\end{longtable}
}
\end{center}
}

\subsection*{Vertex Detector and Tracker}

System cost is estimated based on the experience of the \babar\
detector. A detailed estimate is provided for the cost of the
main detector (layers-1 to 5), while the layer-0 is analyzed
separately, with two different estimates provided, corresponding
to the striplets and the monolithic pixel option. The total cost is obtained summing the main detector cost to the MAPS layer0 cost.

\subsection*{Drift Chamber}
The DCH costing is based on an extrapolation of the costs for the existing \babar\ chamber, assuming the \superb\ design to be comparable. In particular, assuming the number of cells will be similar, many other requirements, such as the length of wire, number of feedthroughs, duration of wire stringing, \etc, can be reliably estimated. Given the more challenging luminosity related backgrounds, we assume that the endplates will be conical in shape and fabricated from carbon fiber composites. This will add a significant period of R\&D to develop the relevant fabrication techniques, include engineering support, but will probably not result in significantly larger production costs for the final endplates.

\subsection*{Particle Identification}
Barrel PID costs and replacement values are derived from \babar\ costs as extrapolated to 2007, except that photon detector costs are taken directly from tube manufacturers' preliminary quotes. A number of different options for Barrel DIRC upgrades are discussed in the detector section. The base option costed uses new pixilated tubes to allow a small, robust detection box for the Barrel DIRC without water coupling, and a forward TOF PID system. A focusing DIRC option with better performance in the barrel is estimated separately, but not summed into the total. The forward TOF PID system cost is expected to be dominated by the cost of the MCP photon detectors, as estimated by the manufacturer (Burle/Photonis).

\subsection*{Electromagnetic Calorimeter}
Electromagnetic calorimeter costs use as their basis expenditures made in construction of the \babar barrel EMC. The reuse value of the barrel calorimeter is based on the actual cost of the barrel escalated for inflation from the time of construction to the current year. Manpower estimates for the barrel construction were obtained by using the costs for ED\&I and Labor, knowledge of the mix of engineers and technicians who contributed to the design and fabrication of individual components, and knowledge of the their salaries. Manpower and costs for engineering and tooling required for the removal and transport of the barrel EMC from SLAC are engineering estimates. The costs and manpower for barrel construction, combined with knowledge of current prices for cost-driver materials, are used to estimate the costs of and manpower needs for forward and backward endcap construction. Cost drivers are crystal procurement, light sensors, and support structures. For the crystals, a strawman layout provides the volume of material needed. Current fabrication prices for full sized crystals are used to obtain materials costs based on this volume. Ancillary costs/manpower needs for vendor development, facility preparation, and crystal Q/C are obtained by scaling barrel costs for the same items, typically by crystal count. Light sensor costs are obtained by use of \babar\ sensor costs escalated for inflation, as well as costs of more recent vendor cost estimates. Ancillary costs and manpower estimates are again obtained from scaling the \babar\ barrel actuals by crystal count in the barrel and strawman endcap. Estimates for costs and manpower for structures are also obtained from barrel calorimeter experience: for the crystal support structure, barrel module actuals are scaled by crystal count to obtain endcap crystal module estimates; for the overall endcap support structure, scaling is more closely tied to relative solid angle. Backward endcap EMC cost and manpower estimates are, because of the sketchy nature of design concepts, less reliable than those for the forward endcap, though the methodology used for the backward endcap is similar to that of the forward endcap.

\subsection*{Instrumented Flux Return}
The  Instrumented Flux Return costs are based on the experience developed for the upgrade of the \babar\ barrel IFR subdetector in 2002 when a scintillator option for \babar\ was developed.

\subsection*{Electronics}

The cost for the Electronics subsystems were estimated with a
combination of scaling from the \babar experience and from direct
estimates.  Infrastructure, high and low voltage, and other items
expected to be similar to those used in \babar were estimated by
scaling the costs from \babar. The readout systems for the DCH, EMC
and DIRC, where the higher data rates require redesigned electronics,
were estimated from the number of different ASICs and printed circuit
boards, and the associated chip and board count. The block diagram
design of the readout electronics for these detectors was based on the
\babar experience, and the cost estimates also reflects that basis.
While any new design would aim to make the readout systems as
uniform as possible across these systems, no associated cost savings
was assumed.

\subsection*{Trigger and DAQ}
The labor and EDIA costs for online computing and DAQ are largely based on extrapolation from the actual \babar\ experience, with some modifications based on "lessons learned" -- in particular, shifting some work on feature extraction from detector subsystems to the core online computing group.

The hardware cost estimates for online computing are based on the current prices of hardware necessary to build the system, with the assumption that Moore's Law will result in future systems with the same unit costs but higher performance (except in the case of networking equipment, where current performance is sufficient for the design).

The cost estimates for DAQ are based on estimates from the SLAC PPA electronics group of the current costs for the "CE" electronics building blocks they are now building for other SLAC projects, which we have assumed as the core elements of the \superb\ DAQ.

A more detailed discussion of considerations in the estimation of online computing costs can be found in Section~\ref{sec:det:TDAQ}




\subsection*{Computing}

Costing the (offline) computing system is different from the
other detector subsystems for two reasons:
\begin{enumerate}
 \item There is no precise ``construction phase''. Instead
computing is best regarded as a continuous operating expense,
through the life of the experiment. As additional data is accumulated,
additional computing resources are required. This is true as well for
the engineering and support required as the system evolves in time.
 \item The cost per unit of computing is a strong function of
time, due to the ``Moore's law'' effect in computing performance.
Assumptions on what year the experiment starts have a significant impact
on cost estimates.
\end{enumerate}
In addition, the computing model is distributed, and it may be assumed
that many of the costs are also borne in a distributed fashion.
Nevertheless, substantial computing resources are required, and it
is important to give an indication of the overall costs.
It should also be noted that data recording and monitoring infrastructure
needs to be localized at the accelerator site and must be fully
functional when data taking begins.

Table~\ref{tab:compCost} shows estimated annual costs for two
different assumptions for the year of first data. All numbers are given in
2007 kEuros. It is assumed that the first major computing investment occurs
in the year prior to first data. Tape silos and drives are constant cost per
unit items; the differences  in cost for these lines in the two
starting year models is due to the
evolving tape density.

\begin{table}[h!tb]
\caption{Annual computing costs for two different assumptions
for the starting year. All amounts are in 2007 kEuros. \label{tab:compCost}}
\begin{center}
\begin{tabular}{lrrrrr}
\hline\hline
Luminosity (ab$^{-1}$) & 0 & 2 & 6 & 12 & 12\\
\hline
Year & 2010 & 2011 & 2012 & 2013 & 2014 \\
\hline
Tapes & 19 & 171 & 314 & 677 & 323 \\
Tape drives & 24 & 213 & 393 & 846 & 404 \\
Tape silos & 270 & 270 & 270 & 539 & 270 \\
Disk & 58 & 345 & 1031 & 1569 & 1410 \\
CPU & 140 & 841 & 1888 & 2232 & 1095 \\
Replacements & 0 & 0 & 0 & 0 & 33 \\[2mm]
Total & 510 & 1840 & 3895 & 5863 & 3533 \\
& \\
\hline
Year & 2011 & 2012 & 2013 & 2014 & 2015 \\
\hline
Tapes & 19 & 85 & 314 & 271 & 323 \\
Tape drives & 24 & 107 & 393 & 338 & 404 \\
Tape silos & 270 & 270 & 270 & 270 & 270 \\
Disk & 38 & 230 & 696 & 1046 & 940 \\
CPU & 93 & 561  & 1274 & 1488 & 730 \\
Replacements & 0 & 0 & 0 & 0 & 22 \\[2mm]
Total & 444 & 1253 & 2947 & 3413 & 2687 \\
\hline
\end{tabular}
\end{center}
\end{table}

\subsection*{Transportation, installation, and commissioning }
Installation and commissioning estimates, including disassembling and reassembling \babar\, are based on the \babar\ experience, using a detail schedule of activities and corresponding manpower requirement.

Although transportation costs are expected to be significant, they are not included in this estimates, because of lack of detailed information at the time of writing.

\section{Schedule}

\begin{figure}[htb]
\centering
\includegraphics[width=1.0\textwidth]{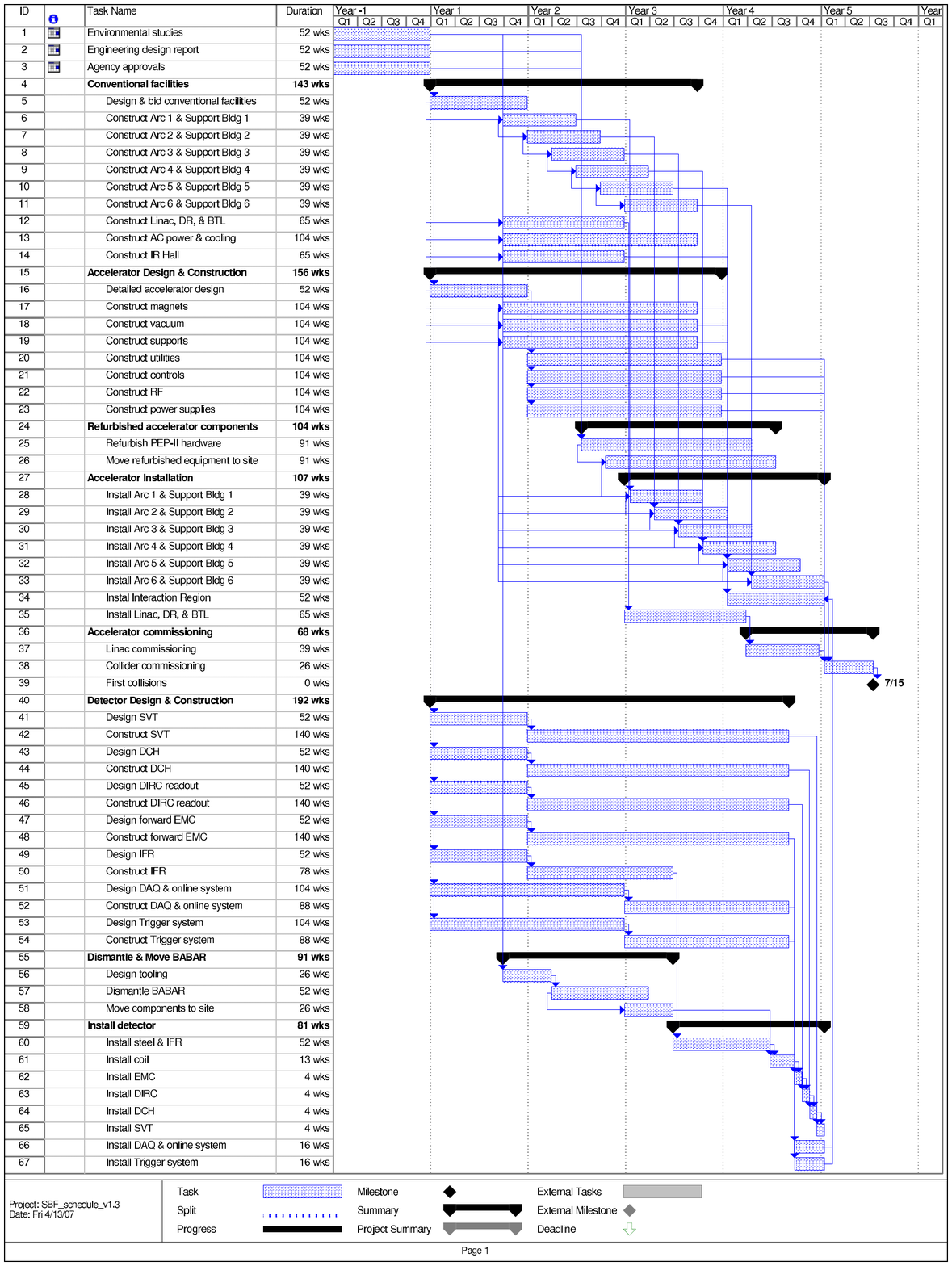}
\caption{Overall schedule for the construction of the \superb\ project.}
\label{fig:schedule}
\end{figure}

The accelerator and detector construction schedule is shown in Fig.~\ref{fig:schedule}. The present \pepii
accelerator (2.2\km with two rings) was built in about four years, but the tunnel already
existed. For the \superb\ schedule, we have added an additional year for tunnel and support building
construction.
      The construction starts with environmental, design finalization, and contract
bidding. The accelerator infrastructure is constructed with a phased approach, moving
around the ring. The accelerator components are installed, again with a phased
approach around the ring followed by system checkout . Finally, beam
commissioning starts, as well as first beam collisions.
The detector schedule is dominated by the time required to disassemble the \babar\ detector,
transport it to the new site, and reassemble it.
The total construction and
commissioning time is estimated to be a little over 5 years.

\afterpage{\clearpage}

\graphicspath{{Appendix/figures/}}
\appendix
\chapter{Appendix}{ }
\label{sec:ph_lattice}
\section{Lattice QCD Calculations}


\def\simge{\mathrel{\rlap{\raise 0.511ex \hbox{$>$}}{\lower 0.511ex
\hbox{$\sim$}}}}
\def\simle{\mathrel{\rlap{\raise 0.511ex \hbox{$<$}}{\lower 0.511ex
\hbox{$\sim$}}}}

The New Physics discovery potential of a Super $B$ Factory
depends on our
capability to control the theoretical determination of hadronic matrix
elements at a level of accuracy comparable to that to be achieved by the
experimental measurements.
For most of the hadronic parameters relevant to flavour physics
the precision necessary to fulfill such a requirement is at the
level of few percent or better.

Lattice QCD is the theoretical tool of choice to compute hadronic quantities.
Being based on first principles, it does not introduce additional free
parameters besides the fundamental couplings of QCD, namely the strong coupling
constant and the quark masses. In addition, all systematic uncertainties
affecting the results of lattice calculations can be systematically reduced in
time, with the continuously increasing availability of computing power. The
development of new algorithms and of new theoretical techniques further speeds
up the process of improving precision.

In spite of the appealing features of the lattice approach, the accuracy
currently reached in the determination of the hadronic matrix elements is
typically at the level of $10$--$15\%$,
{\it i.e.} by far larger than the percent
precision required to match the experimental accuracy at \superb.
For many years, the lack of sufficient computing power has
prevented the possibility of performing ``full QCD'' simulations,
and forced the introduction of the so-called quenched approximation.
In this approximation an error is introduced
which, besides being process dependent, is also difficult to reliably estimate.
This is the main reason why, for most of the relevant observables, the accuracy
of the lattice results has not improved significantly in the last ten years or
so.

In order to assess the impact of \superb\ on our understanding of quark
flavour physics, in this section we attempt to estimate the precision that will
be reached by lattice QCD calculations at the time when such a machine could be
running and producing results. This estimate is unavoidably affected by some
uncertainties. The dominant sources of errors in lattice QCD calculations have
systematic origin, so that the accuracy of the results does not improve in time
according to simple scaling laws. Predictions in this context are necessarily
based on somehow educated guesses, and it should be taken into account that
their reliability decreases the more we attempt to go further in time.

In the following analysis we are going to neglect the impact of algorithmic
improvements and of the development of new theoretical techniques. Past
experience indicates that this effect is actually far from being negligible.
The role played by theoretical developments in improving the accuracy of
lattice QCD
calculations has been as important, to date, as the increase of
computational power. In the last few years, for instance, the improvement of
Monte Carlo algorithms for unquenched calculations has allowed to speed up the
simulations by order of magnitudes, and to simulate light quarks with masses
much closer to their physical values. At the same time, however, the impact of
future theoretical improvements is difficult to predict. For this reason, we
will neglect its effect in this study and we will conservatively only take into
account the increase of precision in lattice QCD calculations which is expected
by the increase of computational power.

\subsection{Estimate of Computational Power}
\begin{figure}[ht]
  \begin{center}
    \includegraphics[scale=0.6]{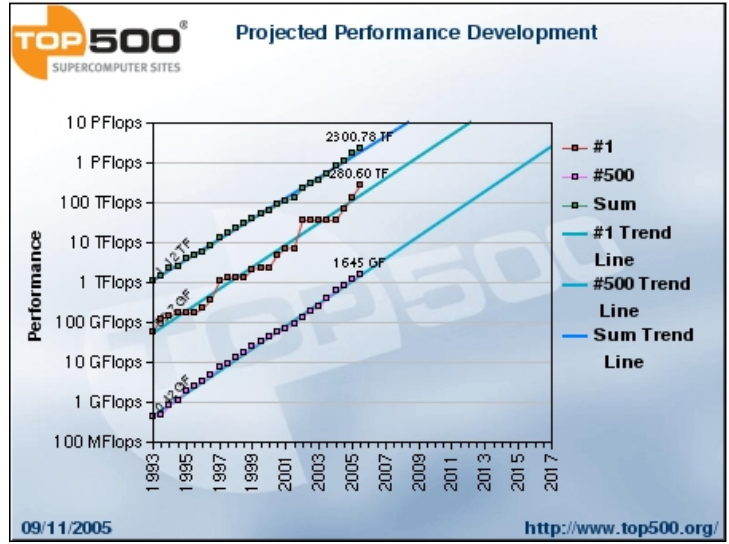}
    \vspace*{2ex}
    \caption{
      Performance development of the most powerful computer systems
      in the world as a function of the year~\cite{top500}.
      The top (green) points correspond to the sum of the
      500 fastest computer systems,
      the middle (red) points to the fastest in the world,
      the bottom (pink) points to the number 500 in the list.
    }
    \label{fig:top500}
  \end{center}
\end{figure}

At variance with other ingredients in the present analysis, the increase with
time of the computational power can be predicted with rather good reliability,
since it is found to follow closely a simple scaling law.
This is illustrated in Fig.~\ref{fig:top500},
which shows the performances of the 500 most powerful
computer systems in the world as a function of the year. From this plot it can
be seen that, at present, the most powerful computer system in the world, an IBM
BlueGene/L system installed at the Lawrence Livermore National Laboratory, has a
sustained speed of 280.6 TFlops, while the 500th in the list has a speed of
about 1.6 TFlops. The performances of typical computer systems available for
lattice QCD calculation today are in the range between 1 and 10 TFlops, thus
entering the lower part of the top-500 list.
Fig.~\ref{fig:top500} then allows us
to derive a prediction for the computer power available to a lattice QCD
collaboration around year 2015.
The typical sustained speed will be between 1 and 10 PFlops,
{\it i.e.}, three orders of magnitude faster than what we have today.

\subsection{Sources of Errors in Lattice QCD Calculations}
Uncertainties in lattice QCD calculations have both statistical and systematic
origin. We now briefly discuss these errors with the aim of understanding their
relevance in future simulations.

\begin{itemize}
\item[-] \underline{Statistical error}:
  this error arises from the fact that the
  functional integral expressing any correlation function in
  quantum field theory is approximated, in numerical simulations,
  by a sum over a finite number of gauge field configurations,
  weighted with their proper Boltzmann factor.
  Typical results of lattice calculations are obtained at present
  by averaging over ${\cal O}(100)$ (or more) independent gauge configurations,
  and the resulting statistical errors are at the level of
  a few percent or better.

\item[-] \underline{Discretization errors}:
  numerical simulations are performed on lattices with
  finite lattice spacing $a$, and an extrapolation to the continuum limit
  $a \to 0$ is required to get the physical result.
  In order to keep the uncertainty due to this extrapolation under control,
  it is important to work with lattices as fine as possible.
  However, for a given spatial or temporal extension $L$ of the lattice,
  the number of sites, $L/a$, increases when approaching the continuum limit,
  thus increasing the overall cost of the simulation.
  Typical values of the lattice spacing used in current simulations
  are around $a \simeq 1$ fm or smaller.

\item[-] \underline{Chiral extrapolation}:
  the inversion of the Dirac operator,
  required to generate the gauge field configurations and to compute quark
  propagators, becomes critical when approaching the chiral limit.
  This is due to presence of exact zero modes in this limit,
  the pseudoscalar Goldstone bosons of QCD.
  For this reason, the cost of numerical simulation increases with
  decreasing values of the light quark masses.
  The values of light quark masses reached in current simulations
  are in the range $1/6 \simle \hat m/m_s \simle 1/2$,
  where $m_s$ is the physical strange quark mass and
  $\hat m$ is the simulated average up-down quark mass.
  Eventually, a chiral extrapolation is thus required in order to bring
  the simulated light quark masses to agreement with their physical values,
  $\hat m \simeq m_s/25$.
  The chiral extrapolation introduces a systematic uncertainty
  which is obviously smaller for lighter values of the simulated quark masses.

\item[-] \underline{Heavy quark extrapolation}:
  the mass of the $b$ quark, $m_b$,
  is larger than the UV cutoff $\Lambda \sim 1/a$ in current lattice
  simulations, since $a^{-1} \simeq 2 \div 3$ GeV.
  Thus discretization effects,
  which are controlled by powers of $a\, m_b \gg 1$,
  prevent the possibility of simulating directly the $b$ quark on the lattice.
  Different approaches have been considered to circumvent this problem.
  A viable method passes through the introduction of effective field theories,
  either HQET or NRQCD, in which the
  heavy quark degrees of freedom are explicitly integrated out.
  Another possibility consists in considering specific discretizations of the
  heavy quark action on the lattice,
  constructed in such a way that discretization errors,
  of ${\cal O}((a m_b)^n)$, have coefficients suppressed by additional powers
  of the lattice spacing~\cite{El-Khadra:1996mp,Aoki:2001ra,Christ:2006us}.
  Finally, one can simulate heavy quarks with masses around the
  charm quark mass, for which $a m_H \simeq a m_c < 1$,
  and extrapolate the results to the $b$ quark mass.
  Each of these approaches has its advantages and drawbacks.
  A reliable estimate of the associated systematic uncertainties
  can only be achieved by comparing the results obtained
  using different approaches.
  The accuracy can be also increased by combining different approaches.
  For instance, the extrapolation of the results obtained in
  the charm quark mass region can be combined with the HQET determination
  of the same observable in the infinite quark mass limit to interpolate
  to the $b$-quark mass.

\item[-] \underline{Finite volume effects}:
  numerical simulations are performed on the lattice with
  finite spatial and temporal extension.
  Therefore, they are affected by finite size effects that
  must be kept under control.
  For a fixed value of the lattice spacing, the number of lattice sites
  increases with the physical size of the lattice,
  and with this number also increases the computational cost of the simulation.
  Since finite volume effects are larger for smaller values of the
  hadron masses (because the Compton wavelength of a particle increases
  with the inverse of its mass), the cost limitations on the lattice size
  also reflect on the values of the lightest quark masses that can be
  considered in a simulation.

\item[-] \underline{Renormalization procedure}:
  the direct outcomes of a lattice calculation are bare values of
  correlation functions which require to be renormalized in order
  to be related to the physical observables.
  The asymptotic freedom of QCD allows the computation of the
  relevant renormalization constants,
  which are connected to the UV properties of the theory,
  using the perturbative expansion.
  However, also because of the technical difficulties of
  lattice perturbation theory, the perturbative series are truncated
  at a typically low order (first or second),
  thus introducing a systematic uncertainties which is often
  far from being negligible. This uncertainty can be however removed using
  various non-perturbative renormalization techniques,
  which have been developed in the last ten years and can be applied
  in most of the relevant cases.
  At present, the typical accuracy reached in the non-perturbative
  determination of lattice renormalization constants is already
  at the level of 1\% or better.
  One expects that this accuracy will be further improved,
  so that it is unlikely that the renormalization procedure will
  represent one of the relevant limiting factors in improving the
  precision of lattice results in the next years.
\end{itemize}

\subsection{Uncertainties in Future Lattice Calculations}
Having discussed the various sources of systematic uncertainties in lattice
QCD calculations, we now proceed to derive the constraints on the parameters of
a numerical simulation (values of the lattice spacing, quark masses and lattice
size) which have to be satisfied in order to keep the systematic errors at the
level of accuracy required to match the experimental precision of a Super $B$
Factory. For definitiveness, we will require such a precision to be 1\% for the
simplest hadronic quantities ({\it e.g.} pseudoscalar decay constants and bag
parameters).

\subsubsection{Minimum Lattice Spacing}
The result for a given observable $Q_{\rm latt}$,
obtained on the lattice at a fixed value of the lattice spacing,
is related to its continuum counterpart $Q_{\rm cont}$
by an expansion of the form
\beq
Q_{\rm latt} = Q_{\rm cont} \left[ 1 + \left(a \Lambda_2\right)^2 +
  \left(a \Lambda_n\right)^n + \ldots \right] \, ,
\label{eq:aexp}
\eeq
where the parameters $\Lambda_2$, $\Lambda_n$, $\ldots$, for observables
involving only light quarks, are of the order of the hadronic scale
$\Lambda_{\rm QCD}$.
In the presence of heavy quarks, instead, the dominant
discretization effects are determined by the heavy quark mass,
{\it i.e.} $\Lambda_2 \sim \Lambda_n \sim m_H$.
In Eq.~(\ref{eq:aexp}) we have assumed that the fermionic action chosen
in the simulation belongs to the class of the
so-called ${\cal O}(a)$-improved actions,
implying the absence in the expansion of the leading discretization
effect of ${\cal O}(a)$.
As for the power $n$ of the subleading correction in Eq.~(\ref{eq:aexp}),
its value depends again on the choice of the action,
and one has $n=3$ for ${\cal O}(a)$-improved Wilson quarks
and $n=4$ for maximally twisted, staggered and Ginsparg-Wilson fermions.

In order to estimate the size of discretization effects left over in a lattice
calculation after the continuum extrapolation has been performed, we will
consider an heuristic argument~\cite{Sharpe} which, besides being very simple,
it is also expected to provide a conservative estimate.
In other words, we expect that the error evaluated in this way
can possibly be overestimated but it is unlikely to be underestimated.

The argument assumes that lattice simulations are performed at only two values
of the lattice spacing, namely $a_{\rm min}$ and $\sqrt{2}\, a_{\rm min}$,
and that the results for $Q_{\rm latt}$ are linearly extrapolated in $a^2$
to get a determination of the physical observable $Q_{\rm cont}$.
Eq.~(\ref{eq:aexp}) can be then used to
evaluate the error introduced in the extrapolation due to the fact that the
next-to-leading order term in the series, $\left(a \Lambda_n\right)^n$,
has been neglected.
A simple calculation gives:
\beq
\varepsilon \equiv \delta Q_{\rm cont}/Q_{\rm cont} \simeq (2^{n/2}-2) \,
\left(a_{\rm min} \Lambda_n \right)^n \, .
\eeq
Requiring the precision of the calculation to be at the level of 1\%,
{\it i.e.} imposing $\varepsilon=0.01$,
provides the minimum value of the lattice spacing that has to be considered
in the simulation. In the presence of light quarks
only, by conservatively assuming $\Lambda_n \simeq 0.8$ GeV, one finds
\beq
\left\{
  \begin{array}{ll}
    a_{\rm min} = 0.056 \ {\rm fm} \quad , \quad n=3 \\
    a_{\rm min} = 0.065 \ {\rm fm} \quad , \quad n=4
  \end{array}
\right. \qquad \rm{(light \ quarks \ only)} \ .
\label{eq:amin1}
\eeq
For studies with heavy quarks with masses around the charm quark mass,
$\Lambda_n \simeq m_c \simeq 1.5$ GeV and one gets
\beq
\left\{
  \begin{array}{ll}
    a_{\rm min} = 0.030 \ {\rm fm} \quad , \quad n=3 \\
    a_{\rm min} = 0.035 \ {\rm fm} \quad , \quad n=4
  \end{array}
\right. \qquad \rm{(heavy \ quarks)} \ .
\label{eq:amin2}
\eeq
For comparison, we recall that typical values of lattice spacing
used in current lattice simulations are in the range
$a ~\sim 0.06 \div 0.10$ fm.

\subsubsection{Minimum Quark Mass}
An argument similar to the one given above can be used to estimate the minimum
value of the light quark masses to be considered in lattice simulations
in order to keep the uncertainty associated with the chiral extrapolation
at the required level of precision~\cite{Sharpe}.

In QCD, the dependence of the physical quantities on the light quark masses is
predicted by chiral perturbation theory. The chiral expansion involves both
analytic (local) and non-analytic terms, the latter being generated by the
contribution of pion loops. To keep the argument simple, we neglect in the
following the presence of the non-analytic contributions and write the chiral
expansion of a given quantity computed on the lattice in the form
\beq
Q_{\rm latt} = Q_{\rm phys}
\left[
  1 + c_1 \left(m_P/m_V \right)^2 + c_2 \left(m_P/m_V \right)^4 + \ldots
\right] \, ,
\label{eq:mexp}
\eeq
where $m_P$ and $m_V$ represent the masses of the pseudoscalar and vector mesons
as obtained with the values of the quark masses considered in the simulation
($m_P=m_\pi$ and $m_V=m_\rho$ at the physical point).
The squared mass $m_P^2$ is proportional to the light quark masses,
while $m_V \simeq m_\rho$ provides the typical scale
entering the chiral expansion. In this way, the coefficients
$c_1$, $c_2$,~$\ldots$ are expected to be of ${\cal O}(1)$.
We are not going to consider that, in most of the cases,
these coefficients are actually predicted by chiral perturbation theory,
so that one is not forced to estimate their value by fitting the results
of the lattice calculation. In this approach, we are providing
again an estimate of the error which is presumably conservative.

By performing the calculation at two values of the light quark masses,
corresponding to $\left(m_P/m_V \right)_{\rm min}$ and
$\sqrt{2}\, \left(m_P/m_V \right)_{\rm min}$ respectively,
and by linearly extrapolating in $\left(m_P/m_V \right)^2$
to the physical point, one introduces an error in the determination
of $Q_{\rm phys}$ which is given, according to Eq.~(\ref{eq:mexp}), by
\beq
\varepsilon \equiv \delta Q_{\rm phys}/Q_{\rm phys} \simeq
2 \, c_2 \left(m_P/m_V\right)_{\rm min}^4 \, .
\eeq
The request of keeping this error at the percent level,
{\it i.e.} $\varepsilon=0.01$, then provides for $c_2=1$
the condition $\left(m_P/m_V\right)_{\rm min} \simeq 0.27$.
Expressed in terms of the light quark masses, this condition implies that
the minimal value of the light quark mass to be considered in the simulation,
in units of the physical strange quark mass, is given by
\beq
\left(\hat m/m_s\right)_{\rm min} \simeq 1/12 \, .
\label{eq:mqmin}
\eeq
Note that this ratio is still about twice larger than its physical value,
$\left(\hat m/m_s\right)_{\rm phys} \simeq 1/25$.
The lightest values of quark masses that have been reached instead
in present lattice calculations are about twice larger than that
indicated by Eq.~(\ref{eq:mqmin}),
namely $\left(\hat m/m_s\right) \simeq 1/6$.

\subsubsection{Minimum Box Size}
Finite volume effects are important to be kept into account in lattice QCD
simulations when aiming at percent accuracy. Since the distortion introduced
by a finite box is related to the infrared behaviour of the theory, its effect
can be estimated in QCD by using chiral perturbation
theory~\cite{Gasser:1986vb}.
One finds that, for correlation functions which are
free of physical singularities (thus, in particular, for all matrix elements
which contain at most one stable particle in the initial and final states)
finite size effects are exponentially suppressed. The dominant contribution
comes from the propagation of virtual pions and can be expressed as
\beq
\varepsilon \equiv \delta Q_{\rm phys}/Q_{\rm phys}
\sim C_Q(m_\pi,L)\, \exp(-m_\pi L)
\eeq
where $L$ is the one-dimensional size of the lattice and the function $C_Q$,
which depends on the observable $Q$, is typically a quantity of ${\cal O}(1)$
that can be explicitly computed.
Assuming $C_Q=1$ and requiring $\varepsilon =0.01$, one gets the requirement
\beq
m_\pi L \simeq 4.5 \, .
\label{eq:Lmin1}
\eeq

For values of the light quark masses as small as indicated
by Eq.~(\ref{eq:mqmin}) the corresponding pion mass turns out
to be of the order of 200 MeV. Therefore, the conditions
of Eq.~(\ref{eq:mqmin}) and Eq.~(\ref{eq:Lmin1}),
combined together, imply that the minimum spatial extension of the lattice
required to keep finite volume effects at the level of 1\% is about
\beq
L \simeq 4.5 \ {\rm fm}\, .
\label{eq:Lmin2}
\eeq
With a lattice spacing of the order of $a = 0.033$ fm, as required by
Eq.~(\ref{eq:amin2}), one also finds that the number of lattice sites in the
four space-time directions is about $136^3 \times 270$, to be compared with
typical lattices of $32^3 \times 64$ sites used in current simulations.

\subsubsection{Treatment of Heavy Quarks}
We will realize below that simulations on lattices so fine for the condition $a
m_b\gg 1$ to be satisfied will be too expensive even with the PFlop machines
that will be presumably available in 2015. Therefore, the treatment of the $b$
quark on the lattice will still require one of the dedicated approaches
discussed in the previous section: use of effective theories, specific
heavy-quark lattice actions or extrapolation of the results obtained in the
charm quark mass region to the $b$-quark mass.

\begin{figure}
  \begin{center}
    \includegraphics[scale=0.6]{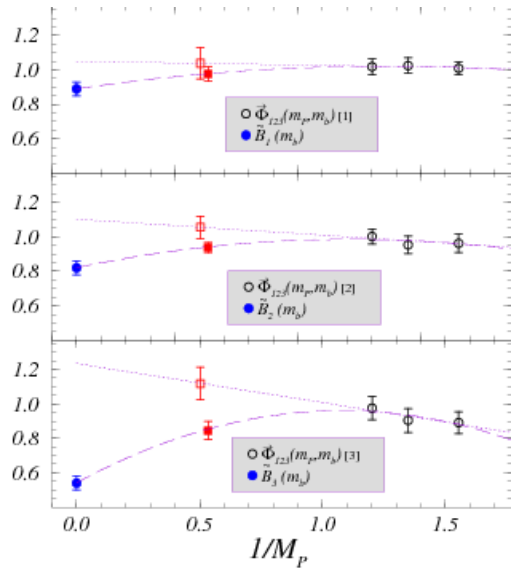}
    \caption{
      Interpolation of lattice results for three parameters
      relevant to $B$ meson mixing~\cite{Becirevic:2001xt}.
      The black (empty) circles are the results obtained in
      the charm quark mass region,
      the blue (filled) circles the HQET determination in the static limit,
      and the red (filled) squares the results obtained after
      interpolation to the $B$ meson
      with the dashed lines showing the interpolation.
      The red (empty) squares show, for comparison,
      the results that are obtained by linearly extrapolating the data
      from the charm mass region (the dotted lines showing the extrapolation),
      without using the information coming from HQET.
    }
    \label{fig:hqextra}
  \end{center}
\end{figure}

In Fig.~\ref{fig:hqextra}, taken from~\cite{Becirevic:2001xt}, we show as an
example the lattice determination of three parameters relevant for $B$ meson
mixing. In the plot, the results obtained in correspondence of meson masses
around the $D$ meson mass are combined with the static (HQET) determination of
the same quantities.
In this way, the matrix elements of interest, corresponding
to the $B$ mesons, are reached through an interpolation rather than an
extrapolation of the lattice data.
The message that can be learnt from this specific example is that, since the
relativistic and static results have comparable precision, the interpolation to
the $B$ meson mass does not increase significantly the uncertainty. One could
also think of further improving the accuracy of the interpolation by computing
within the HQET not only the result in the static limit
but also the sub-leading contribution,
{\it i.e.} the slope in the $1/M$ expansion~\cite{Heitger:2003nj}.

\subsection{Cost of the Target Simulations}
The previous discussion allows to identify the constraints on the relevant
parameters of a ``target'' lattice simulation aiming at the 1\% level of
precision in the determination of the hadronic matrix elements.
These parameters are collected in Table~\ref{tab:target},
where we consider two different
set-ups, the first one for a simulation dedicated to light quark physics only,
the second one for a simulation also involving heavy quarks.
\begin{table}
  \caption{
    Values of the lattice parameters for simulations aiming at the 1\%
    level of precision in the determination of the hadronic matrix elements.
    The left column refers to a calculation dedicated to
    light quark physics only,
    the right column to a calculation also involving heavy quarks.
  }
  \begin{center}
    \begin{tabular}{cc}
      \hline
      \hline
      Light quark physics & Heavy quark physics \\
      \hline
      $N_{\rm conf} = 120$ & $N_{\rm conf} = 120$ \\[2mm]
      $a = 0.05$ fm & $a = 0.033$ fm \\
      $\left[1/a=3.9\ \gev\right]$ & $\left[1/a=6.0\ \gev\right]$ \\[2mm]
      $\hat m/m_s = 1/12$ & $\hat m/m_s = 1/12$ \\
      $\left[m_\pi=200\ \mev\right]$ & $\left[m_\pi=200\ \mev\right]$ \\[2mm]
      $L_s=4.5$ fm & $L_s = 4.5$ fm \\
      $\left[N_{\rm sites}=90^3\times 180\right]$ &
      $\left[N_{\rm sites}=136^3\times 270\right]$ \\[2mm]
      \hline
    \end{tabular}
  \end{center}
  \label{tab:target}
\end{table}

We now estimate the computational cost of the two target simulations indicated
in Table~\ref{tab:target}.
An empirical formula can be used, that approximately expresses
the CPU cost of a numerical simulation as a function of the simulation
parameters: number of independent gauge configurations ($N_{\rm conf}$),
space-time extension of the lattice ($L_s^3 \times L_t$),
value of the average up-down quark mass ($\hat m/m_s$)
and value of the lattice spacing ($a$).
With present algorithms, and for Wilson-like fermions with $N_f=2$ flavours
of dynamical quarks, this formula
reads~\cite{DelDebbio:2006cn}~\footnote{
  Eq.~(\ref{eq:cpucost}) is based on the study of the DD-HMC
  (domain decomposition - Hybrid-Monte-Carlo) algorithm.
  Other recent algorithms, like the HMC with mass preconditioning
  and multiple time scale integration, provide similar performances.
}
\beq
{\rm TFlops-years} \simeq
0.03
\left(\frac{N_{\rm conf}}{100}\right)
\left(\frac{L_s}{3\ {\rm fm}}\right)^5
\left(\frac{L_t}{2 L_s}\right)
\left(\frac{0.2}{\hat m/m_s}\right)
\left(\frac{0.1 \ {\rm fm}}{a}\right)^6 \, ,
\label{eq:cpucost}
\eeq
where the cost, expressed in TFlops-years, represents the number of years of run
required to perform the simulation on a 1-TFlop machine. The overall factor,
which is 0.03 for standard Wilson fermions, becomes approximately 0.05 for the
${\cal O}(a)$-improved theory. The cost of a simulation performed with
Ginsparg-Wilson fermions (domain wall and overlap), which possess better chiral
properties than Wilson fermions, is typically 10-30 times larger than that
indicated by Eq.~(\ref{eq:cpucost}).

\begin{figure}
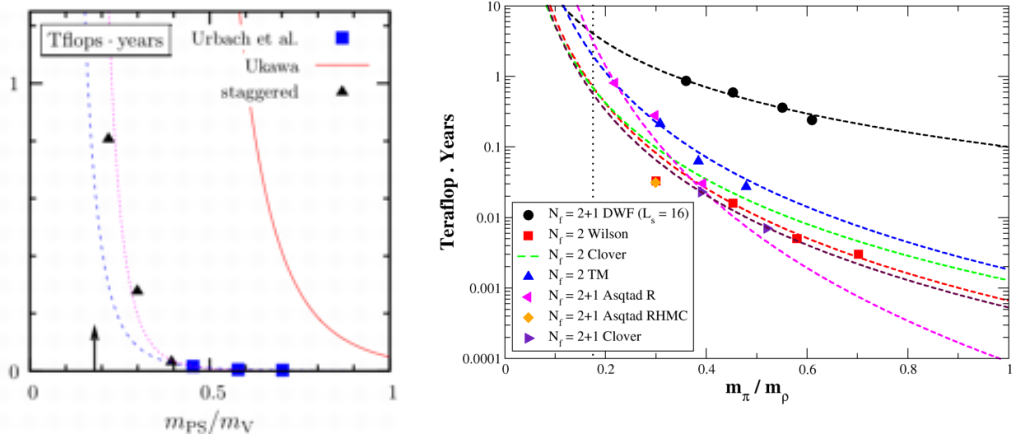

  \begin{center}
    \includegraphics[scale=0.8]{berlinwall2.png}
    \hspace{0.15cm}
    \includegraphics[scale=0.31]{berlin.png}
    \vspace{2ex}
    \caption{
      Cost to generate 1000 independent gauge configurations for
      various fermionic actions and different algorithms,
      as a function of the ratio between the pseudoscalar and
      vector meson masses. On the left plot the cost estimate presented
      by Ukawa at the Lattice 2001 conference is also shown.
      The plots are taken from~\cite{Clark:2006wq}.
    }
    \label{fig:berlinwall}
  \end{center}
\end{figure}

It is interesting to compare the present cost expressed by
Eq.~(\ref{eq:cpucost}) to the analogous formula presented by A.Ukawa at the
Lattice 2001 conference~\cite{Ukawa},
\beq
{\rm TFlops-years} \simeq 3.1 \left(\frac{N_{\rm conf}}{100}\right)
\left(\frac{L_s}{3\ {\rm fm}}\right)^5 \left(\frac{L_t}{2 L_s}\right)
\left(\frac{0.2}{\hat m/m_s}\right)^3 \left(\frac{0.1\ {\rm fm}}{a}\right)^7\,.
\label{eq:berlinwall}
\eeq
Besides the overall factor which is decreased by a factor 100, and the
additional power gained in the dependence on the inverse lattice spacing, the
crucial improvement of current algorithms concerns the scaling of the CPU cost
with the light quark masses, which is reduced from a $1/\hat m^3$ dependence to
$1/\hat m$. The predictions of Eq.~(\ref{eq:berlinwall}) are shown in
Fig.~\ref{fig:berlinwall} (left plot), where they are also compared with the
cost of recent simulations (left and right plots)~\footnote{
  Since the Lattice 2001 conference was held in Berlin,
  it has become common, within the lattice community,
  to refer to the steep increase of the cost with the quark mass as
  the ``Berlin wall''.
}. This comparison provides an important example of how
theoretical and algorithmic developments, which are not taken into account in
our analysis aimed to predict the accuracy of lattice QCD calculations for the
next years, may actually have a significant impact.

Applying Eq.~(\ref{eq:cpucost}) to the target simulations indicated in
Table~\ref{tab:target} one finds a cost of about 0.07 PFlop-years for the
simulation with light quarks only (left column of table~\ref{tab:target}) and
0.9 PFlop-years for the simulation involving also heavy quarks. The overhead
required to perform the calculation at larger values of quark masses and lattice
spacing, necessary to perform the continuum and chiral extrapolations, or to
perform simulations with $N_f=2+1$ (rather than $N_f=2$) dynamical fermions, can
be estimated to be about a factor 3. In both cases, therefore, the expected
costs turn out to be safely affordable with the PFlop machines that are expected
to be available to lattice QCD collaborations around year 2015,
when a Super $B$ Factory could be running and producing results.
This suggests that the required level of precision on the determination
of hadronic parameters can be presumably reached at that time.
In the case of the cheaper simulation involving only light quarks,
the use of the Ginsparg-Wilson fermionic actions, which would
allow to achieve an improved theoretical control on several sources of
systematic uncertainty, increases the computational cost of the simulation at
the level of 1-2 PFlop-years, which is still within the reach of the PFlop
computers.

\subsection{Predicted Accuracy of Lattice QCD Calculations}  
We conclude the analysis of the estimated accuracy of future lattice QCD
calculation by illustrating the differences that should be taken into account
when considering, more specifically, the determination of different hadronic
parameters.

{ \setlength{\tabcolsep}{4pt}  
\begin{table}[tbh]
  \caption{
    Prediction of the accuracy that can be reached in lattice QCD
    determinations of various hadronic parameters assuming the availability
    of a computational power of about 6 TFlops (4th column),
    60 TFlops (5th column) and 1-10 PFlops (6th column).
    The predictions given for the 6 TFlops and 60 TFlops cases
    have been presented by S.~Sharpe in~\cite{Sharpe}.
    The accuracy reached at present in the determination of the various
    parameters is also shown (3rd column).
  }
  \begin{center}
    \setlength{\extrarowheight}{2pt}
    \begin{tabular}{lccccc}
      \hline
      \hline
      Measurement &
      \begin{tabular}{c} Hadronic \\ Parameter \end{tabular} &
      \begin{tabular}{c} Present \\ Error \end{tabular} &
      6 TFlops & 60 TFlops &
      \begin{tabular}{c} 1-10 PFlops \\ (Year 2015) \end{tabular} \\
      \hline
      $K \to \pi\, l\, \nu$   & $f_+^{K\pi}(0)$
      & 0.9\,\% & 0.7\,\% & 0.4\,\% & $<0.1\,\%$ \\
      $\varepsilon_K$ & $\hat B_K$
      & 11\,\%  &  5\,\%  &  3\,\%  & 1\,\% \\
      $B \to l\, \nu$  & $f_B$
      & 14\,\% & 3.5-4.5\,\% & 2.5-4.0\,\% &  1.0-1.5\,\% \\
      $\dmd$ & $f_{Bs}\sqrt{B_{B_s}}$
      & 13\,\% & 4-5\,\% & 3-4\,\% & 1-1.5\,\% \\
      $\dmd/\dms$ & $\xi $
      & 5\,\% & 3\,\% & 1.5-2\,\% & 0.5-0.8\,\% \\
      $B\to D/D^*\,l\,\nu$& ${\cal F}_{B\to D/D^*}$
      & 4\,\%  &  2\,\%  & 1.2\,\% & 0.5\,\% \\
      $B\to \pi/\rho \,l\,\nu$ & $f_+^{B\pi},\ldots$
      & 11\,\% & 5.5-6.5\,\% & 4-5\,\% & 2-3\,\% \\
      $B\to K^*/\rho \,(\gamma, l^+l^-)$ & $T_1^{B\to K^*/\rho}$
      & 13\,\% & ------ & ------ & 3-4\,\% \\
      \hline
    \end{tabular}
  \end{center}
  \label{tab:lattice}
\end{table}
}
A list of observables relevant to studies of flavour physics
is collected in the first column of Table~\ref{tab:lattice},
together with the corresponding hadronic parameters (second column).
In the third column of the table we show the
precision currently reached in the lattice QCD determination of these
quantities. In the fourth and fifth columns we present the accuracy predicted
for future lattice calculations assuming the availability of a computing power
of about 6 and 60 TFlops respectively. These estimates have been presented by
S.~Sharpe~\cite{Sharpe}, on the basis of an error analysis similar to that
performed in the present study. Finally, we give in the last column of
Table~\ref{tab:lattice} the estimates of the accuracy that is expected to be
reached by lattice QCD calculations around year 2015, when computers with
performances of 1-10 PFlops should be available to lattice QCD collaborations.
Some comments on the estimates given in the table are now in order.

The arguments of the previous sections support the conclusion that an accuracy
at the level of 1\% can be reached on the lattice in the determination of the
simplest quantities. These include decay constants, that are obtained from
simple 2-point correlation functions, and $B$ parameters, since the latter are
extracted from more precise ratios of matrix elements. Thus, in the last column
of Table~\ref{tab:lattice} a 1\% error is quoted on $B_K$ and somewhat larger
uncertainties are predicted for $f_{B(s)}$ and $B_{Bd(s)}$, which involve a
treatment of the $b$ quark. The uncertainty on the $B \to \pi/\rho$ semileptonic
form factors is estimated to be larger by a factor of two with respect to the
simplest quantities, both because they are obtained from noisier 3-point (rather
than 2-point) correlation functions and because of the uncertainty associated
with the study of their momentum dependence. For radiative decays, $B\to
K^*/\rho\,\gamma$, an even larger error is estimated, because HQET is of no help
in this case in the kinematical region of interest, {\it i.e.} close to $q^2=0$. On
the other hand, the errors on the hadronic parameter $\xi$ and on the form
factors of $K \to \pi$ and $B \to D/D^*$ semileptonic decays will be most likely
reduced below 1\%, since in these cases one actually measures on the lattice
their differences from 1, {\it i.e.} the value predicted for these parameters in the
SU(3) (for $\xi$ and $f_+^{K\pi}(0)$) and the infinite heavy quark mass (for
${\cal F}_{B\to D/D^*}$) limits. For instance, the accuracy at the level of
0.1\% quoted for $f_+^{K\pi}(0)$ corresponds to an uncertainty of about 2.4\% on
$1 - f_+^{K\pi}(0)$. A slightly larger uncertainty has been quoted for the
determination of $1-{\cal F}_{B\to D/D^*}$, because in this case the
contribution of $(1/m_c-1/m_b)^2$ corrections is presumably larger than the
second order SU(3)-breaking corrections entering the vector form factor of
$K_{\ell 3}$ decays.

The entries in the last column of Table~\ref{tab:lattice}
are the main result of the present analysis,
and can be used to assess the accuracy
reachable at \superb\ in the studies of quark flavour physics.


%
\afterpage{\clearpage}

\newpage
\thispagestyle{empty}
\phantom{test}
\newpage

\graphicspath{{covers/}}
\thispagestyle{empty}
\pagenumbering{none}
\begin{figure}[!h]
\vskip-32.8mm
\hskip-29.9mm
\includegraphics[height=279.4mm]{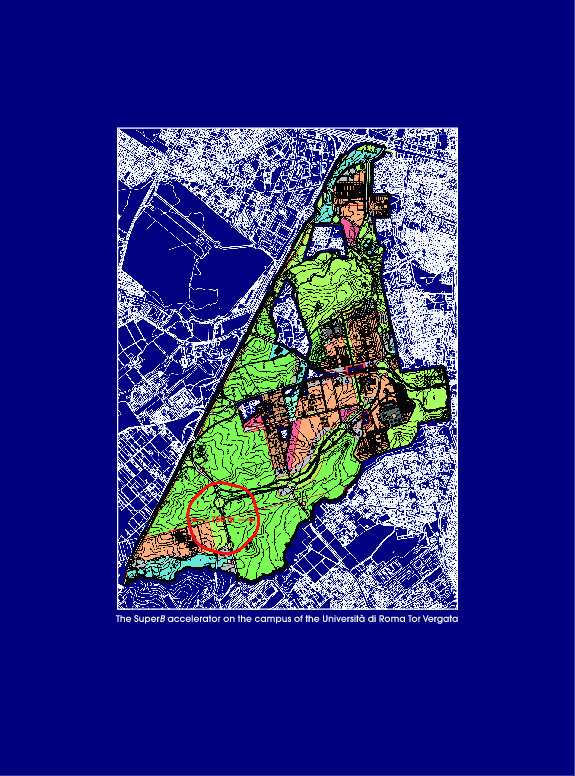}
\end{figure}
\clearpage

\end{document}